\documentclass[a4paper,11pt]{report}
\usepackage{graphicx,amssymb,amstext,amsmath}
\usepackage{subfigure}
\usepackage[retainorgcmds]{IEEEtrantools}
\usepackage[section] {placeins}
\usepackage[cm]{fullpage}
\usepackage{booktabs}
\usepackage{multirow}
\usepackage{hyperref}  
\hypersetup{bookmarks=true, pdfauthor={Pei Zhang},colorlinks=true,linkcolor=black,citecolor=blue,breaklinks=true}
\newcommand{\nocontentsline}[3]{}
\newcommand{\tocless}[2]{\bgroup\let\addcontentsline=\nocontentsline#1{#2}\egroup}
\graphicspath{{./tmp/}}
\begin{document}
\title{\textbf{Higher order mode spectra and the dependence of localized dipole modes on the transverse beam position in third harmonic superconducting cavities at FLASH}}

\author{Pei~Zhang$^{\dagger \ddagger \ast}$, Nicoleta~Baboi$^\ddagger$, Roger~M.~Jones$^{\dagger \ast}$\\
\mbox{$^\dagger$The University of Manchester, Manchester, U.K.}\\
\mbox{$^\ddagger$DESY, Hamburg, Germany}\\
\mbox{$^\ast$The Cockcroft Institute, Daresbury, U.K.}}

\maketitle
\begin{abstract}
An electron beam entering an accelerating cavity excites a wakef{}ield. This wakef{}ield can be decomposed into a series of multi-poles or modes. The dominant component of the transverse wakef{}ield is dipole. This report summarizes the higher order mode (HOM) signals of the third harmonic cavities of FLASH measured at various stages: transmission measurements in the single cavity test stand at Fermilab, at CMTB (Cryo-Module Test Bench) and at FLASH, and beam-excited measurements at FLASH. Modes in the f{}irst two dipole bands and the f{}ifth dipole band have been identif{}ied using a global Lorentzian f{}it technique. The beam-pipe modes at approximately 4~GHz and some modes in the f{}ifth dipole band have been observed as localized modes, while the f{}irst two dipole bands, containing some strong coupling cavity modes, propagate.

This report also presents the dependence of the localized dipole modes on the transverse beam position. Linear dependence for various modes has been observed. This makes them suitable for beam position diagnostics. These modes, together with some propagating, strong coupling modes, have been considered in the design of a dedicated electronics for beam diagnostics with HOMs for the third harmonic cavities. 

\end{abstract}

\renewcommand{\abstractname}{Acknowledgements}

\tableofcontents
\chapter{Introduction}\label{intro}
Wakef{}ields excited by electron bunches in accelerating cavities may adversely af{}fect the beam quality and, in the worst case, result in a beam-break-up instability \cite{rwake}. It is therefore important to ensure that these f{}ields are well suppressed by extracting energy through special couplers. Indeed, for the acceleration of high intensity particle beams, suppressing the wakef{}ields in both superconducting \cite{rsc} and normal conducting \cite{rnc} cavities is a common requirement. On the other hand, since the transverse wakef{}ields depend on the transverse of{}fset of the excitation bunch, they can also be used for beam diagnostics in the cavity without additional vacuum components \cite{rhombpm-1,rhombpm-2,rhombpm-3}. It is this aspect that is focused on in this work.

At FLASH (Free-electron Laser in Hamburg) \cite{rflash}, we plan to make use of the higher order modes (HOM) for beam diagnostics in third harmonic superconducting cavities \cite{racc39-p1,racc39-p2,racc39-p3,racc39-p4}. HOMs are components of the wakef{}ield. The aim is to provide beam position information from the energy radiated to the HOM couplers. In order to achieve this, special electronics is required \cite{racc39-hom}, and is currently under design. Prior to developing electronics, it is essential to characterize the HOMs and understand their behavior relating to the beam of{}fset. For this purpose, HOM measurements presented in this report, both with and without beam-excitations, were conducted \cite{rhommeas-2,rhommeas-3,rhommeas-4,rhommeas-5}. Simulations of the cavities, both with and without couplers, were also performed using various techniques \cite{rhommeas-2,racc39-1,rgsm-2,race3p,rcsc-1}.  

This chapter presents an overview of the FLASH facility and the third harmonic superconducting cavities, followed by a brief description of the principle of beam position diagnostics using dipole modes in accelerating cavities. It proceeds in Chapter~\ref{hom-meas} with the HOM signal measured at several stages after the fabrication of the third harmonic cavities both with and without beam-excitations. HOMs are identif{}ied and the inter-cavity coupling ef{}fects are observed. Detailed studies of HOM dependence on the beam of{}fset for various modes are described in Chapter~\ref{hom-dep}.

\section{FLASH and the Third Harmonic Cavities}\label{intro:flash}
FLASH \cite{rflash} is a free-electron laser facility at DESY providing ultra-short electron bunches with high peak current to generate coherent light with unprecedented brilliance. It is also a test facility for various accelerator studies.
\begin{figure}[h]\center
\includegraphics[width=0.9\textwidth]{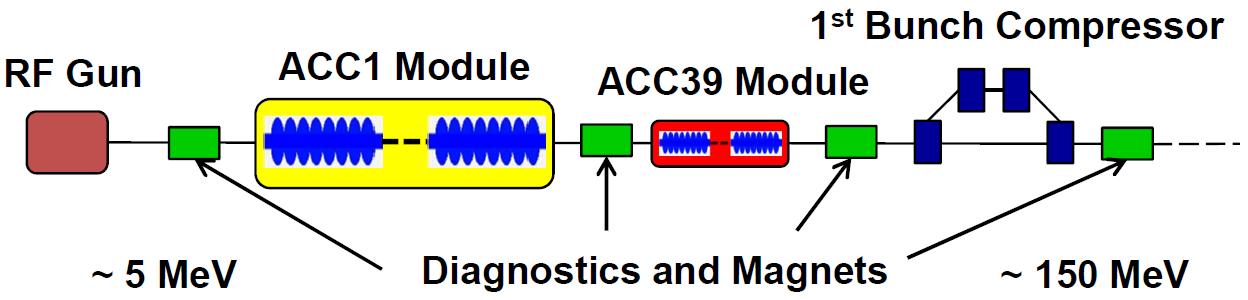}
\caption{Schematic of injector section of the FLASH facility.}
\label{injector-layout}
\end{figure}

\begin{figure}[h]\center
\subfigure[ACC39 module]{
\includegraphics[width=0.45\textwidth]{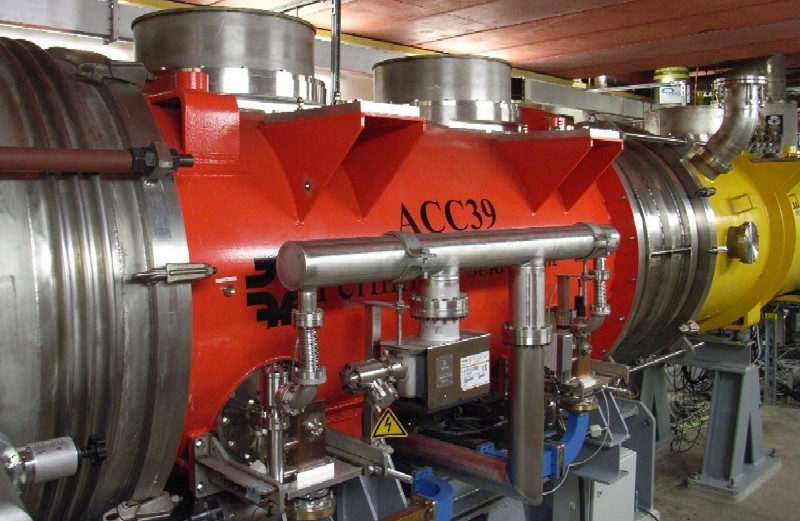}
\label{acc39}
}
\quad
\subfigure[1.3~GHz and 3.9~GHz cavities]{
\includegraphics[width=0.45\textwidth]{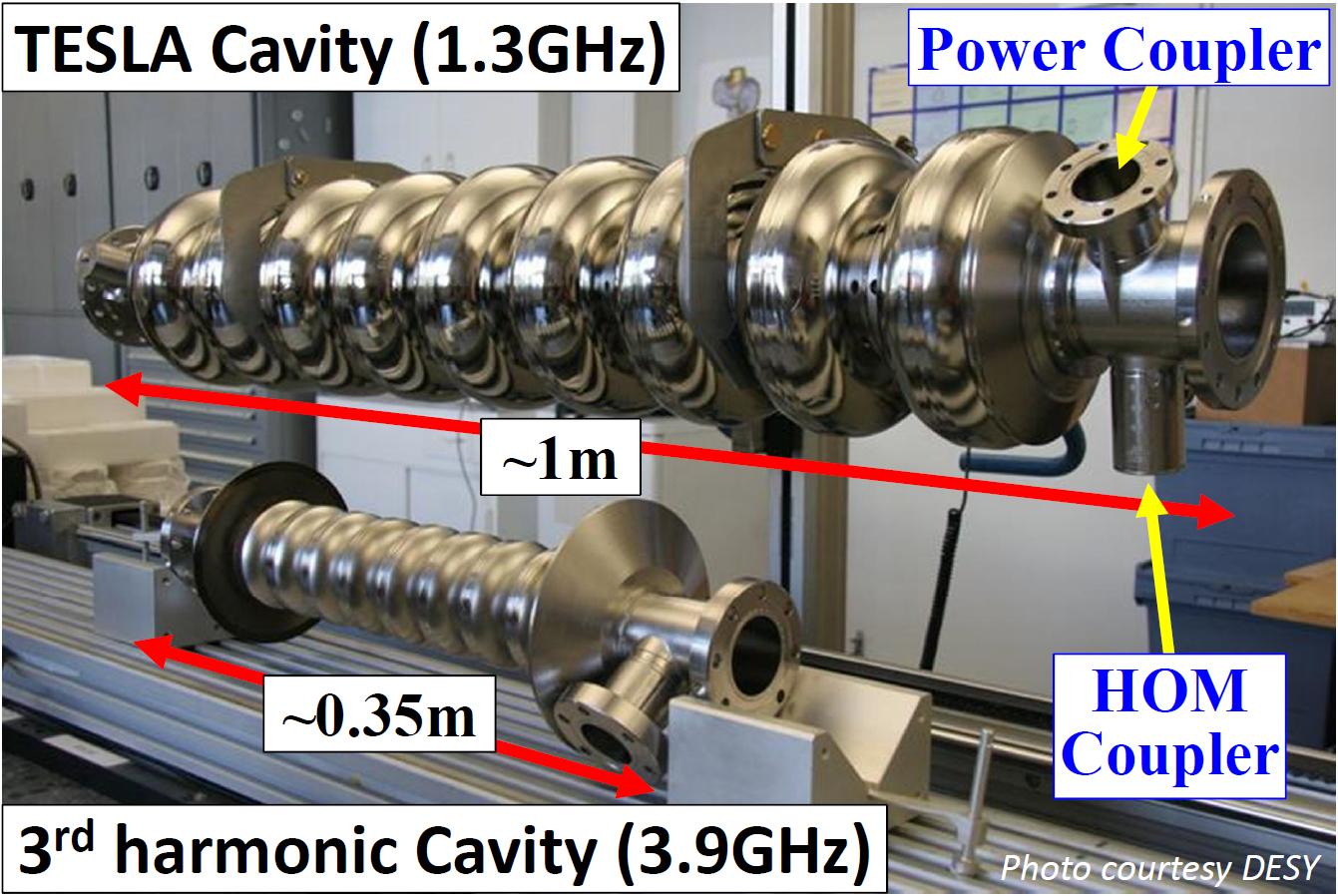}
\label{cavity-acc1}
}
\caption{(a) ACC39 module in FLASH. The module in yellow, on the right, is ACC1. The direction of travel of the multi-bunch beam, is from right to left. (b) A TESLA style cavity operating at 1.3~GHz (top) and the corresponding third harmonic cavity (bottom).}
\label{acc39-cavity}
\end{figure}

Fig.~\ref{injector-layout} shows schematically the injector section of FLASH, which is relevant to our studies. The electron beam generated by a photoelectric gun is accelerated of{}f-crest by eight superconducting 1.3~GHz TESLA cavities \cite{rtesla-1,rtesla-2} in cryo-module ACC1 and compressed by the f{}irst magnetic chicane. Due to the length of the electron bunch in the millimeter range before the f{}irst bunch compressor, and the sinusoidal 1.3~GHz RF f{}ield, a curvature in the energy-phase plane develops, which leads to a long bunch tail and the reduction of peak current in the bunch compression. To linearize the energy spread of the bunch, harmonics of the fundamental accelerating frequency (1.3~GHz) of the main linac are added by third harmonic superconducting cavities operating at 3.9~GHz \cite{racc39-p1,racc39-p2}. Four such cavities are placed in the ACC39 module following ACC1, which were designed and built by Fermilab in collaboration with DESY \cite{racc39-p3,racc39-p4}. Fig.~\ref{acc39} shows a photo of ACC39 in the FLASH beam line.

\begin{figure}[h]\center
\includegraphics[width=0.95\textwidth]{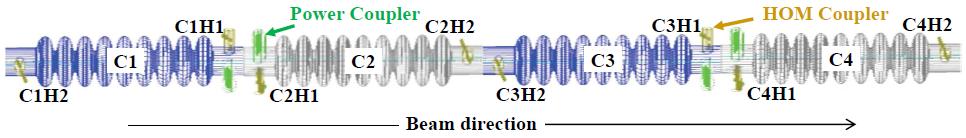}
\caption{Schematic of four cavities within ACC39 module. The power couplers (green) are placed downstream for C1 and C3, and upstream for C2 and C4. The HOM couplers (brown) located on the same side of the power couplers are named H1, while the other ones H2.}
\label{cavity-cartoon}
\end{figure}

The ACC39 module is composed of four cavities namely C1 through C4 (illustrated in Fig.~\ref{cavity-cartoon}). Each cavity is equipped with two higher order mode (HOM) couplers namely H1 and H2. By extracting HOMs through these couplers, the wakef\mbox{}ields are well-suppressed (external $Q$ of the HOMs is required to be less than $10^5$ for a beam-break-up limit) \cite{racc39-4,racc39-fnal-1,racc39-fnal-2} and their deleterious ef{}fects on the beam are minimized. There are two distinct coupler designs: a 1-leg-coupler design (C1 and C3) and a 2-leg-coupler design (C2 and C4) \cite{racc39-4}. Wakef{}ields excited in the cavities are extracted from these couplers and guided by long cables to HOM board rack outside the tunnel. After the module has been installed at FLASH, all the measurements were conducted from this rack as will be discussed in Chapter~\ref{hom-meas}.

By design, the 3.9~GHz cavity has two features in terms of wakef{}ields, compared to the 1.3~GHz cavity. F{}irst, the wakef{}ields in the 3.9~GHz cavity are larger than those in the 1.3~GHz cavity as the iris radius is signif{}icantly smaller: 15~mm in the 3.9~GHz cavity compared to 35~mm in the 1.3~GHz cavity (see Fig.~\ref{cavity-acc1}). From scaling considerations, it can be shown that wakef{}ields per meter in the cavity grow as \cite{rscale-law}
\begin{subequations}
\label{eq:all-scale-law}
\begin{eqnarray}
W_\parallel\sim a^{-2},\label{eq:scale-law-a}
\\
W_\perp\sim a^{-3},\label{eq:scale-law-b}
\end{eqnarray}
\end{subequations}
where $W_\parallel$ and $W_\perp$ are longitudinal and transverse wakef{}ield respectively, and $a$ is the iris radius. Second, the HOM spectrum is signif\mbox{}icantly more complex than that of the 1.3~GHz cavity. The main reason for this is that, unlike the TESLA cavity case, the majority of the modes are above the cutof{}f frequencies of the beam pipes. This allows most of the modes from each independent cavity to propagate through to adjacent cavities. In this case, most modes reach all eight HOM couplers. This facilitates the damping of HOMs to be distributed.

\section{Dipole Dependence on Transverse Beam Of{}fset}\label{intro:dep}
Consider only the long-range wakef\mbox{}ield due to HOMs, the wakef\mbox{}ield excited by an ultra-relativistic particle of velocity $c$ in a periodic, cylindrical symmetric structure can be decomposed by multi-pole expansions into monopole, dipole, quadrupole, etc. Each longitudinal and transverse multi-pole component can be written as \cite{rtesla-2,rscale-law}
\begin{equation}
\mathbf{W_{\parallel m}}=\hat{z}\left(r'\right)^m r^m cosm\theta \sum_n^\infty \omega_{mn} \left(\frac{R}{Q}\right)_{mn} cos\frac{\omega_{mn}s}{c} H(s)
\label{eq:multipole-1}
\end{equation}
and for $m>0$
\begin{IEEEeqnarray}{rCl}
\mathbf{W_{\perp m}}=m\left(r'\right)^m r^{m-1} \left(\hat{r}cosm\theta-\hat{\theta}sinm\theta\right)
\IEEEeqnarraynumspace \nonumber \\
\cdot c \sum_n^\infty  \left(\frac{R}{Q}\right)_{mn} sin\frac{\omega_{mn}s}{c}H(s),
\IEEEeqnarraynumspace
\label{eq:multipole-2}
\end{IEEEeqnarray}
where $H(s)$ is the Heaviside step function \cite{rheav}, $m$=0, 1, 2 modes are the monopole, dipole and quadrupole modes respectively, $\mathbf{W_{\parallel m}}$ is the $m^{th}$ pole component of the longitudinal wake potential, $\mathbf{W_{\perp m}}$ is the $m^{th}$ pole component of the transverse wake potential. The transverse position of the excitation particle is denoted as $(r',\theta')$ where $\theta'$ has been set to $0$ because of the cylindrical symmetry. The distance behind the excitation particle in longitudinal direction is $s$ and $(r,\theta,s)$ denotes a position in 3D space where the wakef\mbox{}ield is examined. Also $\left(\frac{R}{Q}\right)_{mn}$ and $\omega_{mn}/2\pi$ are the $R/Q$ and the frequency of each eigenmode $mn$, $\hat{r}$, $\hat{\theta}$ and $\hat{z}$ are the unit vectors in $r$, $\theta$ and $z$ directions. The total wakef\mbox{}ield, $\mathbf{W_\parallel}$ and $\mathbf{W_\perp}$, can be obtained by summing all the $m$-pole components generated by the excitation particle as
\begin{subequations}
\label{eq:sum-wake}
\begin{eqnarray}
\mathbf{W_\parallel}=\sum_m \mathbf{W_{\parallel m}}, \label{eq:sum-wake-1}
\\
\mathbf{W_\perp}=\sum_m \mathbf{W_{\perp m}}. \label{eq:sum-wake-2}
\end{eqnarray}
\end{subequations}
Each third harmonic cavity has nine cells, therefore each group of nine modes forms a band. The f\mbox{}ield distributions of the modes in the same band are similar in transverse direction but dif\mbox{}ferent in longitudinal direction. The f\mbox{}irst monopole band contains the accelerating mode at 3.9~GHz, and is often referred to as the fundamental band. Modes in other bands have higher frequencies than 3.9~GHz, therefore are naturally represented as higher order modes (HOM).

During normal operations, bunches are contained in the near-axis region, meaning that the $r'$ term is small. Thus monopole modes ($m$=0) dominate the longitudinal wakef\mbox{}ield, while dipole modes ($m$=1) dominate the transverse wakef\mbox{}ield. Therefore, those wakef\mbox{}ields can be approximated as
\begin{subequations}
\label{eq:prox-wake}
\begin{eqnarray}
\mathbf{W_\parallel} \simeq \hat{z} \sum_n^\infty \omega_{0n} \left(\frac{R}{Q}\right)_{0n} cos\frac{\omega_{0n}s}{c}H(s),
\label{eq:prox-wake-1}
\\
\mathbf{W_\perp} \simeq \hat{x} r' c \sum_n^\infty \left(\frac{R}{Q}\right)_{1n} sin\frac{\omega_{1n}s}{c}H(s),
\label{eq:prox-wake-2}
\end{eqnarray}
\end{subequations}
where $\hat{x}=(\hat{r}cos\theta - \hat{\theta}sin\theta)$ is the unit vector in the transverse direction. The $\hat{x}$ term describes the polarization of the dipole mode in the transverse plane. There are two polarizations of each dipole eigenmode perpendicular to each other in the cylindrical symmetric structure. The dipole transverse wakef\mbox{}ield is of particular interest, since it has a linear dependence on the transverse of\mbox{}fset $r'$ of the excitation particle, regardless of the observation transverse position $r$. Therefore, by monitoring beam-excited dipole modes, one can determine the beam position within the cavity \mbox{\cite{rhombpm-1,rhombpm-2,rhombpm-3}}. On the other hand, quadrupole modes have a quadratic relation to the transverse of\mbox{}fset $r'$, therefore are not very sensitive in the near-axis region. They have not been considered in this report.

\chapter{HOM Spectra}\label{hom-meas}
In order to understand the HOM spectra of the third harmonic cavities, modal characterization is needed at all stages after the fabrication of the cavities: transmission spectra between the two HOM couplers measured for each single cavity are presented in Section~\ref{hom-meas:fnal}, module-based transmission measurements in Section~\ref{hom-meas:cmtb} and beam-excited HOM measurements in Section~\ref{hom-meas:beam}. The signals were measured from all eight HOM couplers. Not all measured signals are shown in this chapter. The selection is based on a representative sample of the typical behavior of the modes. HOM signals measured from all eight HOM couplers can be found in Appendix~\ref{app-spec}. 

\section{Transmission Spectra of A Single Cavity Measured at Fermilab}\label{hom-meas:fnal}
\begin{figure}[h]\center
\subfigure[Measured across C1 from C1H1 to C1H2]{
\includegraphics[width=0.98\textwidth]{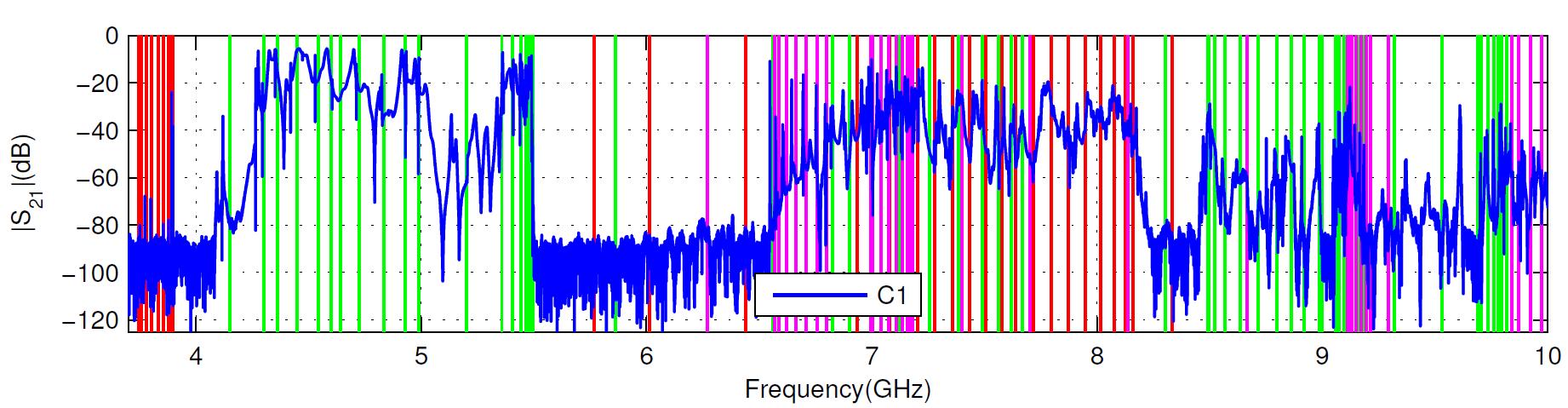}
\label{full-trans-fnal}
}
\subfigure[The fundamental band]{
\includegraphics[width=0.31\textwidth]{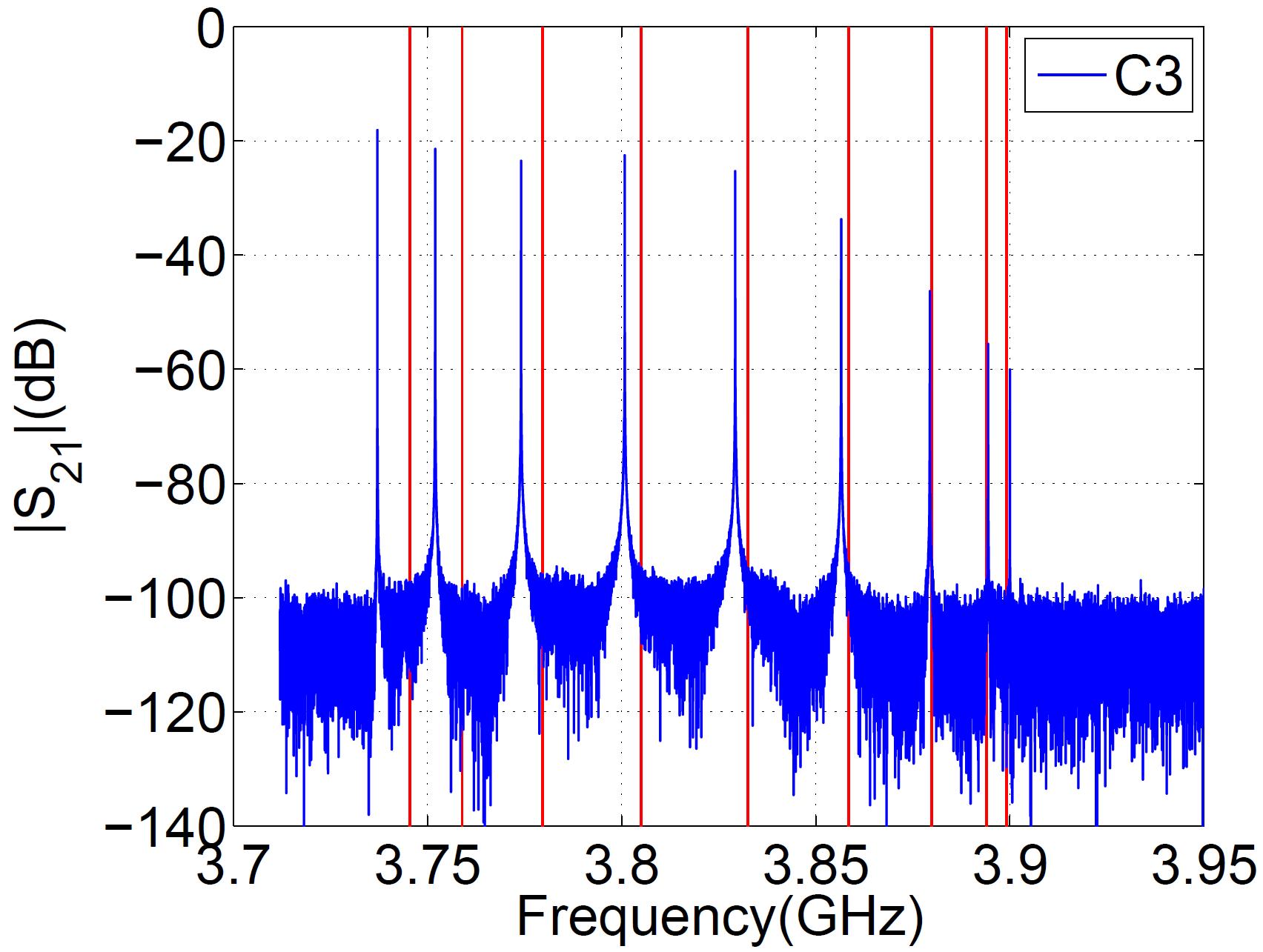}
\label{mono-trans-fnal}
}
\subfigure[The f\mbox{}irst two dipole bands]{
\includegraphics[width=0.31\textwidth]{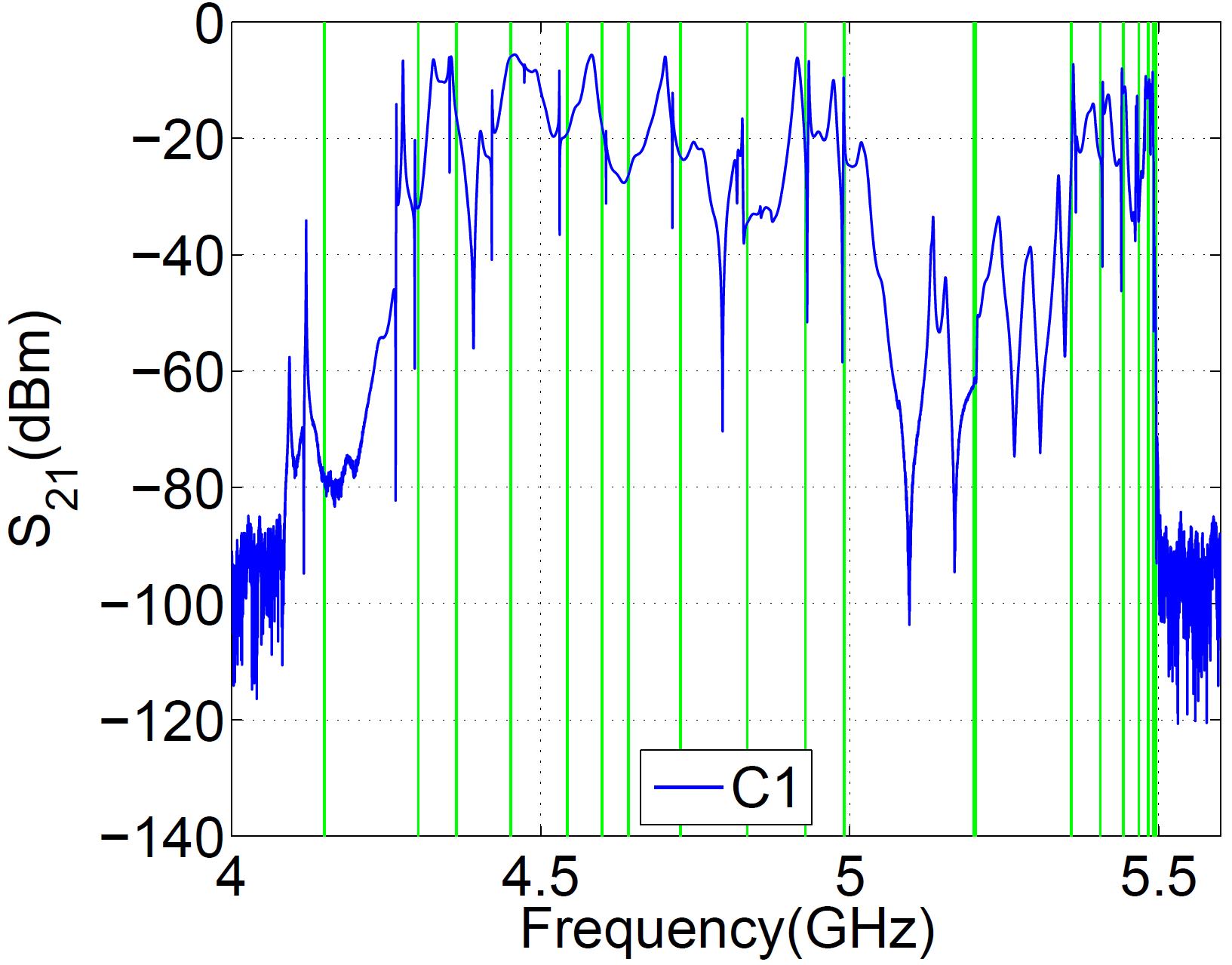}
\label{D1D2-trans-fnal}
}
\subfigure[The f\mbox{}ifth dipole band]{
\includegraphics[width=0.31\textwidth]{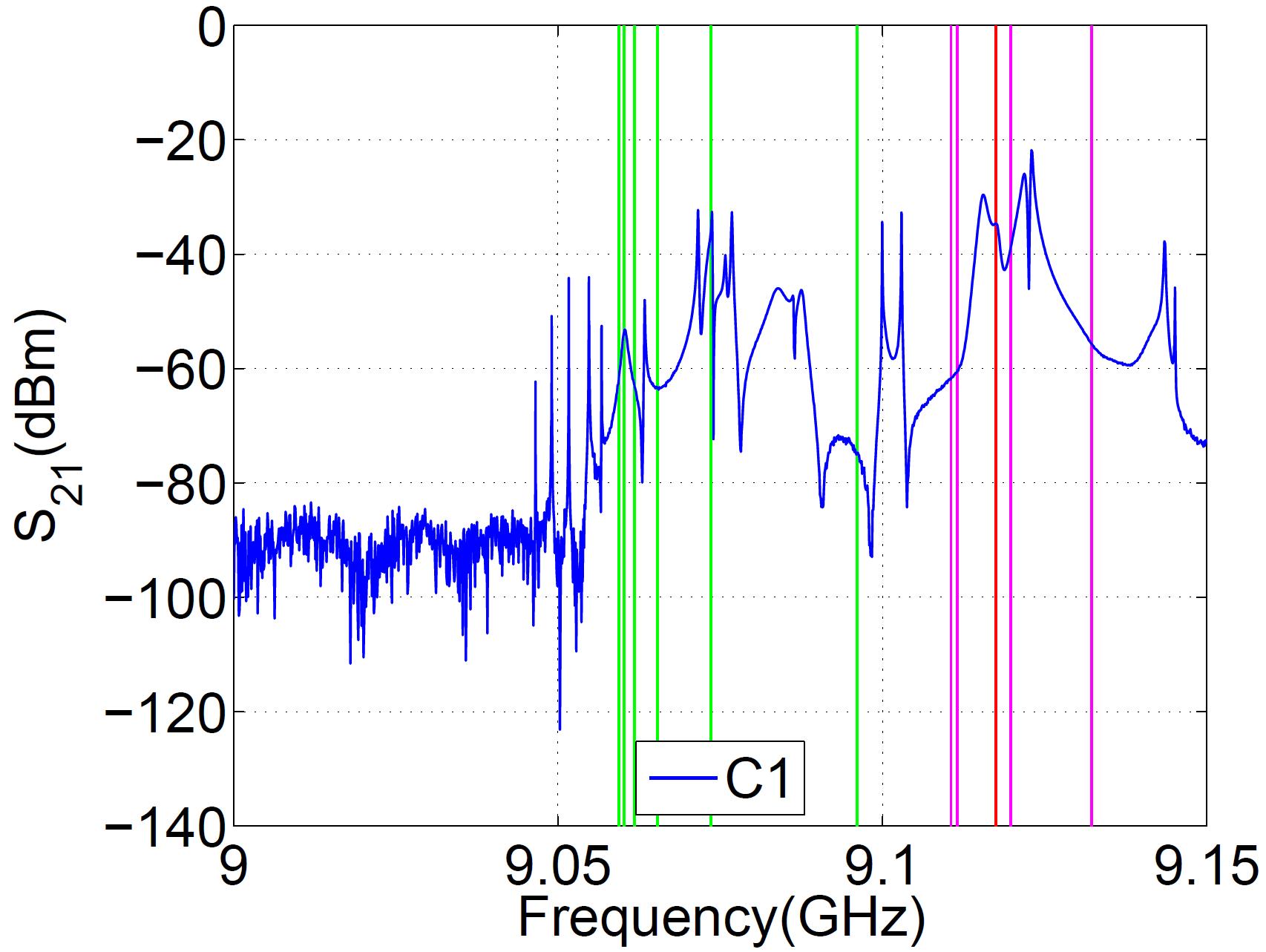}
\label{D5-trans-fnal}
}
\caption{Typical transmission spectrum (S$_{21}$) of a single isolated cavity. The vertical lines indicate simulation results of the eigenmodes \cite{racc39-1}. The colors red, green and magenta represent monopole, dipole and quadrupole modes, respectively.}
\label{spec-trans-fnal}
\end{figure}

After fabrication, the cavities were mounted on the test stand at Fermilab and cooled down \cite{racc39-fnal-1,racc39-fnal-2}. The RF transmission spectrum (S$_{21}$) was subsequently measured for each single cavity.\footnote{Data are kindly provided by T.~Khabibouline from Fermilab.} A typical one is shown in Fig.~\ref{full-trans-fnal} along with simulations. The simulations are performed on an ideal single cavity without power coupler and HOM couplers. The band structure of a single cavity is depicted in terms of monopole, dipole and quadrupole modes from 3.7 to 10~GHz frequency span. Specif{}ic regions are respectively the fundamental band (Fig.~\ref{mono-trans-fnal}), the f{}irst two dipole bands (Fig.~\ref{D1D2-trans-fnal}) and the f{}ifth dipole band (Fig.~\ref{D5-trans-fnal}). The nine modes in the fundamental band can be identif{}ied in Fig.~\ref{mono-trans-fnal}. The accelerating mode is the last peak at 3.9~GHz. Unlike the simulated cavity, the actual cavity has couplers, which breaks the symmetry of the structure. This accounts for the dif{}ferences between simulations and measurements. In addition, other sources such as fabrication errors and cavity tuning can also contribute to the dif{}ferences to the simulations. The dipole modes are not easy to identify anymore as shown in Fig.~\ref{D1D2-trans-fnal} and Fig.~\ref{D5-trans-fnal}. 

\begin{figure}[h]\center
\includegraphics[width=0.5\textwidth]{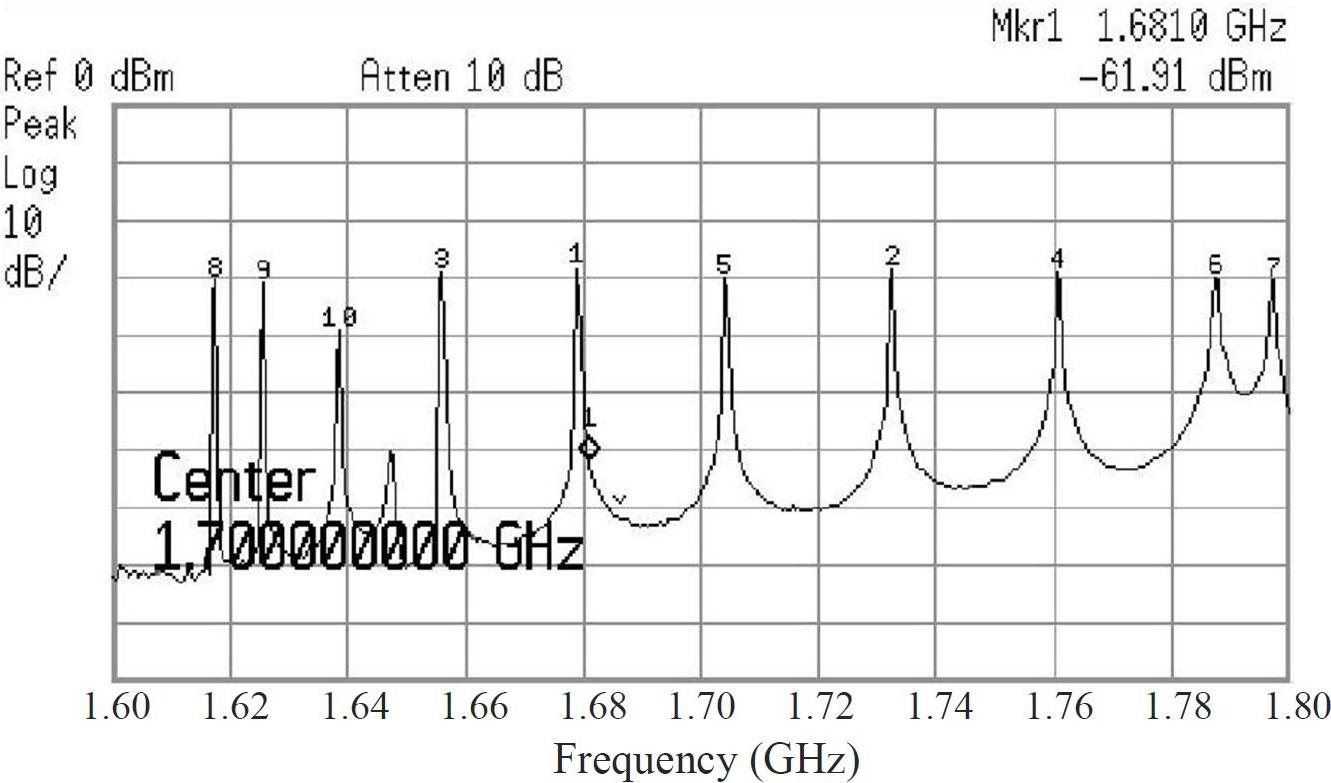}
\caption{The beam-excited spectrum of the f{}irst dipole band of the TESLA cavity operating at 1.3~GHz. Each peak shown in this plot is in fact two peaks close to each other representing the two polarizations of each dipole mode.}
\label{D1-acc1}
\end{figure}

For comparison, a typical beam-excited spectrum of the f{}irst dipole band of the TESLA cavity operating at 1.3~GHz is shown in Fig.~\ref{D1-acc1}. Here one can identify nine peaks \footnote{The last peak in the plot belongs to the second dipole band according to simulations.}, clearly separated from each other. The coupling of a specif{}ic mode to the beam is characterized by the quantity $R/Q$ \cite{rtesla-2}. Comparing with simulations, peaks marked as \#5 and \#2 in Fig.~\ref{D1-acc1} are identif{}ied as modes which have strong coupling to the beam as calculated from simulations. Therefore, these modes are more sensitive to the transverse beam of{}fset, which leads to a better resolution in the diagnostics electronics. Mode \#5 at approximately 1.7~GHz was used for HOM electronics of the 1.3~GHz cavity \cite{rhombpm-3}. The $R/Q$ of this mode is 5.54~$\Omega/$cm$^2$ \cite{rtesla-2}. However, the modal spectrum is more complicated for the 3.9~GHz cavity (see Fig.~\ref{D1D2-trans-fnal}). Mode identif{}ication is by no means straightforward.

\begin{figure}[h]\center
\subfigure[The f\mbox{}irst dipole band of C1]{
\includegraphics[width=0.48\textwidth]{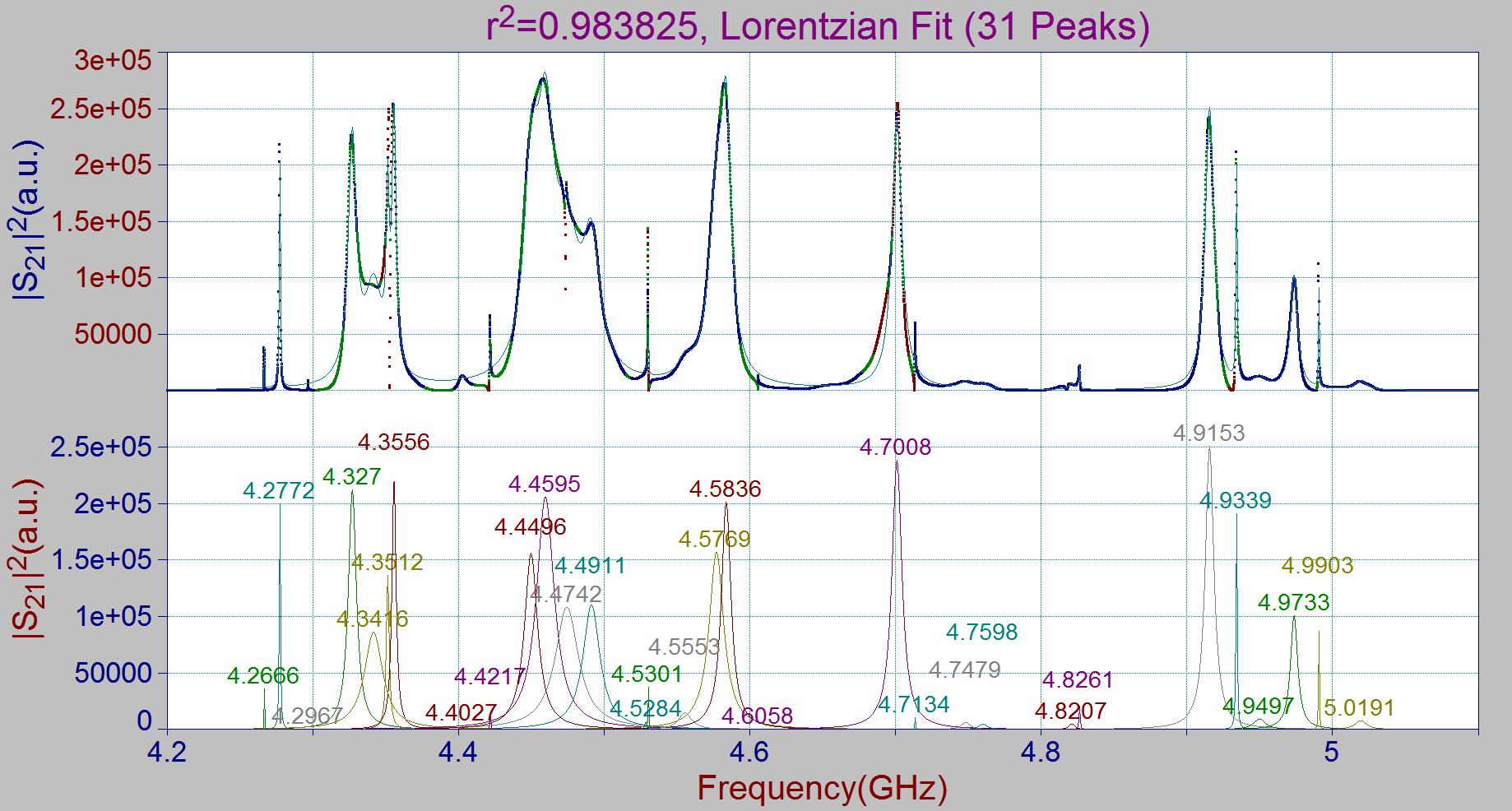}
\label{lorfit-D1}
}
\subfigure[The second dipole band of C1]{
\includegraphics[width=0.48\textwidth]{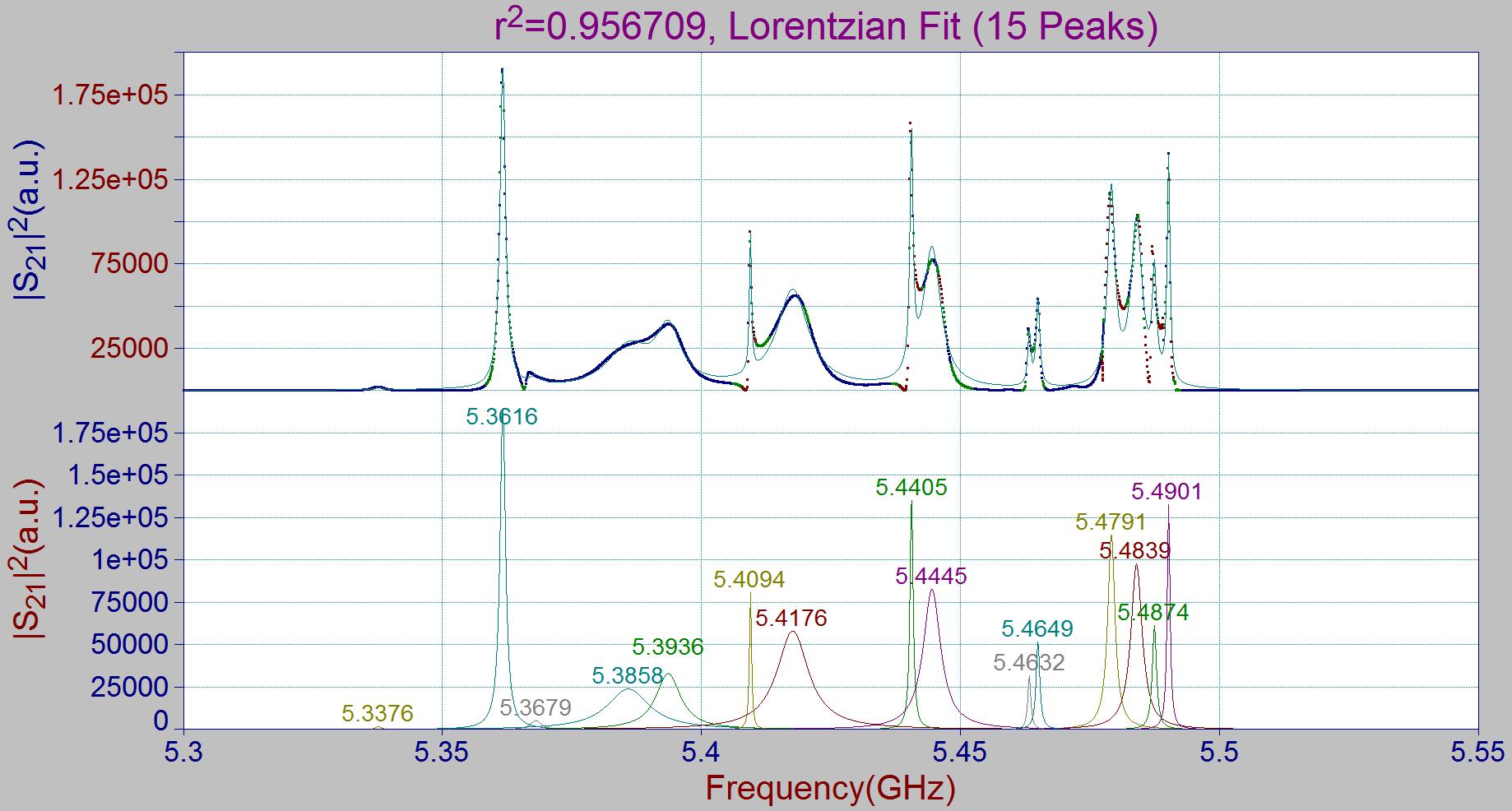}
\label{lorfit-D2}
}
\caption{F{}it of the f{}irst two dipole bands of C1 (from the single cavity measurement) as Lorentzian distributions.}
\label{lorfit-D1D2}
\end{figure}

\begin{figure}[h]\center
\subfigure[part 1]{
\includegraphics[width=0.45\textwidth]{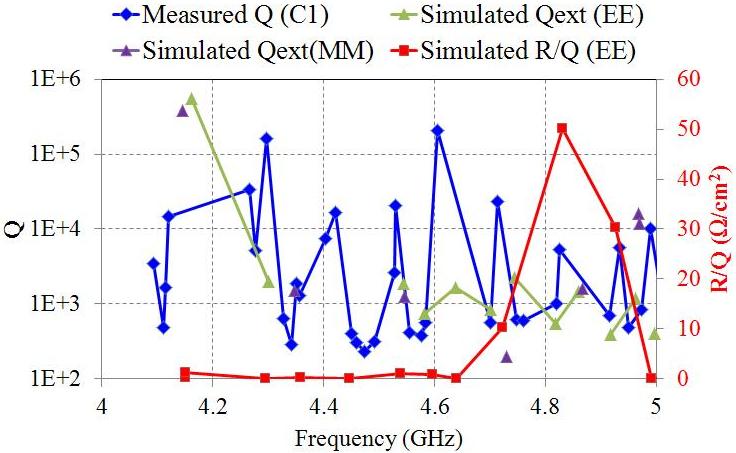}
\label{lorfit-D1D2-fvsQ-part1}
}
\quad
\subfigure[part 2]{
\includegraphics[width=0.45\textwidth]{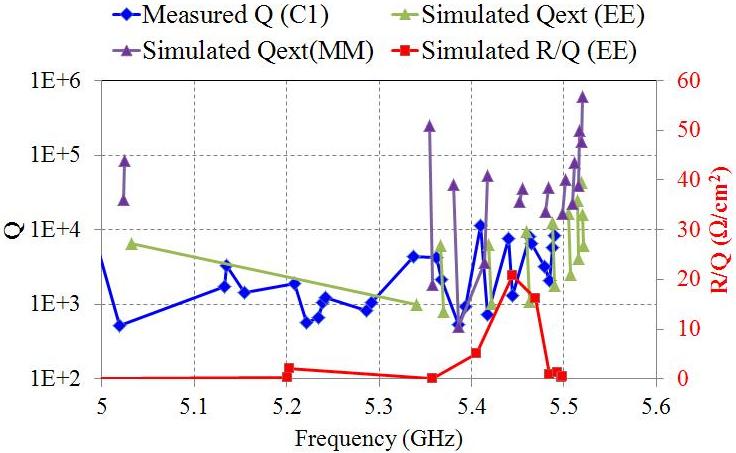}
\label{lorfit-D1D2-fvsQ-part2}
}
\caption{The simulation \cite{racc39-1} and the single cavity measurement of the dipole beam-pipe modes and the f{}irst two dipole bands of C1.}
\label{lorfit-D1D2-fvsQ}
\end{figure}

Each mode shown in the spectrum in Fig.~\ref{spec-trans-fnal} is a resonant peak having a Lorentzian distribution \cite{rstat-3}:
\begin{equation}
y=\frac{y_0}{1+\left(\frac{f-f_0}{\Delta f}\right)^2}.
\label{eq:lorfit}
\end{equation}
Here $y_0$ is the amplitude, $f_0$ is the center frequency and $\Delta f$ is the half-width at half-amplitude (HWHM). In order to identify the modes, due to the complexity of the spectrum, all modes in a dipole band are f{}it simultaneously rather than f{}itting individual peak. The frequency ($f_0$) and the quality factor $Q$ ($Q=f_0/(2\Delta f)$) is then obtained for each mode. Fig.~\ref{lorfit-D1D2} shows the f\mbox{}it results of the f{}irst two dipole bands of one cavity using the code PeakFit \cite{rpeakfit}. The goodness of f{}it is measured by the coef{}f{}icient of determination $r^2$ ($r^2=1$, perfect f{}it; $r^2=0$, poor f{}it) \cite{rstat-1}. Each peak with a frequency marked on top denotes a mode. Hidden peaks are revealed. The $Q$ of each mode is shown in Fig.~\ref{lorfit-D1D2-fvsQ} along with the coupling strength to the beam for each mode from simulations (described by the $R/Q$ value) \cite{racc39-1}. The $Q_{ext}$ is also shown in Fig.~\ref{lorfit-D1D2-fvsQ} for each mode simulated on an cavity with power and HOM couplers \cite{racc39-1}. The polarization split of dipole modes and the frequency shift from ideal simulations can be observed. By comparing the modal frequencies between simulations and measurements, modes can be approximately identif{}ied, but this is by no means as straightforward or precise as it was with the TESLA-style cavities.    

\section{Module-Based Transmission Spectra}\label{hom-meas:cmtb}
\begin{figure}[h]
\subfigure[The f\mbox{}irst dipole band of C2]{
\includegraphics[width=0.47\textwidth]{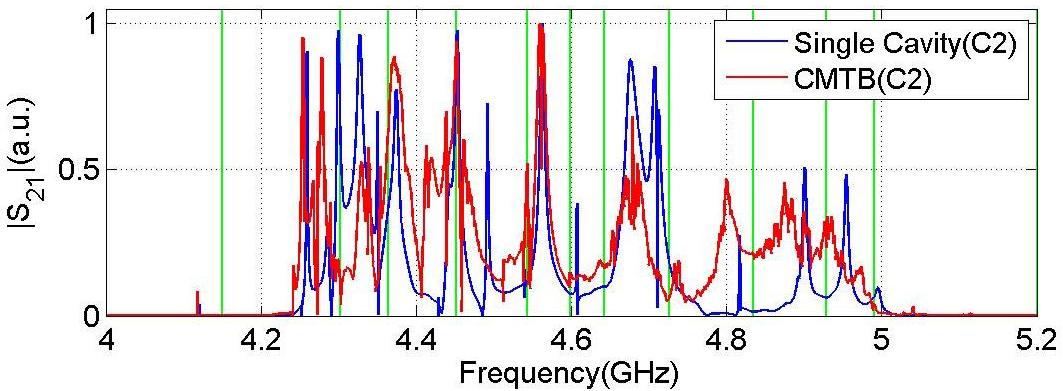}
\label{D1-fnal-cmtb}
}
\subfigure[The second dipole band of C2]{
\includegraphics[width=0.47\textwidth]{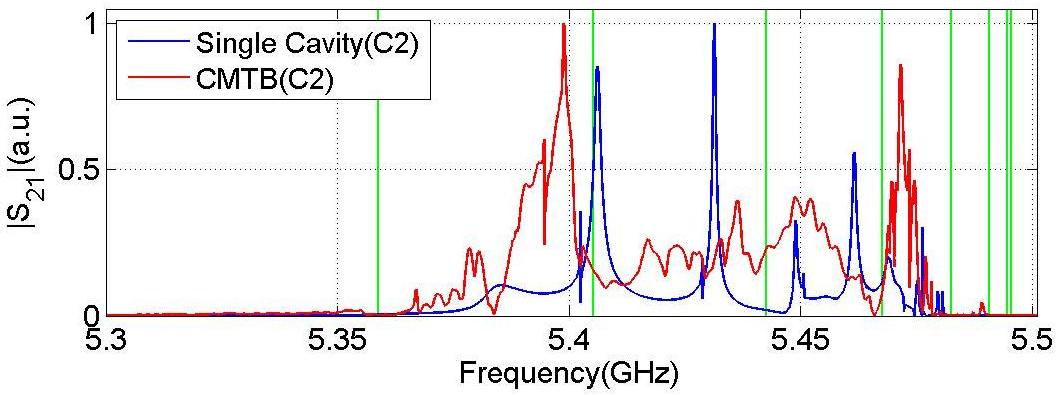}
\label{D2-fnal-cmtb}
}
\caption{Spectra of C2 from single cavity measurement (blue) and CMTB (red). The vertical lines are from simulations \cite{racc39-1}.}
\label{D1D2-fnal-cmtb}
\end{figure}

After the assembly of all four cavities into the cryo-module ACC39, this was installed in the Cryo-Module Test Bench (CMTB) at DESY. RF measurements were conducted in the CMTB tunnel without beam-excitation.\footnote{Data are kindly provided by T.~Khabibouline from Fermilab.} The measurements were repeated after the installation in FLASH from the HOM board rack outside the tunnel\footnote{H.W.~Glock along with T.~Flisgen from the University of Rostock lead the measurements.} \cite{rhommeas-2}. The HOM signals are stronger in CMTB than in FLASH because additional amplif{}iers were used and measurements were conducted inside the tunnel. The two measurements are otherwise similar. Therefore CMTB results are shown regarding most of the module-based measurements without beam-excitation. The measurements at FLASH are shown in Appendix~\ref{app-spec:flash}. From the recorded spectra, monopole, dipole and quadrupole bands are identif{}ied. Instead of the usual discrete modes in single cavity spectrum described in the previous section, more modes arise due to the coupling of inter-connected cavities. These coupling ef{}fects can be clearly seen in Fig.~\ref{D1D2-fnal-cmtb} where the $S_{21}$ parameter of an isolated cavity is compared with the one measured at CMTB. 

\begin{figure}[h]
\subfigure[Spectrum from 3.7 to 9~GHz]{
\includegraphics[width=0.47\textwidth]{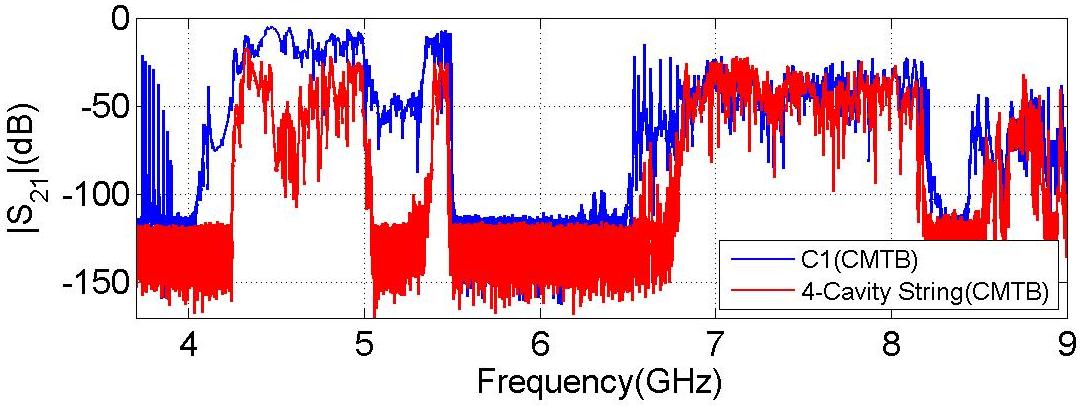}
\label{cutoff-full}
}
\subfigure[The f\mbox{}ifth dipole band]{
\includegraphics[width=0.47\textwidth]{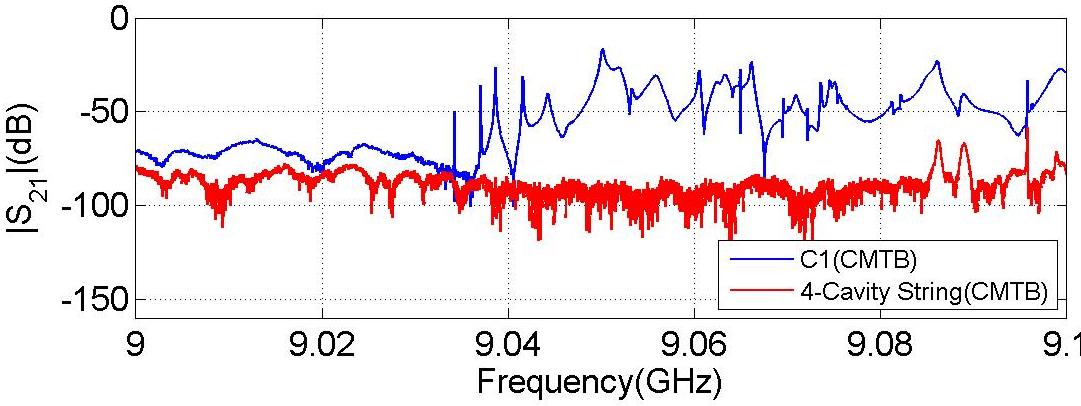}
\label{cutoff-D5}
}
\caption{Coupling ef{}fects of inter-connected cavities. The measurements were made at CMTB, while the spectra measured across C1 (from C1H1 to C1H2) is in blue, and the spectra measured across the entire four-cavity string (from C1H2 to C4H2) is in red.}
\label{cutoff-full-D5}
\end{figure}

To characterize the coupling of these modes, transmission spectra were measured along the entire four-cavity string (from C1H2 to C4H2), and then compared with the spectra measured between upstream and downstream couplers of each cavity within the cryo-module. Most HOMs couple to adjacent cavities through attached beam pipes. This can be observed in Fig.~\ref{cutoff-full-D5}, where the usual discrete modes which are below the cutof{}f frequencies of the attached beam pipes (peaks present only in the cavity C1 spectrum but absent from the string spectrum). The theoretical cutof{}f frequency of the beam pipe is 4.39~GHz for dipole TE$_{11}$ modes, while the measurement shows an even lower value at approximately 4.25~GHz. Dipole beam-pipe modes at approximately 4.1~GHz and the f{}ifth dipole band at approximately 9.05~GHz are clearly visible as localized within each beam pipe or cavity. Simulations were made and showed the non-propagating characteristics of those modes as well \cite{racc39-hfss,racc39-cst}.

\section{Beam-Excited HOM Spectra}\label{hom-meas:beam}
\begin{figure}[h]\center
\subfigure[C2H1]{
\includegraphics[width=0.4\textwidth]{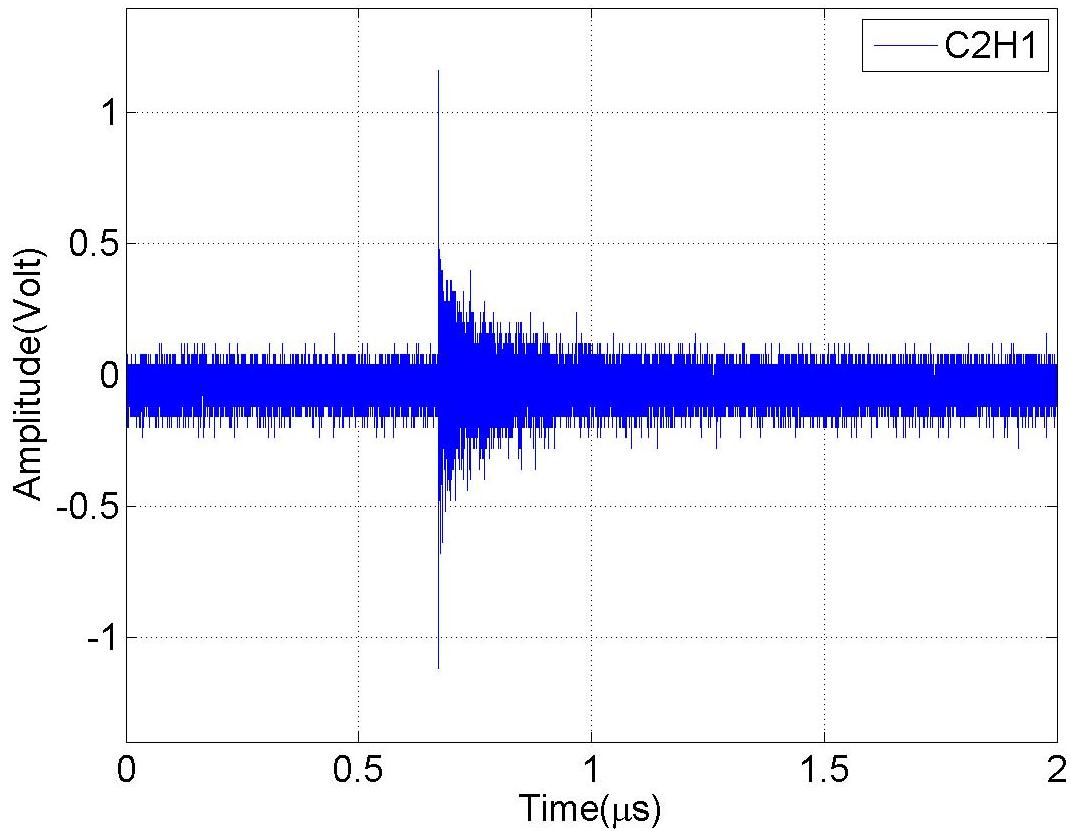}
\label{wfm-1}
}
\quad
\subfigure[C2H2]{
\includegraphics[width=0.4\textwidth]{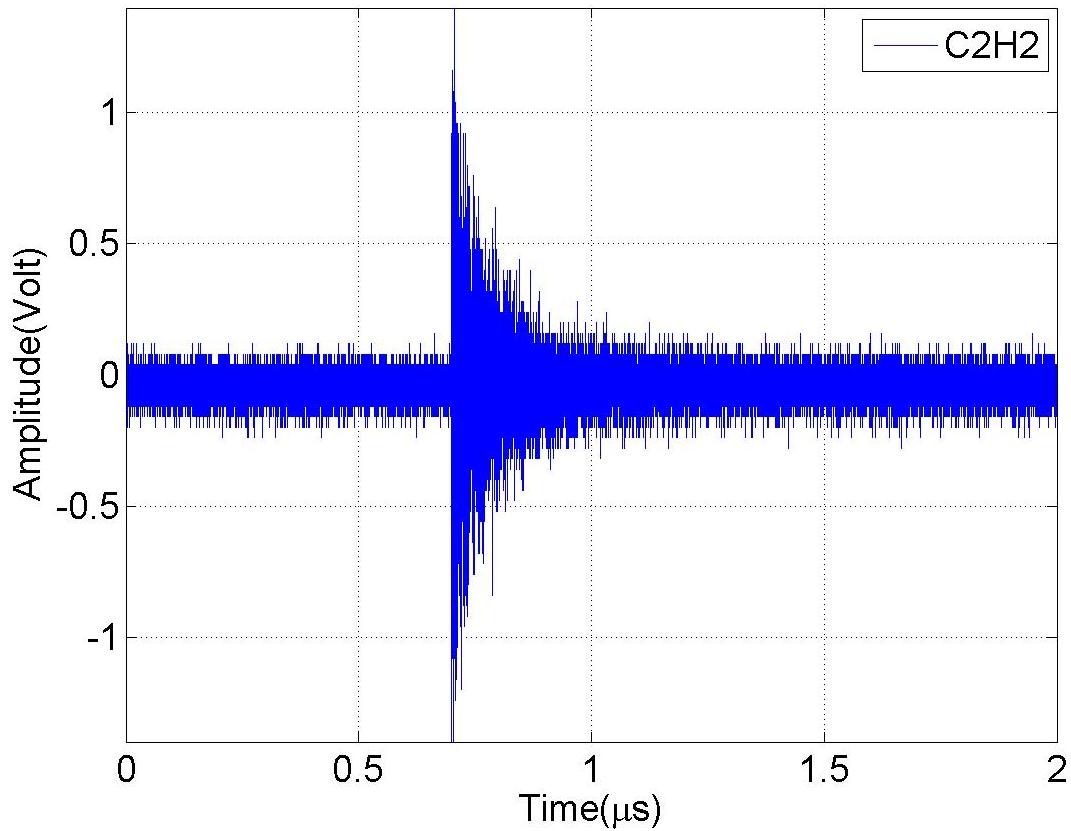}
\label{wfm-2}
}
\caption{Waveforms excited by a single electron bunch measured from the upstream and the downstream HOM coupler of C2.}
\label{wfm}
\end{figure}

The beam-excited measurements were conducted using a Tektronix Oscilloscope (scope) with a bandwidth of 6~GHz and a Tektronix Real-time Spectrum Analyzer (RSA) \cite{rrtsa}. HOM signals were taken from both HOM couplers of all four cavities at ACC39 HOM board rack outside the tunnel. A 10~dB external attenuator was connected to each HOM coupler to reduce the power of the beam-excited signals radiated to the coupler. Time-domain waveforms and real-time spectra were recorded. Each waveform was sampled with 20~GS/s and 200,000 points were recorded in a time window of 10~$\mu$s. Example waveforms from the upstream and the downstream couplers of C2 are shown in Fig.~\ref{wfm}. Each waveform was excited by a single electron bunch with a charge of approximately 0.5~nC. The scope was triggered synchronously with the beam pulse. 

\begin{figure}[h]\center
\subfigure[HOM spectrum]{
\includegraphics[width=0.9\textwidth]{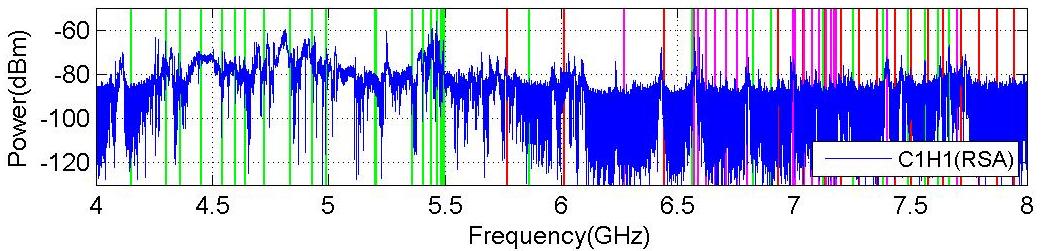}
\label{rsa-full-spec}
}
\subfigure[The fundamental band]{
\includegraphics[width=0.45\textwidth]{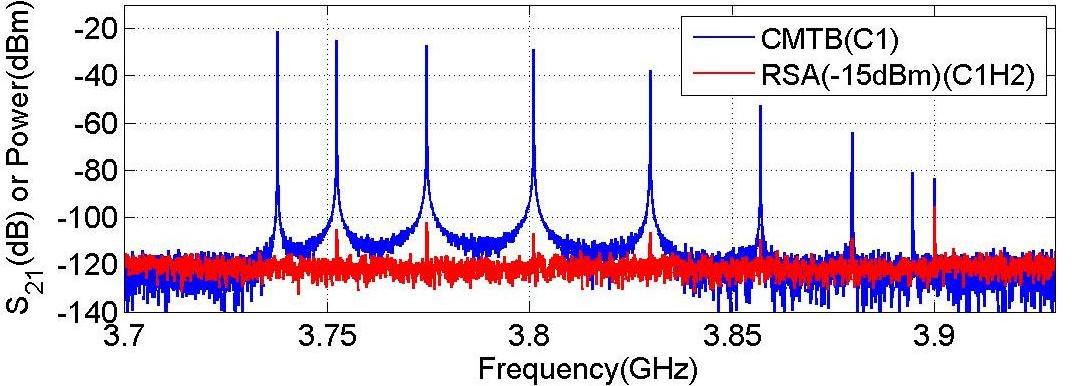}
\label{mono-rsa-nwa}
}
\subfigure[The f{}irst two dipole bands]{
\includegraphics[width=0.45\textwidth]{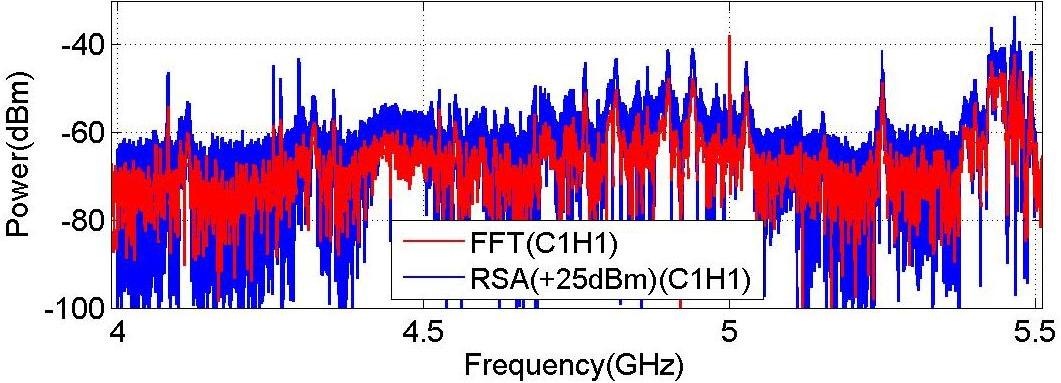}
\label{D1D2-rsa-fft}
}
\subfigure[The f{}ifth dipole band]{
\includegraphics[width=0.45\textwidth]{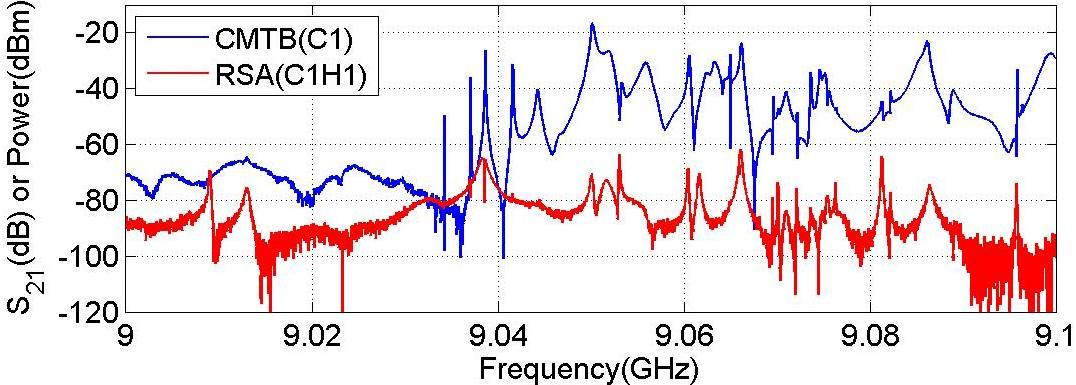}
\label{D5-rsa-nwa}
}
\caption{Beam-excited spectrum measured from HOM coupler C1H1. The vertical lines indicate simulation results of the eigenmodes \cite{racc39-1}. The colors red, green and magenta represent monopole, dipole and quadrupole modes, respectively.}
\label{rsa-spec}
\end{figure}

The same trigger was also used for the spectrum measurements by RSA. The spectrum ranging from 3.7 to 10~GHz was recorded from each HOM coupler. Each 50~MHz of the spectrum was excited by a single electron bunch. A frequency step of 10~kHz and a resolution bandwidth of 22.5~kHz were used. Fig.~\ref{rsa-full-spec} shows the HOM spectrum ranging from 4--8~GHz measured from coupler C1H1. Fig.~\ref{mono-rsa-nwa} shows the beam-excited spectrum of the fundamental band of one cavity compared with transmission spectrum measured at CMTB without beam-excitation. After a FFT (\textit{Fast Fourier Transform}) applied on the time-domain waveform (Fig.~\ref{wfm}), HOM signals measured by scope and RSA are compared in Fig.~\ref{D1D2-rsa-fft}. They show very good consistency. The spectrum of the f{}ifth dipole band is presented in Fig.~\ref{D5-rsa-nwa} (without the 10~dB external attenuator). 

\chapter{HOM Dependence on Transverse Beam Of{}fset}\label{hom-dep}
After the modal characterizations of the third harmonic cavities and the ACC39 module, experiments were conducted at FLASH to study the correlation of dipole modes to the transverse beam position. Localized beam-pipe modes and trapped cavity modes are presented here. Lorentzian f{}it technique was used in this chapter.

\section{Measurement Setup}\label{hom-dep:setup}
\begin{figure}[h]\center
\includegraphics[width=0.9\textwidth]{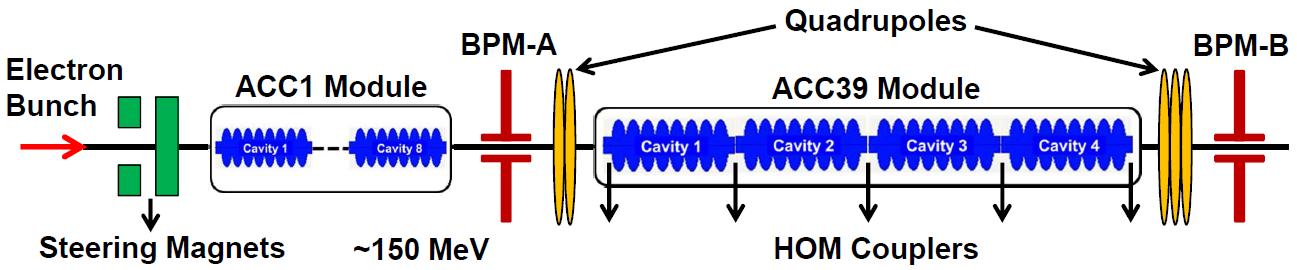}
\caption{Schematic of measurement setup for HOM dependence study (not to scale, cavities in ACC1 module are approximately three times larger than those in ACC39 module).}
\label{hom-setup}
\end{figure}

The schematic of the measurement setup is shown in Fig.~\ref{hom-setup}. An electron bunch of approximately 0.5~nC is accelerated on-crest by ACC1 before entering the ACC39 module. Two steering magnets located upstream of ACC1 can def{}lect the beam horizontally and vertically respectively. These are used to produce transverse of{}fsets of the electron bunch in ACC39. Two beam position monitors (BPM-A and BPM-B) are used to record transverse beam positions before and after ACC39. Switching of{}f the accelerating f{}ield in ACC39 and all quadruples close to ACC39, a straight line trajectory of the electron bunch is produced between those two BPMs. Therefore, the transverse of{}fset of the electron bunch in each cavity can be determined by interpolating the readouts of the two BPMs.

\begin{figure}[h]\center
\subfigure[Cross-like move]{
\includegraphics[width=0.226\textwidth]{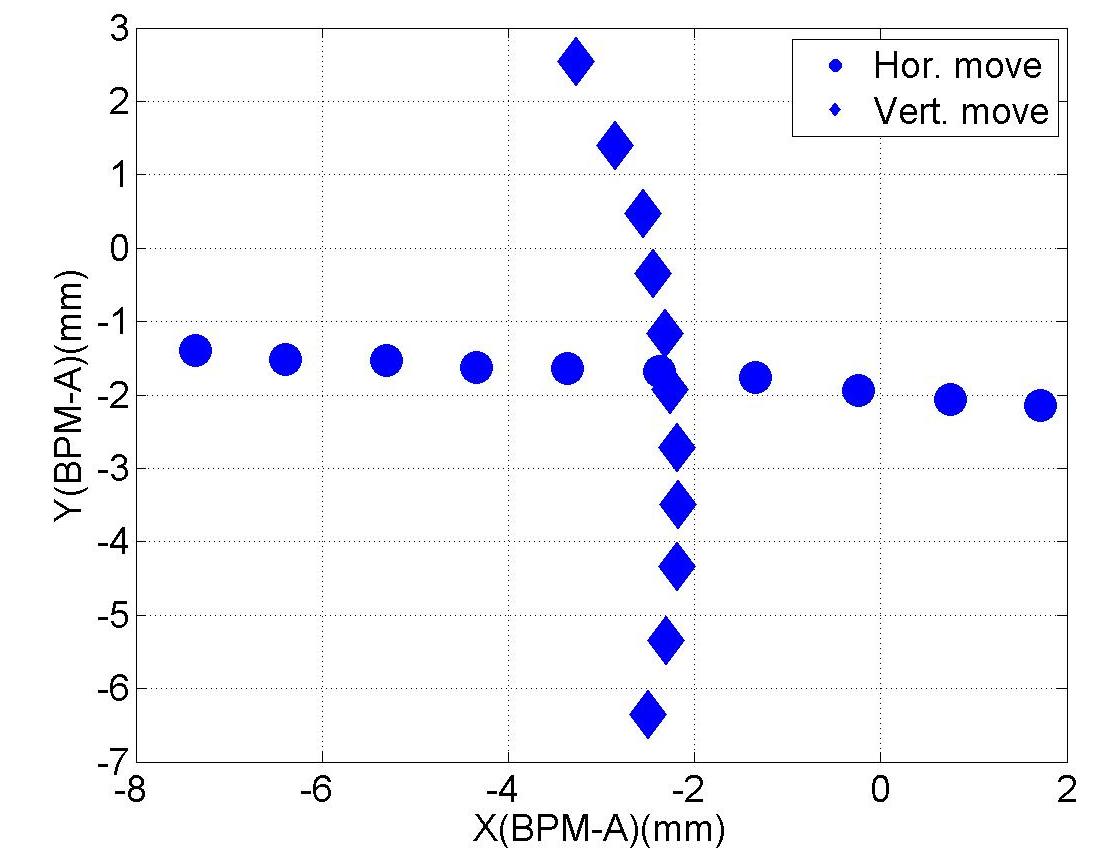}
\label{4D-9ACC1-cross-BP}
}
\subfigure[Cross-like move]{
\includegraphics[width=0.226\textwidth]{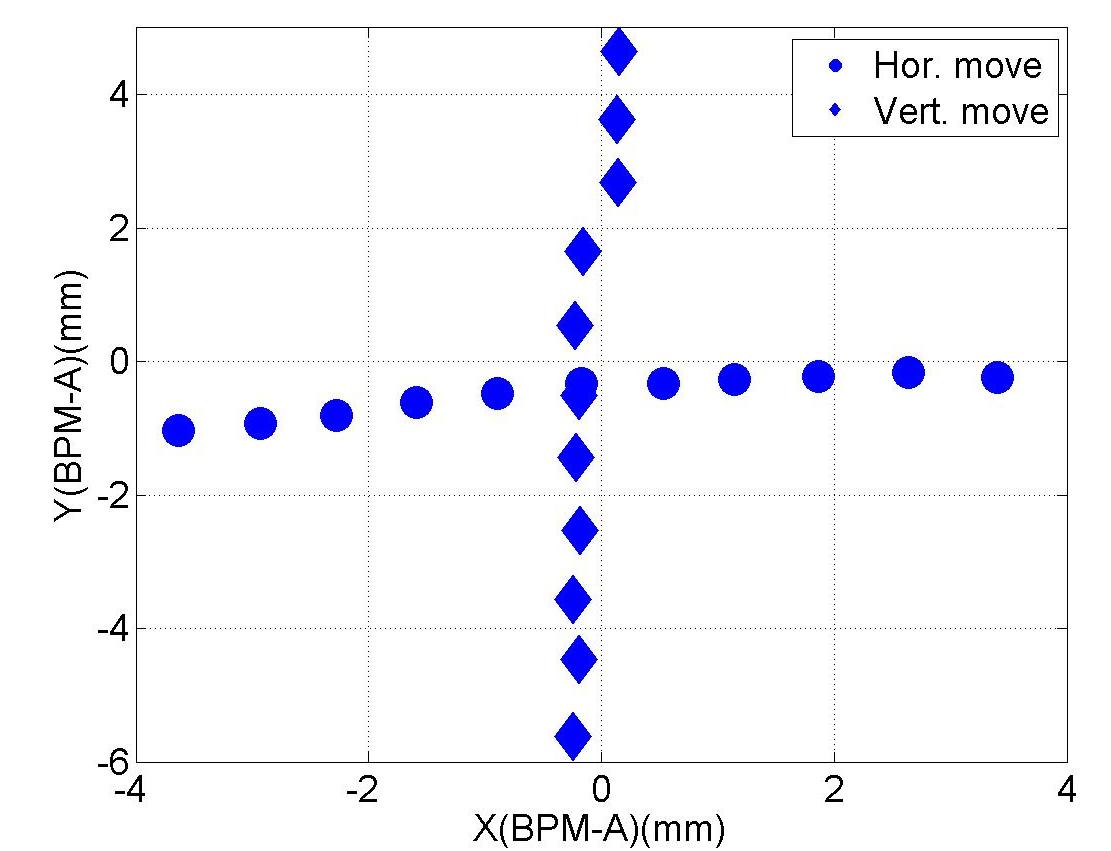}
\label{4D-9ACC1-cross-D5}
}
\subfigure[Grid-like move]{
\includegraphics[width=0.226\textwidth]{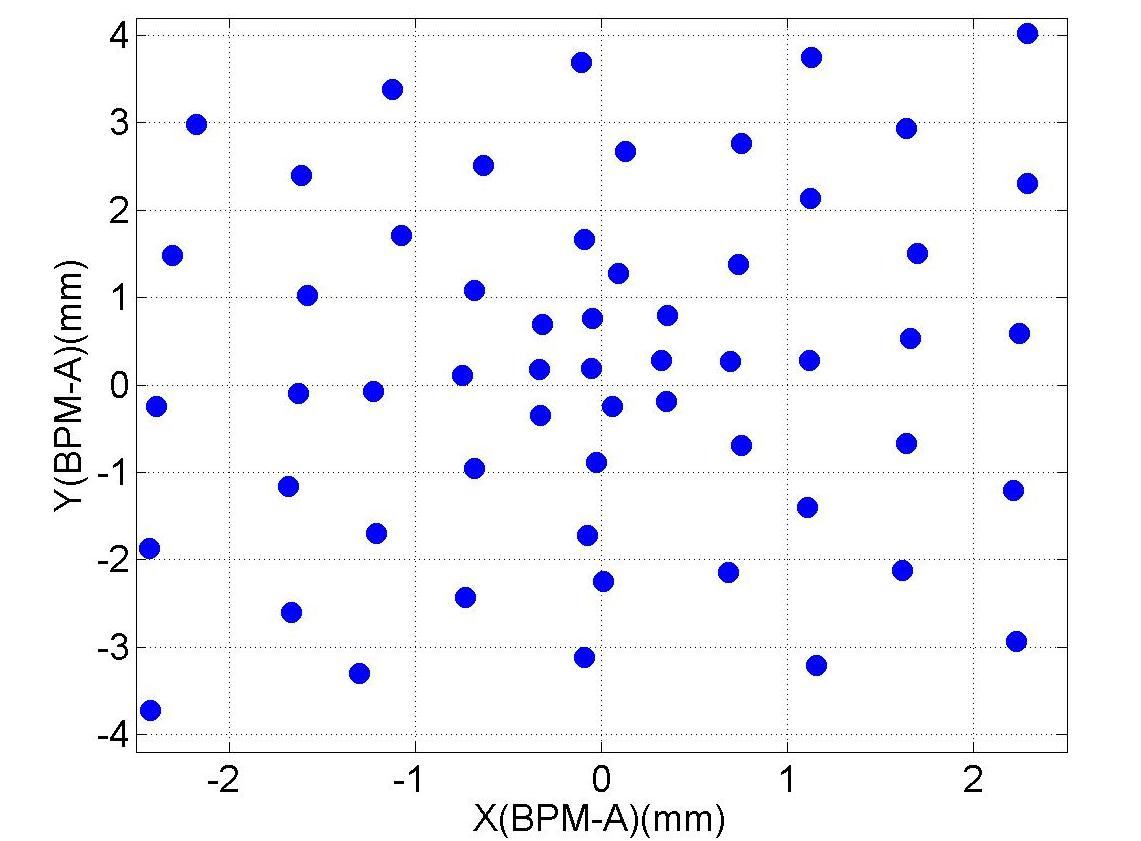}
\label{4D-9ACC1-grid-D1D2}
}
\subfigure[Grid-like move]{
\includegraphics[width=0.226\textwidth]{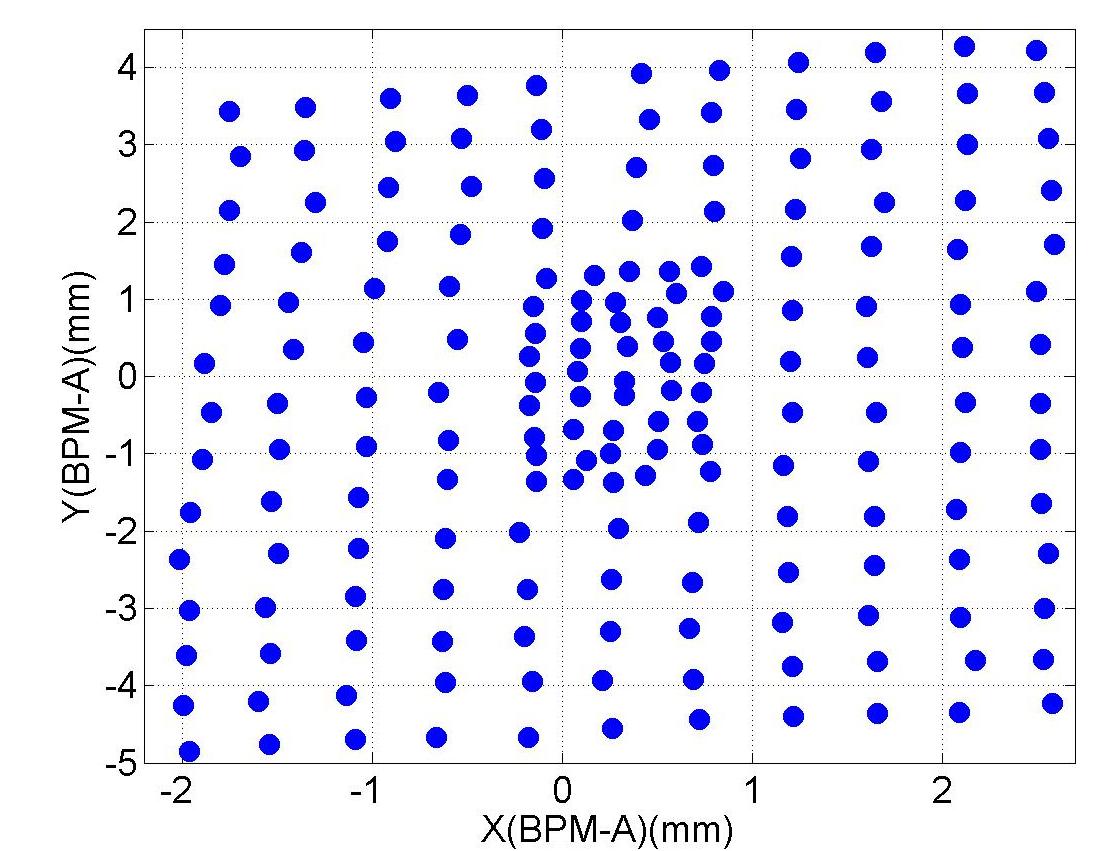}
\label{4D-9ACC1-grid-D5}
}
\caption{Cross-like beam scan during the measurements of the dipole beam-pipe modes (a) and the f{}ifth dipole band (b). Grid-like beam scan during the measurements of the dipole beam-pipe modes (c) and the f{}ifth dipole band (d). The quadrupoles were still on for the beam scan of (a).}
\label{2D-cross-grid}
\end{figure}

For each beam position, along with HOM signals, beam information is also recorded synchronously by reading nearby toroids, BPMs and currents of steering magnets. The electron bunch is moved in two-dimensional (2D) cross manner and 2D grid manner. The readings from BPM-A are shown in Fig.~\ref{2D-cross-grid}. Position interpolations are applicable throughout this paper except the 2D cross scan used for measuring beam-pipe modes (Fig.~\ref{4D-9ACC1-cross-BP}). In this case, the beam trajectory is no longer a straight line because the quadrupoles between BPM-A and BPM-B were still on. One notices some tilt of the position readings during both scans. This is due to the coupling between $x$ and $y$ plane caused partially by the ACC1 module and partially by BPM-A itself. The quadrupoles can also contribute to the nonlinearity for the readout of BPM-B. The details of various beam scans for the data presented in this note are summarized in Appendix~\ref{app-move}.

\section{The Localized Dipole Beam-pipe Modes}\label{hom-dep:bp}
\begin{figure}[h]
\subfigure[Mode \#1 (x)]{
\includegraphics[width=0.226\textwidth]{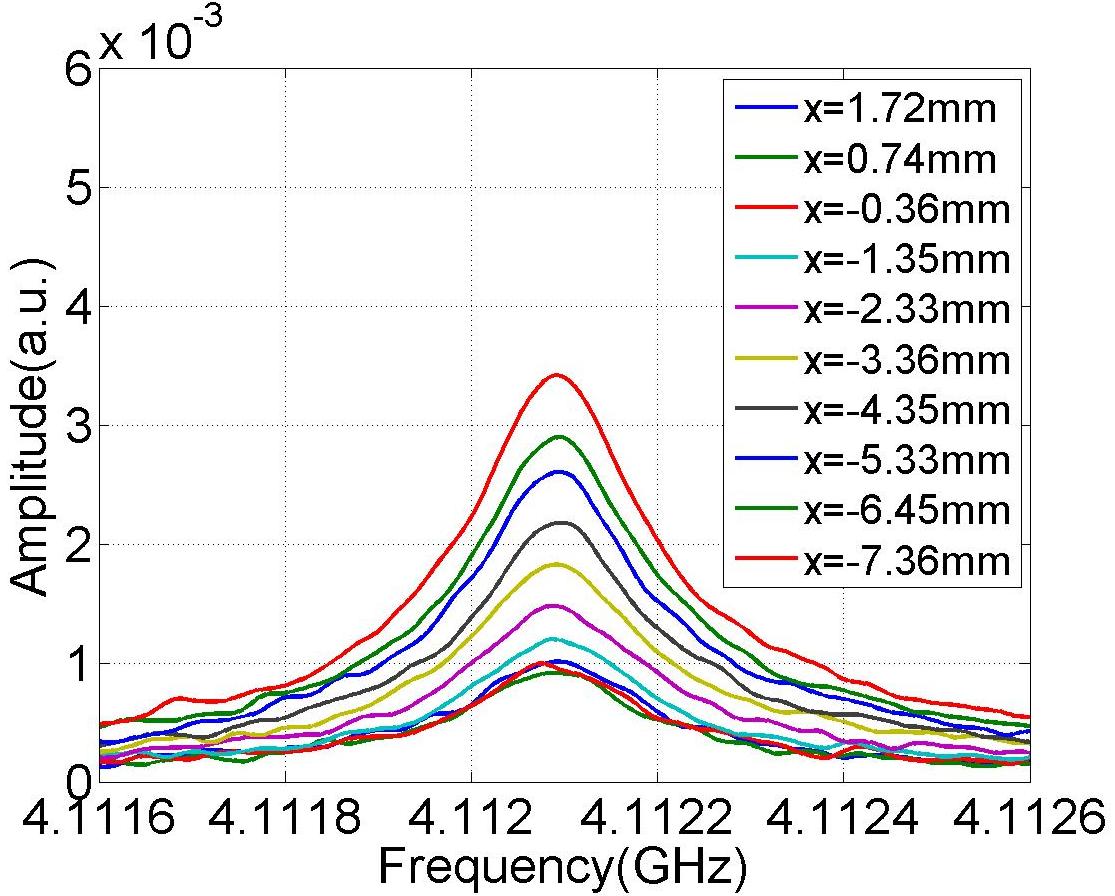}
\label{amp-cross-x-bp1}
}
\subfigure[Mode \#1 (y)]{
\includegraphics[width=0.226\textwidth]{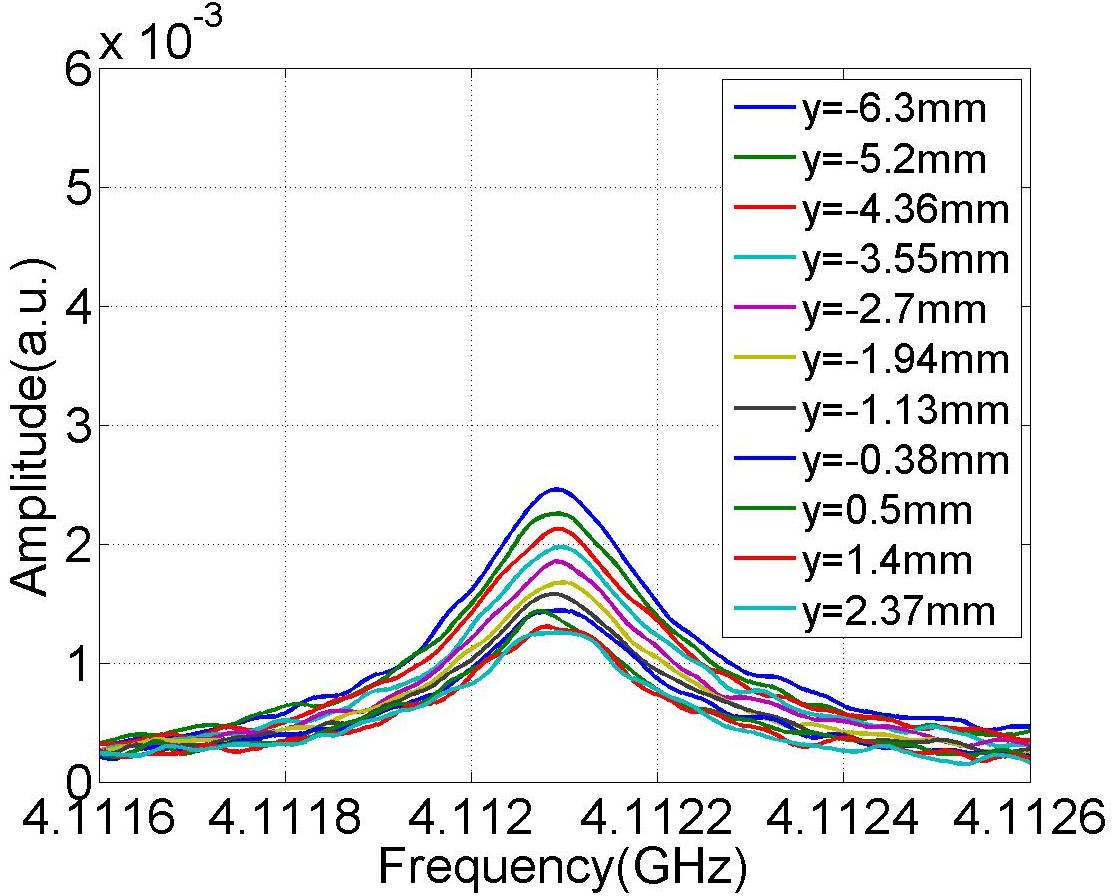}
\label{amp-cross-y-bp1}
}
\subfigure[Mode \#2 (x)]{
\includegraphics[width=0.226\textwidth]{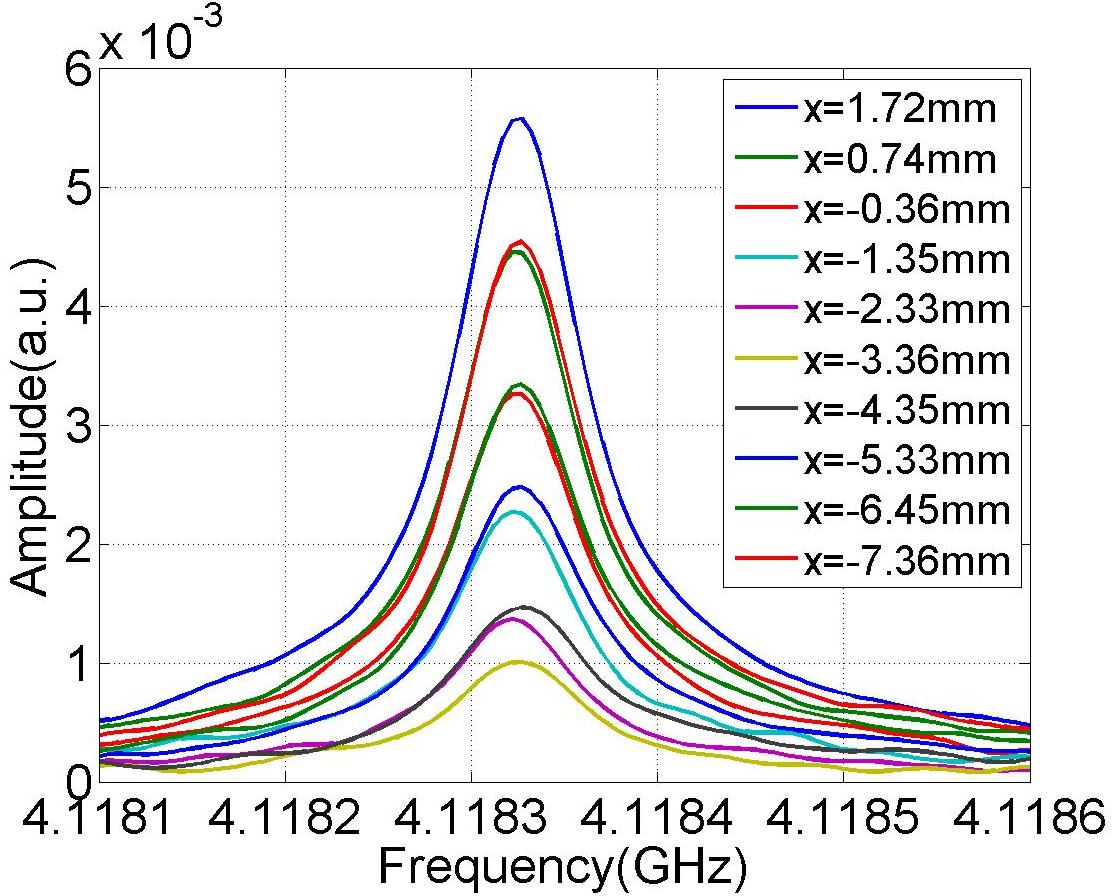}
\label{amp-cross-x-bp2}
}
\subfigure[Mode \#2 (y)]{
\includegraphics[width=0.226\textwidth]{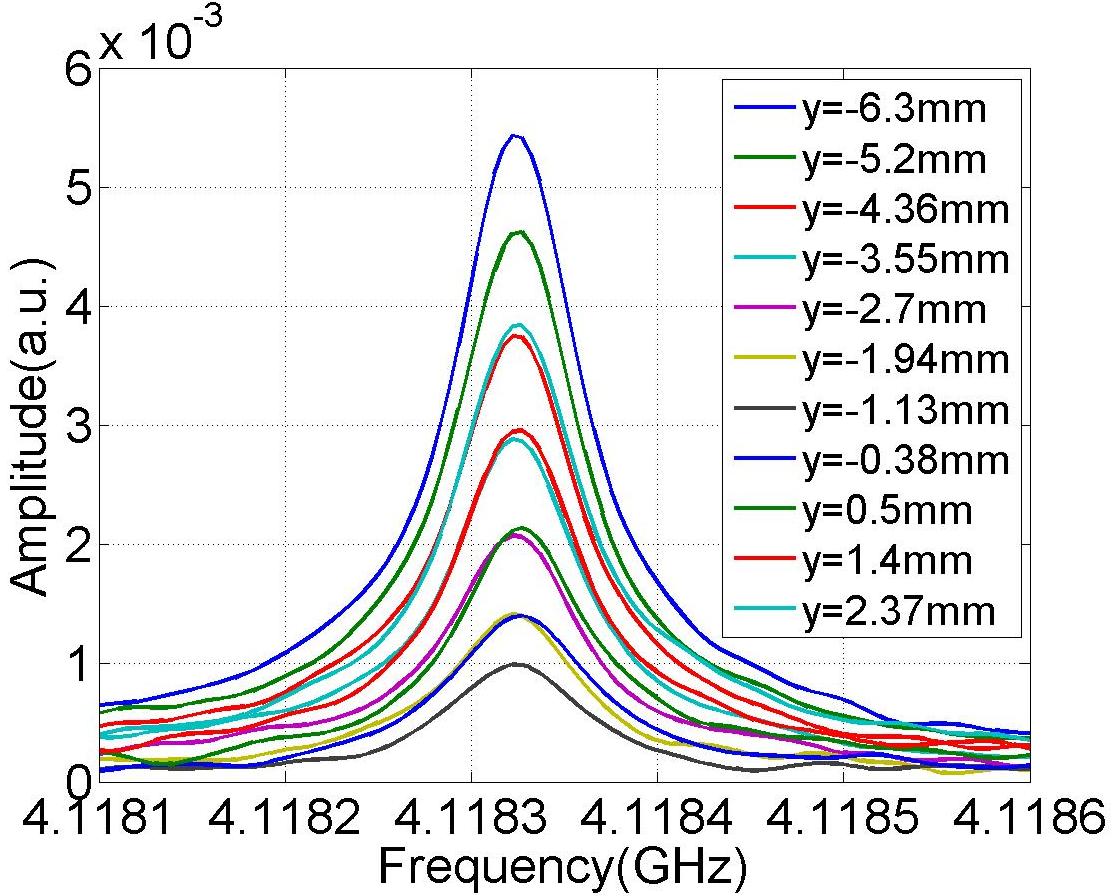}
\label{amp-cross-y-bp2}
}
\subfigure[Mode \#1 (x)]{
\includegraphics[width=0.226\textwidth]{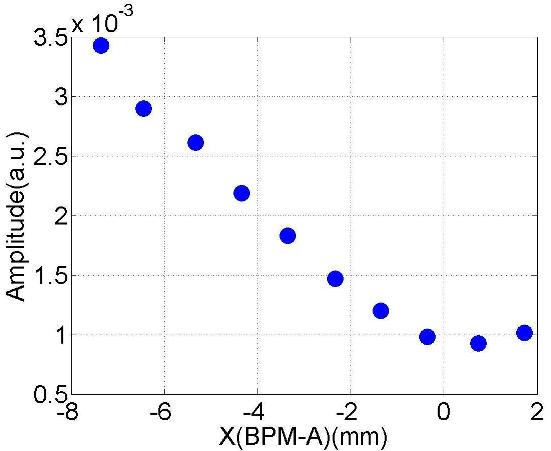}
\label{dep-cross-x-bp1}
}
\subfigure[Mode \#1 (y)]{
\includegraphics[width=0.226\textwidth]{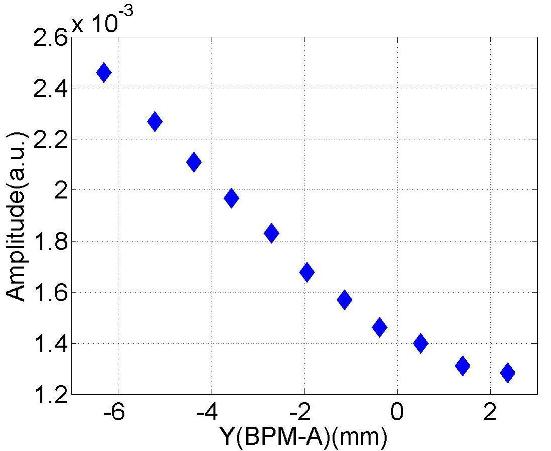}
\label{dep-cross-y-bp1}
}
\subfigure[Mode \#2 (x)]{
\includegraphics[width=0.226\textwidth]{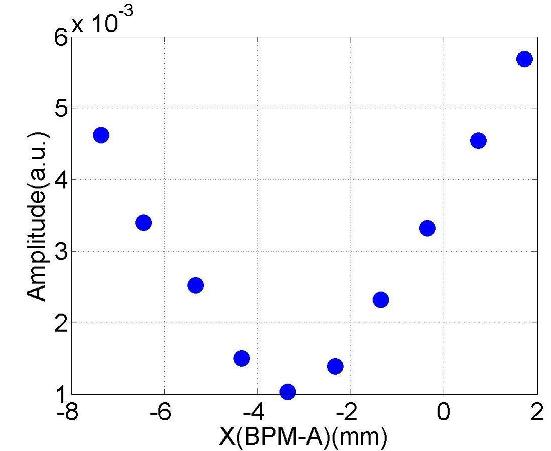}
\label{dep-cross-x-bp2}
}
\subfigure[Mode \#2 (y)]{
\includegraphics[width=0.226\textwidth]{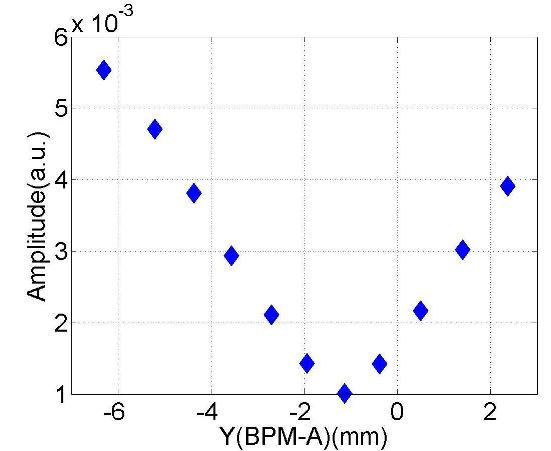}
\label{dep-cross-y-bp2}
}
\caption{Amplitude of dipole beam-pipe modes varies with transverse beam position read from BPM-A. The spectra were measured from HOM coupler C2H2.}
\label{amp-dep-cross-bp}
\end{figure}

The spectra of two modes at approximately 4.1~GHz measured from C2H2 at ten dif{}ferent horizontal beam positions (dots in Fig.~\ref{4D-9ACC1-cross-BP}) are shown in Fig.~\ref{amp-cross-x-bp1} and Fig.~\ref{amp-cross-x-bp2}, while the modal spectra of eleven dif{}ferent vertical beam positions (diamonds in Fig.~\ref{4D-9ACC1-cross-BP}) are shown in Fig.~\ref{amp-cross-y-bp1} and Fig.~\ref{amp-cross-y-bp2}. According to the previous studies, they are identif{}ied as dipole beam-pipe modes. The vertical position reading from BPM-A varied by $\pm$0.24~mm during the horizontal scan, and the horizontal position varied by $\pm$0.33~mm during the vertical scan. Variations of the mode amplitude with respect to the horizontal beam position can be observed. The amplitudes have been normalized to 1~nC of beam charge. The amplitudes do not reach zero because of an of{}fset in the vertical direction during the horizontal scan and it is probably the case that there is an angular displacement in the beam trajectory.

By performing Lorentzian f\mbox{}its (Eq.~\ref{eq:lorfit}), one can obtain the mode amplitude from each spectrum. Fig.~\ref{amp-dep-cross-bp}(e)-(h) show the amplitude of each mode as a function of the transverse beam position read from BPM-A for both horizontal and vertical beam scans (Fig.~\ref{4D-9ACC1-cross-BP}). A linear dependence of the mode amplitude on the transverse beam position can be observed, which indicates a dipole-like behavior. 

\begin{figure}[h]\center
\subfigure[Mode \#1]{
\includegraphics[width=0.4\textwidth]{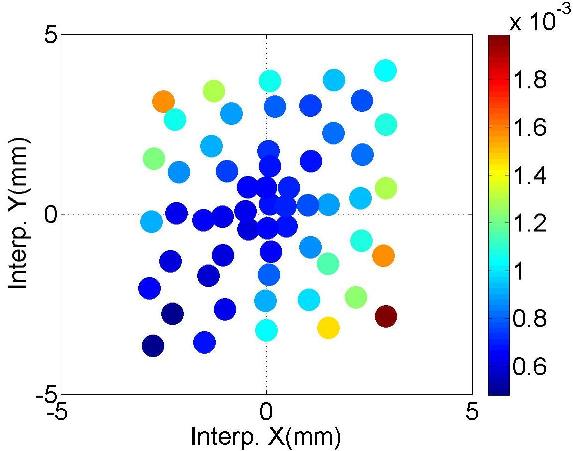}
\label{polar-bp1}
}
\quad
\subfigure[Mode \#2]{
\includegraphics[width=0.4\textwidth]{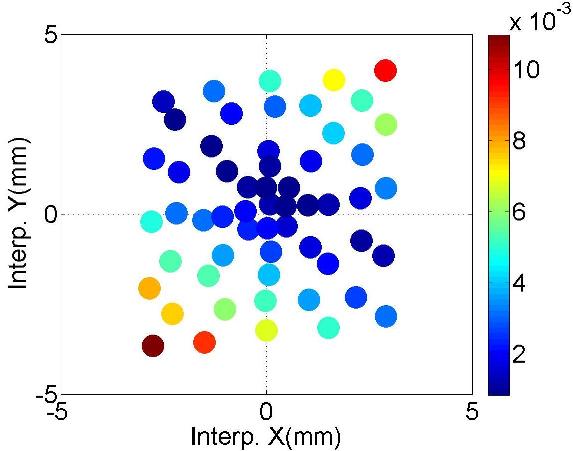}
\label{polar-bp2}
}
\caption{Amplitude of dipole beam-pipe modes as a function of the transverse beam position interpolated in the cavity (C2). The color varies according to amplitude value. The signals were measured from HOM coupler C2H2.}
\label{polar-bp}
\end{figure}

The polarization of each dipole beam-pipe mode is shown in Fig.~\ref{polar-bp} where the amplitude was measured in the 2D grid scan (Fig.~\ref{4D-9ACC1-grid-D1D2}). Position interpolations from the two BPM readouts (BPM-A and BPM-B) are applied to get the transverse beam positions in the related beam pipe. These two modes are polarized perpendicularly to each other, which indicates potential splitting of the mode degeneracy. This is also representative of a dipole-like behavior. The symmetry of the ideal cylindrical structure is broken by the HOM couplers installed on the connecting beam pipes. This causes the frequency split of the two polarizations along with the inevitable manufacturing tolerances. There are three beam pipes inter-connecting cavities in ACC39 (see Fig.~\ref{cavity-cartoon}), but clear polarizations can only be observed from the beam pipe connecting C2 and C3. The power couplers installed on the other two beam pipes may account for this. The linear dependencies and polarizations of modes within the frequency range of 4.0 - 4.15~GHz measured from all eight HOM couplers are shown in Appendix~\ref{app-bp}.

\section{Trapped Cavity Modes in the Fifth Dipole Band}\label{hom-dep:d5}
\begin{figure}[h]\center
\subfigure[Mode amplitudes (C2H2)]{
\includegraphics[width=0.31\textwidth]{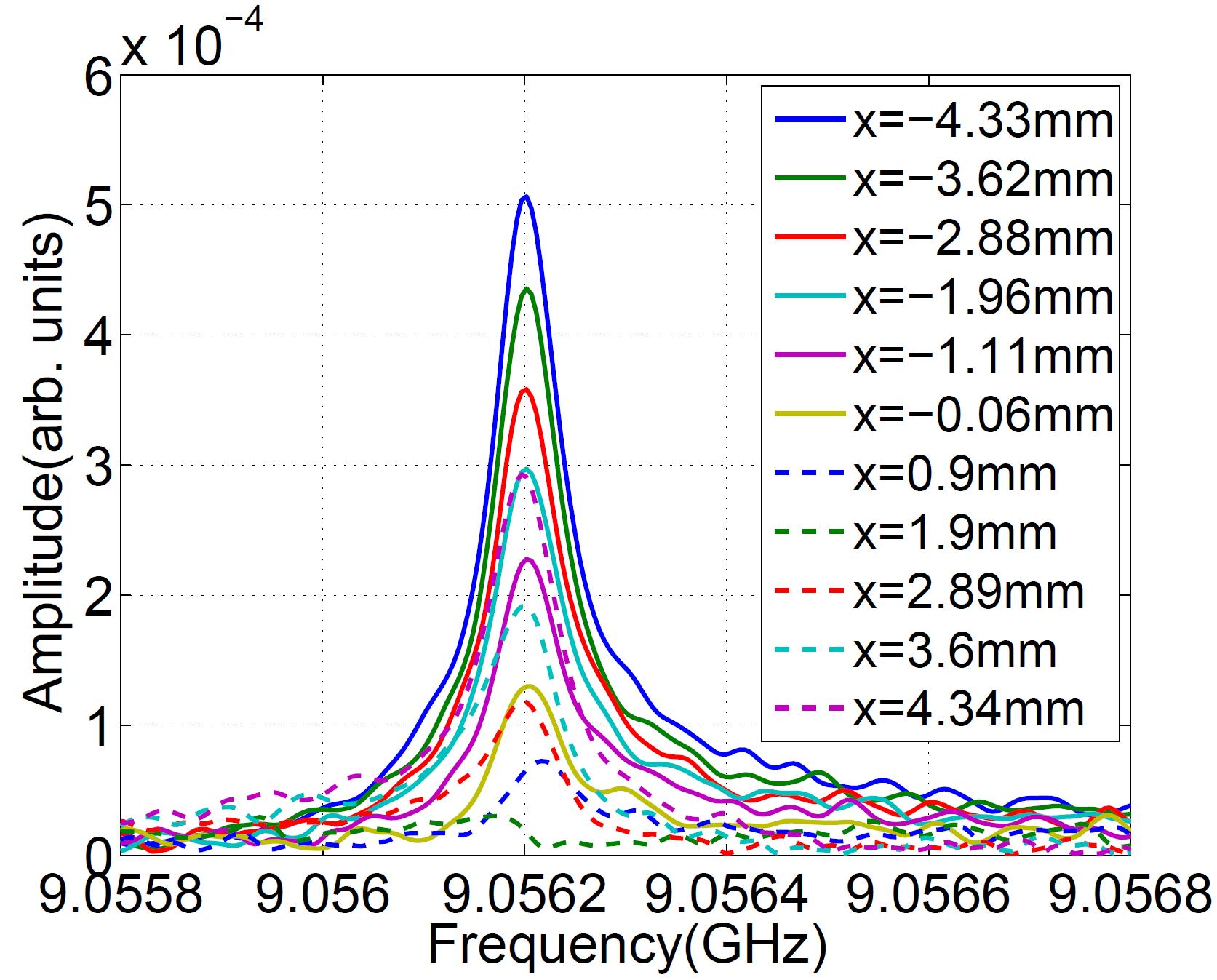}
\label{amp-cross-x-D5}
}
\subfigure[Linear dependence]{
\includegraphics[width=0.31\textwidth]{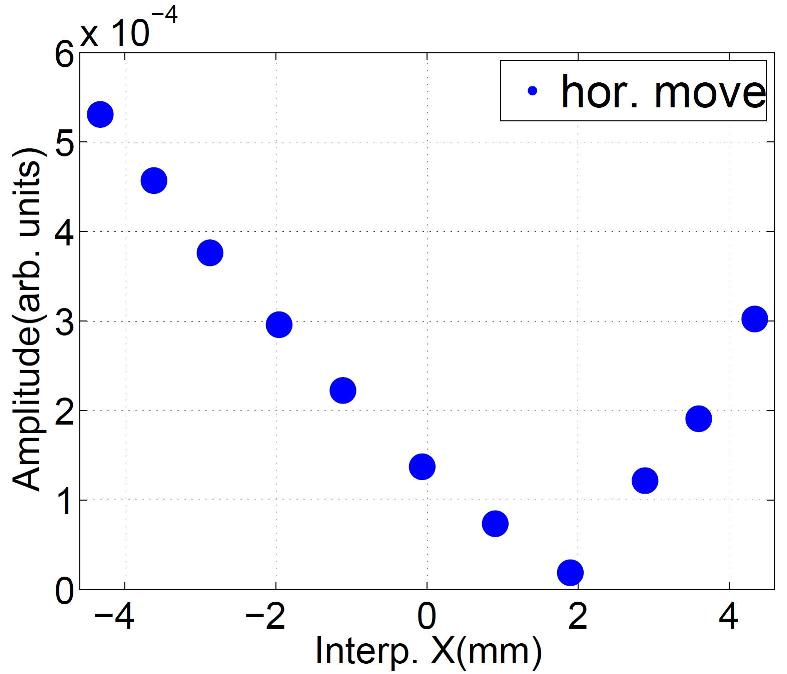}
\label{dep-cross-x-D5}
}
\subfigure[Polarization (C2H2)]{
\includegraphics[width=0.32\textwidth]{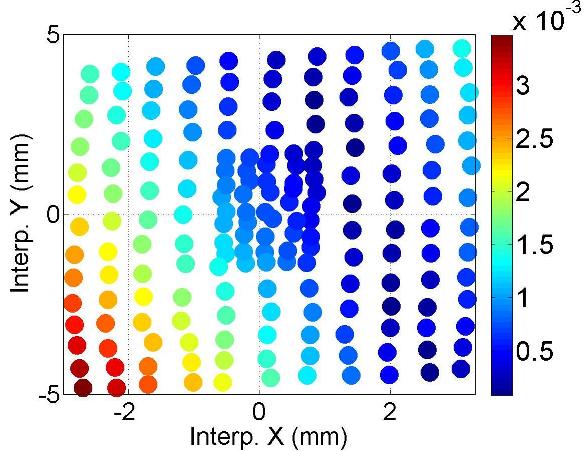}
\label{polar-D5}
}
\caption{Amplitudes of one dipole mode as a function of the transverse beam position interpolated in the cavity (C2). The signals were measured from HOM coupler C2H2.}
\label{amp-dep-cross-D5}
\end{figure}

As explained in Chapter~\ref{hom-meas}, there are trapped modes in the f{}ifth dipole band. By moving the beam horizontally (dots in Fig.~\ref{4D-9ACC1-cross-D5}), the variation of the amplitude of a mode at approximately 9.0562~GHz can be clearly seen in Fig.~\ref{amp-cross-x-D5}. The vertical position in C2 varied by $\pm0.29$~mm during the horizontal scan. Using a Lorentzian f{}it to obtain the amplitudes of the modes and plot them against horizontal beam positions interpolated from BPM readouts, the linear dependence can be clearly seen in Fig.~\ref{dep-cross-x-D5}, which indicates a dipole-like behavior. 

In general, modes in the f{}ifth dipole band have small amplitude, therefore a dedicated grid-like beam scan was conducted without the 10~dB external attenuators. The BPM readouts for this scan are shown in Fig.~\ref{4D-9ACC1-grid-D5}. The amplitude of the same mode at approximately 9.0562~GHz is again obtained by a Lorentzian f{}it for each beam position, and plotted in Fig.~\ref{polar-D5}. The color denotes the amplitude magnitude. The polarization of this mode can be observed. The linear dependencies and polarizations of modes within the frequency range of 9.0 - 9.1~GHz measured from all eight HOM couplers are shown in Appendix~\ref{app-d5}.

\chapter{Conclusions}
Modal characterizations of the third harmonic cavities have been performed on HOM measurements made at several stages: isolated cavities, module-based and beam-based during FLASH operations. This is the f{}irst time for third harmonic cavities that the dependencies of HOMs on transverse beam positions have been observed. Beam-pipe and cavity modes have been characterized. Beam-pipe modes and some cavity modes in the f{}ifth dipole band are localized in segments of the module. Their analysis has been presented in this report. They allow for the beam position to be determined for each individual component (beam pipe or cavity). We have identif{}ied modes and in some cases polarizations of two beam-pipe modes have been observed. From transmission measurements, cavity modes in the f{}irst two dipole bands propagate. These were reported in other papers. Dedicated electronics for HOM-based beam diagnostics are under design. Various potential dipole modes for the electronics are being studied.
\begin{abstract}
We would like to thank T.~Khabibouline for providing the data of the single cavity measurements at Fermilab and module-based measurements at CMTB. The contribution of these two gentlemen from the University of Rostock cannot be neglected to the module-based measurements at FLASH: Dr.~H.W.~Glock and T.~Flisgen. We also acknowledge the contribution of B.~Lorbeer from DESY and Dr.~I.R.R.~Shinton from the University of Manchester to the measurements. We thank the FLASH crew for supporting the measurements, members of MEW group from the University of Manchester for enlightening ideas and colleagues from EuCARD WP10.5 for useful discussions. We are also grateful to Dr.~Rainer~Wanzenberg for carefully reading this manuscript. This work received support from the European Commission under the FP7 Research Infrastructures grant agreement No.227579.
\end{abstract}
\addcontentsline{toc}{chapter}{\numberline{}Bibliography}

\appendix
\chapter{HOM Signal}\label{app-spec}
\section{HOM Spectra from Single Cavity Measurements at Fermilab}\label{app-spec:fnal}
\begin{figure}[h]\center
\includegraphics[width=0.9\textwidth]{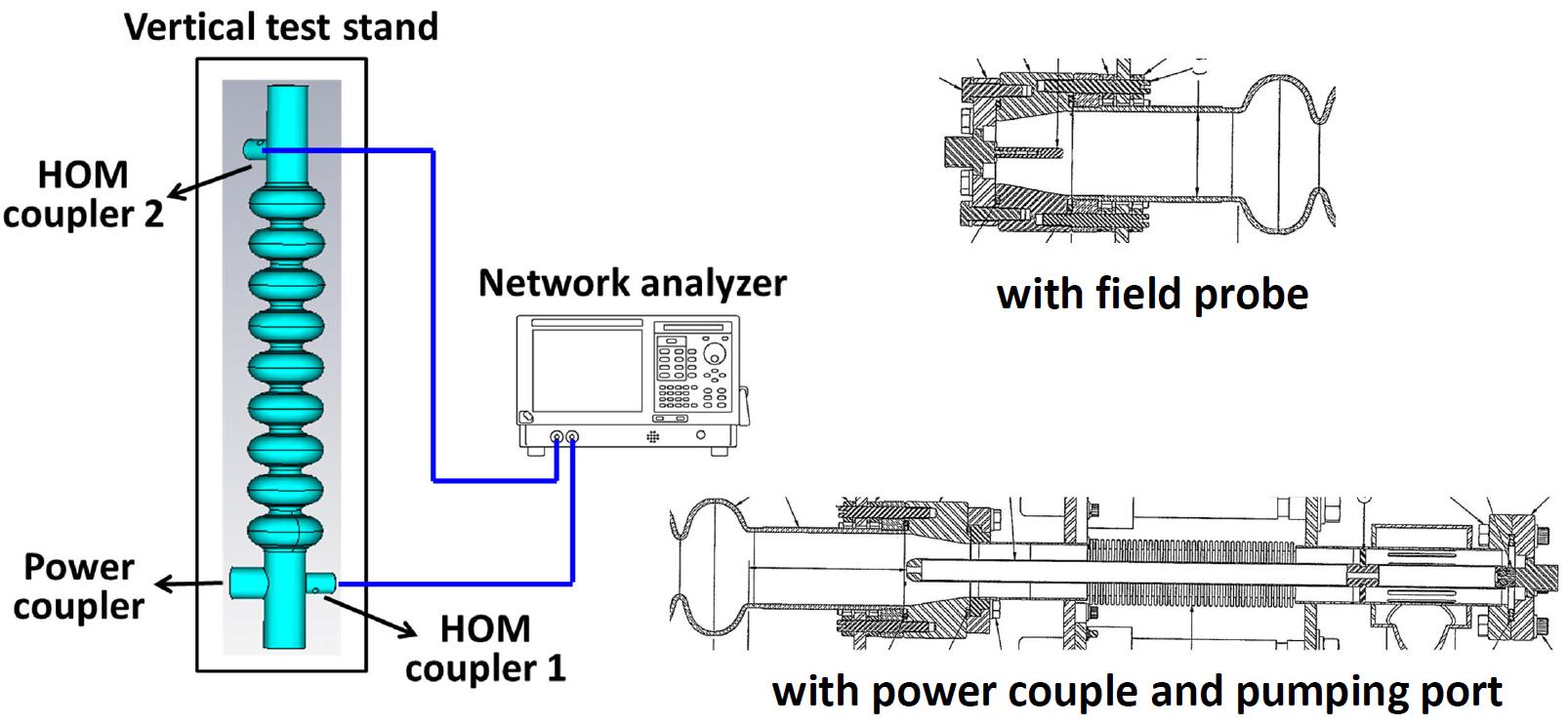}
\caption{The schematic setup of the single cavity RF transmission measurement.}
\label{setup-fnal}
\end{figure}

\begin{figure}
\subfigure[C1 (from C1H1 to C1H2)]{
\includegraphics[width=1\textwidth]{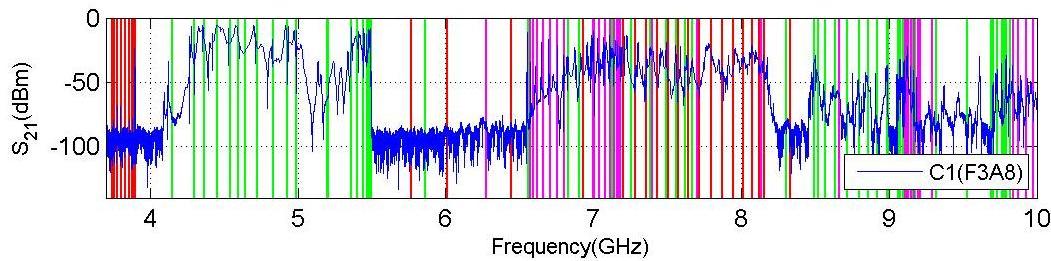}
\label{fnal-full-spec-C1}
}
\subfigure[C2 (from C2H1 to C2H2)]{
\includegraphics[width=1\textwidth]{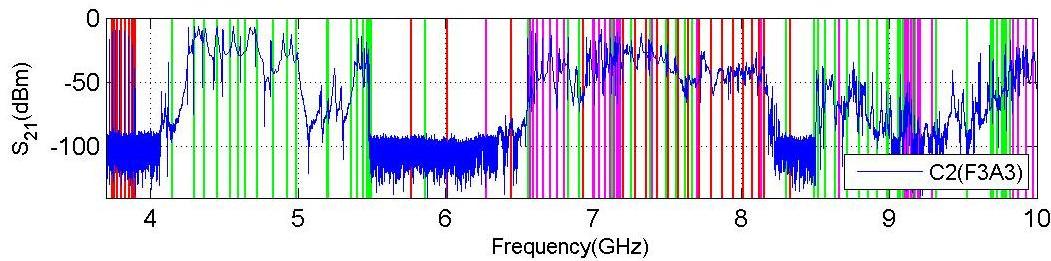}
\label{fnal-full-spec-C2}
}
\subfigure[C3 (from C3H1 to C3H2)]{
\includegraphics[width=1\textwidth]{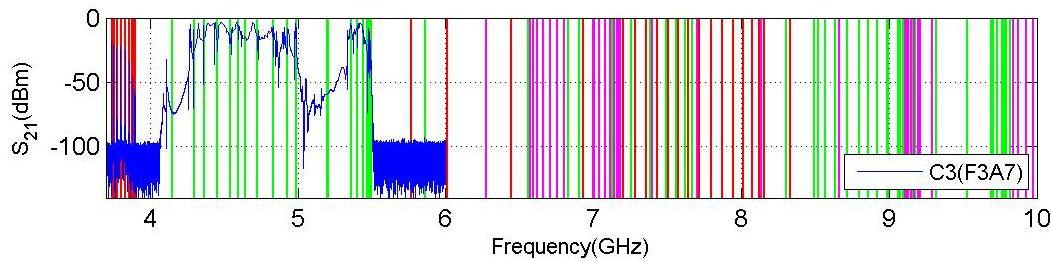}
\label{fnal-full-spec-C3}
}
\subfigure[C4 (from C4H1 to C4H2)]{
\includegraphics[width=1\textwidth]{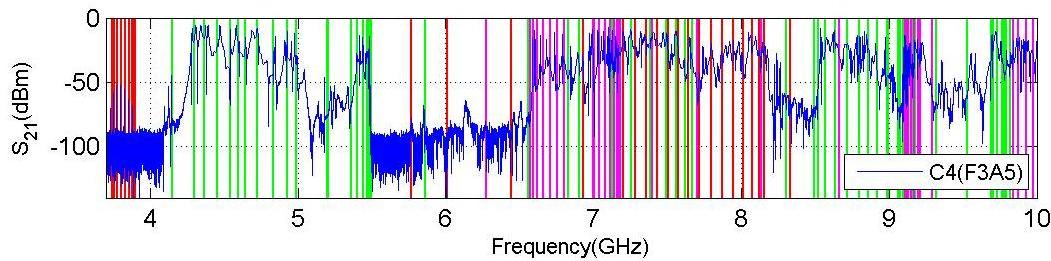}
\label{fnal-full-spec-C4}
}
\caption{Transmission scattering parameter $S_{21}$ measured across each single cavity at Fermilab. The vertical lines indicate the simulation results. The colors red, green and magenta represent monopole, dipole and quadrupole modes respectively, and this applies in Fig.~\ref{fnal-fund-all}--\ref{fnal-D5-all}. Details of the spectra are shown in the following f{}igures.}
\label{fnal-full-spec-all}
\end{figure}

\begin{figure}
\subfigure[C1 (from C1H1 to C1H2)]{
\includegraphics[width=1\textwidth]{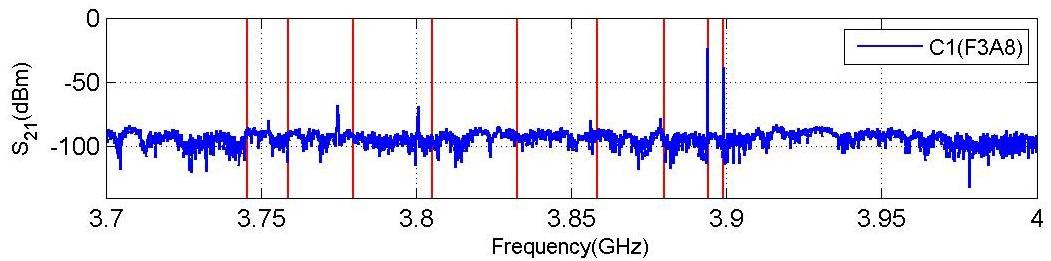}
\label{fnal-fund-C1}
}
\subfigure[C2 (from C2H1 to C2H2)]{
\includegraphics[width=1\textwidth]{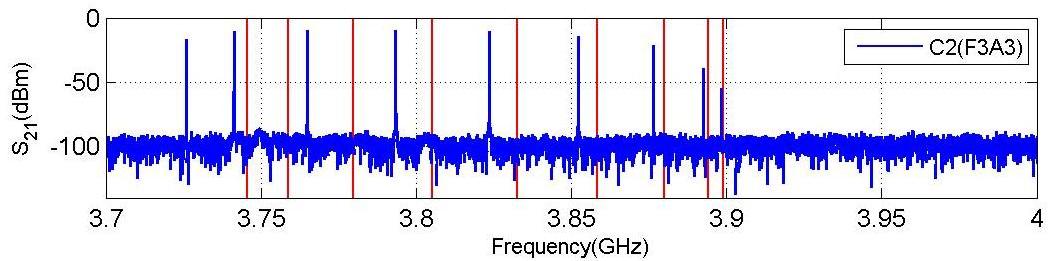}
\label{fnal-fund-C2}
}
\subfigure[C3 (from C3H1 to C3H2)]{
\includegraphics[width=1\textwidth]{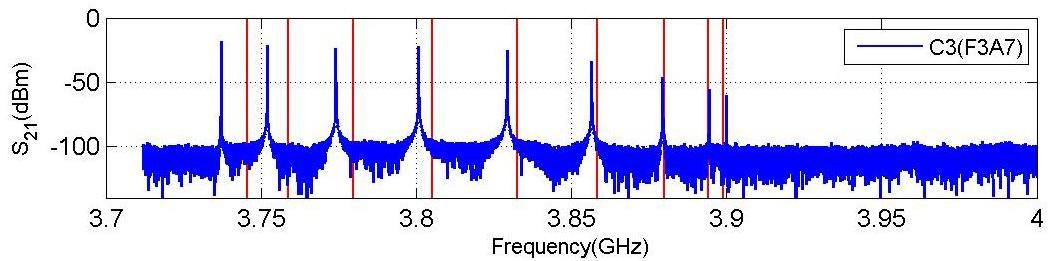}
\label{fnal-fund-C3}
}
\subfigure[C4 (from C4H1 to C4H2)]{
\includegraphics[width=1\textwidth]{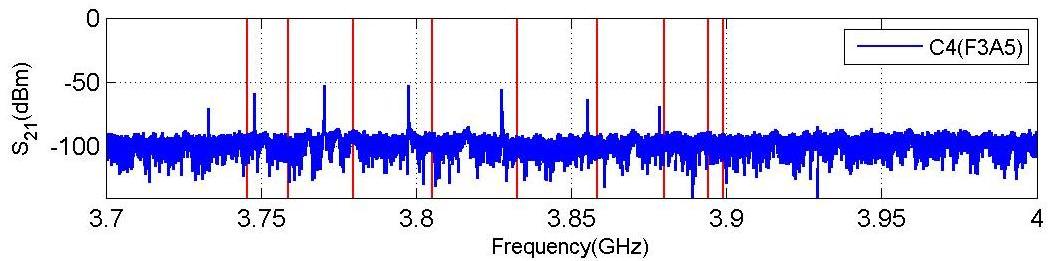}
\label{fnal-fund-C4}
}
\caption{Details of plots in Fig.~\ref{fnal-full-spec-all}: the fundamental band.}
\label{fnal-fund-all}
\end{figure}

\begin{figure}
\subfigure[C1 (from C1H1 to C1H2)]{
\includegraphics[width=1\textwidth]{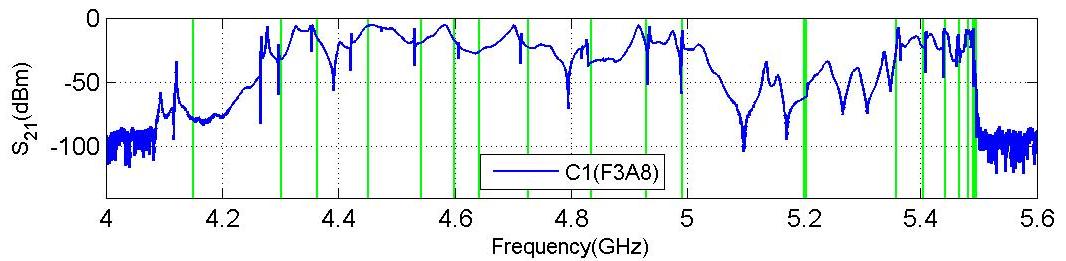}
\label{fnal-D1D2-C1}
}
\subfigure[C2 (from C2H1 to C2H2)]{
\includegraphics[width=1\textwidth]{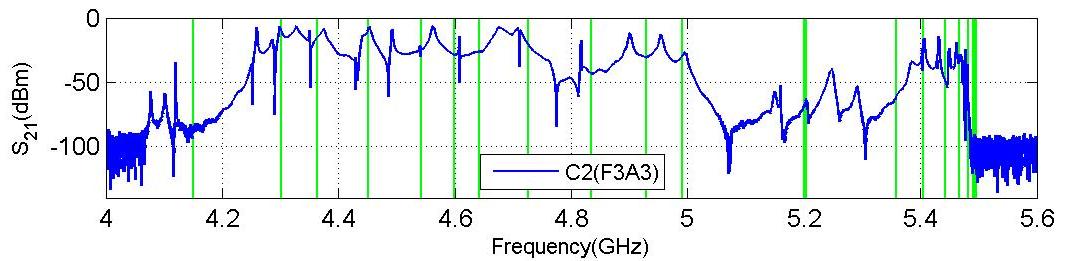}
\label{fnal-D1D2-C2}
}
\subfigure[C3 (from C3H1 to C3H2)]{
\includegraphics[width=1\textwidth]{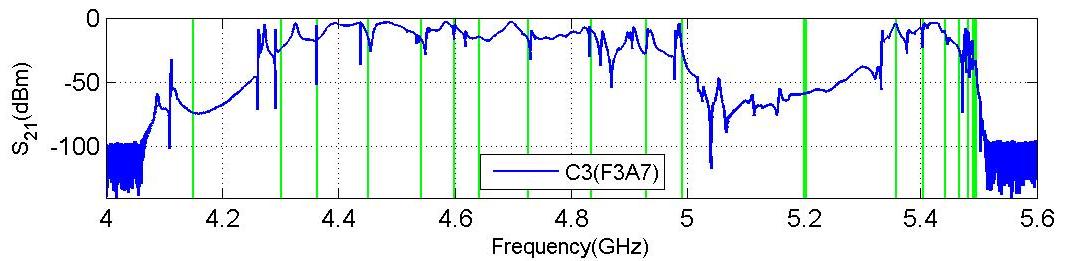}
\label{fnal-D1D2-C3}
}
\subfigure[C4 (from C4H1 to C4H2)]{
\includegraphics[width=1\textwidth]{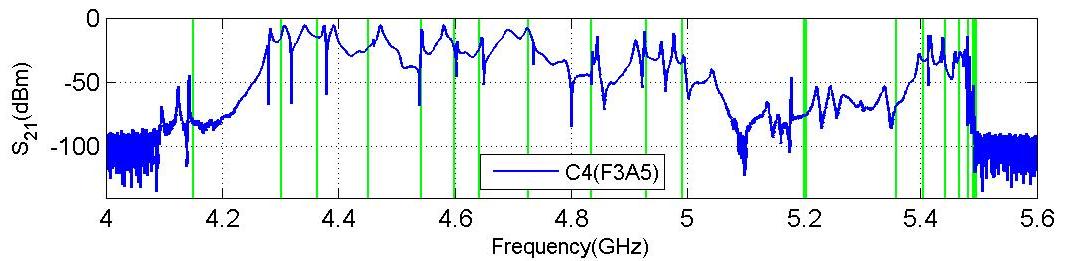}
\label{fnal-D1D2-C4}
}
\caption{Details of plots in Fig.~\ref{fnal-full-spec-all}: the f\mbox{}irst and second dipole band.}
\label{fnal-D1D2-all}
\end{figure}

\begin{figure}
\subfigure[C1 (from C1H1 to C1H2)]{
\includegraphics[width=1\textwidth]{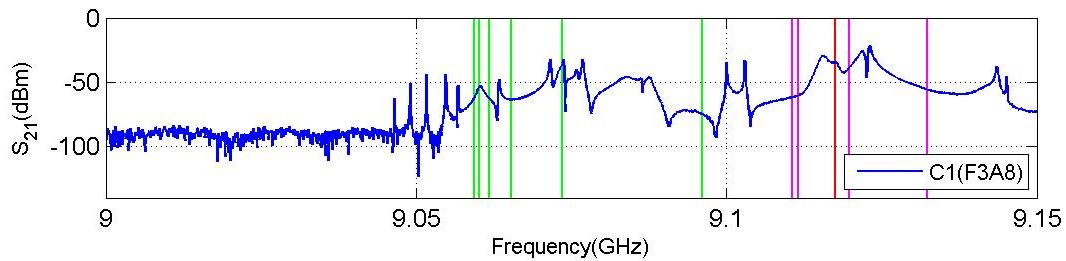}
\label{fnal-D5-C1}
}
\subfigure[C2 (from C2H1 to C2H2)]{
\includegraphics[width=1\textwidth]{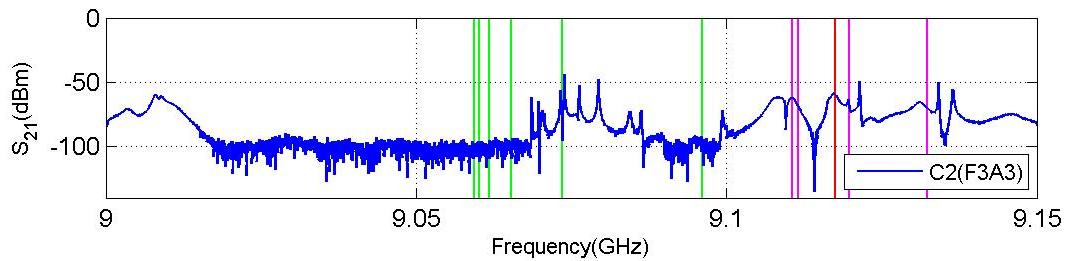}
\label{fnal-D5-C2}
}
\subfigure[C3 (from C3H1 to C3H2)]{
\includegraphics[width=1\textwidth]{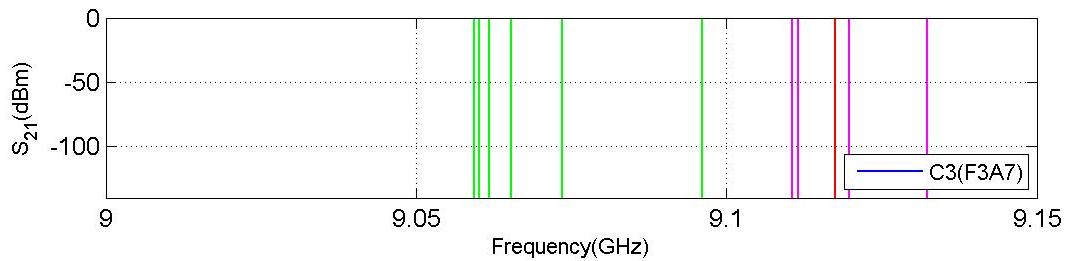}
\label{fnal-D5-C3}
}
\subfigure[C4 (from C4H1 to C4H2)]{
\includegraphics[width=1\textwidth]{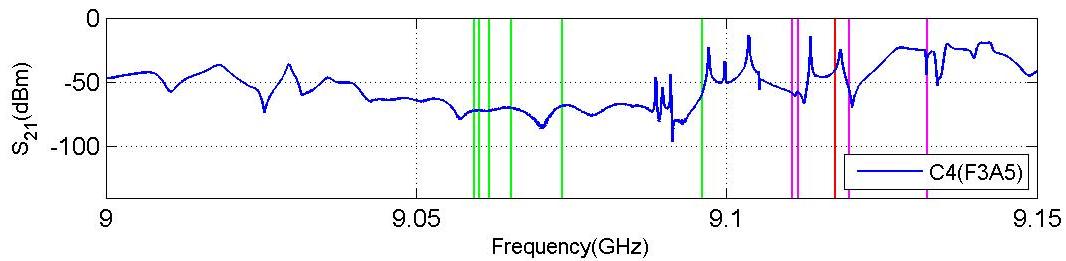}
\label{fnal-D5-C4}
}
\caption{Details of plots in Fig.~\ref{fnal-full-spec-all}: the f\mbox{}ifth dipole band.}
\label{fnal-D5-all}
\end{figure}

After fabrication, each 3.9~GHz cavity was mounted on the test stand at Fermilab and cooled down. The measurements were made on a vertical test stand for C1, C2 and C4, while C3 was on a horizontal test stand. C3 has been tuned to the accelerating mode 3.9~GHz, while C1 and C2 have not. The tuning status of C4 is not clear. In the vertical test stand, to both beam pipe f{}langes conical transitions ($\sim$40~mm long) were attached in order to form a diameter of 40~mm of the cavity beam pipe to the diameter of 30~mm. A f{}lange was attached to the end with HOM coupler 2. A pumping port was attached to the other end with power coupler. In the horizontal test stand, a blank f{}lange was attached to the end with HOM coupler 2, while a vacuum pumping port was attached to the other end with power coupler. Fig.~\ref{setup-fnal} shows schematically how the transmission measurement was set up on a vertical test stand. The transmission scattering parameter $S_{21}$ was subsequently measured for each single cavity. C1, C2 and C4 have been measured up to 10~GHz, while only the spectra data of up to 6~GHz is available for C3 \cite{racc39-fnal-1,racc39-fnal-2}. Measurement data are kindly provided by T.~Khabibouline from Fermilab. Fig.~\ref{fnal-full-spec-all}--\ref{fnal-D5-all} are transmission spectra from single cavity measurements at Fermilab. 


\FloatBarrier
\section{Mode Identif{}ication in the Single Cavity Spectra}\label{app-spec:pk}
The modes in each spectrum are f{}it globally. The software used for f{}itting is PeakFit\textregistered \cite{rpeakfit}. In some cases, more than two peaks have been found for a peak-shaped piece of spectrum, not all of these peaks are necessarily modes. We believe it comes from the fact that these peaks tend to f{}it the shape of the spectrum, which might be biased. The results are presented in the following f{}igures and tables. The simulated $Q_{ext}$ are from \cite{racc39-1}.

\begin{figure}\center
\subfigure[The f\mbox{}irst dipole band (C1)]{
\includegraphics[width=0.8\textwidth]{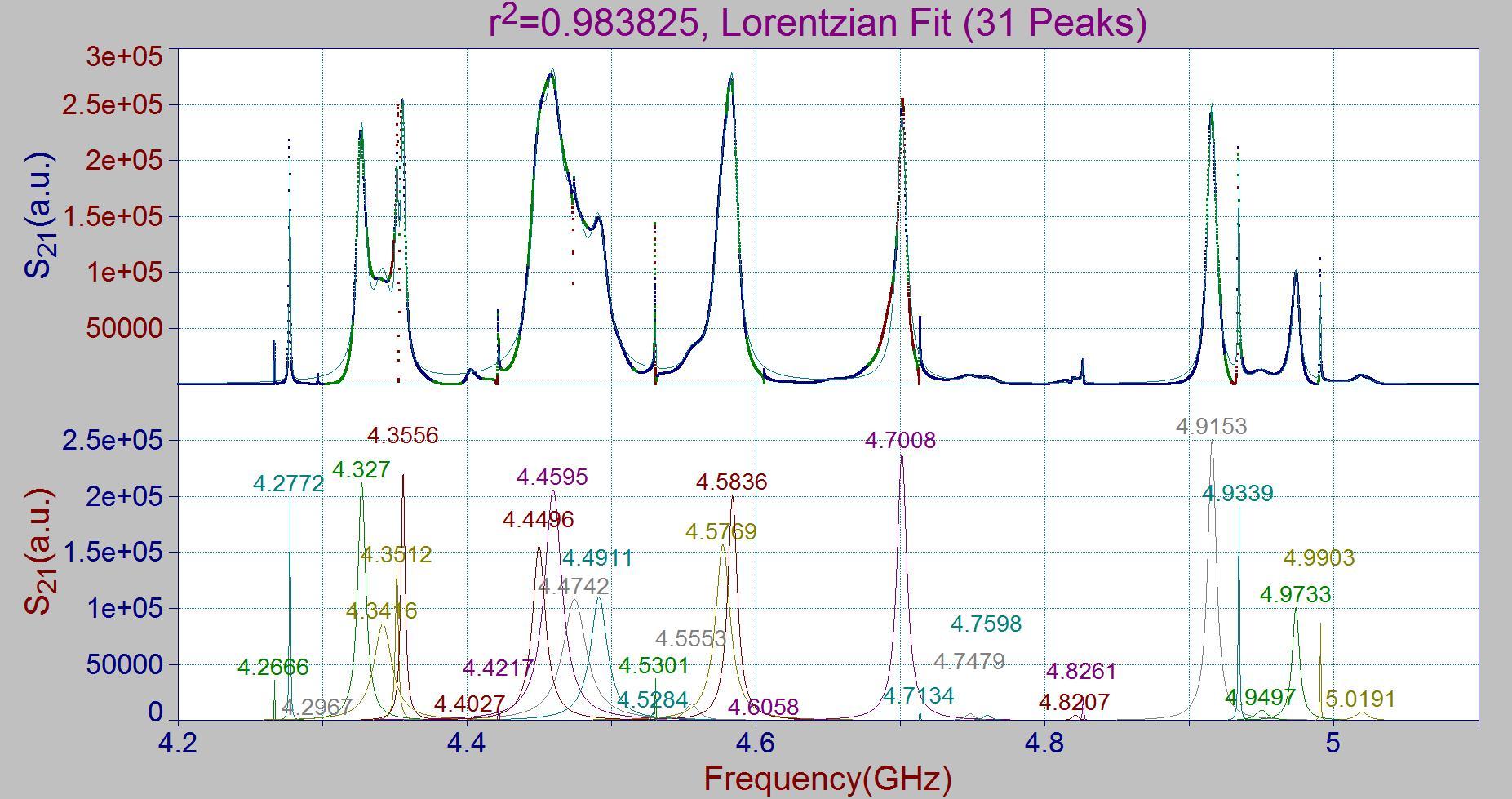}
\label{fnalc1_D1_PF}
}
\subfigure[The second dipole band (C1)]{
\includegraphics[width=0.8\textwidth]{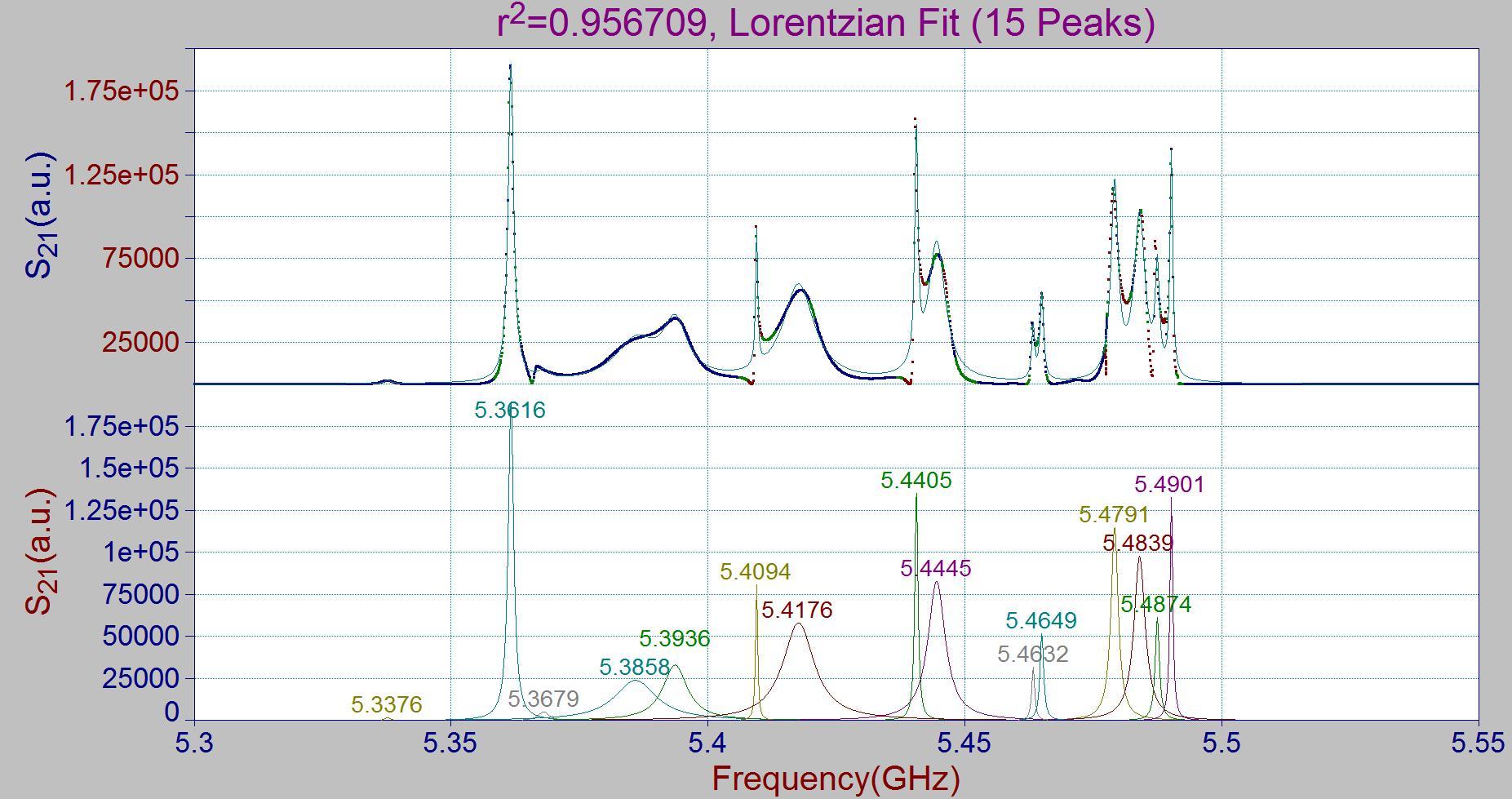}
\label{fnalc1_D2_PF}
}
\subfigure[The f\mbox{}irst dipole band and beampipe modes (C1)]{
\includegraphics[width=0.45\textwidth]{simu-fnal-C1-part1}
\label{simu-fnal-C1-part1}
}
\quad
\subfigure[The second dipole band and beampipe modes (C1)]{
\includegraphics[width=0.45\textwidth]{simu-fnal-C1-part2}
\label{simu-fnal-C1-part2}
}
\caption{Lorentzian f\mbox{}it of the f\mbox{}irst two dipole bands and beampipe modes of C1 from single cavity measurement at Fermilab.}
\label{fnalc1_D1D2_PF_simu}
\end{figure}
\begin{figure}\center
\includegraphics[width=0.8\textwidth]{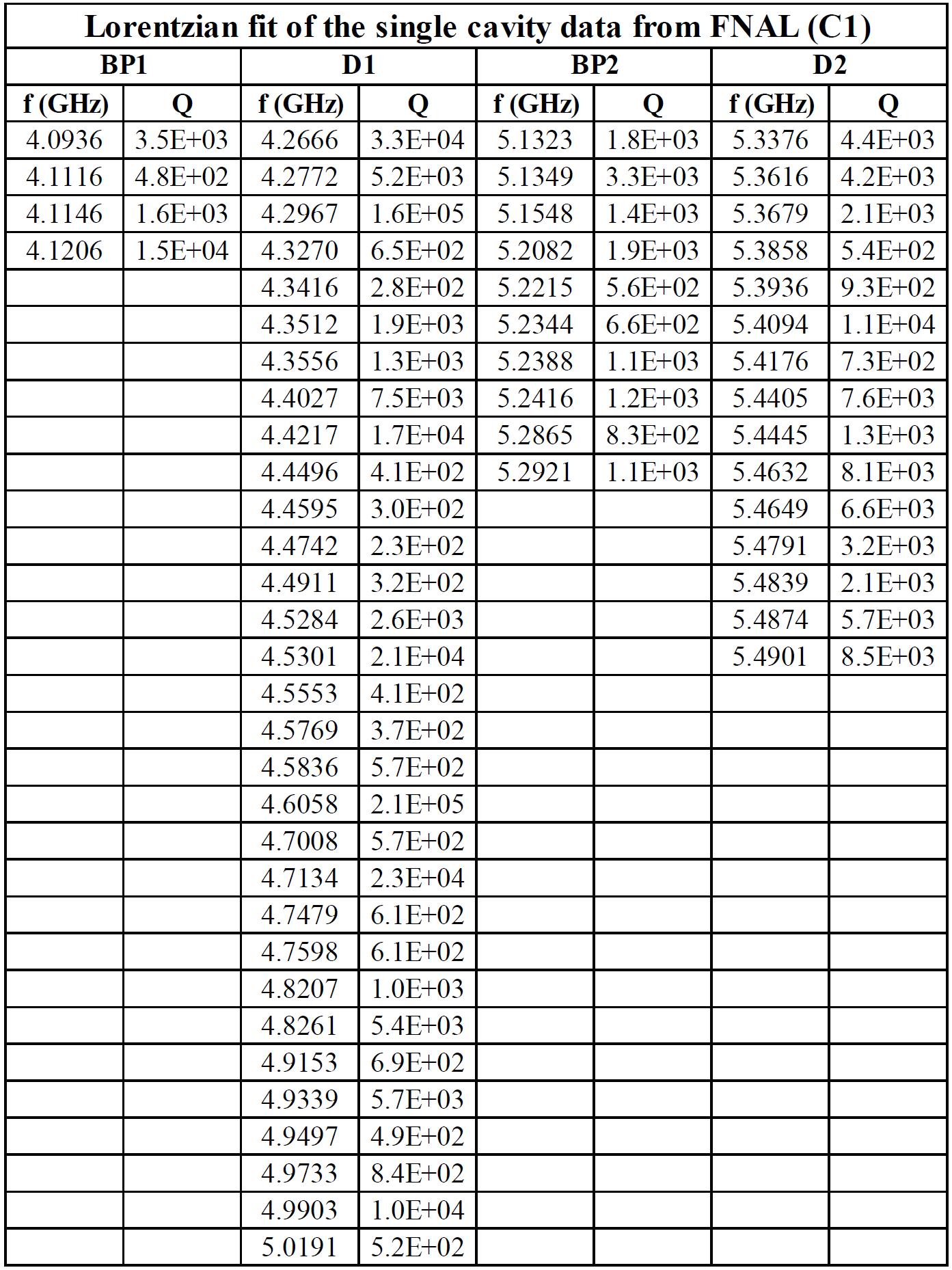}
\caption{Table of peaks in Fig.~\ref{fnalc1_D1D2_PF_simu}.}
\label{simu-fnal-C1-table}
\end{figure}

\begin{figure}\center
\subfigure[The f\mbox{}irst dipole band (C2)]{
\includegraphics[width=0.8\textwidth]{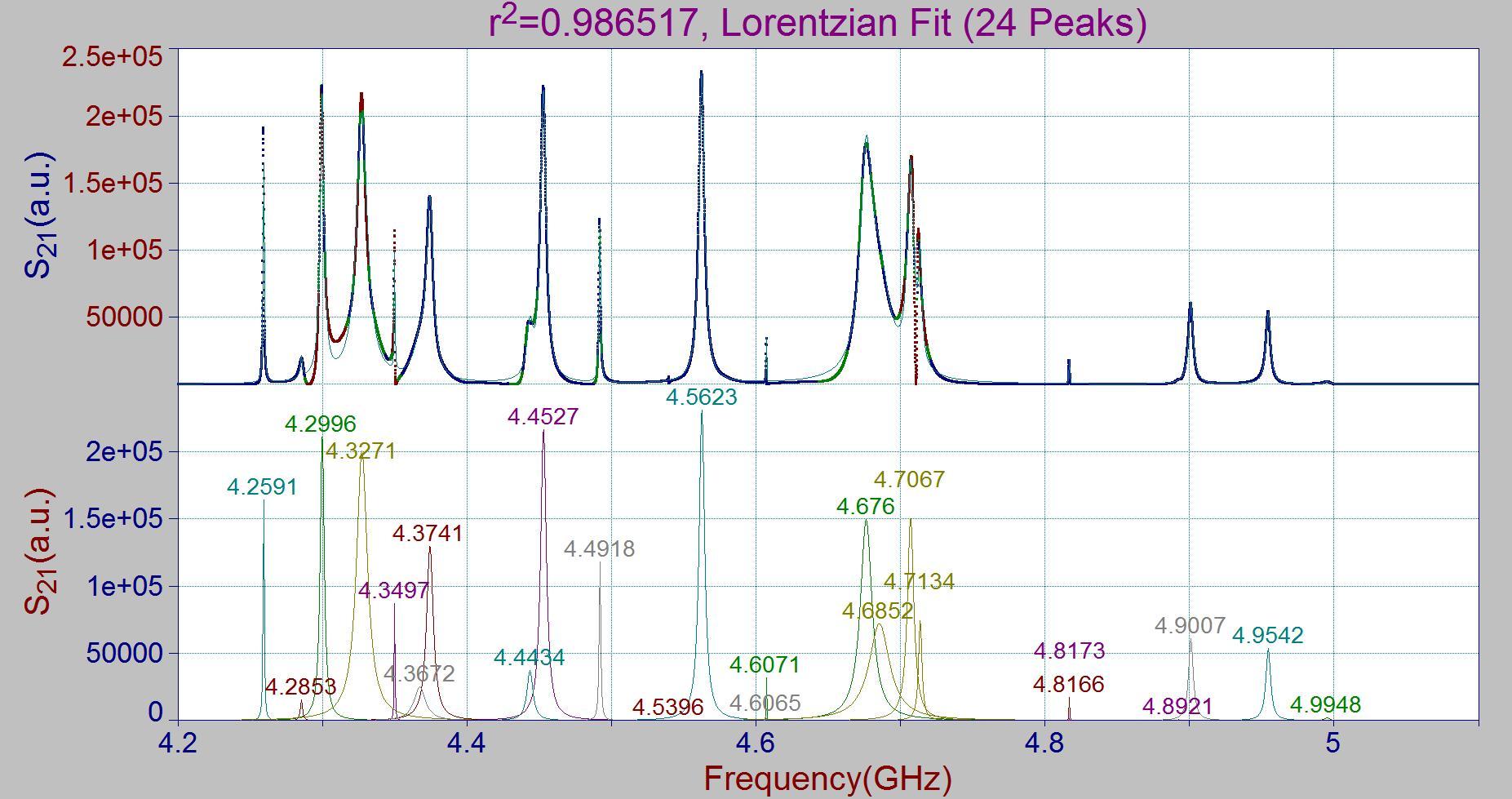}
\label{fnalc2_D1_PF}
}
\subfigure[The second dipole band (C2)]{
\includegraphics[width=0.8\textwidth]{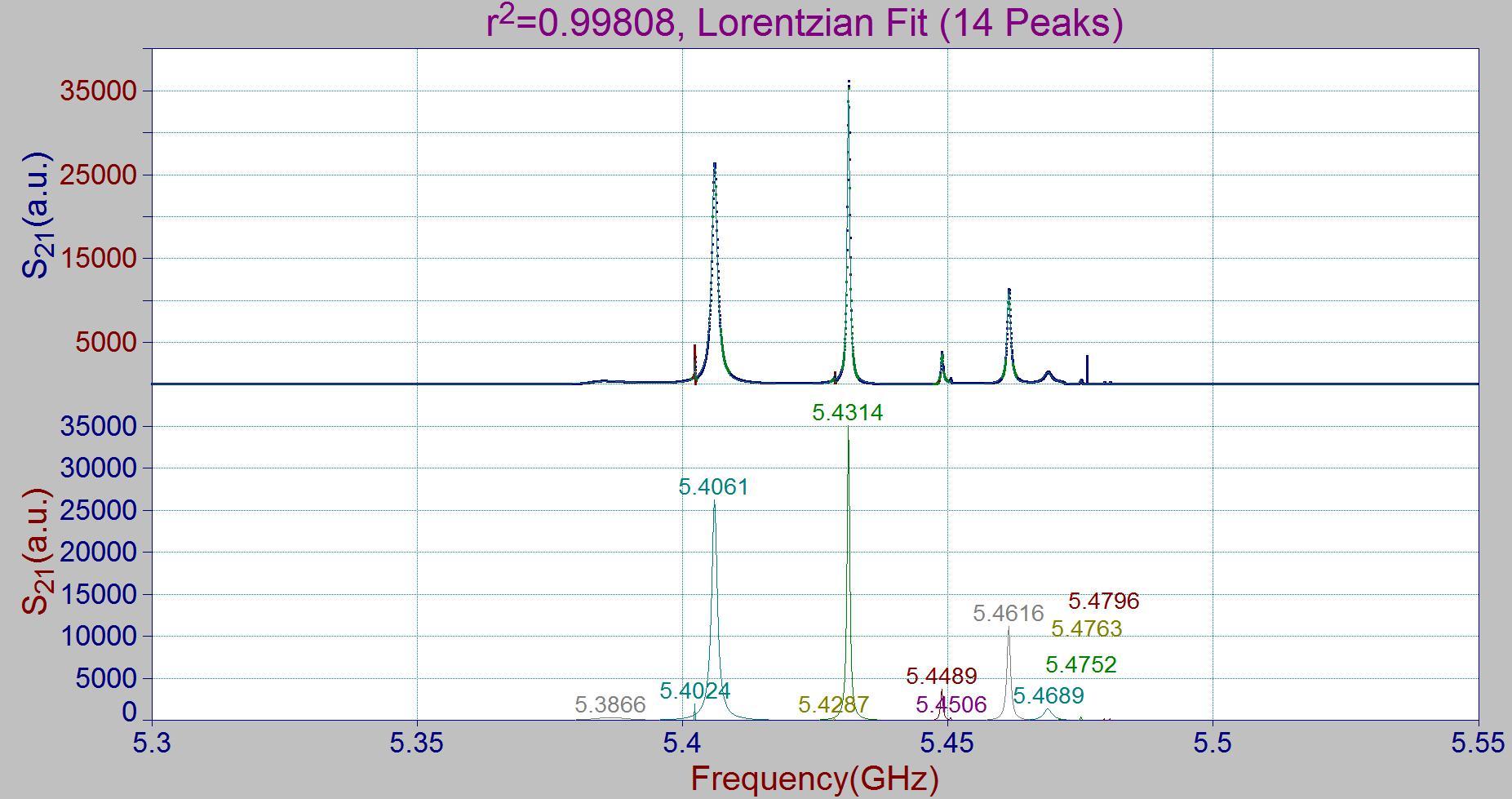}
\label{fnalc2_D2_PF}
}
\subfigure[The f\mbox{}irst dipole band and beampipe modes (C2)]{
\includegraphics[width=0.45\textwidth]{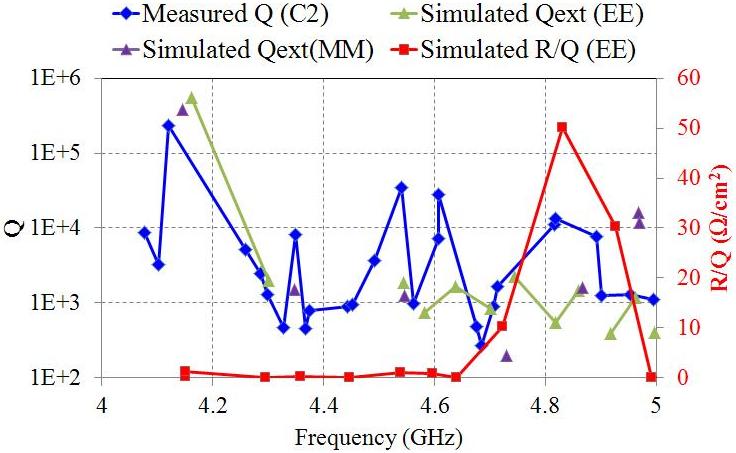}
\label{simu-fnal-C2-part1}
}
\quad
\subfigure[The second dipole band and beampipe modes (C2)]{
\includegraphics[width=0.45\textwidth]{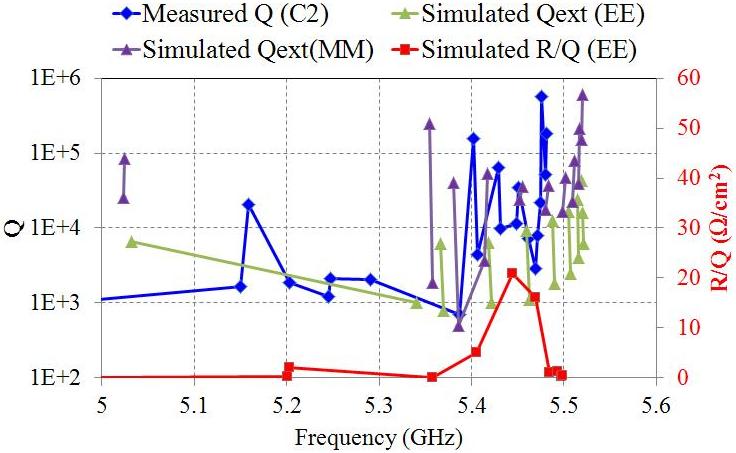}
\label{simu-fnal-C2-part2}
}
\caption{Lorentzian f\mbox{}it of the f\mbox{}irst two dipole bands and beampipe modes of C2 from single cavity measurement at Fermilab.}
\label{fnalc2_D1D2_PF_simu}
\end{figure}
\begin{figure}\center
\includegraphics[width=0.8\textwidth]{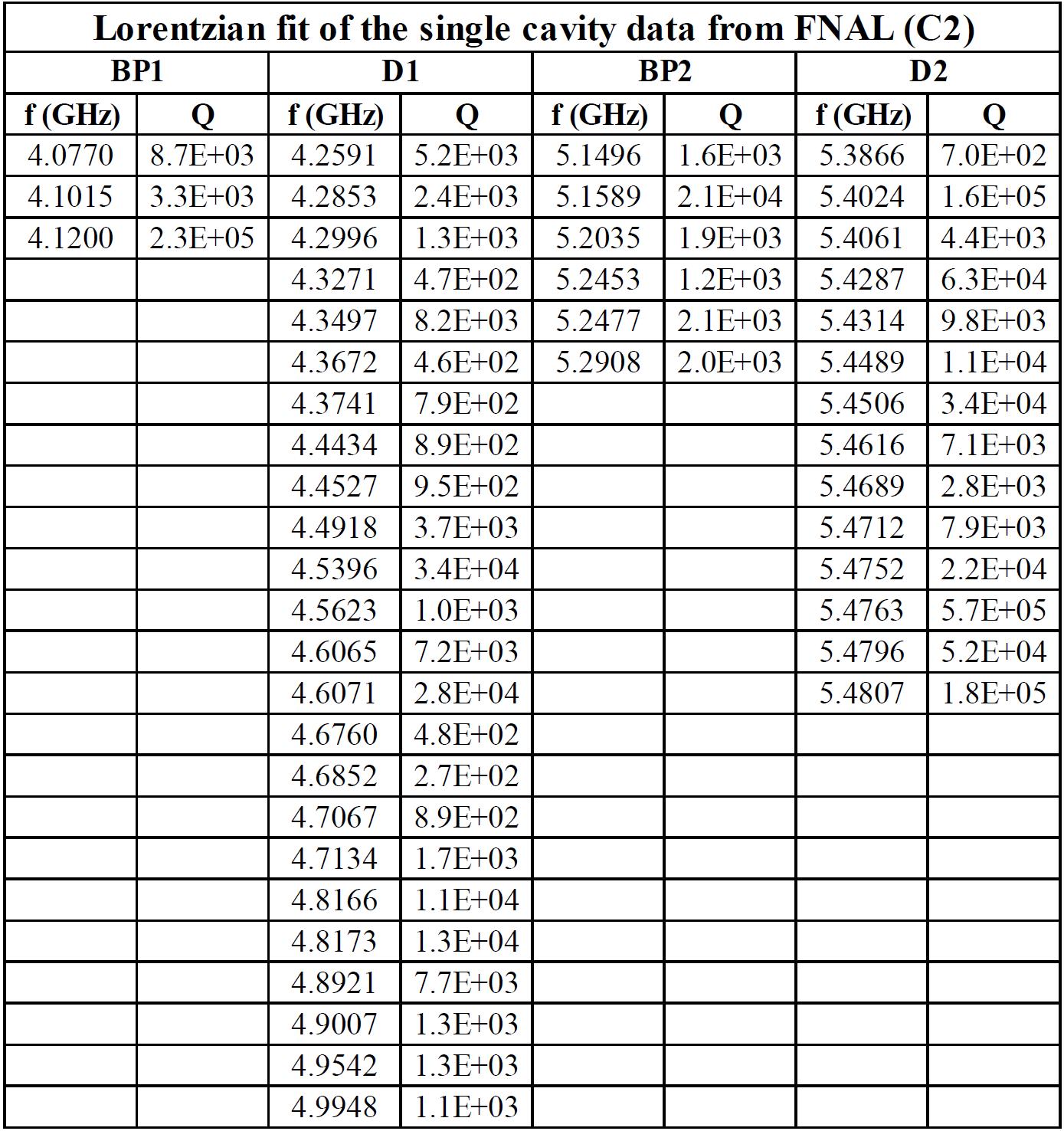}
\caption{Table of peaks in Fig.~\ref{fnalc2_D1D2_PF_simu}.}
\label{simu-fnal-C2-table}
\end{figure}

\begin{figure}\center
\subfigure[The f\mbox{}irst dipole band]{
\includegraphics[width=0.8\textwidth]{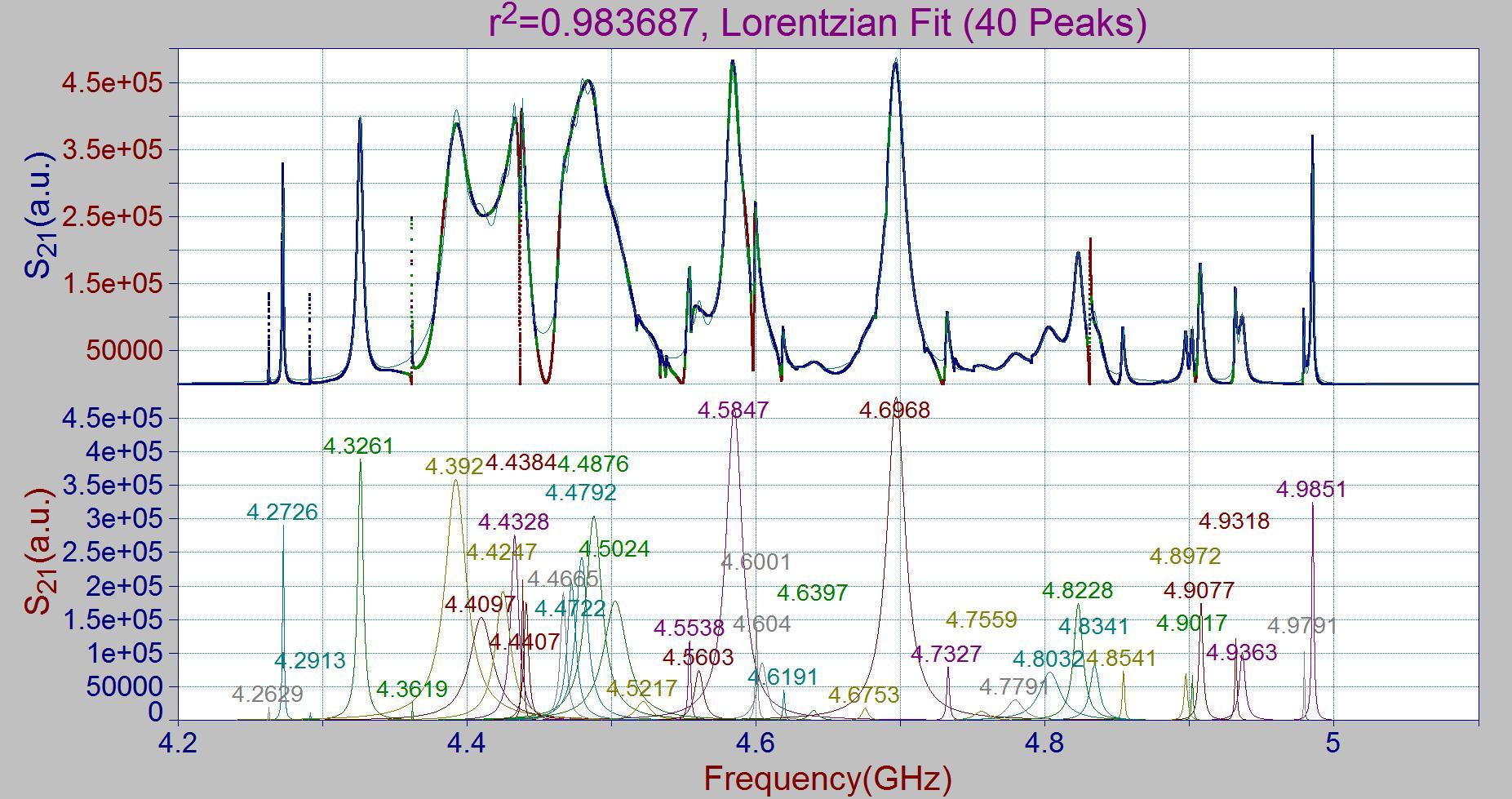}
\label{fnalc3_D1_PF}
}
\subfigure[The second dipole band]{
\includegraphics[width=0.8\textwidth]{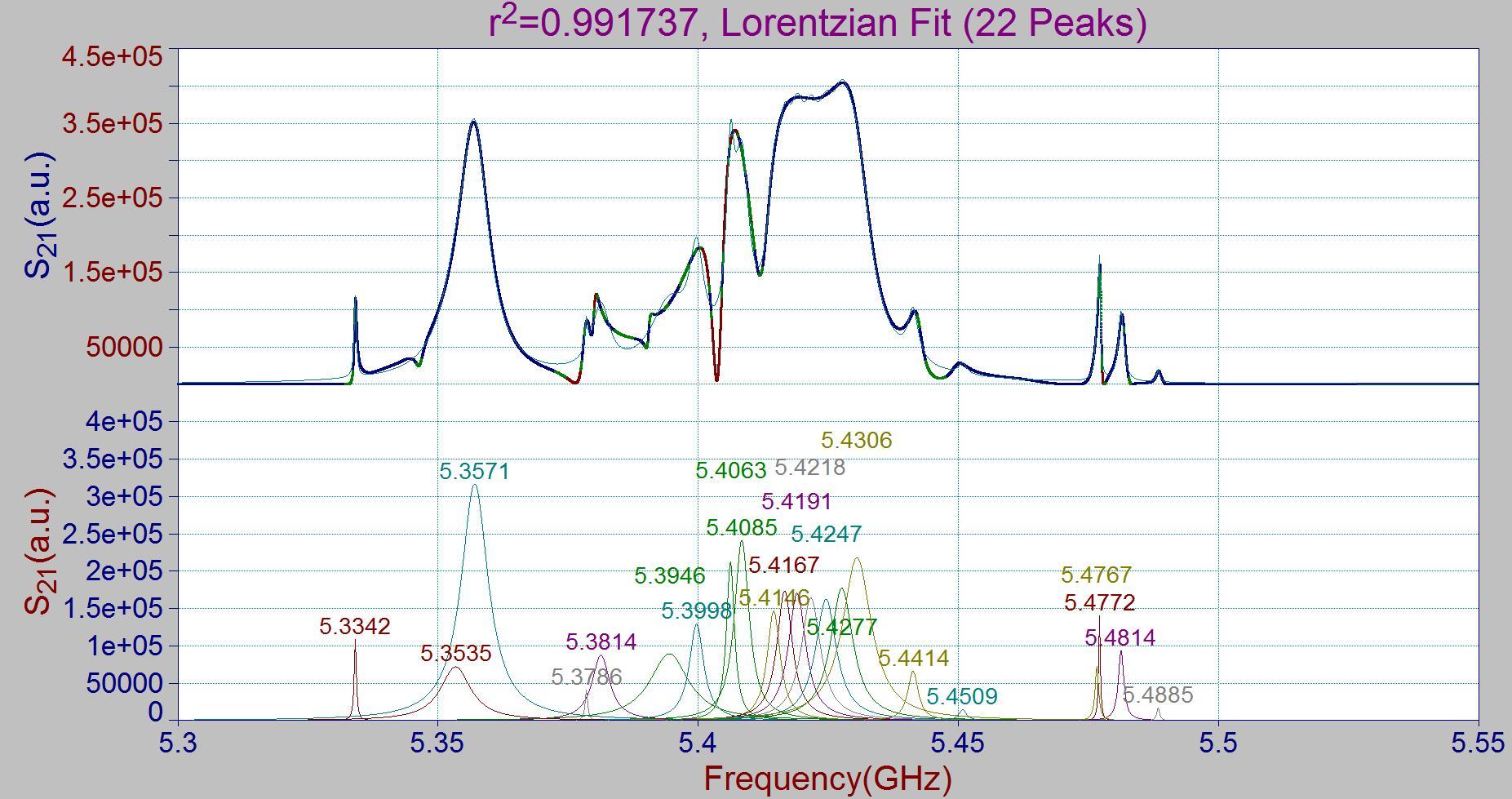}
\label{fnalc3_D2_PF}
}
\subfigure[The f\mbox{}irst dipole band and beampipe modes (C3)]{
\includegraphics[width=0.45\textwidth]{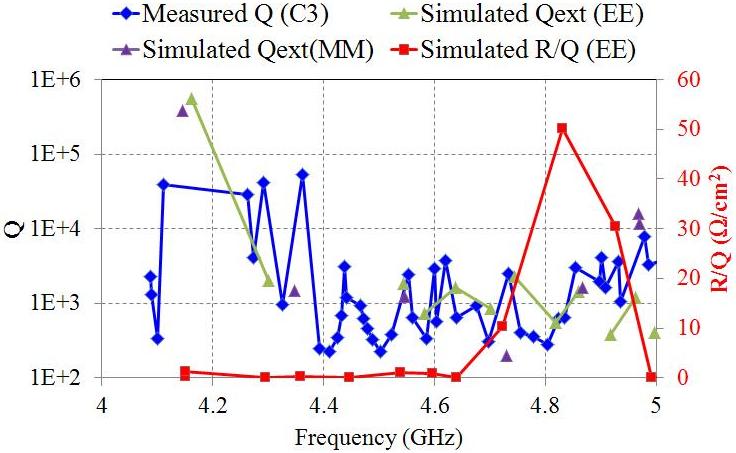}
\label{simu-fnal-C3-part1}
}
\quad
\subfigure[The second dipole band and beampipe modes (C3)]{
\includegraphics[width=0.45\textwidth]{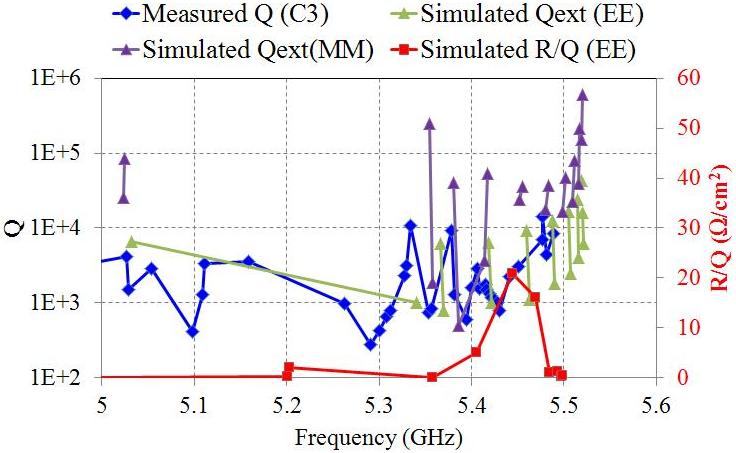}
\label{simu-fnal-C3-part2}
}
\caption{Lorentzian f\mbox{}it of the f\mbox{}irst two dipole bands and beampipe modes of C3 from single cavity measurement at Fermilab.}
\label{fnalc3_D1D2_PF_simu}
\end{figure}
\begin{figure}\center
\includegraphics[width=0.8\textwidth]{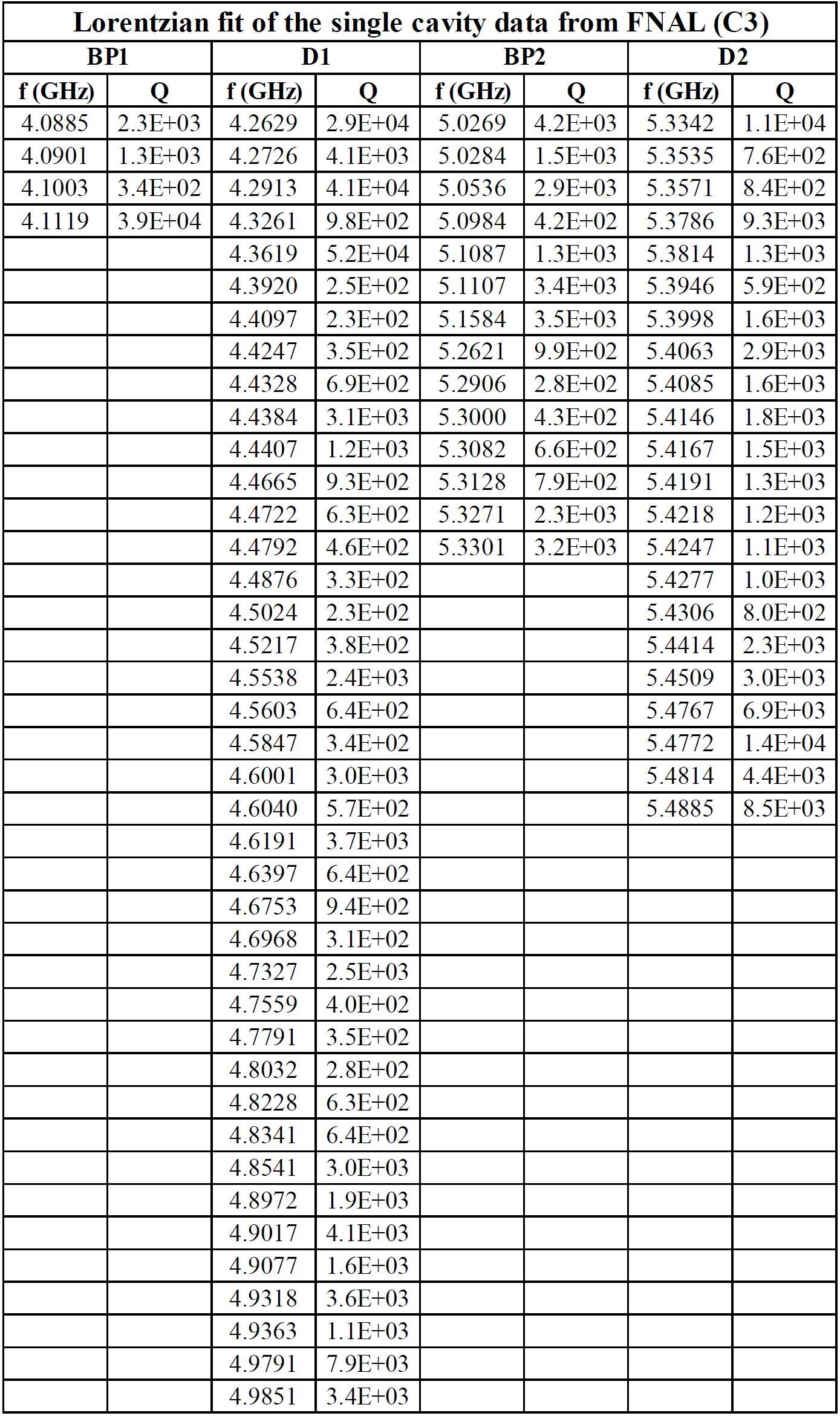}
\caption{Table of peaks in Fig.~\ref{fnalc3_D1D2_PF_simu}.}
\label{simu-fnal-C3-table}
\end{figure}

\begin{figure}\center
\subfigure[The f\mbox{}irst dipole band (C4)]{
\includegraphics[width=0.8\textwidth]{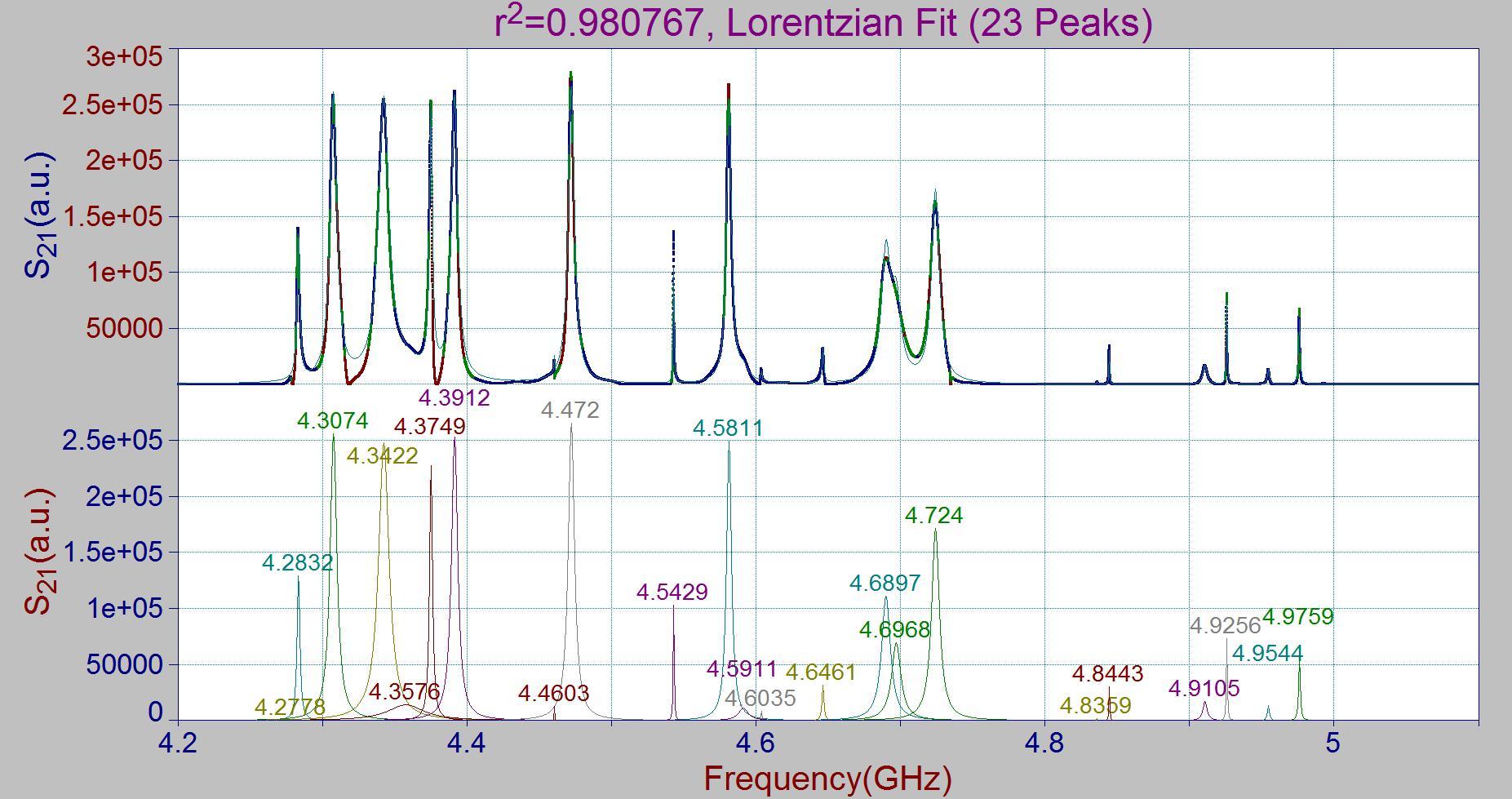}
\label{fnalc4_D1_PF}
}
\subfigure[The second dipole band (C4)]{
\includegraphics[width=0.8\textwidth]{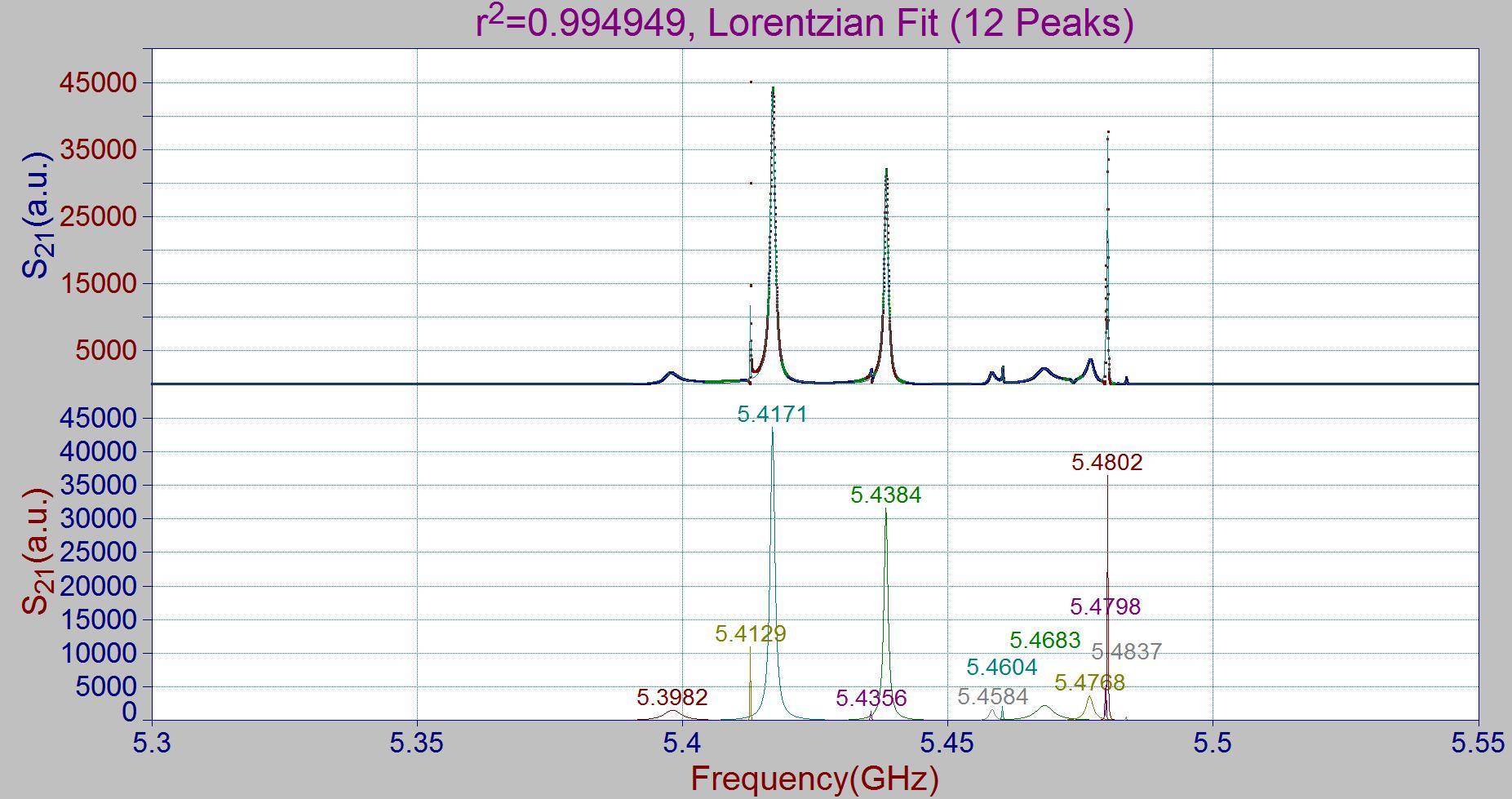}
\label{fnalc4_D2_PF}
}
\subfigure[The f\mbox{}irst dipole band and beampipe modes (C4)]{
\includegraphics[width=0.45\textwidth]{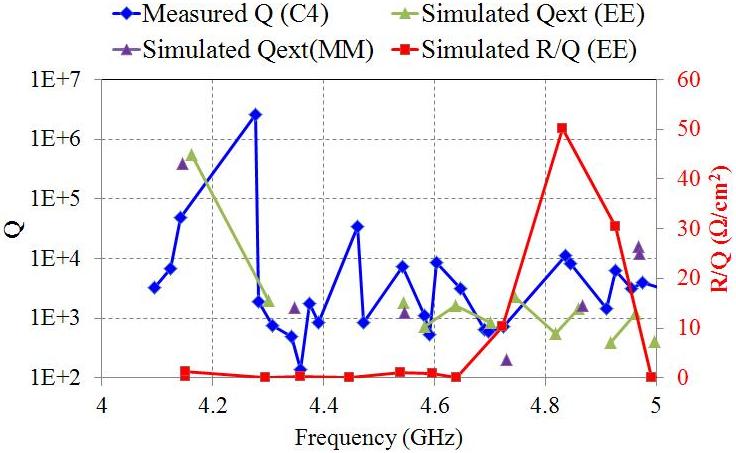}
\label{simu-fnal-C4-part1}
}
\quad
\subfigure[The second dipole band and beampipe modes (C4)]{
\includegraphics[width=0.45\textwidth]{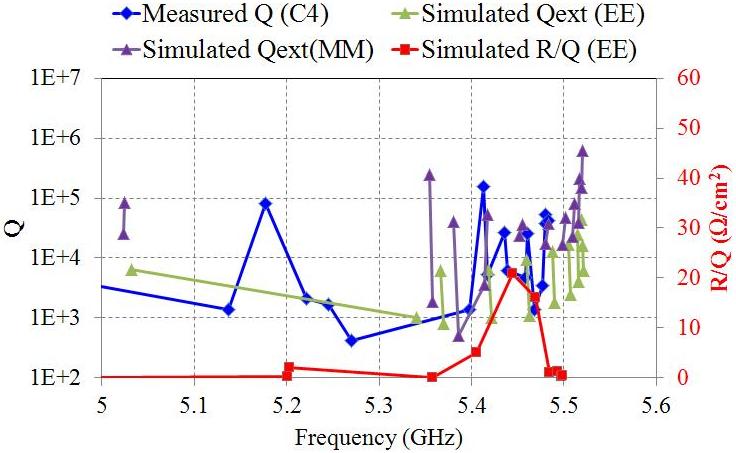}
\label{simu-fnal-C4-part2}
}
\caption{Lorentzian f\mbox{}it of the f\mbox{}irst two dipole bands and beampipe modes of C4 from single cavity measurement at Fermilab.}
\label{fnalc4_D1D2_PF_simu}
\end{figure}
\begin{figure}\center
\includegraphics[width=0.8\textwidth]{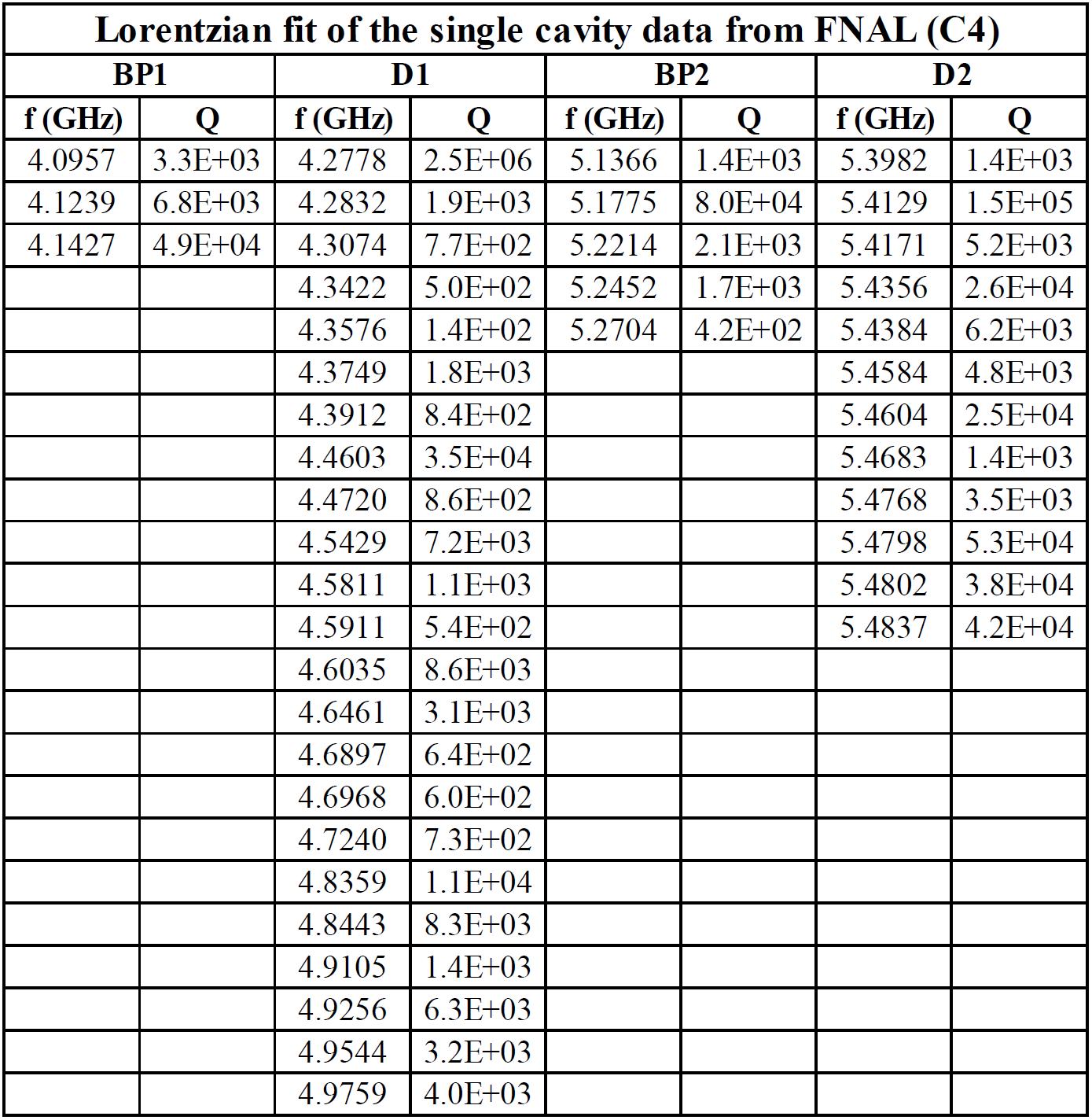}
\caption{Table of peaks in Fig.~\ref{fnalc4_D1D2_PF_simu}.}
\label{simu-fnal-C4-table}
\end{figure}

\begin{figure}\center
\subfigure[The f\mbox{}ifth dipole band (C1) (part1)]{
\includegraphics[width=0.8\textwidth]{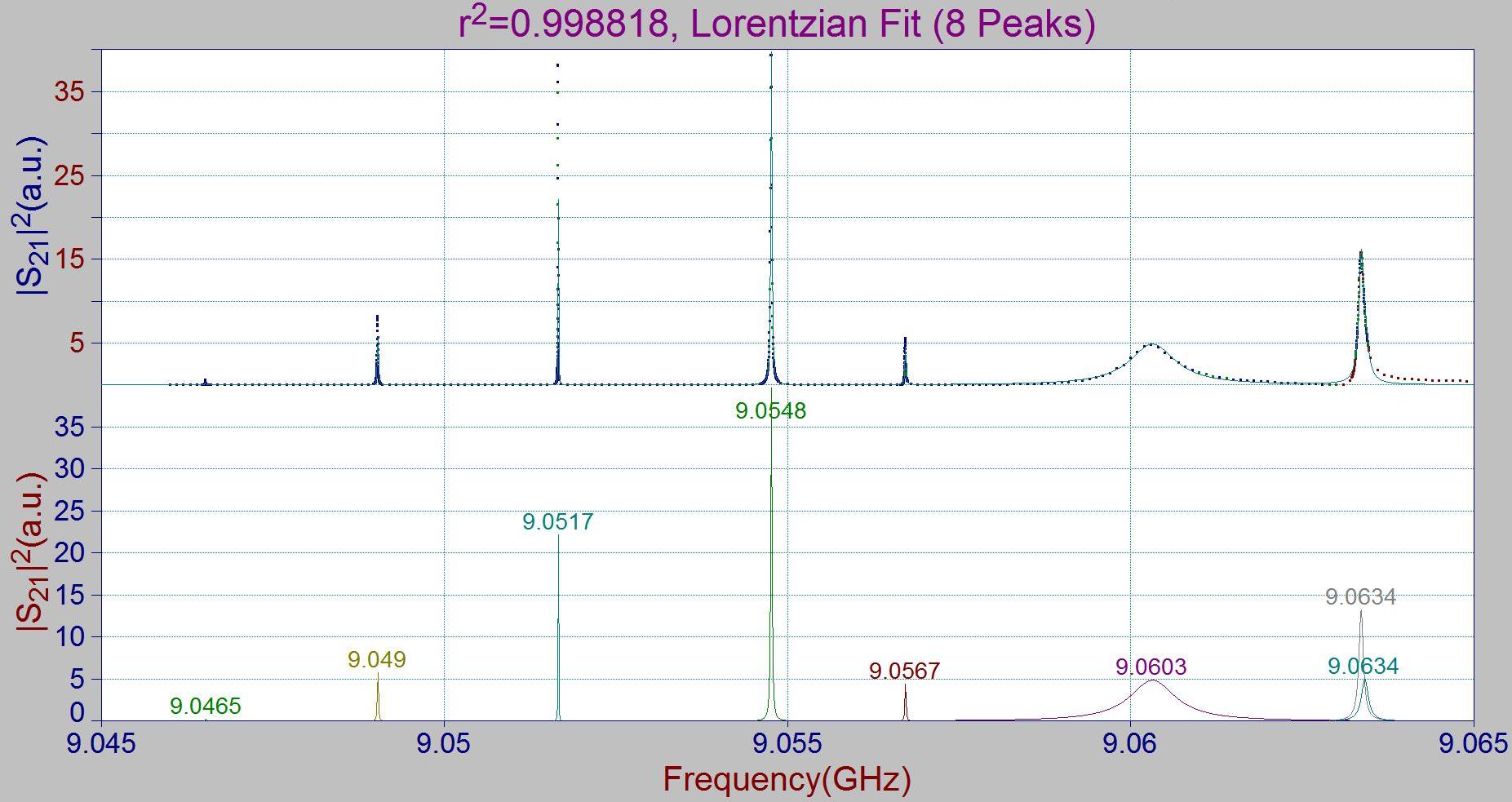}
\label{fnalc1_D5_1_PF}
}
\subfigure[The f\mbox{}ifth dipole band (C1) (part2)]{
\includegraphics[width=0.8\textwidth]{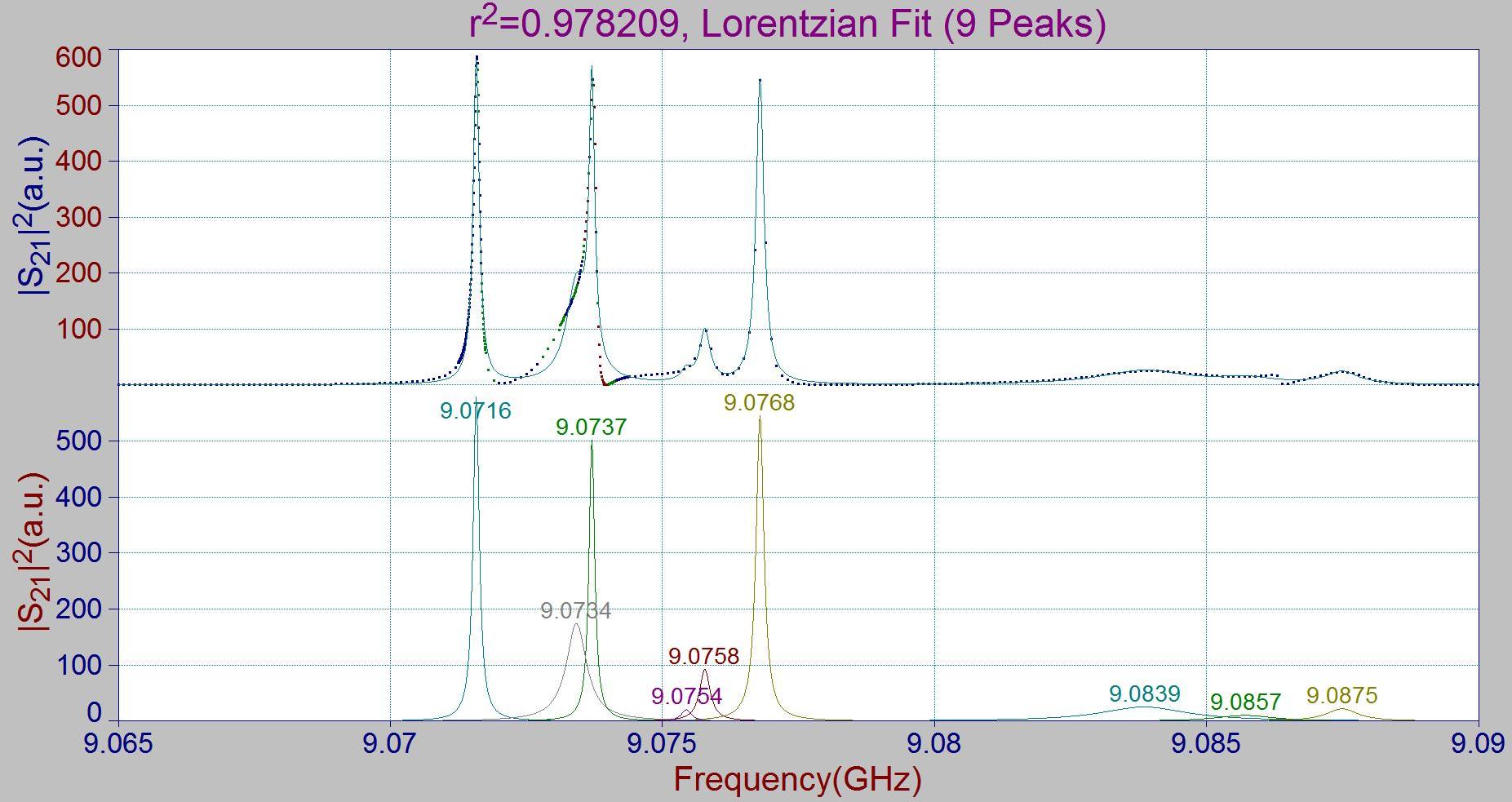}
\label{fnalc1_D5_2_PF}
}
\subfigure[The f\mbox{}ifth dipole band (C1)]{
\includegraphics[width=0.7\textwidth]{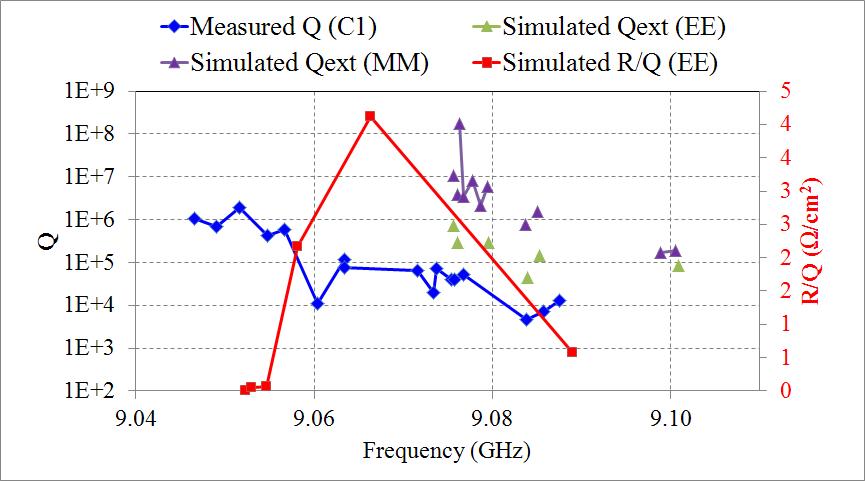}
\label{simu-fnal-C1-D5}
}
\caption{Lorentzian f\mbox{}it of the f\mbox{}ifth dipole band of C1 from single cavity measurement at Fermilab.}
\label{fnalc1_D5_PF_simu}
\end{figure}
\begin{figure}\center
\includegraphics[width=0.5\textwidth]{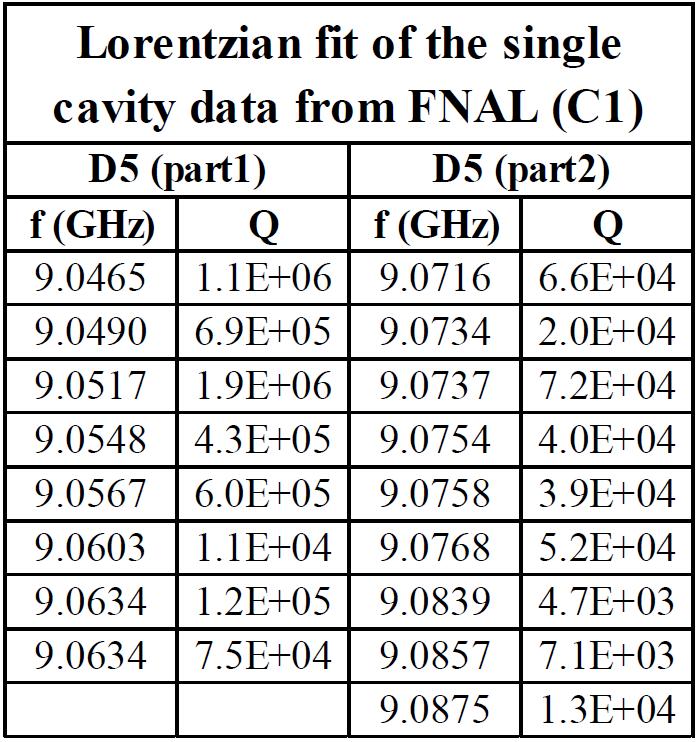}
\caption{Table of peaks in Fig.~\ref{fnalc1_D5_PF_simu}.}
\label{simu-fnal-C1-D5-table}
\end{figure}

\begin{figure}\center
\subfigure[The f\mbox{}ifth dipole band (C2) (part1)]{
\includegraphics[width=0.45\textwidth]{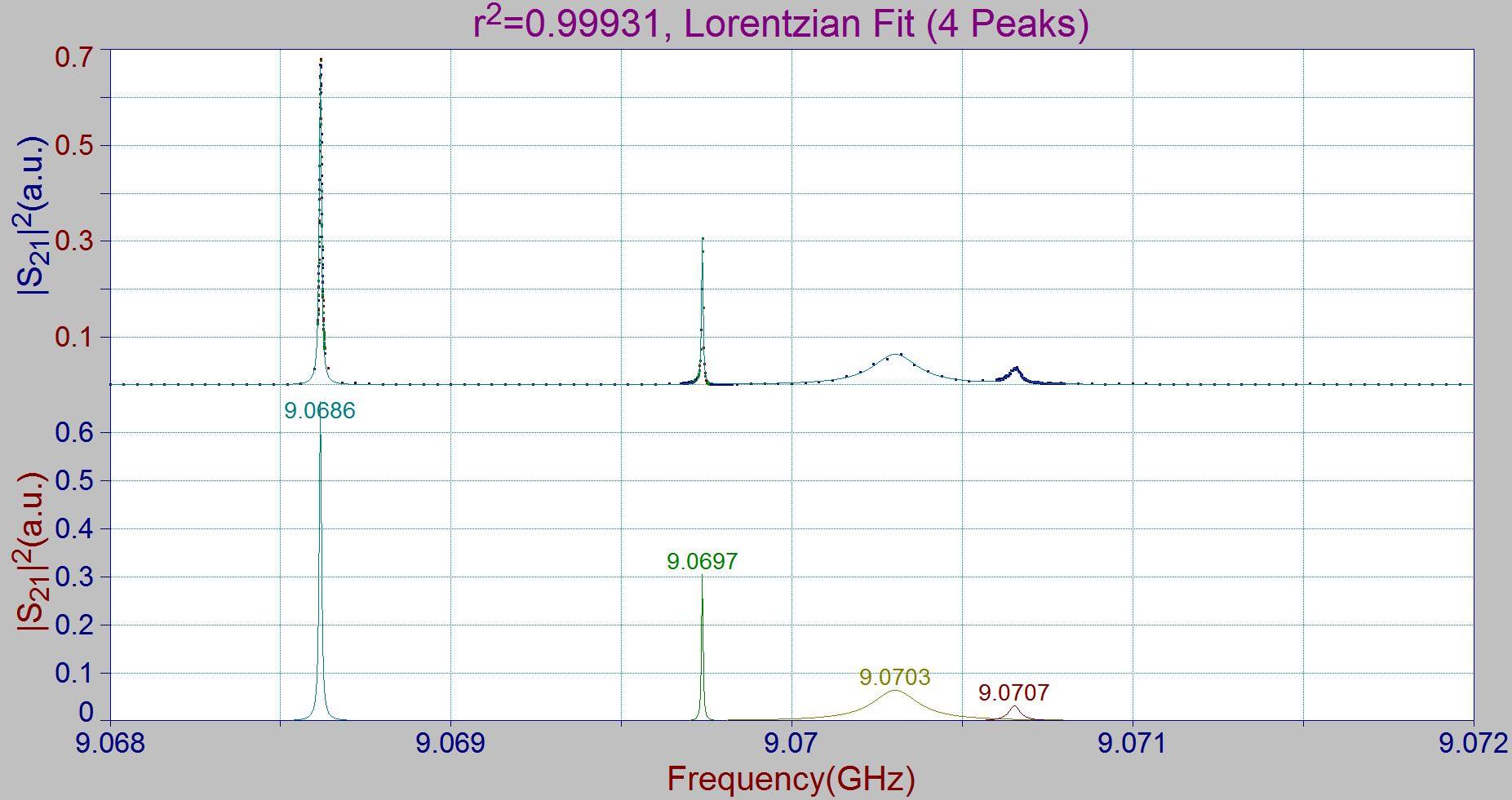}
\label{fnalc2_D5_1_PF}
}
\subfigure[The f\mbox{}ifth dipole band (C2) (part2)]{
\includegraphics[width=0.45\textwidth]{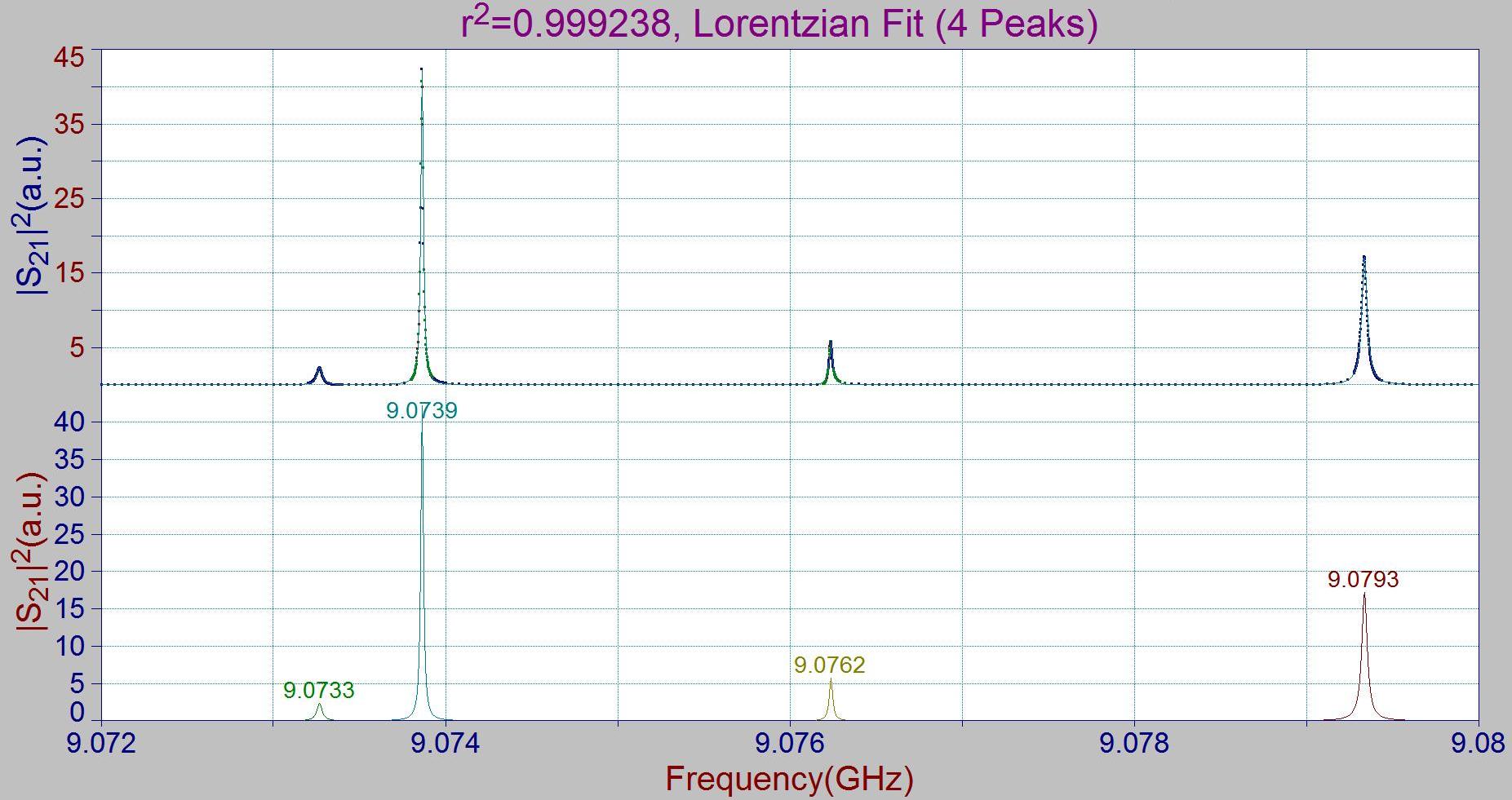}
\label{fnalc2_D5_2_PF}
}
\subfigure[The f\mbox{}ifth dipole band (C2) (part3)]{
\includegraphics[width=0.85\textwidth]{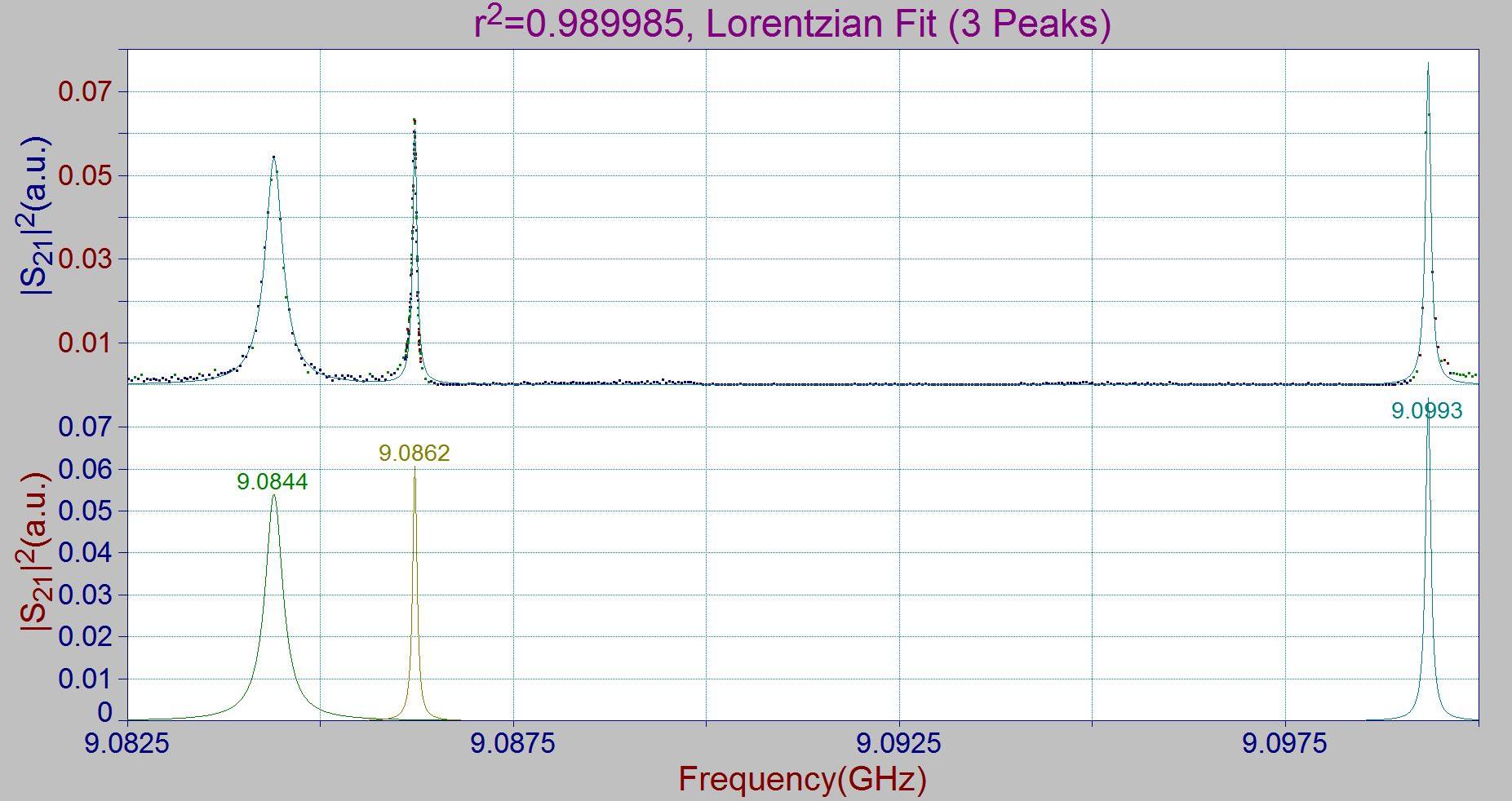}
\label{fnalc2_D5_3_PF}
}
\subfigure[The f\mbox{}ifth dipole band (C2)]{
\includegraphics[width=0.8\textwidth]{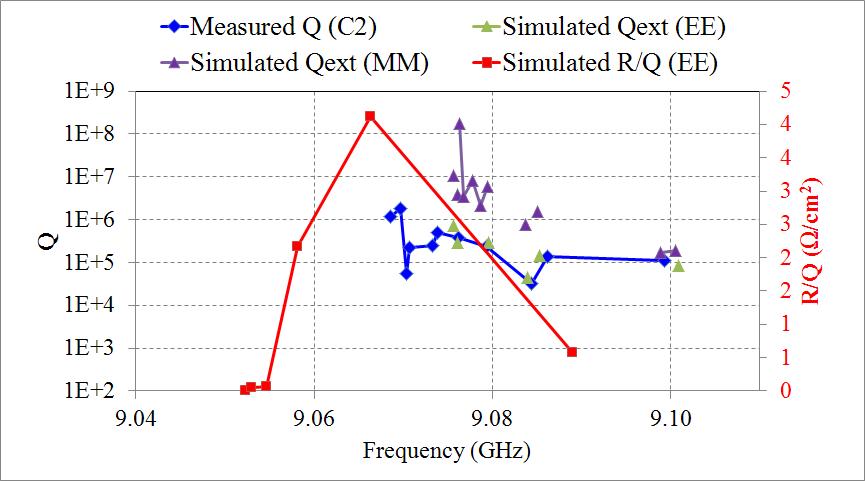}
\label{simu-fnal-C2-D5}
}
\caption{Lorentzian f\mbox{}it of the f\mbox{}ifth dipole band of C2 from single cavity measurement at Fermilab.}
\label{fnalc2_D5_PF_simu}
\end{figure}
\begin{figure}\center
\includegraphics[width=0.65\textwidth]{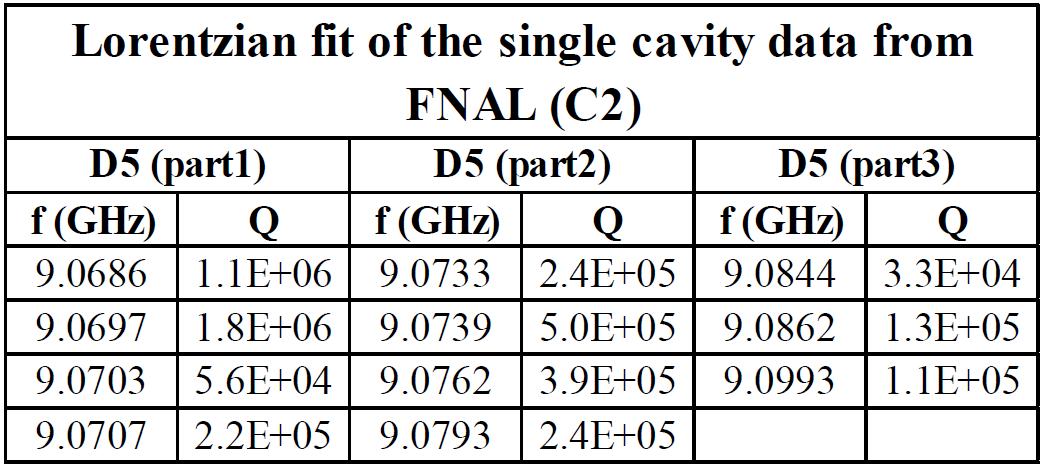}
\caption{Table of peaks in Fig.~\ref{fnalc2_D5_PF_simu}.}
\label{simu-fnal-C2-D5-table}
\end{figure}

\begin{figure}\center
\subfigure[The f\mbox{}ifth dipole band (C4) (part1)]{
\includegraphics[width=0.85\textwidth]{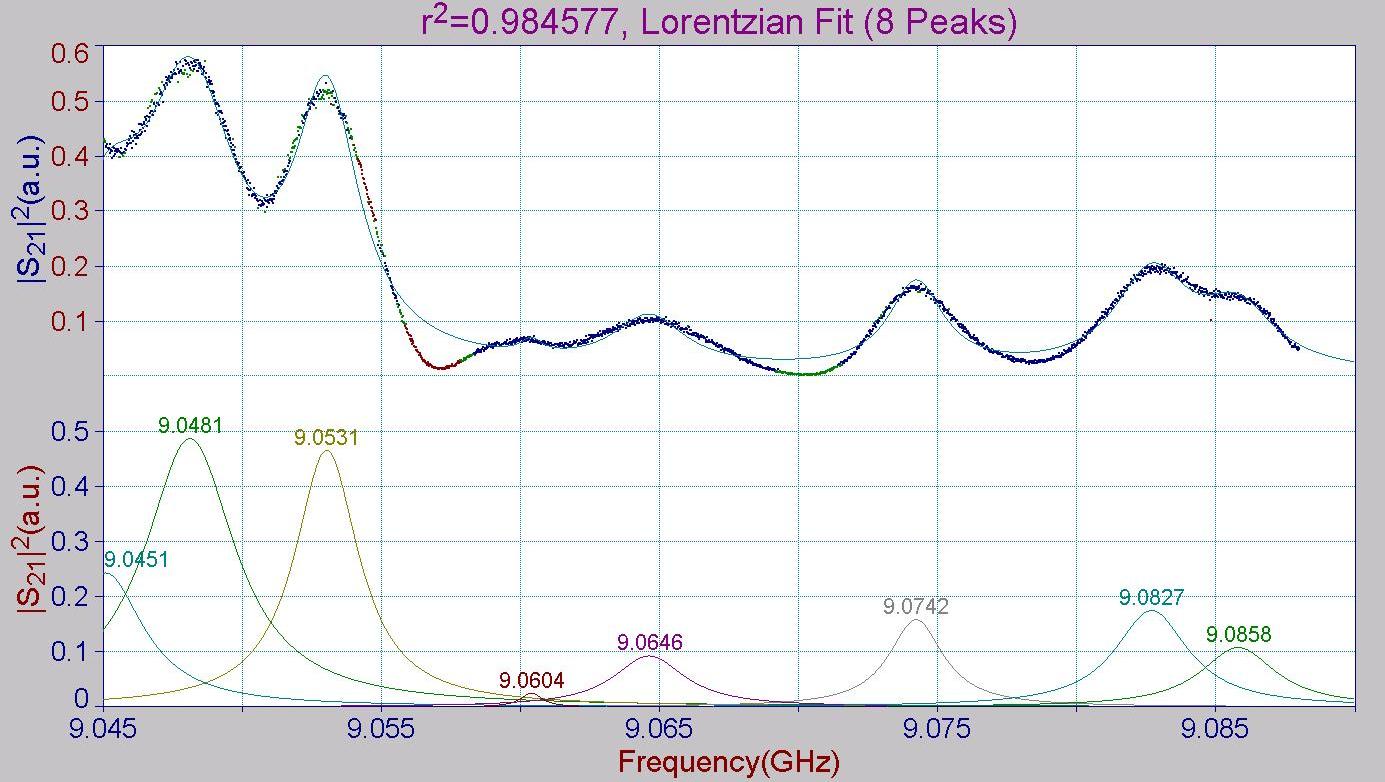}
\label{fnalc4_D5_1_PF}
}
\subfigure[The f\mbox{}ifth dipole band (C4) (part2)]{
\includegraphics[width=0.45\textwidth]{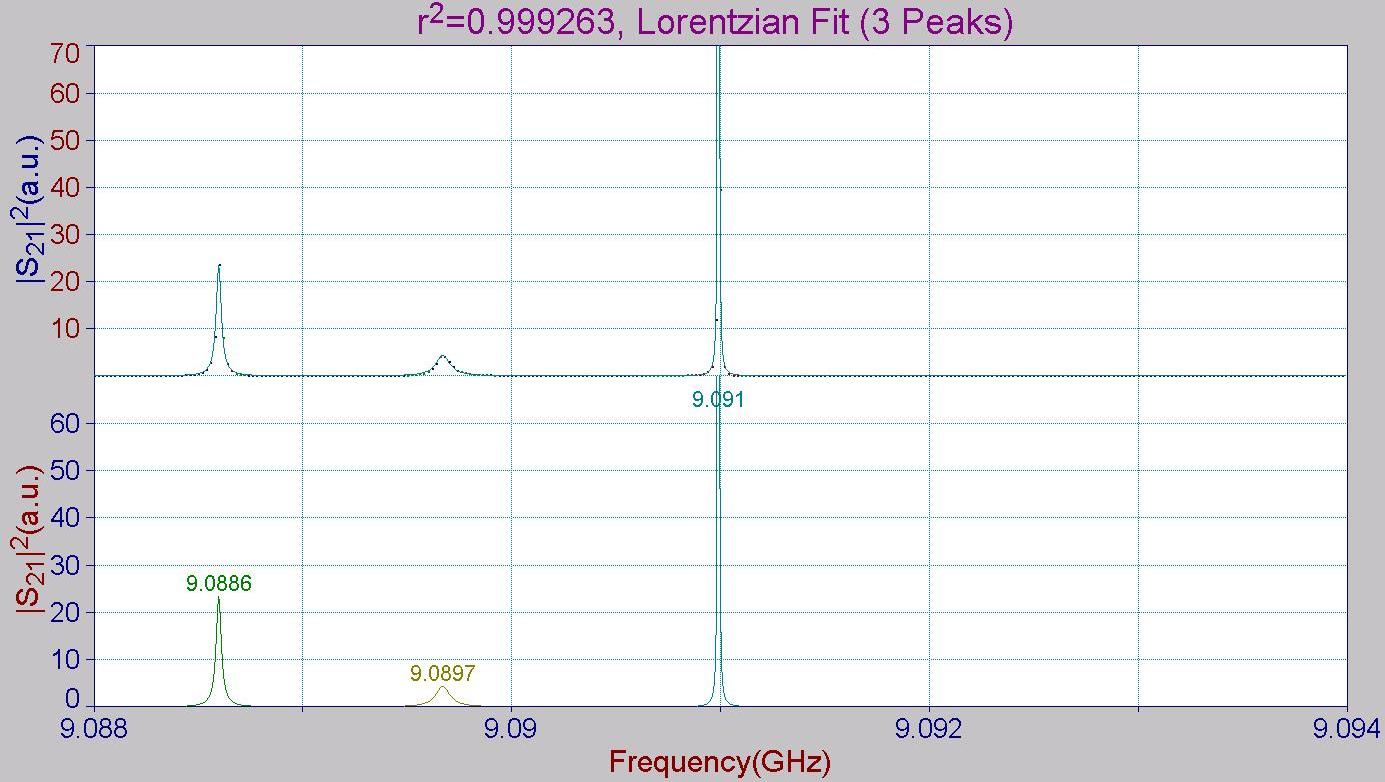}
\label{fnalc4_D5_2_PF}
}
\subfigure[The f\mbox{}ifth dipole band (C4) (part3)]{
\includegraphics[width=0.45\textwidth]{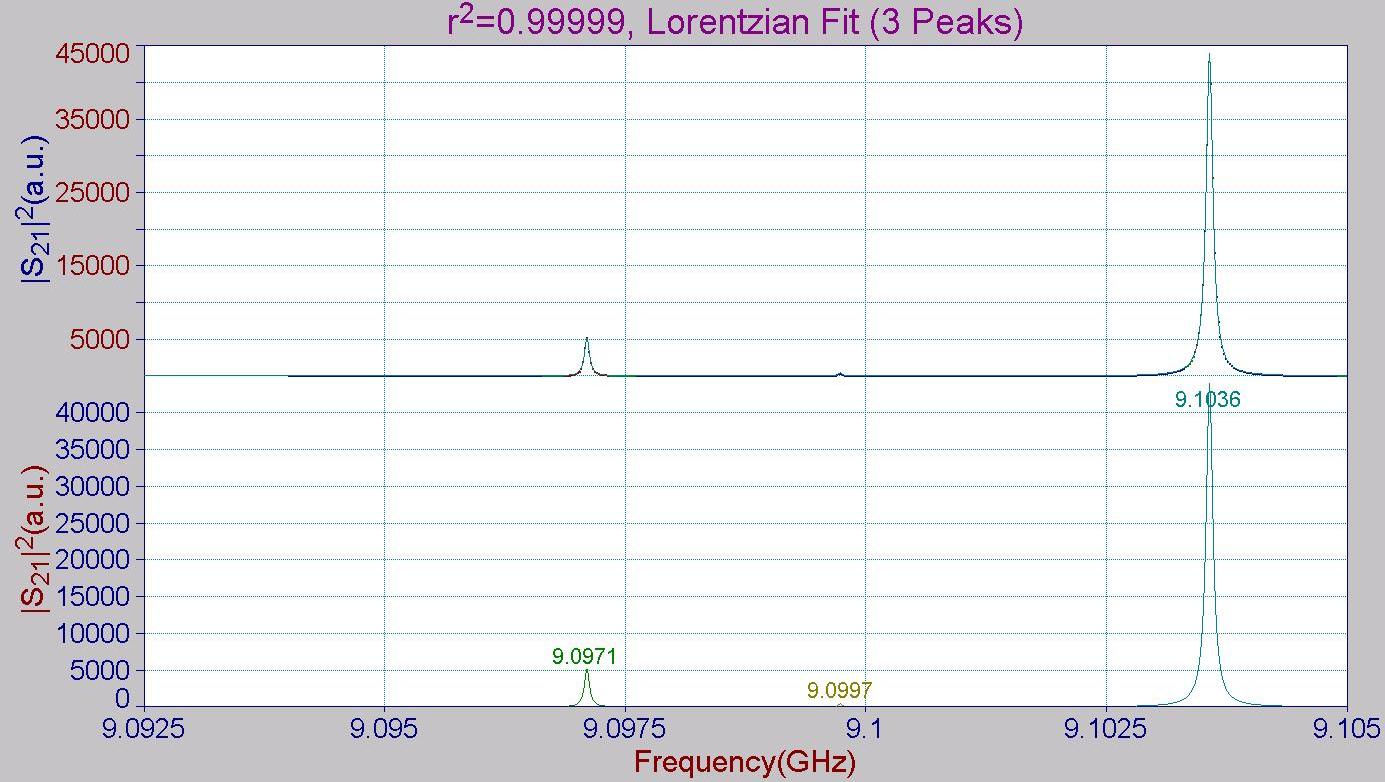}
\label{fnalc4_D5_3_PF}
}
\subfigure[The f\mbox{}ifth dipole band (C4)]{
\includegraphics[width=0.8\textwidth]{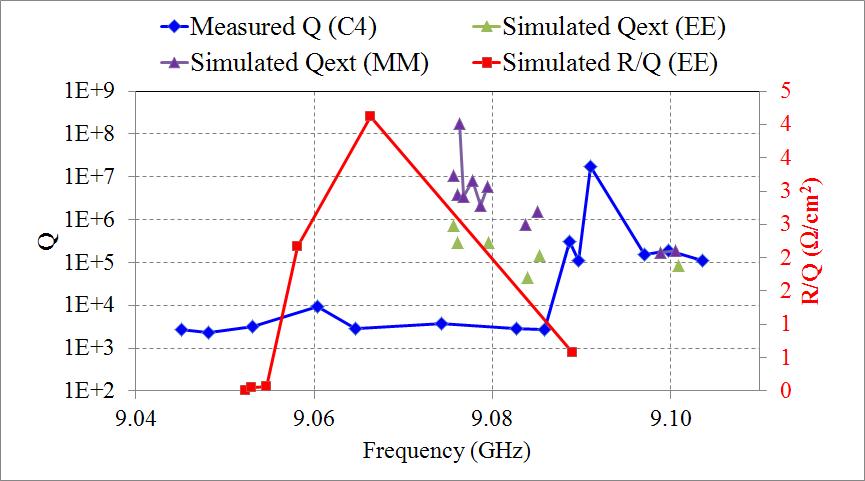}
\label{simu-fnal-C4-D5}
}
\caption{Lorentzian f\mbox{}it of the f\mbox{}ifth dipole band of C4 from single cavity measurement at Fermilab.}
\label{fnalc4_D5_PF_simu}
\end{figure}
\begin{figure}\center
\includegraphics[width=0.65\textwidth]{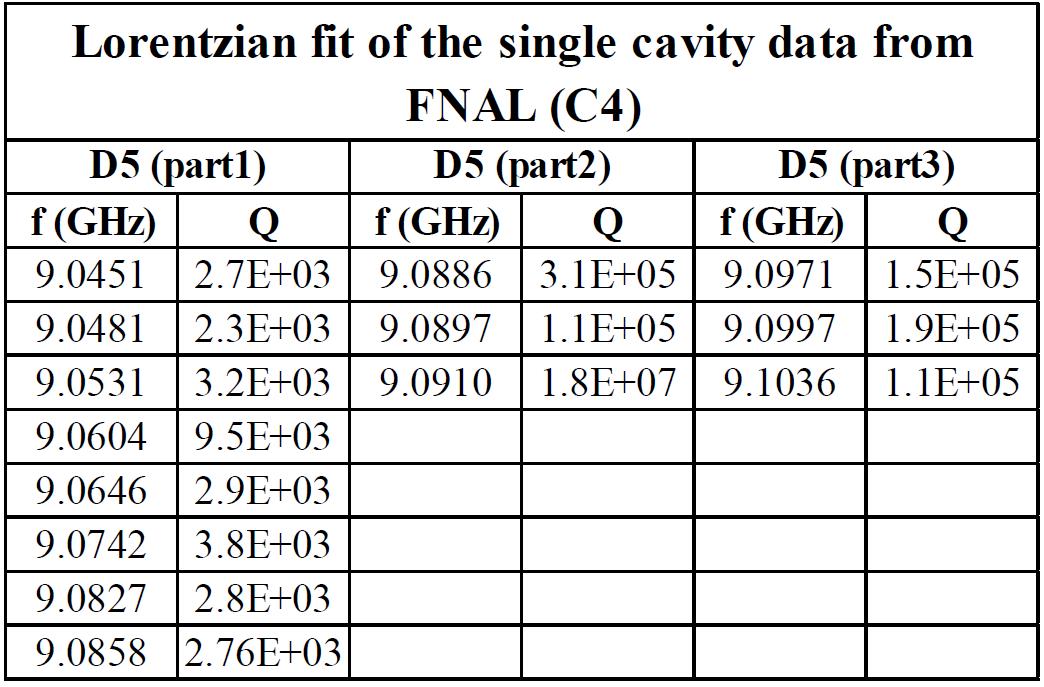}
\caption{Table of peaks in Fig.~\ref{fnalc1_D5_PF_simu}.}
\label{simu-fnal-C4-D5-table}
\end{figure}

\FloatBarrier
\section{Module-Based Transmission Spectra Measured at CMTB}\label{app-spec:cmtb}
\begin{figure}[h]\center
\includegraphics[width=0.8\textwidth]{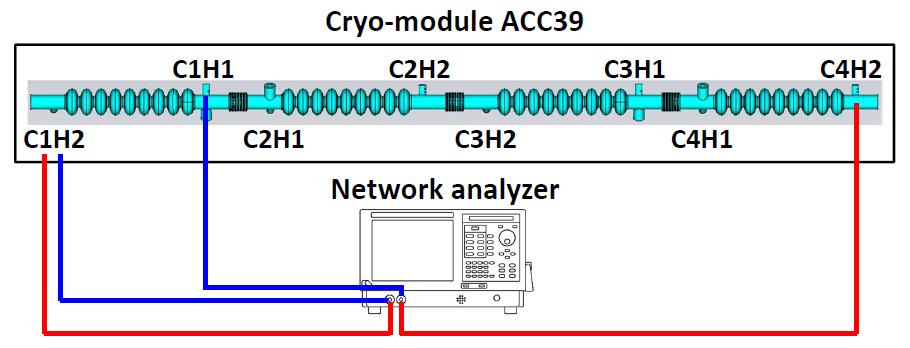}
\caption{The schematic setup of the module-based RF transmission measurement.}
\label{cmtb-setup}
\end{figure}

\begin{figure}
\subfigure[C1 (from C1H1 to C1H2)]{
\includegraphics[width=1\textwidth]{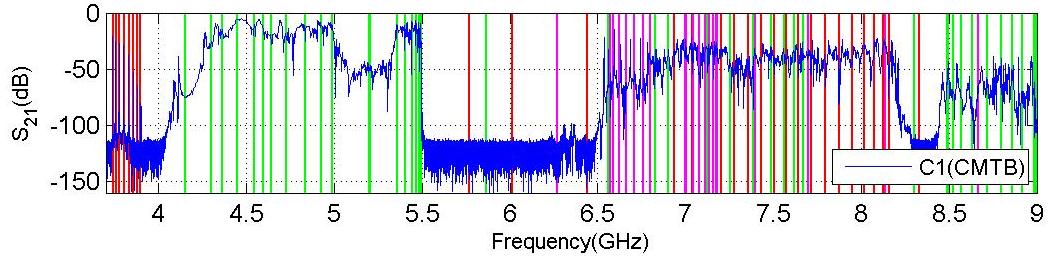}
\label{cmtb-full-spec-C1}
}
\subfigure[C2 (from C2H1 to C2H2)]{
\includegraphics[width=1\textwidth]{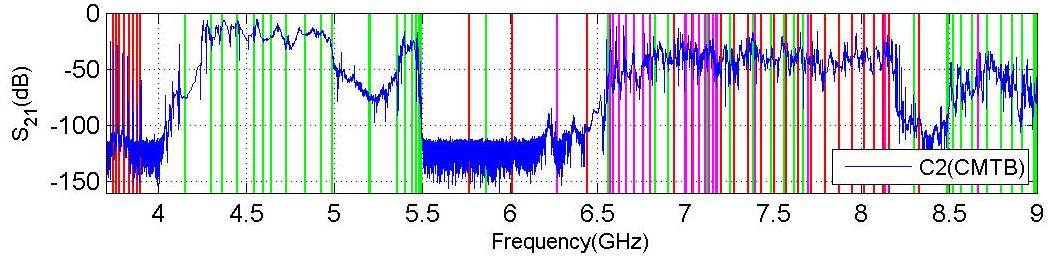}
\label{cmtb-full-spec-C2}
}
\subfigure[C3 (from C3H1 to C3H2)]{
\includegraphics[width=1\textwidth]{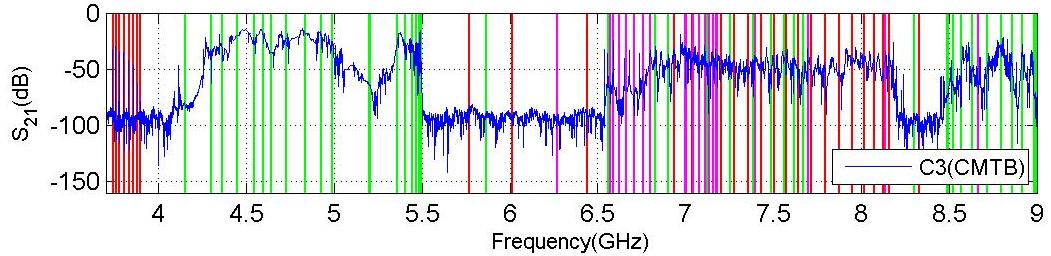}
\label{cmtb-full-spec-C3}
}
\subfigure[C4 (from C4H1 to C4H2)]{
\includegraphics[width=1\textwidth]{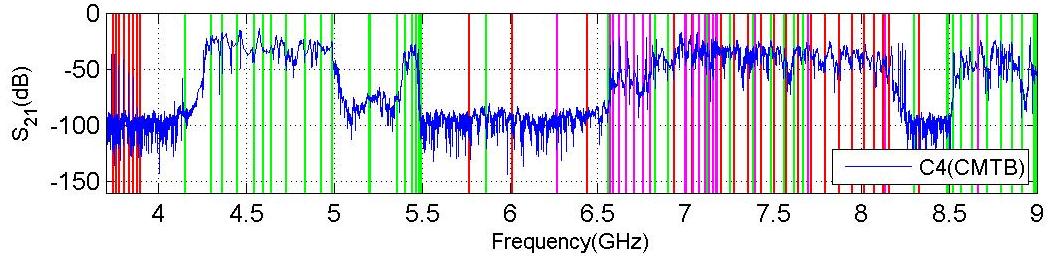}
\label{cmtb-full-spec-C4}
}
\caption{Transmission spectrum measured across each cavity at CMTB without beam excitations. The vertical lines indicate the simulation results. The colors red, green and magenta represent monopole, dipole and quadrupole modes, respectively.}
\label{cmtb-full-spec-all}
\end{figure}

\begin{figure}
\subfigure[The f\mbox{}irst dipole band of C1]{
\includegraphics[width=1\textwidth]{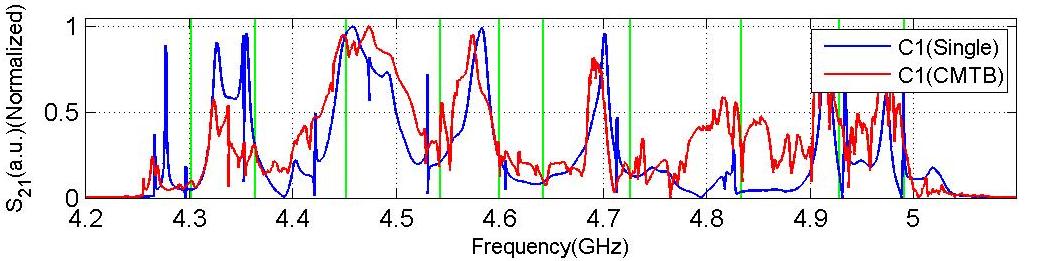}
\label{cmtb-fnal-D1-C1}
}
\subfigure[The f\mbox{}irst dipole band of C2]{
\includegraphics[width=1\textwidth]{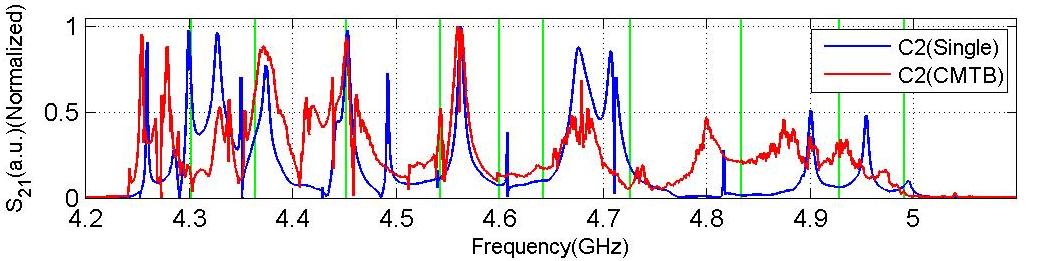}
\label{cmtb-fnal-D1-C2}
}
\subfigure[The f\mbox{}irst dipole band of C3]{
\includegraphics[width=1\textwidth]{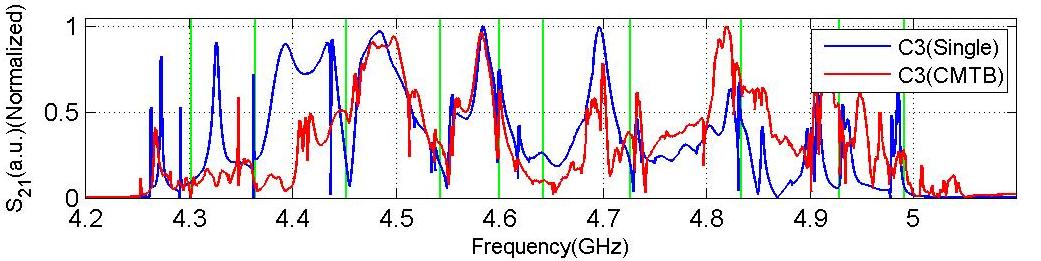}
\label{cmtb-fnal-D1-C3}
}
\subfigure[The f\mbox{}irst dipole band of C4]{
\includegraphics[width=1\textwidth]{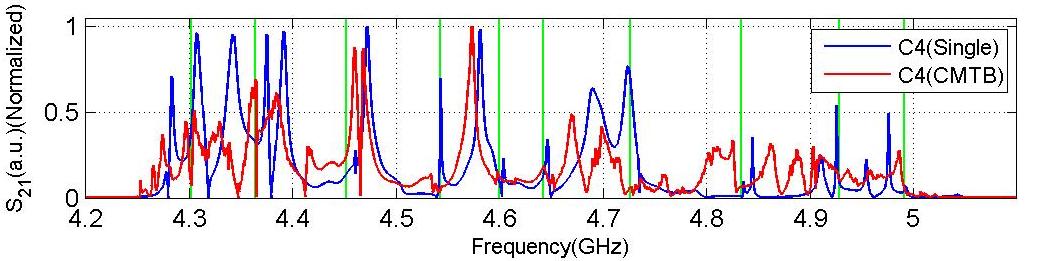}
\label{cmtb-fnal-D1-C4}
}
\caption{Comparison of single cavity measurement (blue) with CMTB (red) for the f{}irst dipole band. The vertical lines are simulation results.}
\label{cmtb-fnal-D1-all}
\end{figure}

\begin{figure}
\subfigure[The second dipole band of C1]{
\includegraphics[width=1\textwidth]{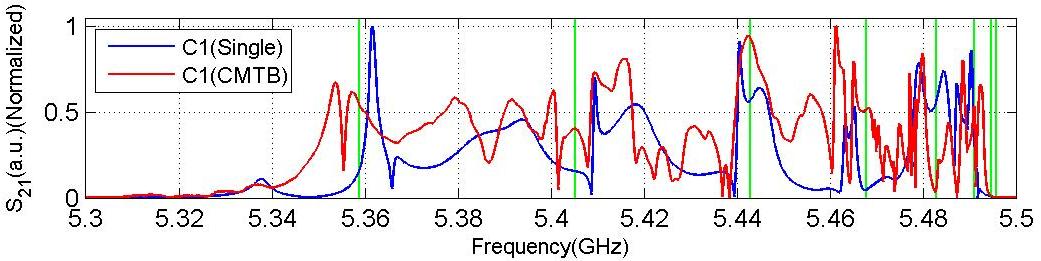}
\label{cmtb-fnal-D2-C1}
}
\subfigure[The second dipole band of C2]{
\includegraphics[width=1\textwidth]{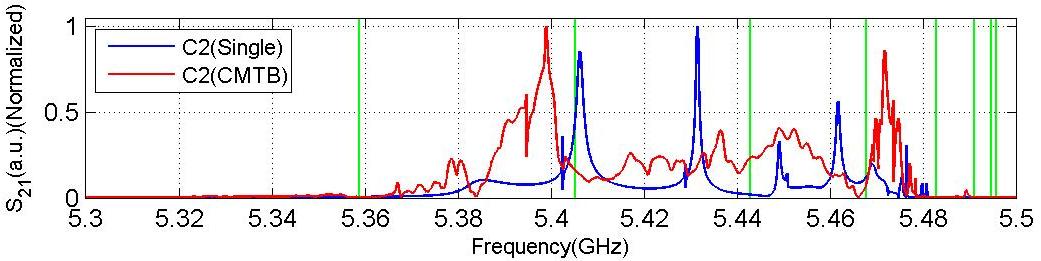}
\label{cmtb-fnal-D2-C2}
}
\subfigure[The second dipole band of C3]{
\includegraphics[width=1\textwidth]{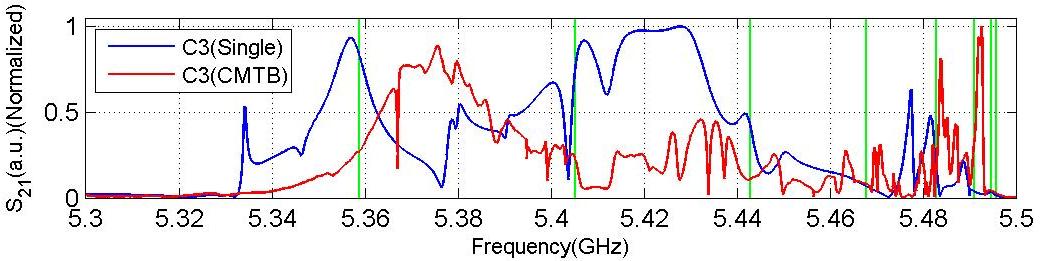}
\label{cmtb-fnal-D2-C3}
}
\subfigure[The second dipole band of C4]{
\includegraphics[width=1\textwidth]{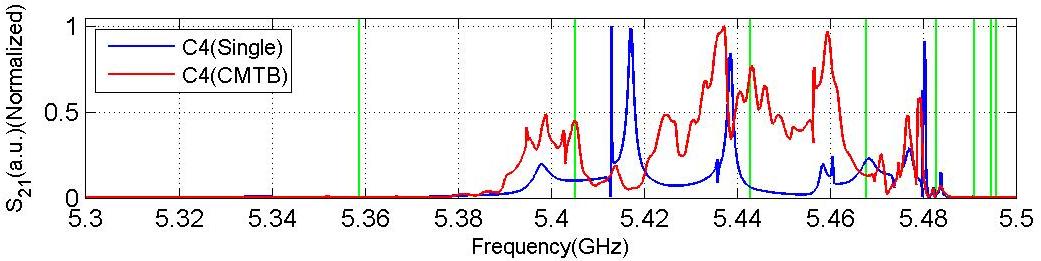}
\label{cmtb-fnal-D2-C4}
}
\caption{Comparison of single cavity measurement (blue) with CMTB (red) for the second dipole band. The vertical lines are simulation results.}
\label{cmtb-fnal-D2-all}
\end{figure}

\begin{figure}
\subfigure[C1 (from C1H1 to C1H2) with four-cavity string (from C1H2 to C4H2)]{
\includegraphics[width=1\textwidth]{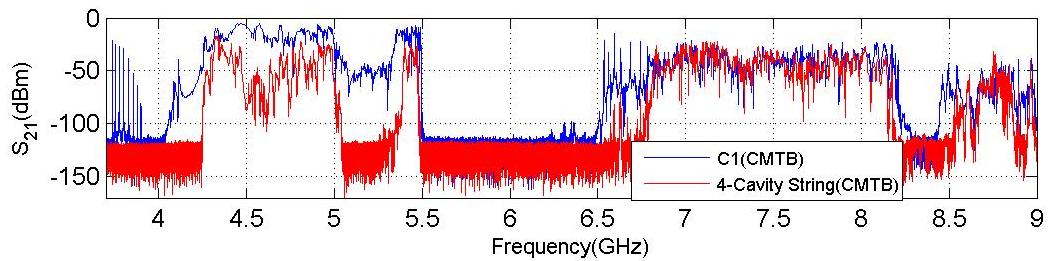}
\label{cutoff-full-spec-cmtb-C1}
}
\subfigure[C2 (from C2H1 to C2H2) with four-cavity string (from C1H2 to C4H2)]{
\includegraphics[width=1\textwidth]{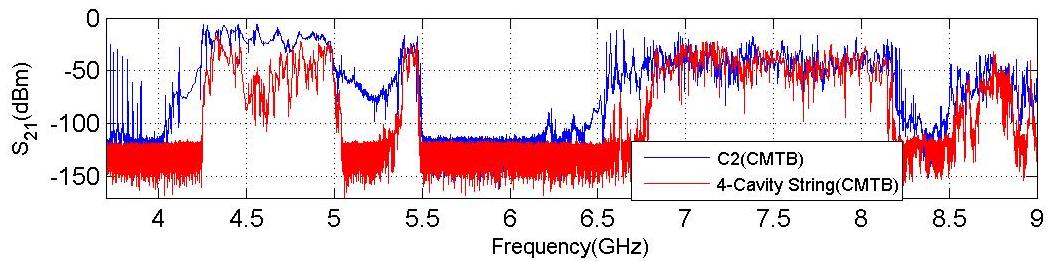}
\label{cutoff-full-spec-cmtb-C2}
}
\subfigure[C3 (from C3H1 to C3H2) with four-cavity string (from C1H2 to C4H2)]{
\includegraphics[width=1\textwidth]{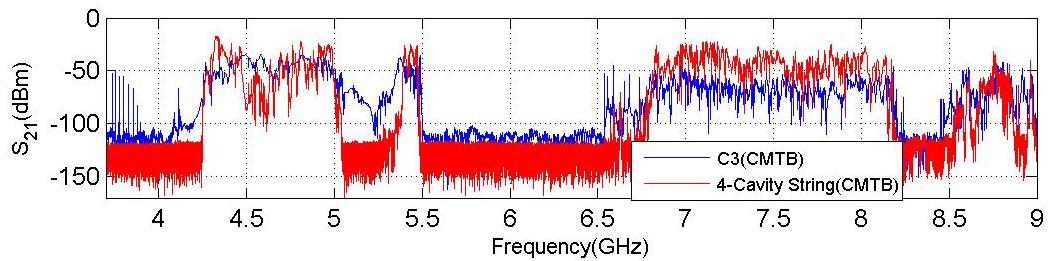}
\label{cutoff-full-spec-cmtb-C3}
}
\subfigure[C4 (from C4H1 to C4H2) with four-cavity string (from C1H2 to C4H2)]{
\includegraphics[width=1\textwidth]{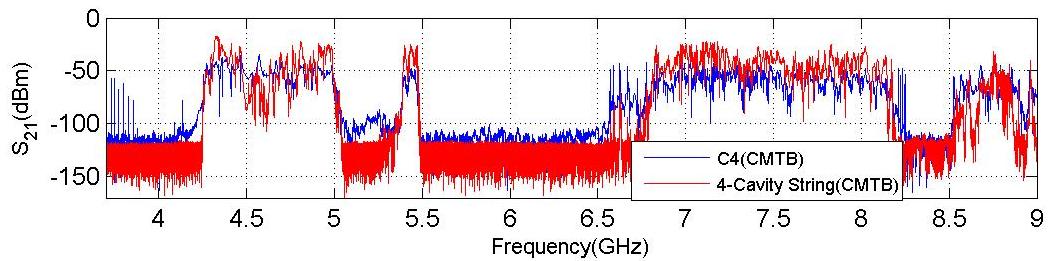}
\label{cutoff-full-spec-cmtb-C4}
}
\caption{Coupling ef\mbox{}fects of inter-connected cavities. The measurement is done at CMTB, while the spectra measured across each cavity (from coupler 1 to coupler 2) is in blue, and the spectra measured across the entire four-cavity string (from C1H2 to C4H2) is in red.}
\label{cutoff-full-spec-cmtb-all}
\end{figure}

\begin{figure}
\subfigure[C1 (from C1H1 to C1H2) with four-cavity string (from C1H2 to C4H2)]{
\includegraphics[width=1\textwidth]{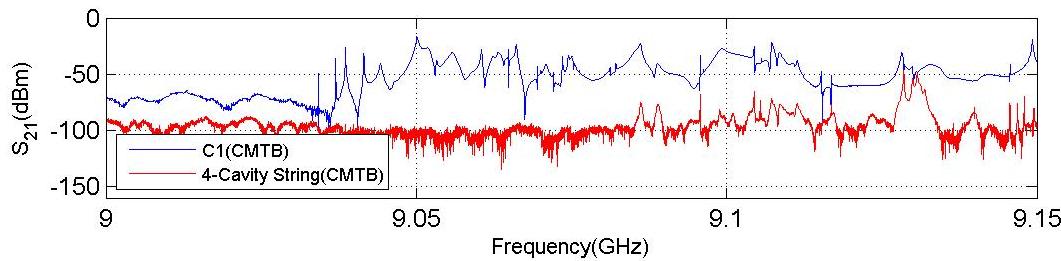}
\label{cutoff-D5-cmtb-C1}
}
\subfigure[C2 (from C2H1 to C2H2) with four-cavity string (from C1H2 to C4H2)]{
\includegraphics[width=1\textwidth]{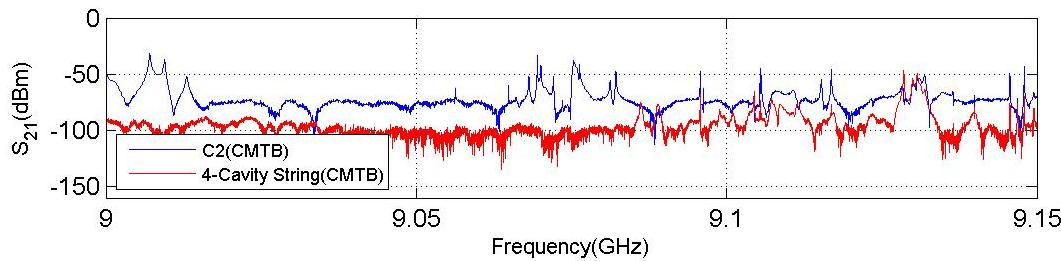}
\label{cutoff-D5-cmtb-C2}
}
\subfigure[C3 (from C3H1 to C3H2) with four-cavity string (from C1H2 to C4H2)]{
\includegraphics[width=1\textwidth]{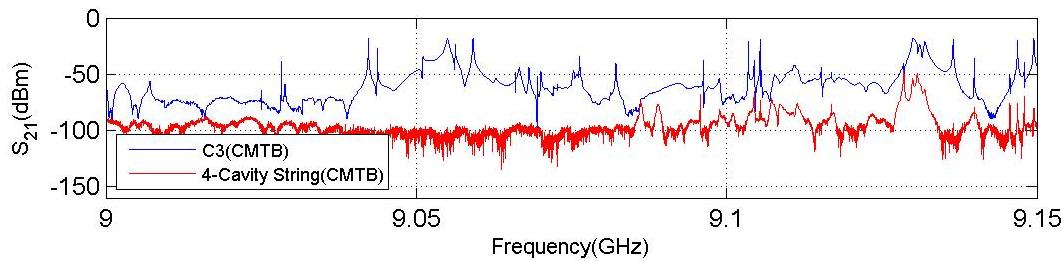}
\label{cutoff-D5-cmtb-C3}
}
\subfigure[C4 (from C4H1 to C4H2) with four-cavity string (from C1H2 to C4H2)]{
\includegraphics[width=1\textwidth]{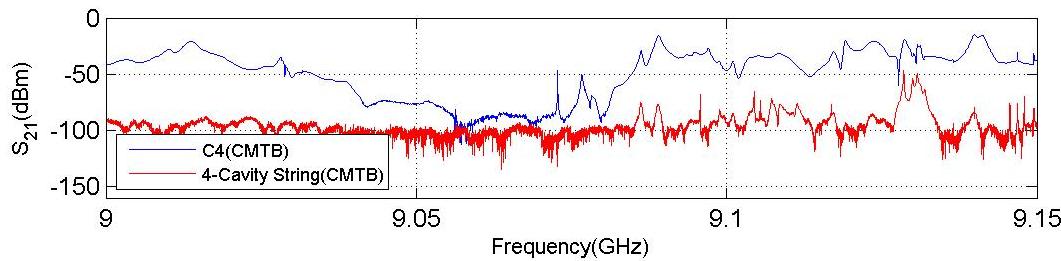}
\label{cutoff-D5-cmtb-C4}
}
\caption{Coupling ef\mbox{}fects of inter-connected cavities. The measurement is done at CMTB, while the spectra measured across each cavity (from coupler 1 to coupler 2) is in blue, and the spectra measured across the entire four-cavity string (from C1H2 to C4H2) is in red.}
\label{cutoff-D5-cmtb-all}
\end{figure}

The transmission parameter $S_{21}$ was measured in CMTB (Cryo-Module Test Bench). The schematic of the measurement setup is shown in Fig.~\ref{cmtb-setup}. The four cavities were connected with bellows of 40~mm diameter. A pumping line with gate valve was attached to one end of the module while another gate valve was attached to the other end \cite{racc39-fnal-2}. Measurement data are kindly provided by T.~Khabibouline from Fermilab. The transmission spectrum through each cavity is shown in Fig.~\ref{cmtb-full-spec-all}. A comparison of transmission spectra between CMTB and single cavity is shown in Fig.~\ref{cmtb-fnal-D1-all} for the f{}irst dipole band and Fig.~\ref{cmtb-fnal-D2-all} for the second dipole band. The propagating characteristics of most HOMs are shown in Fig.~\ref{cutoff-full-spec-cmtb-all} and Fig.~\ref{cutoff-D5-cmtb-all}.

\FloatBarrier
\section{Module-Based Transmission Spectra Measured at FLASH}\label{app-spec:flash}
\begin{figure}\center
\subfigure[C1 (from C1H1 to C1H2)]{
\includegraphics[width=1\textwidth]{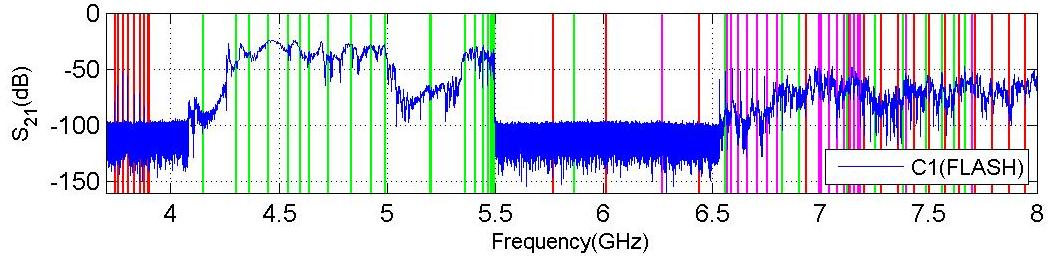}
\label{flash-full-spec-C1}
}
\subfigure[C2 (from C2H1 to C2H2)]{
\includegraphics[width=1\textwidth]{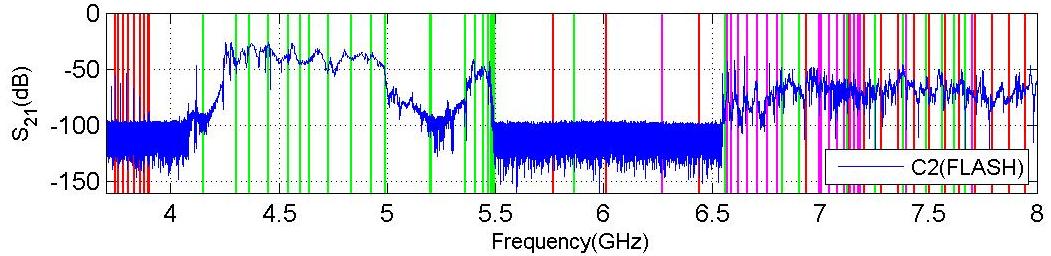}
\label{flash-full-spec-C2}
}
\subfigure[C3 (from C3H1 to C3H2)]{
\includegraphics[width=1\textwidth]{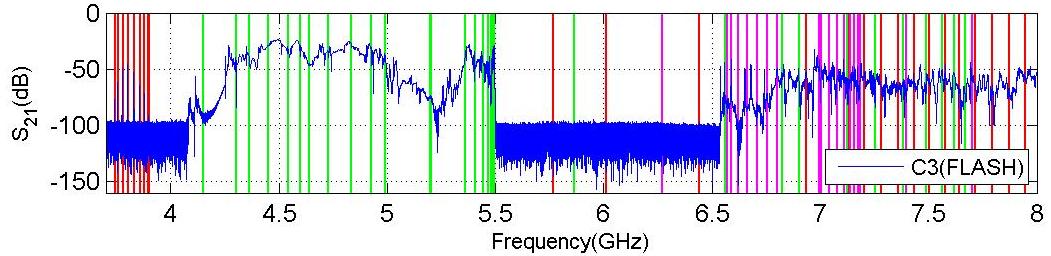}
\label{flash-full-spec-C3}
}
\subfigure[C4 (from C4H1 to C4H2)]{
\includegraphics[width=1\textwidth]{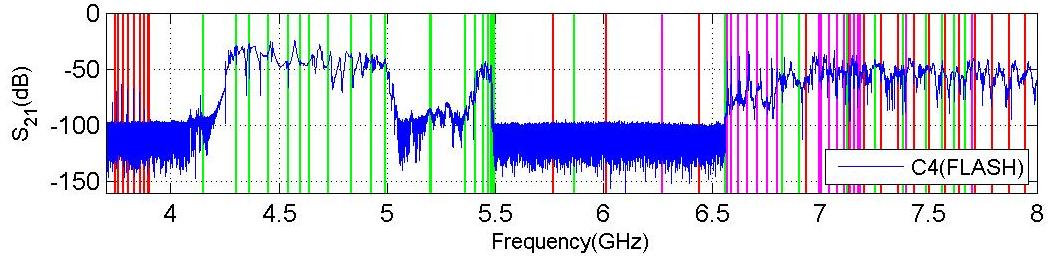}
\label{flash-full-spec-C4}
}
\caption{Transmission spectrum measured across each cavity at FLASH without beam excitations. The vertical lines indicate the simulation results. The colors red, green and magenta represent monopole, dipole and quadrupole modes, respectively.}
\label{flash-full-spec-all}
\end{figure}

\begin{figure}\center
\subfigure[C1 (from C1H1 to C1H2) with four-cavity string (from C1H2 to C4H2)]{
\includegraphics[width=1\textwidth]{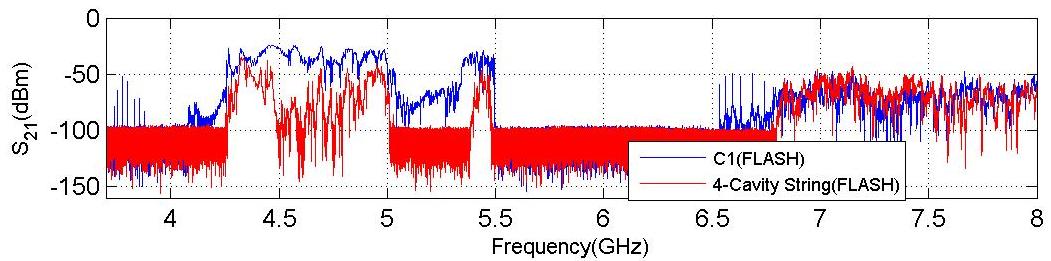}
\label{cutoff-full-spec-flash-C1}
}
\subfigure[C2 (from C2H1 to C2H2) with four-cavity string (from C1H2 to C4H2)]{
\includegraphics[width=1\textwidth]{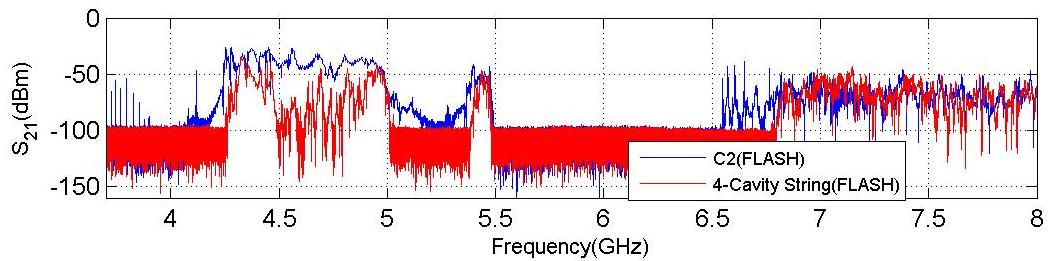}
\label{cutoff-full-spec-flash-C2}
}
\subfigure[C3 (from C3H1 to C3H2) with four-cavity string (from C1H2 to C4H2)]{
\includegraphics[width=1\textwidth]{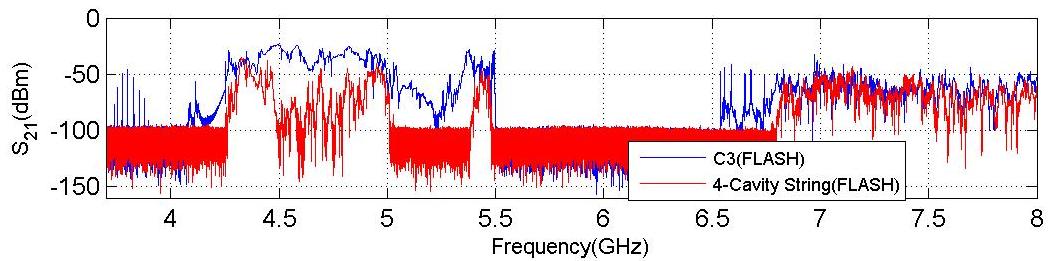}
\label{cutoff-full-spec-flash-C3}
}
\subfigure[C4 (from C4H1 to C4H2) with four-cavity string (from C1H2 to C4H2)]{
\includegraphics[width=1\textwidth]{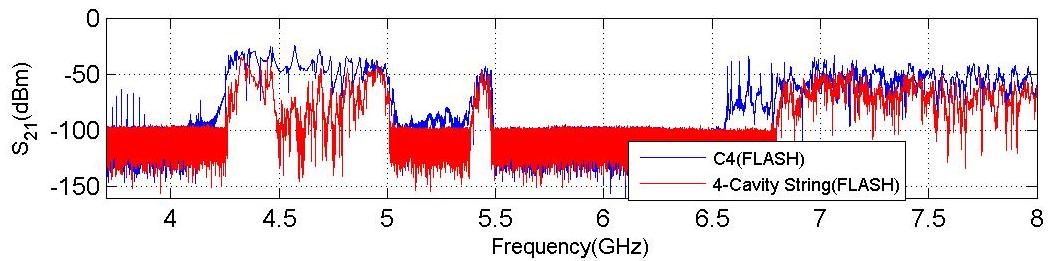}
\label{cutoff-full-spec-flash-C4}
}
\caption{Coupling ef\mbox{}fects of inter-connected cavities. The measurement is done at FLASH, while the spectra measured across each cavity (from coupler 1 to coupler 2) is in blue, and the spectra measured across the entire four-cavity string (from C1H2 to C4H2) is in red.}
\label{cutoff-full-spec-flash-all}
\end{figure}

\begin{figure}\center
\subfigure[C1 (from C1H1 to C1H2)]{
\includegraphics[width=0.9\textwidth]{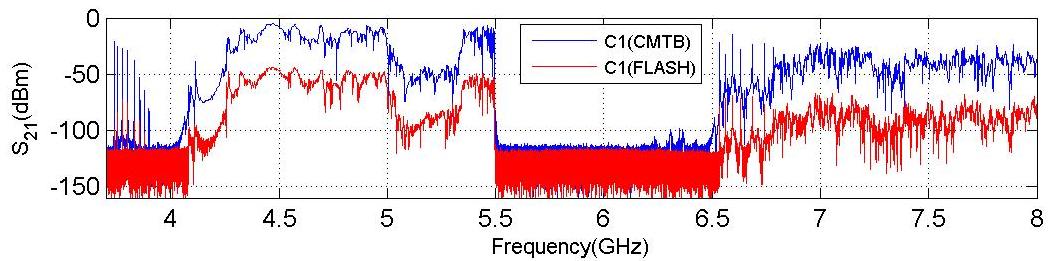}
\label{flash-cmtb-full-spec-C1}
}
\subfigure[C2 (from C2H1 to C2H2)]{
\includegraphics[width=0.9\textwidth]{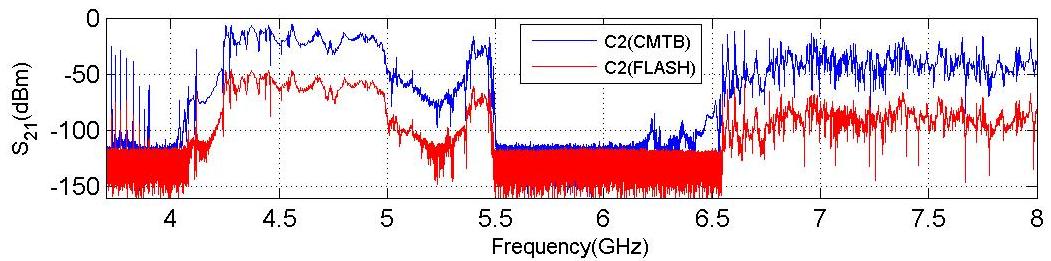}
\label{flash-cmtb-full-spec-C2}
}
\subfigure[C3 (from C3H1 to C3H2)]{
\includegraphics[width=0.9\textwidth]{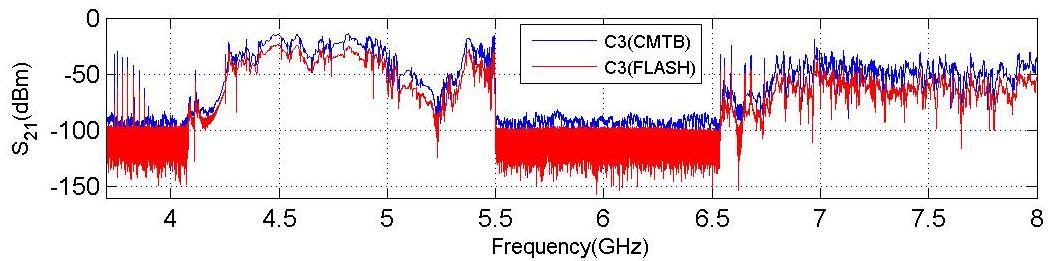}
\label{flash-cmtb-full-spec-C3}
}
\subfigure[C4 (from C4H1 to C4H2)]{
\includegraphics[width=0.9\textwidth]{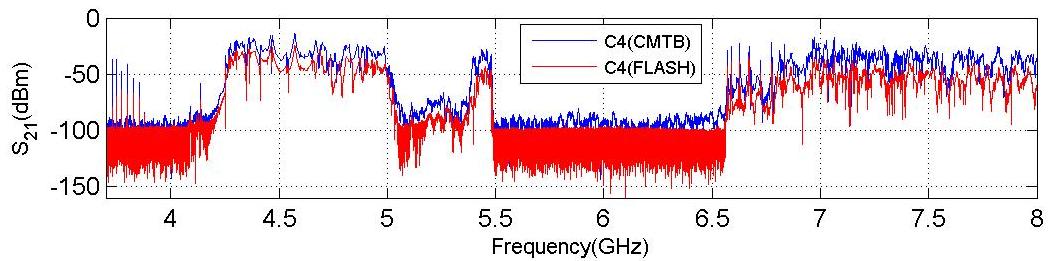}
\label{flash-cmtb-full-spec-C4}
}
\subfigure[Four-cavity string (from C1H2 to C4H2)]{
\includegraphics[width=0.9\textwidth]{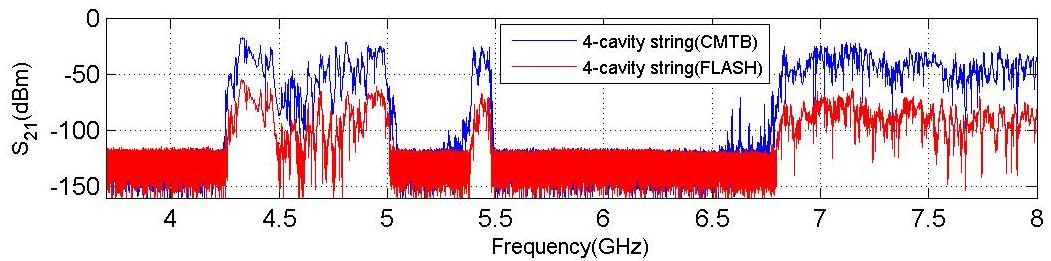}
\label{flash-cmtb-full-spec-4-cavity}
}
\caption{Comparison of spectra measured at CMTB (blue) with those obtained at FLASH (red).}
\label{flash-cmtb-full-spec-all}
\end{figure}

The transmission measurement was made after the module been installed in FLASH. The setup is similar to that of CMTB (Fig.~\ref{cmtb-setup}). The transmission spectrum though each cavity is shown in Fig.~\ref{flash-full-spec-all}. The coupling ef{}fect is shown in Fig.~\ref{cutoff-full-spec-flash-all}. A direct comparison of transmission spectra between CMTB and FLASH measurement is shown in Fig.~\ref{flash-cmtb-full-spec-all}. The spectra are similar except the CMTB data has larger dynamic range since the measurement was conducted in the tunnel. 

\FloatBarrier
\section{Beam-Excited HOM Spectra Measured at FLASH}\label{app-spec:beam}
\begin{figure}\center
\subfigure[C1H1]{
\includegraphics[width=0.37\textwidth]{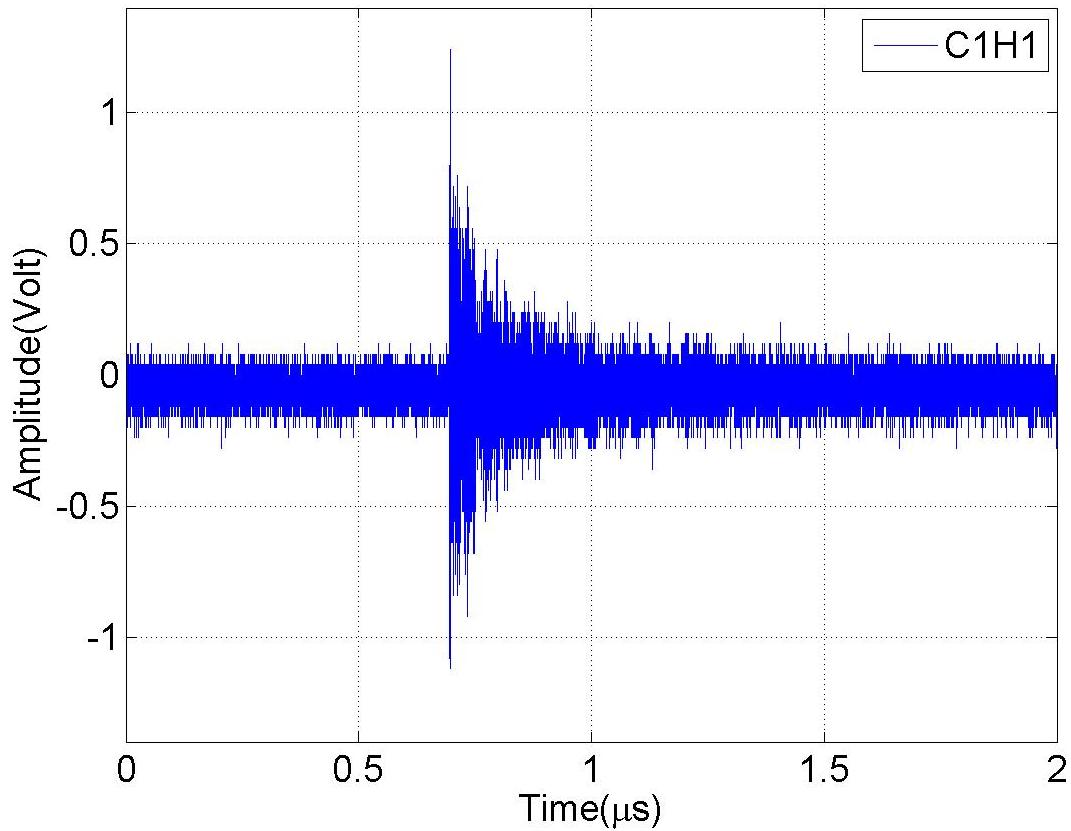}
\label{wfm-C1H1}
}
\quad\quad
\subfigure[C1H2]{
\includegraphics[width=0.37\textwidth]{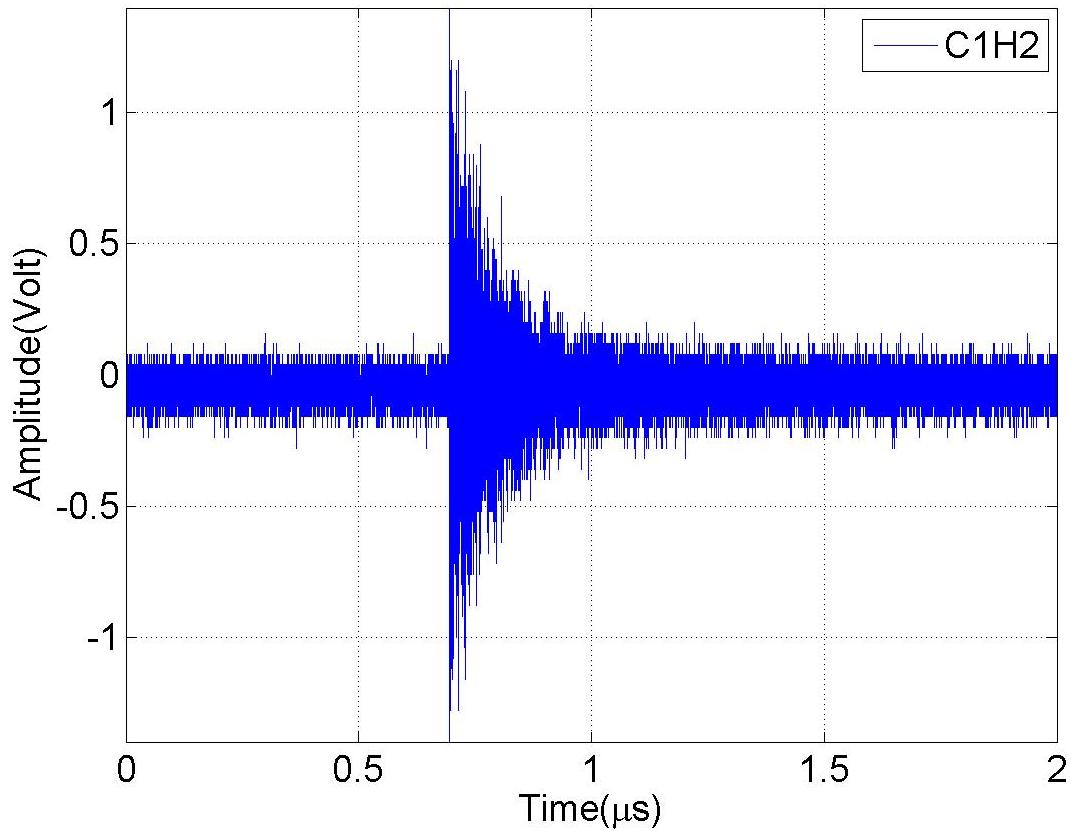}
\label{wfm-C1H2}
}
\subfigure[C2H1]{
\includegraphics[width=0.37\textwidth]{scope-G2move01-C2H1}
\label{wfm-C2H1}
}
\quad\quad
\subfigure[C2H2]{
\includegraphics[width=0.37\textwidth]{scope-G2move01-C2H2}
\label{wfm-C2H2}
}
\subfigure[C3H1]{
\includegraphics[width=0.37\textwidth]{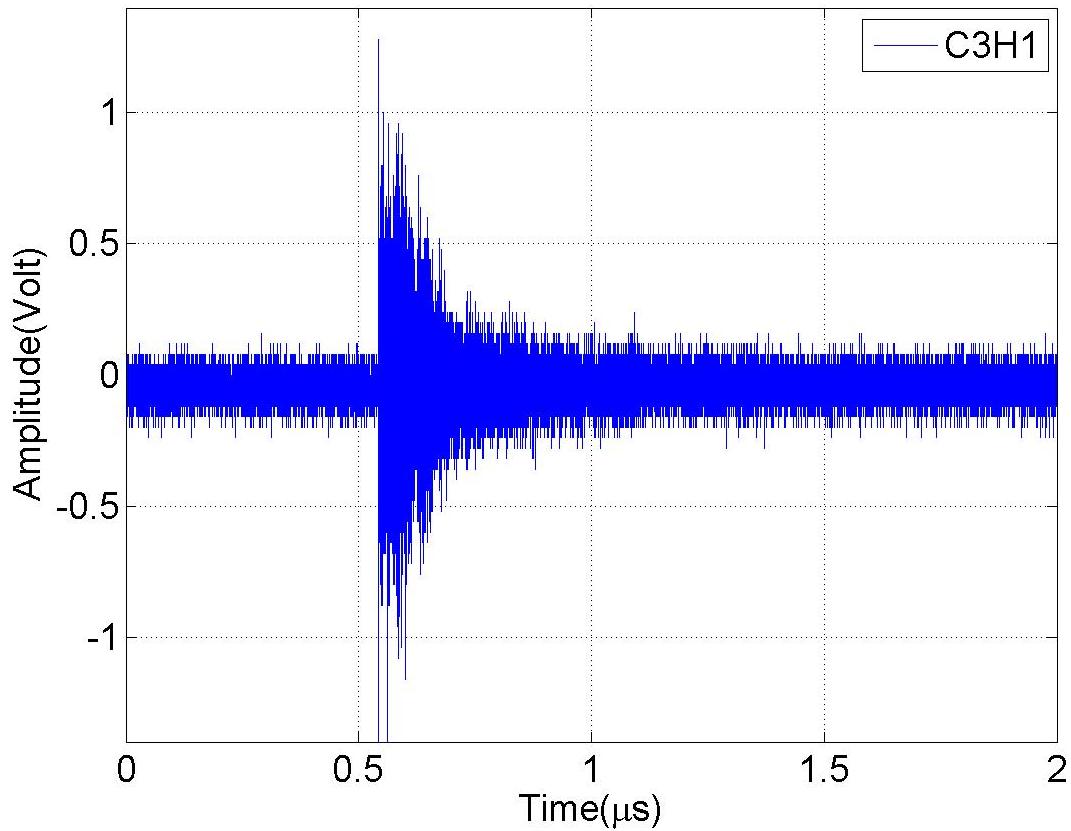}
\label{wfm-C3H1}
}
\quad\quad
\subfigure[C3H2]{
\includegraphics[width=0.37\textwidth]{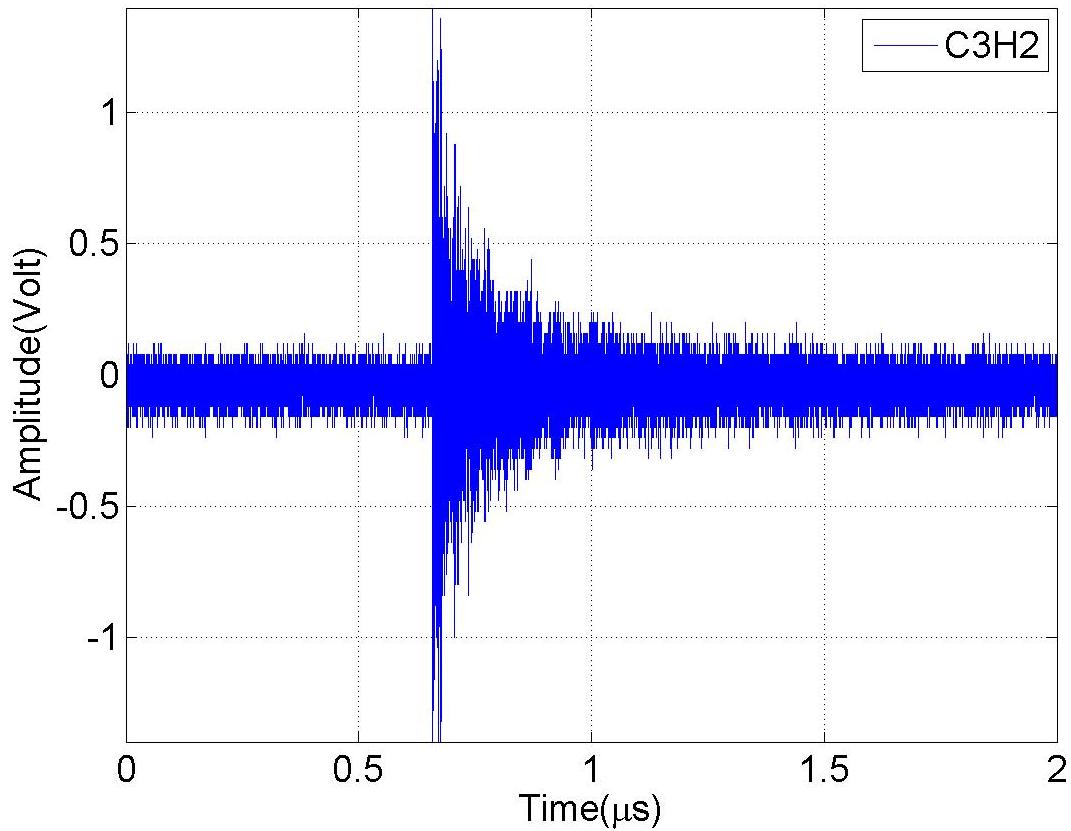}
\label{wfm-C3H2}
}
\subfigure[C4H1]{
\includegraphics[width=0.37\textwidth]{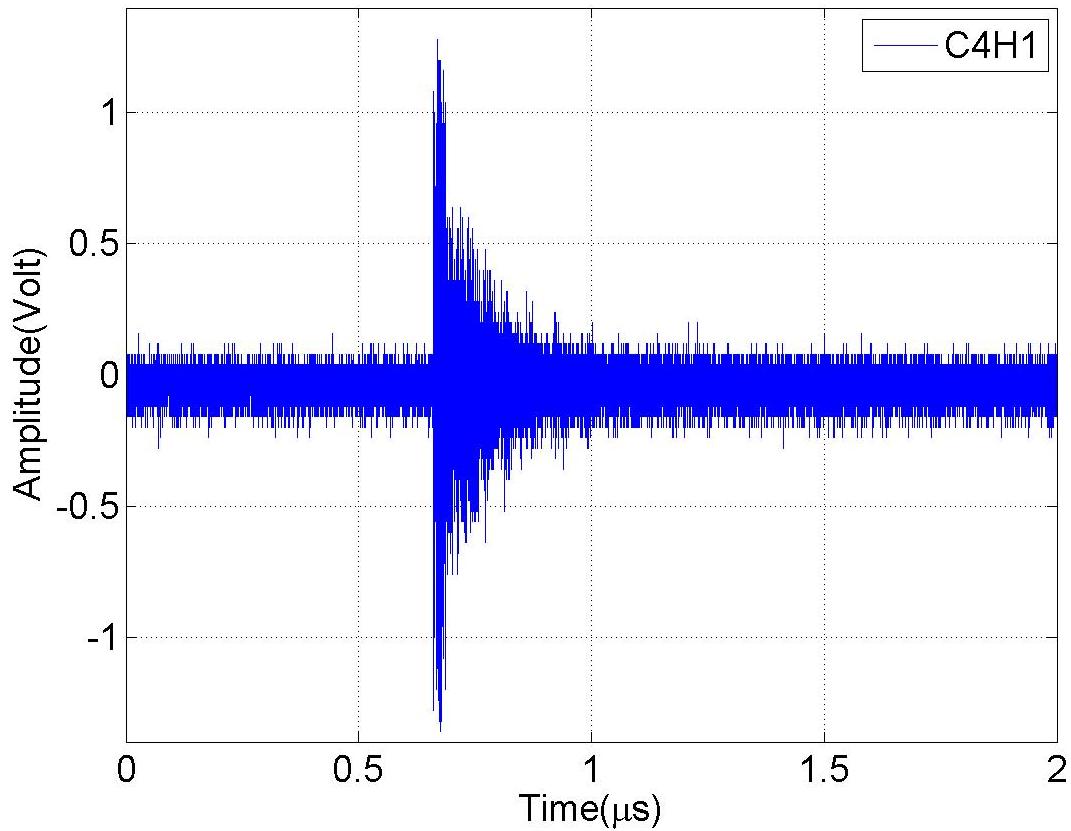}
\label{wfm-C4H1}
}
\quad\quad
\subfigure[C4H2]{
\includegraphics[width=0.37\textwidth]{scope-G2move01-C1H1}
\label{wfm-C4H2}
}
\caption{Waveform excited by a single electron bunch measured from each HOM coupler.}
\label{wfm-all}
\end{figure}

\begin{figure}
\subfigure[C1H1]{
\includegraphics[width=1\textwidth]{rsa-full-spec-C1H1-2in1}
\label{rsa-full-spec-C1H1}
}
\subfigure[C1H2)]{
\includegraphics[width=1\textwidth]{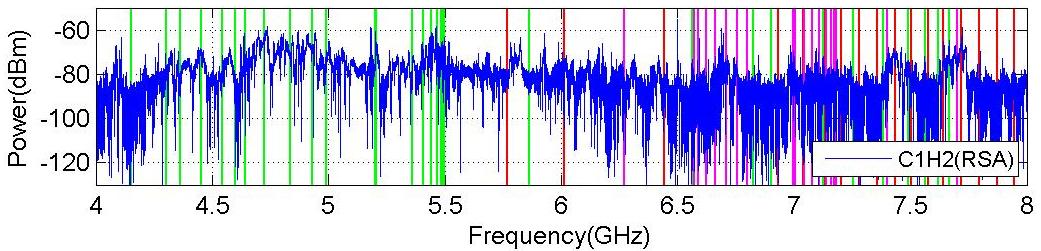}
\label{rsa-full-spec-C1H2}
}
\subfigure[C2H1]{
\includegraphics[width=1\textwidth]{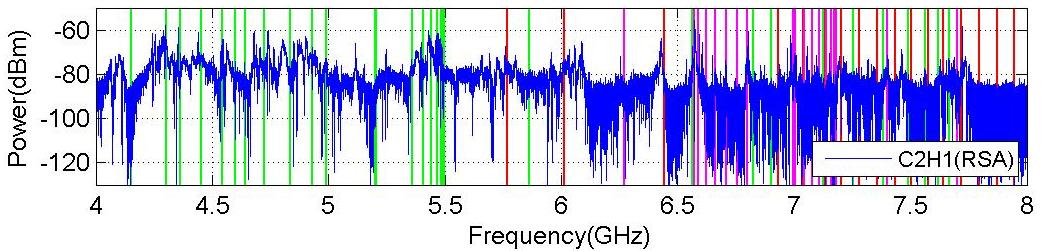}
\label{rsa-full-spec-C2H1}
}
\subfigure[C2H2]{
\includegraphics[width=1\textwidth]{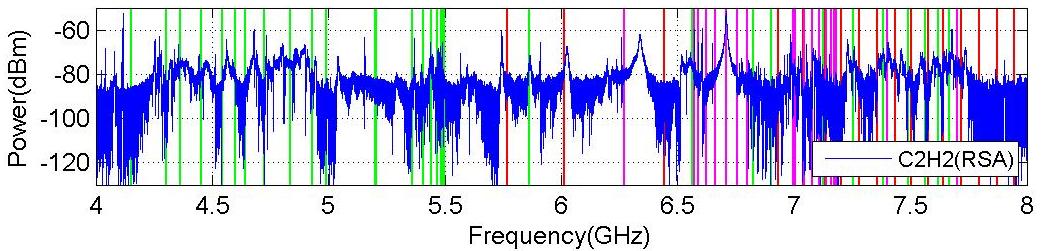}
\label{rsa-full-spec-C2H2}
}
\caption{Beam-excited spectra measured by RSA. Each 50~MHz is excited by a single electron bunch. The vertical lines indicate the simulation results. The colors red, green and magenta represent monopole, dipole and quadrupole modes, respectively.}
\label{rsa-full-spec-C1C2}
\end{figure}

\begin{figure}
\subfigure[C3H1]{
\includegraphics[width=1\textwidth]{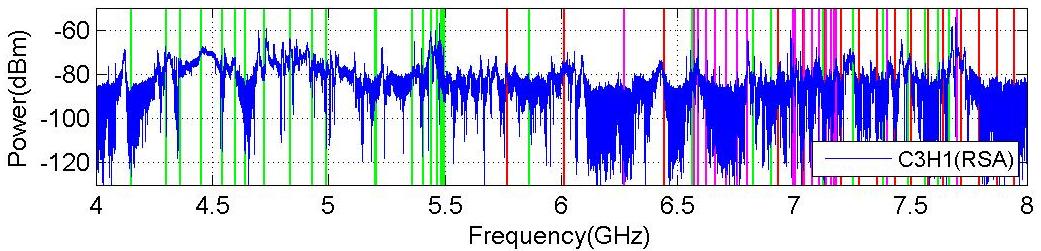}
\label{rsa-full-spec-C3H1}
}
\subfigure[C3H2)]{
\includegraphics[width=1\textwidth]{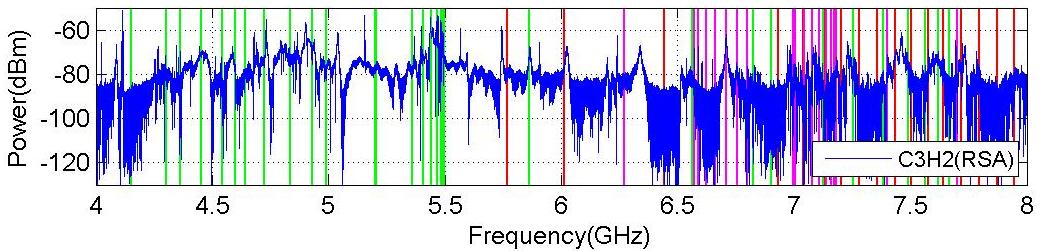}
\label{rsa-full-spec-C3H2}
}
\subfigure[C4H1]{
\includegraphics[width=1\textwidth]{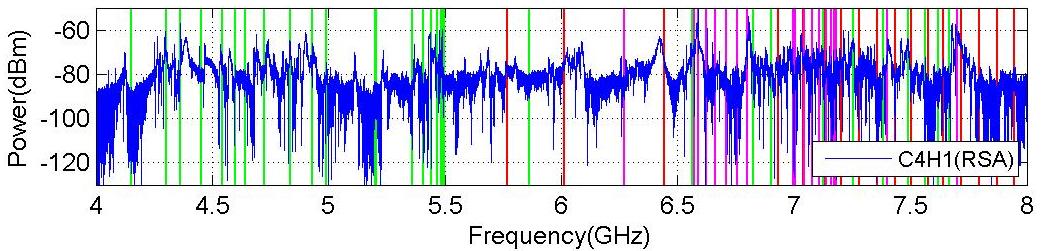}
\label{rsa-full-spec-C4H1}
}
\subfigure[C4H2]{
\includegraphics[width=1\textwidth]{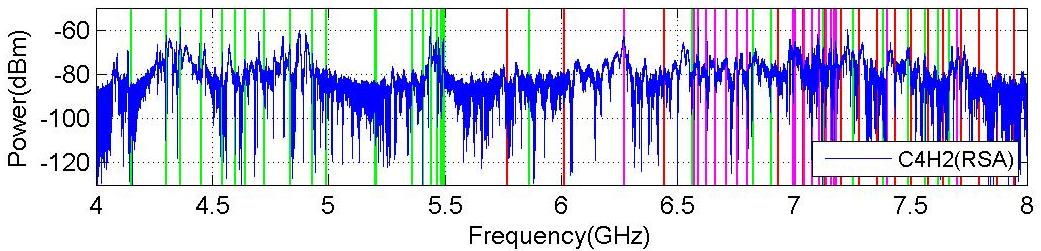}
\label{rsa-full-spec-C4H2}
}
\caption{Beam-excited spectra measured by RSA. Each 50~MHz is excited by a single electron bunch. The vertical lines indicate the simulation results. The colors red, green and magenta represent monopole, dipole and quadrupole modes, respectively.}
\label{rsa-full-spec-C3C4}
\end{figure}

\begin{figure}
\subfigure[C1]{
\includegraphics[width=1\textwidth]{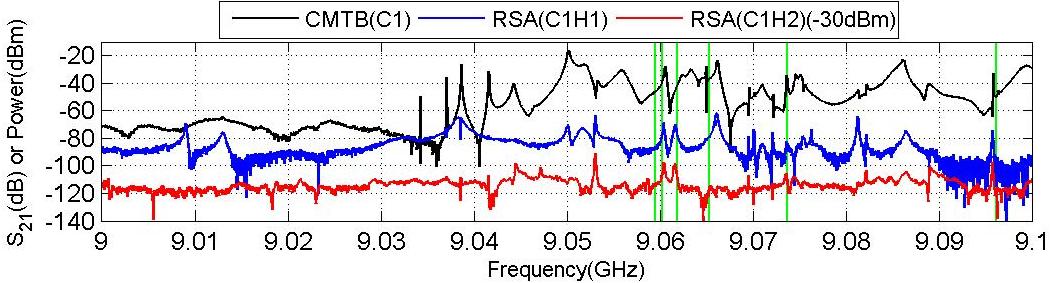}
\label{rsa-cmtb-D5-C1}
}
\subfigure[C2]{
\includegraphics[width=1\textwidth]{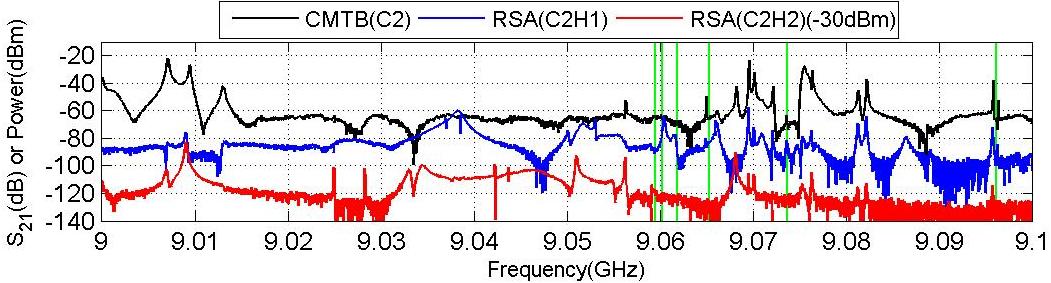}
\label{rsa-cmtb-D5-C2}
}
\subfigure[C3]{
\includegraphics[width=1\textwidth]{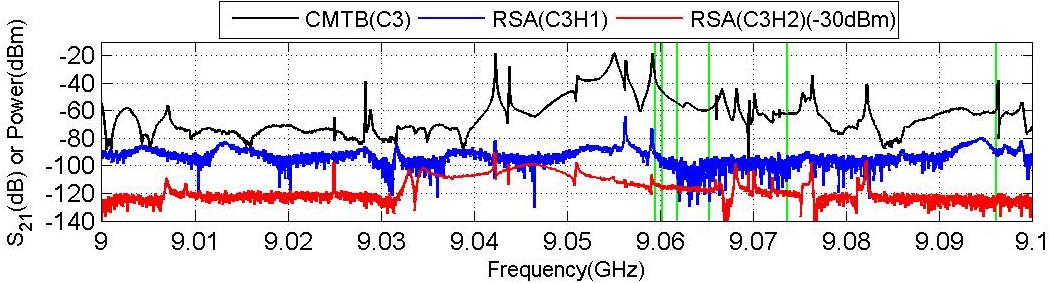}
\label{rsa-cmtb-D5-C3}
}
\subfigure[C4]{
\includegraphics[width=1\textwidth]{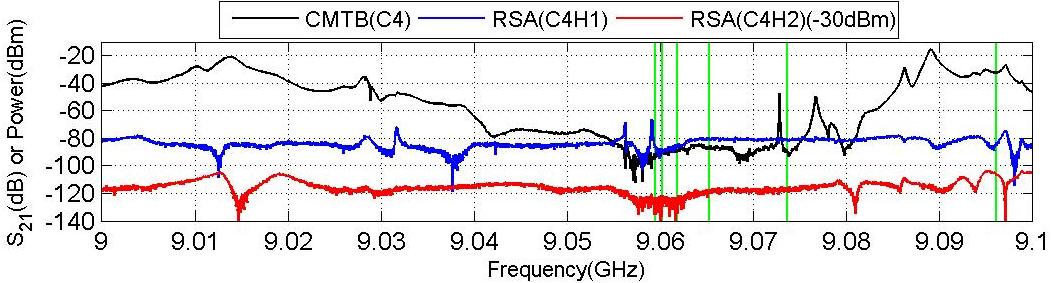}
\label{rsa-cmtb-D5-C4}
}
\caption{Beam-excited spectra measured by RSA (blue for H1 and red for H2) compared with CMTB measurement. Each 50~MHz is excited by a single electron bunch. The vertical lines in red indicate the simulation results.}
\label{rsa-cmtb-D5-all}
\end{figure}

The single bunch excited HOM spectrum was measured from each coupler. The measurement details are described in Section~\ref{hom-meas:beam}. The time-domain signals are shown in Fig.~\ref{wfm-all}. The spectra are shown in Fig.~\ref{rsa-full-spec-C1C2} and Fig.~\ref{rsa-full-spec-C3C4}. A direct comparison between real-time spectra and f{}f{}t-applied time-domain waveforms are shown in Fig.~\ref{rsa-full-spec-C1C2} and Fig.~\ref{rsa-full-spec-C3C4} for the f{}irst two dipole bands. The beam-excited spectra of the f{}ifth dipole band are shown in Fig.~\ref{rsa-cmtb-D5-all} along with the CMTB spectra as a comparison.

\begin{figure}
\subfigure[C1H1]{
\includegraphics[width=1\textwidth]{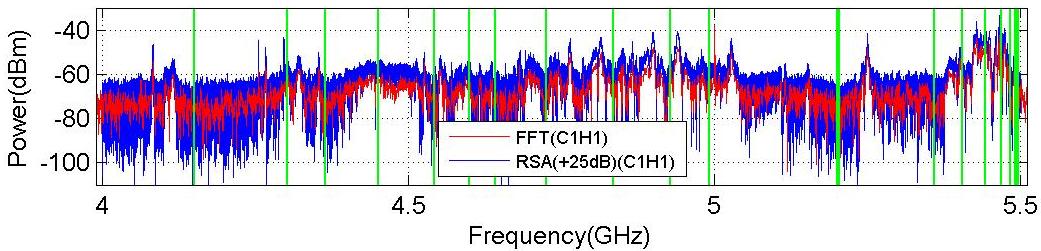}
\label{fft-rsa-C1H1}
}
\subfigure[C1H2]{
\includegraphics[width=1\textwidth]{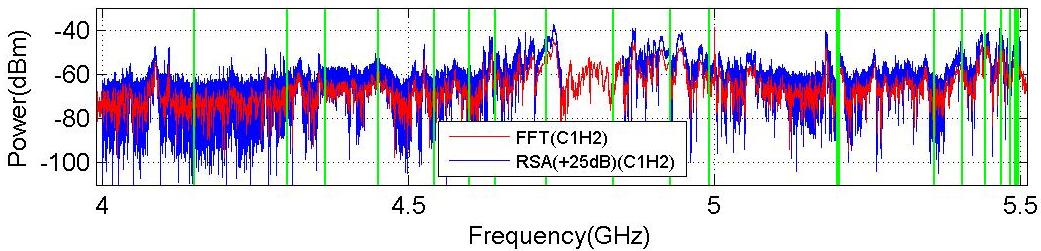}
\label{fft-rsa-C1H2}
}
\subfigure[C2H1]{
\includegraphics[width=1\textwidth]{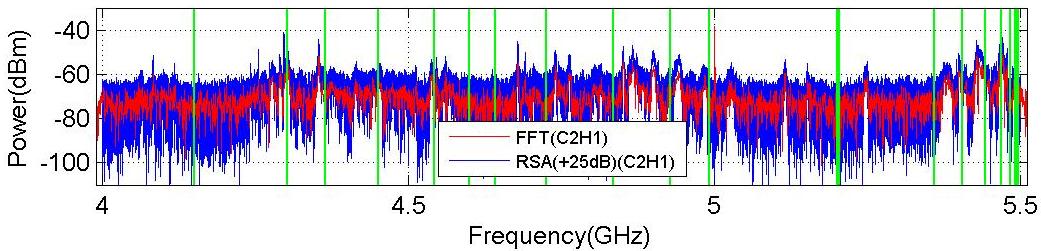}
\label{fft-rsa-C2H1}
}
\subfigure[C2H2]{
\includegraphics[width=1\textwidth]{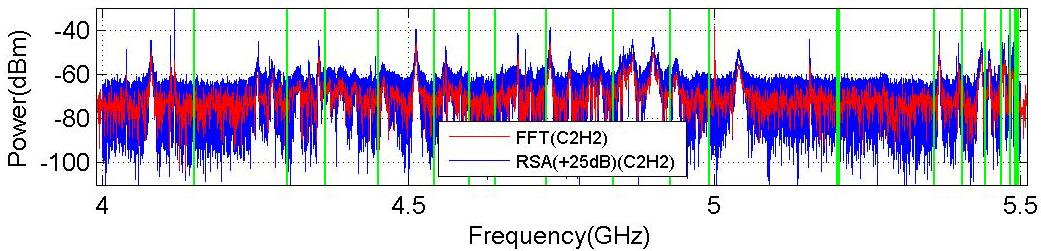}
\label{fft-rsa-C2H2}
}
\caption{Beam-excited spectrum of the f\mbox{}irst and second dipole band measured from each HOM coupler using RSA (blue) and Scope (after FFT with 25 dB added, red). The vertical lines in green are simulation results.}
\label{fft-rsa-C1C2}
\end{figure}
\begin{figure}
\subfigure[C3H1]{
\includegraphics[width=1\textwidth]{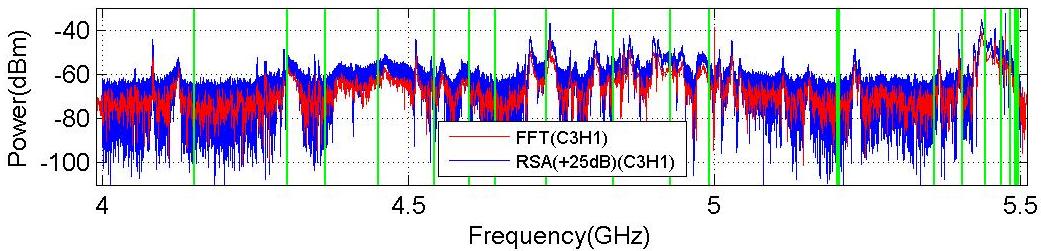}
\label{fft-rsa-C3H1}
}
\subfigure[C3H2]{
\includegraphics[width=1\textwidth]{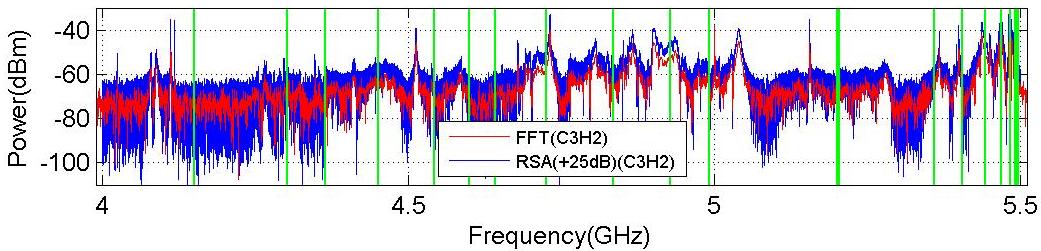}
\label{fft-rsa-C3H2}
}
\subfigure[C4H1]{
\includegraphics[width=1\textwidth]{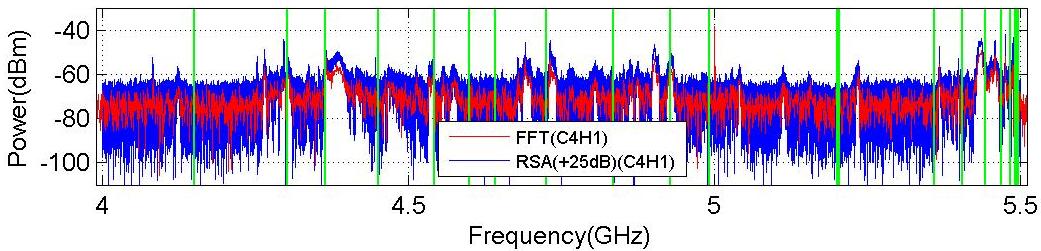}
\label{fft-rsa-C4H1}
}
\subfigure[C4H2]{
\includegraphics[width=1\textwidth]{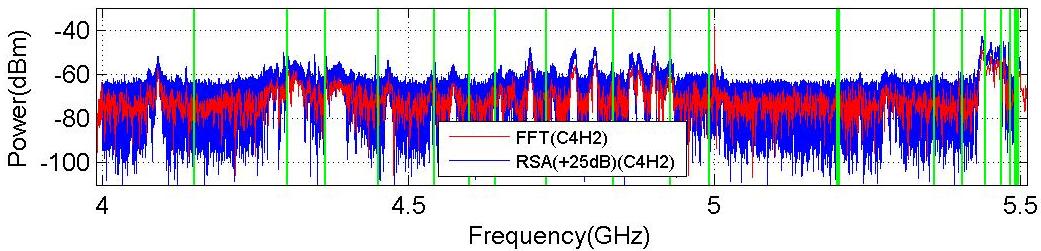}
\label{fft-rsa-C4H2}
}
\caption{Beam-excited spectrum of the f\mbox{}irst and second dipole band measured from each HOM coupler using RSA (blue) and Scope (after FFT with 25 dB added, red). The vertical lines in green are simulation results.}
\label{fft-rsa-C3C4}
\end{figure}

\chapter{Various Beam Scans}\label{app-move}
In order to study dipole dependence of the modes on the transverse beam position, we have moved the beam in various ways by using a pair of steering magnets as shown in Fig.~\ref{hom-setup}. The various scans used for the data shown in this note are summarized here. Figures shown in this chapter are extensions of Chapter~\ref{hom-dep:setup}. 

\begin{figure}\center
\subfigure[Steerer (2GUN)]{
\includegraphics[width=0.37\textwidth]{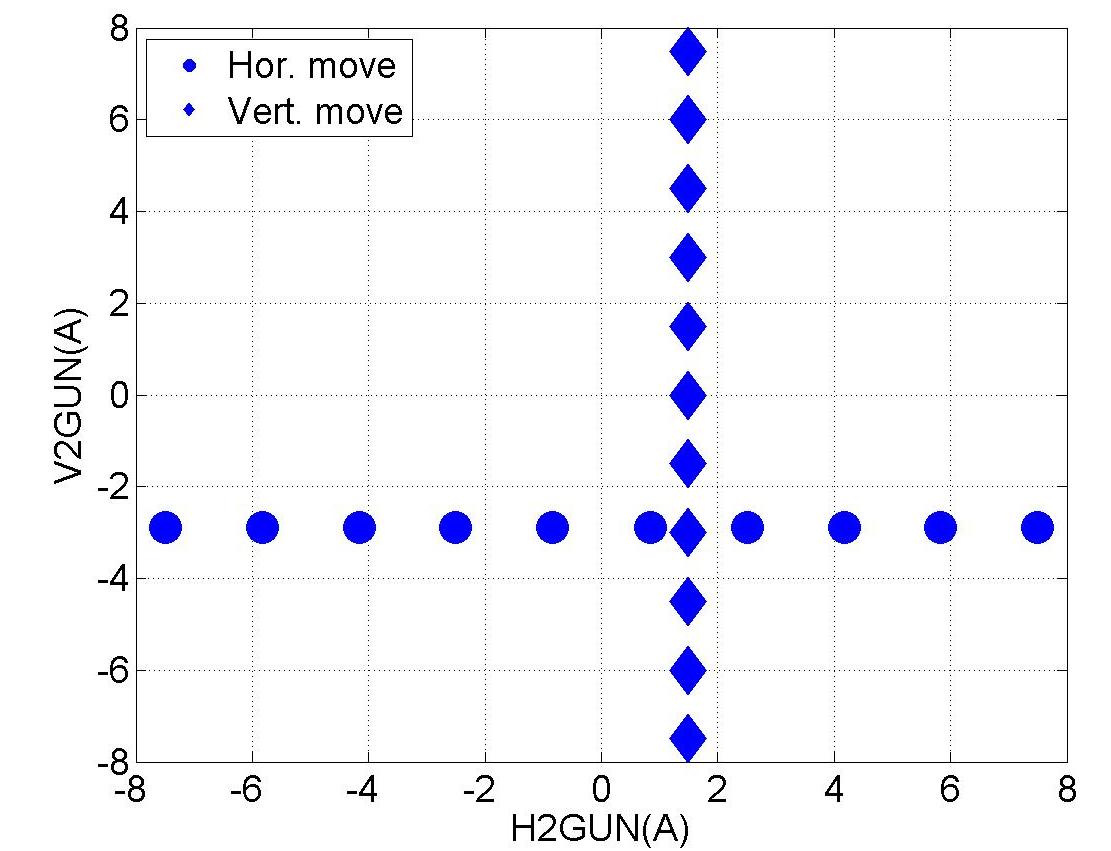}
\label{4D-HV2GUN-cross}
}\\
\subfigure[$x$ vs. $y$ (BPM-A)]{
\includegraphics[width=0.37\textwidth]{4D-9ACC1-cross}
\label{4D-9ACC1-cross}
}
\quad\quad
\subfigure[$x$ vs. $y$ (BPM-B)]{
\includegraphics[width=0.37\textwidth]{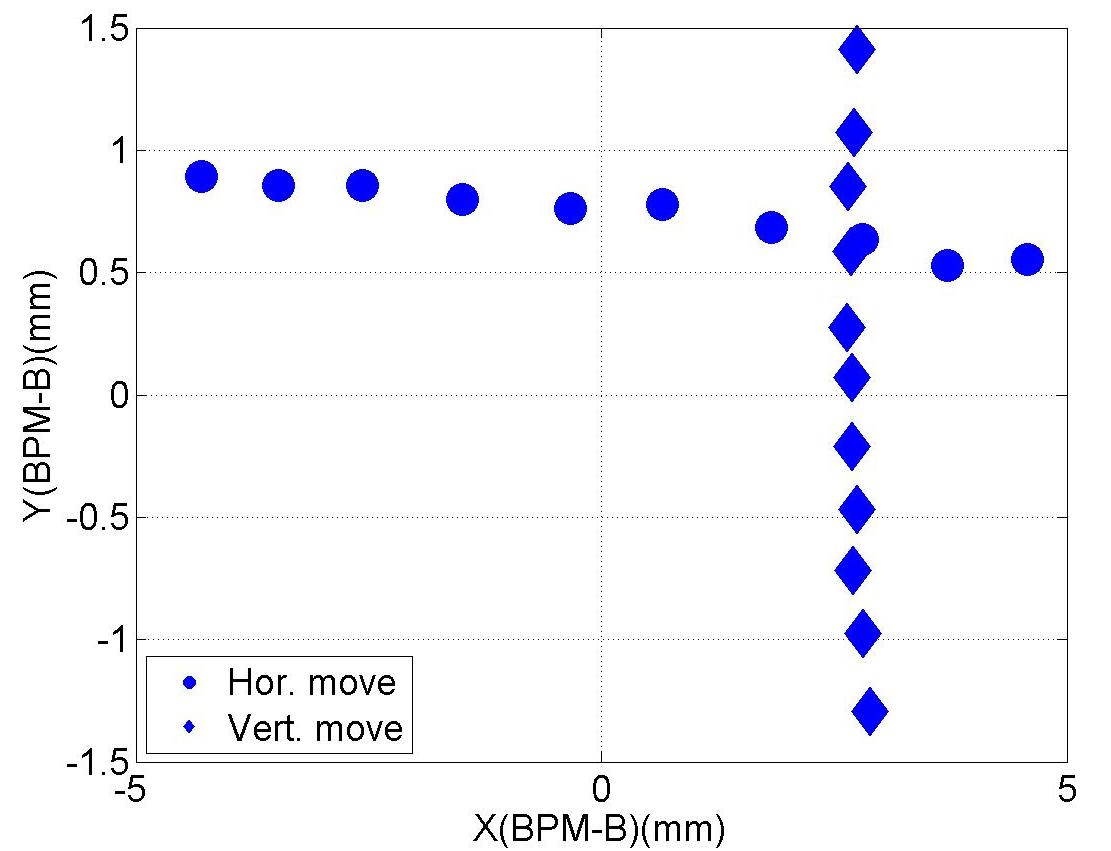}
\label{4D-2UBC2-cross}
}
\subfigure[$x$ vs. $x'$ (BPM-A)]{
\includegraphics[width=0.37\textwidth]{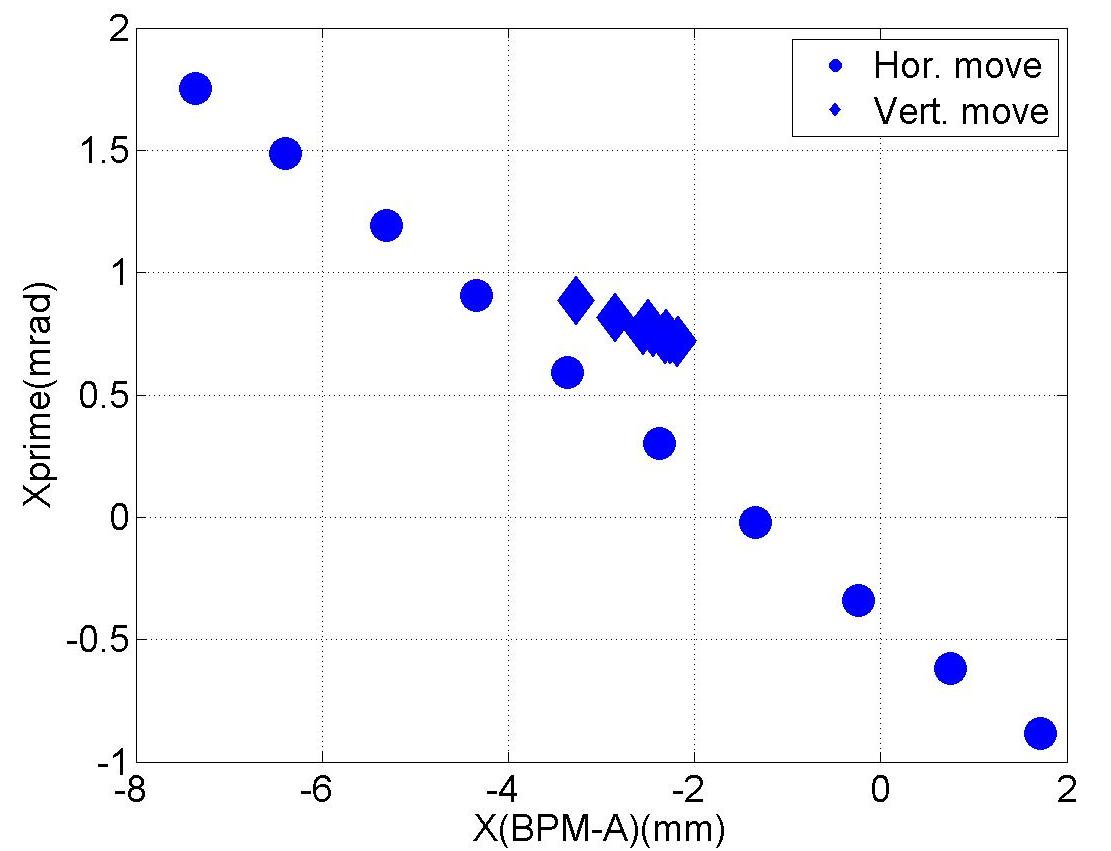}
\label{4D-9ACC1-x-xp-cross}
}
\quad\quad
\subfigure[$y$ vs. $y'$ (BPM-A)]{
\includegraphics[width=0.37\textwidth]{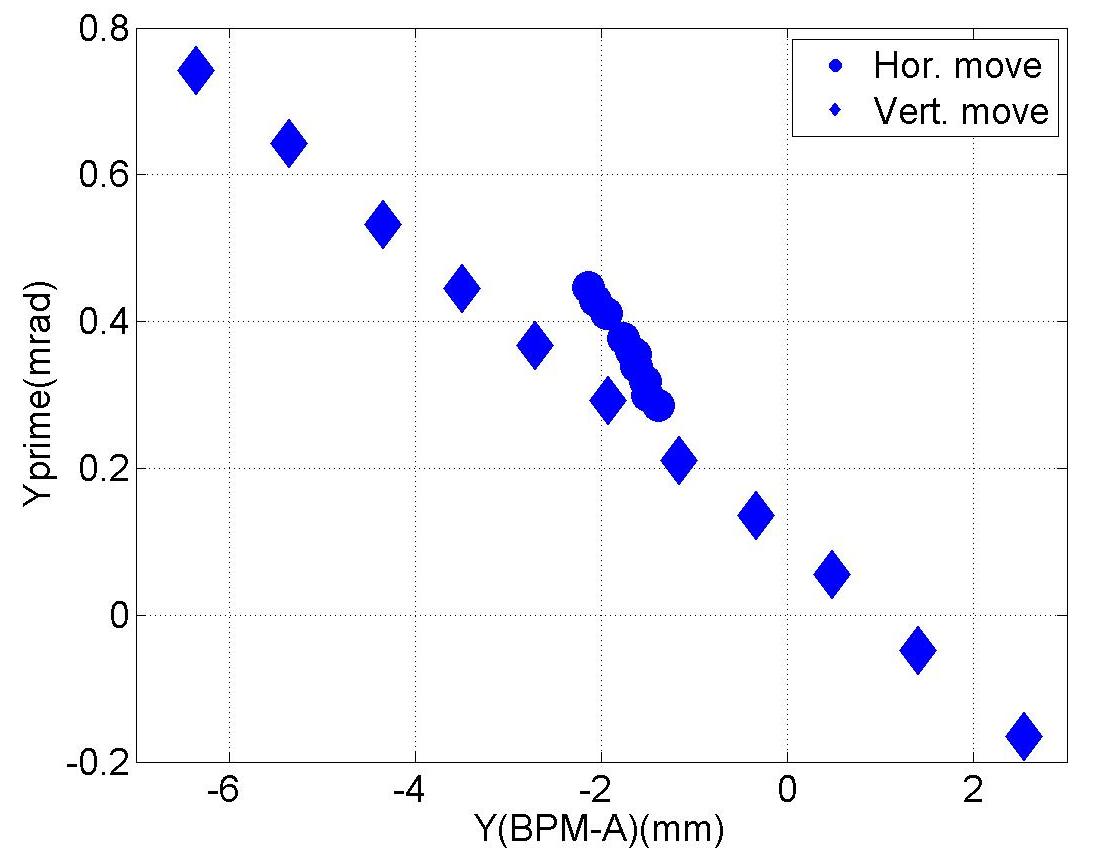}
\label{4D-9ACC1-y-yp-cross}
}
\subfigure[$x$ vs. $x'$ (BPM-B)]{
\includegraphics[width=0.37\textwidth]{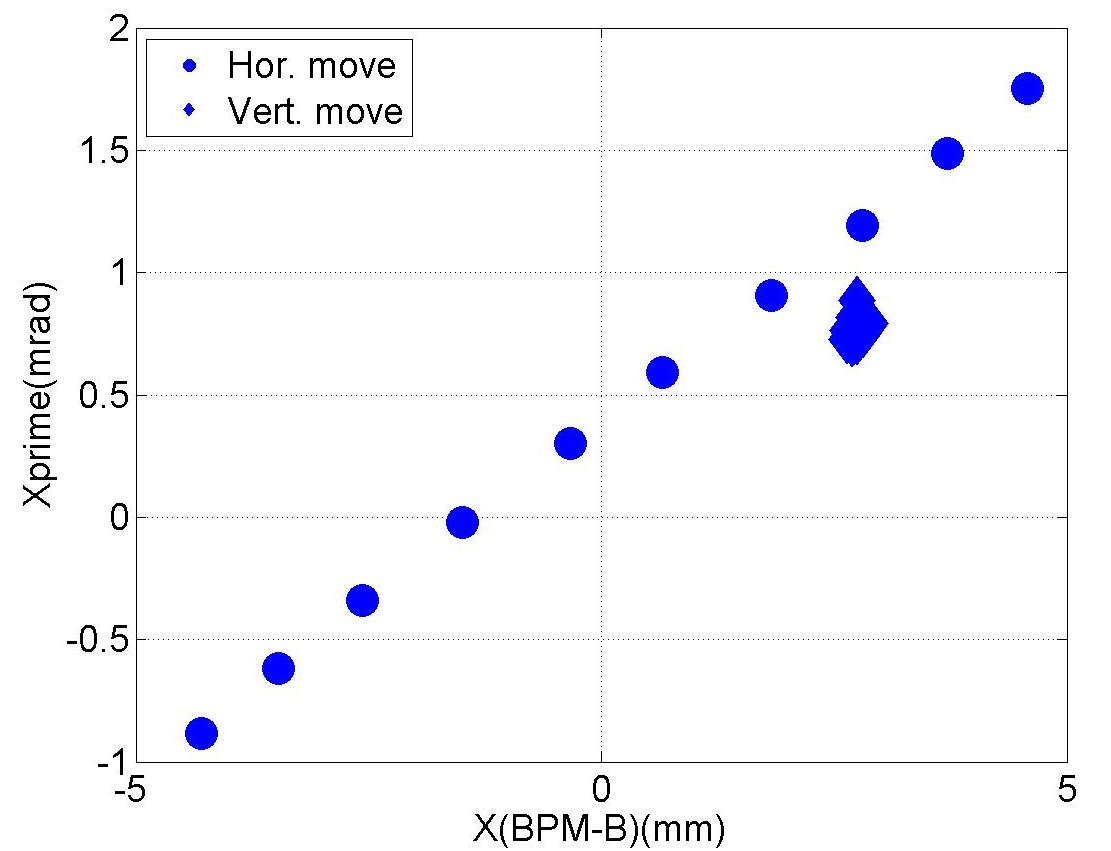}
\label{4D-2UBC2-x-xp-cross}
}
\quad\quad
\subfigure[$y$ vs. $y'$ (BPM-B)]{
\includegraphics[width=0.37\textwidth]{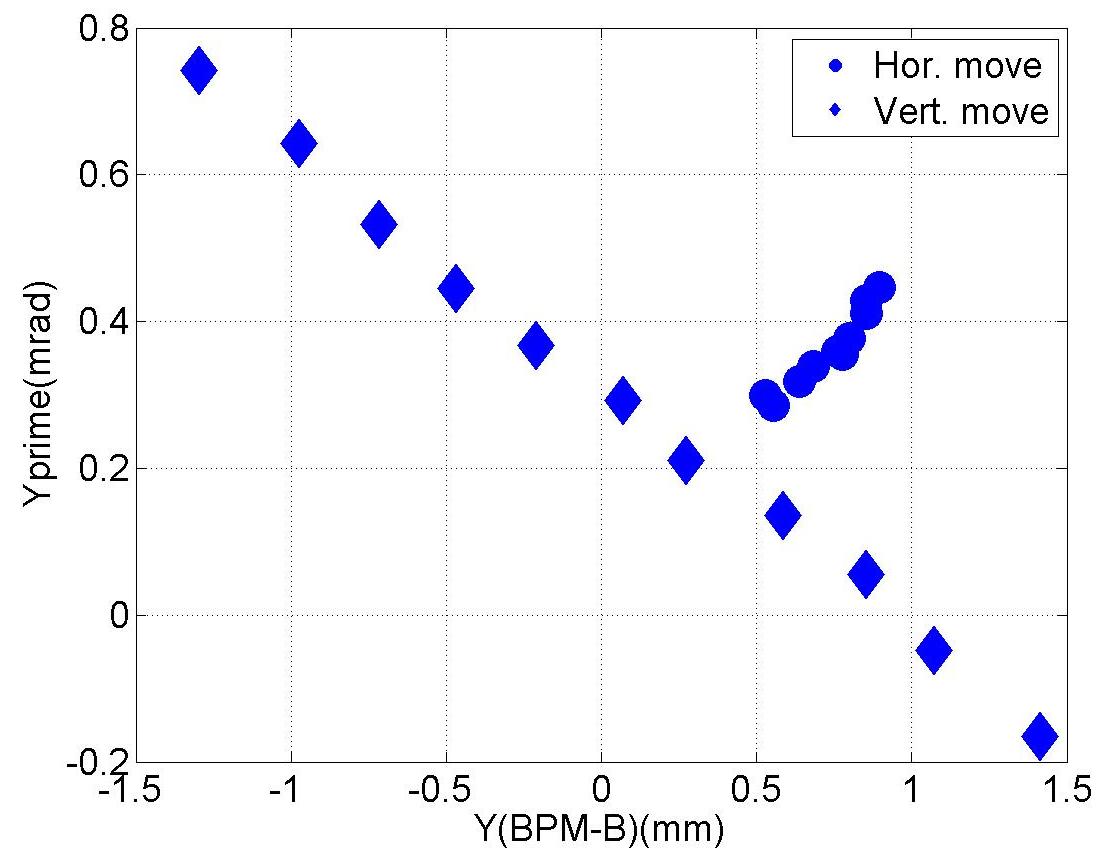}
\label{4D-2UBC2-y-yp-cross}
}
\caption{BPM readouts during 2D cross movement for measuring the dipole beam-pipe modes. The quadrupoles were still on during the scan.}
\label{4D-cross}
\end{figure}
\begin{figure}\center
\subfigure[Steerer (2GUN)]{
\includegraphics[width=0.3\textwidth]{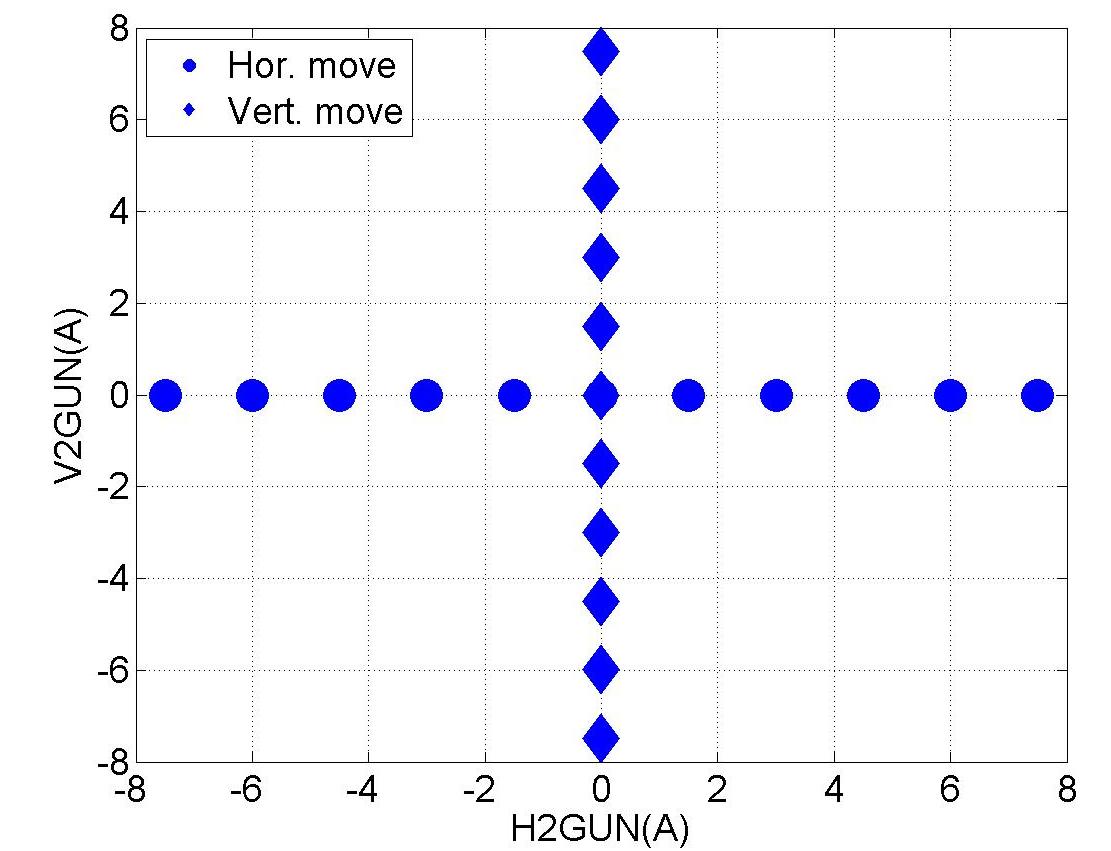}
\label{4D-HV2GUN-cross}
}
\subfigure[$x$ vs. $y$ (BPM-A)]{
\includegraphics[width=0.3\textwidth]{4D-9ACC1-cross-D5}
\label{4D-9ACC1-cross}
}
\subfigure[$x$ vs. $y$ (BPM-B)]{
\includegraphics[width=0.3\textwidth]{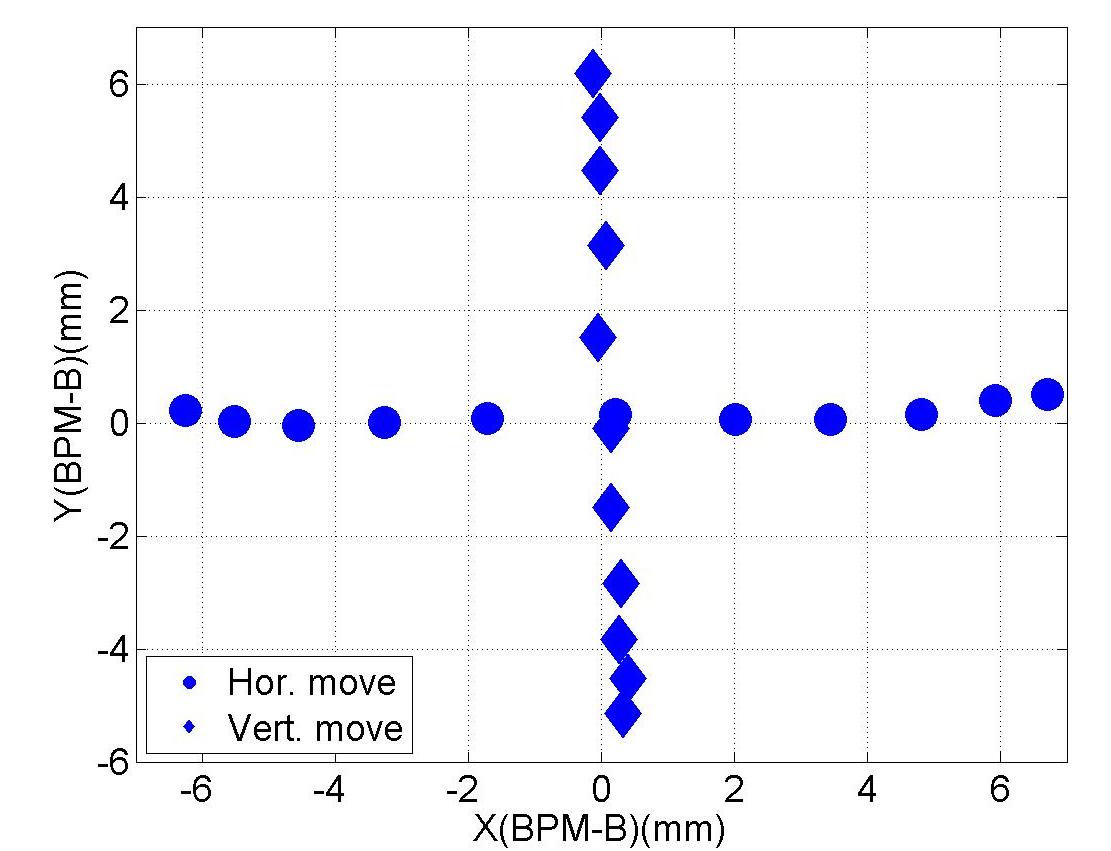}
\label{4D-2UBC2-cross}
}
\subfigure[$x$ vs. $y$ (Interpolated into C1)]{
\includegraphics[width=0.3\textwidth]{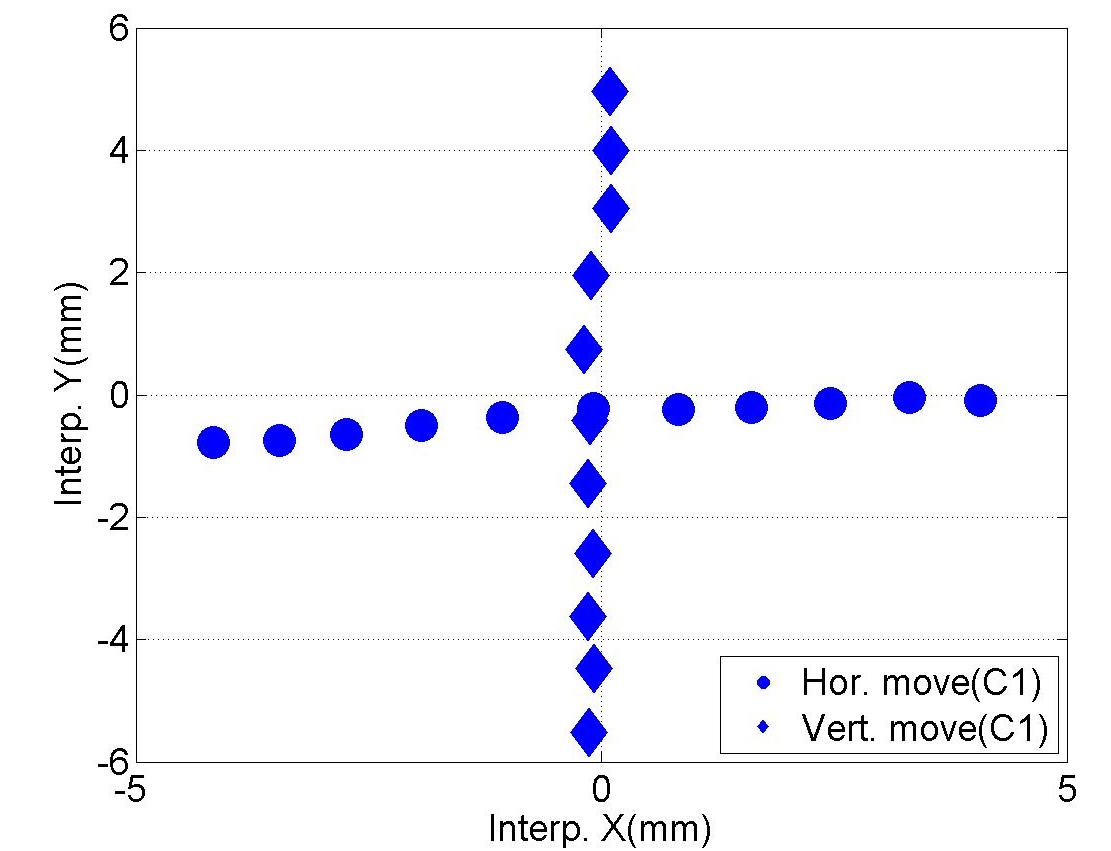}
\label{4D-interp-C1-D5}
}
\subfigure[$x$ vs. $x'$ (Interpolated into C1)]{
\includegraphics[width=0.3\textwidth]{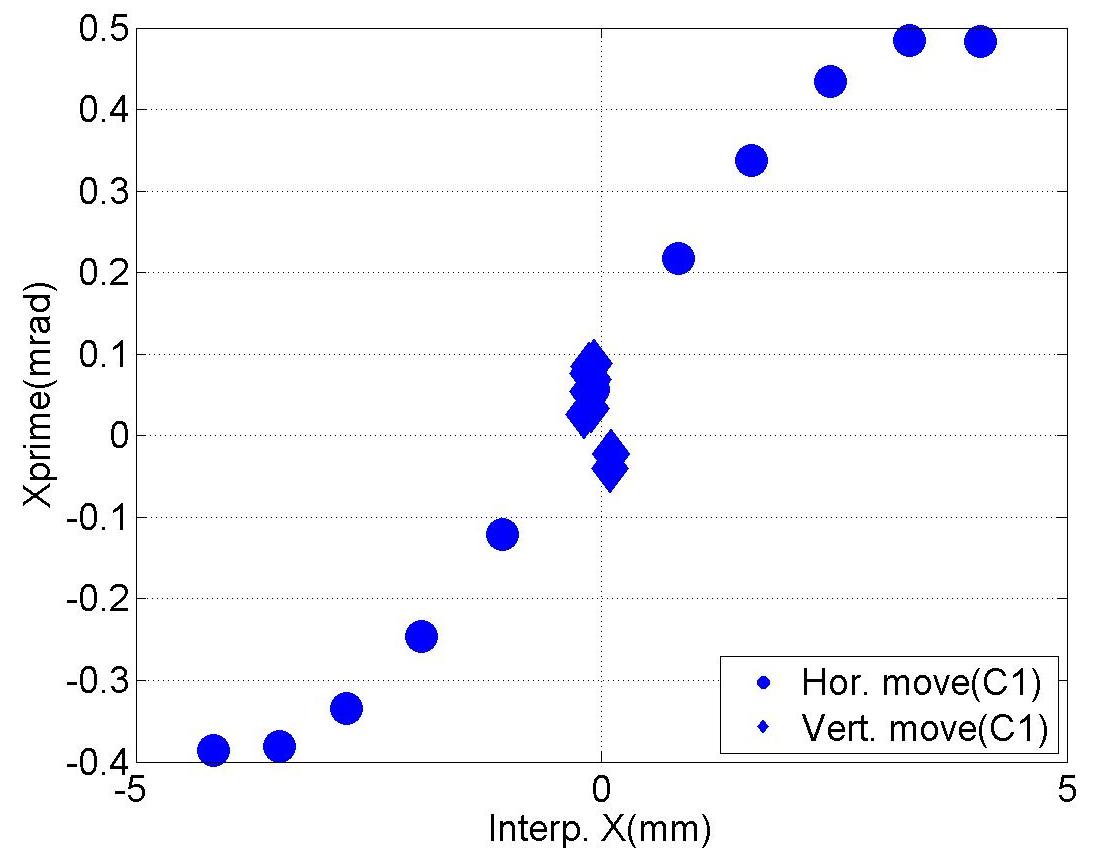}
\label{4D-interp-x-xp-C1-D5}
}
\subfigure[$y$ vs. $y'$ (Interpolated into C1)]{
\includegraphics[width=0.3\textwidth]{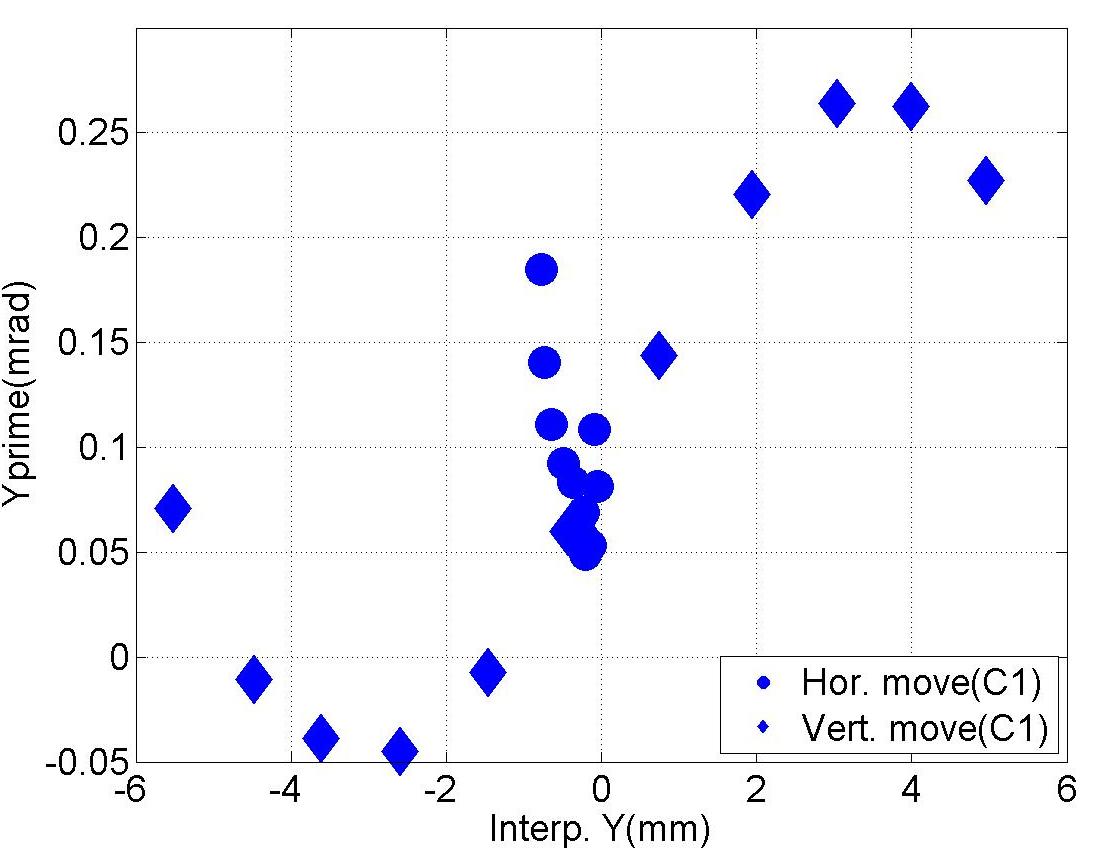}
\label{4D-interp-y-yp-C1-D5}
}
\subfigure[$x$ vs. $y$ (Interpolated into C2)]{
\includegraphics[width=0.3\textwidth]{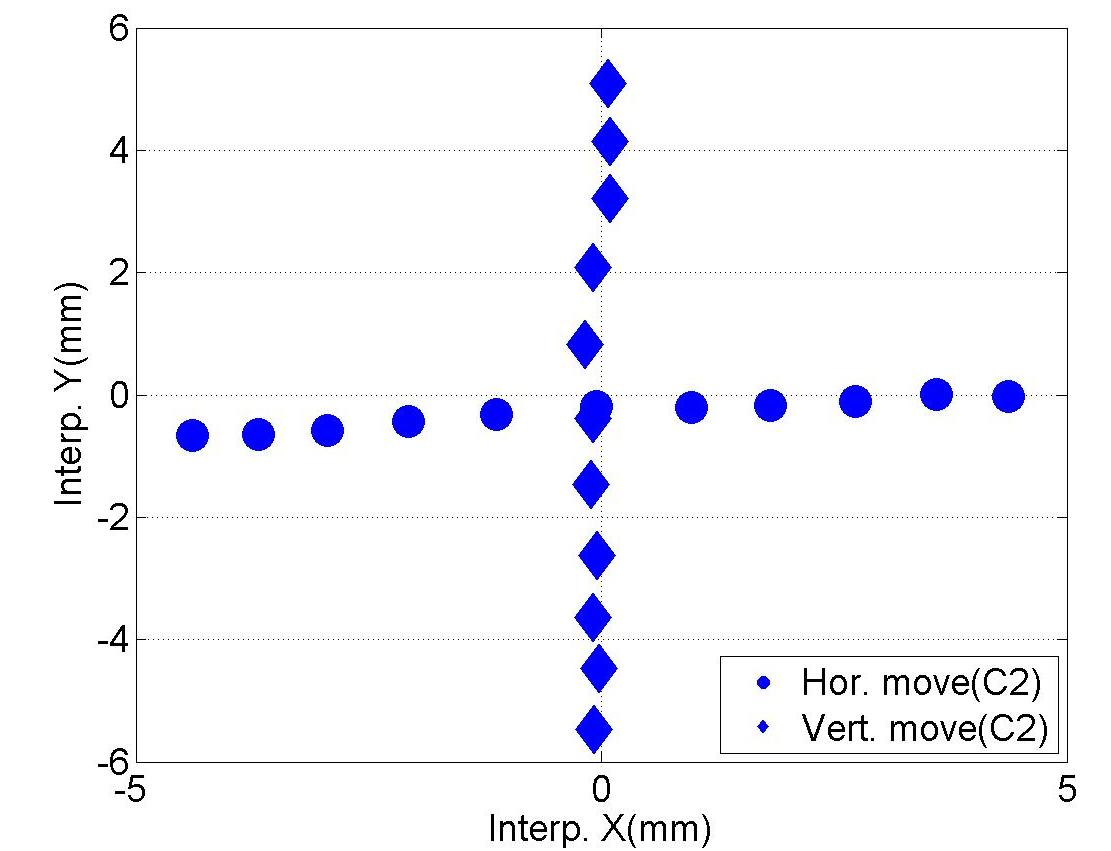}
\label{4D-interp-C2-D5}
}
\subfigure[$x$ vs. $x'$ (Interpolated into C2)]{
\includegraphics[width=0.3\textwidth]{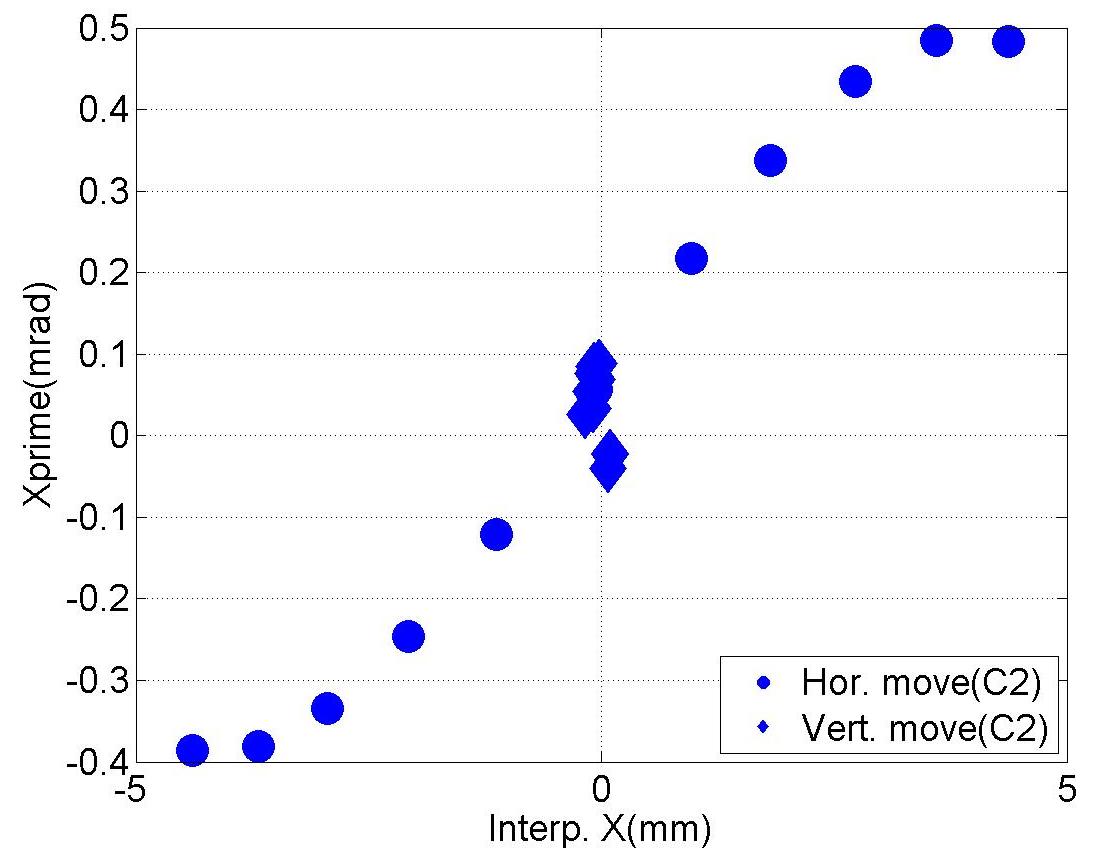}
\label{4D-interp-x-xp-C2-D5}
}
\subfigure[$y$ vs. $y'$ (Interpolated into C2)]{
\includegraphics[width=0.3\textwidth]{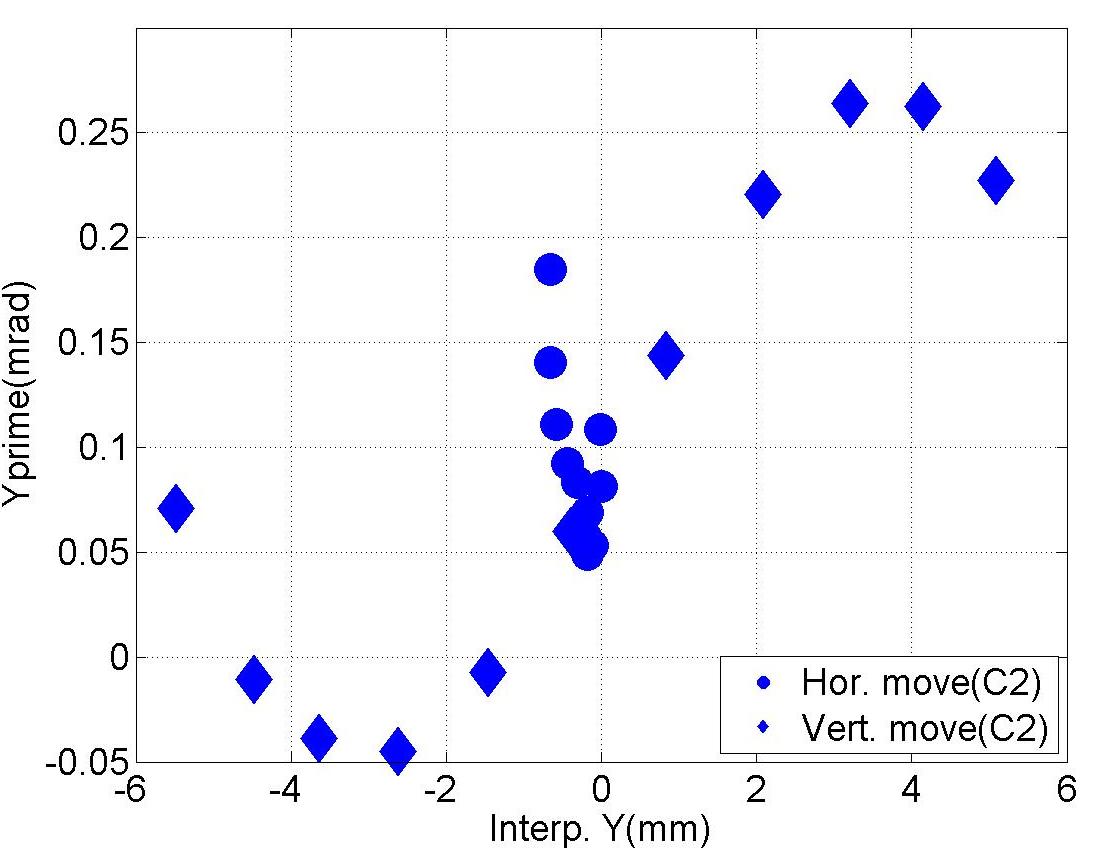}
\label{4D-interp-y-yp-C2-D5}
}
\subfigure[$x$ vs. $y$ (Interpolated into C3)]{
\includegraphics[width=0.3\textwidth]{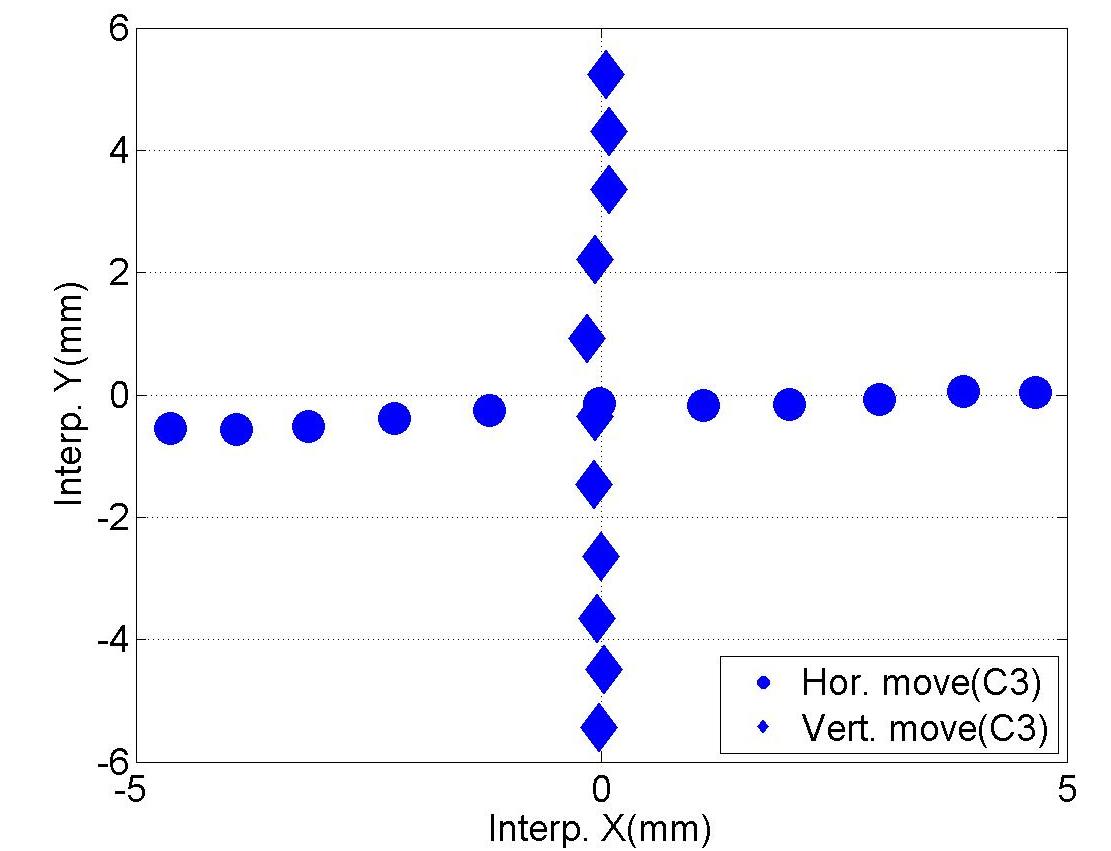}
\label{4D-interp-C3-D5}
}
\subfigure[$x$ vs. $x'$ (Interpolated into C3)]{
\includegraphics[width=0.3\textwidth]{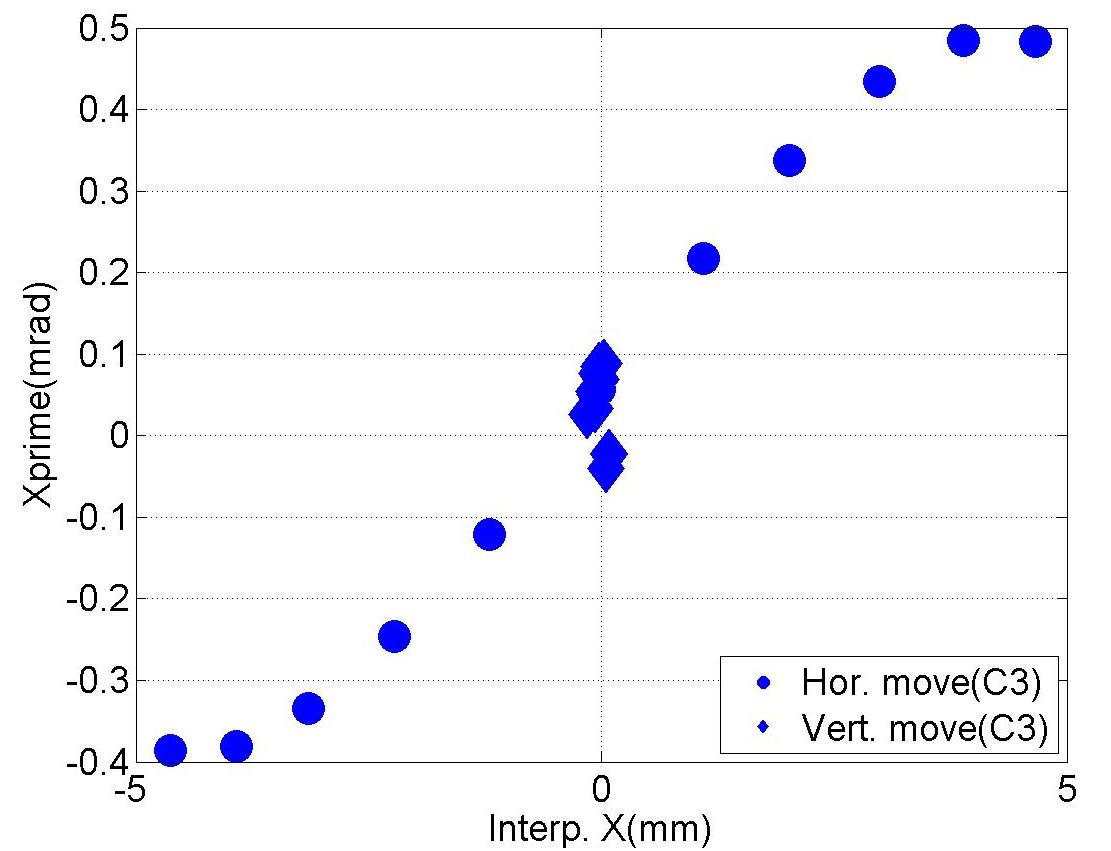}
\label{4D-interp-x-xp-C3-D5}
}
\subfigure[$y$ vs. $y'$ (Interpolated into C3)]{
\includegraphics[width=0.3\textwidth]{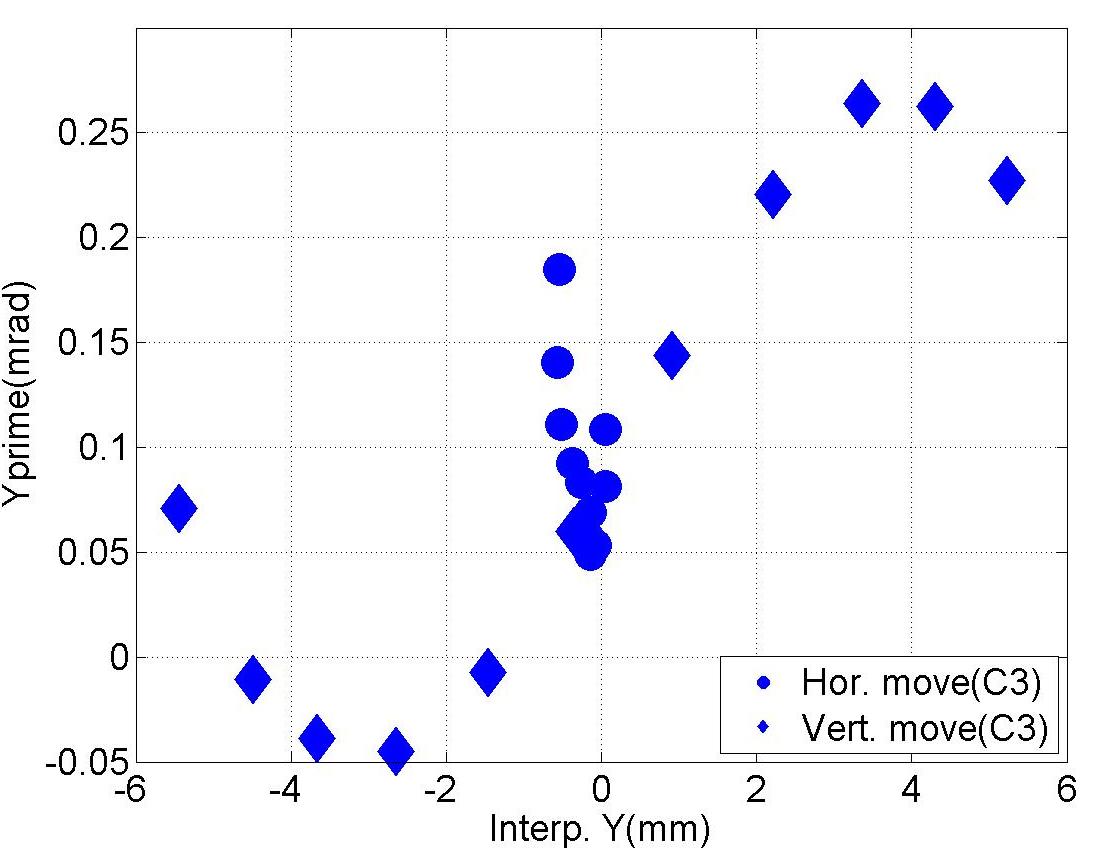}
\label{4D-interp-y-yp-C3-D5}
}
\subfigure[$x$ vs. $y$ (Interpolated into C4)]{
\includegraphics[width=0.3\textwidth]{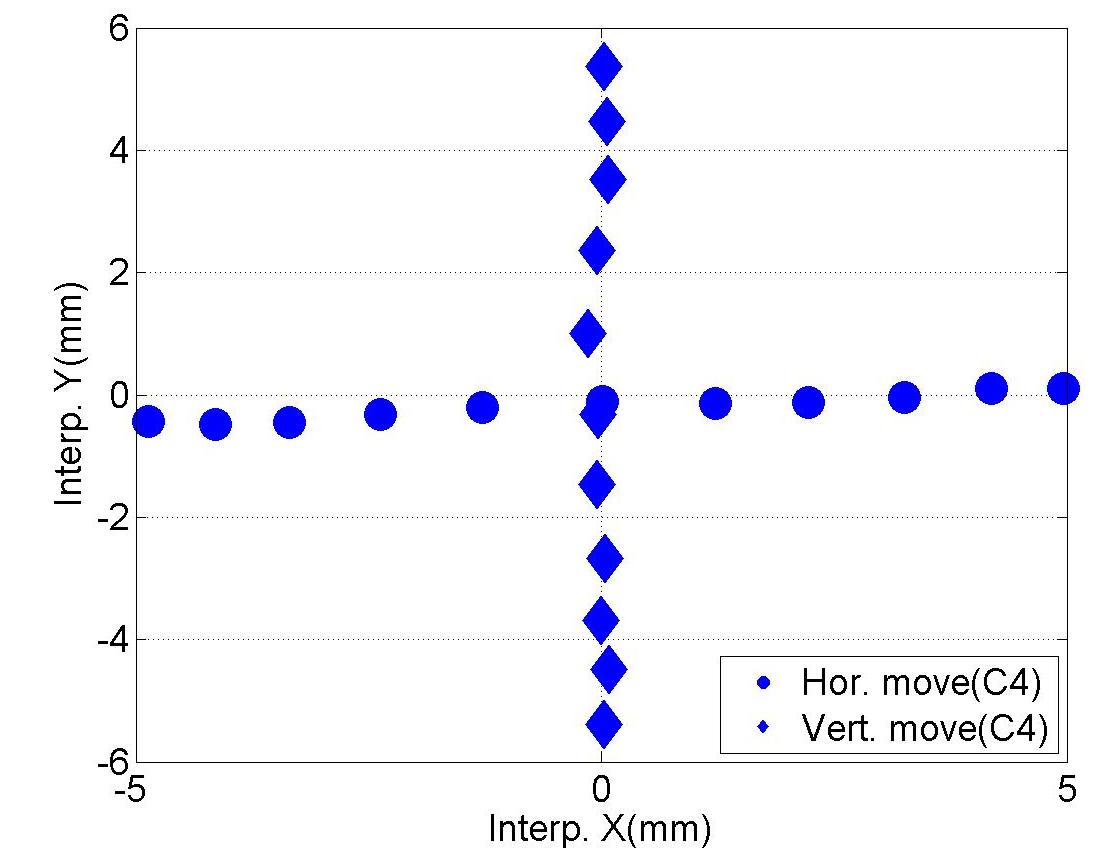}
\label{4D-interp-C4-D5}
}
\subfigure[$x$ vs. $x'$ (Interpolated into C4)]{
\includegraphics[width=0.3\textwidth]{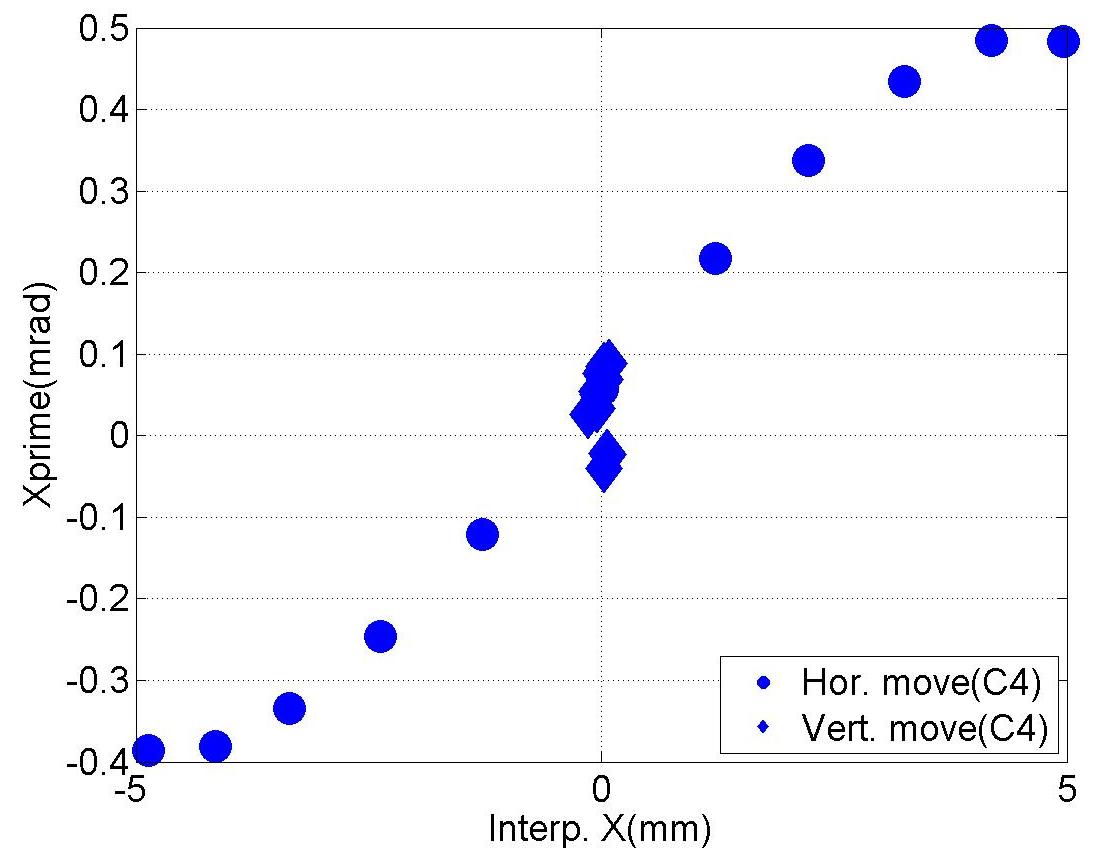}
\label{4D-interp-x-xp-C4-D5}
}
\subfigure[$y$ vs. $y'$ (Interpolated into C4)]{
\includegraphics[width=0.3\textwidth]{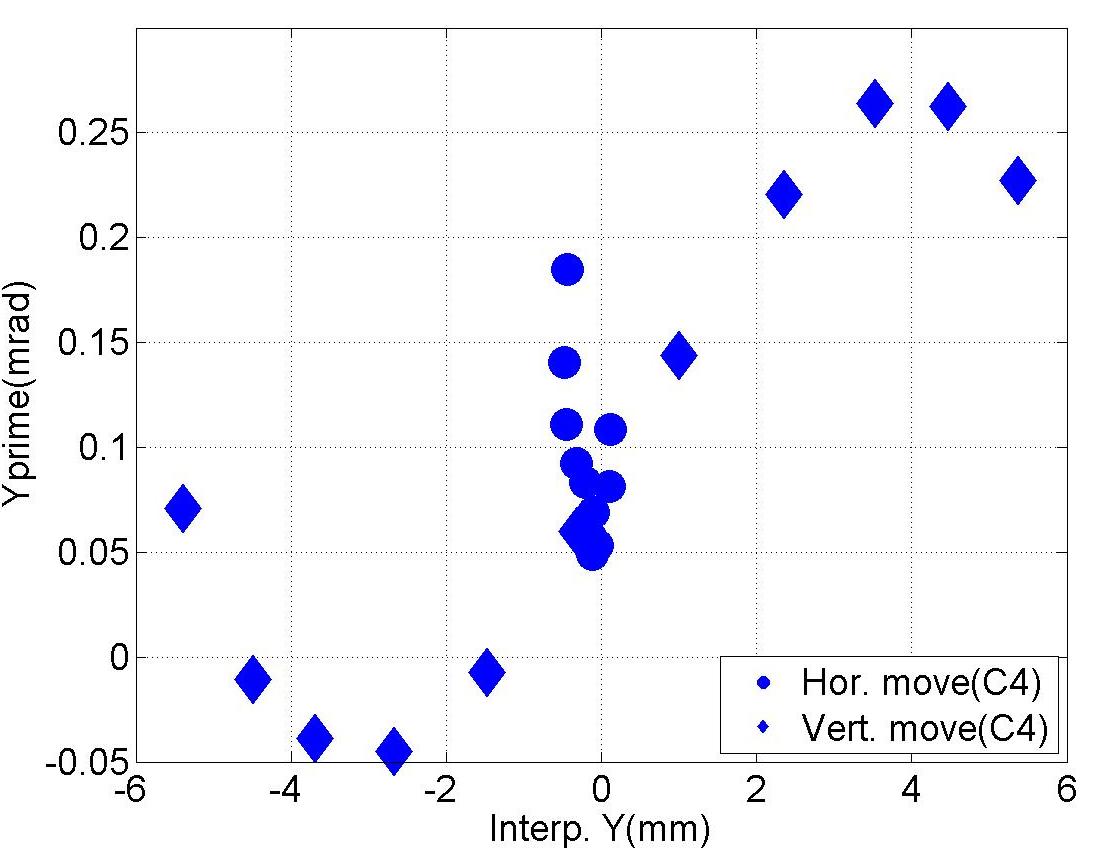}
\label{4D-interp-y-yp-C4-D5}
}
\caption{BPM readouts during 2D cross movement for measuring the f{}ifth dipole band.}
\label{4D-cross}
\end{figure}
\begin{figure}\center
\subfigure[Steerer (2GUN)]{
\includegraphics[width=0.37\textwidth]{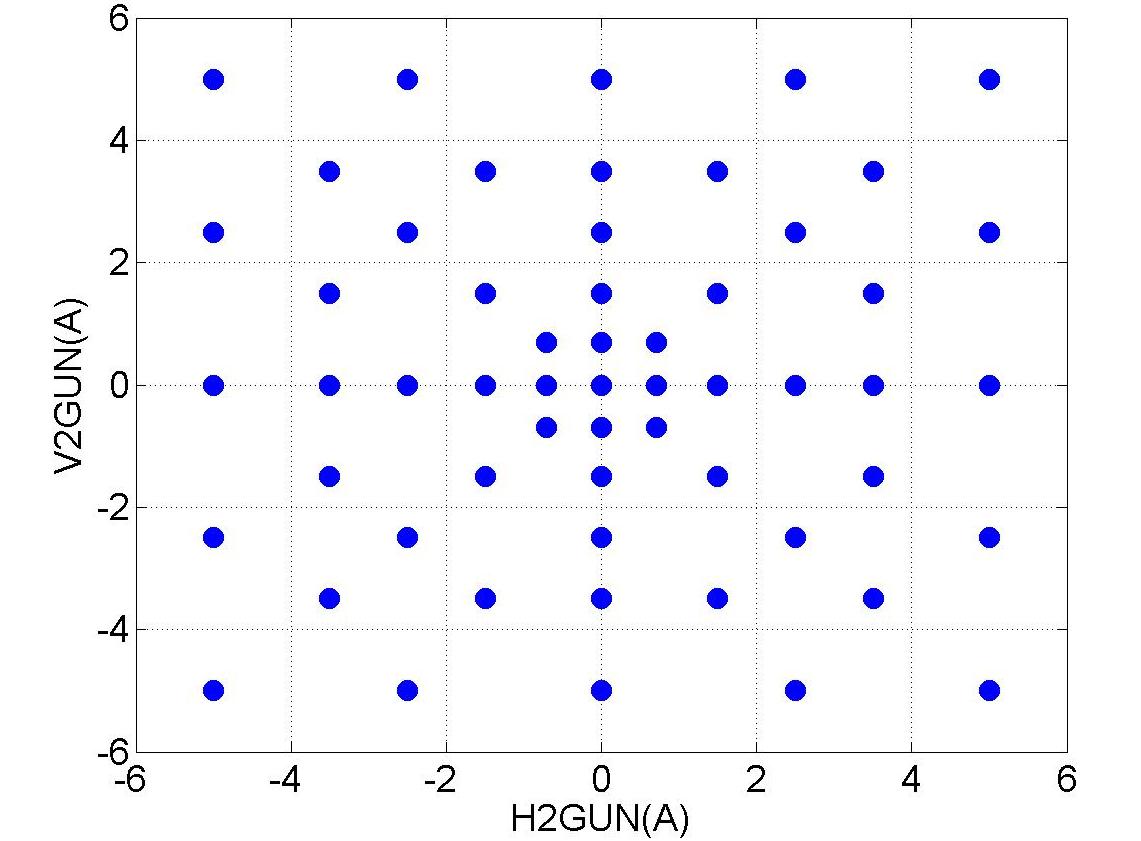}
\label{4D-HV2GUN-grid}
}\\
\subfigure[$x$ vs. $y$ (BPM-A)]{
\includegraphics[width=0.37\textwidth]{4D-9ACC1-grid}
\label{4D-9ACC1-grid}
}
\quad\quad
\subfigure[$x$ vs. $y$ (BPM-B)]{
\includegraphics[width=0.37\textwidth]{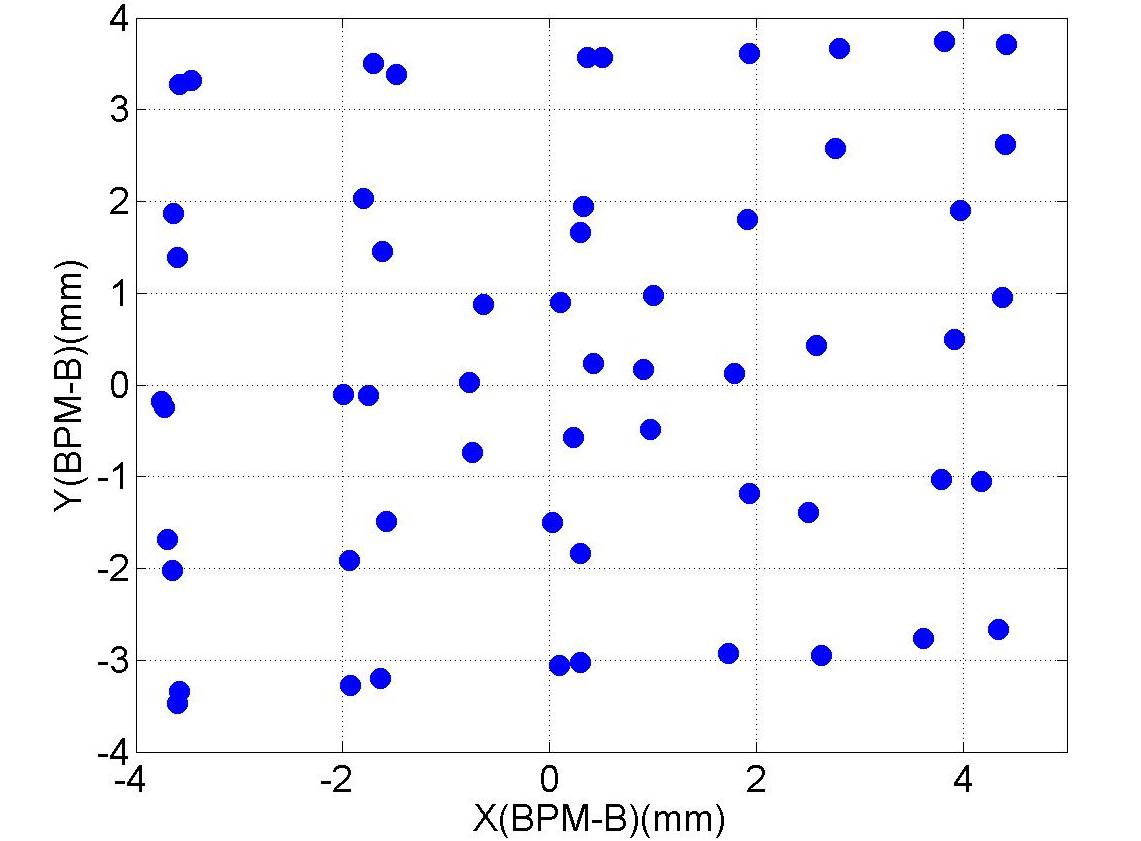}
\label{4D-2UBC2-grid}
}
\subfigure[$x$ vs. $x'$ (BPM-A)]{
\includegraphics[width=0.37\textwidth]{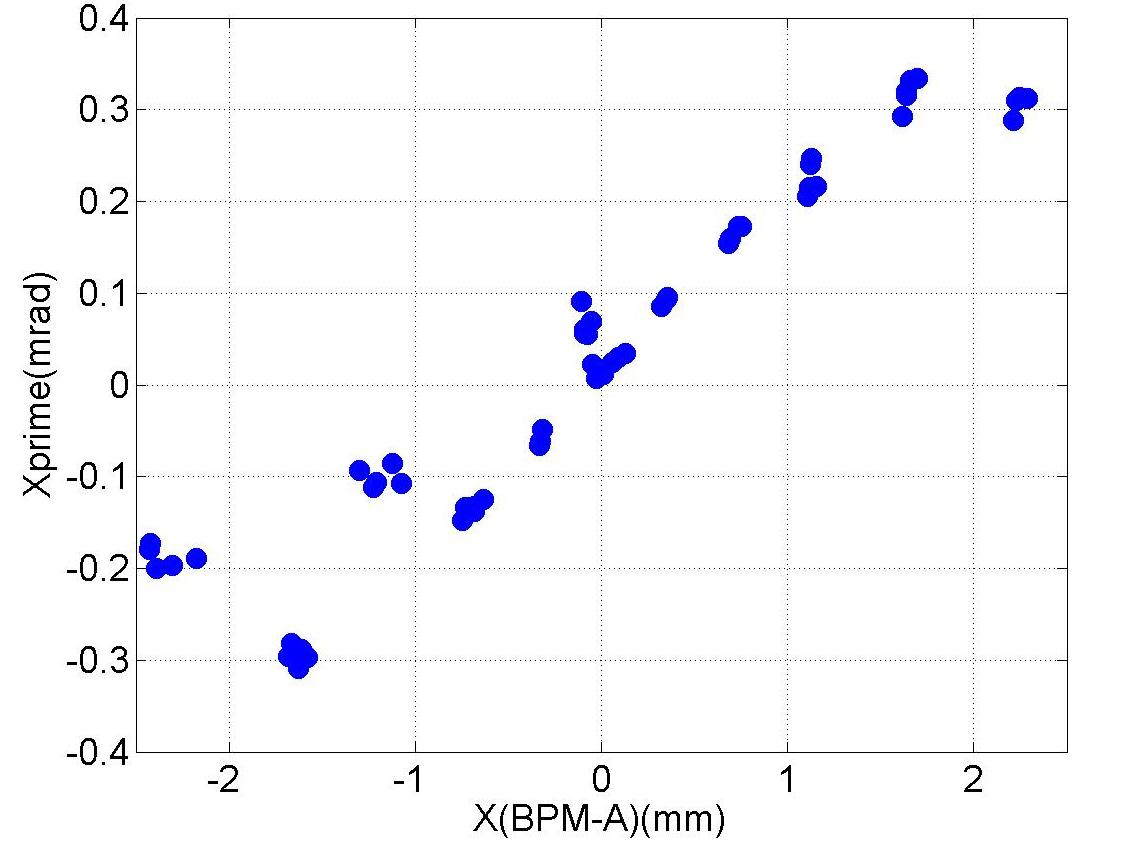}
\label{4D-9ACC1-x-xp-grid}
}
\quad\quad
\subfigure[$y$ vs. $y'$ (BPM-A)]{
\includegraphics[width=0.37\textwidth]{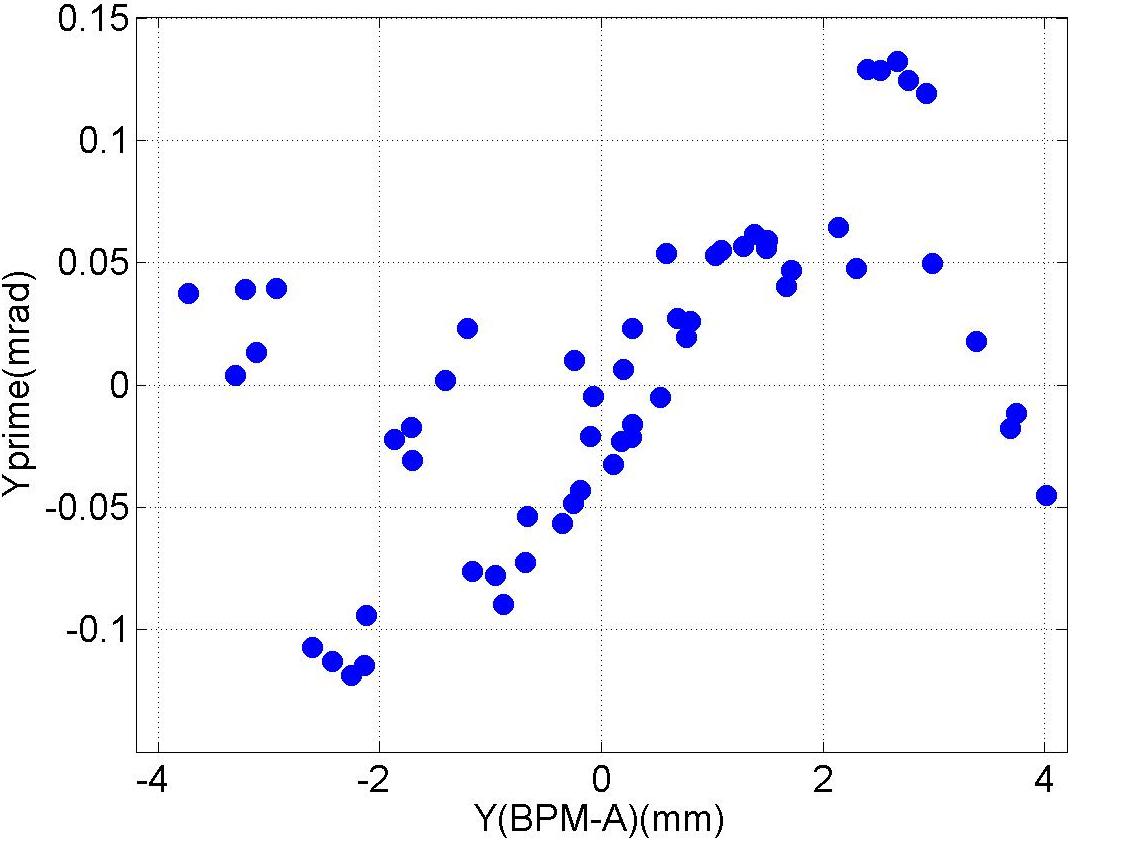}
\label{4D-9ACC1-y-yp-grid}
}
\subfigure[$x$ vs. $x'$ (BPM-B)]{
\includegraphics[width=0.37\textwidth]{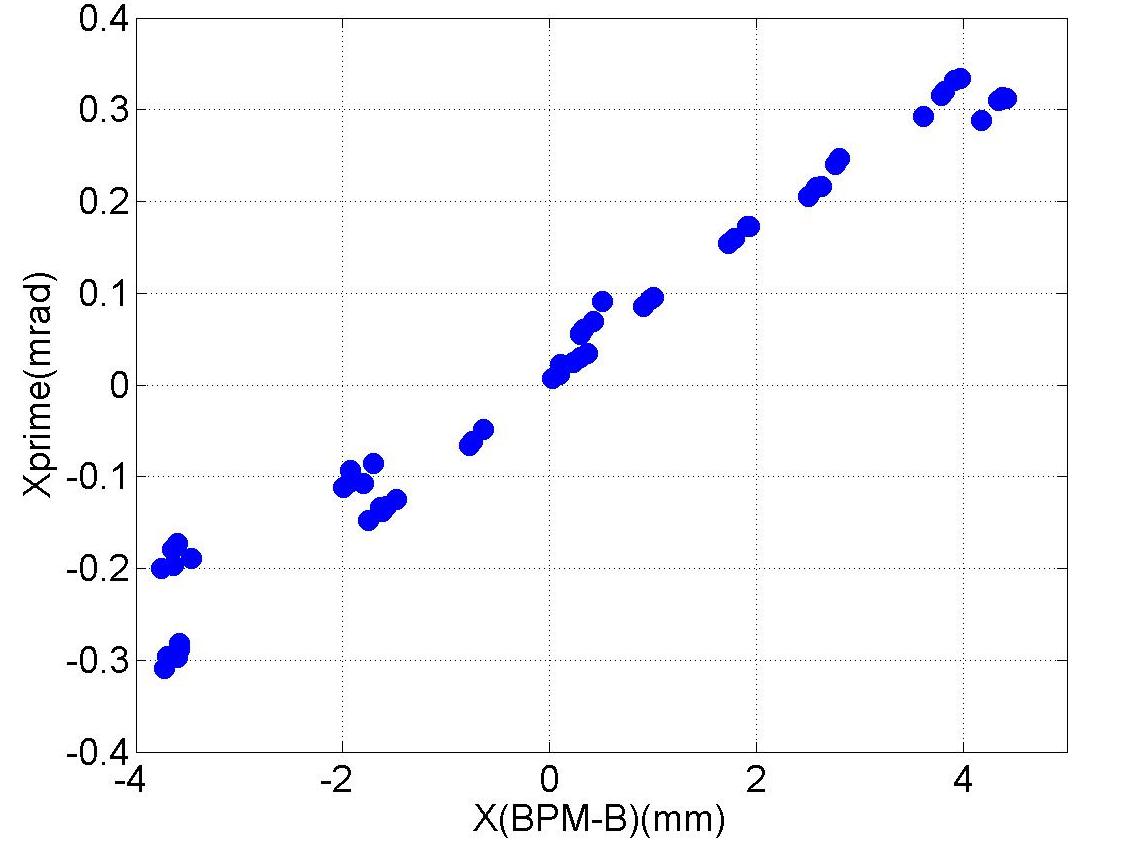}
\label{4D-2UBC2-x-xp-grid}
}
\quad\quad
\subfigure[$y$ vs. $y'$ (BPM-B)]{
\includegraphics[width=0.37\textwidth]{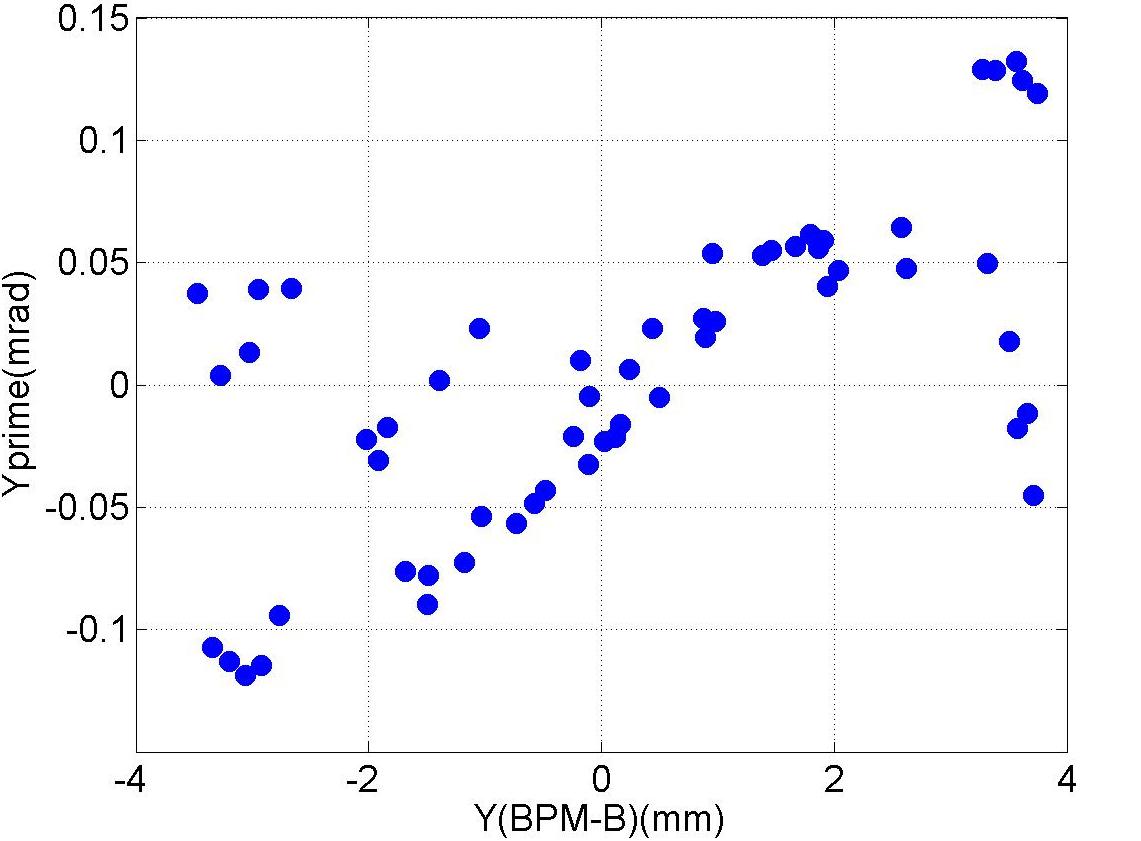}
\label{4D-2UBC2-y-yp-grid}
}
\caption{BPM readouts during 2D grid movement for measuring the dipole beam-pipe modes.}
\label{4D-grid}
\end{figure}
\begin{figure}\center
\subfigure[Steerer (2GUN)]{
\includegraphics[width=0.37\textwidth]{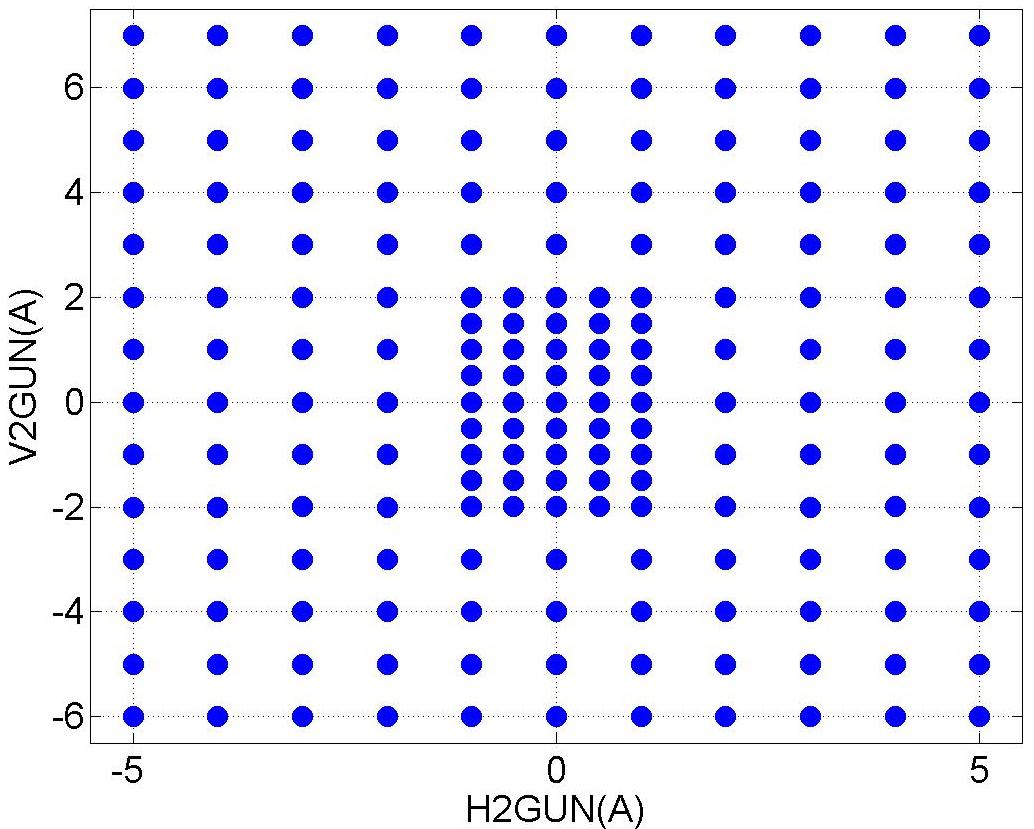}
\label{4D-HV2GUN-grid-D5}
}\\
\subfigure[$x$ vs. $y$ (BPM-A)]{
\includegraphics[width=0.37\textwidth]{4D-9ACC1-grid-D5}
\label{4D-9ACC1-grid-D5}
}
\quad\quad
\subfigure[$x$ vs. $y$ (BPM-B)]{
\includegraphics[width=0.37\textwidth]{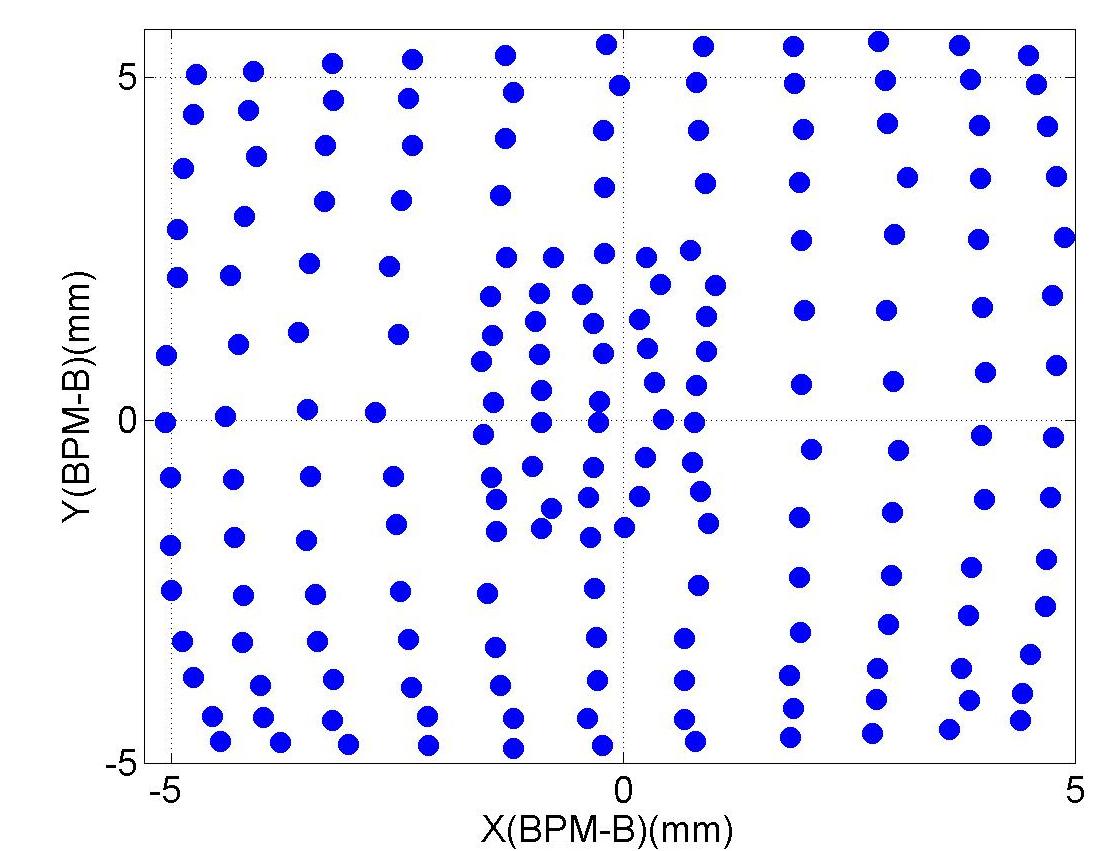}
\label{4D-2UBC2-grid-D5}
}
\subfigure[$x$ vs. $x'$ (BPM-A)]{
\includegraphics[width=0.37\textwidth]{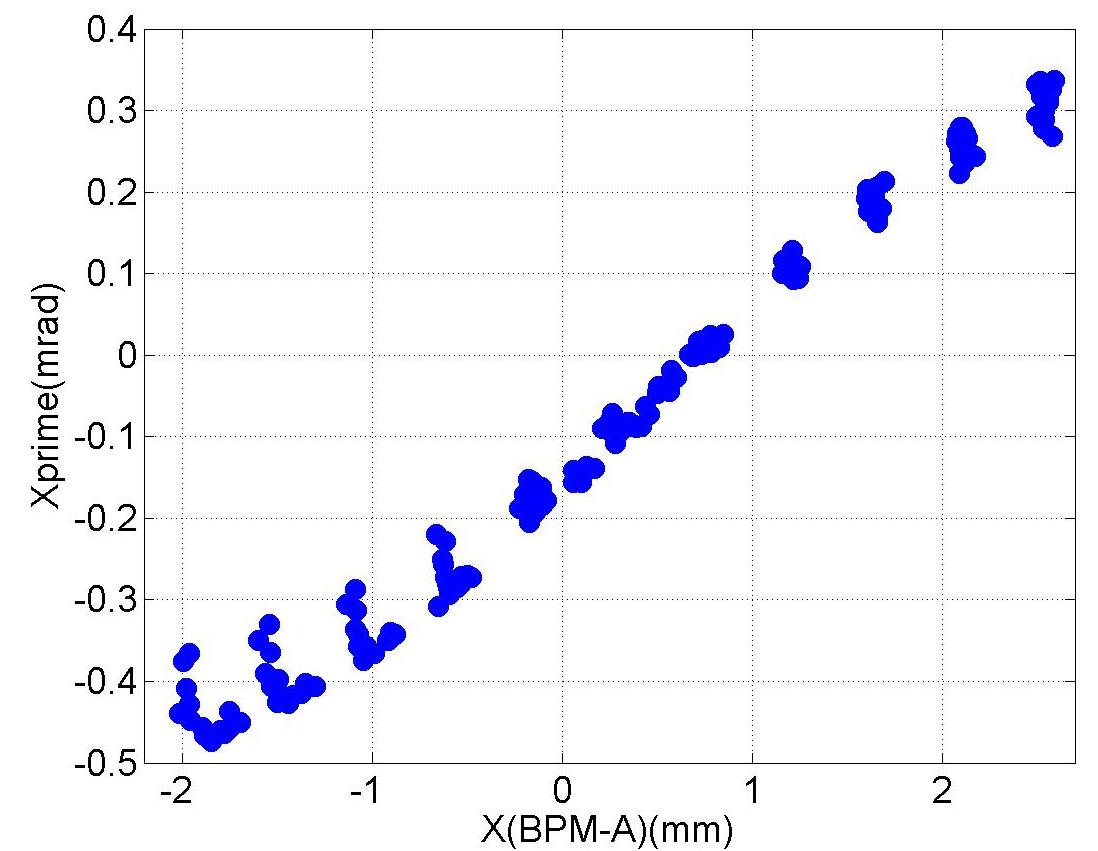}
\label{4D-9ACC1-x-xp-grid-D5}
}
\quad\quad
\subfigure[$y$ vs. $y'$ (BPM-A)]{
\includegraphics[width=0.37\textwidth]{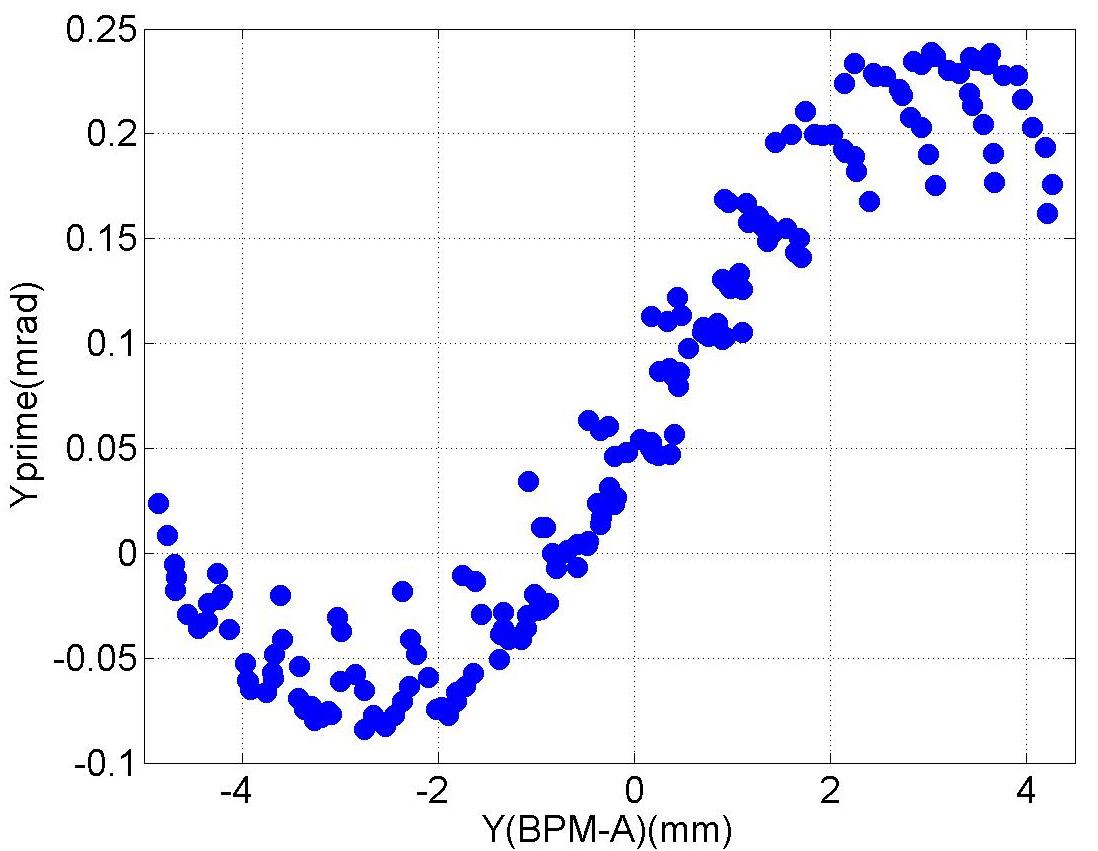}
\label{4D-9ACC1-y-yp-grid-D5}
}
\subfigure[$x$ vs. $x'$ (BPM-B)]{
\includegraphics[width=0.37\textwidth]{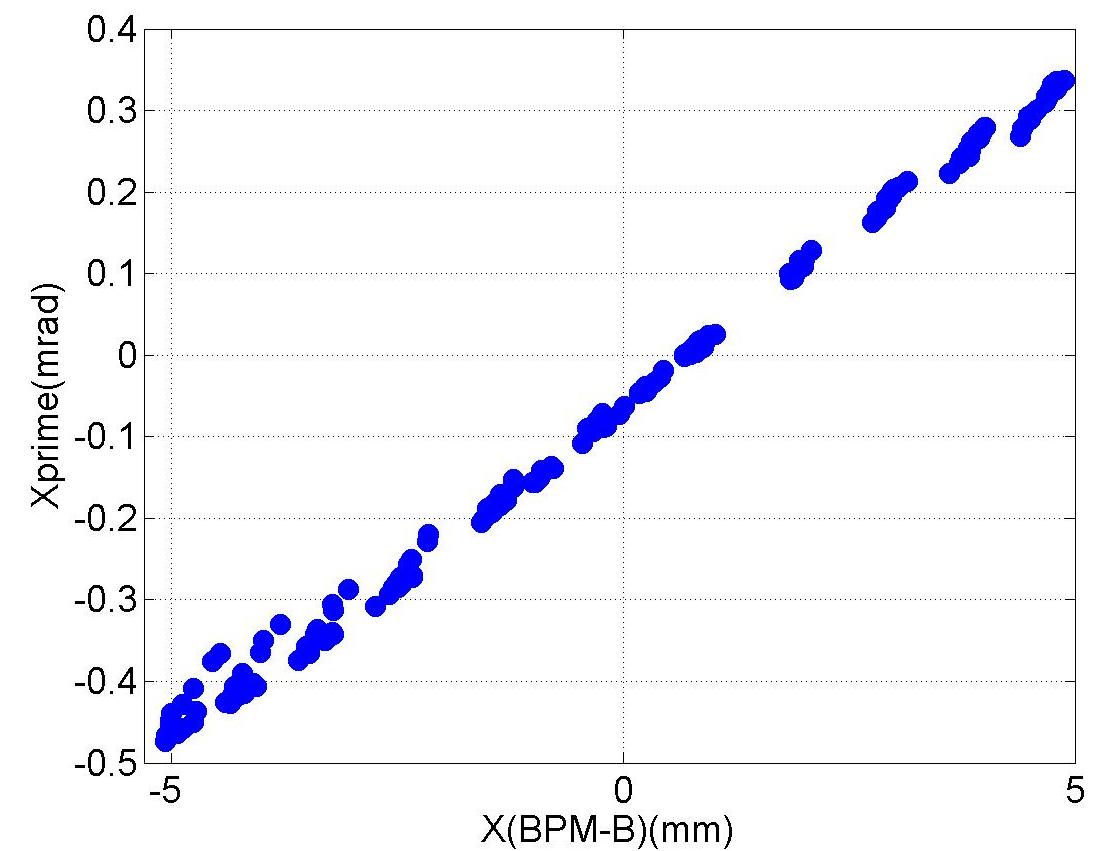}
\label{4D-2UBC2-x-xp-grid-D5}
}
\quad\quad
\subfigure[$y$ vs. $y'$ (BPM-B)]{
\includegraphics[width=0.37\textwidth]{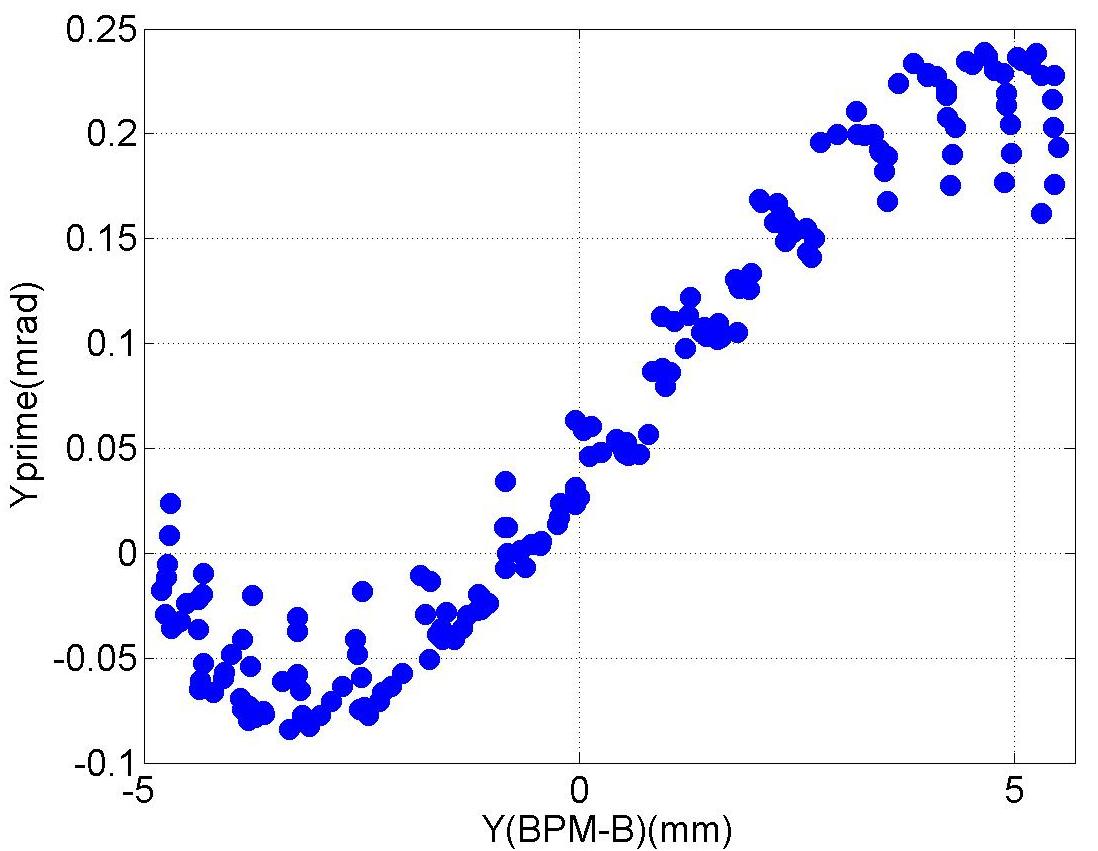}
\label{4D-2UBC2-y-yp-grid-D5}
}
\caption{BPM readouts during 2D grid movement for measuring the f{}ifth dipole bands.}
\label{4D-grid-D5}
\end{figure}

\chapter{Dipole Dependence: The Beam-pipe Modes}\label{app-bp}
Each peak shown in the spectrum ranging from 4-4.15~GHz is f{}itted with Lorentzian distribution (Eq.~\ref{eq:lorfit}). The f{}itted mode amplitude is plotted against $x$ and $y$ readings from BPM-A (Fig.~\ref{hom-setup}) for each coupler. The position interpolation is not applicable for the cross movement as explained in Chapter~\ref{hom-dep:setup}. The mode polarization obtained from a 2D scan is also plotted versus the beam position interpolated into each cavity. This chapter is an extension of Chapter~\ref{hom-dep:bp}. 

\section{BP: HOM Coupler C1H1}
\begin{figure}[h]
\subfigure[Spectrum (C1H1)]{
\includegraphics[width=1\textwidth]{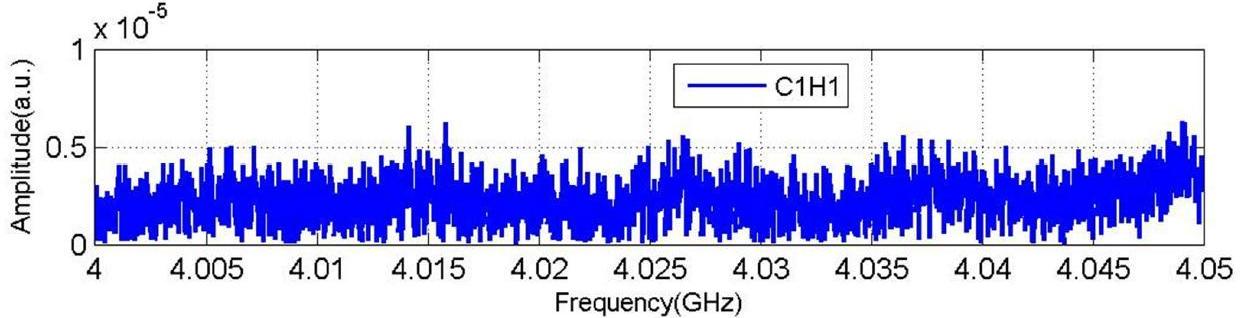}
\label{spec-C1H1-X-1-BP}
}
\subfigure[Spectrum (C1H1)]{
\includegraphics[width=1\textwidth]{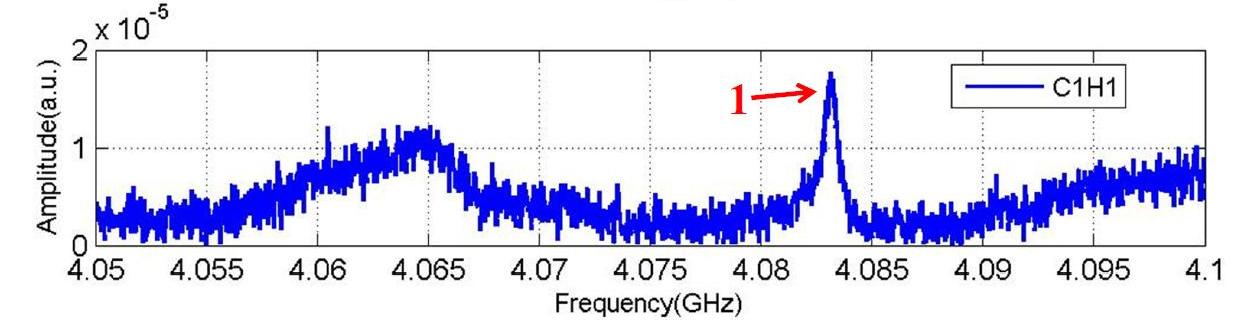}
\label{spec-C1H1-X-2-BP}
}
\subfigure[\#1 ($f$:4.0831GHz; Q:10$^3$)]{
\includegraphics[width=0.31\textwidth]{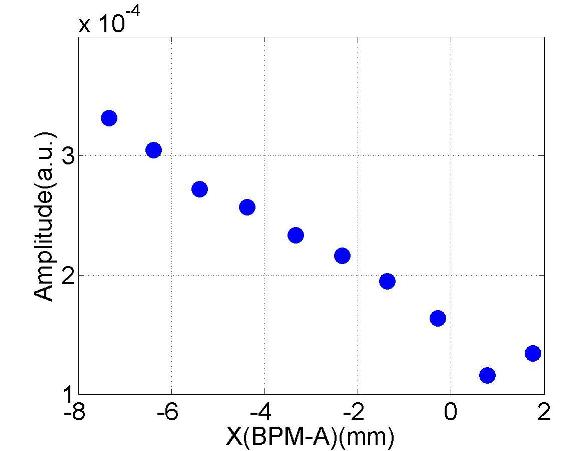}
\label{dep-C1H1-X-1-BP}
}
\subfigure[\#1 ($f$:4.0832GHz; Q:10$^3$)]{
\includegraphics[width=0.31\textwidth]{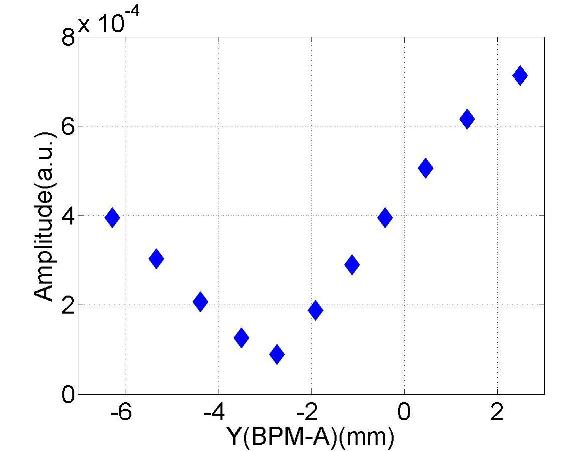}
\label{dep-C1H1-Y-1-BP}
}
\subfigure[\#1 ($f$:4.0831GHz; Q:10$^3$)]{
\includegraphics[width=0.31\textwidth]{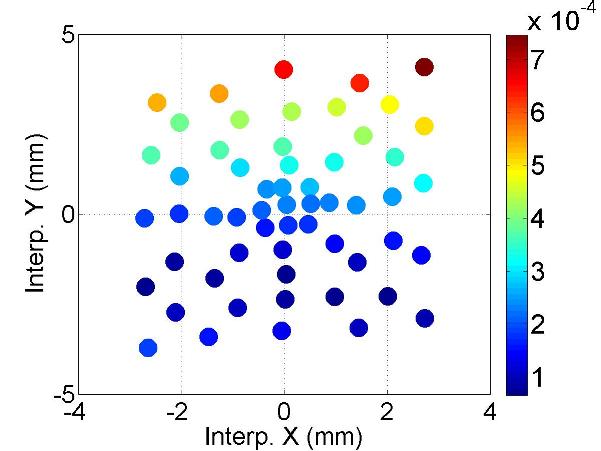}
\label{polar-C1H1-1-BP}
}
\caption{Depedence of the mode amplitude on the transverse beam of{}fset.}
\label{spec-dep-C1H1-XY-1-2-BP}
\end{figure}
\begin{figure}[h]\center
\subfigure[Spectrum (C1H1)]{
\includegraphics[width=1\textwidth]{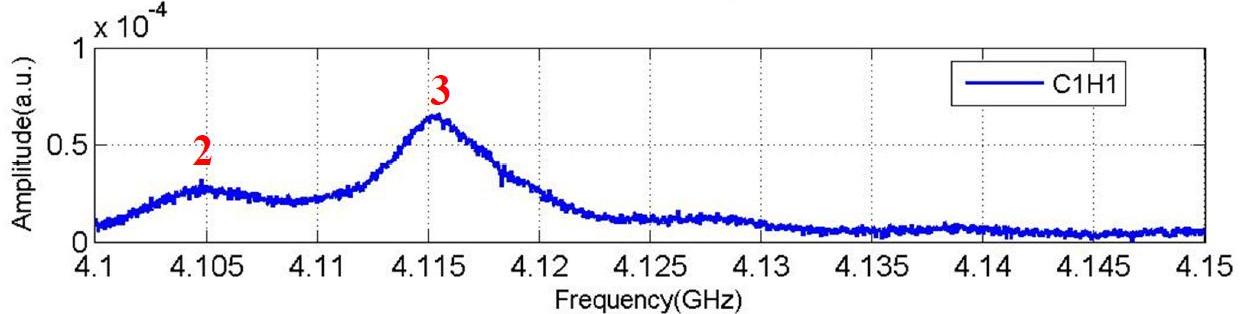}
\label{spec-C1H1-X-3-BP}
}
\subfigure[\#2 ($f$:4.1048GHz; Q:10$^2$)]{
\includegraphics[width=0.31\textwidth]{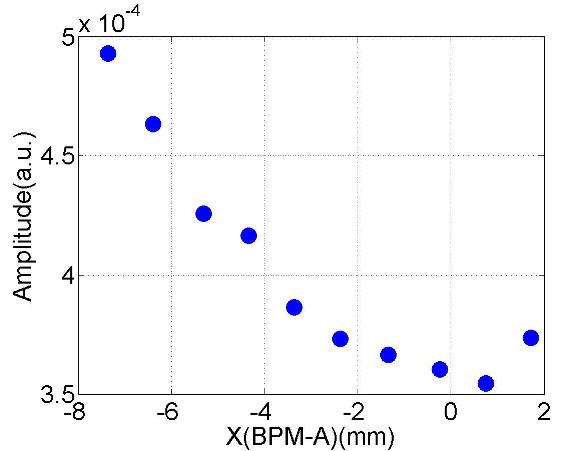}
\label{dep-C1H1-X-2-BP}
}
\subfigure[\#3 ($f$:4.1153GHz; Q:10$^3$)]{
\includegraphics[width=0.31\textwidth]{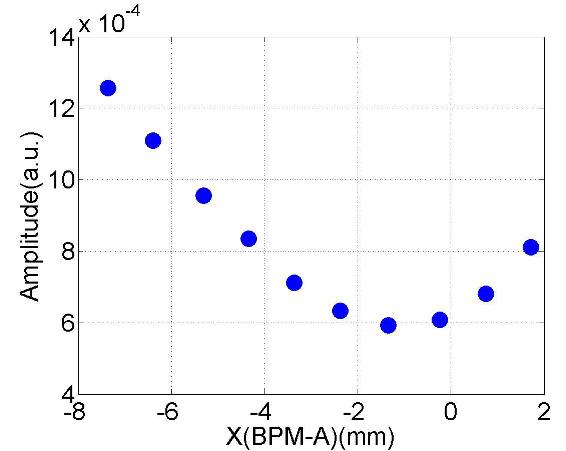}
\label{dep-C1H1-X-3-BP}
}\\
\subfigure[\#2 ($f$:4.1047GHz; Q:10$^2$)]{
\includegraphics[width=0.31\textwidth]{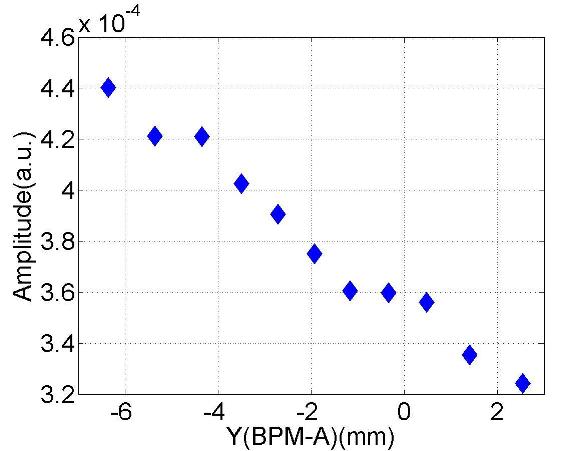}
\label{dep-C1H1-Y-2-BP}
}
\subfigure[\#3 ($f$:4.1152GHz; Q:10$^3$)]{
\includegraphics[width=0.31\textwidth]{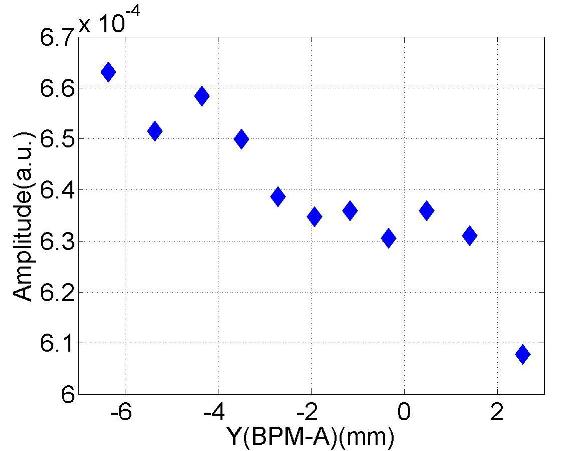}
\label{dep-C1H1-Y-3-BP}
}\\
\subfigure[\#2 ($f$:4.1048GHz; Q:10$^2$)]{
\includegraphics[width=0.31\textwidth]{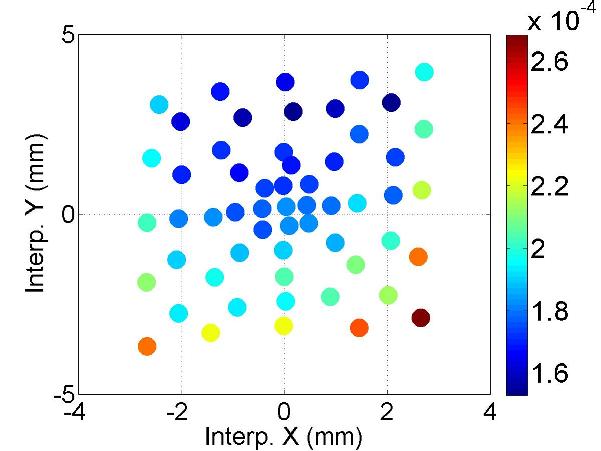}
\label{polar-C1H1-2-BP}
}
\subfigure[\#3 ($f$:4.1152GHz; Q:10$^3$)]{
\includegraphics[width=0.31\textwidth]{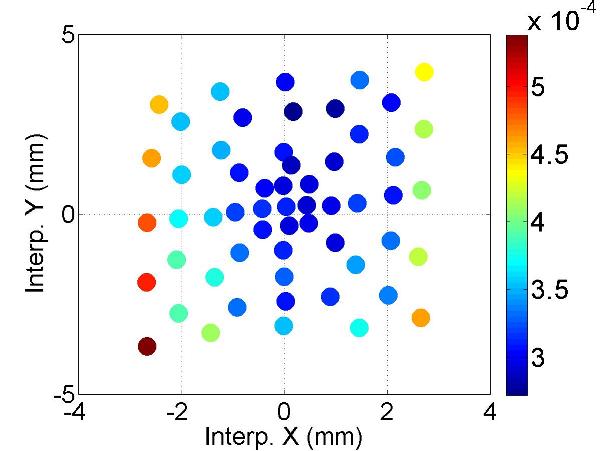}
\label{polar-C1H1-3-BP}
}
\caption{Depedence of the mode amplitude on the transverse beam of{}fset.}
\label{spec-dep-C1H1-XY-3-BP}
\end{figure}

\FloatBarrier
\section{BP: HOM Coupler C1H2}
\begin{figure}[h]\center
\subfigure[Spectrum (C1H2)]{
\includegraphics[width=1\textwidth]{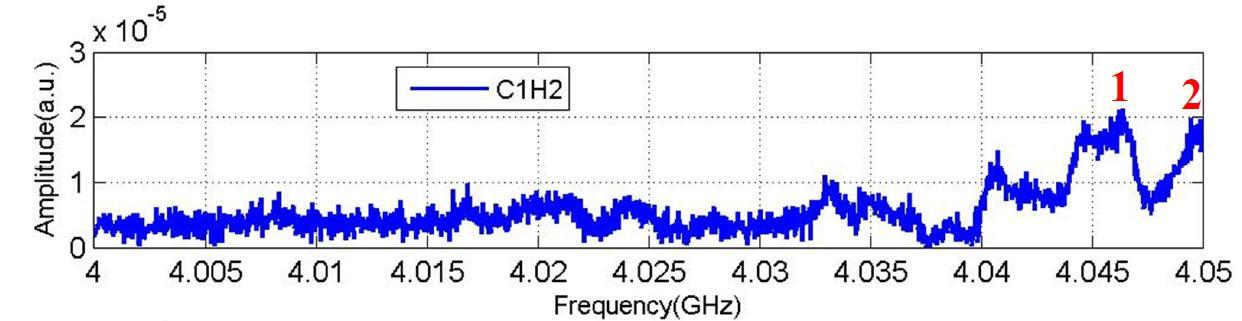}
\label{spec-C1H2-X-1-BP}
}
\subfigure[\#1 ($f$:4.0463GHz; Q:10$^3$)]{
\includegraphics[width=0.3\textwidth]{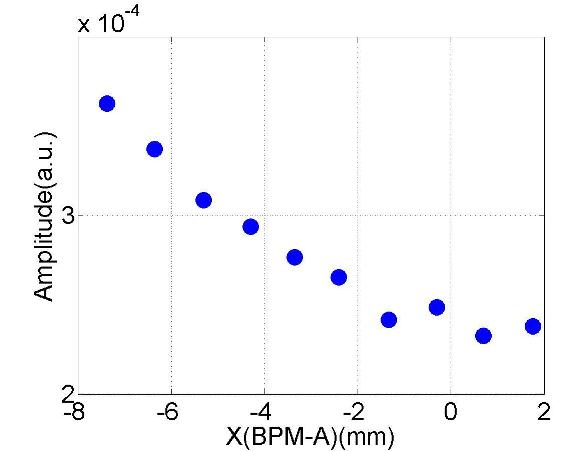}
\label{dep-C1H2-X-1-BP}
}
\subfigure[\#1 ($f$:4.0463GHz; Q:10$^3$)]{
\includegraphics[width=0.3\textwidth]{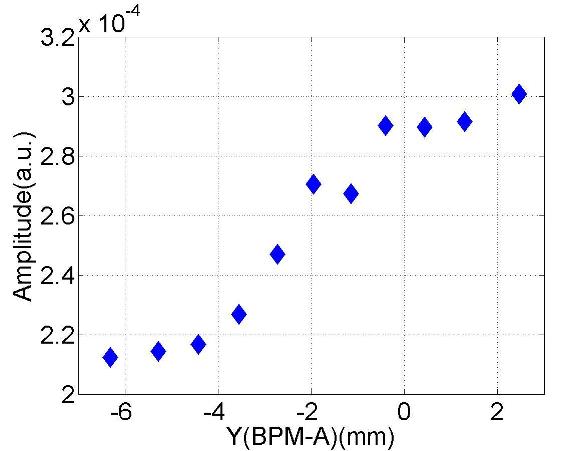}
\label{dep-C1H2-Y-1-BP}
}
\subfigure[\#1 ($f$:4.0463GHz; Q:10$^3$)]{
\includegraphics[width=0.31\textwidth]{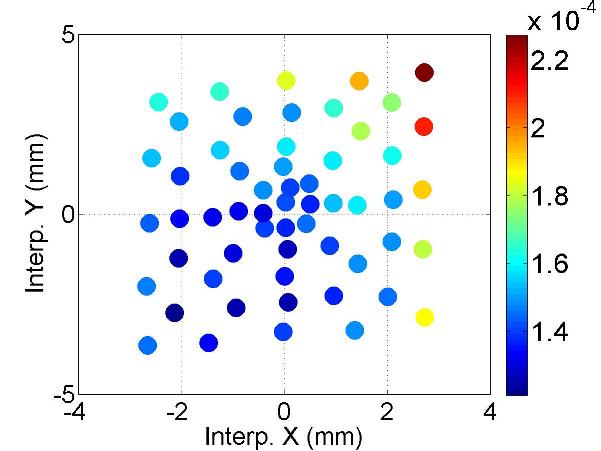}
\label{polar-C1H2-1-BP}
}\\
\subfigure[\#2 ($f$:4.0498GHz; Q:10$^3$)]{
\includegraphics[width=0.3\textwidth]{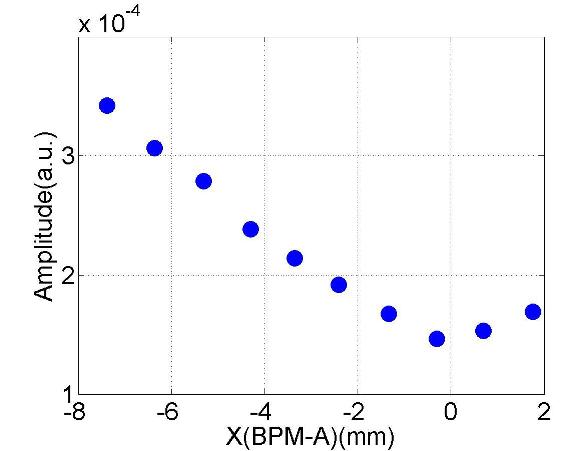}
\label{dep-C1H2-X-2-BP}
}
\subfigure[\#2 ($f$:4.0496GHz; Q:10$^3$)]{
\includegraphics[width=0.3\textwidth]{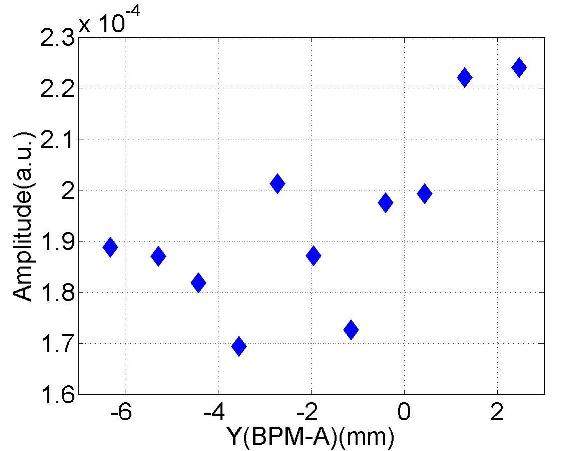}
\label{dep-C1H2-Y-2-BP}
}
\subfigure[\#2 ($f$:4.0496GHz; Q:10$^3$)]{
\includegraphics[width=0.31\textwidth]{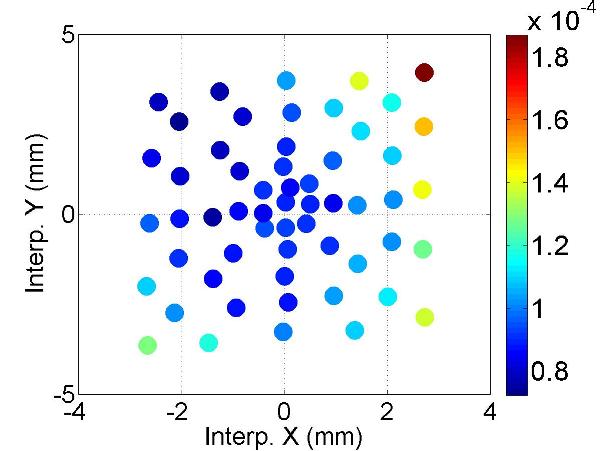}
\label{polar-C1H2-2-BP}
}
\caption{Depedence of the mode amplitude on the transverse beam of{}fset.}
\label{spec-dep-C1H2-XY-1-BP}
\end{figure}
\begin{figure}[h]
\subfigure[Spectrum (C1H2)]{
\includegraphics[width=1\textwidth]{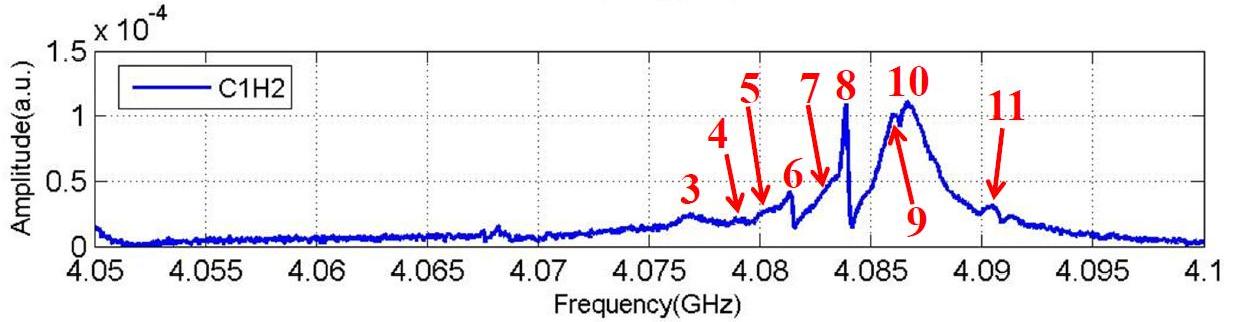}
\label{spec-C1H2-X-2-BP}
}
\subfigure[\#3 ($f$:4.0768GHz; Q:10$^3$)]{
\includegraphics[width=0.31\textwidth]{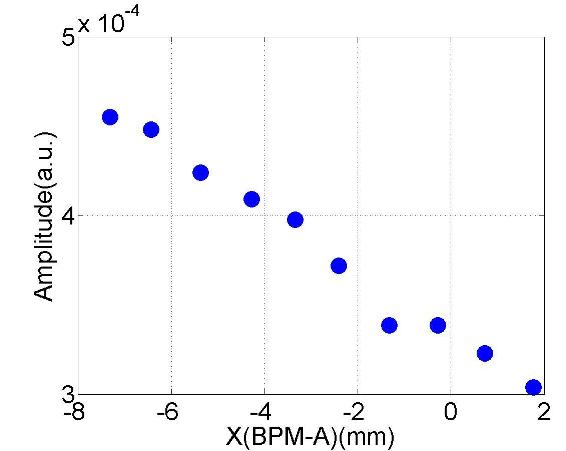}
\label{dep-C1H2-X-3-BP}
}
\subfigure[\#4 ($f$:4.0791GHz; Q:10$^3$)]{
\includegraphics[width=0.31\textwidth]{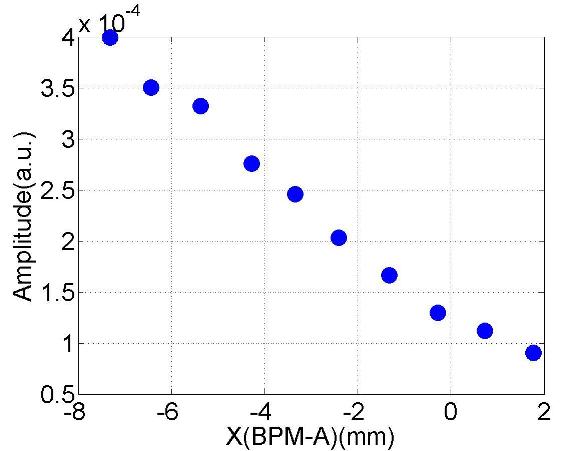}
\label{dep-C1H2-X-4-BP}
}
\subfigure[\#5 ($f$:4.0805GHz; Q:10$^3$)]{
\includegraphics[width=0.31\textwidth]{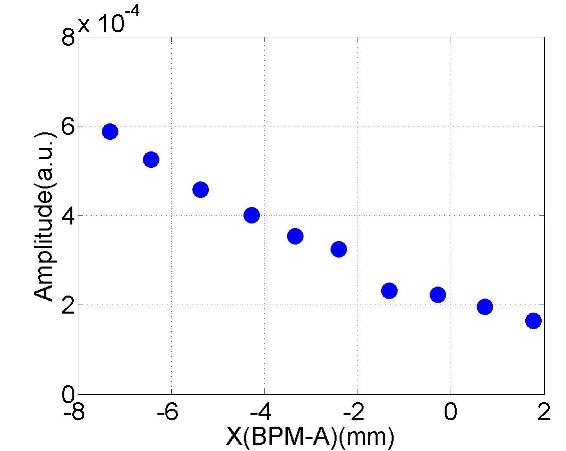}
\label{dep-C1H2-X-5-BP}
}
\subfigure[\#6 ($f$:4.0813GHz; Q:10$^4$)]{
\includegraphics[width=0.31\textwidth]{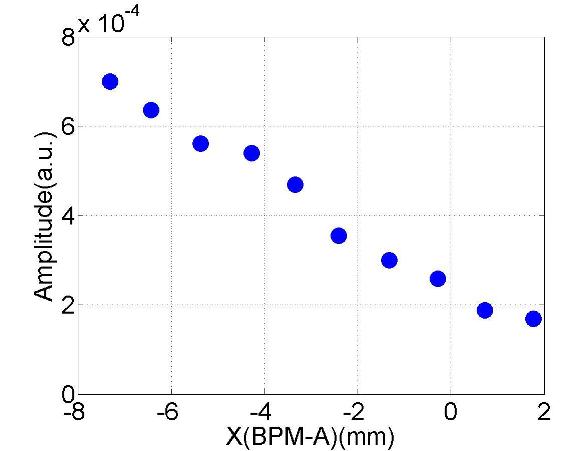}
\label{dep-C1H2-X-6-BP}
}
\subfigure[\#7 ($f$:4.0832GHz; Q:10$^3$)]{
\includegraphics[width=0.31\textwidth]{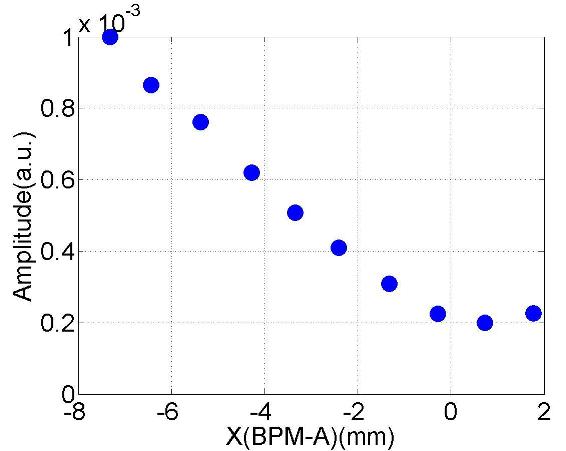}
\label{dep-C1H2-X-7-BP}
}
\subfigure[\#8 ($f$:4.0838GHz; Q:10$^4$)]{
\includegraphics[width=0.31\textwidth]{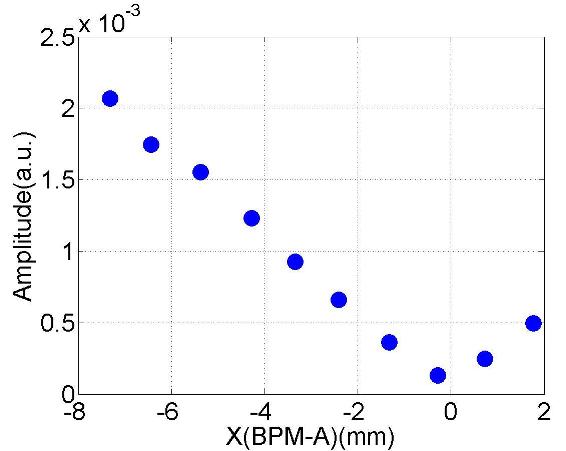}
\label{dep-C1H2-X-8-BP}
}
\subfigure[\#9 ($f$:4.0859GHz; Q:10$^3$)]{
\includegraphics[width=0.31\textwidth]{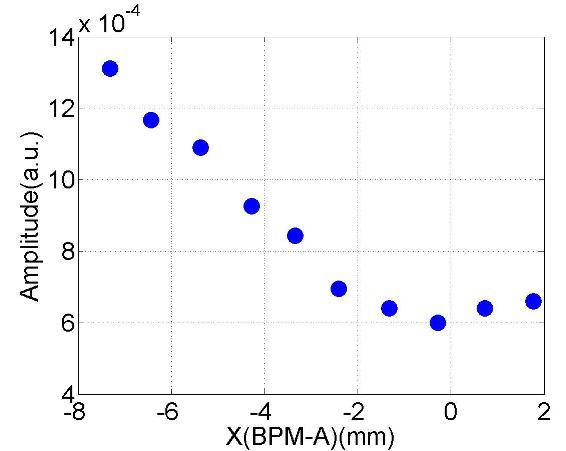}
\label{dep-C1H2-X-9-BP}
}
\subfigure[\#10 ($f$:4.0867GHz; Q:10$^3$)]{
\includegraphics[width=0.31\textwidth]{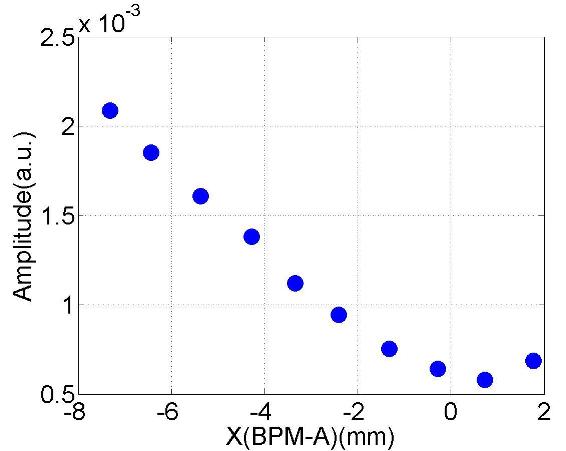}
\label{dep-C1H2-X-10-BP}
}
\subfigure[\#11 ($f$:4.0902GHz; Q:10$^3$)]{
\includegraphics[width=0.31\textwidth]{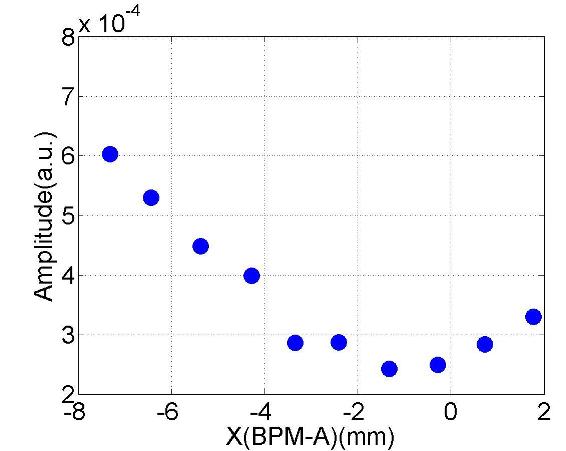}
\label{dep-C1H2-X-11-BP}
}
\caption{Depedence of the mode amplitude on the horizontal beam of{}fset.}
\label{spec-dep-C1H2-X-2-BP}
\end{figure}
\begin{figure}[h]
\subfigure[Spectrum (C1H2)]{
\includegraphics[width=1\textwidth]{Xmove-Spec-C1H2-2}
\label{spec-C1H2-X-2-BP}
}
\subfigure[\#3 ($f$:4.0768GHz; Q:10$^3$)]{
\includegraphics[width=0.31\textwidth]{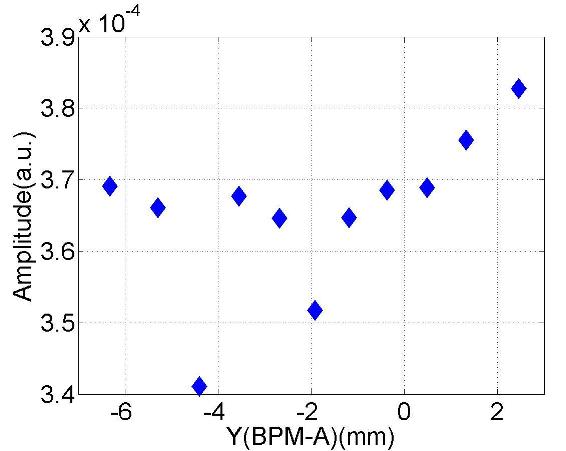}
\label{dep-C1H2-Y-3-BP}
}
\subfigure[\#4 ($f$:4.0791GHz; Q:10$^3$)]{
\includegraphics[width=0.31\textwidth]{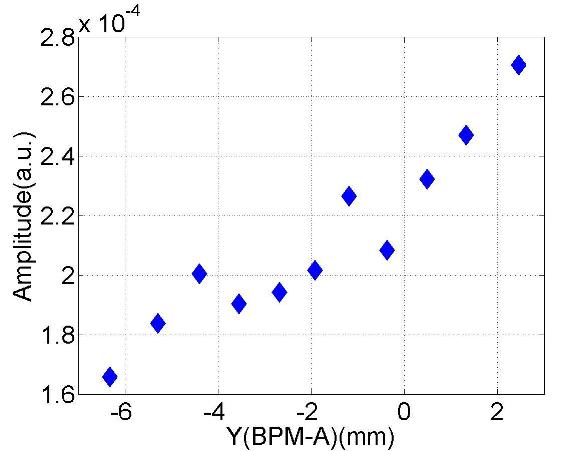}
\label{dep-C1H2-Y-4-BP}
}
\subfigure[\#5 ($f$:4.0805GHz; Q:10$^3$)]{
\includegraphics[width=0.31\textwidth]{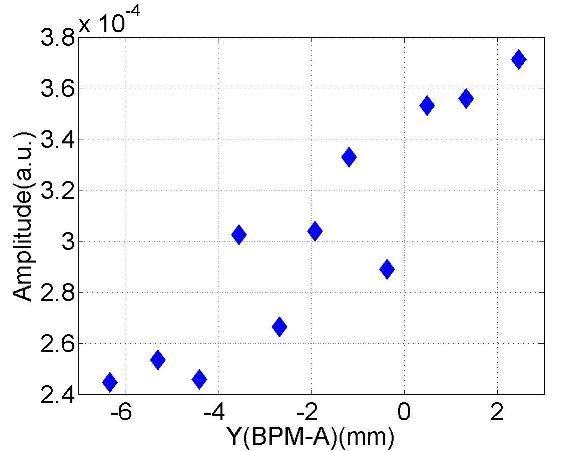}
\label{dep-C1H2-Y-5-BP}
}
\subfigure[\#6 ($f$:4.0813GHz; Q:10$^4$)]{
\includegraphics[width=0.31\textwidth]{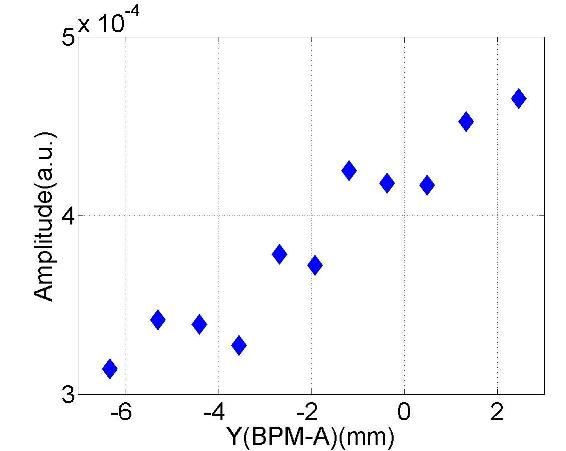}
\label{dep-C1H2-Y-6-BP}
}
\subfigure[\#7 ($f$:4.0833GHz; Q:10$^3$)]{
\includegraphics[width=0.31\textwidth]{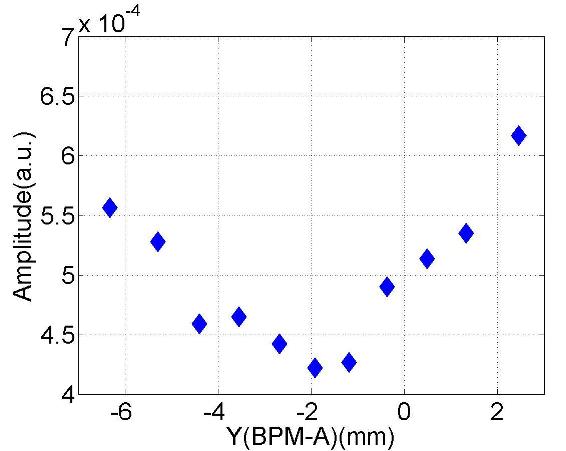}
\label{dep-C1H2-Y-7-BP}
}
\subfigure[\#8 ($f$:4.0838GHz; Q:10$^4$)]{
\includegraphics[width=0.31\textwidth]{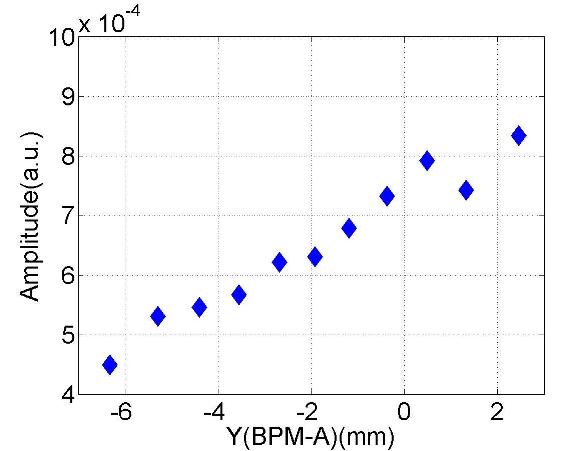}
\label{dep-C1H2-Y-8-BP}
}
\subfigure[\#9 ($f$:4.0859GHz; Q:10$^3$)]{
\includegraphics[width=0.31\textwidth]{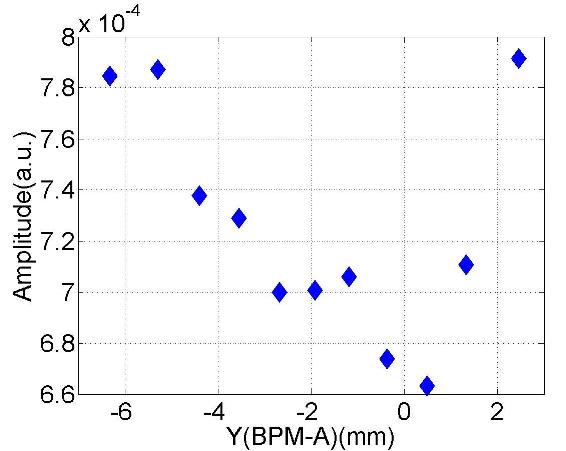}
\label{dep-C1H2-Y-9-BP}
}
\subfigure[\#10 ($f$:4.0867GHz; Q:10$^3$)]{
\includegraphics[width=0.31\textwidth]{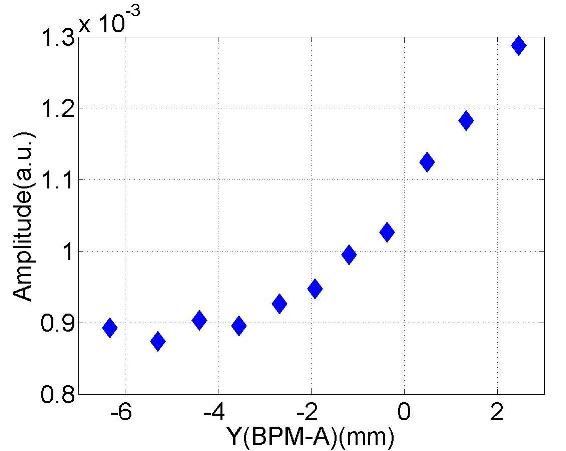}
\label{dep-C1H2-Y-10-BP}
}
\subfigure[\#11 ($f$:4.0903GHz; Q:10$^3$)]{
\includegraphics[width=0.31\textwidth]{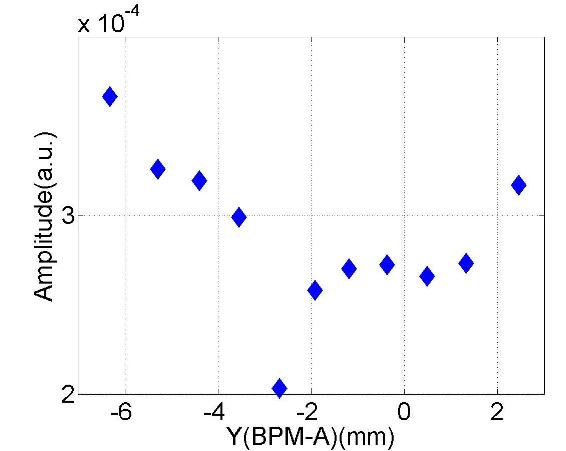}
\label{dep-C1H2-Y-11-BP}
}
\caption{Depedence of the mode amplitude on the vertical beam of{}fset.}
\label{spec-dep-C1H2-Y-2-BP}
\end{figure}
\begin{figure}[h]
\subfigure[Spectrum (C1H2)]{
\includegraphics[width=1\textwidth]{Xmove-Spec-C1H2-2}
\label{spec-C1H2-X-2-BP}
}
\subfigure[\#3 ($f$:4.0767GHz; Q:10$^3$)]{
\includegraphics[width=0.31\textwidth]{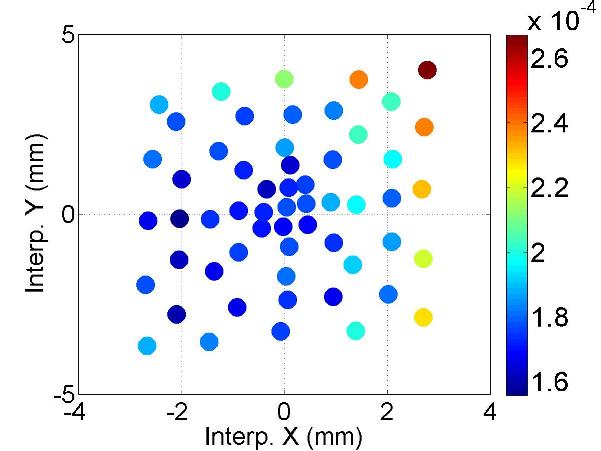}
\label{polar-C1H2-3-BP}
}
\subfigure[\#4 ($f$:4.0791GHz; Q:10$^3$)]{
\includegraphics[width=0.31\textwidth]{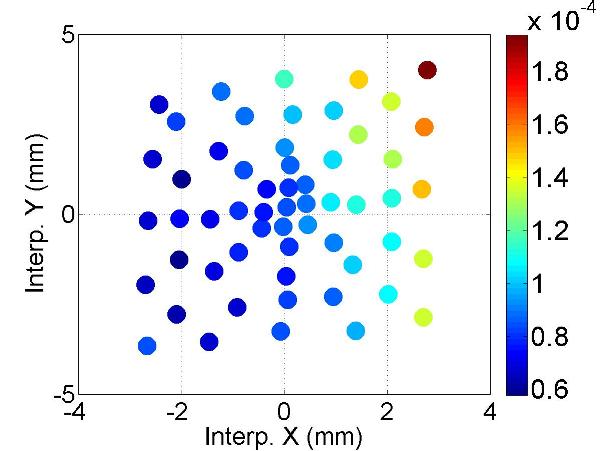}
\label{polar-C1H2-4-BP}
}
\subfigure[\#5 ($f$:4.0805GHz; Q:10$^3$)]{
\includegraphics[width=0.31\textwidth]{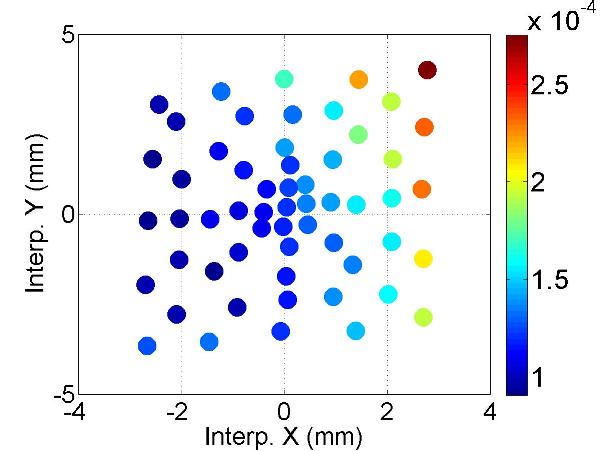}
\label{polar-C1H2-5-BP}
}
\subfigure[\#6 ($f$:4.0813GHz; Q:10$^4$)]{
\includegraphics[width=0.31\textwidth]{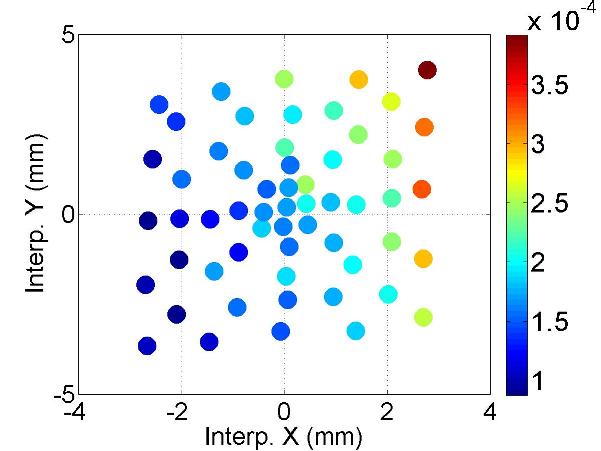}
\label{polar-C1H2-6-BP}
}
\subfigure[\#7 ($f$:4.0830GHz; Q:10$^3$)]{
\includegraphics[width=0.31\textwidth]{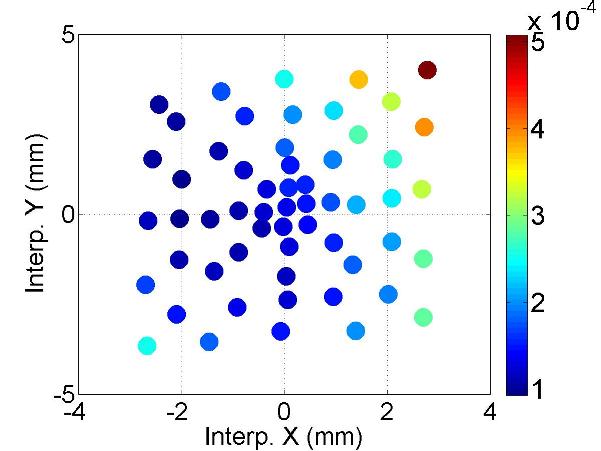}
\label{polar-C1H2-7-BP}
}
\subfigure[\#8 ($f$:4.0838GHz; Q:10$^4$)]{
\includegraphics[width=0.31\textwidth]{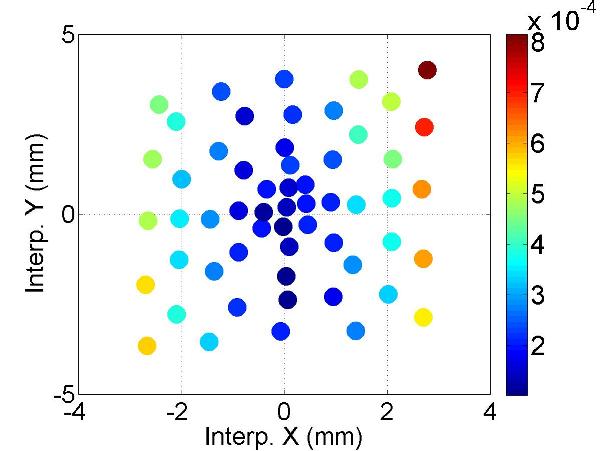}
\label{polar-C1H2-8-BP}
}
\subfigure[\#9 ($f$:4.0858GHz; Q:10$^3$)]{
\includegraphics[width=0.31\textwidth]{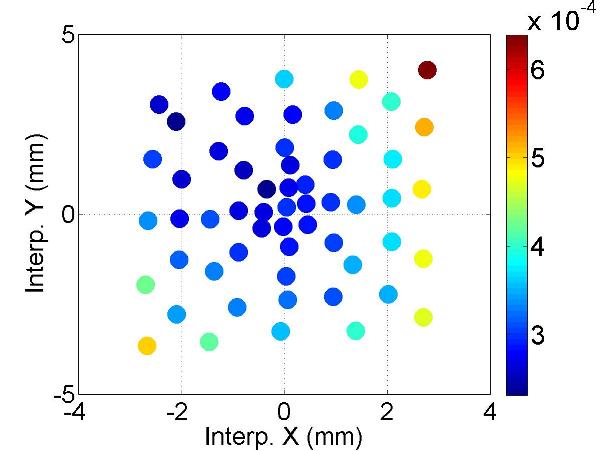}
\label{polar-C1H2-9-BP}
}
\subfigure[\#10 ($f$:4.0867GHz; Q:10$^3$)]{
\includegraphics[width=0.31\textwidth]{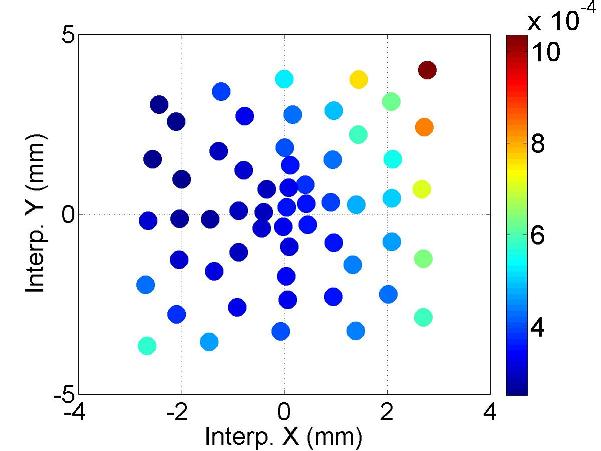}
\label{polar-C1H2-10-BP}
}
\subfigure[\#11 ($f$:4.0904GHz; Q:10$^3$)]{
\includegraphics[width=0.31\textwidth]{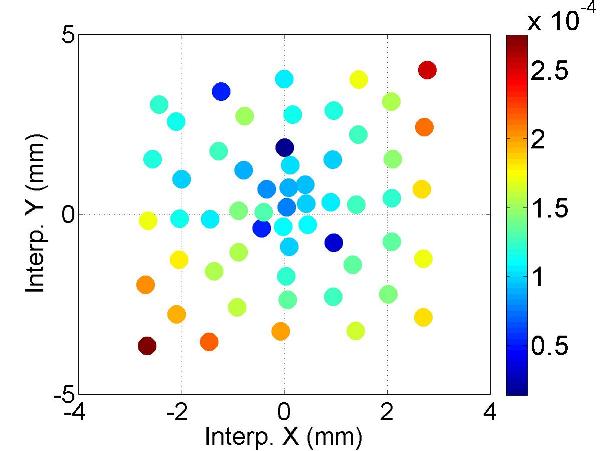}
\label{polar-C1H2-11-BP}
}
\caption{Depedence of the mode amplitude on the beam of{}fset in the cavity.}
\label{spec-polar-C1H2-2-BP}
\end{figure}
\begin{figure}[h]
\subfigure[Spectrum (C1H2)]{
\includegraphics[width=1\textwidth]{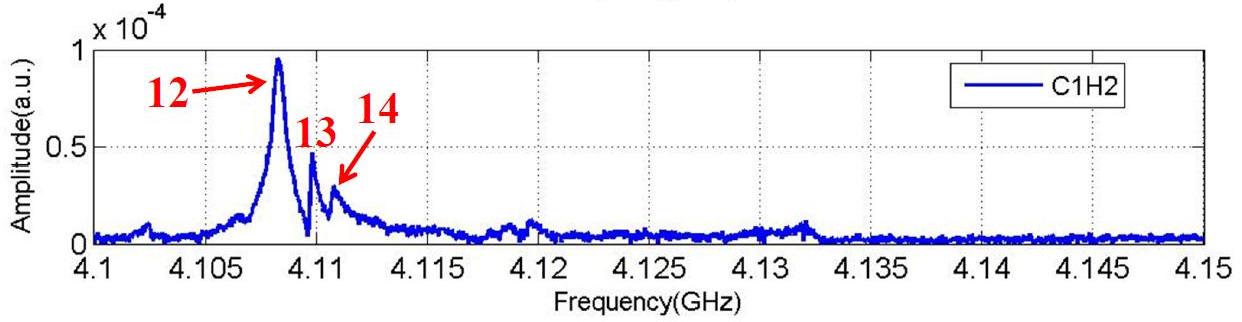}
\label{spec-C1H2-X-3-BP}
}
\subfigure[\#12 ($f$:4.1083GHz; Q:10$^3$)]{
\includegraphics[width=0.31\textwidth]{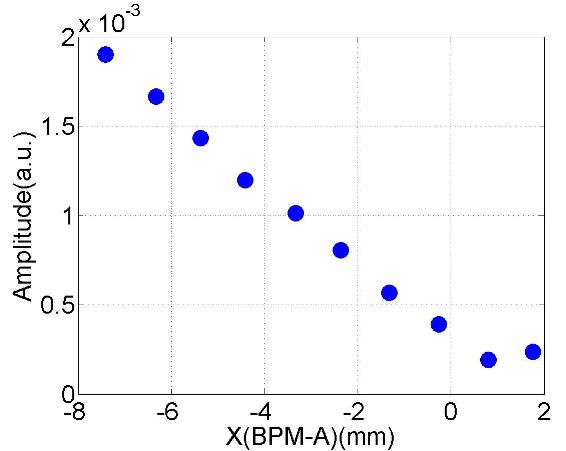}
\label{dep-C1H2-X-12-BP}
}
\subfigure[\#13 ($f$:4.1099GHz; Q:10$^4$)]{
\includegraphics[width=0.31\textwidth]{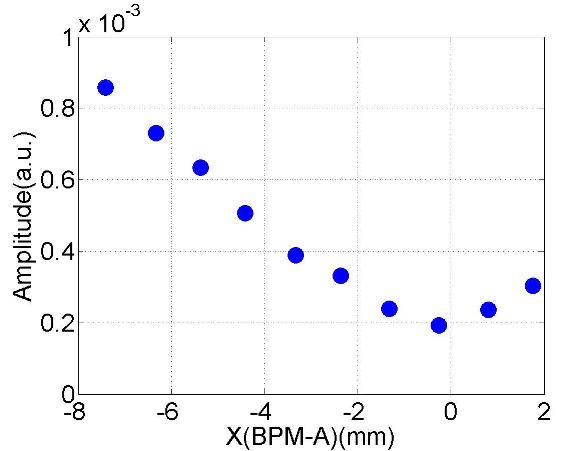}
\label{dep-C1H2-X-13-BP}
}
\subfigure[\#14 ($f$:4.1108GHz; Q:10$^4$)]{
\includegraphics[width=0.31\textwidth]{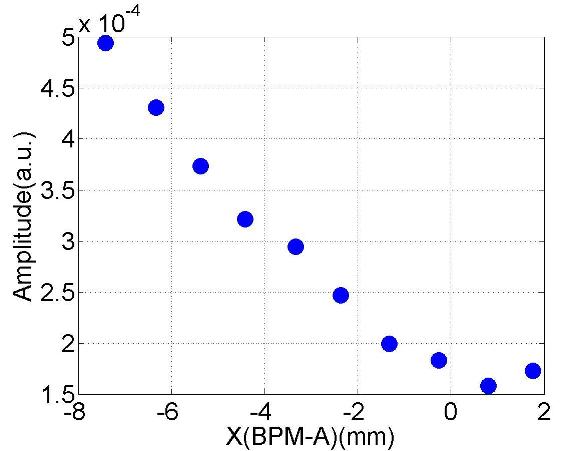}
\label{dep-C1H2-X-14-BP}
}
\subfigure[\#12 ($f$:4.1083GHz; Q:10$^3$)]{
\includegraphics[width=0.31\textwidth]{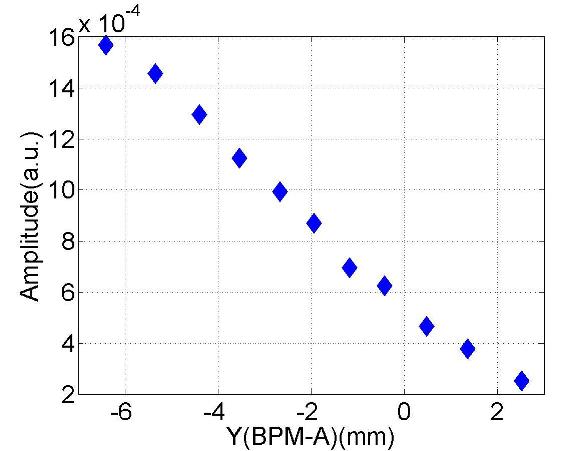}
\label{dep-C1H2-Y-12-BP}
}
\subfigure[\#13 ($f$:4.1099GHz; Q:10$^4$)]{
\includegraphics[width=0.31\textwidth]{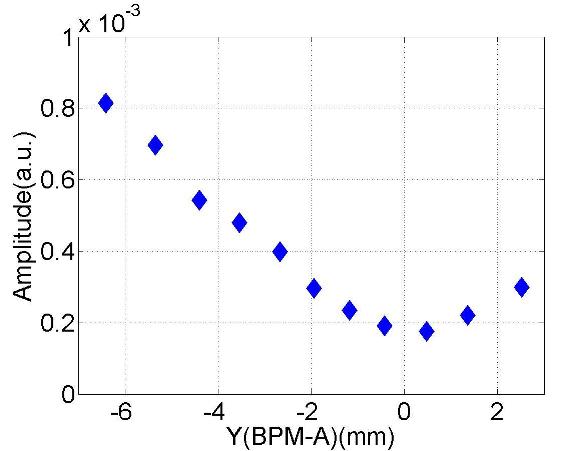}
\label{dep-C1H2-Y-13-BP}
}
\subfigure[\#14 ($f$:4.1109GHz; Q:10$^4$)]{
\includegraphics[width=0.31\textwidth]{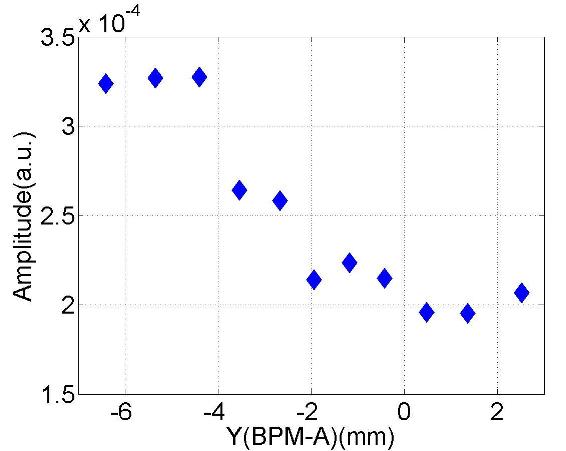}
\label{dep-C1H2-Y-14-BP}
}
\subfigure[\#12 ($f$:4.1083GHz; Q:10$^3$)]{
\includegraphics[width=0.31\textwidth]{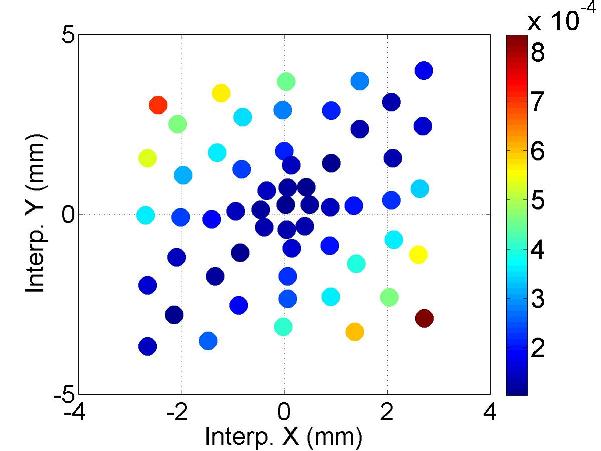}
\label{polar-C1H2-12-BP}
}
\subfigure[\#13 ($f$:4.1098GHz; Q:10$^4$)]{
\includegraphics[width=0.31\textwidth]{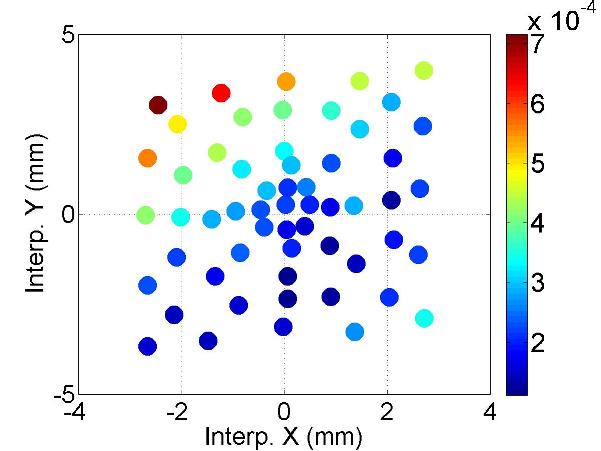}
\label{polar-C1H2-13-BP}
}
\subfigure[\#14 ($f$:4.1107GHz; Q:10$^3$)]{
\includegraphics[width=0.31\textwidth]{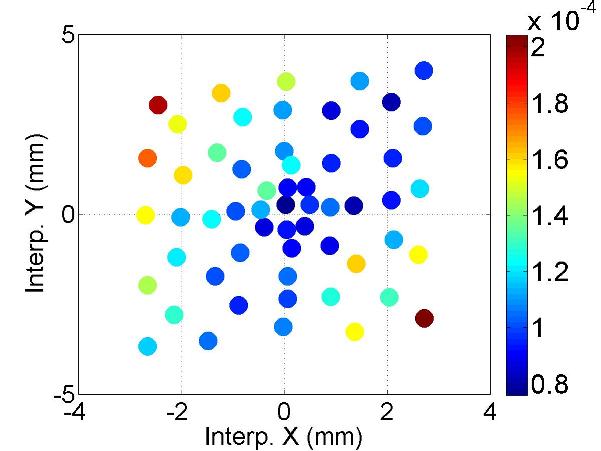}
\label{polar-C1H2-14-BP}
}
\caption{Depedence of the mode amplitude on the transverse beam of{}fset.}
\label{spec-dep-C1H2-XY-3-BP}
\end{figure}

\FloatBarrier
\section{BP: HOM Coupler C2H1}
\begin{figure}[h]\center
\subfigure[Spectrum (C2H1)]{
\includegraphics[width=0.85\textwidth]{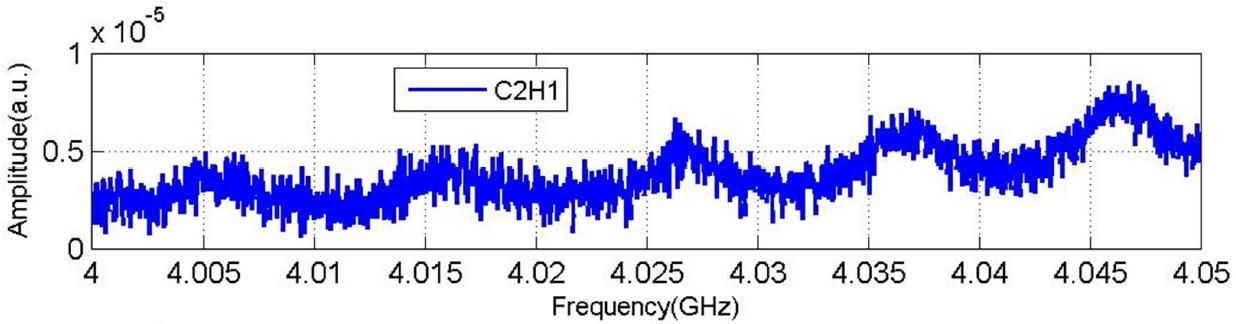}
\label{spec-C2H1-X-1-BP}
}
\subfigure[Spectrum (C2H1)]{
\includegraphics[width=0.85\textwidth]{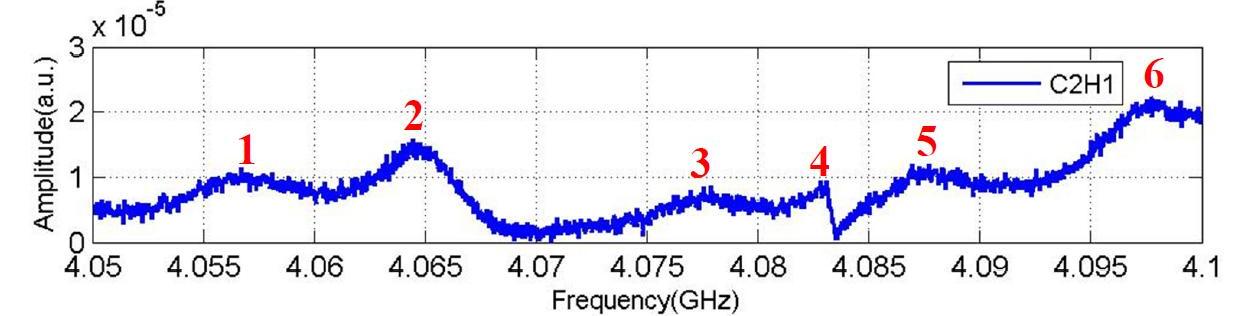}
\label{spec-C2H1-X-2-BP}
}
\subfigure[\#1 ($f$:4.0569GHz; Q:10$^2$)]{
\includegraphics[width=0.26\textwidth]{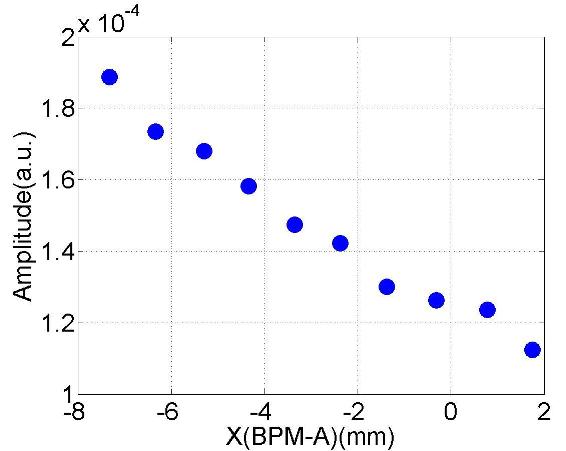}
\label{dep-C2H1-X-1-BP}
}
\subfigure[\#2 ($f$:4.0647GHz; Q:10$^3$)]{
\includegraphics[width=0.26\textwidth]{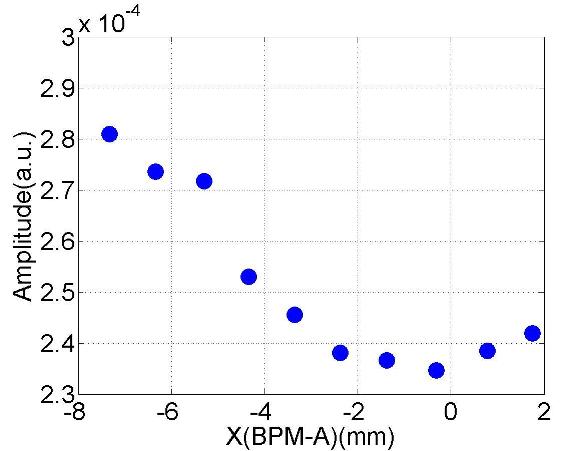}
\label{dep-C2H1-X-2-BP}
}
\subfigure[\#3 ($f$:4.0780GHz; Q:10$^2$)]{
\includegraphics[width=0.26\textwidth]{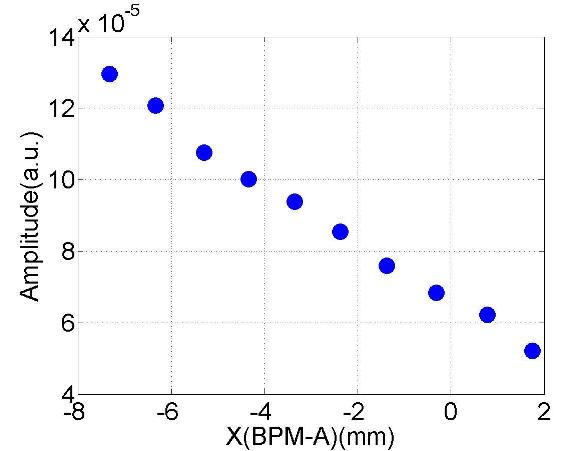}
\label{dep-C2H1-X-3-BP}
}
\subfigure[\#4 ($f$:4.0822GHz; Q:10$^3$)]{
\includegraphics[width=0.26\textwidth]{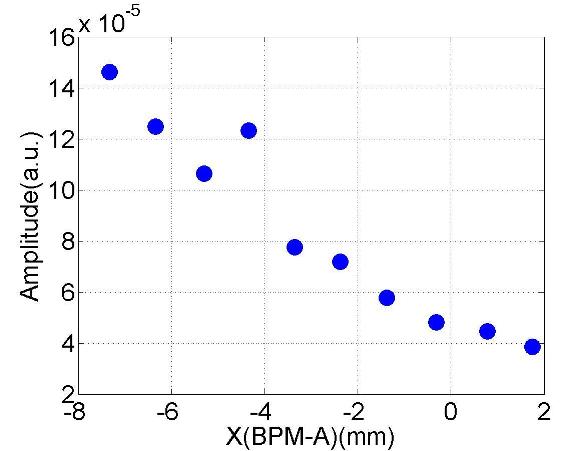}
\label{dep-C2H1-X-4-BP}
}
\subfigure[\#5 ($f$:4.0877GHz; Q:10$^3$)]{
\includegraphics[width=0.26\textwidth]{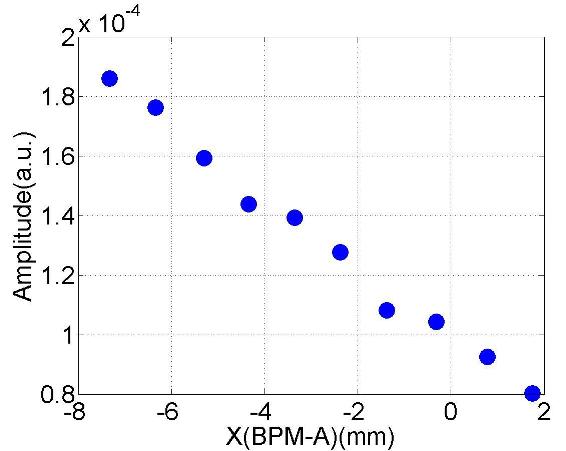}
\label{dep-C2H1-X-5-BP}
}
\subfigure[\#6 ($f$:4.0982GHz; Q:10$^2$)]{
\includegraphics[width=0.26\textwidth]{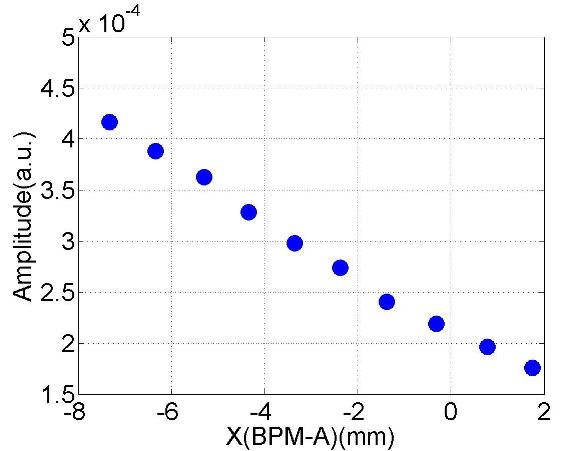}
\label{dep-C2H1-X-6-BP}
}
\caption{Depedence of the mode amplitude on the horizontal beam of{}fset.}
\label{spec-dep-C2H1-X-1-2-BP}
\end{figure}
\begin{figure}[h]
\subfigure[Spectrum (C2H1)]{
\includegraphics[width=1\textwidth]{Xmove-Spec-C2H1-1}
\label{spec-C2H1-X-1-BP}
}
\subfigure[Spectrum (C2H1)]{
\includegraphics[width=1\textwidth]{Xmove-Spec-C2H1-2}
\label{spec-C2H1-X-2-BP}
}
\subfigure[\#1 ($f$:4.0569GHz; Q:10$^2$)]{
\includegraphics[width=0.31\textwidth]{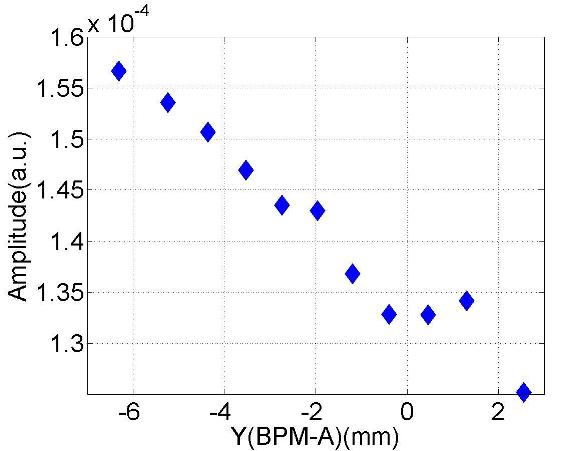}
\label{dep-C2H1-Y-1-BP}
}
\subfigure[\#2 ($f$:4.0647GHz; Q:10$^3$)]{
\includegraphics[width=0.31\textwidth]{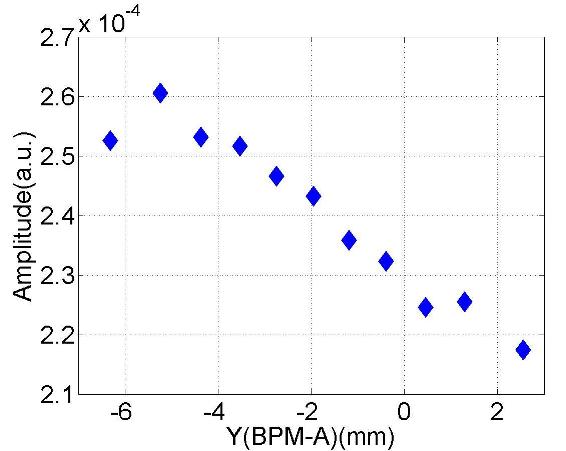}
\label{dep-C2H1-Y-2-BP}
}
\subfigure[\#3 ($f$:4.0784GHz; Q:10$^2$)]{
\includegraphics[width=0.31\textwidth]{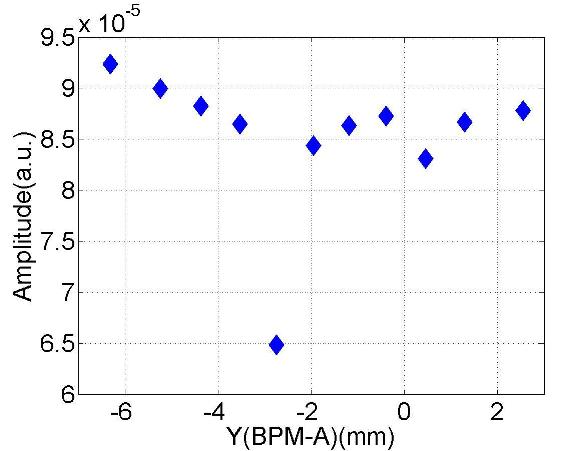}
\label{dep-C2H1-Y-3-BP}
}
\subfigure[\#4 ($f$:4.0829GHz; Q:10$^3$)]{
\includegraphics[width=0.31\textwidth]{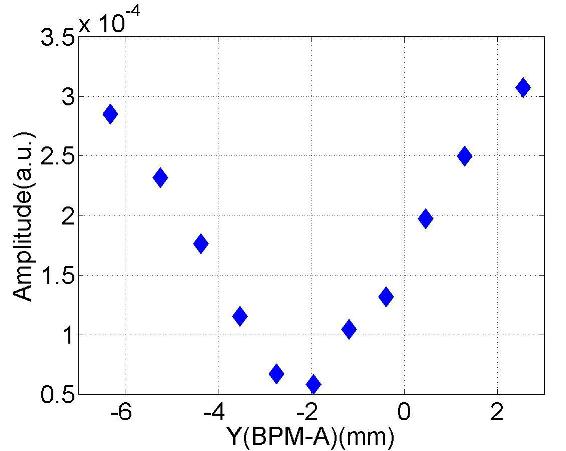}
\label{dep-C2H1-Y-4-BP}
}
\subfigure[\#5 ($f$:4.0876GHz; Q:10$^3$)]{
\includegraphics[width=0.31\textwidth]{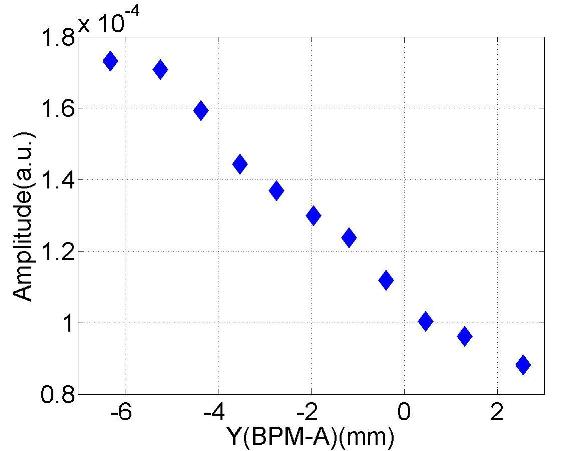}
\label{dep-C2H1-Y-5-BP}
}
\subfigure[\#6 ($f$:4.0983GHz; Q:10$^2$)]{
\includegraphics[width=0.31\textwidth]{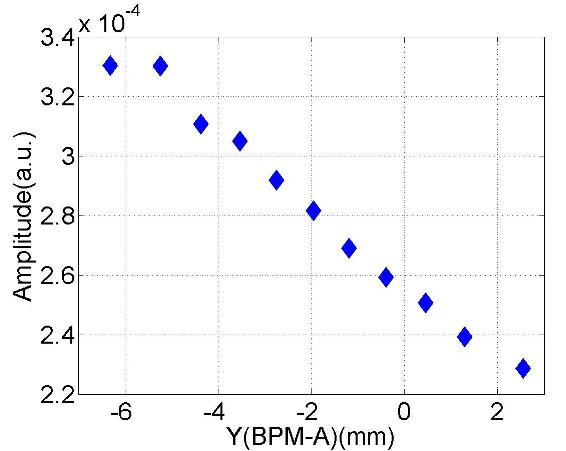}
\label{dep-C2H1-Y-6-BP}
}
\caption{Depedence of the mode amplitude on the vertical beam of{}fset.}
\label{spec-dep-C2H1-Y-1-2-BP}
\end{figure}
\begin{figure}[h]
\subfigure[Spectrum (C2H1)]{
\includegraphics[width=1\textwidth]{Xmove-Spec-C2H1-1}
\label{spec-C2H1-X-1-BP}
}
\subfigure[Spectrum (C2H1)]{
\includegraphics[width=1\textwidth]{Xmove-Spec-C2H1-2}
\label{spec-C2H1-X-2-BP}
}
\subfigure[\#1 ($f$:4.0568GHz; Q:10$^2$)]{
\includegraphics[width=0.31\textwidth]{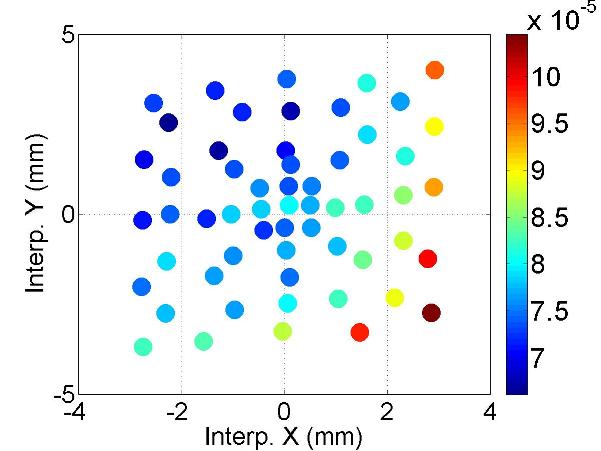}
\label{polar-C2H1-1-BP}
}
\subfigure[\#2 ($f$:4.0649GHz; Q:10$^3$)]{
\includegraphics[width=0.31\textwidth]{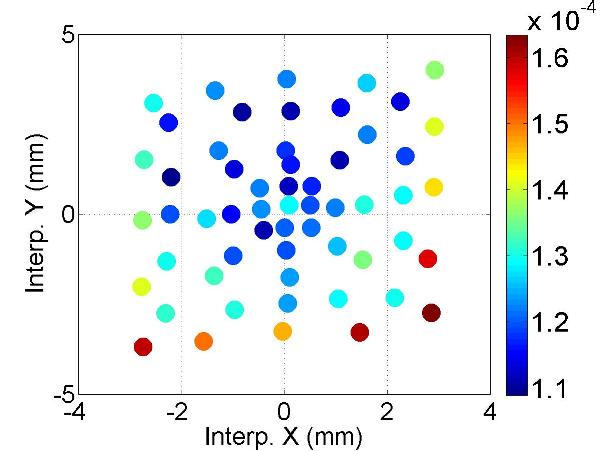}
\label{polar-C2H1-2-BP}
}
\subfigure[\#3 ($f$:4.0787GHz; Q:10$^2$)]{
\includegraphics[width=0.31\textwidth]{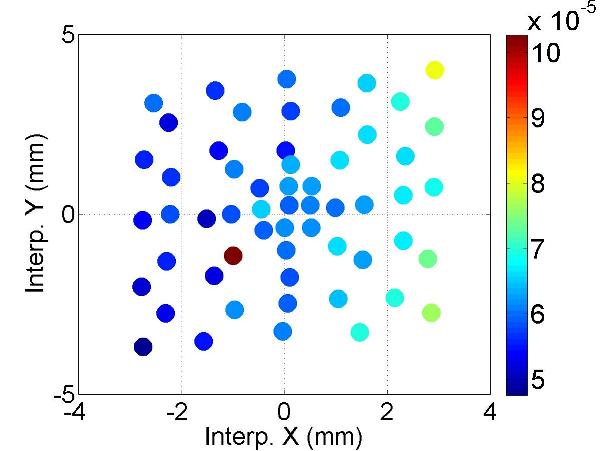}
\label{polar-C2H1-3-BP}
}
\subfigure[\#4 ($f$:4.0830GHz; Q:10$^3$)]{
\includegraphics[width=0.31\textwidth]{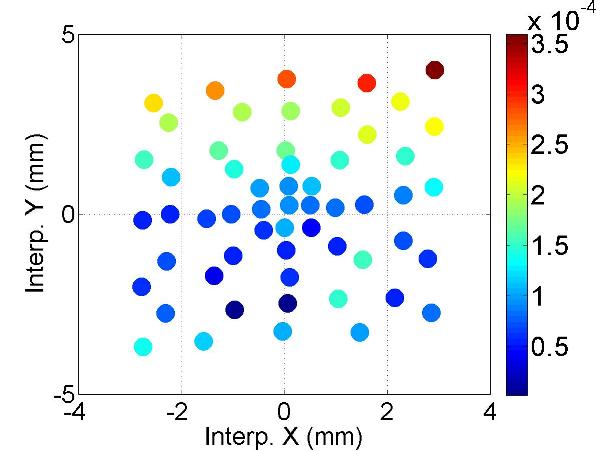}
\label{polar-C2H1-4-BP}
}
\subfigure[\#5 ($f$:4.0865GHz; Q:10$^2$)]{
\includegraphics[width=0.31\textwidth]{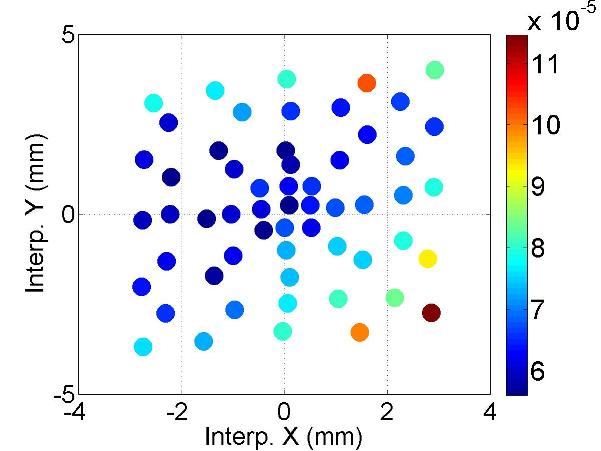}
\label{polar-C2H1-5-BP}
}
\subfigure[\#6 ($f$:4.0983GHz; Q:10$^2$)]{
\includegraphics[width=0.31\textwidth]{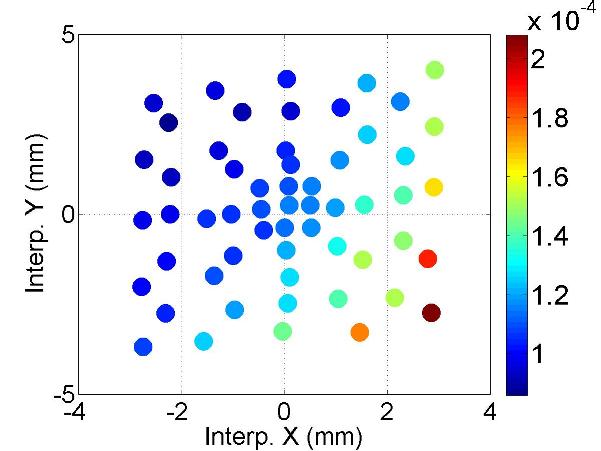}
\label{polar-C2H1-6-BP}
}
\caption{Depedence of the mode amplitude on the beam of{}fset in the cavity.}
\label{spec-polar-C2H1-1-2-BP}
\end{figure}
\begin{figure}[h]
\subfigure[Spectrum (C2H1)]{
\includegraphics[width=1\textwidth]{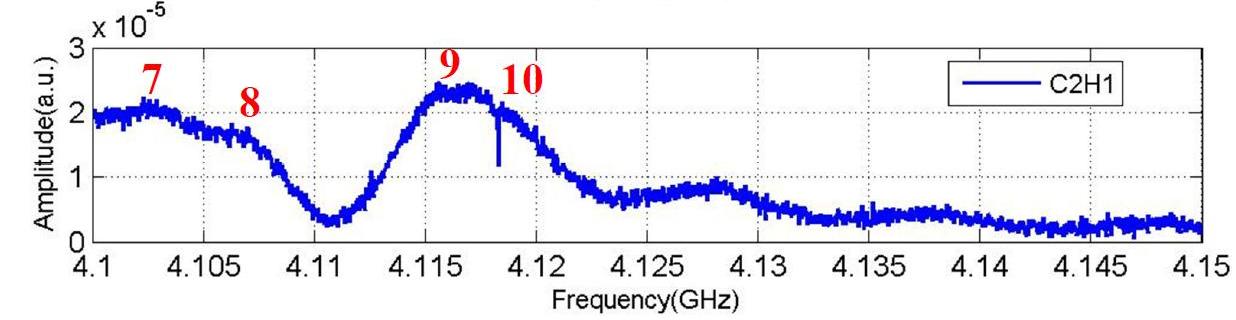}
\label{spec-C2H1-X-3-BP}
}
\subfigure[\#7 ($f$:4.1017GHz; Q:10$^2$)]{
\includegraphics[width=0.23\textwidth]{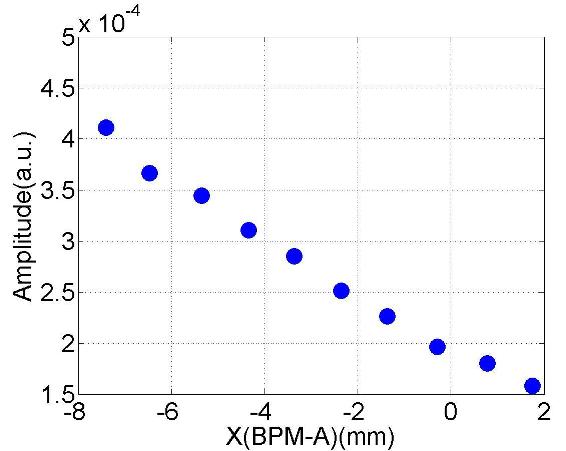}
\label{dep-C2H1-X-7-BP}
}
\subfigure[\#8 ($f$:4.1067GHz; Q:10$^3$)]{
\includegraphics[width=0.23\textwidth]{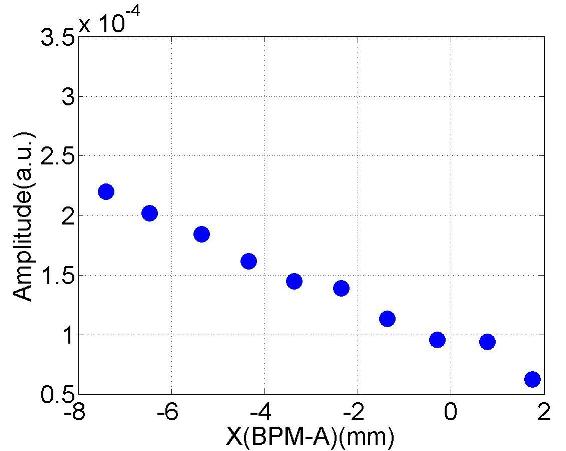}
\label{dep-C2H1-X-8-BP}
}
\subfigure[\#9 ($f$:4.1157GHz; Q:10$^3$)]{
\includegraphics[width=0.23\textwidth]{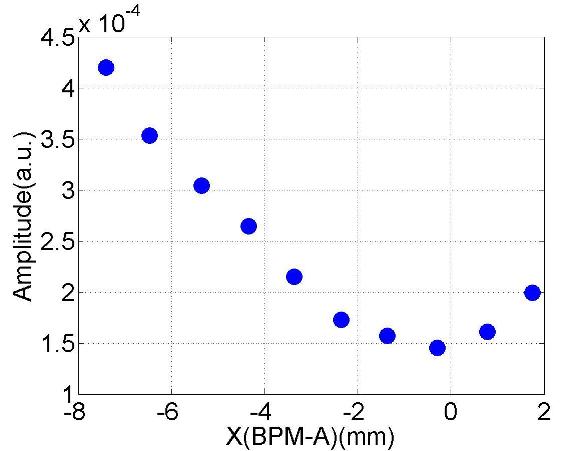}
\label{dep-C2H1-X-9-BP}
}
\subfigure[\#10 ($f$:4.1183GHz; Q:10$^3$)]{
\includegraphics[width=0.23\textwidth]{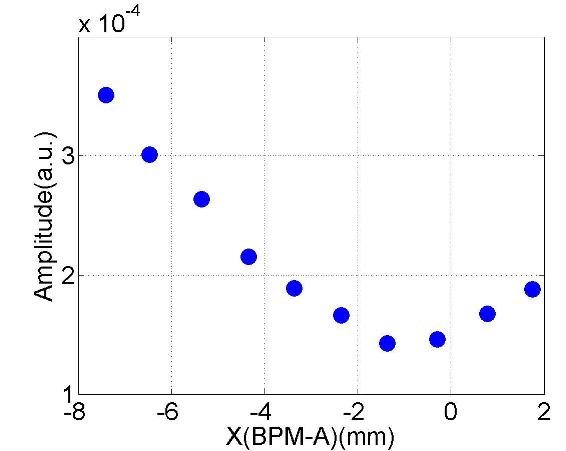}
\label{dep-C2H1-X-10-BP}
}
\subfigure[\#7 ($f$:4.1019GHz; Q:10$^2$)]{
\includegraphics[width=0.23\textwidth]{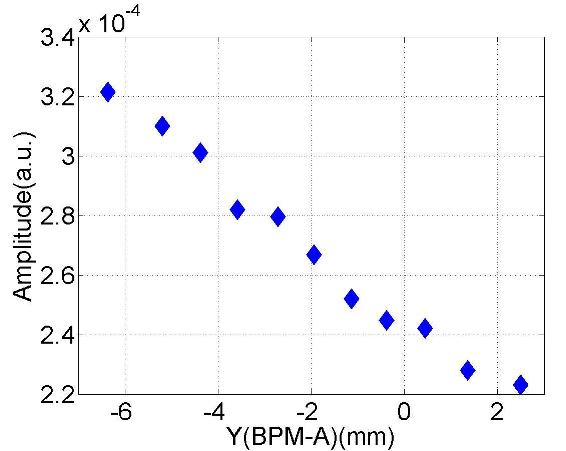}
\label{dep-C2H1-Y-7-BP}
}
\subfigure[\#8 ($f$:4.1068GHz; Q:10$^3$)]{
\includegraphics[width=0.23\textwidth]{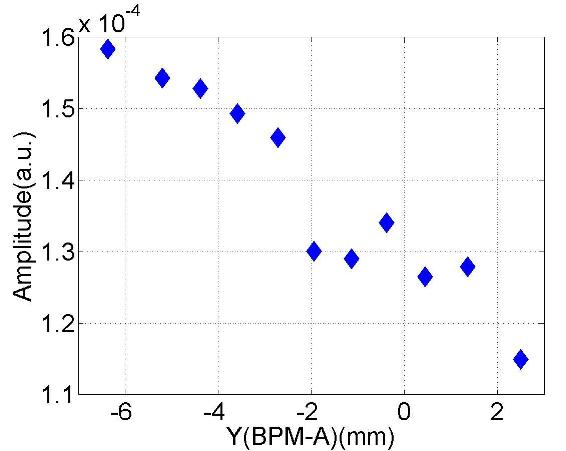}
\label{dep-C2H1-Y-8-BP}
}
\subfigure[\#9 ($f$:4.1159GHz; Q:10$^3$)]{
\includegraphics[width=0.23\textwidth]{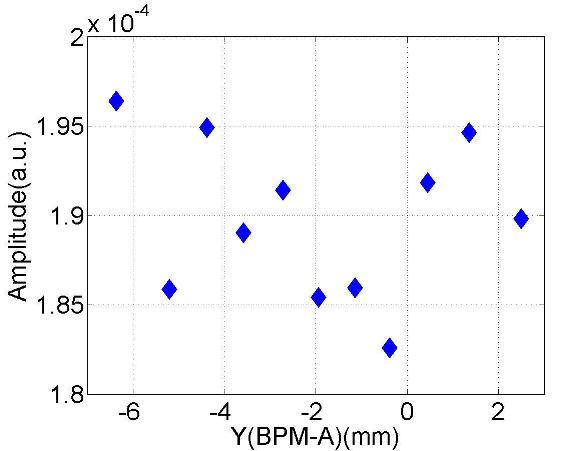}
\label{dep-C2H1-Y-9-BP}
}
\subfigure[\#10 ($f$:4.1183GHz; Q:10$^3$)]{
\includegraphics[width=0.23\textwidth]{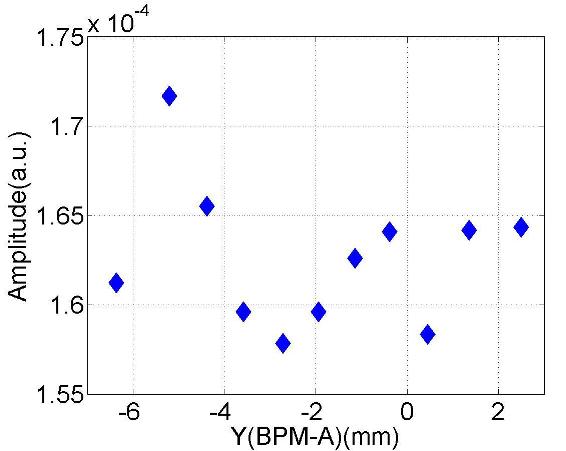}
\label{dep-C2H1-Y-10-BP}
}
\subfigure[\#7 ($f$:4.1013GHz; Q:10$^2$)]{
\includegraphics[width=0.23\textwidth]{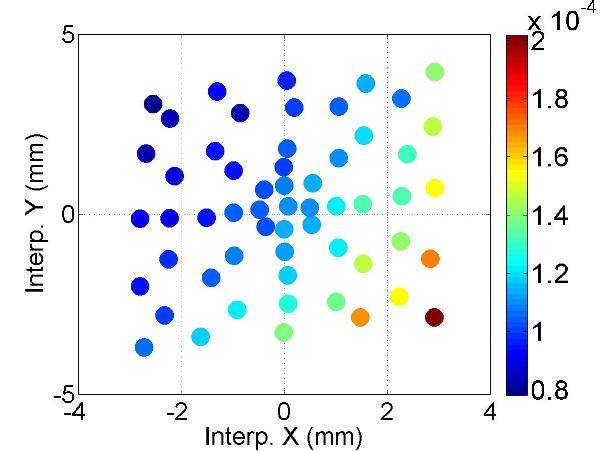}
\label{polar-C2H1-7-BP}
}
\subfigure[\#8 ($f$:4.1070GHz; Q:10$^3$)]{
\includegraphics[width=0.23\textwidth]{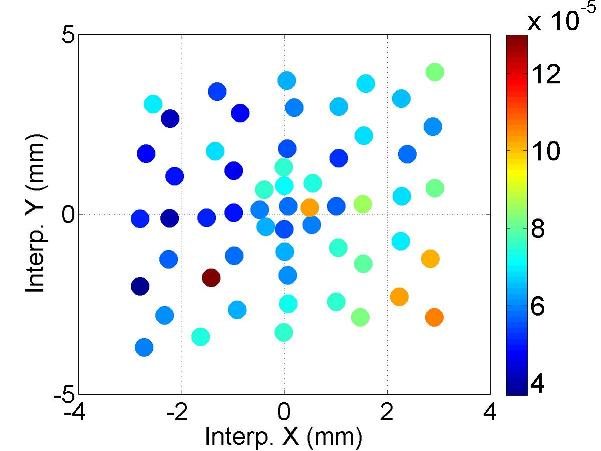}
\label{polar-C2H1-8-BP}
}
\subfigure[\#9 ($f$:4.1161GHz; Q:10$^3$)]{
\includegraphics[width=0.23\textwidth]{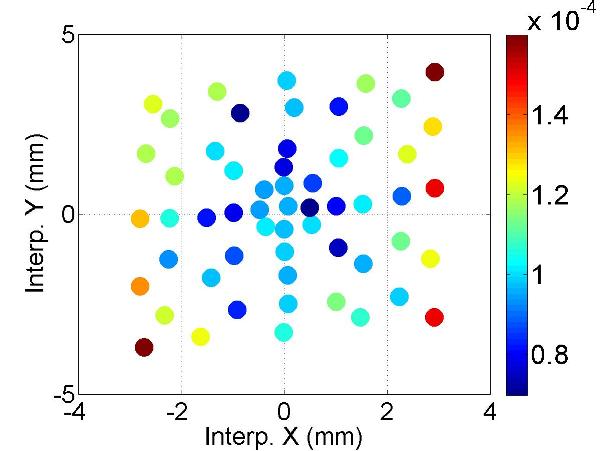}
\label{polar-C2H1-9-BP}
}
\subfigure[\#10 ($f$:4.1188GHz; Q:10$^3$)]{
\includegraphics[width=0.23\textwidth]{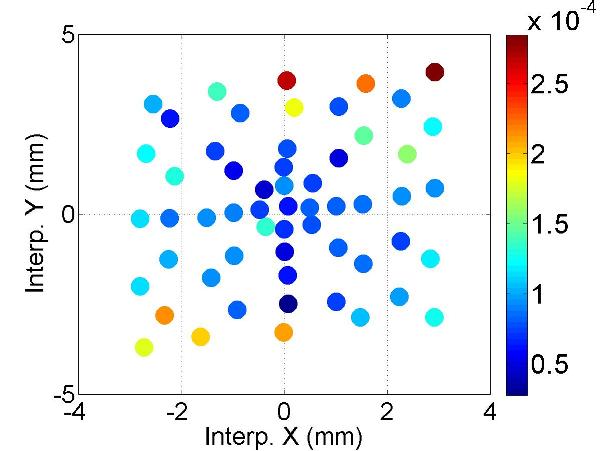}
\label{polar-C2H1-10-BP}
}
\caption{Depedence of the mode amplitude on the transverse beam of{}fset.}
\label{spec-dep-C2H1-XY-3-BP}
\end{figure}

\FloatBarrier
\section{BP: HOM Coupler C2H2}
\begin{figure}[h]\center
\subfigure[Spectrum (C2H2)]{
\includegraphics[width=0.95\textwidth]{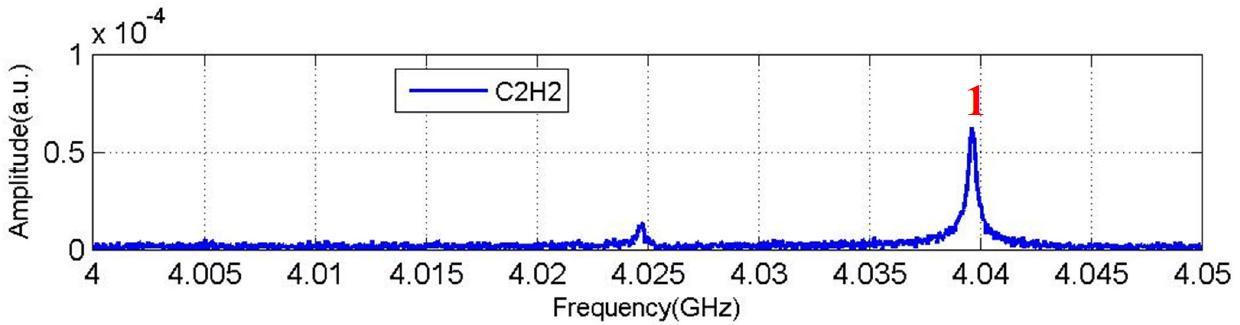}
\label{spec-C2H2-X-1-BP}
}
\subfigure[Spectrum (C2H2)]{
\includegraphics[width=0.95\textwidth]{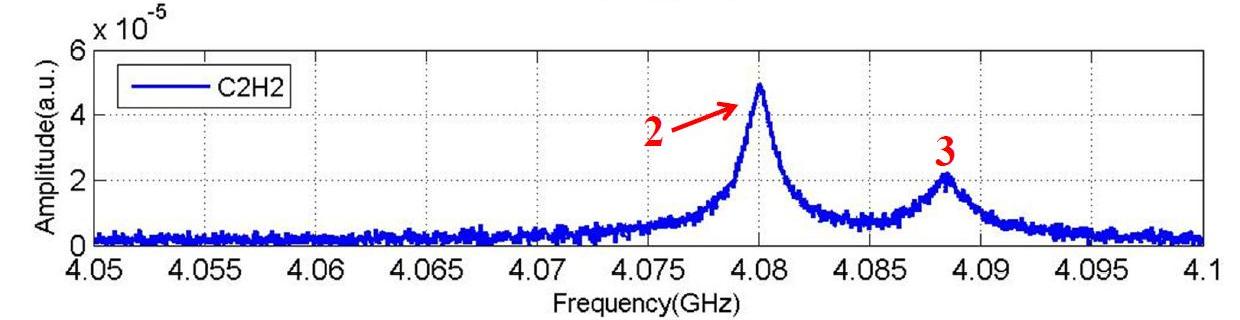}
\label{spec-C2H2-X-2-BP}
}\\
\subfigure[\#1 ($f$:4.0396GHz; Q:10$^4$)]{
\includegraphics[width=0.23\textwidth]{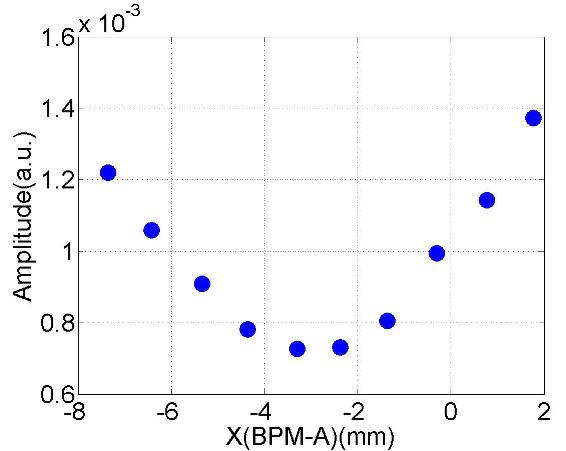}
\label{dep-C2H2-X-1-BP}
}
\subfigure[\#2 ($f$:4.0800GHz; Q:10$^3$)]{
\includegraphics[width=0.23\textwidth]{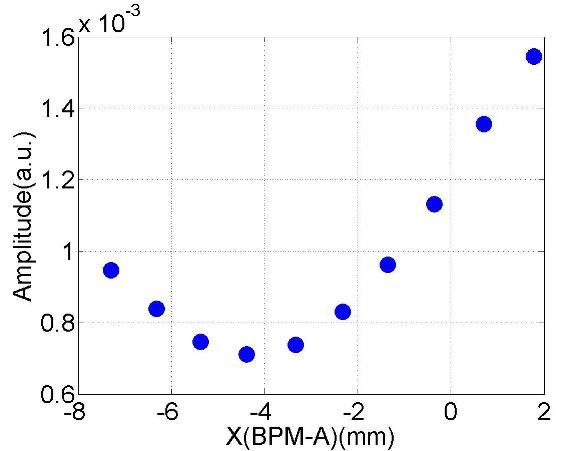}
\label{dep-C2H2-X-2-BP}
}
\subfigure[\#3 ($f$:4.0881GHz; Q:10$^3$)]{
\includegraphics[width=0.23\textwidth]{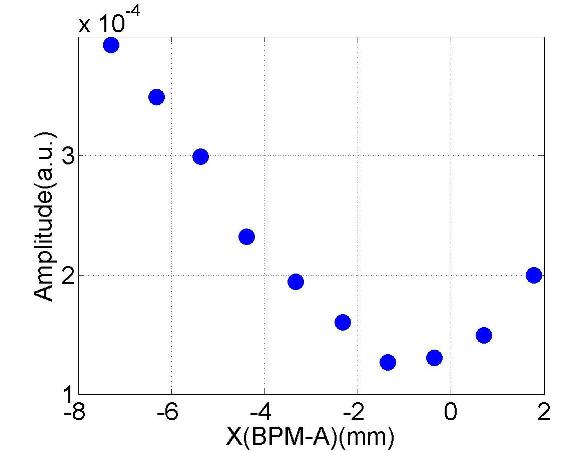}
\label{dep-C2H2-X-3-BP}
}\\
\subfigure[\#1 ($f$:4.0396GHz; Q:10$^4$)]{
\includegraphics[width=0.23\textwidth]{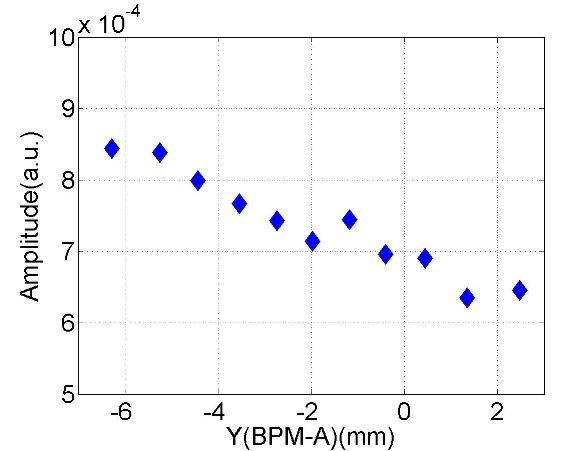}
\label{dep-C2H2-Y-1-BP}
}
\subfigure[\#2 ($f$:4.0800GHz; Q:10$^3$)]{
\includegraphics[width=0.23\textwidth]{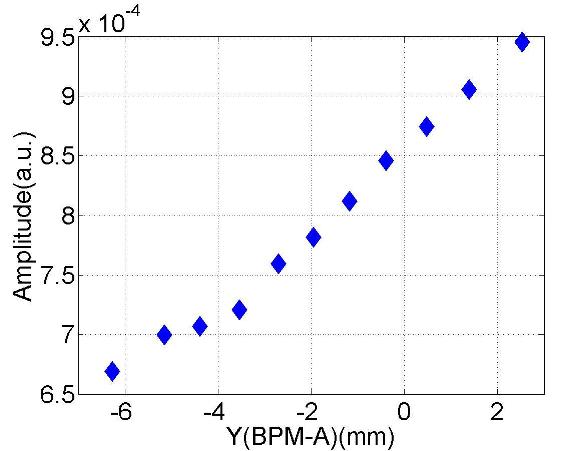}
\label{dep-C2H2-Y-2-BP}
}
\subfigure[\#3 ($f$:4.0881GHz; Q:10$^3$)]{
\includegraphics[width=0.23\textwidth]{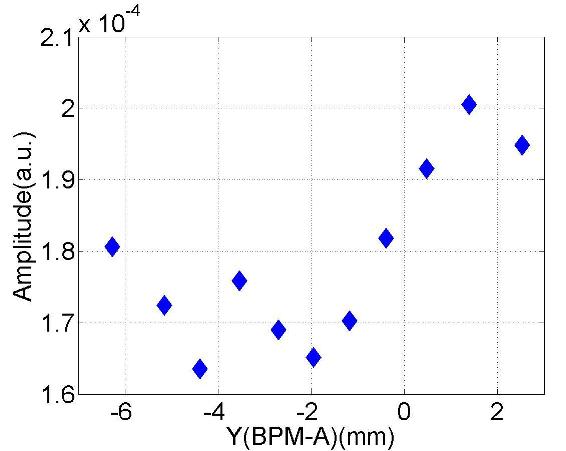}
\label{dep-C2H2-Y-3-BP}
}\\
\subfigure[\#1 ($f$:4.0396GHz; Q:10$^4$)]{
\includegraphics[width=0.23\textwidth]{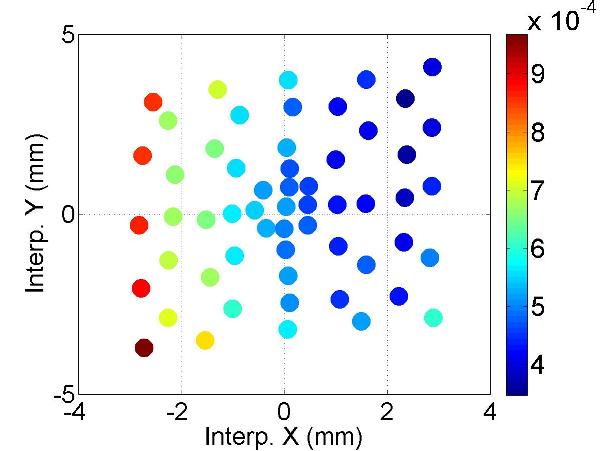}
\label{polar-C2H2-1-BP}
}
\subfigure[\#2 ($f$:4.0800GHz; Q:10$^3$)]{
\includegraphics[width=0.23\textwidth]{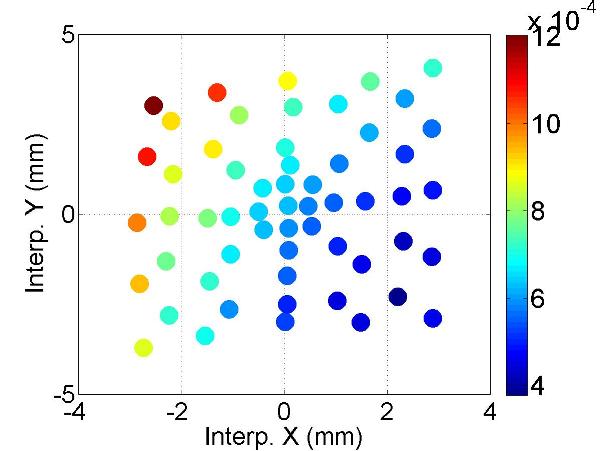}
\label{polar-C2H2-2-BP}
}
\subfigure[\#3 ($f$:4.0876GHz; Q:10$^3$)]{
\includegraphics[width=0.23\textwidth]{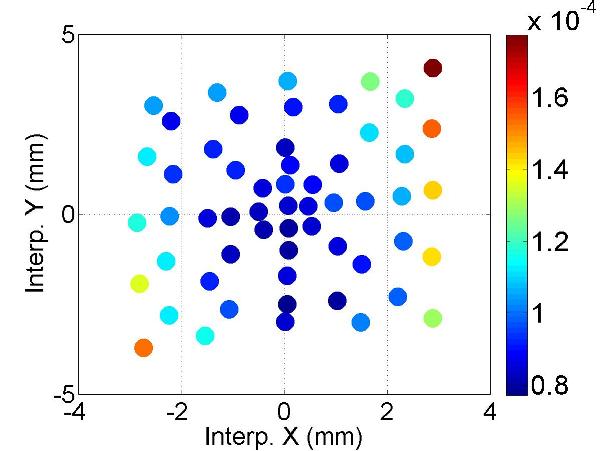}
\label{polar-C2H2-3-BP}
}
\caption{Depedence of the mode amplitude on the transverse beam of{}fset.}
\label{spec-dep-C2H2-XY-1-2-BP}
\end{figure}
\begin{figure}[h]\center
\subfigure[Spectrum (C2H2)]{
\includegraphics[width=1\textwidth]{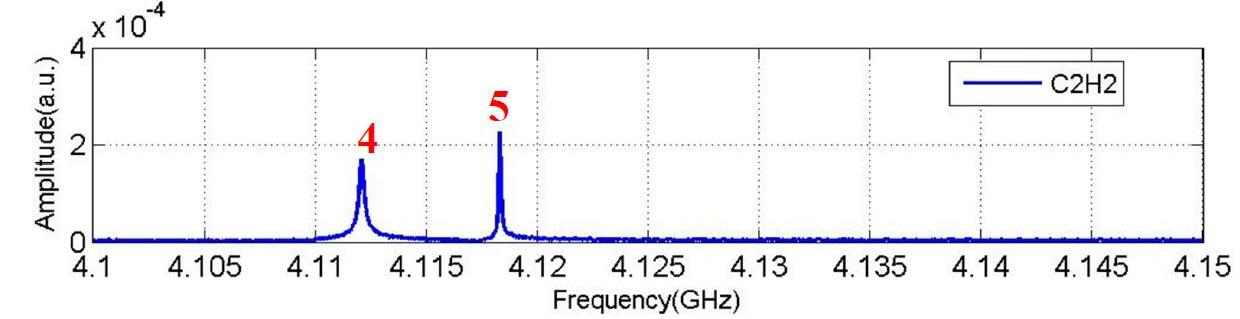}
\label{spec-C2H2-X-3-BP}
}
\subfigure[\#4 ($f$:4.1121GHz; Q:10$^4$)]{
\includegraphics[width=0.31\textwidth]{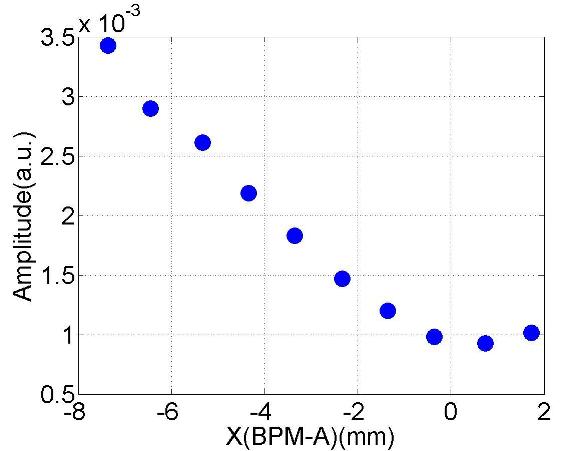}
\label{dep-C2H2-X-4-BP}
}
\subfigure[\#5 ($f$:4.1183GHz; Q:10$^4$)]{
\includegraphics[width=0.31\textwidth]{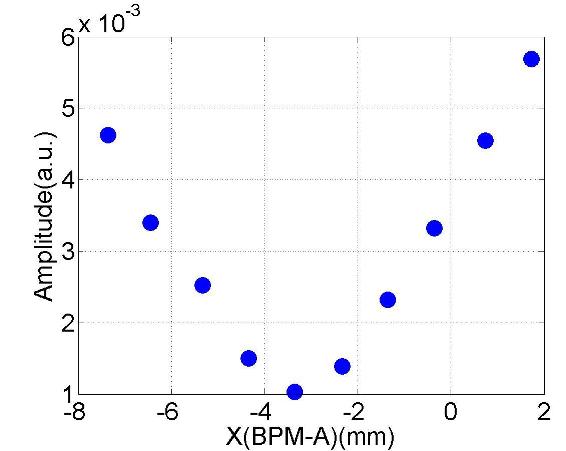}
\label{dep-C2H2-X-5-BP}
}\\
\subfigure[\#4 ($f$:4.1121GHz; Q:10$^4$)]{
\includegraphics[width=0.31\textwidth]{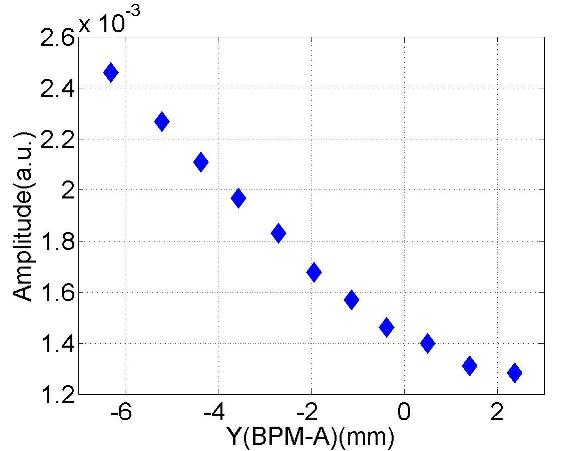}
\label{dep-C2H2-Y-4-BP}
}
\subfigure[\#5 ($f$:4.1183GHz; Q:10$^4$)]{
\includegraphics[width=0.31\textwidth]{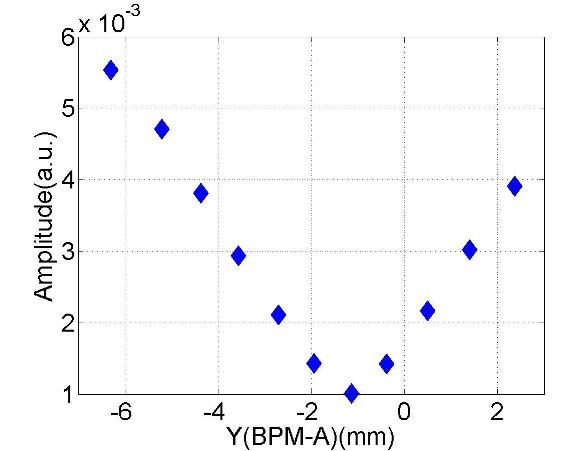}
\label{dep-C2H2-Y-5-BP}
}\\
\subfigure[\#4 ($f$:4.1121GHz; Q:10$^4$)]{
\includegraphics[width=0.31\textwidth]{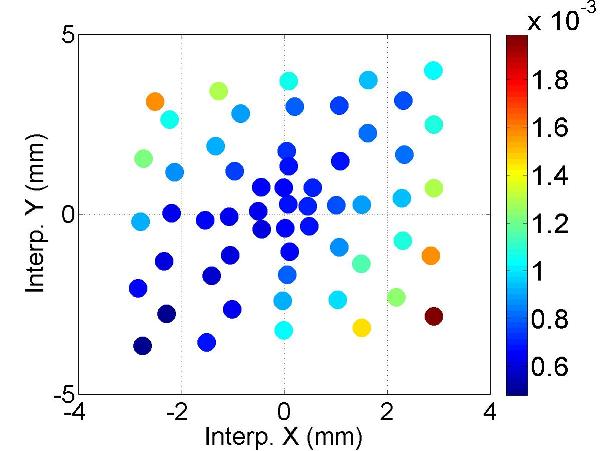}
\label{polar-C2H2-4-BP}
}
\subfigure[\#5 ($f$:4.1183GHz; Q:10$^5$)]{
\includegraphics[width=0.31\textwidth]{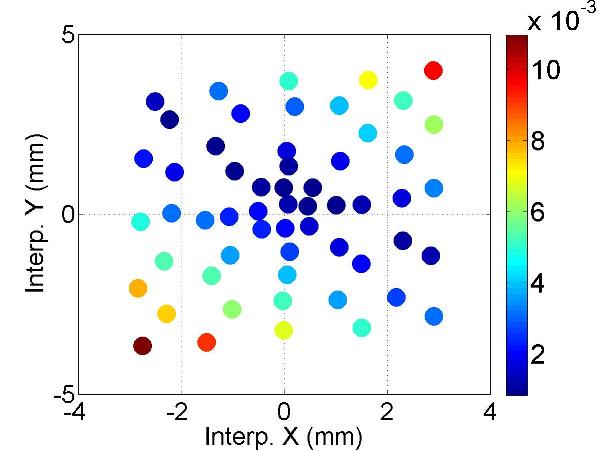}
\label{polar-C2H2-5-BP}
}
\caption{Depedence of the mode amplitude on the transverse beam of{}fset.}
\label{spec-dep-C2H2-XY-3-BP}
\end{figure}

\FloatBarrier
\section{BP: HOM Coupler C3H1}
\begin{figure}[h]
\subfigure[Spectrum (C3H1)]{
\includegraphics[width=1\textwidth]{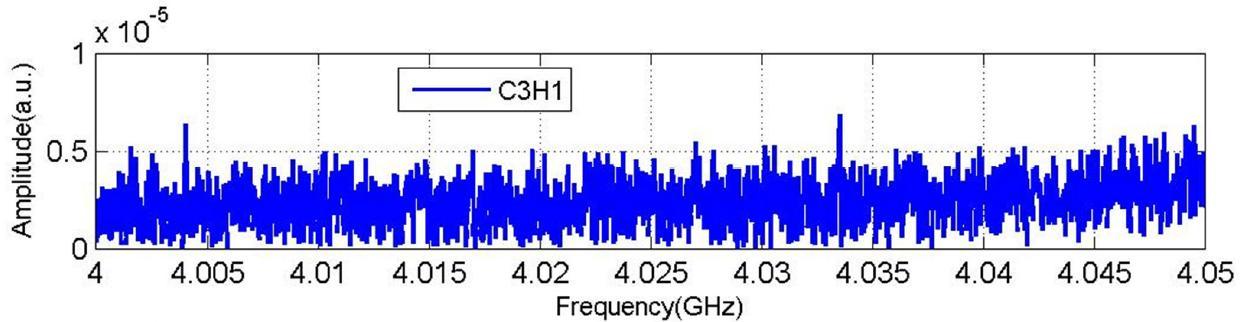}
\label{spec-C3H1-X-1-BP}
}
\subfigure[Spectrum (C3H1)]{
\includegraphics[width=1\textwidth]{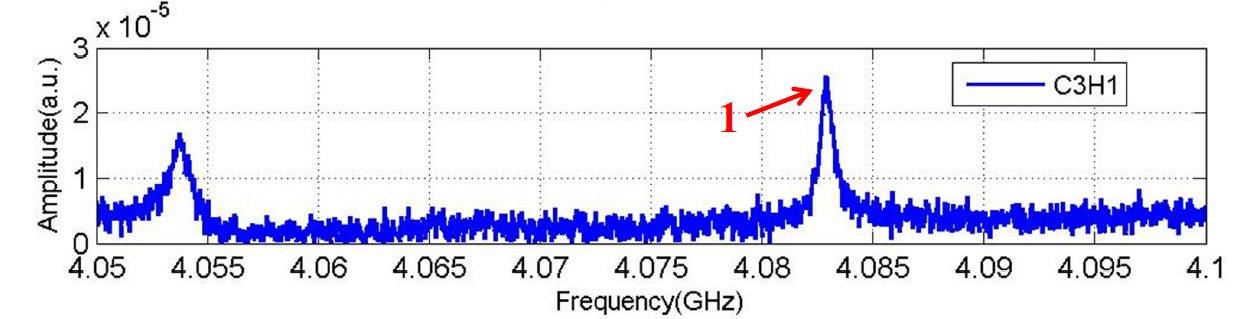}
\label{spec-C3H1-X-2-BP}
}
\subfigure[\#1 ($f$:4.0829GHz; Q:10$^3$)]{
\includegraphics[width=0.31\textwidth]{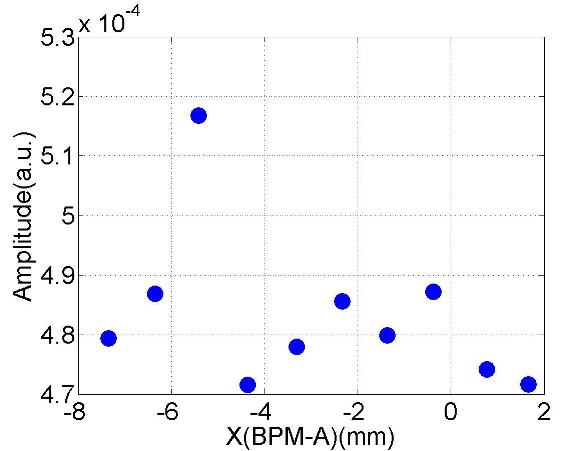}
\label{dep-C3H1-X-1-BP}
}
\subfigure[\#1 ($f$:4.0829GHz; Q:10$^3$)]{
\includegraphics[width=0.31\textwidth]{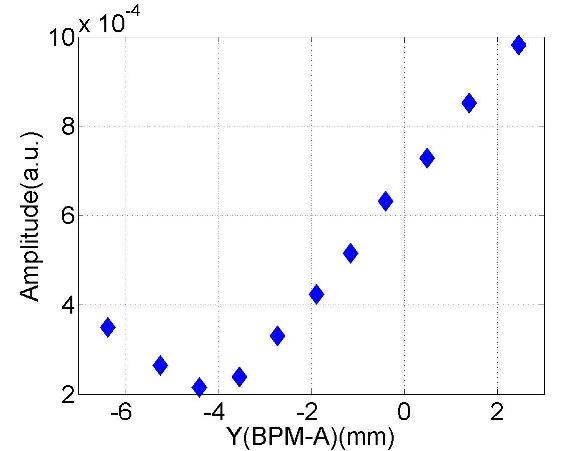}
\label{dep-C3H1-Y-1-BP}
}
\subfigure[\#1 ($f$:4.0829GHz; Q:10$^3$)]{
\includegraphics[width=0.31\textwidth]{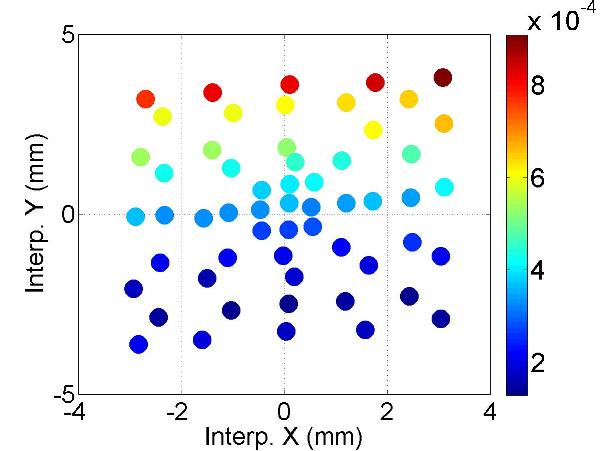}
\label{polar-C3H1-1-BP}
}
\caption{Depedence of the mode amplitude on the transverse beam of{}fset.}
\label{spec-dep-C3H1-XY-1-2-BP}
\end{figure}
\begin{figure}[h]
\subfigure[Spectrum (C3H1)]{
\includegraphics[width=1\textwidth]{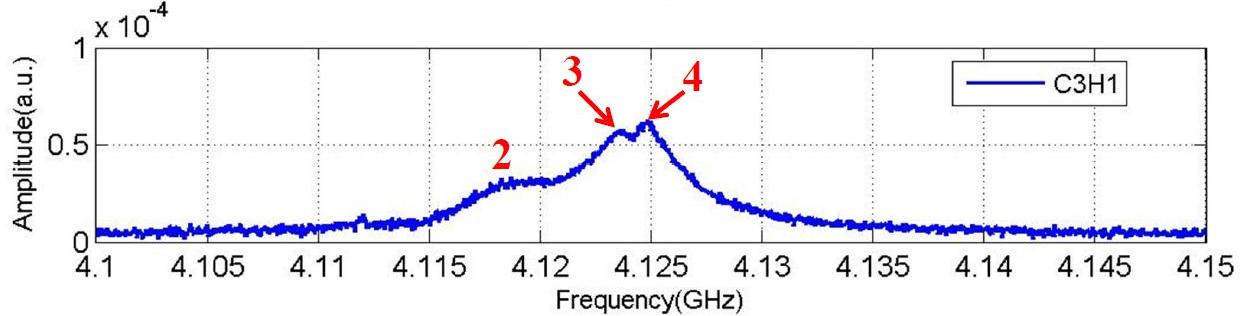}
\label{spec-C3H1-X-3-BP}
}
\subfigure[\#2 ($f$:4.1183GHz; Q:10$^2$)]{
\includegraphics[width=0.31\textwidth]{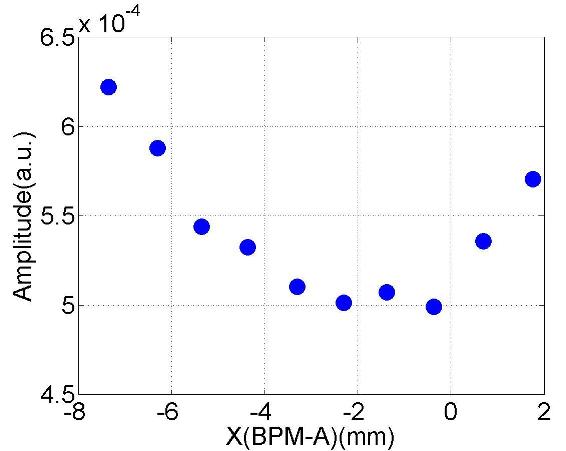}
\label{dep-C3H1-X-2-BP}
}
\subfigure[\#3 ($f$:4.1235GHz; Q:10$^3$)]{
\includegraphics[width=0.31\textwidth]{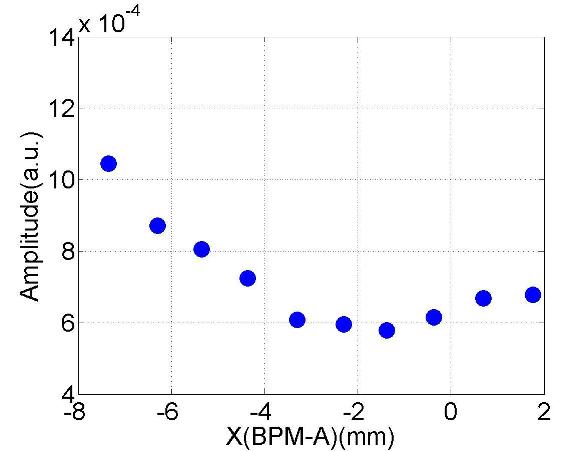}
\label{dep-C3H1-X-3-BP}
}
\subfigure[\#4 ($f$:4.1250GHz; Q:10$^3$)]{
\includegraphics[width=0.31\textwidth]{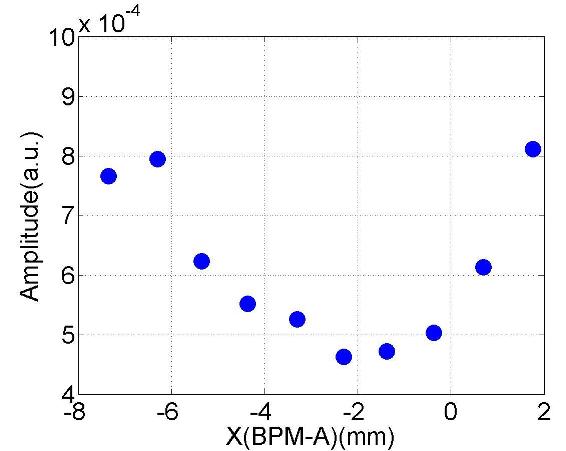}
\label{dep-C3H1-X-4-BP}
}
\subfigure[\#2 ($f$:4.1185GHz; Q:10$^2$)]{
\includegraphics[width=0.31\textwidth]{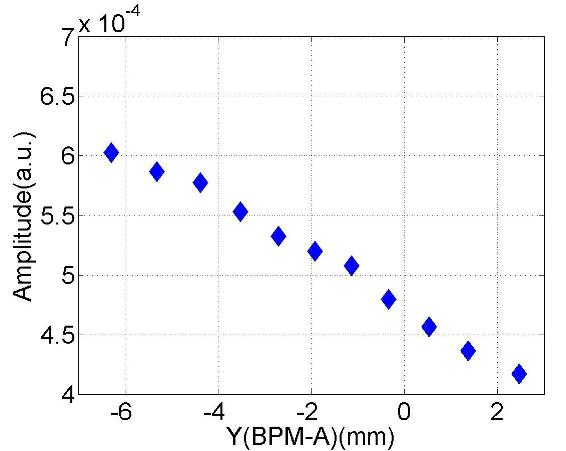}
\label{dep-C3H1-Y-2-BP}
}
\subfigure[\#3 ($f$:4.1235GHz; Q:10$^3$)]{
\includegraphics[width=0.31\textwidth]{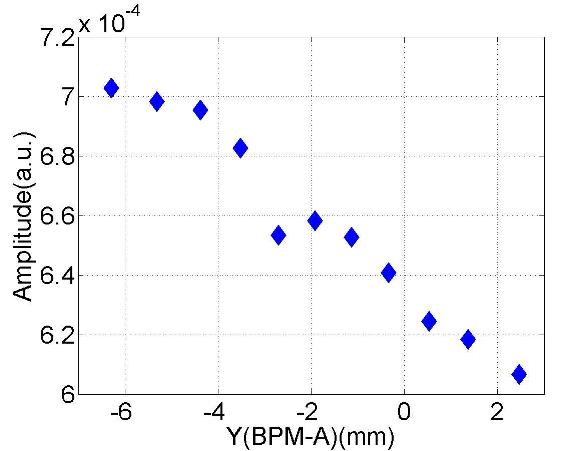}
\label{dep-C3H1-Y-3-BP}
}
\subfigure[\#4 ($f$:4.1252GHz; Q:10$^3$)]{
\includegraphics[width=0.31\textwidth]{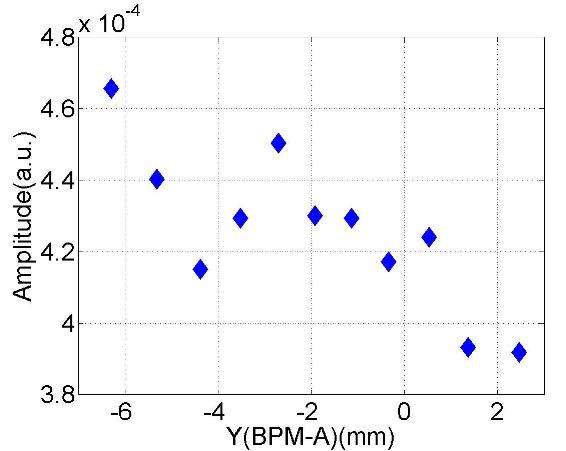}
\label{dep-C3H1-Y-4-BP}
}
\subfigure[\#2 ($f$:4.1181GHz; Q:10$^2$)]{
\includegraphics[width=0.31\textwidth]{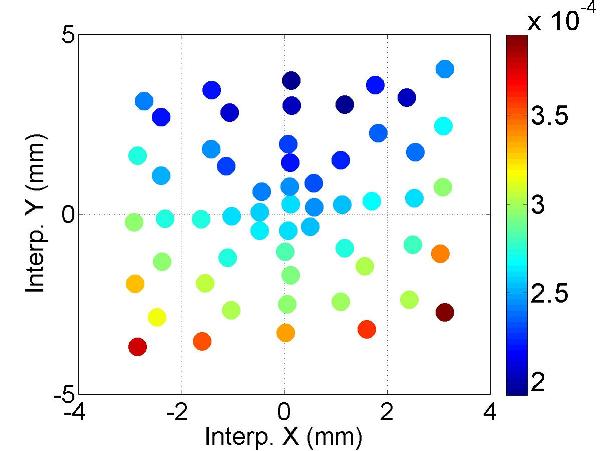}
\label{polar-C3H1-2-BP}
}
\subfigure[\#3 ($f$:4.1236GHz; Q:10$^3$)]{
\includegraphics[width=0.31\textwidth]{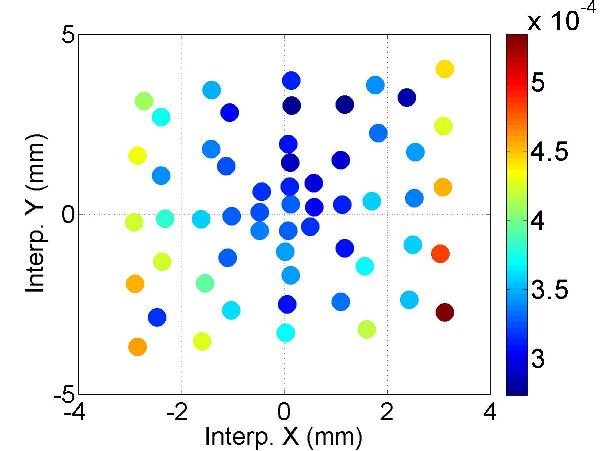}
\label{polar-C3H1-3-BP}
}
\subfigure[\#4 ($f$:4.1252GHz; Q:10$^3$)]{
\includegraphics[width=0.31\textwidth]{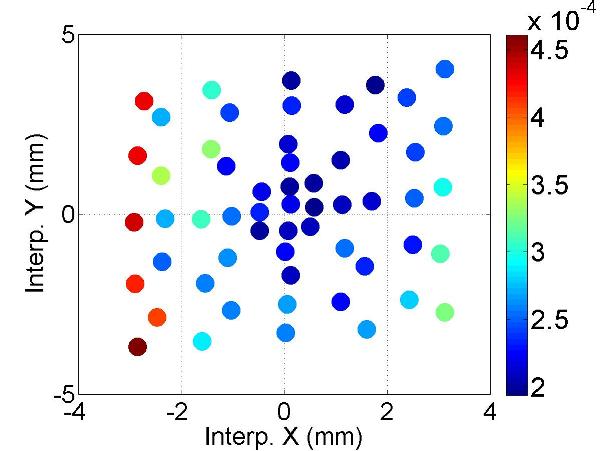}
\label{polar-C3H1-4-BP}
}
\caption{Depedence of the mode amplitude on the transverse beam of{}fset.}
\label{spec-dep-C3H1-XY-3-BP}
\end{figure}

\FloatBarrier
\section{BP: HOM Coupler C3H2}
\begin{figure}[h]\center
\subfigure[Spectrum (C3H2)]{
\includegraphics[width=1\textwidth]{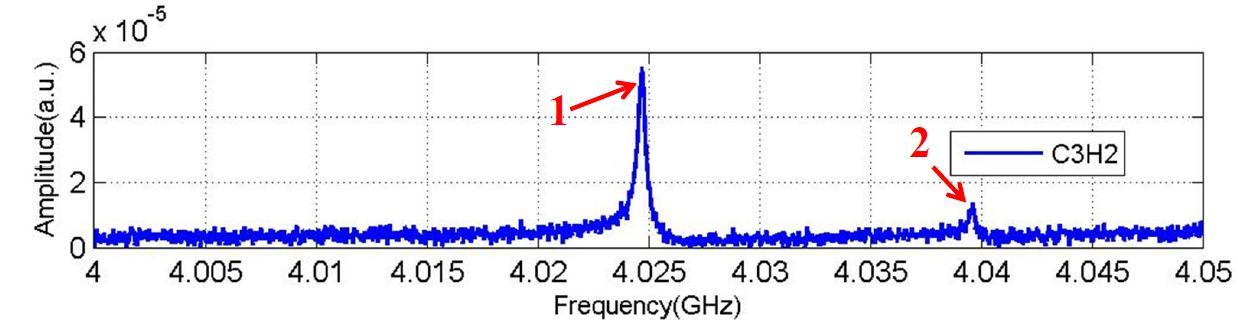}
\label{spec-C3H2-X-1-BP}
}
\subfigure[\#1 ($f$:4.0247GHz; Q:10$^4$)]{
\includegraphics[width=0.3\textwidth]{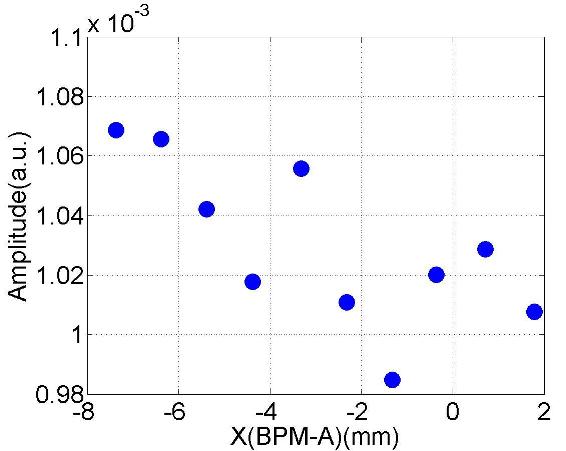}
\label{dep-C3H2-X-1-BP}
}
\subfigure[\#1 ($f$:4.0247GHz; Q:10$^4$)]{
\includegraphics[width=0.3\textwidth]{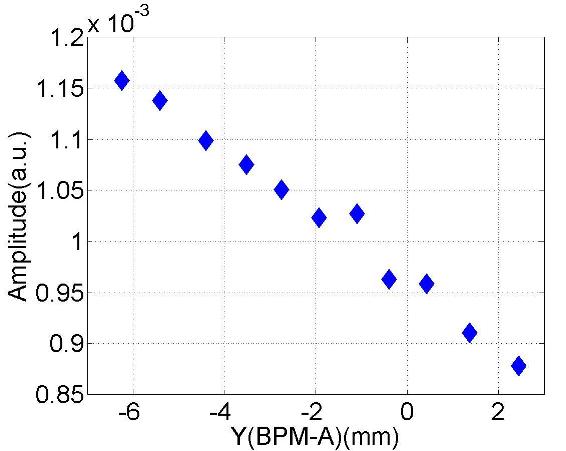}
\label{dep-C3H2-Y-1-BP}
}
\subfigure[\#1 ($f$:4.0247GHz; Q:10$^4$)]{
\includegraphics[width=0.32\textwidth]{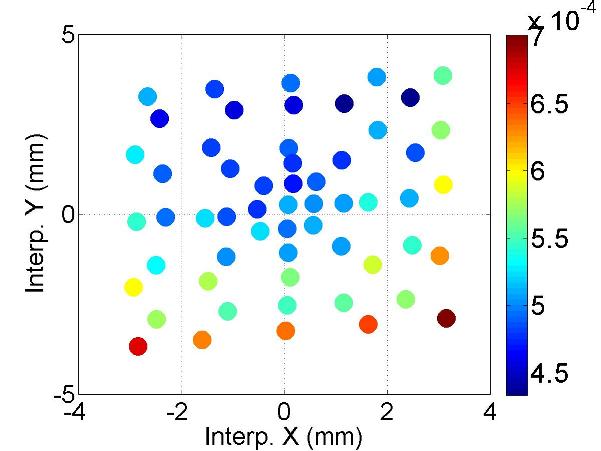}
\label{polar-C3H2-1-BP}
}\\
\subfigure[\#2 ($f$:4.0396GHz; Q:10$^4$)]{
\includegraphics[width=0.3\textwidth]{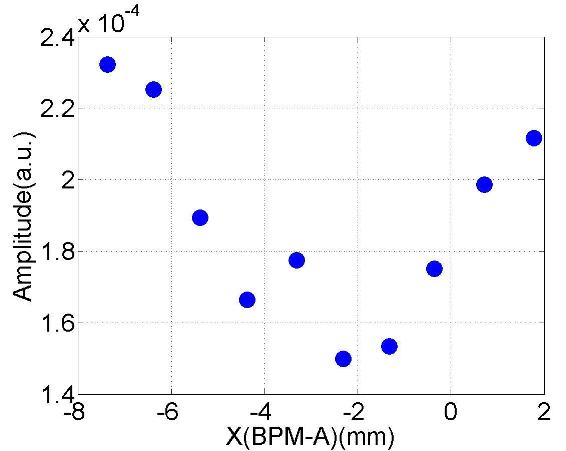}
\label{dep-C3H2-X-2-BP}
}
\subfigure[\#2 ($f$:4.0396GHz; Q:10$^4$)]{
\includegraphics[width=0.3\textwidth]{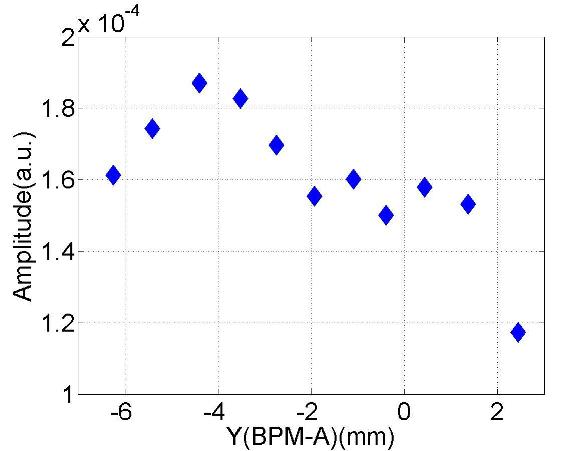}
\label{dep-C3H2-Y-2-BP}
}
\subfigure[\#2 ($f$:4.0396GHz; Q:10$^4$)]{
\includegraphics[width=0.32\textwidth]{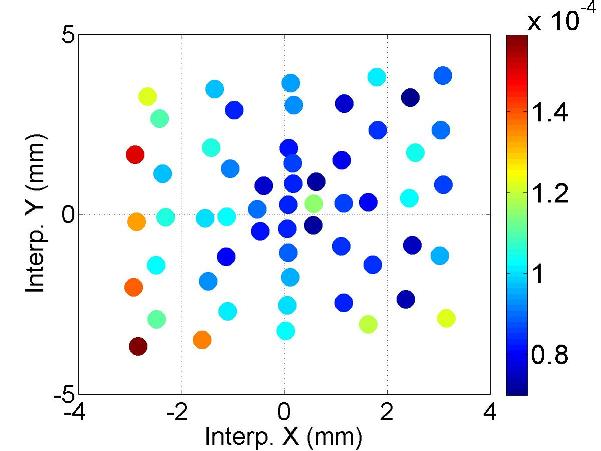}
\label{polar-C3H2-2-BP}
}
\caption{Depedence of the mode amplitude on the transverse beam of{}fset.}
\label{spec-dep-C3H2-XY-1-BP}
\end{figure}
\begin{figure}[h]\center
\subfigure[Spectrum (C3H2)]{
\includegraphics[width=1\textwidth]{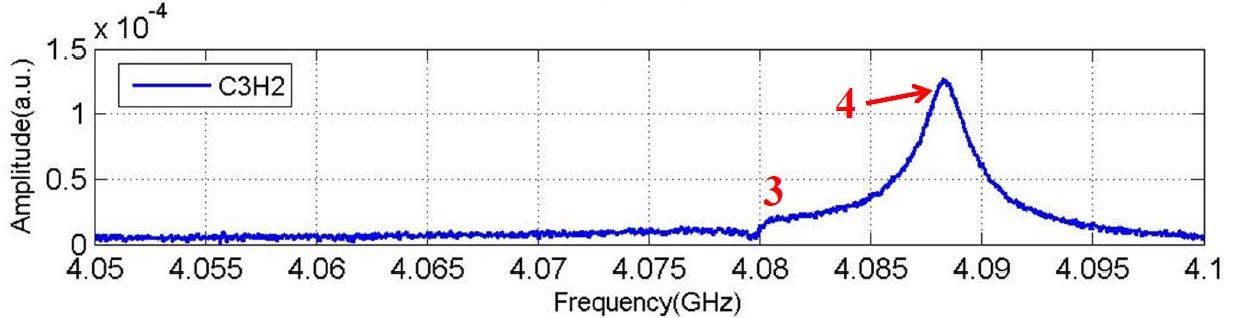}
\label{spec-C3H2-X-2-BP}
}
\subfigure[\#3 ($f$:4.0802GHz; Q:10$^3$)]{
\includegraphics[width=0.31\textwidth]{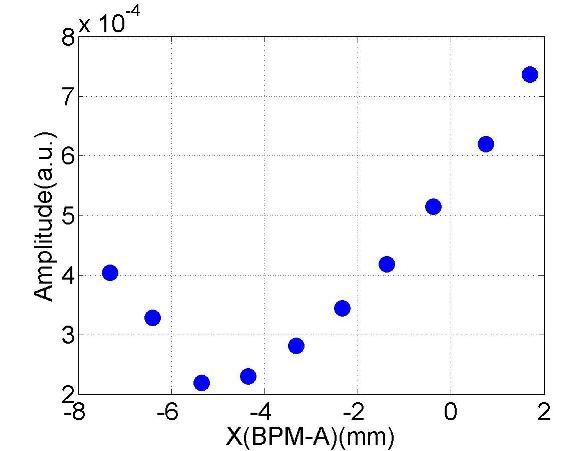}
\label{dep-C3H2-X-3-BP}
}
\subfigure[\#4 ($f$:4.0883GHz; Q:10$^3$)]{
\includegraphics[width=0.31\textwidth]{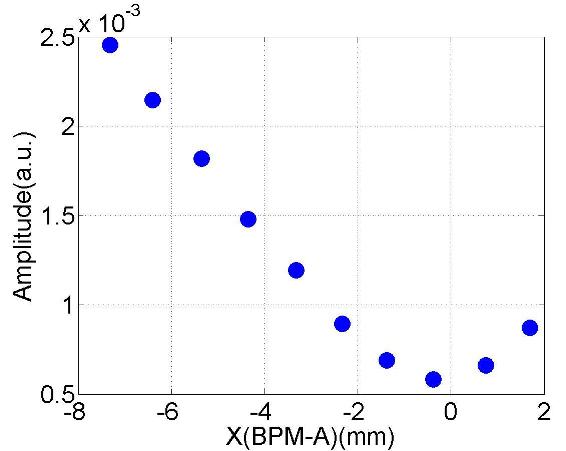}
\label{dep-C3H2-X-4-BP}
}\\
\subfigure[\#3 ($f$:4.0799GHz; Q:10$^3$)]{
\includegraphics[width=0.31\textwidth]{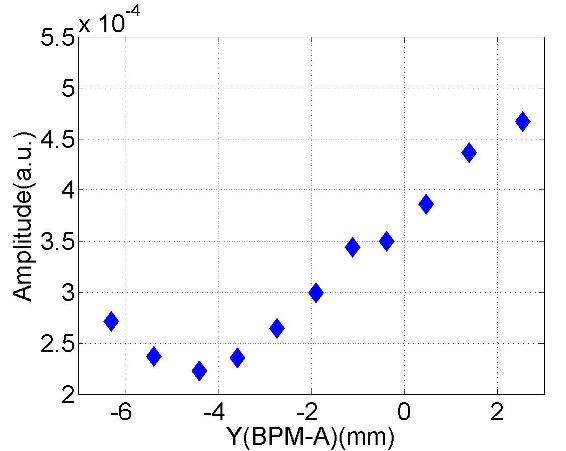}
\label{dep-C3H2-Y-3-BP}
}
\subfigure[\#4 ($f$:4.0883GHz; Q:10$^3$)]{
\includegraphics[width=0.31\textwidth]{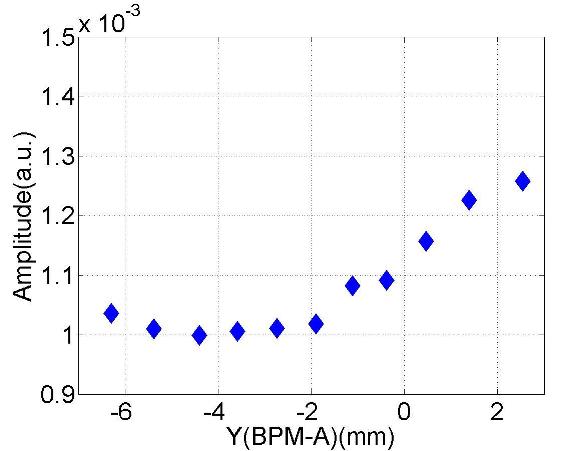}
\label{dep-C3H2-Y-4-BP}
}\\
\subfigure[\#3 ($f$:4.0799GHz; Q:10$^3$)]{
\includegraphics[width=0.31\textwidth]{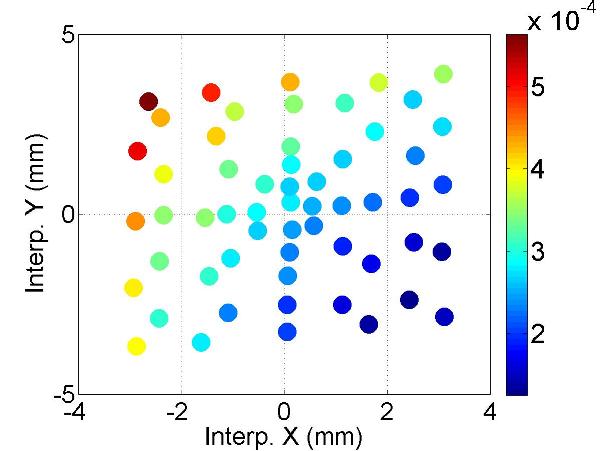}
\label{polar-C3H2-3-BP}
}
\subfigure[\#4 ($f$:4.0883GHz; Q:10$^3$)]{
\includegraphics[width=0.31\textwidth]{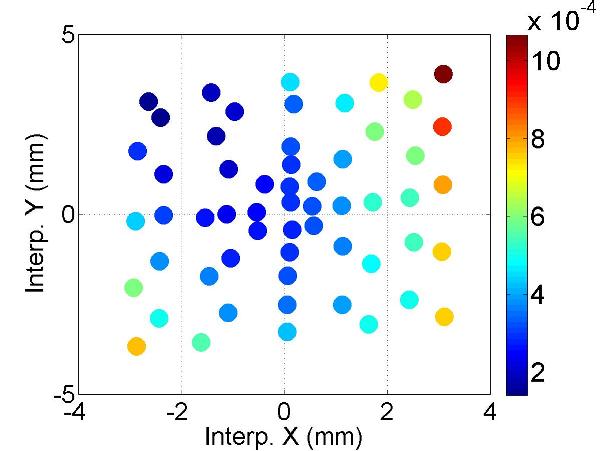}
\label{polar-C3H2-4-BP}
}
\caption{Depedence of the mode amplitude on the transverse beam of{}fset.}
\label{spec-dep-C3H2-XY-2-BP}
\end{figure}
\begin{figure}[h]\center
\subfigure[Spectrum (C3H2)]{
\includegraphics[width=1\textwidth]{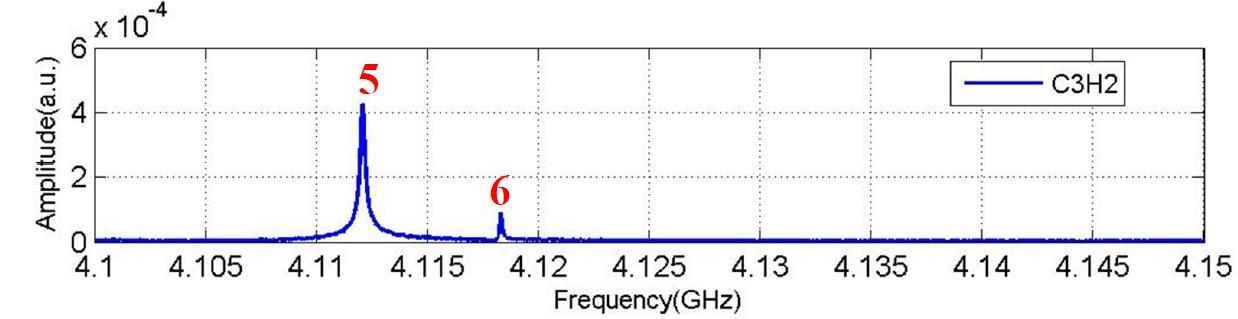}
\label{spec-C3H2-X-3-BP}
}
\subfigure[\#5 ($f$:4.1121GHz; Q:10$^4$)]{
\includegraphics[width=0.31\textwidth]{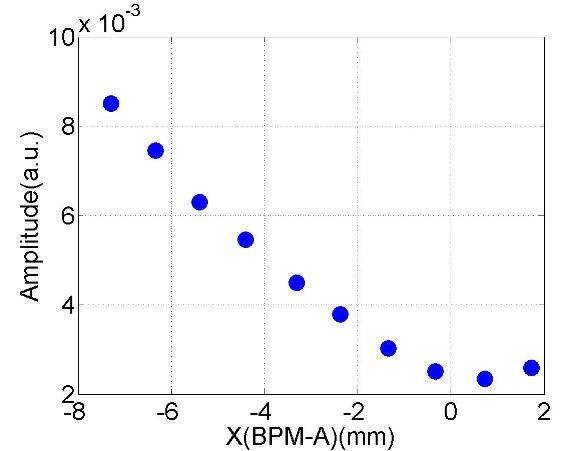}
\label{dep-C3H2-X-5-BP}
}
\subfigure[\#6 ($f$:4.1183GHz; Q:10$^4$)]{
\includegraphics[width=0.31\textwidth]{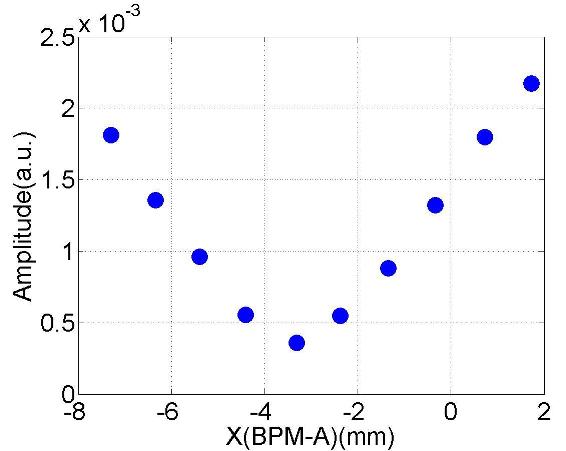}
\label{dep-C3H2-X-6-BP}
}\\
\subfigure[\#5 ($f$:4.1121GHz; Q:10$^4$)]{
\includegraphics[width=0.31\textwidth]{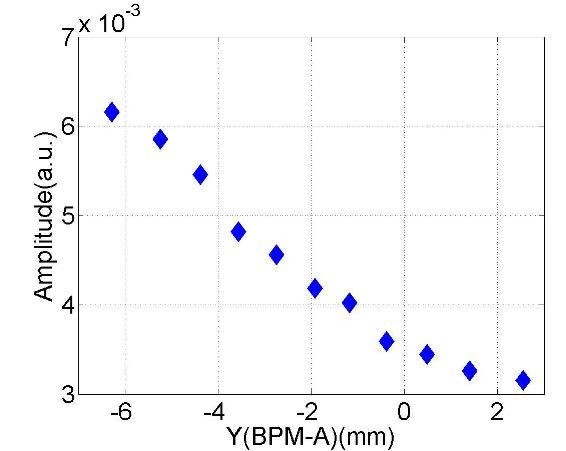}
\label{dep-C3H2-Y-5-BP}
}
\subfigure[\#6 ($f$:4.1183GHz; Q:10$^4$)]{
\includegraphics[width=0.31\textwidth]{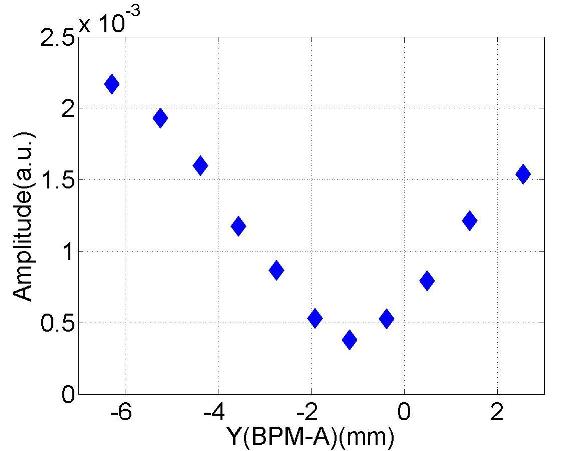}
\label{dep-C3H2-Y-6-BP}
}\\
\subfigure[\#5 ($f$:4.1121GHz; Q:10$^4$)]{
\includegraphics[width=0.31\textwidth]{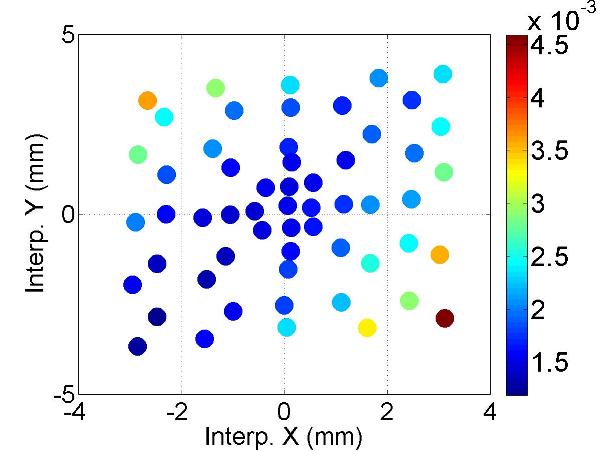}
\label{polar-C3H2-5-BP}
}
\subfigure[\#6 ($f$:4.1183GHz; Q:10$^5$)]{
\includegraphics[width=0.31\textwidth]{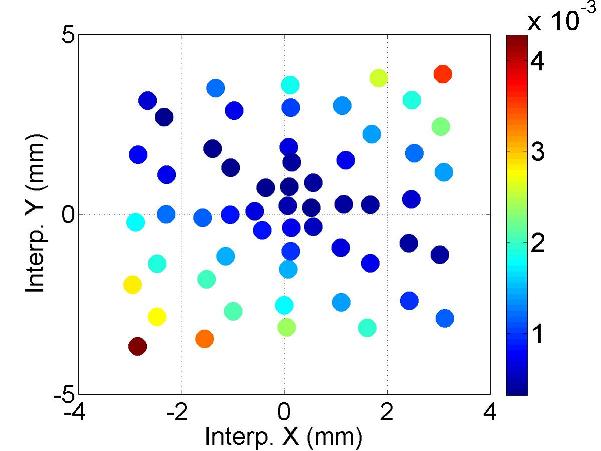}
\label{polar-C3H2-6-BP}
}
\caption{Depedence of the mode amplitude on the transverse beam of{}fset.}
\label{spec-dep-C3H2-XY-3-BP}
\end{figure}

\FloatBarrier
\section{BP: HOM Coupler C4H1}
\begin{figure}[h]\center
\subfigure[Spectrum (C4H1)]{
\includegraphics[width=0.85\textwidth]{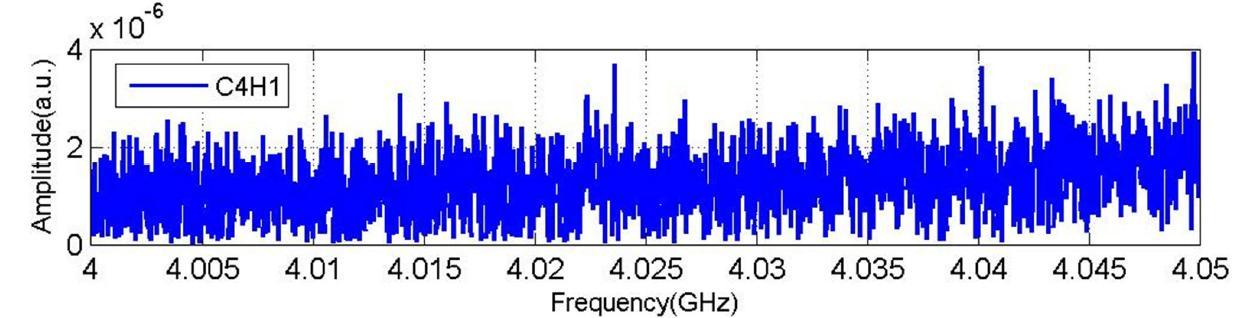}
\label{spec-C4H1-X-1-BP}
}
\subfigure[Spectrum (C4H1)]{
\includegraphics[width=0.85\textwidth]{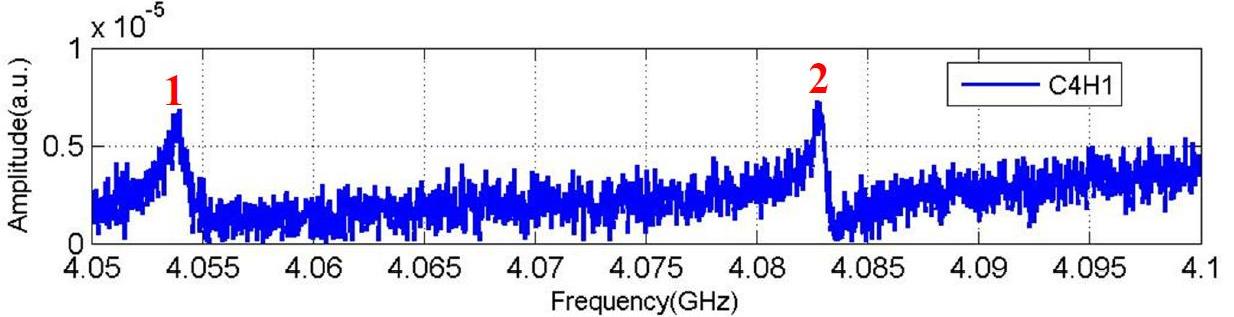}
\label{spec-C4H1-X-2-BP}
}
\subfigure[\#1 ($f$:4.0538GHz; Q:10$^3$)]{
\includegraphics[width=0.26\textwidth]{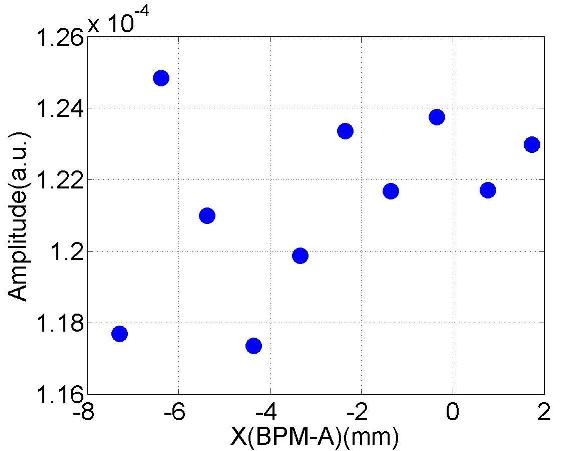}
\label{dep-C4H1-X-1-BP}
}
\subfigure[\#1 ($f$:4.0537GHz; Q:10$^3$)]{
\includegraphics[width=0.26\textwidth]{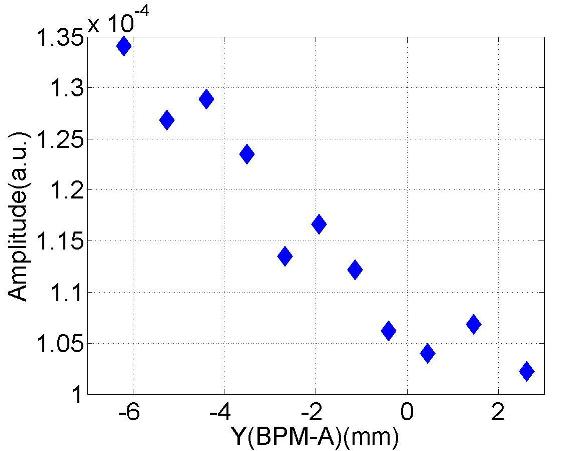}
\label{dep-C4H1-Y-1-BP}
}
\subfigure[\#1 ($f$:4.0537GHz; Q:10$^3$)]{
\includegraphics[width=0.27\textwidth]{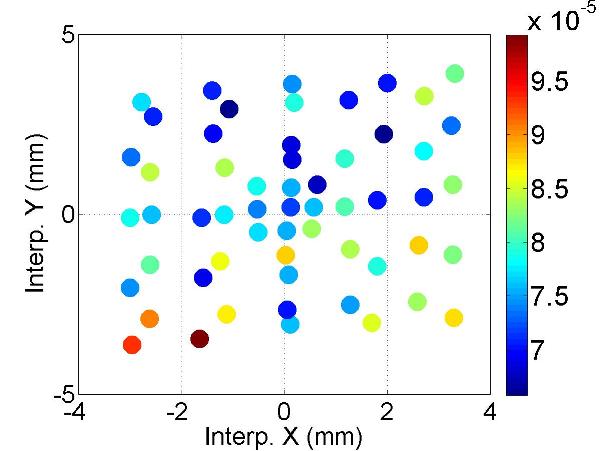}
\label{polar-C4H1-1-BP}
}
\subfigure[\#2 ($f$:4.0828GHz; Q:10$^4$)]{
\includegraphics[width=0.26\textwidth]{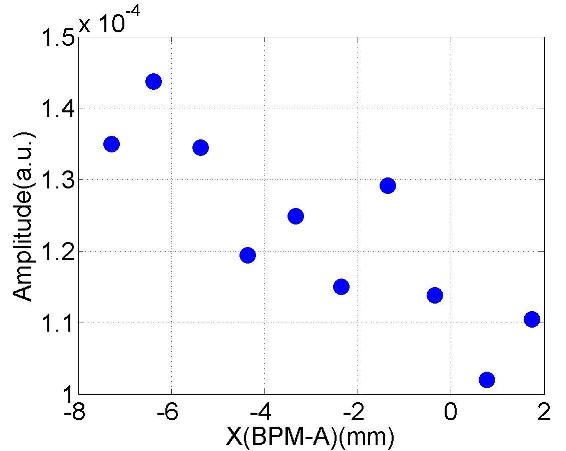}
\label{dep-C4H1-X-2-BP}
}
\subfigure[\#2 ($f$:4.0828GHz; Q:10$^4$)]{
\includegraphics[width=0.26\textwidth]{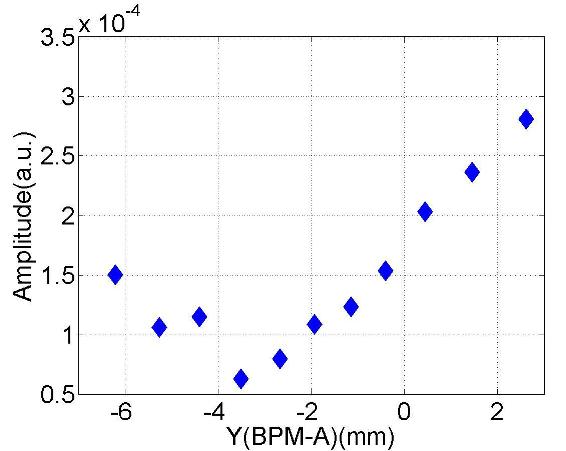}
\label{dep-C4H1-Y-2-BP}
}
\subfigure[\#2 ($f$:4.0828GHz; Q:10$^4$)]{
\includegraphics[width=0.27\textwidth]{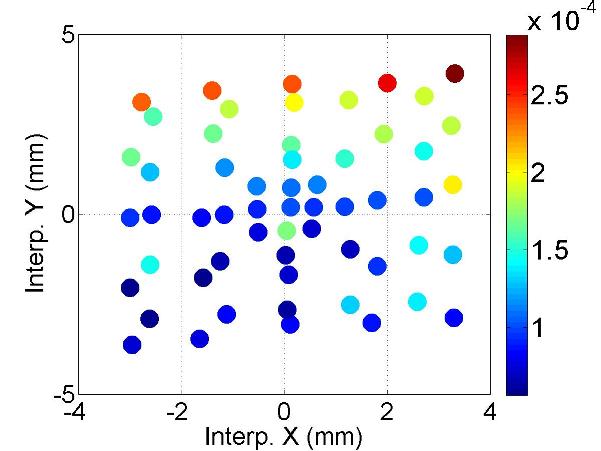}
\label{polar-C4H1-2-BP}
}
\caption{Depedence of the mode amplitude on the transverse beam of{}fset.}
\label{spec-dep-C4H1-XY-1-2-BP}
\end{figure}
\begin{figure}[h]
\subfigure[Spectrum (C4H1)]{
\includegraphics[width=1\textwidth]{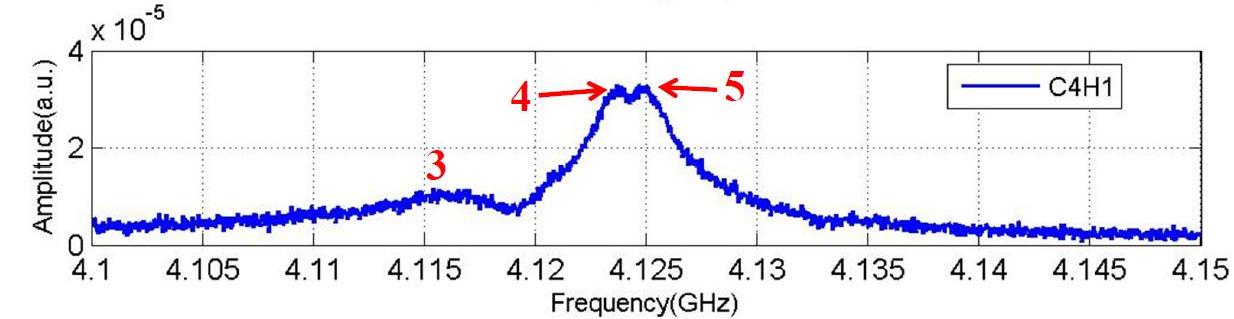}
\label{spec-C4H1-X-3-BP}
}
\subfigure[\#3 ($f$:4.1160GHz; Q:10$^2$)]{
\includegraphics[width=0.31\textwidth]{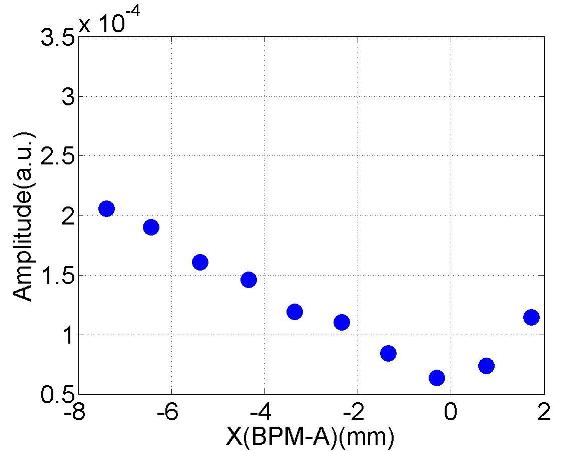}
\label{dep-C4H1-X-3-BP}
}
\subfigure[\#4 ($f$:4.1234GHz; Q:10$^3$)]{
\includegraphics[width=0.31\textwidth]{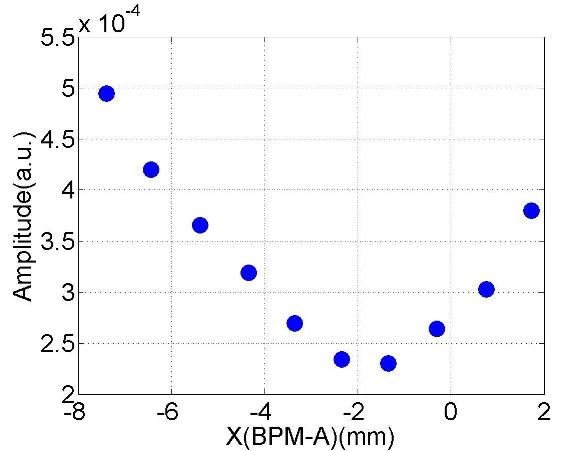}
\label{dep-C4H1-X-4-BP}
}
\subfigure[\#5 ($f$:4.1251GHz; Q:10$^3$)]{
\includegraphics[width=0.31\textwidth]{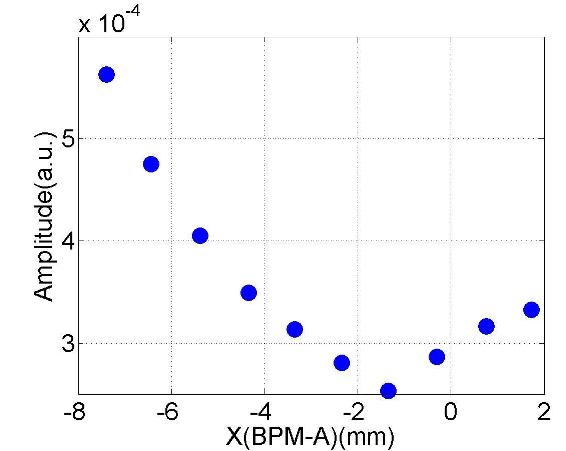}
\label{dep-C4H1-X-5-BP}
}
\subfigure[\#3 ($f$:4.1153GHz; Q:10$^2$)]{
\includegraphics[width=0.31\textwidth]{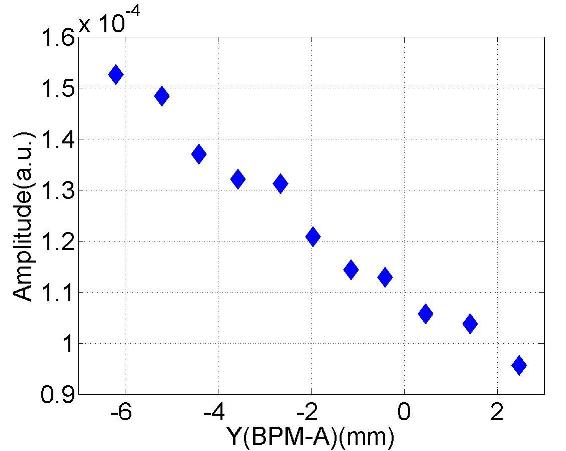}
\label{dep-C4H1-Y-3-BP}
}
\subfigure[\#4 ($f$:4.1236GHz; Q:10$^3$)]{
\includegraphics[width=0.31\textwidth]{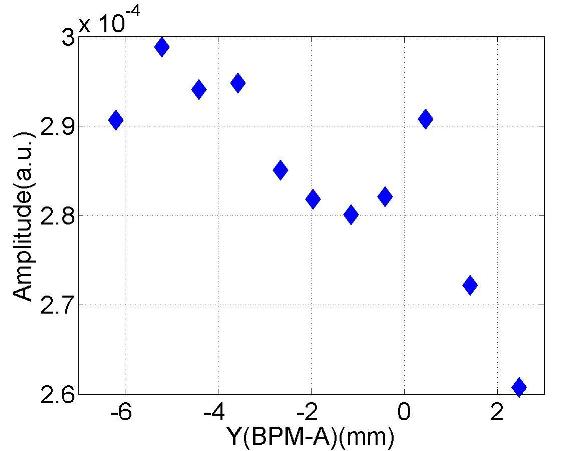}
\label{dep-C4H1-Y-4-BP}
}
\subfigure[\#5 ($f$:4.1252GHz; Q:10$^3$)]{
\includegraphics[width=0.31\textwidth]{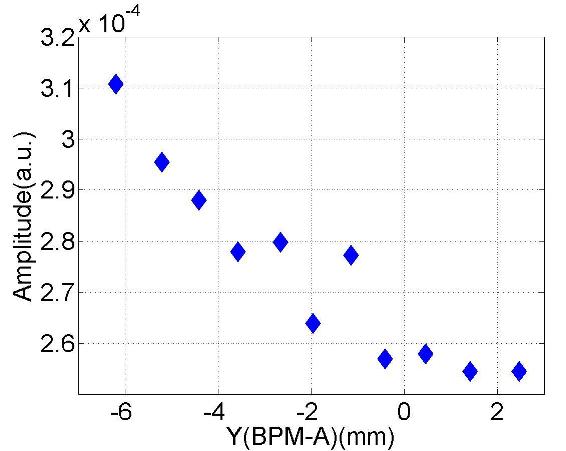}
\label{dep-C4H1-Y-5-BP}
}
\subfigure[\#3 ($f$:4.1155GHz; Q:10$^2$)]{
\includegraphics[width=0.31\textwidth]{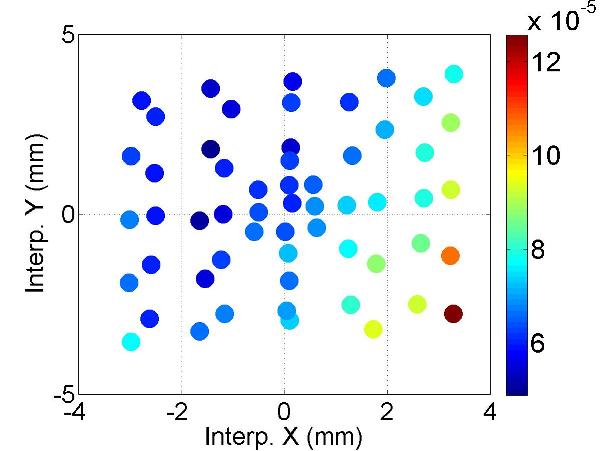}
\label{polar-C4H1-3-BP}
}
\subfigure[\#4 ($f$:4.1235GHz; Q:10$^3$)]{
\includegraphics[width=0.31\textwidth]{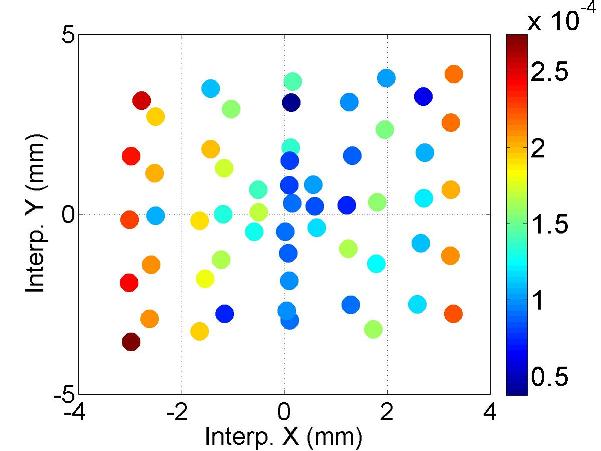}
\label{polar-C4H1-4-BP}
}
\subfigure[\#5 ($f$:4.1249GHz; Q:10$^3$)]{
\includegraphics[width=0.31\textwidth]{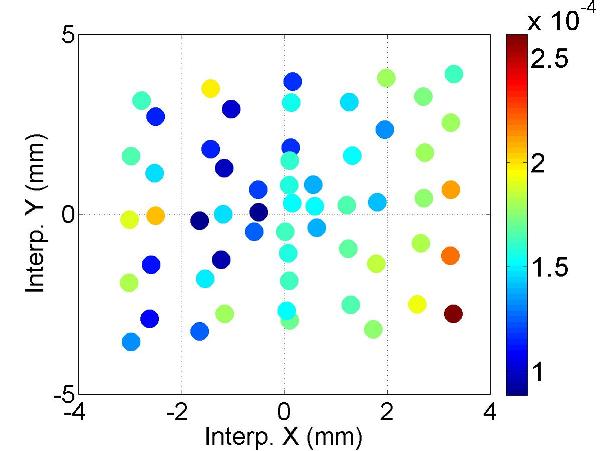}
\label{polar-C4H1-5-BP}
}
\caption{Depedence of the mode amplitude on the transverse beam of{}fset.}
\label{spec-dep-C4H1-XY-3-BP}
\end{figure}

\FloatBarrier
\section{BP: HOM Coupler C4H2}
\begin{figure}[h]\center
\subfigure[Spectrum (C4H2)]{
\includegraphics[width=0.95\textwidth]{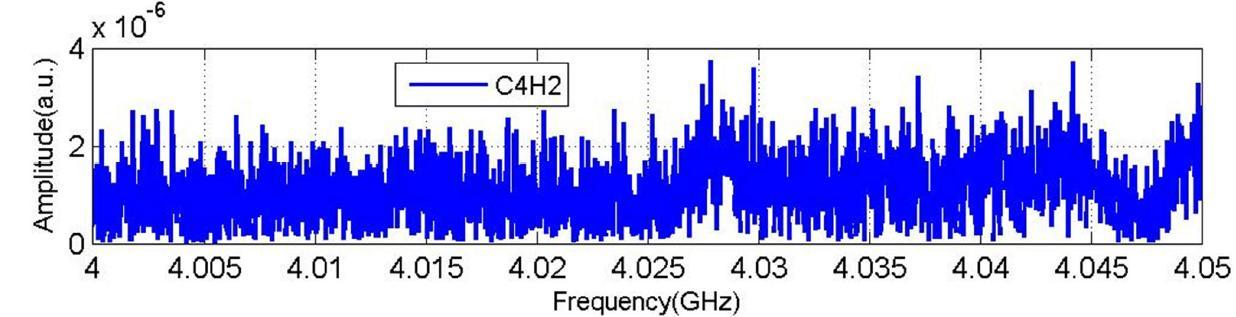}
\label{spec-C4H2-X-1-BP}
}
\subfigure[Spectrum (C4H2)]{
\includegraphics[width=0.95\textwidth]{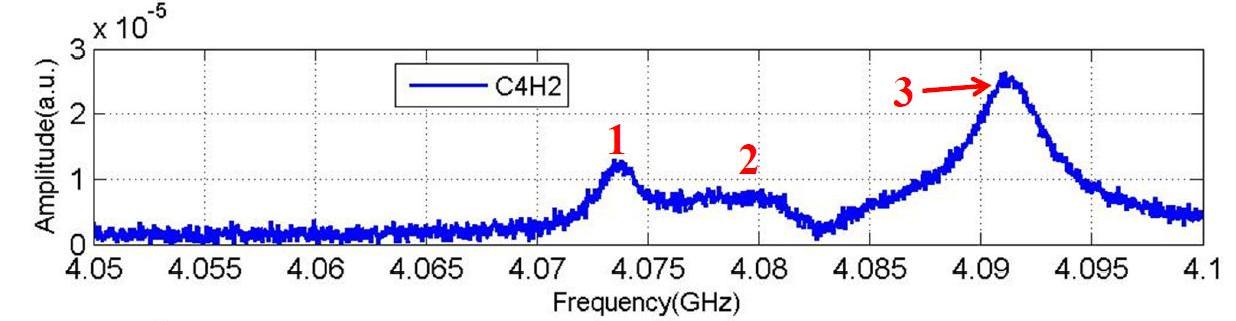}
\label{spec-C4H2-X-2-BP}
}
\subfigure[\#1 ($f$:4.0738GHz; Q:10$^3$)]{
\includegraphics[width=0.24\textwidth]{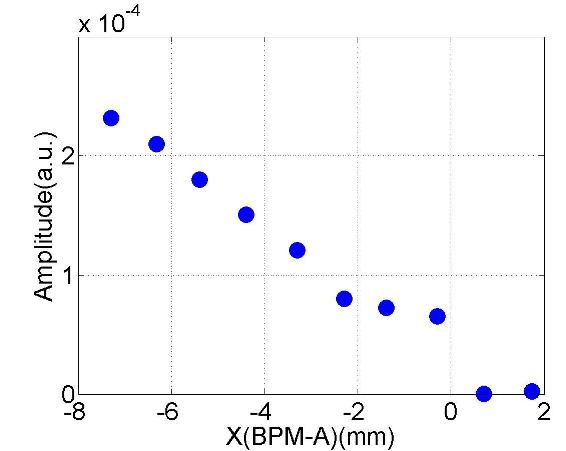}
\label{dep-C4H2-X-1-BP}
}
\subfigure[\#2 ($f$:4.0775GHz; Q:10$^2$)]{
\includegraphics[width=0.24\textwidth]{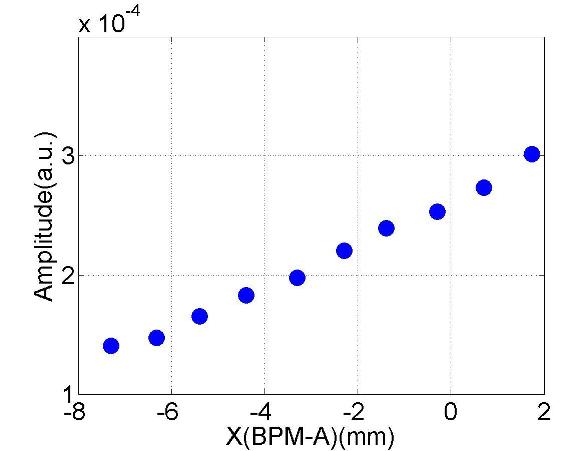}
\label{dep-C4H2-X-2-BP}
}
\subfigure[\#3 ($f$:4.0912GHz; Q:10$^3$)]{
\includegraphics[width=0.24\textwidth]{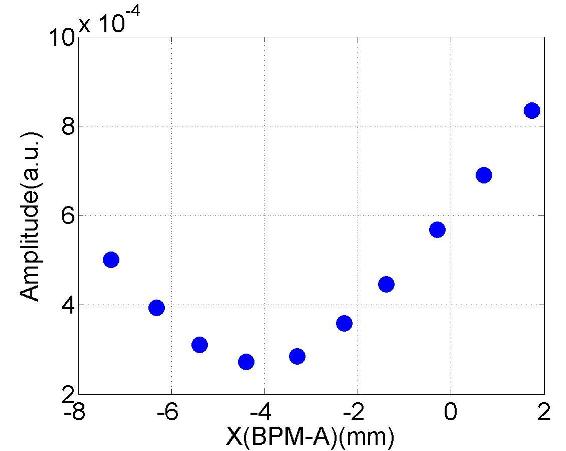}
\label{dep-C4H2-X-3-BP}
}\\
\subfigure[\#1 ($f$:4.0739GHz; Q:10$^3$)]{
\includegraphics[width=0.24\textwidth]{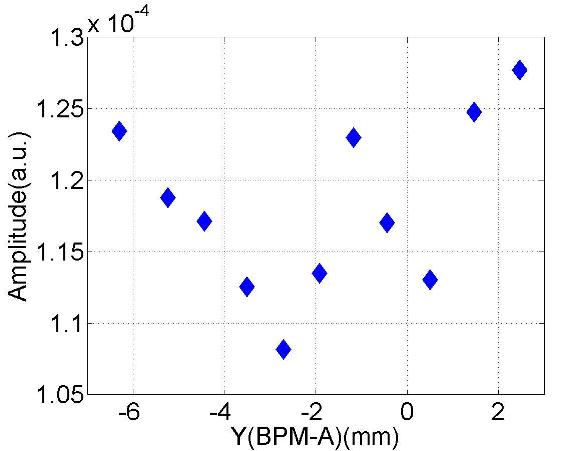}
\label{dep-C4H2-Y-1-BP}
}
\subfigure[\#2 ($f$:4.0771GHz; Q:10$^2$)]{
\includegraphics[width=0.24\textwidth]{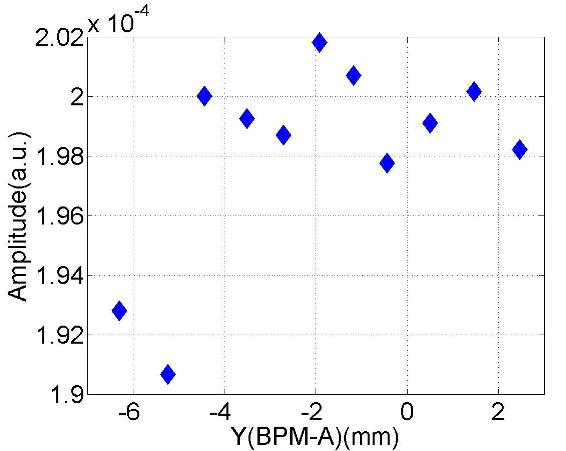}
\label{dep-C4H2-Y-2-BP}
}
\subfigure[\#3 ($f$:4.0911GHz; Q:10$^3$)]{
\includegraphics[width=0.24\textwidth]{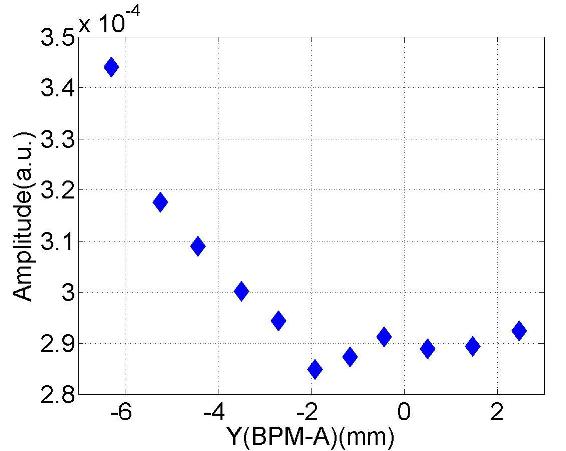}
\label{dep-C4H2-Y-3-BP}
}\\
\subfigure[\#1 ($f$:4.0738GHz; Q:10$^3$)]{
\includegraphics[width=0.25\textwidth]{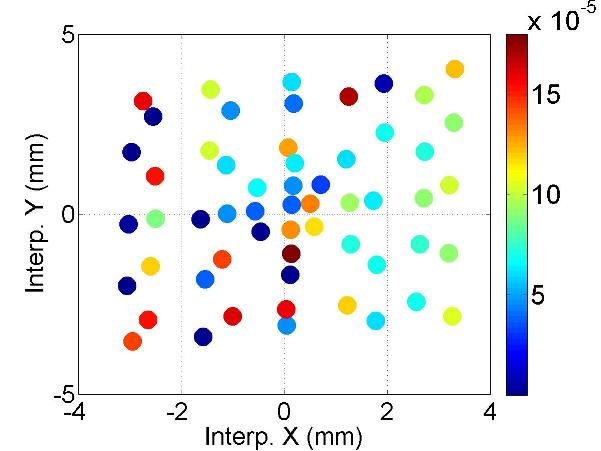}
\label{polar-C4H2-1-BP}
}
\subfigure[\#2 ($f$:4.0772GHz; Q:10$^2$)]{
\includegraphics[width=0.25\textwidth]{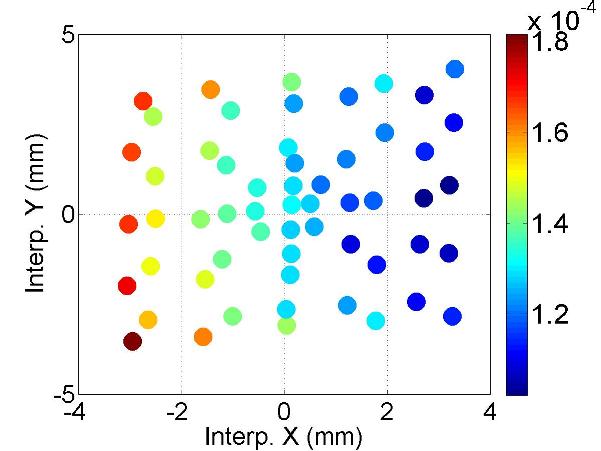}
\label{polar-C4H2-2-BP}
}
\subfigure[\#3 ($f$:4.0911GHz; Q:10$^3$)]{
\includegraphics[width=0.25\textwidth]{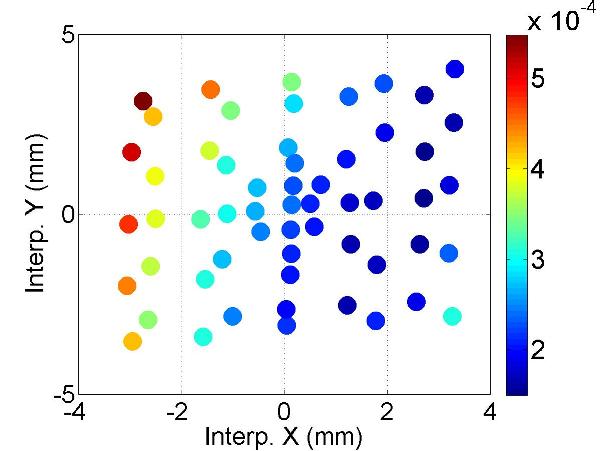}
\label{polar-C4H2-3-BP}
}
\caption{Depedence of the mode amplitude on the transverse beam of{}fset.}
\label{spec-dep-C4H2-XY-1-2-BP}
\end{figure}
\begin{figure}[h]\center
\subfigure[Spectrum (C4H2)]{
\includegraphics[width=1\textwidth]{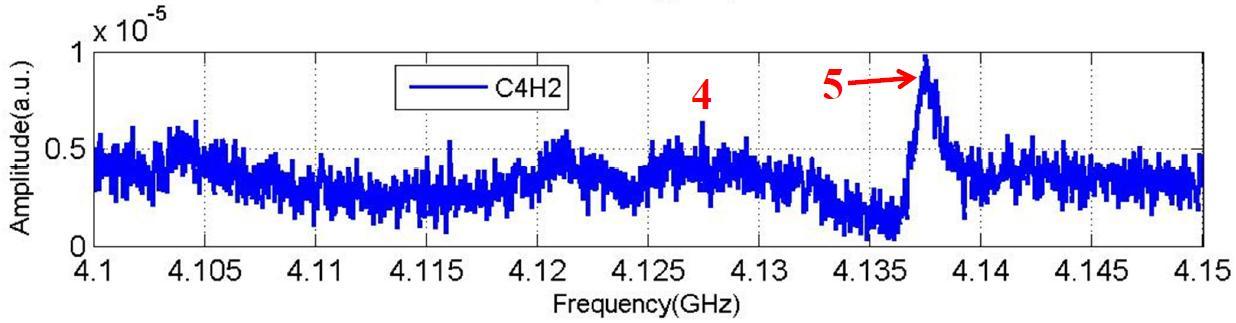}
\label{spec-C4H2-X-3-BP}
}
\subfigure[\#4 ($f$:4.1253GHz; Q:10$^3$)]{
\includegraphics[width=0.31\textwidth]{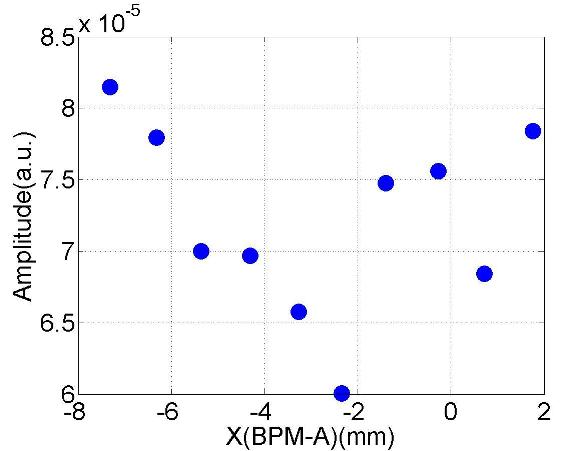}
\label{dep-C4H2-X-4-BP}
}
\subfigure[\#5 ($f$:4.1374GHz; Q:10$^3$)]{
\includegraphics[width=0.31\textwidth]{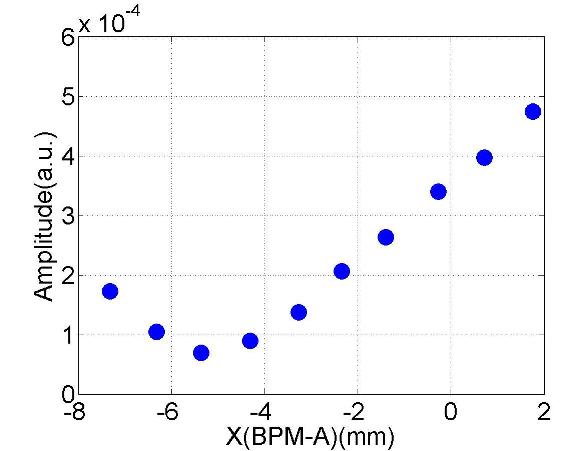}
\label{dep-C4H2-X-5-BP}
}\\
\subfigure[\#4 ($f$:4.1259GHz; Q:10$^3$)]{
\includegraphics[width=0.31\textwidth]{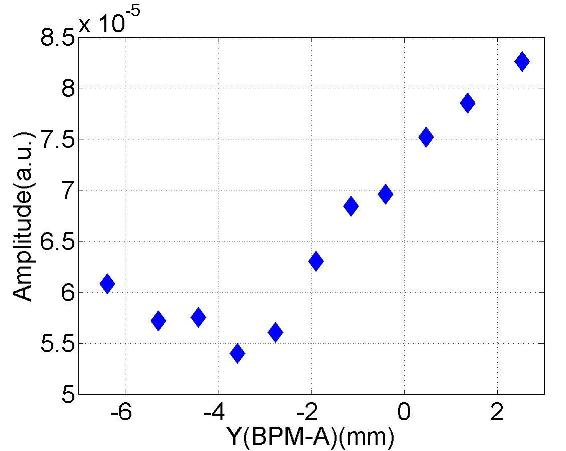}
\label{dep-C4H2-Y-4-BP}
}
\subfigure[\#5 ($f$:4.1373GHz; Q:10$^3$)]{
\includegraphics[width=0.31\textwidth]{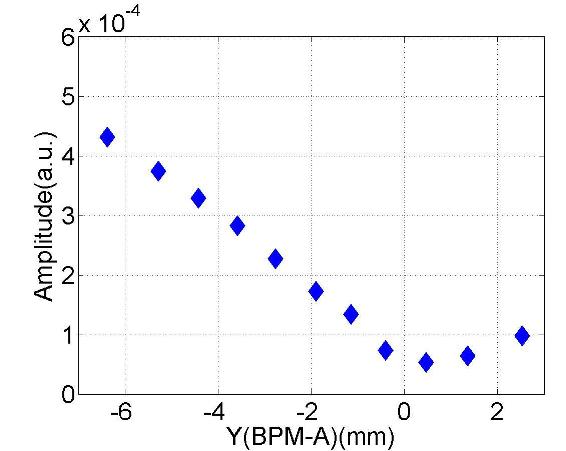}
\label{dep-C4H2-Y-5-BP}
}\\
\subfigure[\#4 ($f$:4.1255GHz; Q:10$^3$)]{
\includegraphics[width=0.32\textwidth]{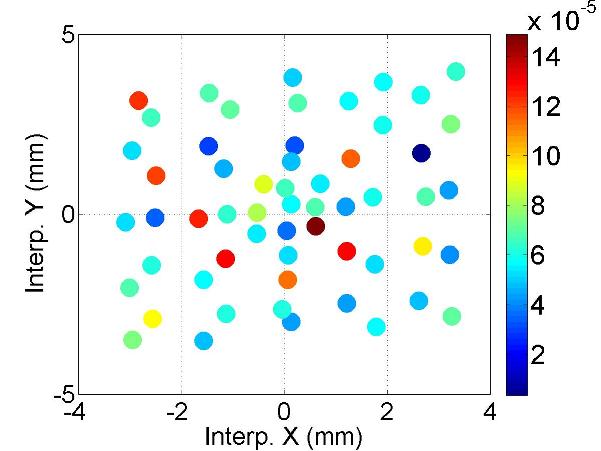}
\label{polar-C4H2-4-BP}
}
\subfigure[\#5 ($f$:4.1374GHz; Q:10$^3$)]{
\includegraphics[width=0.32\textwidth]{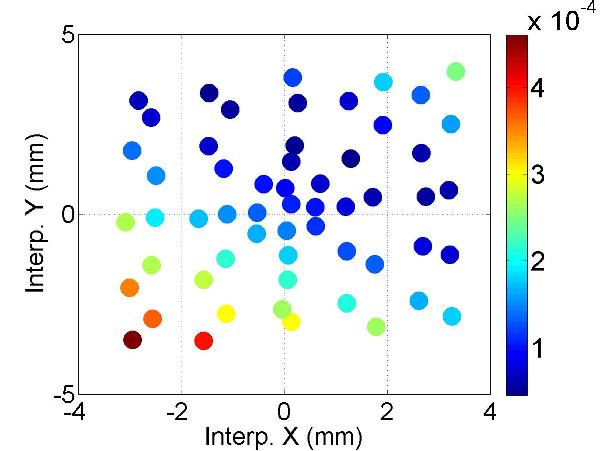}
\label{polar-C4H2-5-BP}
}
\caption{Dependence of the mode amplitude on the transverse beam of{}fset.}
\label{spec-dep-C4H2-XY-3-BP}
\end{figure}

\chapter{Dipole Dependence: The Fifth Dipole Band}\label{app-d5}
Each peak shown in the spectrum ranging from 9-9.1~GHz is f{}itted with Lorentzian distribution (Eq.~\ref{eq:lorfit}). The f{}itted mode amplitude is plotted against interpolated $x$ and $y$ for each cavity. The mode polarization is also plotted with beam position interpolated into each cavity. This chapter is an extension of Chapter~\ref{hom-dep:d5}. 

\section{D5: HOM Coupler C1H1}
\begin{figure}[h]
\subfigure[Spectrum (C1H1)]{
\includegraphics[width=1\textwidth]{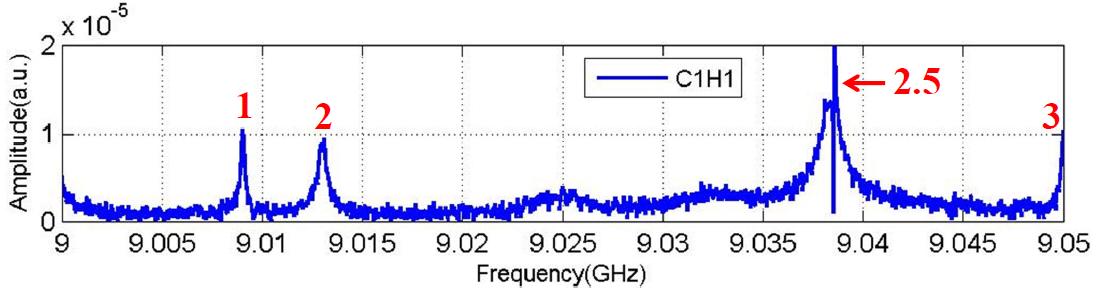}
\label{spec-C1H1-X-1}
}
\subfigure[\#1 ($f$:9.0090GHz; $Q$:10$^4$)]{
\includegraphics[width=0.23\textwidth]{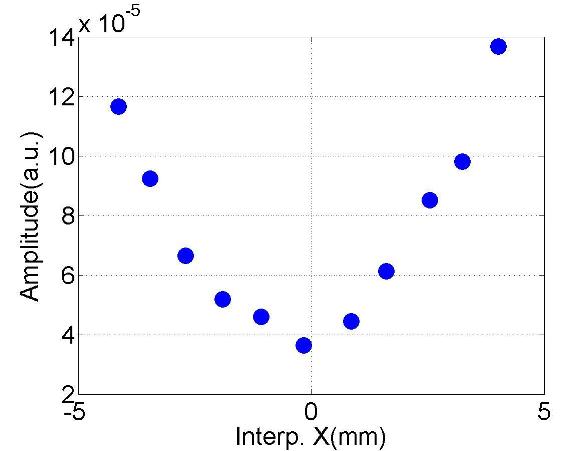}
\label{dep-C1H1-X-1}
}
\subfigure[\#2 ($f$:9.0130GHz; $Q$:10$^4$)]{
\includegraphics[width=0.23\textwidth]{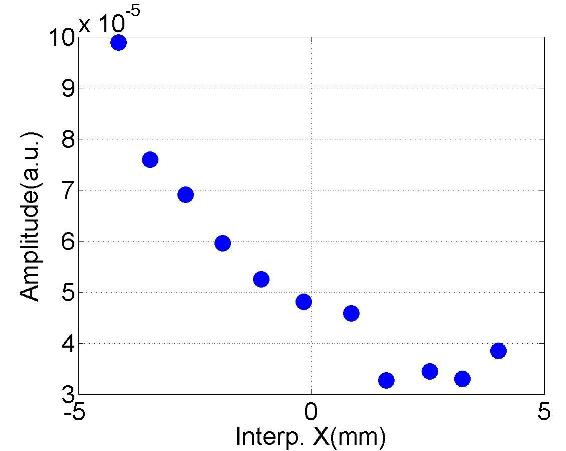}
\label{dep-C1H1-X-2}
}
\subfigure[\#3 ($f$:9.0500GHz; $Q$:10$^4$)]{
\includegraphics[width=0.23\textwidth]{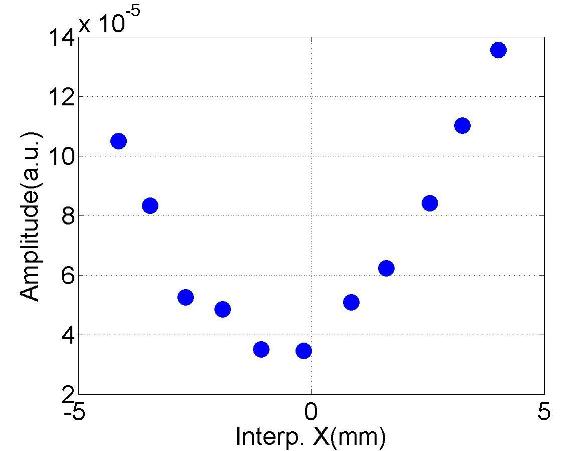}
\label{dep-C1H1-X-3}
}\\
\subfigure[\#1 ($f$:9.0089GHz; $Q$:10$^4$)]{
\includegraphics[width=0.23\textwidth]{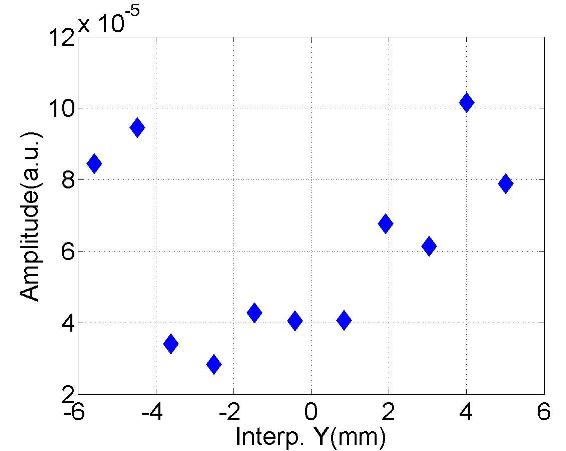}
\label{dep-C1H1-Y-1}
}
\subfigure[\#2 ($f$:9.0130GHz; $Q$:10$^4$)]{
\includegraphics[width=0.23\textwidth]{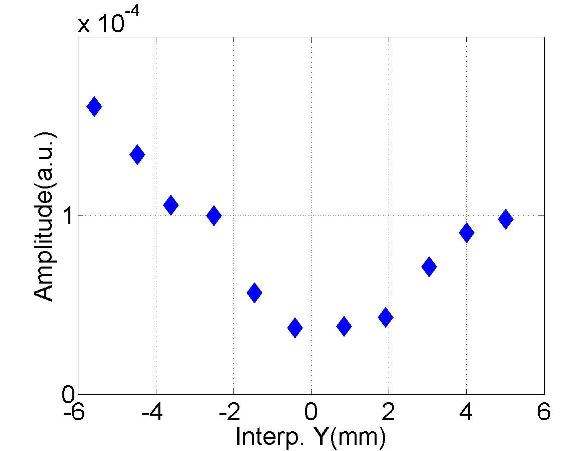}
\label{dep-C1H1-Y-2}
}
\subfigure[\#3 ($f$:9.0499GHz; $Q$:10$^4$)]{
\includegraphics[width=0.23\textwidth]{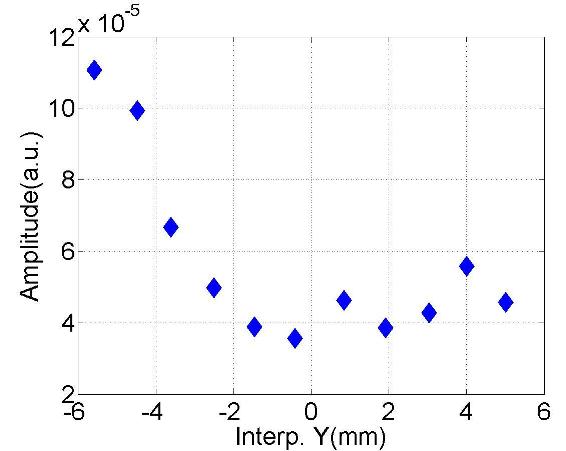}
\label{dep-C1H1-Y-3}
}\\
\subfigure[\#1 ($f$:9.0090GHz; $Q$:10$^4$)]{
\includegraphics[width=0.23\textwidth]{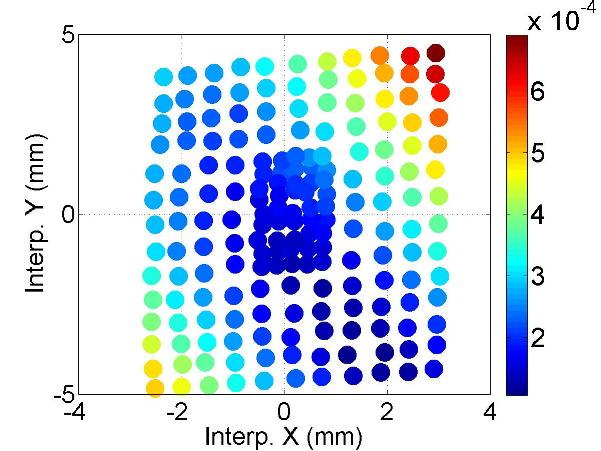}
\label{polar-C1H1-1}
}
\subfigure[\#2 ($f$:9.0130GHz; $Q$:10$^4$)]{
\includegraphics[width=0.23\textwidth]{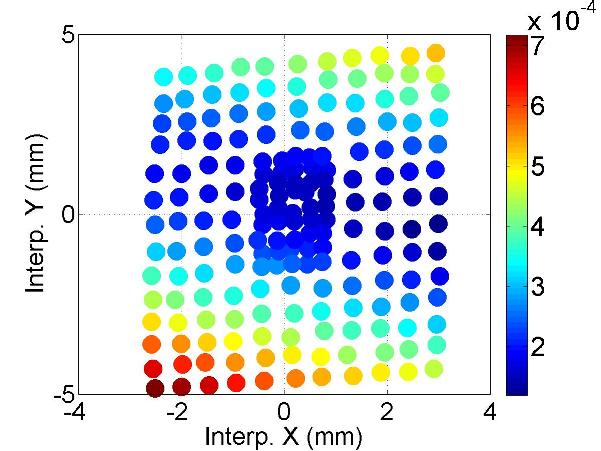}
\label{polar-C1H1-2}
}
\subfigure[\#2.5 ($f$:9.0387GHz; $Q$:10$^5$)]{
\includegraphics[width=0.23\textwidth]{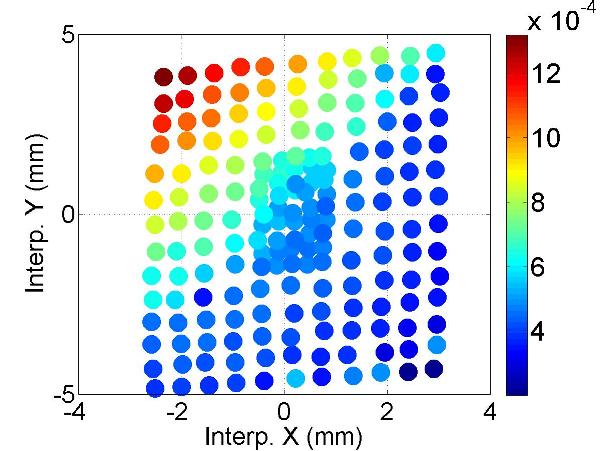}
\label{polar-C1H1-2_5}
}
\subfigure[\#3 ($f$:9.0450GHz; $Q$:10$^4$)]{
\includegraphics[width=0.23\textwidth]{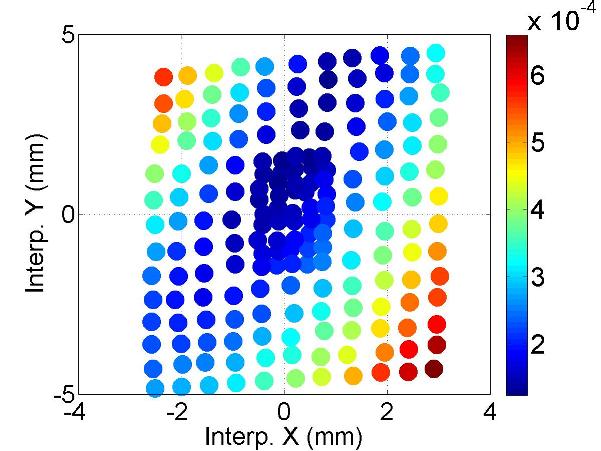}
\label{polar-C1H1-3}
}
\caption{Dependence of the mode amplitude on the transverse beam of{}fset in the cavity.}
\label{spec-dep-C1H1-XY-1}
\end{figure}
\begin{figure}[h]
\subfigure[Spectrum (C1H1)]{
\includegraphics[width=1\textwidth]{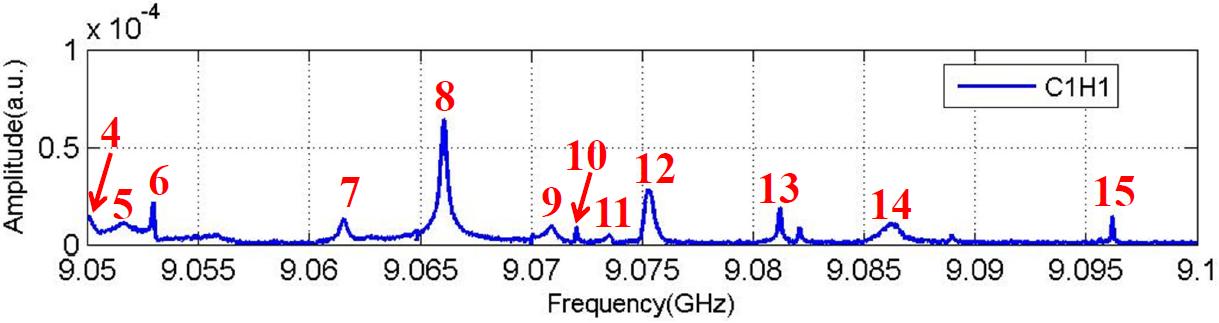}
\label{spec-C1H1-X-2}
}
\subfigure[\#4 ($f$:9.0501GHz; $Q$:10$^4$)]{
\includegraphics[width=0.23\textwidth]{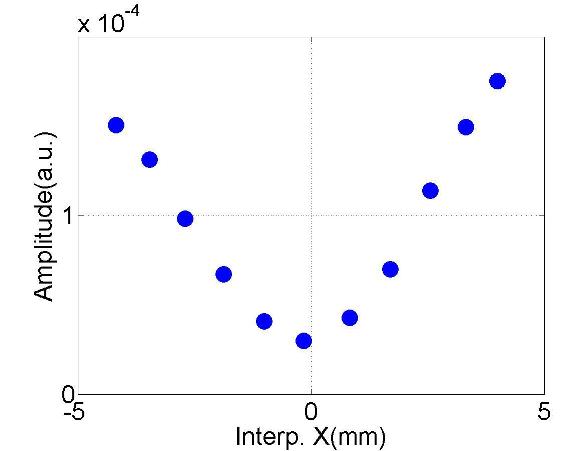}
\label{dep-C1H1-X-4}
}
\subfigure[\#5 ($f$:9.0517GHz; $Q$:10$^3$)]{
\includegraphics[width=0.23\textwidth]{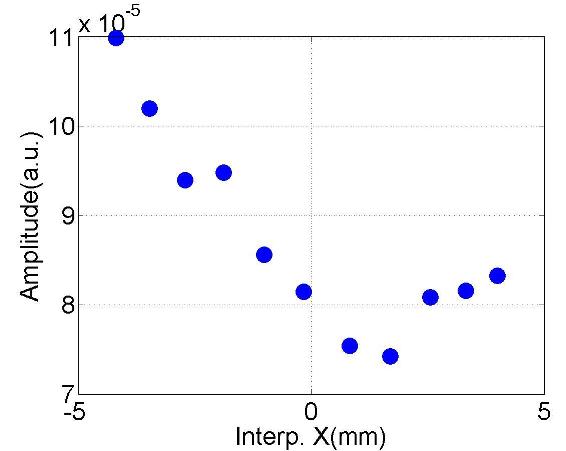}
\label{dep-C1H1-X-5}
}
\subfigure[\#6 ($f$:9.0530GHz; $Q$:10$^5$)]{
\includegraphics[width=0.23\textwidth]{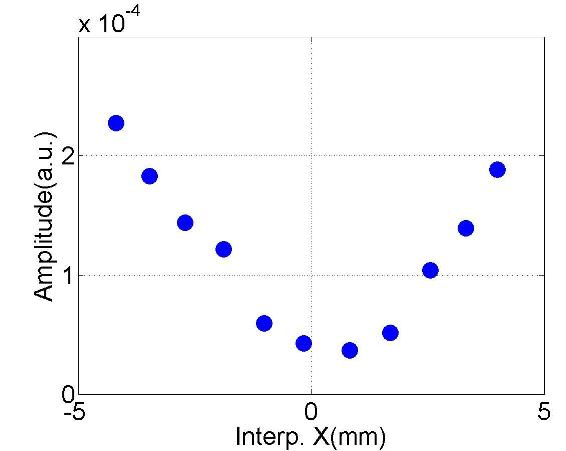}
\label{dep-C1H1-X-6}
}
\subfigure[\#7 ($f$:9.0616GHz; $Q$:10$^4$)]{
\includegraphics[width=0.23\textwidth]{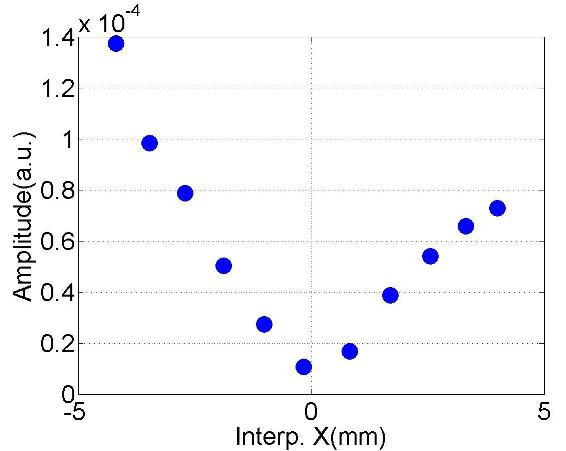}
\label{dep-C1H1-X-7}
}
\subfigure[\#8 ($f$:9.0661GHz; $Q$:10$^4$)]{
\includegraphics[width=0.23\textwidth]{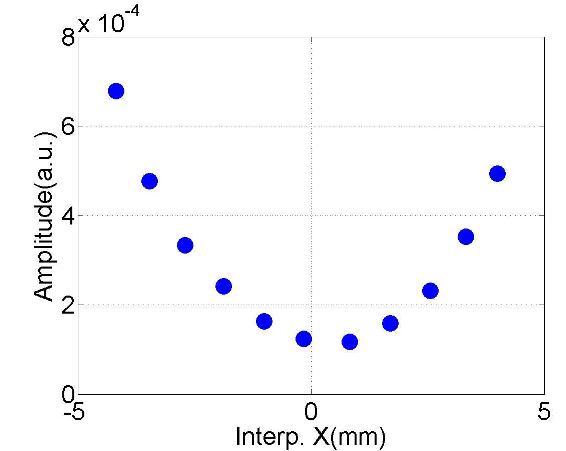}
\label{dep-C1H1-X-8}
}
\subfigure[\#9 ($f$:9.0709GHz; $Q$:10$^4$)]{
\includegraphics[width=0.23\textwidth]{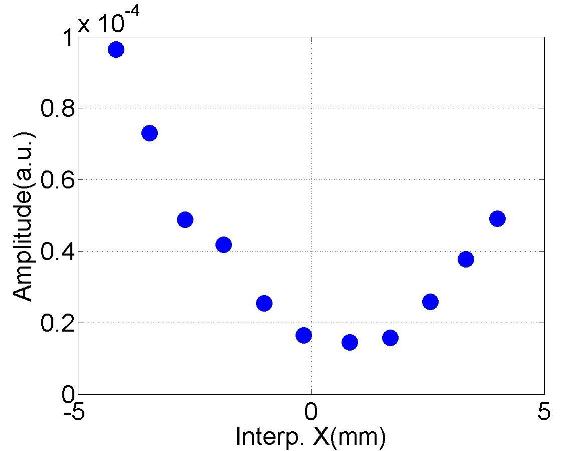}
\label{dep-C1H1-X-9}
}
\subfigure[\#10 ($f$:9.0720GHz; $Q$:10$^4$)]{
\includegraphics[width=0.23\textwidth]{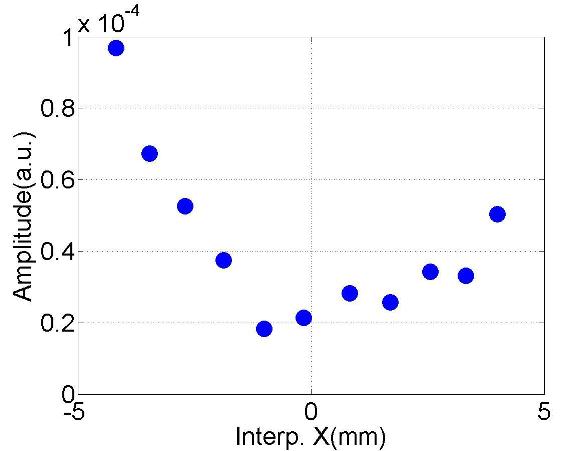}
\label{dep-C1H1-X-10}
}
\subfigure[\#11 ($f$:9.0736GHz; $Q$:10$^4$)]{
\includegraphics[width=0.23\textwidth]{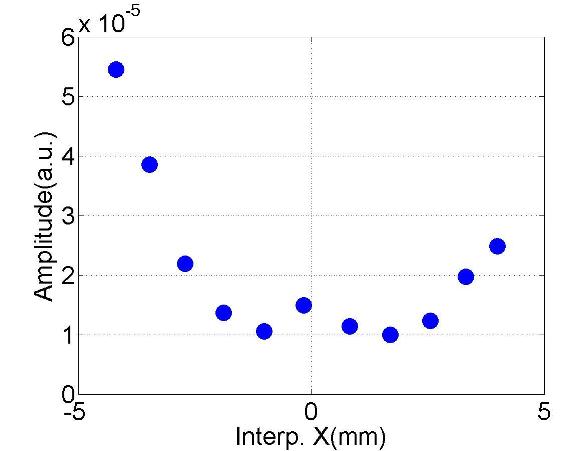}
\label{dep-C1H1-X-11}
}
\subfigure[\#12 ($f$:9.0754GHz; $Q$:10$^4$)]{
\includegraphics[width=0.23\textwidth]{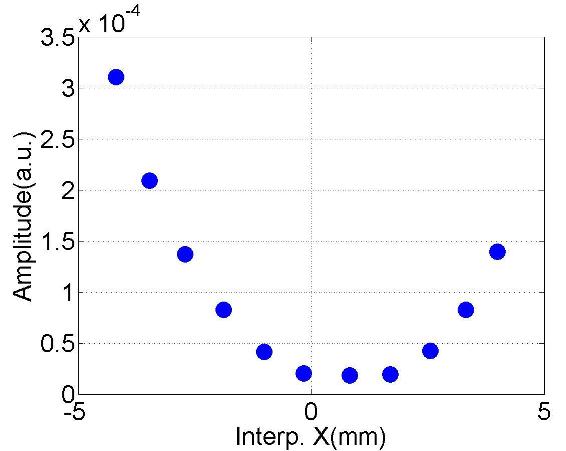}
\label{dep-C1H1-X-12}
}
\subfigure[\#13 ($f$:9.0812GHz; $Q$:10$^5$)]{
\includegraphics[width=0.23\textwidth]{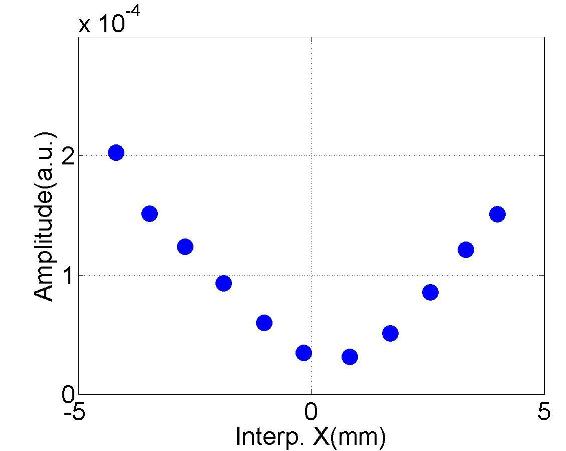}
\label{dep-C1H1-X-13}
}
\subfigure[\#14 ($f$:9.0862GHz; $Q$:10$^4$)]{
\includegraphics[width=0.23\textwidth]{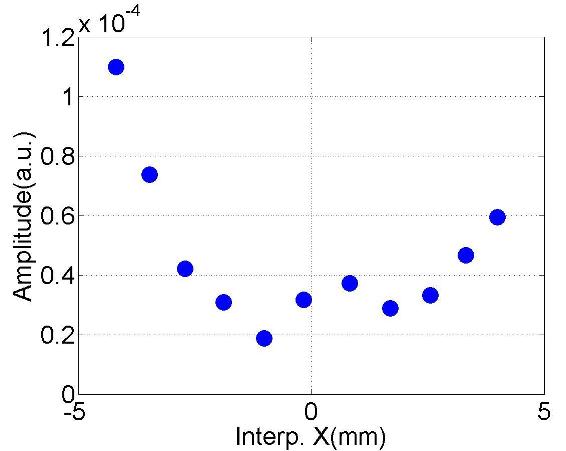}
\label{dep-C1H1-X-14}
}
\subfigure[\#15 ($f$:9.0962GHz; $Q$:10$^5$)]{
\includegraphics[width=0.23\textwidth]{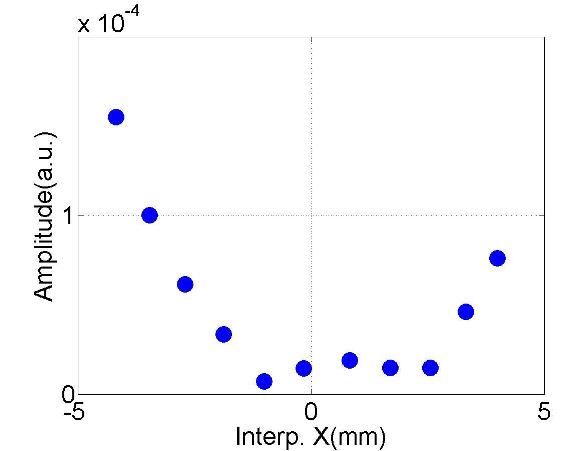}
\label{dep-C1H1-X-15}
}
\caption{Dependence of the mode amplitude on the horizontal beam of{}fset in the cavity.}
\label{spec-dep-C1H1-X-2}
\end{figure}
\begin{figure}[h]
\subfigure[Spectrum (C1H1)]{
\includegraphics[width=1\textwidth]{D5Xmove-Spec-C1H1-2}
\label{spec-C1H1-X-1}
}
\subfigure[\#4 ($f$:9.0501GHz; $Q$:10$^4$)]{
\includegraphics[width=0.23\textwidth]{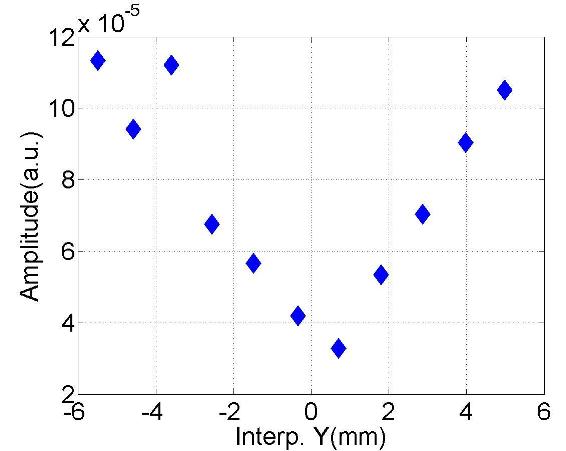}
\label{dep-C1H1-Y-4}
}
\subfigure[\#5 ($f$:9.0517GHz; $Q$:10$^3$)]{
\includegraphics[width=0.23\textwidth]{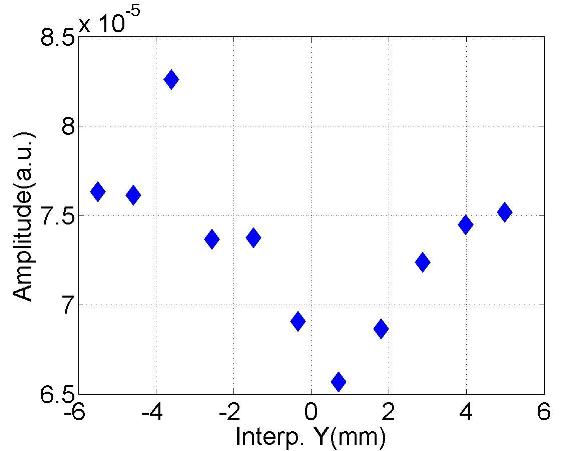}
\label{dep-C1H1-Y-5}
}
\subfigure[\#6 ($f$:9.0530GHz; $Q$:10$^5$)]{
\includegraphics[width=0.23\textwidth]{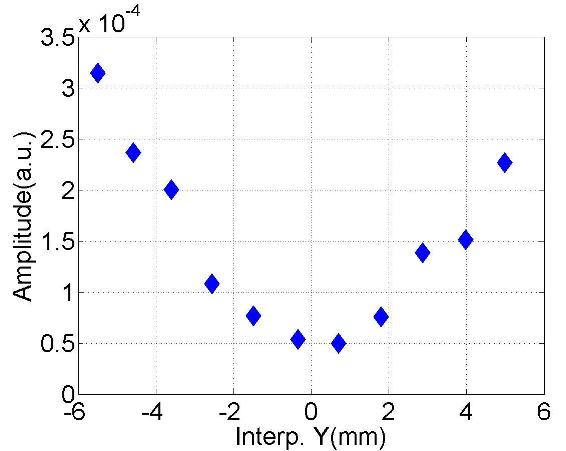}
\label{dep-C1H1-Y-6}
}
\subfigure[\#7 ($f$:9.0614GHz; $Q$:10$^4$)]{
\includegraphics[width=0.23\textwidth]{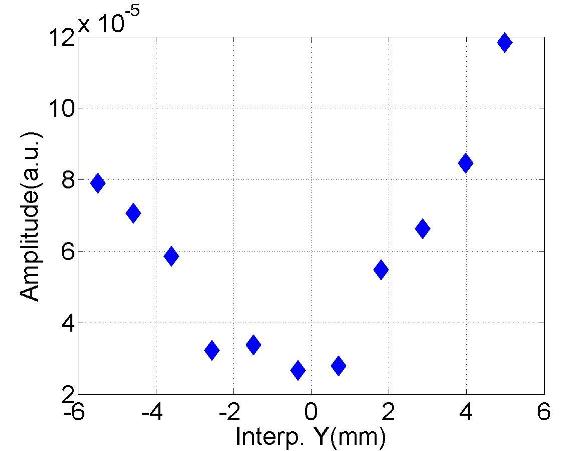}
\label{dep-C1H1-Y-7}
}
\subfigure[\#8 ($f$:9.0661GHz; $Q$:10$^4$)]{
\includegraphics[width=0.23\textwidth]{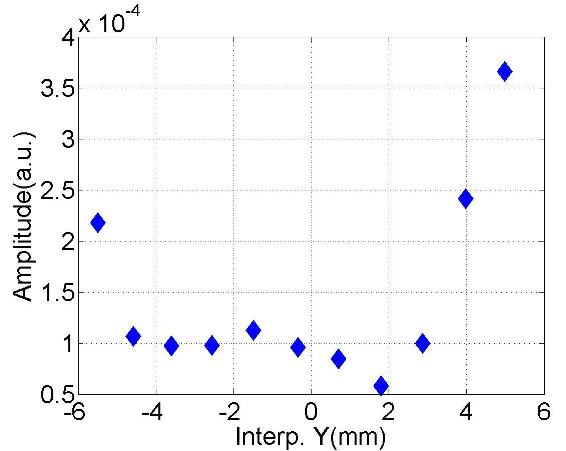}
\label{dep-C1H1-Y-8}
}
\subfigure[\#9 ($f$:9.0709GHz; $Q$:10$^4$)]{
\includegraphics[width=0.23\textwidth]{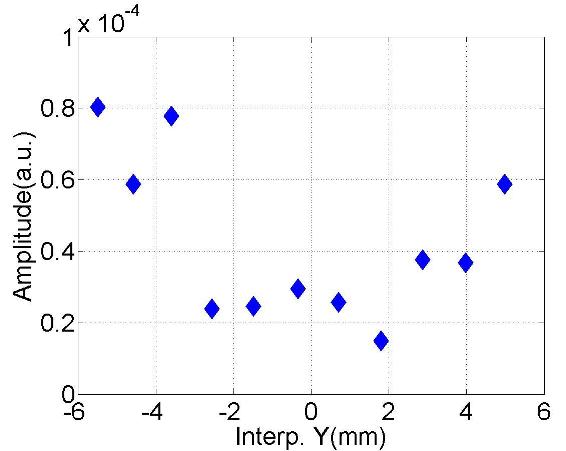}
\label{dep-C1H1-Y-9}
}
\subfigure[\#10 ($f$:9.0722GHz; $Q$:10$^4$)]{
\includegraphics[width=0.23\textwidth]{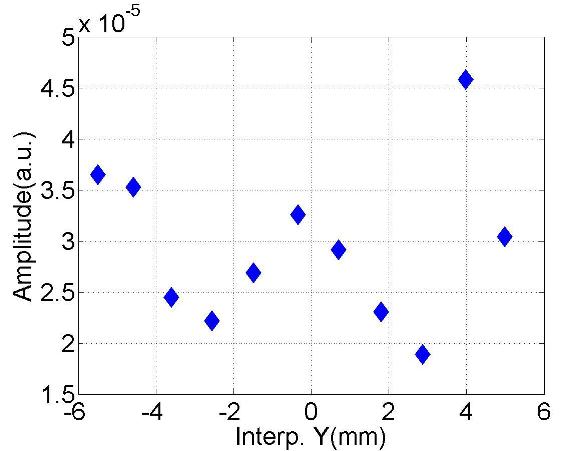}
\label{dep-C1H1-Y-10}
}
\subfigure[\#11 ($f$:9.0735GHz; $Q$:10$^4$)]{
\includegraphics[width=0.23\textwidth]{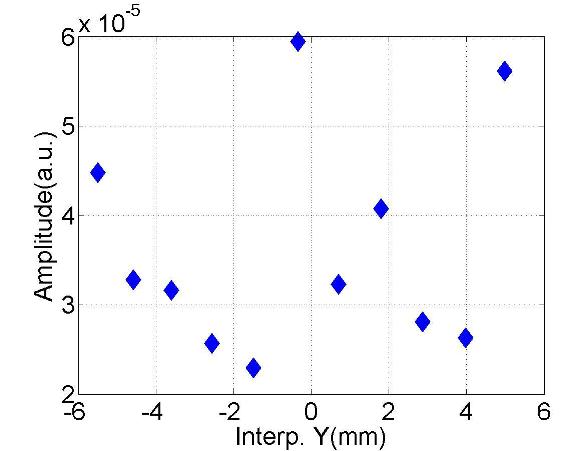}
\label{dep-C1H1-Y-11}
}
\subfigure[\#12 ($f$:9.0754GHz; $Q$:10$^4$)]{
\includegraphics[width=0.23\textwidth]{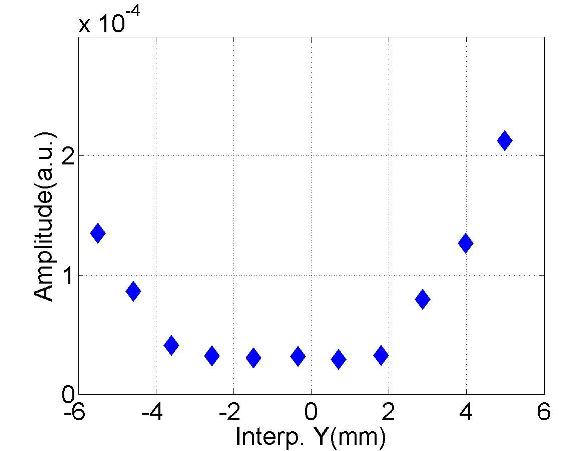}
\label{dep-C1H1-Y-12}
}
\subfigure[\#13 ($f$:9.0812GHz; $Q$:10$^5$)]{
\includegraphics[width=0.23\textwidth]{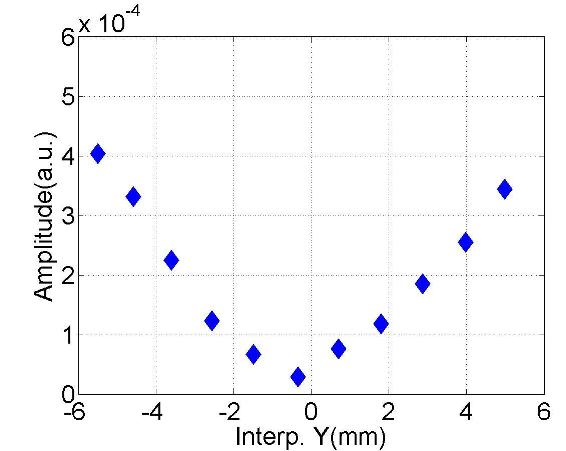}
\label{dep-C1H1-Y-13}
}
\subfigure[\#14 ($f$:9.0865GHz; $Q$:10$^4$)]{
\includegraphics[width=0.23\textwidth]{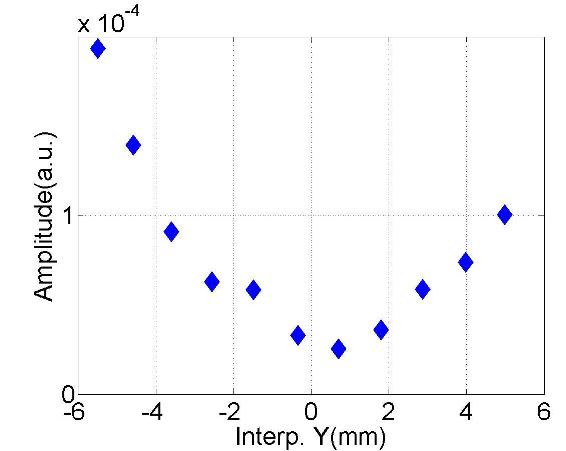}
\label{dep-C1H1-Y-14}
}
\subfigure[\#15 ($f$:9.0962GHz; $Q$:10$^5$)]{
\includegraphics[width=0.23\textwidth]{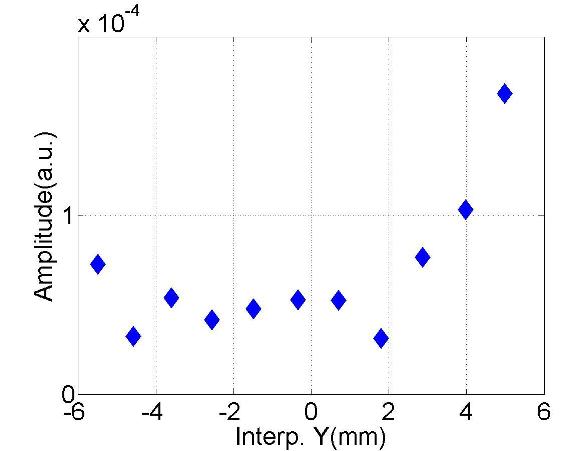}
\label{dep-C1H1-Y-15}
}
\caption{Dependence of the mode amplitude on the vertical beam of{}fset in the cavity.}
\label{spec-dep-C1H1-Y-2}
\end{figure}
\begin{figure}[h]
\subfigure[Spectrum (C1H1)]{
\includegraphics[width=1\textwidth]{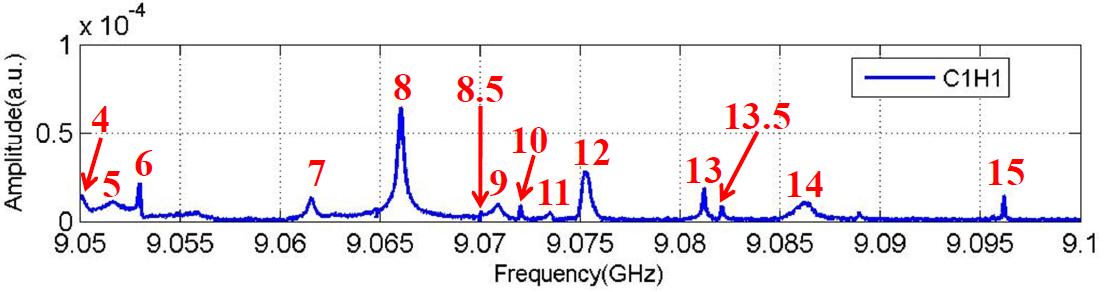}
\label{spec-C1H1-2}
}
\subfigure[\#4 ($f$:9.0501GHz; $Q$:10$^4$)]{
\includegraphics[width=0.23\textwidth]{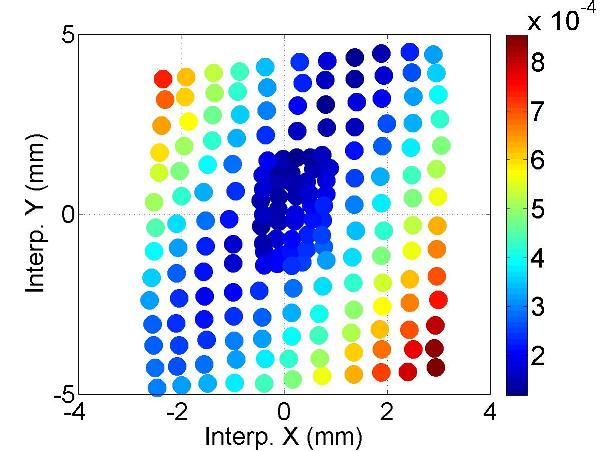}
\label{polar-C1H1-4}
}
\subfigure[\#5 ($f$:9.0516GHz; $Q$:10$^4$)]{
\includegraphics[width=0.23\textwidth]{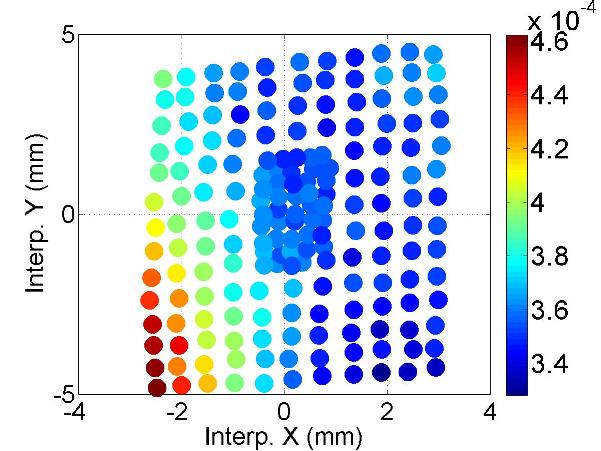}
\label{polar-C1H1-5}
}
\subfigure[\#6 ($f$:9.0530GHz; $Q$:10$^5$)]{
\includegraphics[width=0.23\textwidth]{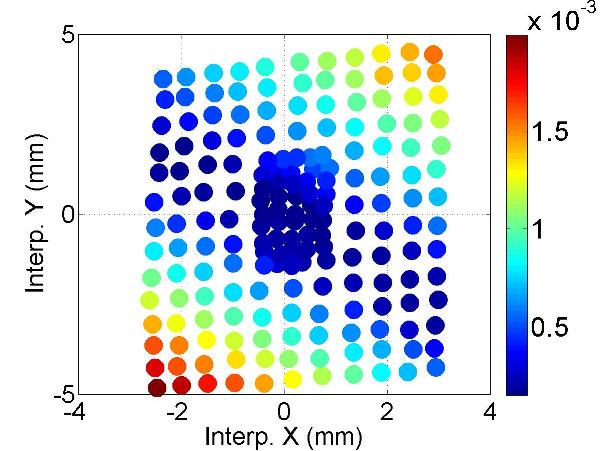}
\label{polar-C1H1-6}
}
\subfigure[\#7 ($f$:9.0616GHz; $Q$:10$^4$)]{
\includegraphics[width=0.23\textwidth]{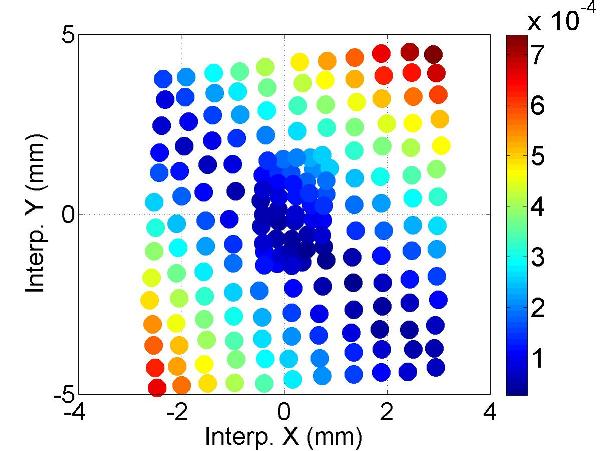}
\label{polar-C1H1-7}
}
\subfigure[\#8 ($f$:9.0661GHz; $Q$:10$^4$)]{
\includegraphics[width=0.23\textwidth]{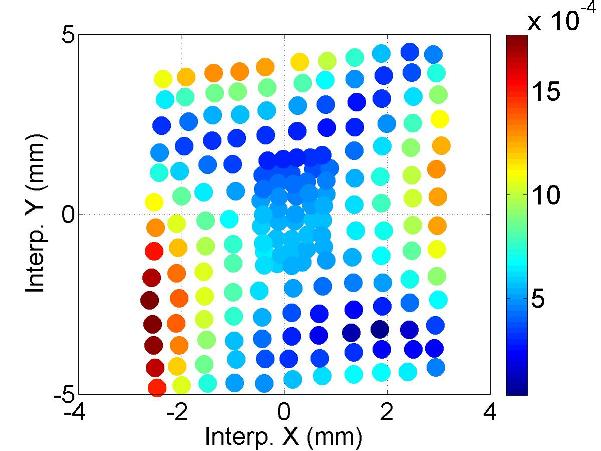}
\label{polar-C1H1-8}
}
\subfigure[\#8.5 ($f$:9.0700GHz; $Q$:10$^5$)]{
\includegraphics[width=0.23\textwidth]{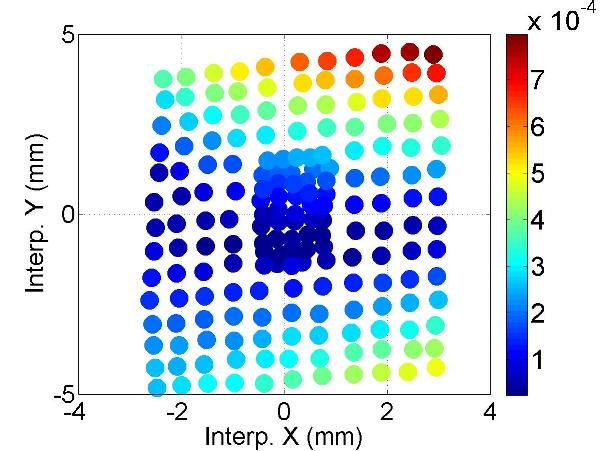}
\label{polar-C1H1-8_5}
}
\subfigure[\#9 ($f$:9.0709GHz; $Q$:10$^4$)]{
\includegraphics[width=0.23\textwidth]{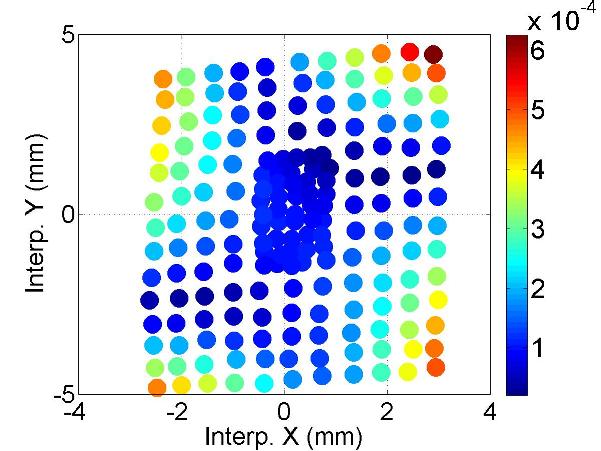}
\label{polar-C1H1-9}
}
\subfigure[\#10 ($f$:9.0720GHz; $Q$:10$^5$)]{
\includegraphics[width=0.23\textwidth]{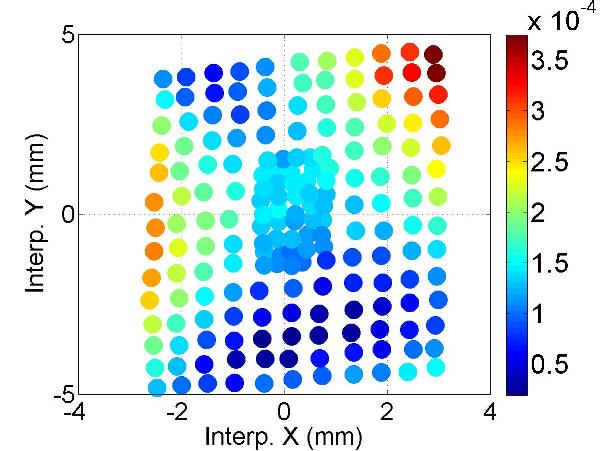}
\label{polar-C1H1-10}
}
\subfigure[\#11 ($f$:9.0737GHz; $Q$:10$^4$)]{
\includegraphics[width=0.23\textwidth]{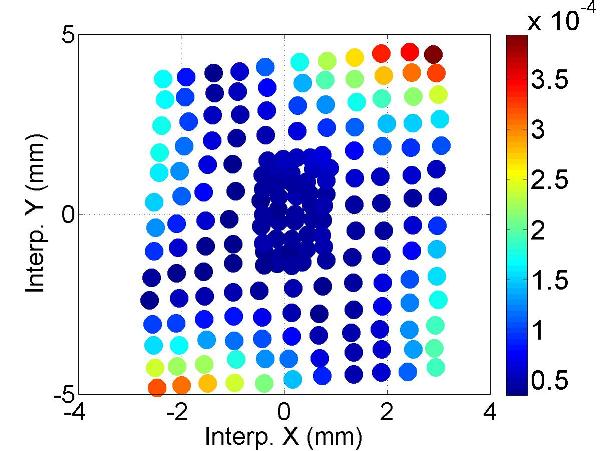}
\label{polar-C1H1-11}
}
\subfigure[\#12 ($f$:9.0753GHz; $Q$:10$^4$)]{
\includegraphics[width=0.23\textwidth]{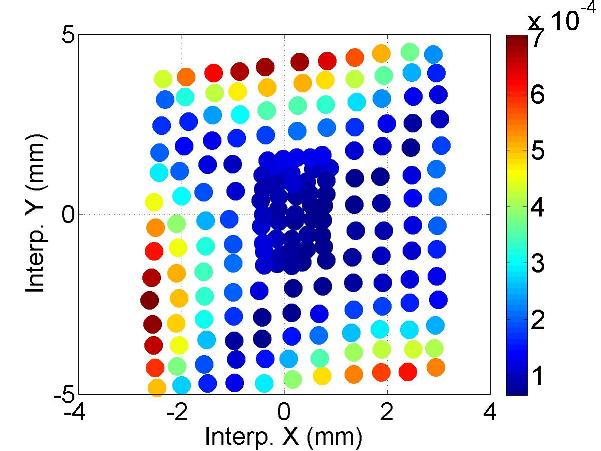}
\label{polar-C1H1-12}
}
\subfigure[\#13 ($f$:9.0812GHz; $Q$:10$^5$)]{
\includegraphics[width=0.23\textwidth]{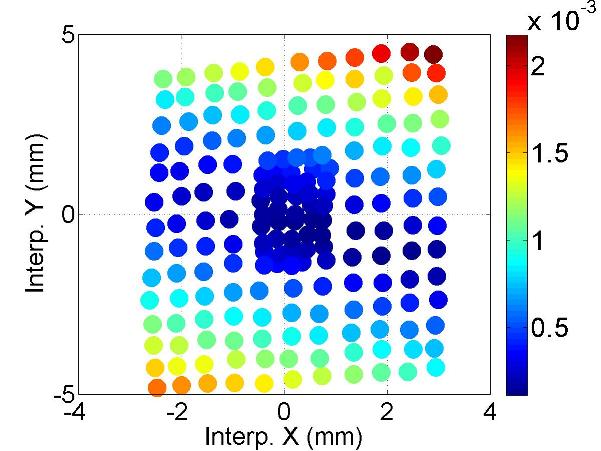}
\label{polar-C1H1-13}
}
\subfigure[\#13.5 ($f$:9.0820GHz; $Q$:10$^5$)]{
\includegraphics[width=0.23\textwidth]{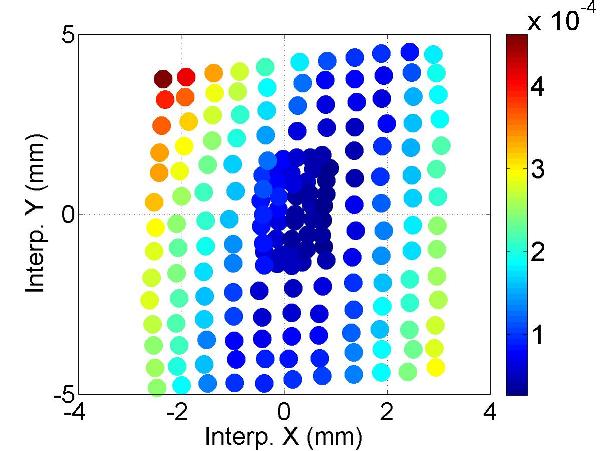}
\label{polar-C1H1-13_5}
}
\subfigure[\#14 ($f$:9.0862GHz; $Q$:10$^4$)]{
\includegraphics[width=0.23\textwidth]{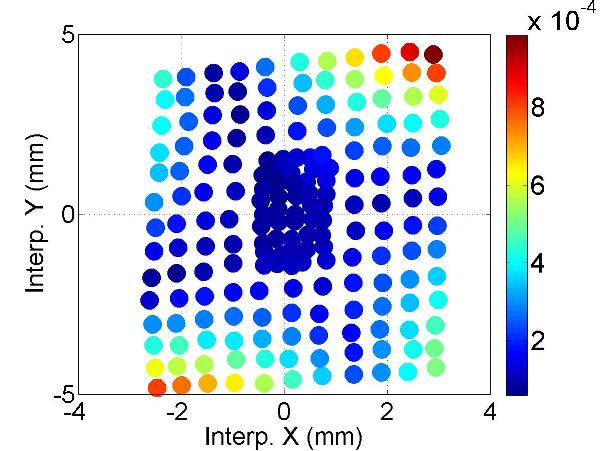}
\label{polar-C1H1-14}
}
\subfigure[\#15 ($f$:9.0962GHz; $Q$:10$^5$)]{
\includegraphics[width=0.23\textwidth]{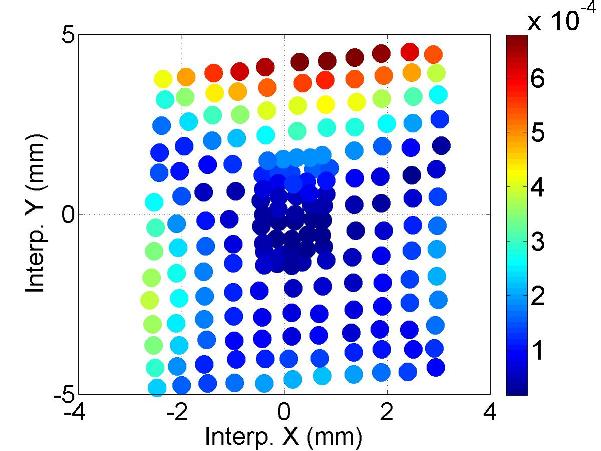}
\label{polar-C1H1-15}
}
\caption{Polarization of the mode.}
\label{spec-polar-C1H1-2}
\end{figure}

\FloatBarrier
\section{D5: HOM Coupler C1H2}
\begin{figure}[h]\center
\subfigure[Spectrum (C1H2)]{
\includegraphics[width=0.85\textwidth]{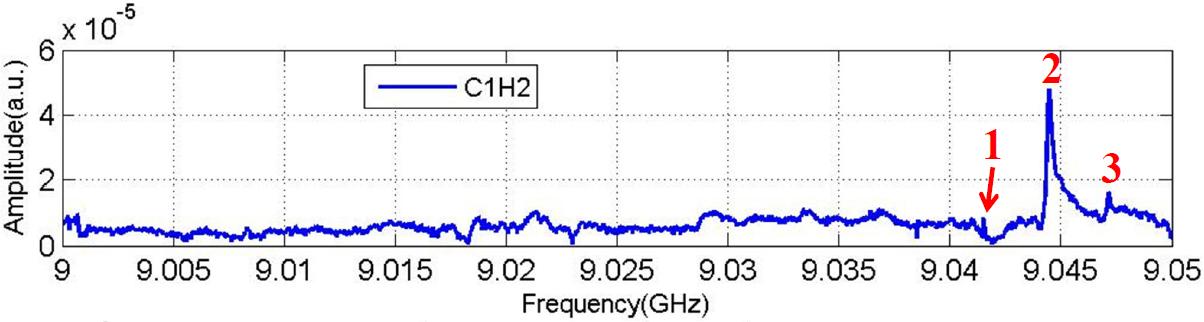}
\label{spec-C1H2-X-1}
}
\subfigure[\#1 ($f$:9.0415GHz; $Q$:10$^5$)]{
\includegraphics[width=0.26\textwidth]{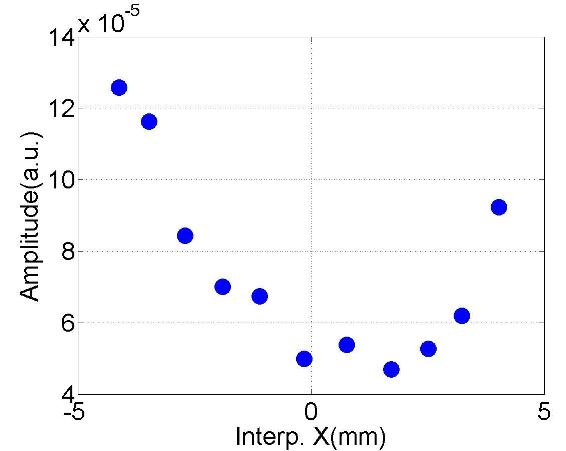}
\label{dep-C1H2-X-1}
}
\subfigure[\#2 ($f$:9.0445GHz; $Q$:10$^4$)]{
\includegraphics[width=0.26\textwidth]{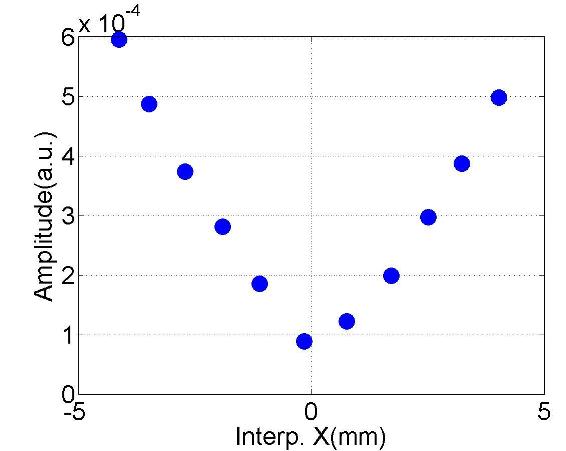}
\label{dep-C1H2-X-2}
}
\subfigure[\#3 ($f$:9.0472GHz; $Q$:10$^4$)]{
\includegraphics[width=0.26\textwidth]{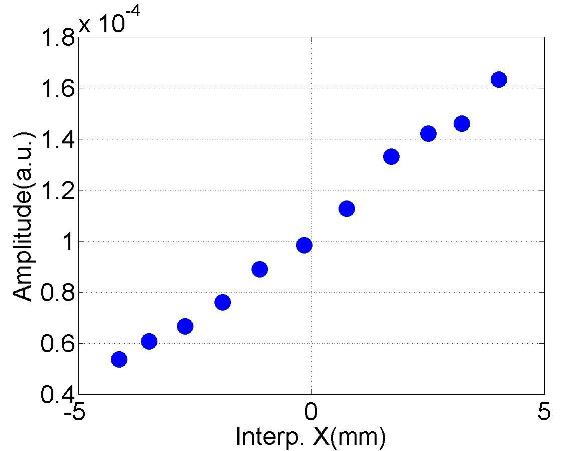}
\label{dep-C1H2-X-3}
}
\subfigure[\#1 ($f$:9.0415GHz; $Q$:10$^5$)]{
\includegraphics[width=0.26\textwidth]{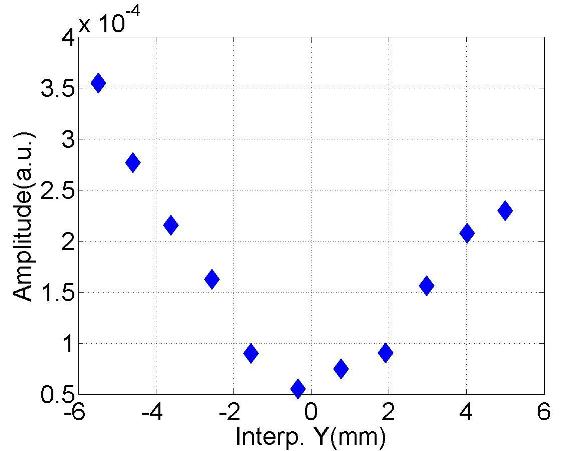}
\label{dep-C1H2-Y-1}
}
\subfigure[\#2 ($f$:9.0446GHz; $Q$:10$^4$)]{
\includegraphics[width=0.26\textwidth]{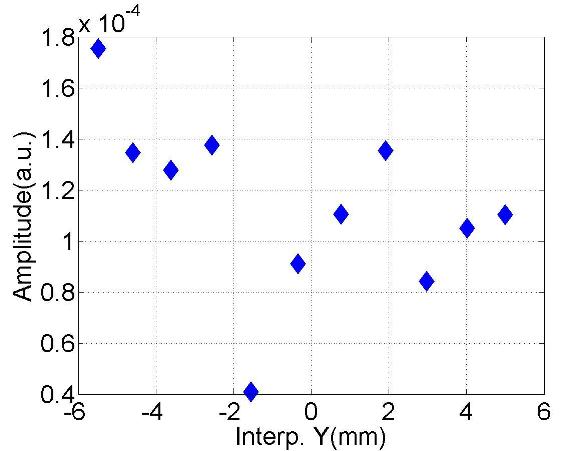}
\label{dep-C1H2-Y-2}
}
\subfigure[\#3 ($f$:9.0472GHz; $Q$:10$^4$)]{
\includegraphics[width=0.26\textwidth]{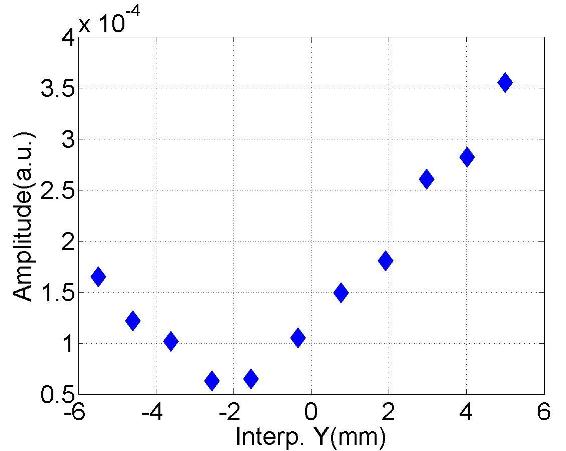}
\label{dep-C1H2-Y-3}
}
\subfigure[\#1 ($f$:9.0415GHz; $Q$:10$^5$)]{
\includegraphics[width=0.27\textwidth]{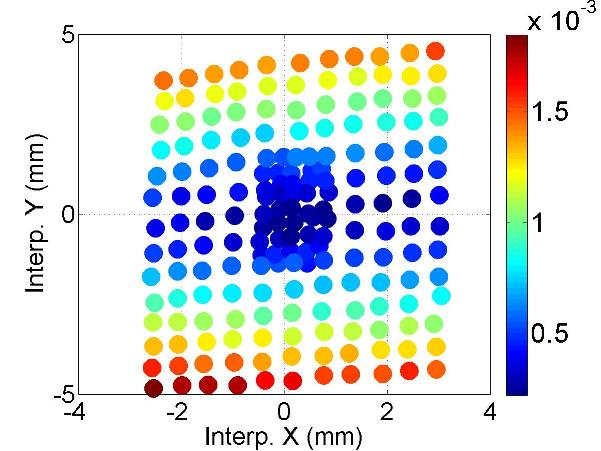}
\label{polar-C1H2-1}
}
\subfigure[\#2 ($f$:9.0446GHz; $Q$:10$^4$)]{
\includegraphics[width=0.27\textwidth]{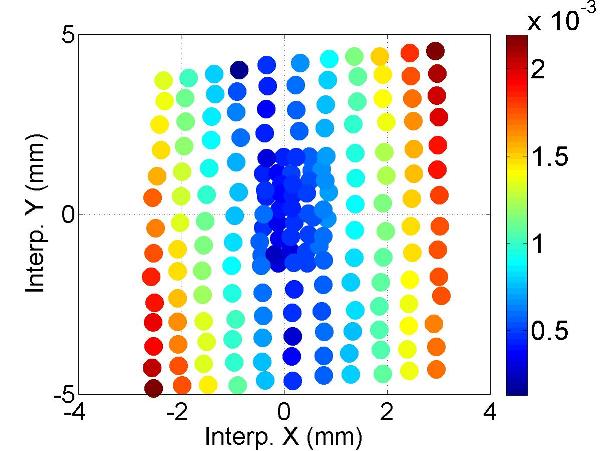}
\label{polar-C1H2-2}
}
\subfigure[\#3 ($f$:9.0471GHz; $Q$:10$^4$)]{
\includegraphics[width=0.27\textwidth]{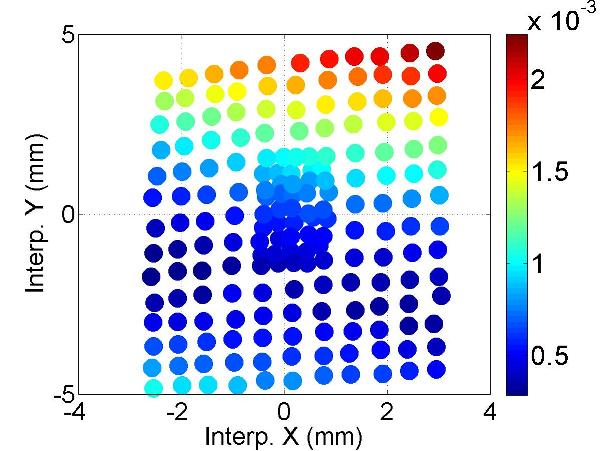}
\label{polar-C1H2-3}
}
\caption{Dependence of the mode amplitude on the transverse beam of{}fset in the cavity.}
\label{spec-dep-C1H2-XY-1}
\end{figure}
\begin{figure}[h]\center
\subfigure[Spectrum (C1H2)]{
\includegraphics[width=0.85\textwidth]{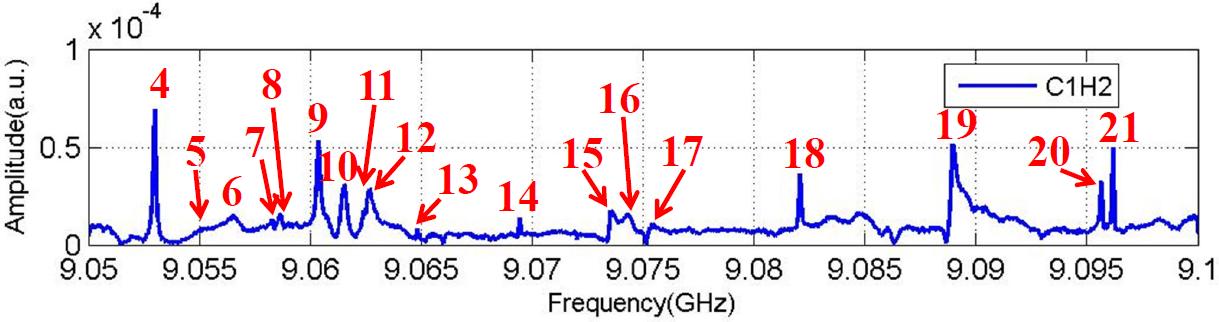}
\label{spec-C1H2-X-2}
}
\subfigure[{\scriptsize\#4 ($f$:9.0530GHz; $Q$:10$^5$)}]{
\includegraphics[width=0.22\textwidth]{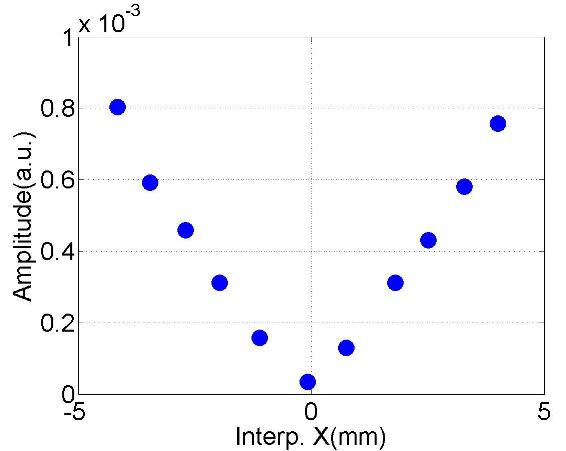}
\label{dep-C1H2-X-4}
}
\subfigure[{\scriptsize\#5 ($f$:9.0553GHz; $Q$:10$^4$)}]{
\includegraphics[width=0.22\textwidth]{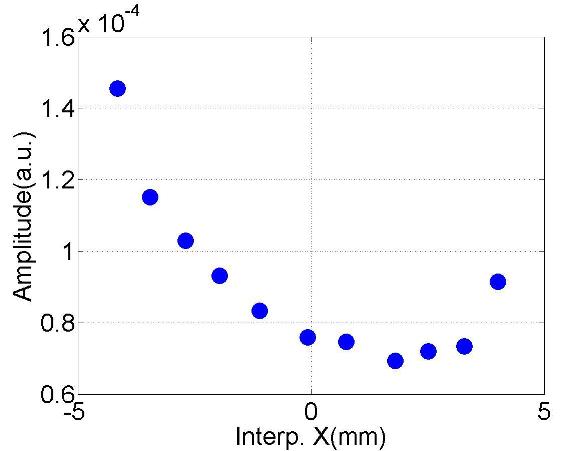}
\label{dep-C1H2-X-5}
}
\subfigure[{\scriptsize\#6 ($f$:9.0565GHz; $Q$:10$^4$)}]{
\includegraphics[width=0.22\textwidth]{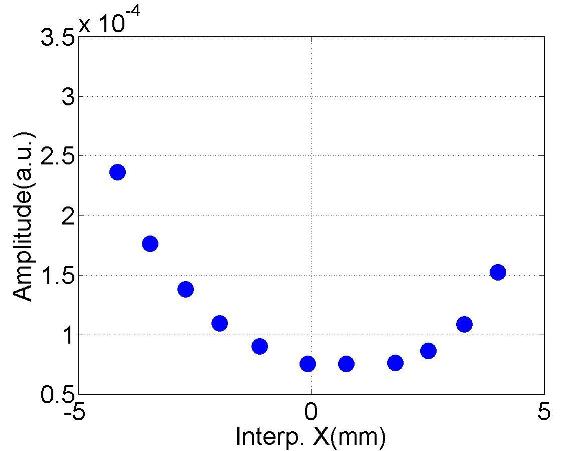}
\label{dep-C1H2-X-6}
}
\subfigure[{\scriptsize\#7 ($f$:9.0583GHz; $Q$:10$^4$)}]{
\includegraphics[width=0.22\textwidth]{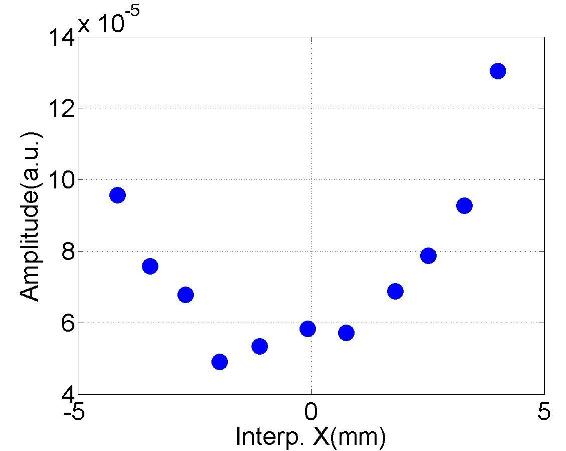}
\label{dep-C1H2-X-7}
}
\subfigure[{\scriptsize\#8 ($f$:9.0586GHz; $Q$:10$^4$)}]{
\includegraphics[width=0.22\textwidth]{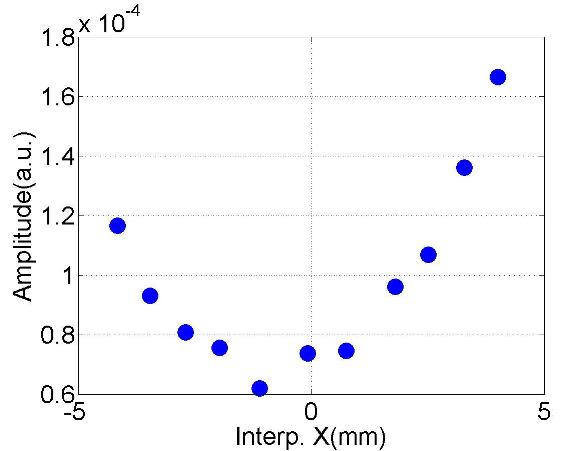}
\label{dep-C1H2-X-8}
}
\subfigure[{\scriptsize\#9 ($f$:9.0604GHz; $Q$:10$^4$)}]{
\includegraphics[width=0.22\textwidth]{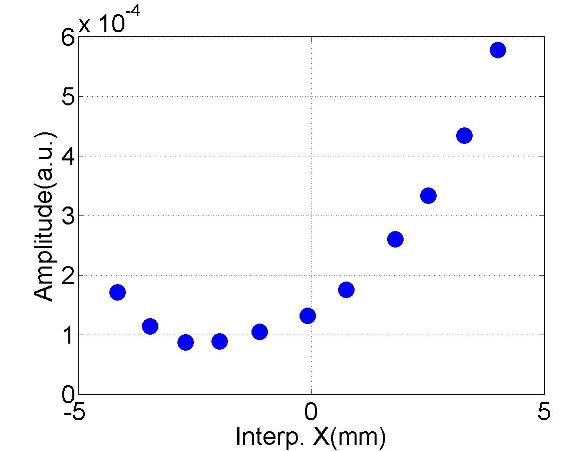}
\label{dep-C1H2-X-9}
}
\subfigure[{\scriptsize\#10 ($f$:9.0616GHz; $Q$:10$^4$)}]{
\includegraphics[width=0.22\textwidth]{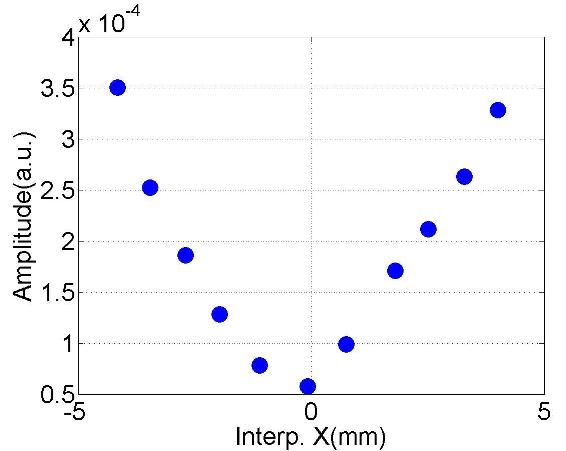}
\label{dep-C1H2-X-10}
}
\subfigure[{\scriptsize\#11 ($f$:9.0623GHz; $Q$:10$^4$)}]{
\includegraphics[width=0.22\textwidth]{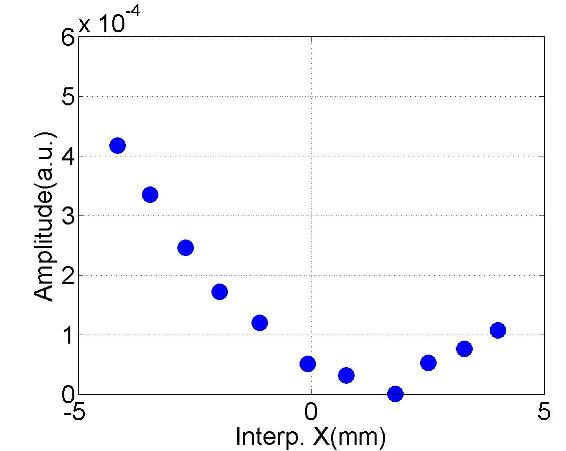}
\label{dep-C1H2-X-11}
}
\subfigure[{\scriptsize\#12 ($f$:9.0627GHz; $Q$:10$^4$)}]{
\includegraphics[width=0.22\textwidth]{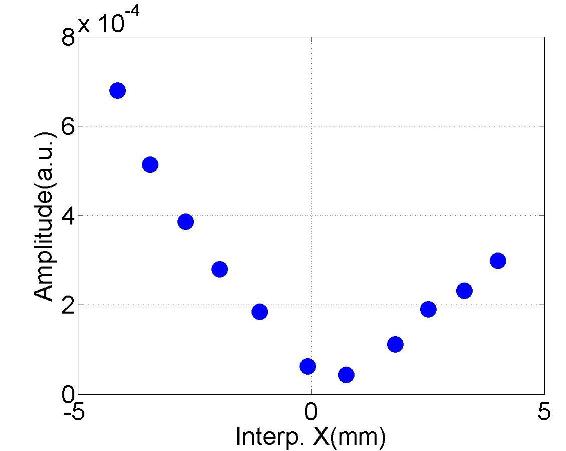}
\label{dep-C1H2-X-12}
}
\subfigure[{\scriptsize\#13 ($f$:9.0648GHz; $Q$:10$^5$)}]{
\includegraphics[width=0.22\textwidth]{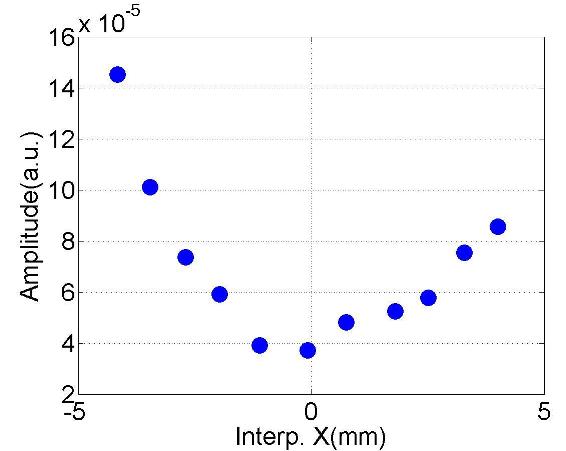}
\label{dep-C1H2-X-13}
}
\subfigure[{\scriptsize\#14 ($f$:9.0695GHz; $Q$:10$^5$)}]{
\includegraphics[width=0.22\textwidth]{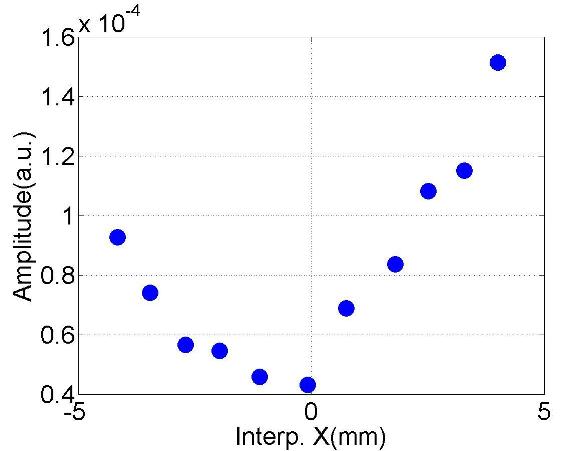}
\label{dep-C1H2-X-14}
}
\subfigure[{\scriptsize\#15 ($f$:9.0735GHz; $Q$:10$^4$)}]{
\includegraphics[width=0.22\textwidth]{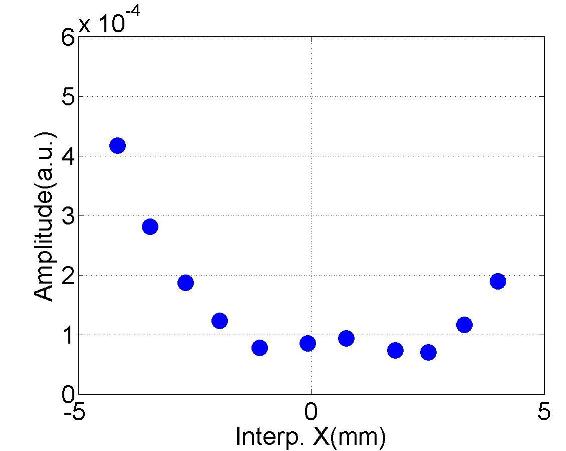}
\label{dep-C1H2-X-15}
}
\subfigure[{\scriptsize\#16 ($f$:9.0745GHz; $Q$:10$^4$)}]{
\includegraphics[width=0.22\textwidth]{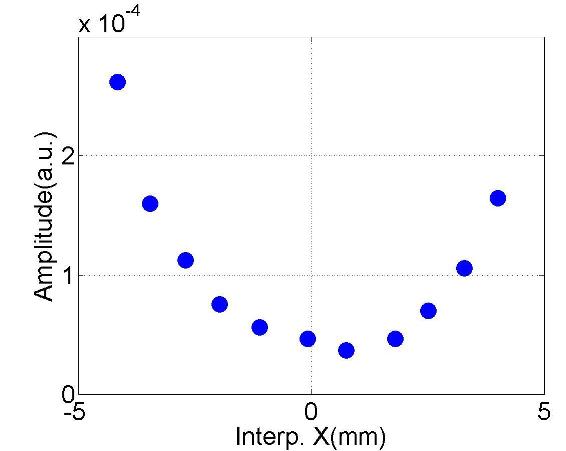}
\label{dep-C1H2-X-16}
}
\subfigure[{\scriptsize\#17 ($f$:9.0755GHz; $Q$:10$^4$)}]{
\includegraphics[width=0.22\textwidth]{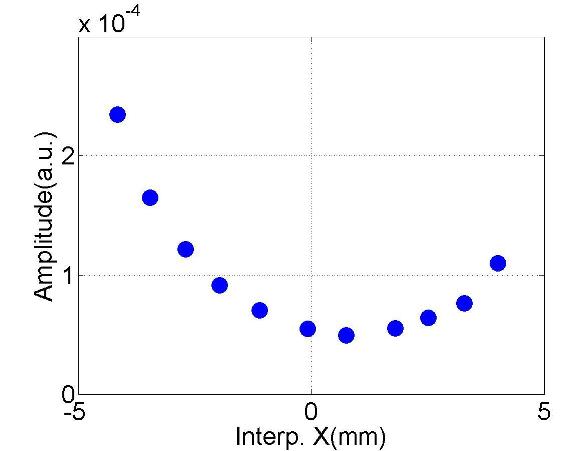}
\label{dep-C1H2-X-17}
}
\subfigure[{\scriptsize\#18 ($f$:9.0821GHz; $Q$:10$^5$)}]{
\includegraphics[width=0.22\textwidth]{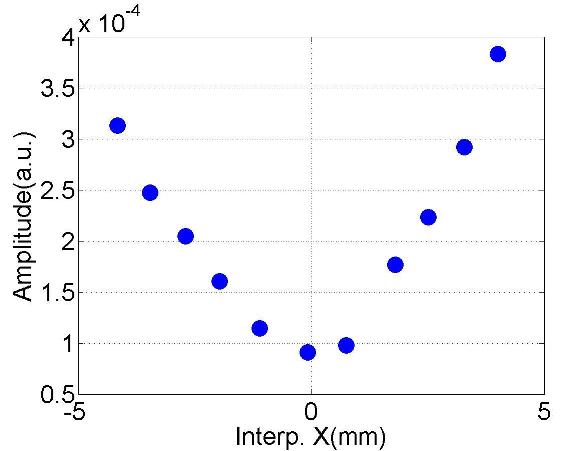}
\label{dep-C1H2-X-18}
}
\subfigure[{\scriptsize\#19 ($f$:9.0892GHz; $Q$:10$^4$)}]{
\includegraphics[width=0.22\textwidth]{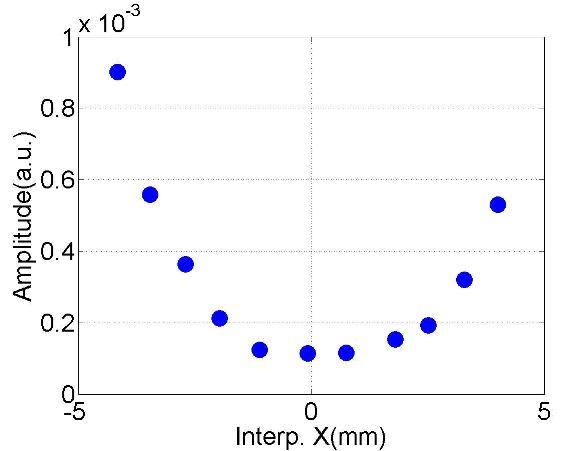}
\label{dep-C1H2-X-19}
}
\subfigure[{\scriptsize\#20 ($f$:9.0957GHz; $Q$:10$^5$)}]{
\includegraphics[width=0.22\textwidth]{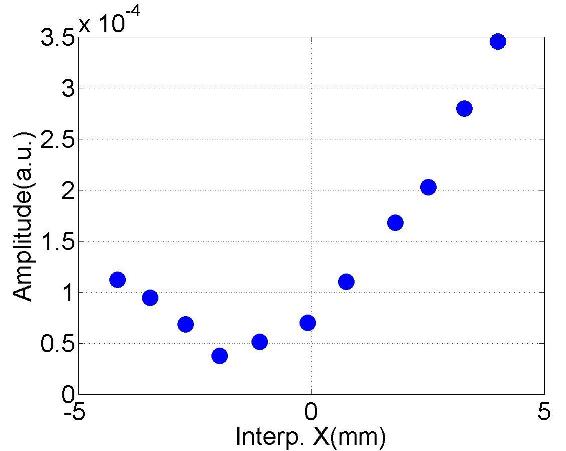}
\label{dep-C1H2-X-20}
}
\subfigure[{\scriptsize\#21 ($f$:9.0962GHz; $Q$:10$^5$)}]{
\includegraphics[width=0.22\textwidth]{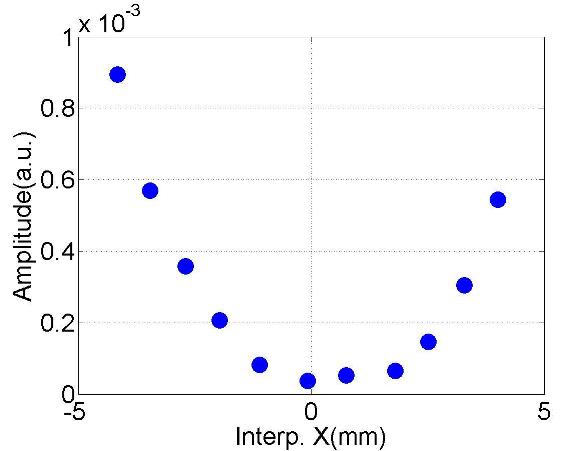}
\label{dep-C1H2-X-21}
}
\caption{Dependence of the mode amplitude on the horizontal beam of{}fset in the cavity.}
\label{spec-dep-C1H2-X-2}
\end{figure}
\begin{figure}[h]\center
\subfigure[Spectrum (C1H2)]{
\includegraphics[width=0.85\textwidth]{D5Xmove-Spec-C1H2-2}
\label{spec-C1H2-Y-2}
}
\subfigure[{\scriptsize\#4 ($f$:9.0530GHz; $Q$:10$^5$)}]{
\includegraphics[width=0.22\textwidth]{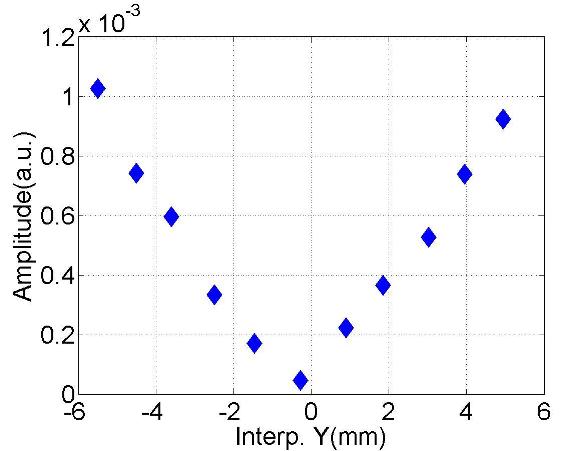}
\label{dep-C1H2-Y-4}
}
\subfigure[{\scriptsize\#5 ($f$:9.0554GHz; $Q$:10$^4$)}]{
\includegraphics[width=0.22\textwidth]{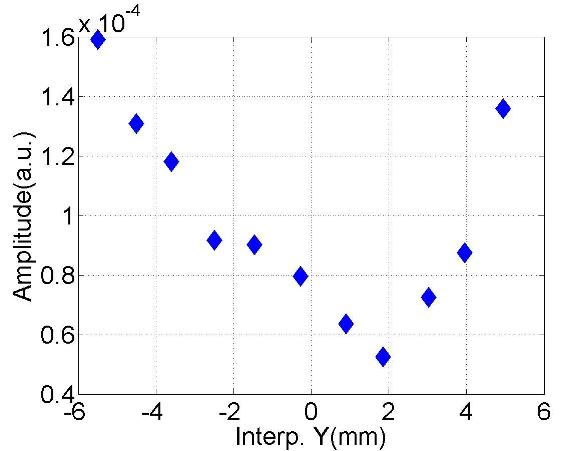}
\label{dep-C1H2-Y-5}
}
\subfigure[{\scriptsize\#6 ($f$:9.0564GHz; $Q$:10$^4$)}]{
\includegraphics[width=0.22\textwidth]{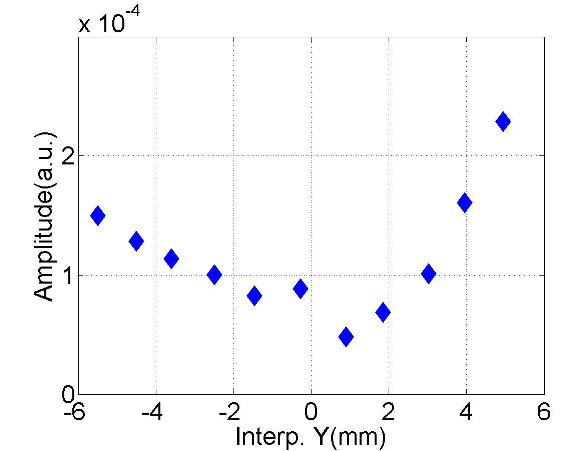}
\label{dep-C1H2-Y-6}
}
\subfigure[{\scriptsize\#7 ($f$:9.0583GHz; $Q$:10$^5$)}]{
\includegraphics[width=0.22\textwidth]{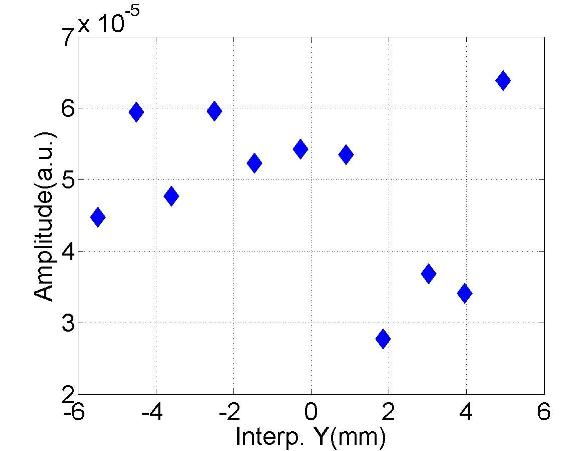}
\label{dep-C1H2-Y-7}
}
\subfigure[{\scriptsize\#8 ($f$:9.0587GHz; $Q$:10$^5$)}]{
\includegraphics[width=0.22\textwidth]{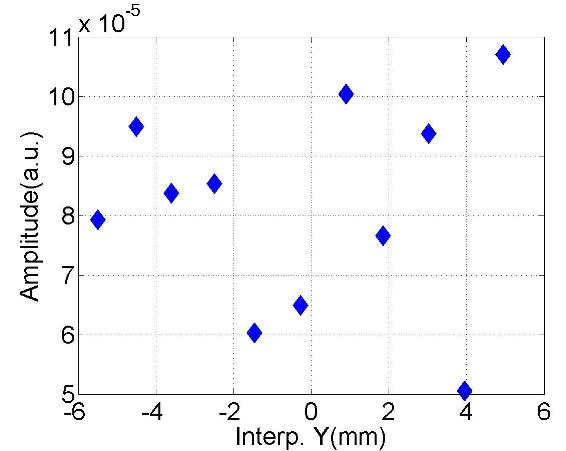}
\label{dep-C1H2-Y-8}
}
\subfigure[{\scriptsize\#9 ($f$:9.0604GHz; $Q$:10$^4$)}]{
\includegraphics[width=0.22\textwidth]{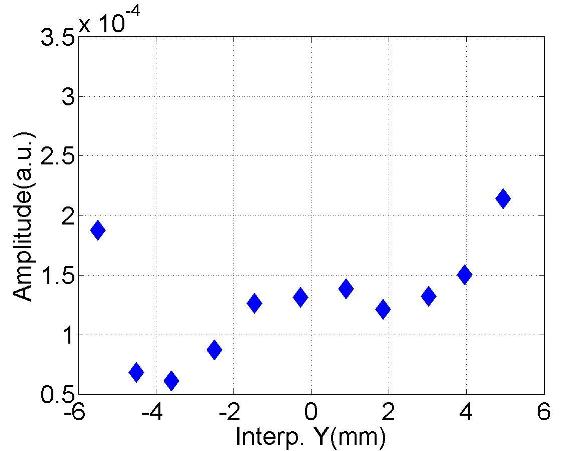}
\label{dep-C1H2-Y-9}
}
\subfigure[{\scriptsize\#10 ($f$:9.0616GHz; $Q$:10$^4$)}]{
\includegraphics[width=0.22\textwidth]{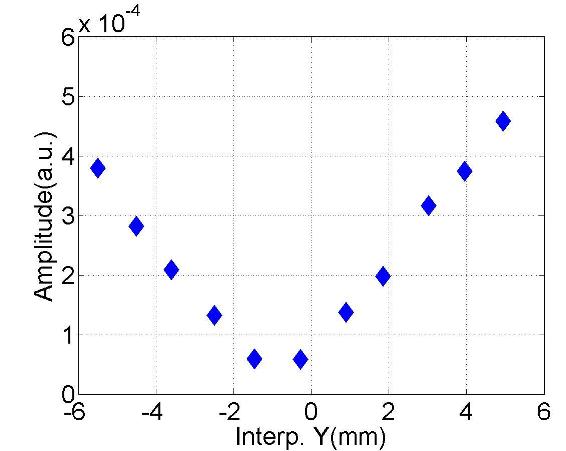}
\label{dep-C1H2-Y-10}
}
\subfigure[{\scriptsize\#11 ($f$:9.0624GHz; $Q$:10$^4$)}]{
\includegraphics[width=0.22\textwidth]{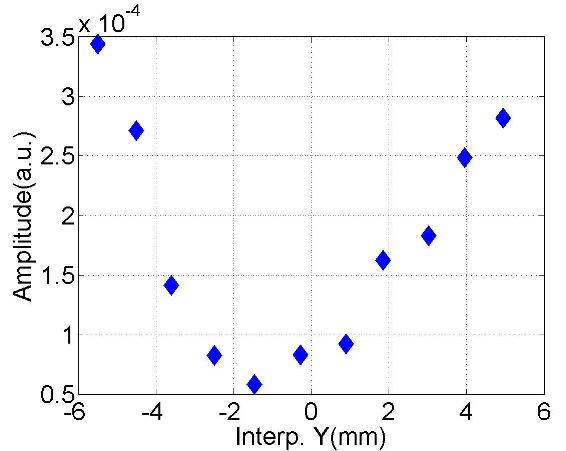}
\label{dep-C1H2-Y-11}
}
\subfigure[{\scriptsize\#12 ($f$:9.0627GHz; $Q$:10$^4$)}]{
\includegraphics[width=0.22\textwidth]{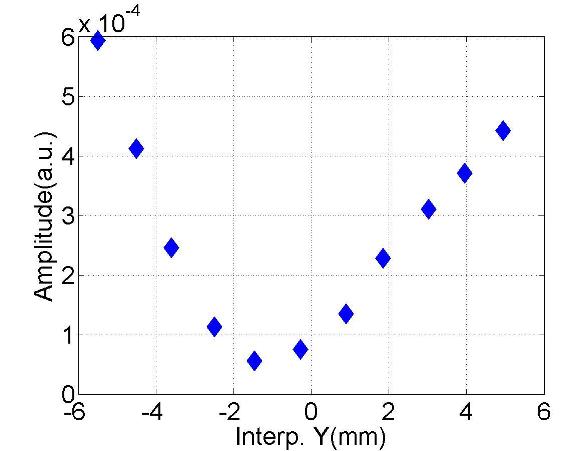}
\label{dep-C1H2-Y-12}
}
\subfigure[{\scriptsize\#13 ($f$:9.0648GHz; $Q$:10$^5$)}]{
\includegraphics[width=0.22\textwidth]{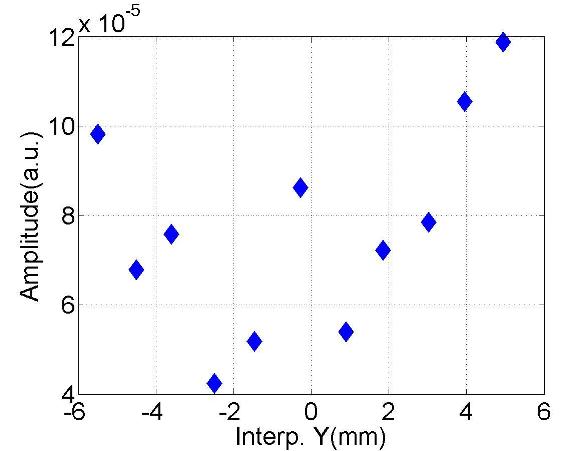}
\label{dep-C1H2-Y-13}
}
\subfigure[{\scriptsize\#14 ($f$:9.0694GHz; $Q$:10$^5$)}]{
\includegraphics[width=0.22\textwidth]{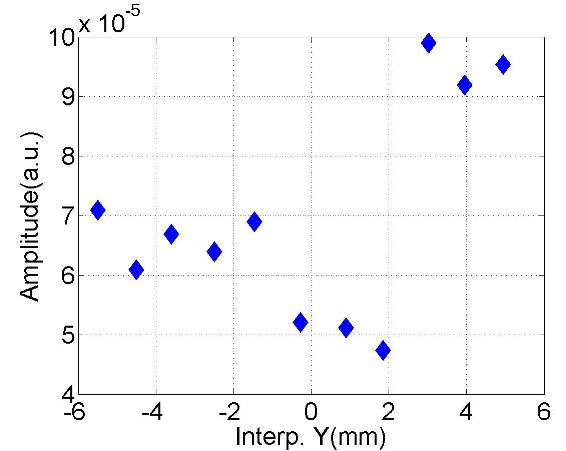}
\label{dep-C1H2-Y-14}
}
\subfigure[{\scriptsize\#15 ($f$:9.0735GHz; $Q$:10$^4$)}]{
\includegraphics[width=0.22\textwidth]{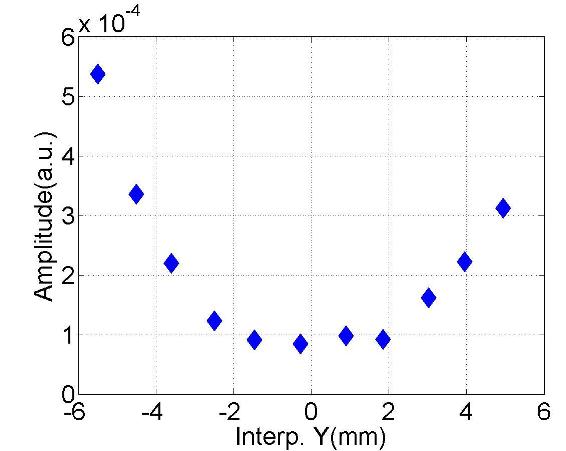}
\label{dep-C1H2-Y-15}
}
\subfigure[{\scriptsize\#16 ($f$:9.0746GHz; $Q$:10$^4$)}]{
\includegraphics[width=0.22\textwidth]{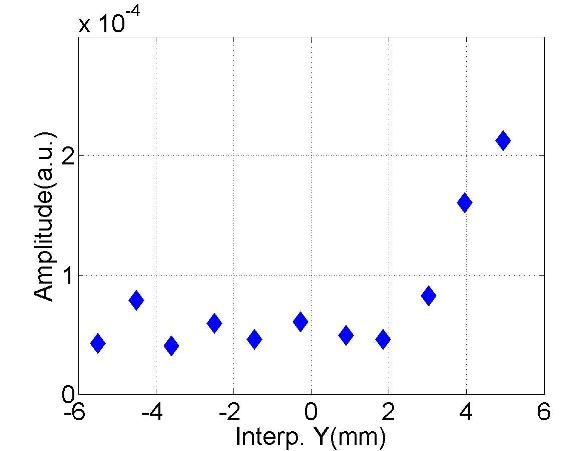}
\label{dep-C1H2-Y-16}
}
\subfigure[{\scriptsize\#17 ($f$:9.0754GHz; $Q$:10$^4$)}]{
\includegraphics[width=0.22\textwidth]{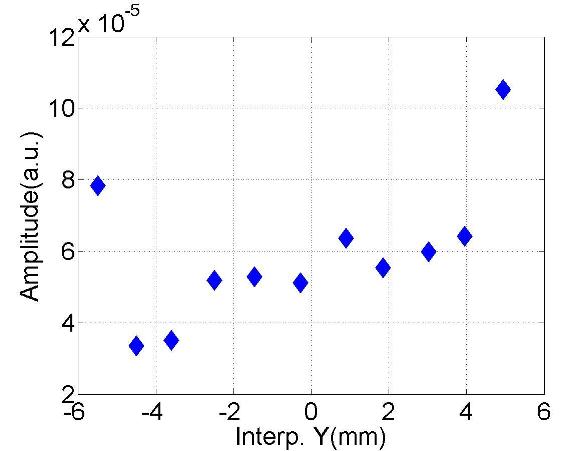}
\label{dep-C1H2-Y-17}
}
\subfigure[{\scriptsize\#18 ($f$:9.0821GHz; $Q$:10$^4$)}]{
\includegraphics[width=0.22\textwidth]{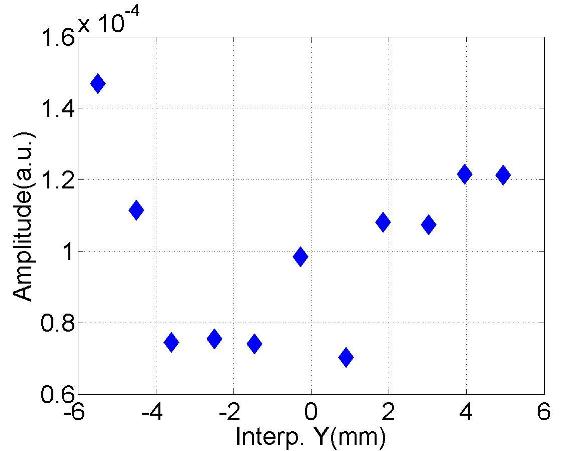}
\label{dep-C1H2-Y-18}
}
\subfigure[{\scriptsize\#19 ($f$:9.0890GHz; $Q$:10$^4$)}]{
\includegraphics[width=0.22\textwidth]{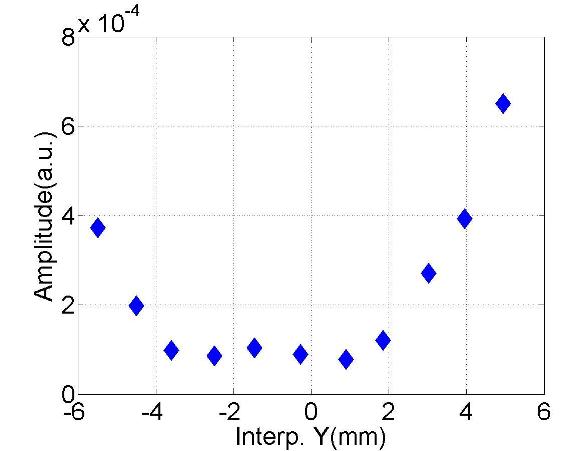}
\label{dep-C1H2-Y-19}
}
\subfigure[{\scriptsize\#20 ($f$:9.0956GHz; $Q$:10$^5$)}]{
\includegraphics[width=0.22\textwidth]{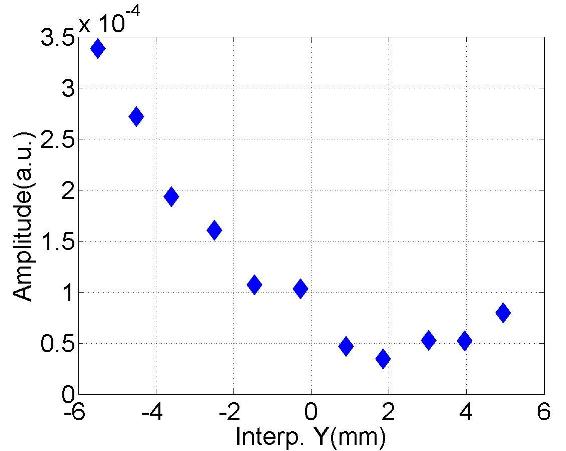}
\label{dep-C1H2-Y-20}
}
\subfigure[{\scriptsize\#21 ($f$:9.0962GHz; $Q$:10$^5$)}]{
\includegraphics[width=0.22\textwidth]{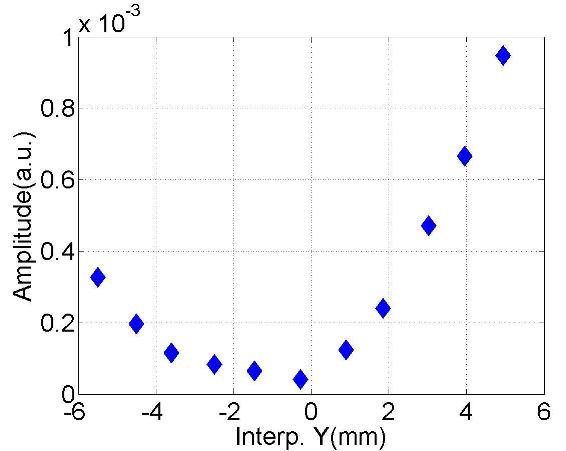}
\label{dep-C1H2-Y-21}
}
\caption{Dependence of the mode amplitude on the horizontal beam of{}fset in the cavity.}
\label{spec-dep-C1H2-X-2}
\end{figure}
\begin{figure}[h]\center
\subfigure[Spectrum (C1H2)]{
\includegraphics[width=0.9\textwidth]{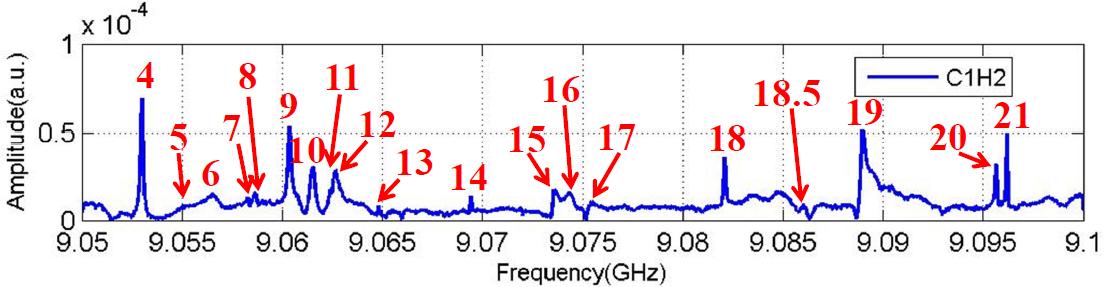}
\label{spec-C1H2-2}
}
\subfigure[{\scriptsize\#4 ($f$:9.0530GHz; $Q$:10$^5$)}]{
\includegraphics[width=0.22\textwidth]{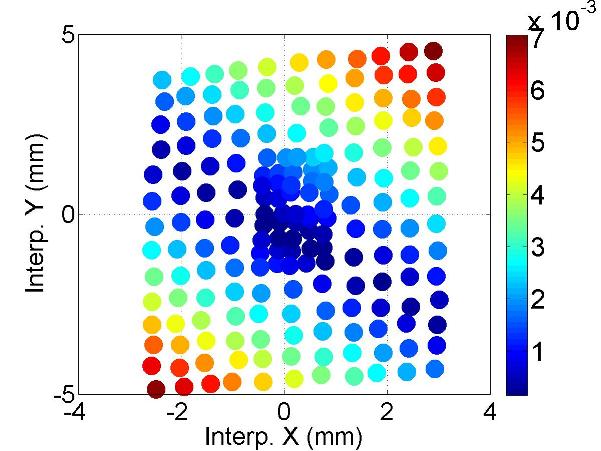}
\label{polar-C1H2-4}
}
\subfigure[{\scriptsize\#5 ($f$:9.0553GHz; $Q$:10$^3$)}]{
\includegraphics[width=0.22\textwidth]{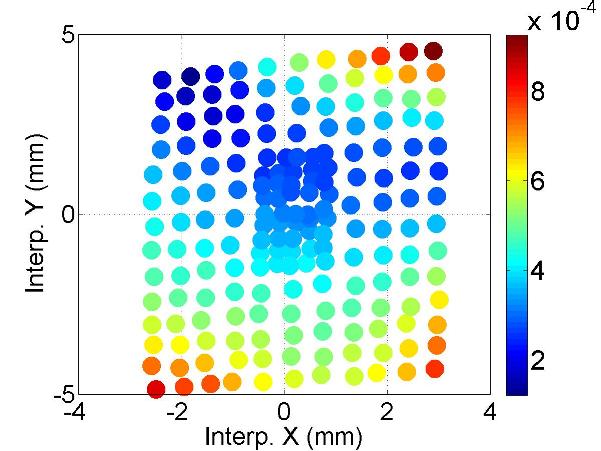}
\label{polar-C1H2-5}
}
\subfigure[{\scriptsize\#6 ($f$:9.0563GHz; $Q$:10$^4$)}]{
\includegraphics[width=0.22\textwidth]{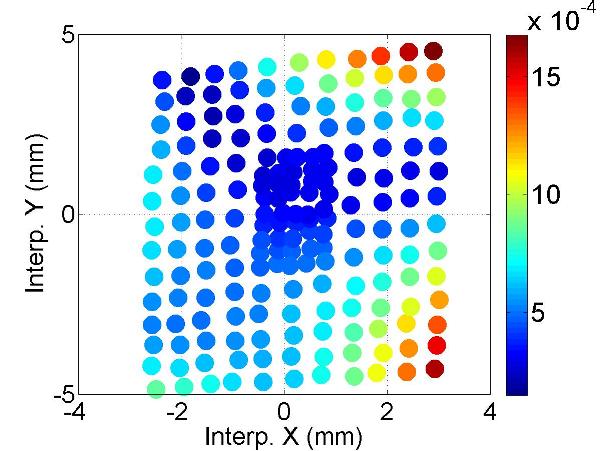}
\label{polar-C1H2-6}
}
\subfigure[{\scriptsize\#7 ($f$:9.0583GHz; $Q$:10$^4$)}]{
\includegraphics[width=0.22\textwidth]{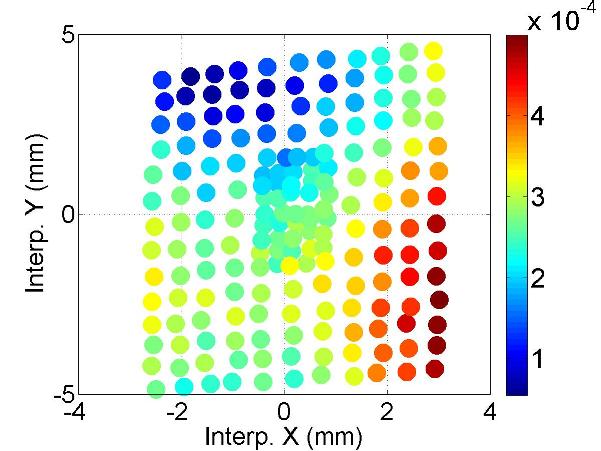}
\label{polar-C1H2-7}
}
\subfigure[{\scriptsize\#8 ($f$:9.0587GHz; $Q$:10$^4$)}]{
\includegraphics[width=0.22\textwidth]{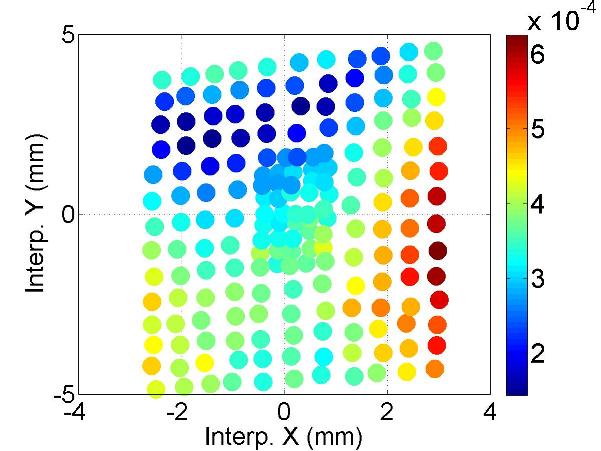}
\label{polar-C1H2-8}
}
\subfigure[{\scriptsize\#9 ($f$:9.0604GHz; $Q$:10$^4$)}]{
\includegraphics[width=0.22\textwidth]{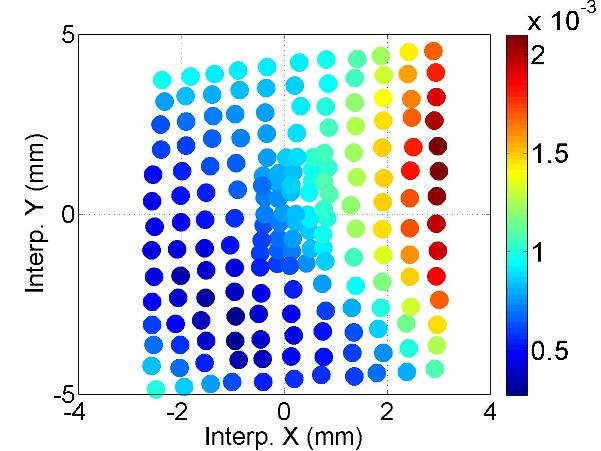}
\label{polar-C1H2-9}
}
\subfigure[{\scriptsize\#10 ($f$:9.0616GHz; $Q$:10$^4$)}]{
\includegraphics[width=0.22\textwidth]{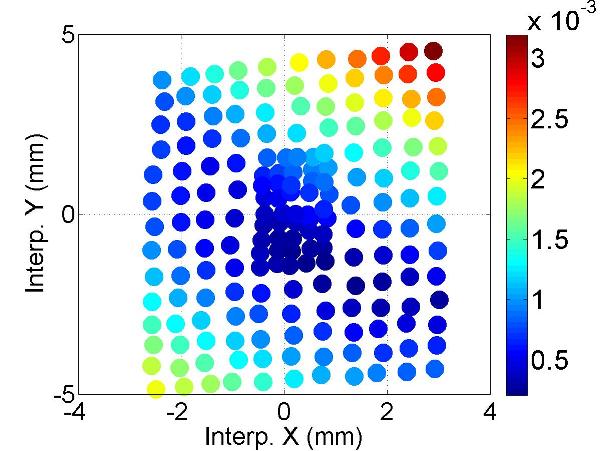}
\label{polar-C1H2-10}
}
\subfigure[{\scriptsize\#11 ($f$:9.0624GHz; $Q$:10$^4$)}]{
\includegraphics[width=0.22\textwidth]{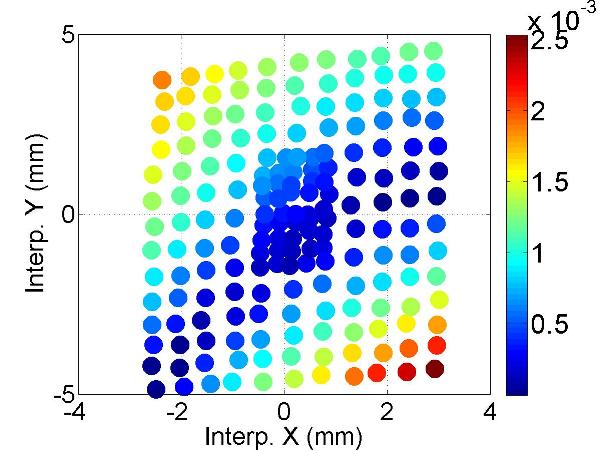}
\label{polar-C1H2-11}
}
\subfigure[{\scriptsize\#12 ($f$:9.0627GHz; $Q$:10$^4$)}]{
\includegraphics[width=0.22\textwidth]{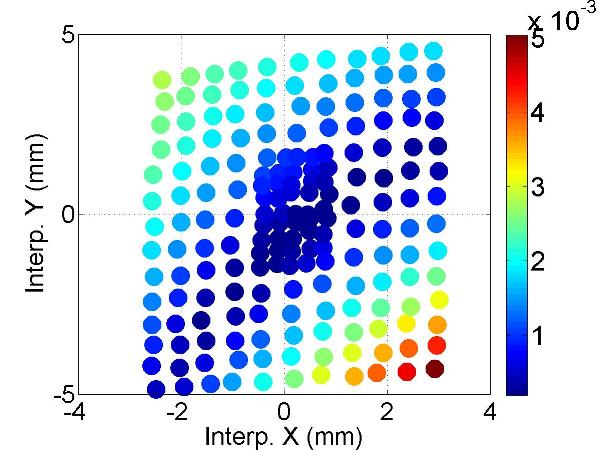}
\label{polar-C1H2-12}
}
\subfigure[{\scriptsize\#13 ($f$:9.0648GHz; $Q$:10$^4$)}]{
\includegraphics[width=0.22\textwidth]{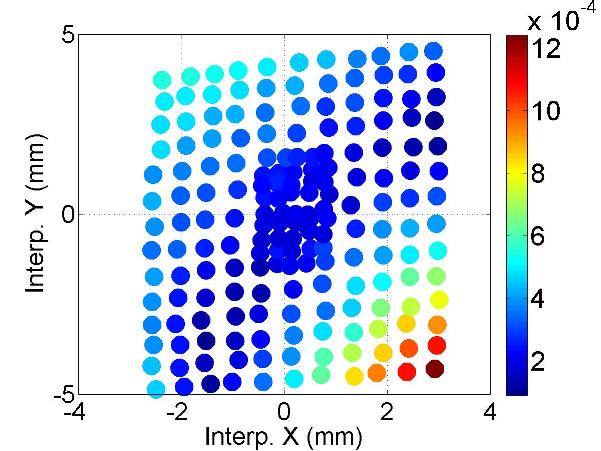}
\label{polar-C1H2-13}
}
\subfigure[{\scriptsize\#14 ($f$:9.0695GHz; $Q$:10$^5$)}]{
\includegraphics[width=0.22\textwidth]{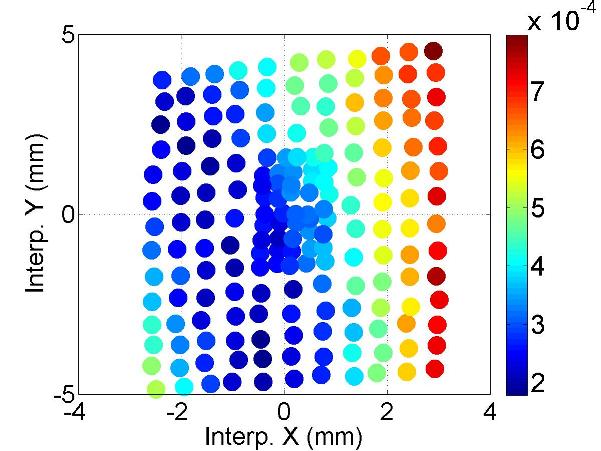}
\label{polar-C1H2-14}
}
\subfigure[{\scriptsize\#15 ($f$:9.0735GHz; $Q$:10$^4$)}]{
\includegraphics[width=0.22\textwidth]{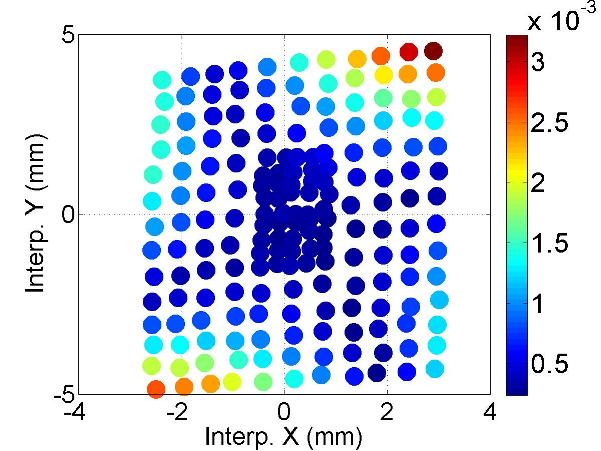}
\label{polar-C1H2-15}
}
\subfigure[{\scriptsize\#16 ($f$:9.0746GHz; $Q$:10$^4$)}]{
\includegraphics[width=0.22\textwidth]{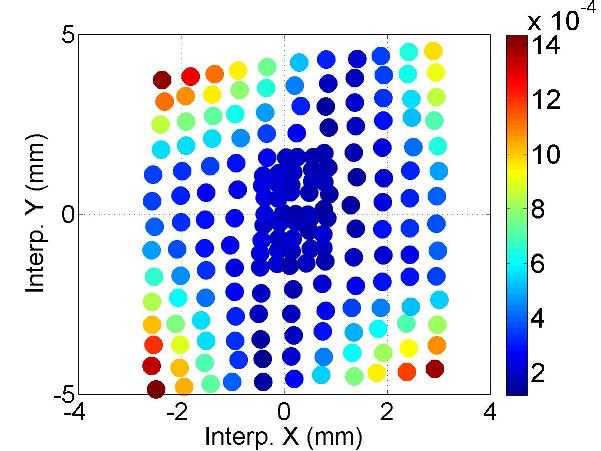}
\label{polar-C1H2-16}
}
\subfigure[{\scriptsize\#17 ($f$:9.0755GHz; $Q$:10$^4$)}]{
\includegraphics[width=0.22\textwidth]{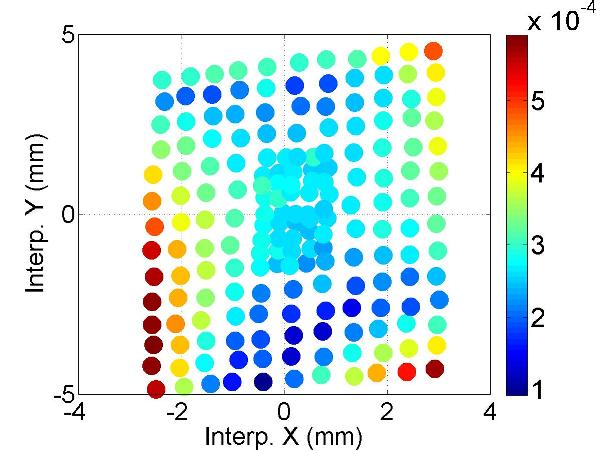}
\label{polar-C1H2-17}
}
\subfigure[{\scriptsize\#18 ($f$:9.0821GHz; $Q$:10$^5$)}]{
\includegraphics[width=0.22\textwidth]{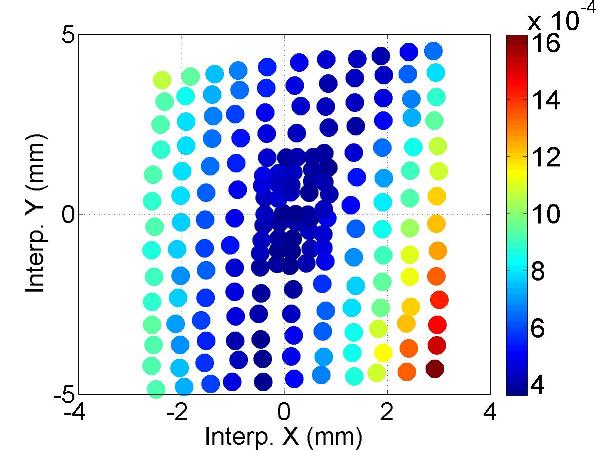}
\label{polar-C1H2-18}
}
\subfigure[{\scriptsize\#18.5 ($f$:9.0865GHz; $Q$:10$^4$)}]{
\includegraphics[width=0.22\textwidth]{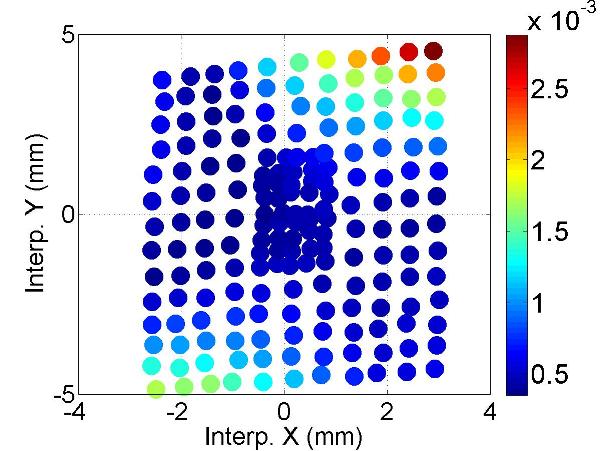}
\label{polar-C1H2-18_5}
}
\subfigure[{\scriptsize\#19 ($f$:9.0891GHz; $Q$:10$^4$)}]{
\includegraphics[width=0.22\textwidth]{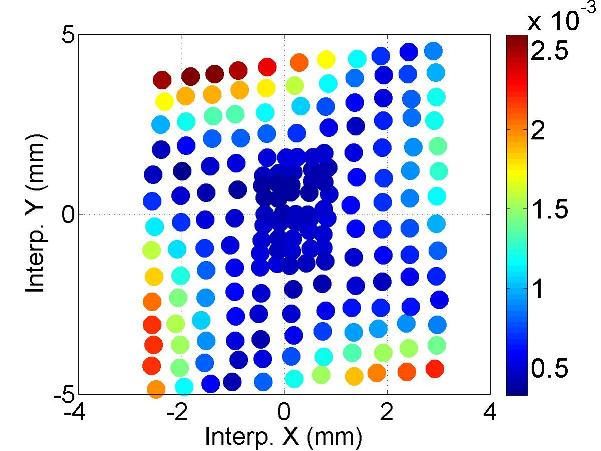}
\label{polar-C1H2-19}
}
\subfigure[{\scriptsize\#20 ($f$:9.0956GHz; $Q$:10$^4$)}]{
\includegraphics[width=0.22\textwidth]{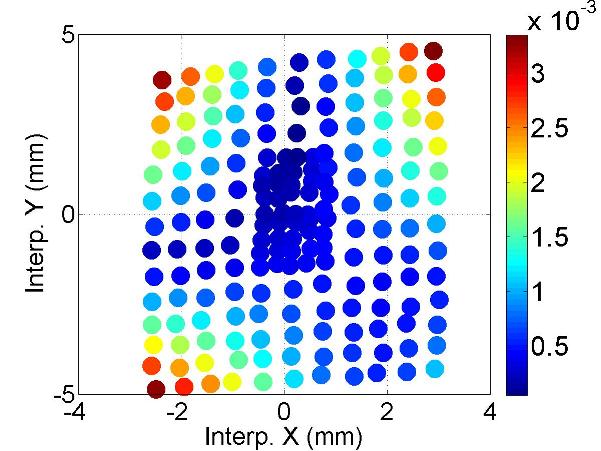}
\label{polar-C1H2-20}
}
\subfigure[{\scriptsize\#21 ($f$:9.0962GHz; $Q$:10$^5$)}]{
\includegraphics[width=0.22\textwidth]{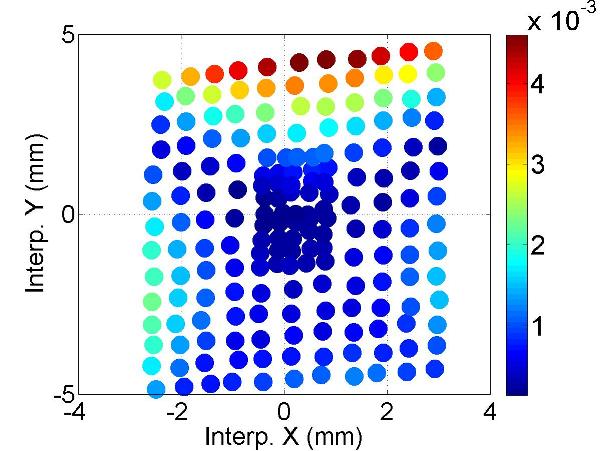}
\label{polar-C1H2-21}
}
\caption{Polarization of the mode.}
\label{spec-polar-C1H2-2}
\end{figure}

\FloatBarrier
\section{D5: HOM Coupler C2H1}
\begin{figure}[h]\center
\subfigure[Spectrum (C2H1)]{
\includegraphics[width=0.85\textwidth]{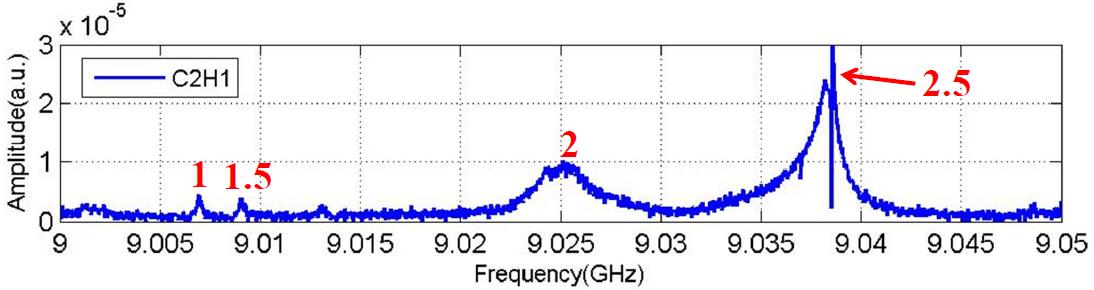}
\label{spec-C2H1-X-1}
}
\subfigure[Mode \#1 ($f$:9.0069GHz; $Q$:10$^4$)]{
\includegraphics[width=0.26\textwidth]{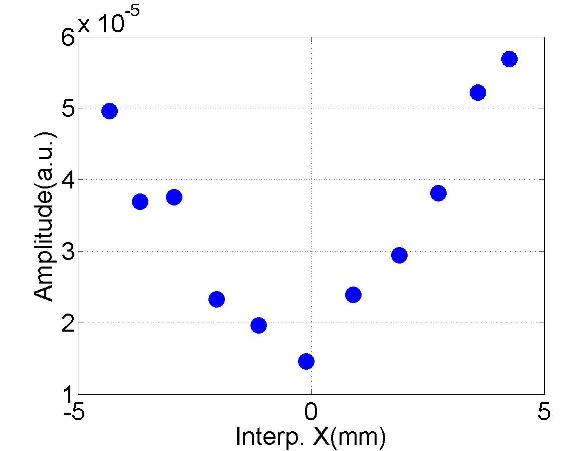}
\label{dep-C2H1-X-1}
}
\subfigure[Mode \#1 ($f$:9.00070GHz; $Q$:10$^4$)]{
\includegraphics[width=0.26\textwidth]{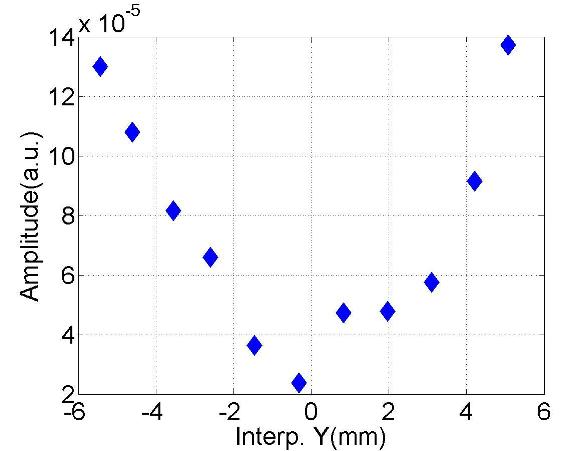}
\label{dep-C2H1-Y-1}
}
\subfigure[\#1 ($f$:9.0069GHz; $Q$:10$^4$)]{
\includegraphics[width=0.26\textwidth]{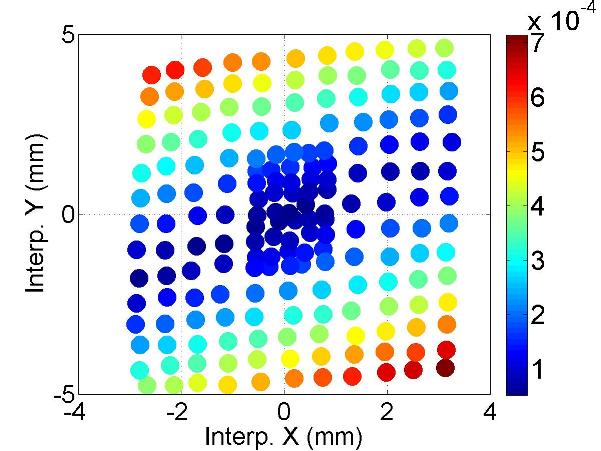}
\label{polar-C2H1-1}
}
\subfigure[\#2 ($f$:9.0257GHz; $Q$:10$^3$)]{
\includegraphics[width=0.26\textwidth]{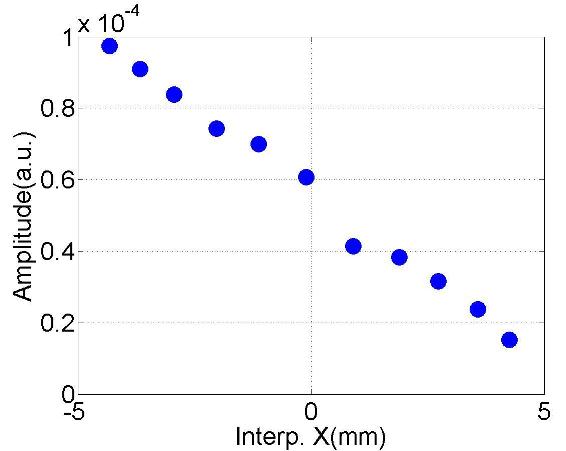}
\label{dep-C2H1-X-2}
}
\subfigure[\#2 ($f$:9.0257GHz; $Q$:10$^3$)]{
\includegraphics[width=0.26\textwidth]{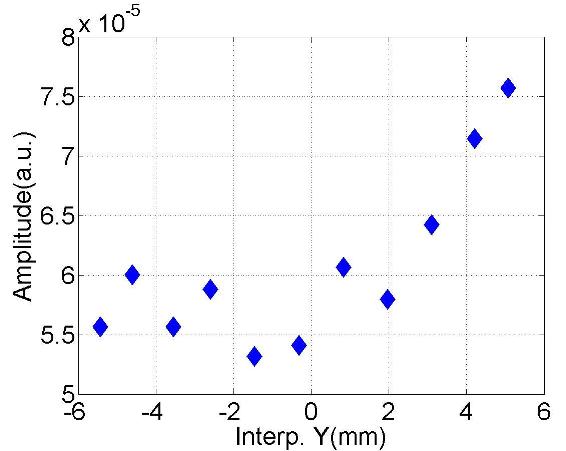}
\label{dep-C2H1-Y-2}
}
\subfigure[ \#2 ($f$:9.0257GHz; $Q$:10$^3$)]{
\includegraphics[width=0.26\textwidth]{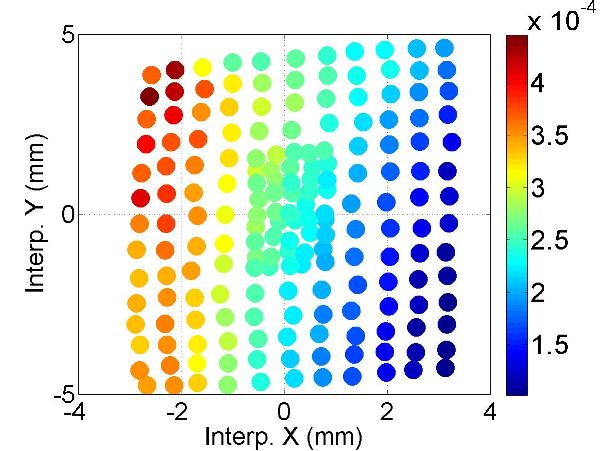}
\label{polar-C2H1-2}
}
\subfigure[\#1.5 ($f$:9.0090GHz; $Q$:10$^4$)]{
\includegraphics[width=0.26\textwidth]{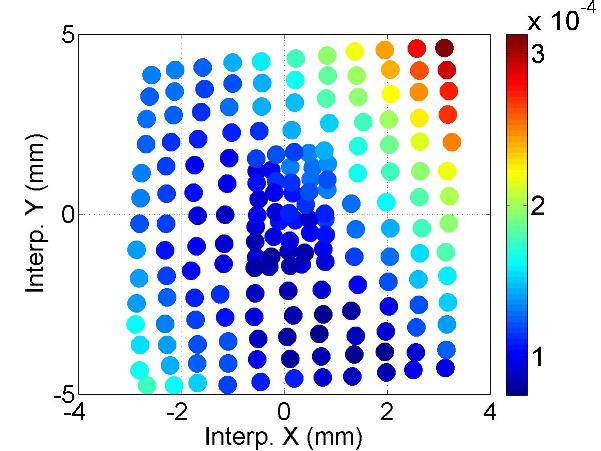}
\label{polar-C2H1-1_5}
}
\subfigure[\#2.5 ($f$:9.0387GHz; $Q$:10$^5$)]{
\includegraphics[width=0.26\textwidth]{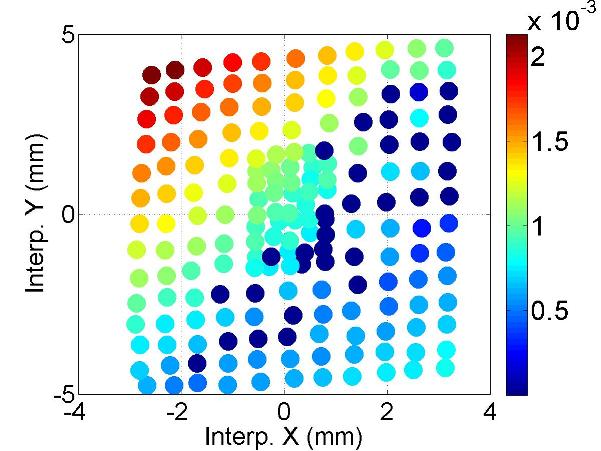}
\label{polar-C2H1-2_5}
}
\caption{Dependence of the mode amplitude on the transverse beam of{}fset in the cavity.}
\label{spec-dep-C2H1-XY-1}
\end{figure}
\begin{figure}[h]
\subfigure[Spectrum (C2H1)]{
\includegraphics[width=1\textwidth]{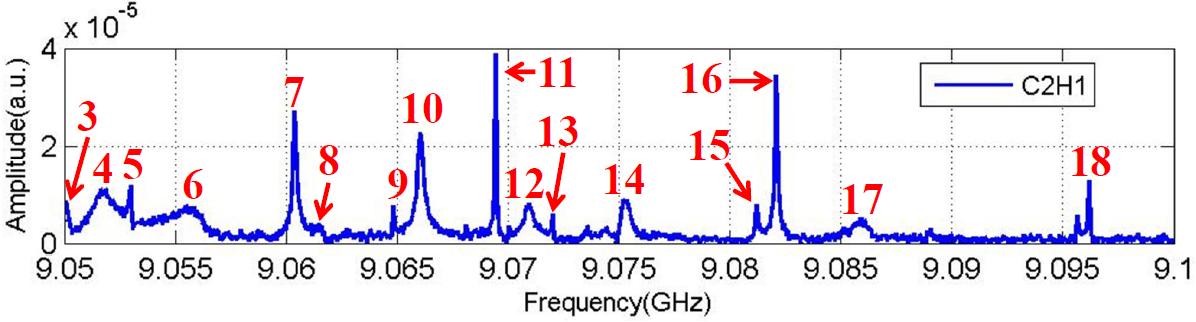}
\label{spec-C2H1-X-2}
}
\subfigure[\#3 ($f$:9.0501GHz; $Q$:10$^4$)]{
\includegraphics[width=0.23\textwidth]{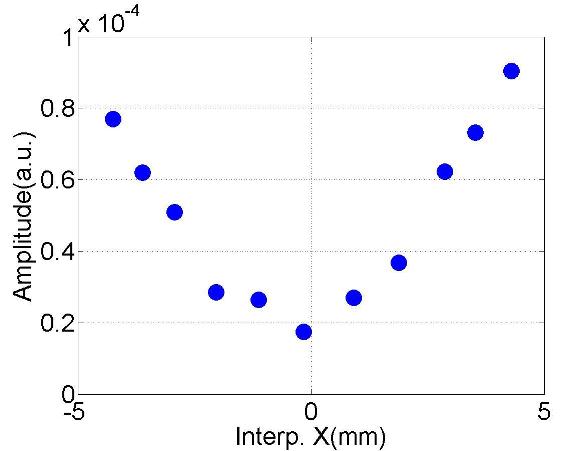}
\label{dep-C2H1-X-3}
}
\subfigure[\#4 ($f$:9.0517GHz; $Q$:10$^3$)]{
\includegraphics[width=0.23\textwidth]{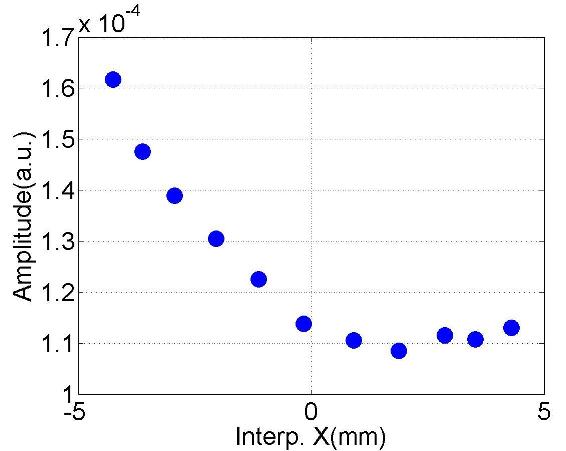}
\label{dep-C2H1-X-4}
}
\subfigure[\#5 ($f$:9.0530GHz; $Q$:10$^4$)]{
\includegraphics[width=0.23\textwidth]{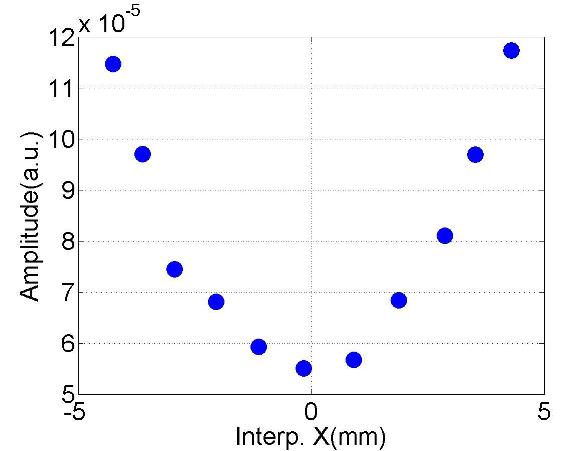}
\label{dep-C2H1-X-5}
}
\subfigure[\#6 ($f$:9.0552GHz; $Q$:10$^3$)]{
\includegraphics[width=0.23\textwidth]{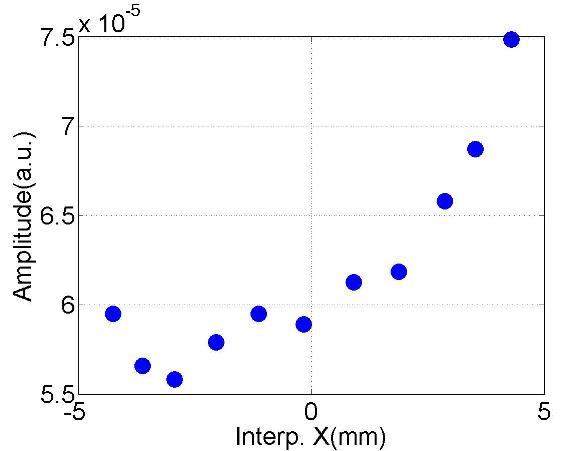}
\label{dep-C2H1-X-6}
}
\subfigure[\#7 ($f$:9.0604GHz; $Q$:10$^4$)]{
\includegraphics[width=0.23\textwidth]{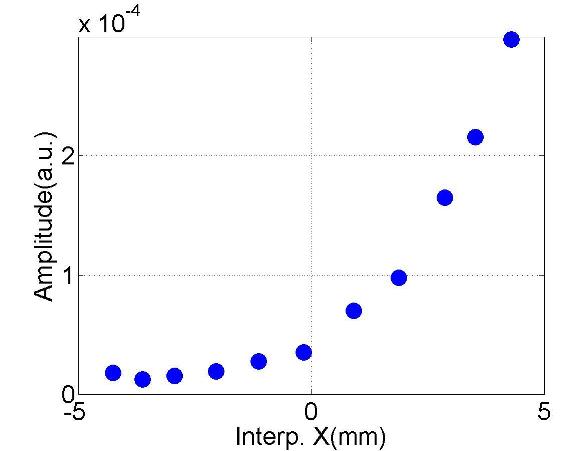}
\label{dep-C2H1-X-7}
}
\subfigure[\#8 ($f$:9.0615GHz; $Q$:10$^4$)]{
\includegraphics[width=0.23\textwidth]{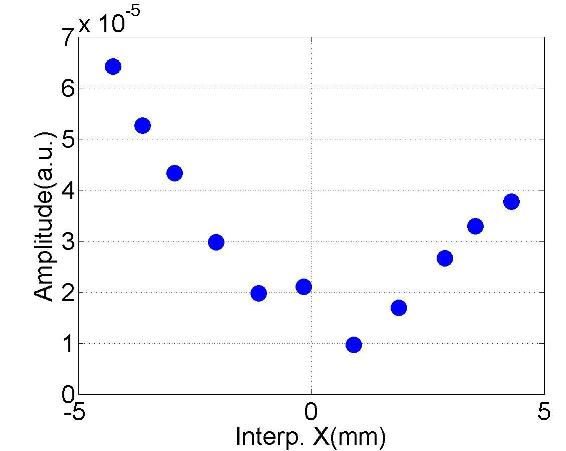}
\label{dep-C2H1-X-8}
}
\subfigure[\#9 ($f$:9.0648GHz; $Q$:10$^5$)]{
\includegraphics[width=0.23\textwidth]{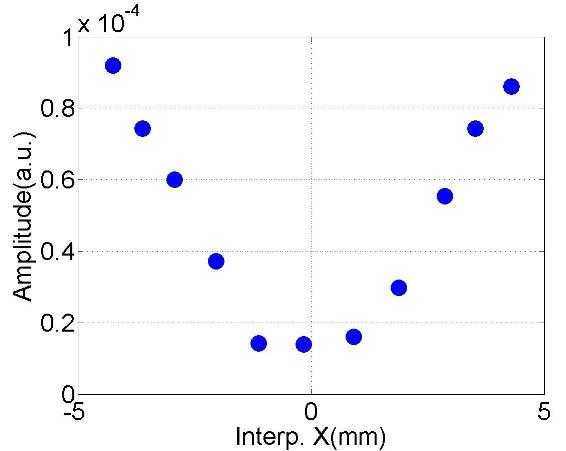}
\label{dep-C2H1-X-9}
}
\subfigure[\#10 ($f$:9.0660GHz; $Q$:10$^4$)]{
\includegraphics[width=0.23\textwidth]{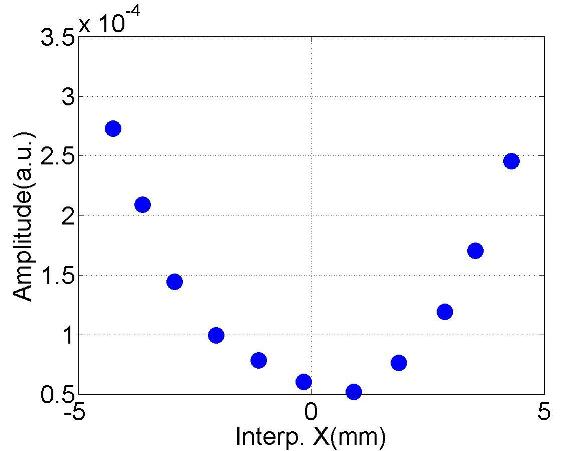}
\label{dep-C2H1-X-10}
}
\subfigure[\#11 ($f$:9.0694GHz; $Q$:10$^5$)]{
\includegraphics[width=0.23\textwidth]{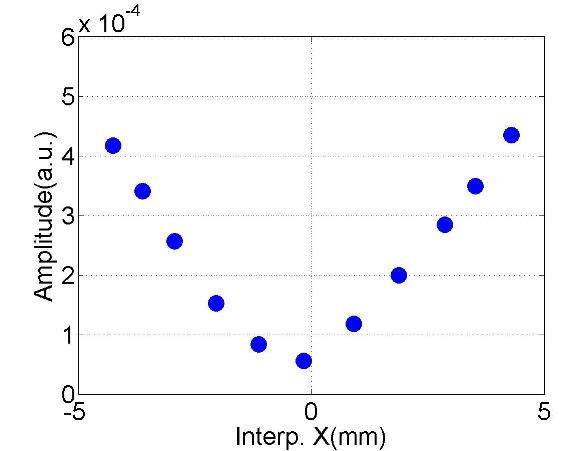}
\label{dep-C2H1-X-11}
}
\subfigure[\#12 ($f$:9.0709GHz; $Q$:10$^4$)]{
\includegraphics[width=0.23\textwidth]{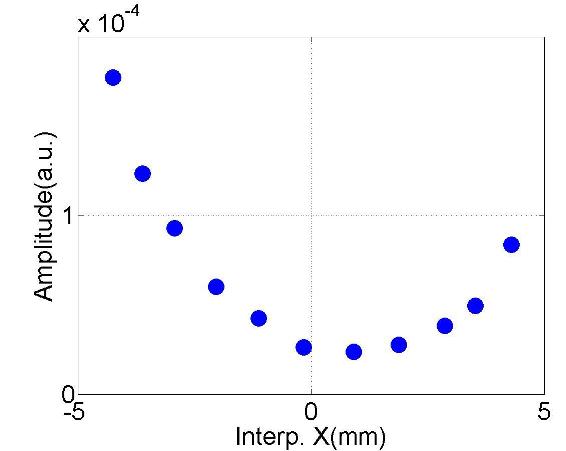}
\label{dep-C2H1-X-12}
}
\subfigure[\#13 ($f$:9.0720GHz; $Q$:10$^5$)]{
\includegraphics[width=0.23\textwidth]{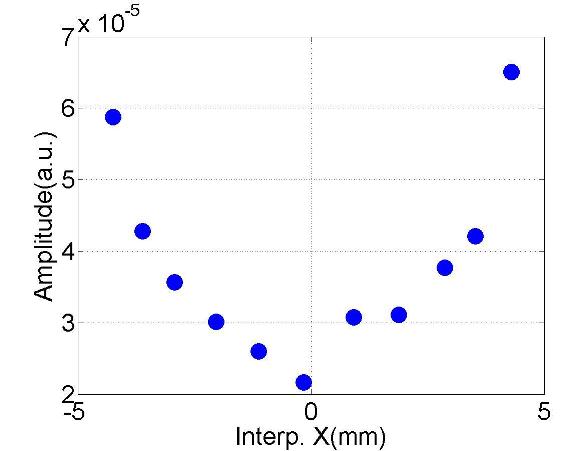}
\label{dep-C2H1-X-13}
}
\subfigure[\#14 ($f$:9.0755GHz; $Q$:10$^4$)]{
\includegraphics[width=0.23\textwidth]{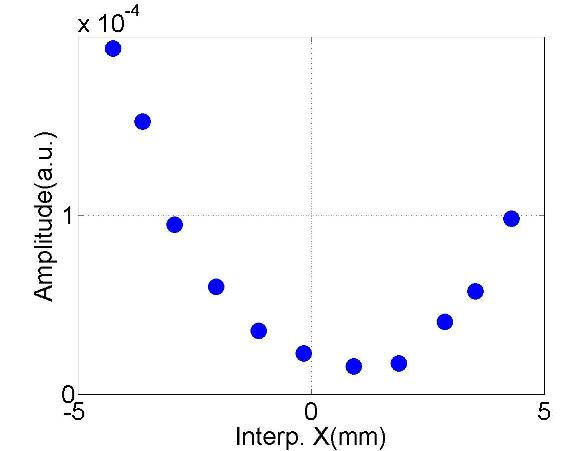}
\label{dep-C2H1-X-14}
}
\subfigure[\#15 ($f$:9.0812GHz; $Q$:10$^5$)]{
\includegraphics[width=0.23\textwidth]{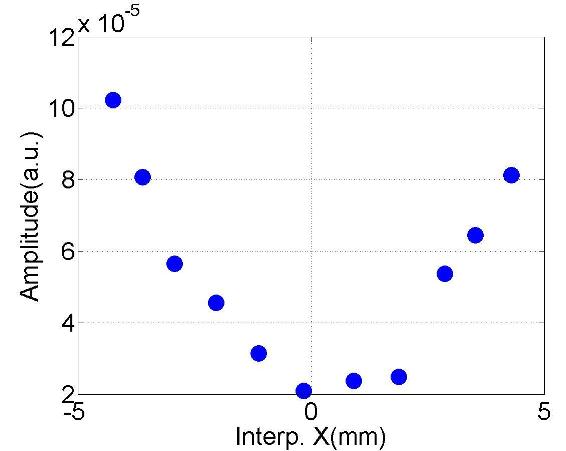}
\label{dep-C2H1-X-15}
}
\subfigure[\#16 ($f$:9.0821GHz; $Q$:10$^5$)]{
\includegraphics[width=0.23\textwidth]{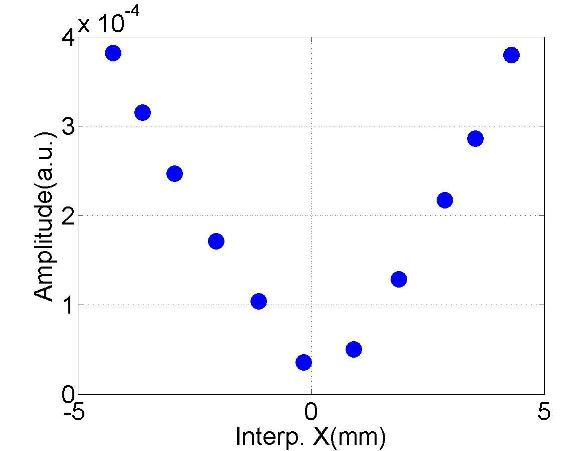}
\label{dep-C2H1-X-16}
}
\subfigure[\#17 ($f$:9.0861GHz; $Q$:10$^4$)]{
\includegraphics[width=0.23\textwidth]{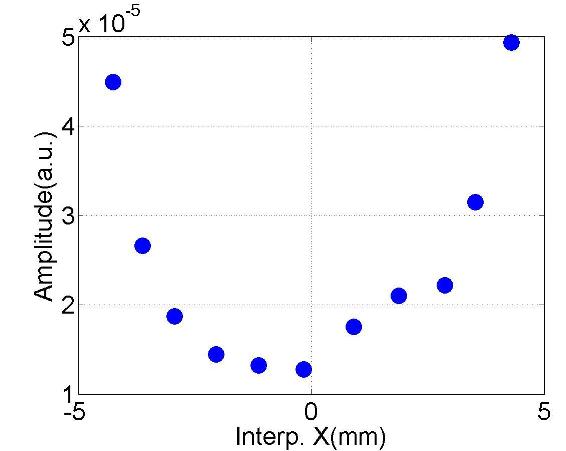}
\label{dep-C2H1-X-17}
}
\subfigure[\#18 ($f$:9.0962GHz; $Q$:10$^5$)]{
\includegraphics[width=0.23\textwidth]{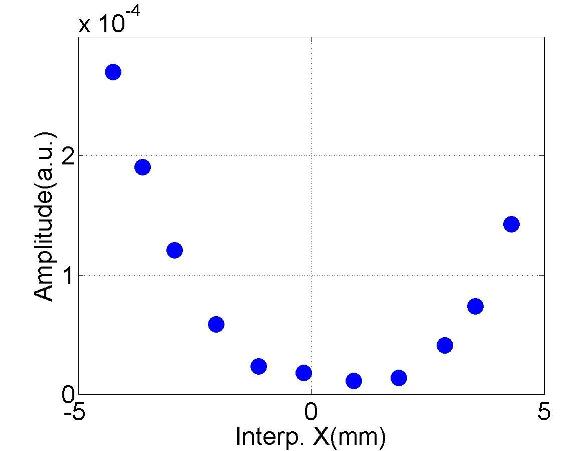}
\label{dep-C2H1-X-18}
}
\caption{Dependence of the mode amplitude on the horizontal beam of{}fset in the cavity.}
\label{spec-dep-C2H1-X-2}
\end{figure}
\begin{figure}[h]
\subfigure[Spectrum (C2H1)]{
\includegraphics[width=1\textwidth]{D5Xmove-Spec-C2H1-2}
\label{spec-C2H1-X-2}
}
\subfigure[\#3 ($f$:9.0501GHz; $Q$:10$^5$)]{
\includegraphics[width=0.23\textwidth]{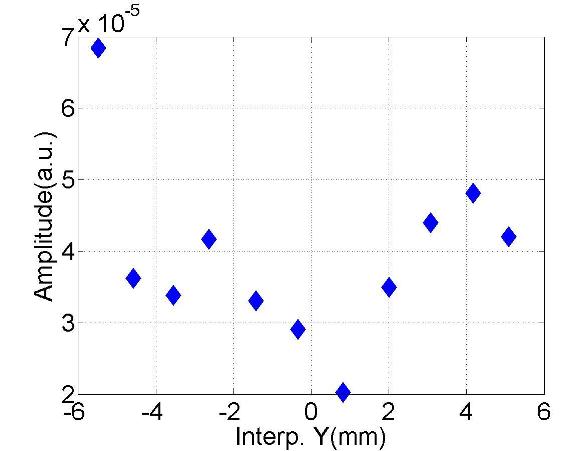}
\label{dep-C2H1-Y-3}
}
\subfigure[\#4 ($f$:9.0518GHz; $Q$:10$^3$)]{
\includegraphics[width=0.23\textwidth]{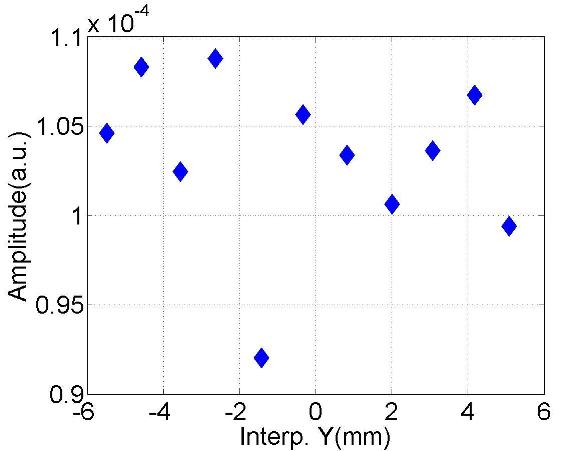}
\label{dep-C2H1-Y-4}
}
\subfigure[\#5 ($f$:9.0530GHz; $Q$:10$^4$)]{
\includegraphics[width=0.23\textwidth]{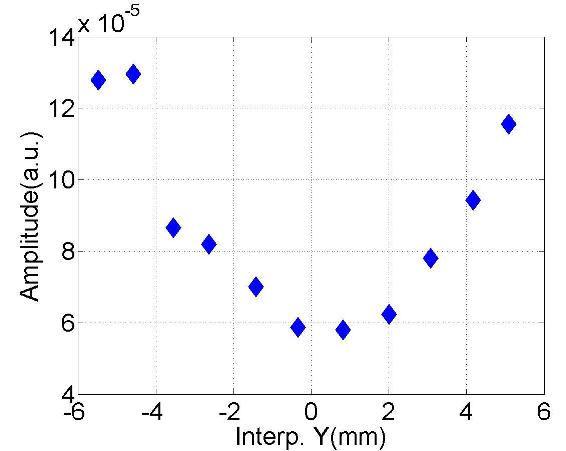}
\label{dep-C2H1-Y-5}
}
\subfigure[\#6 ($f$:9.0552GHz; $Q$:10$^3$)]{
\includegraphics[width=0.23\textwidth]{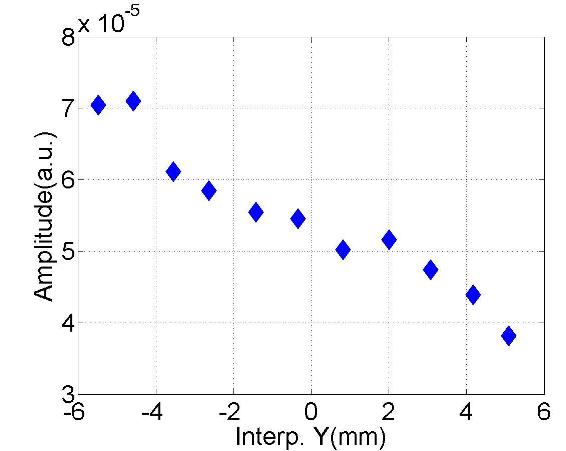}
\label{dep-C2H1-Y-6}
}
\subfigure[\#7 ($f$:9.0604GHz; $Q$:10$^4$)]{
\includegraphics[width=0.23\textwidth]{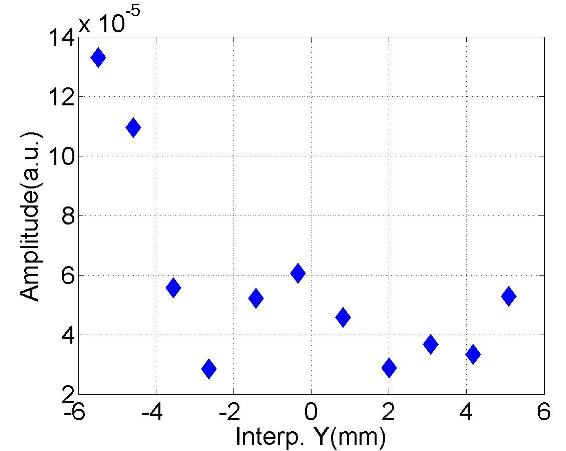}
\label{dep-C2H1-Y-7}
}
\subfigure[\#8 ($f$:9.0616GHz; $Q$:10$^4$)]{
\includegraphics[width=0.23\textwidth]{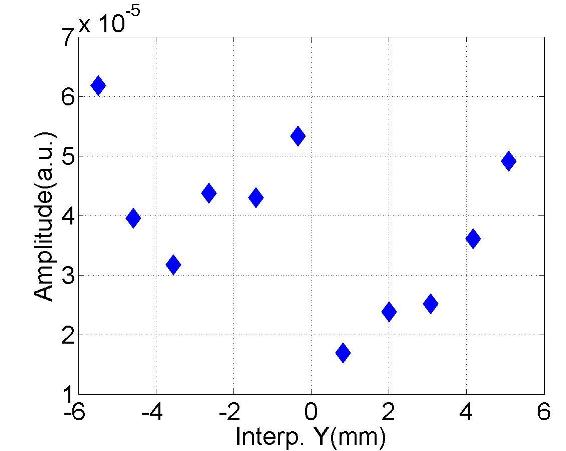}
\label{dep-C2H1-Y-8}
}
\subfigure[\#9 ($f$:9.0648GHz; $Q$:10$^5$)]{
\includegraphics[width=0.23\textwidth]{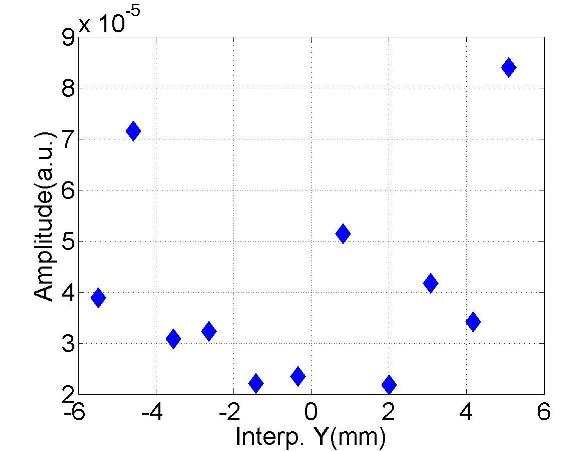}
\label{dep-C2H1-Y-9}
}
\subfigure[\#10 ($f$:9.0661GHz; $Q$:10$^4$)]{
\includegraphics[width=0.23\textwidth]{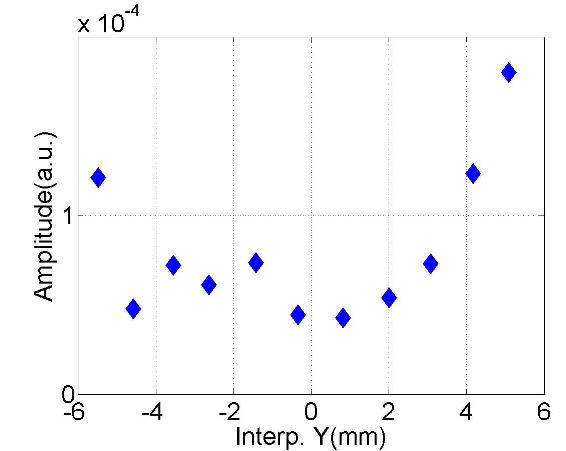}
\label{dep-C2H1-Y-10}
}
\subfigure[\#11 ($f$:9.0694GHz; $Q$:10$^5$)]{
\includegraphics[width=0.23\textwidth]{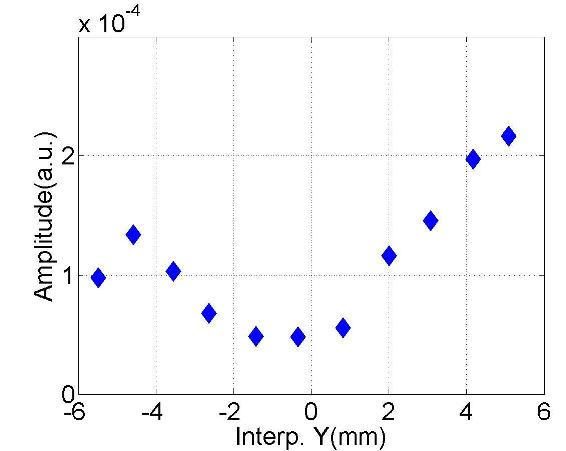}
\label{dep-C2H1-Y-11}
}
\subfigure[\#12 ($f$:9.0709GHz; $Q$:10$^4$)]{
\includegraphics[width=0.23\textwidth]{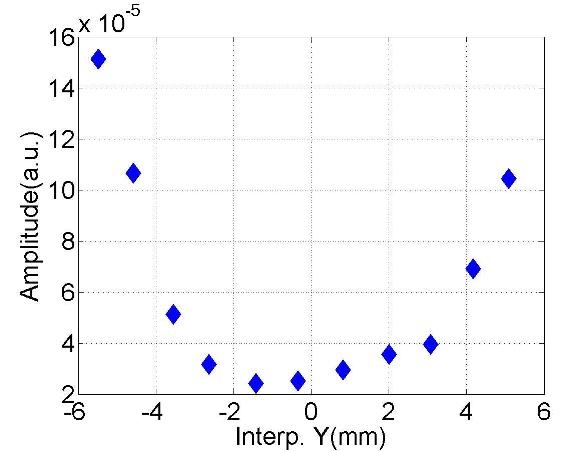}
\label{dep-C2H1-Y-12}
}
\subfigure[\#13 ($f$:9.0720GHz; $Q$:10$^5$)]{
\includegraphics[width=0.23\textwidth]{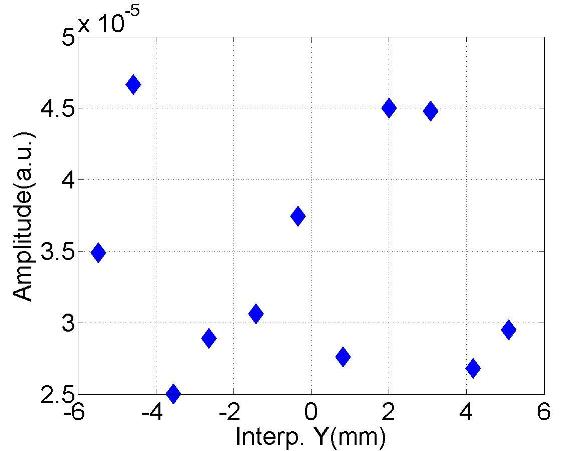}
\label{dep-C2H1-Y-13}
}
\subfigure[\#14 ($f$:9.0754GHz; $Q$:10$^4$)]{
\includegraphics[width=0.23\textwidth]{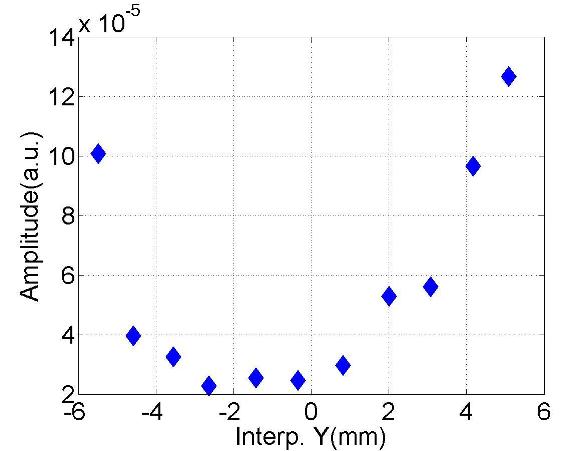}
\label{dep-C2H1-Y-14}
}
\subfigure[\#15 ($f$:9.0812GHz; $Q$:10$^5$)]{
\includegraphics[width=0.23\textwidth]{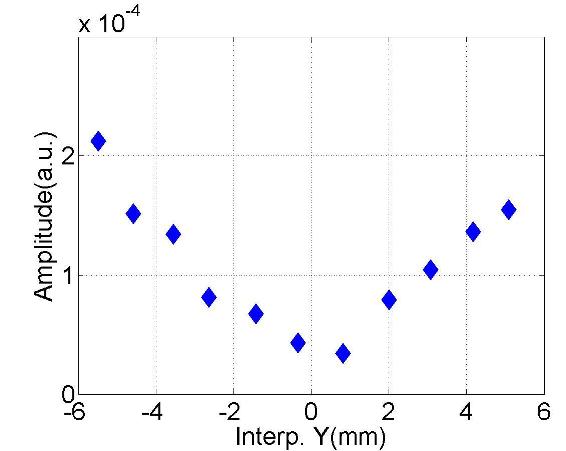}
\label{dep-C2H1-Y-15}
}
\subfigure[\#16 ($f$:9.0821GHz; $Q$:10$^5$)]{
\includegraphics[width=0.23\textwidth]{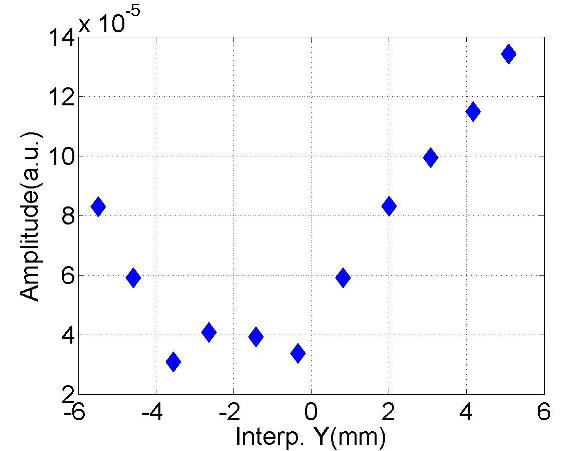}
\label{dep-C2H1-Y-16}
}
\subfigure[\#17 ($f$:9.0864GHz; $Q$:10$^4$)]{
\includegraphics[width=0.23\textwidth]{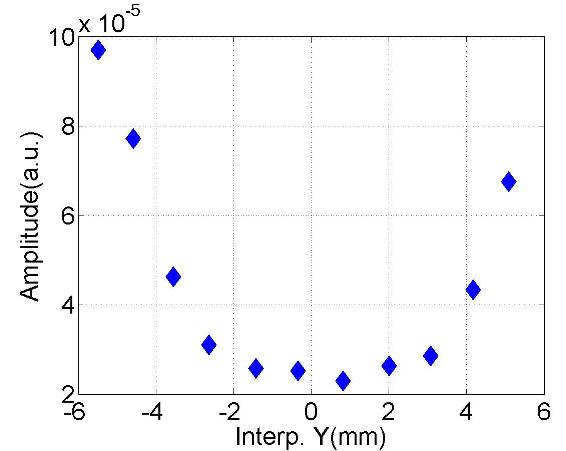}
\label{dep-C2H1-Y-17}
}
\subfigure[\#18 ($f$:9.0962GHz; $Q$:10$^5$)]{
\includegraphics[width=0.23\textwidth]{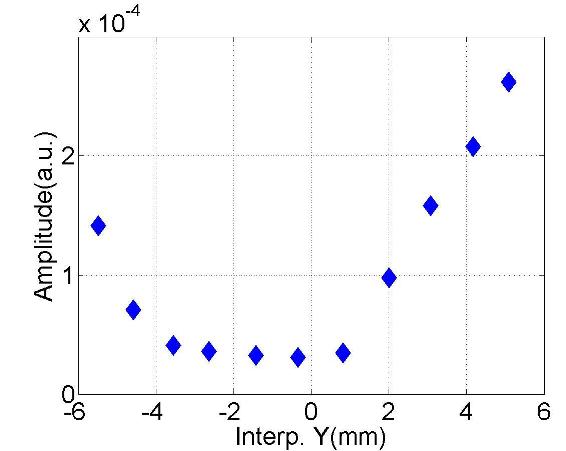}
\label{dep-C2H1-Y-18}
}
\caption{Dependence of the mode amplitude on the vertical beam of{}fset in the cavity.}
\label{spec-dep-C2H1-Y-2}
\end{figure}
\begin{figure}[h]
\subfigure[Spectrum (C2H1)]{
\includegraphics[width=1\textwidth]{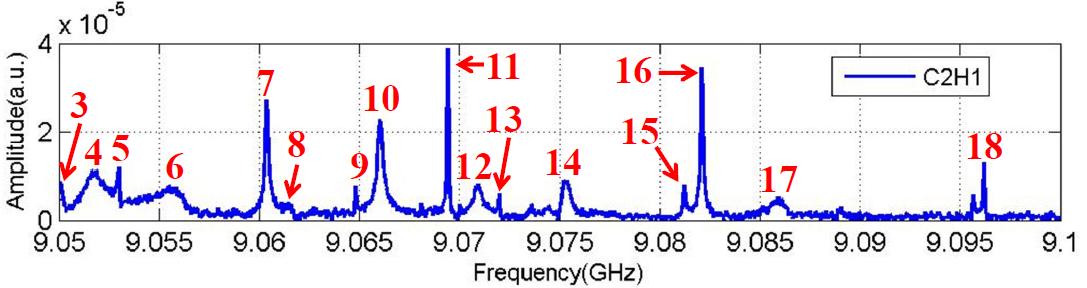}
\label{spec-C2H1-2}
}
\subfigure[\#3 ($f$:9.0501GHz; $Q$:10$^4$)]{
\includegraphics[width=0.23\textwidth]{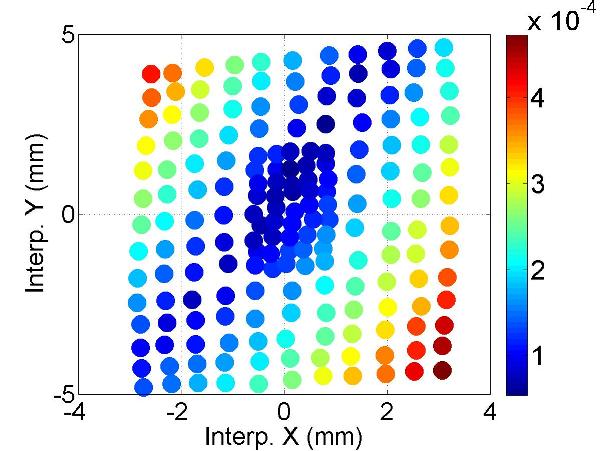}
\label{polar-C2H1-3}
}
\subfigure[\#4 ($f$:9.0517GHz; $Q$:10$^3$)]{
\includegraphics[width=0.23\textwidth]{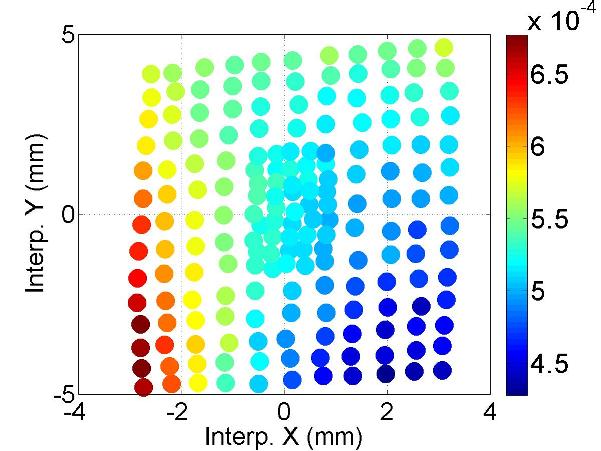}
\label{polar-C2H1-4}
}
\subfigure[\#5 ($f$:9.0530GHz; $Q$:10$^5$)]{
\includegraphics[width=0.23\textwidth]{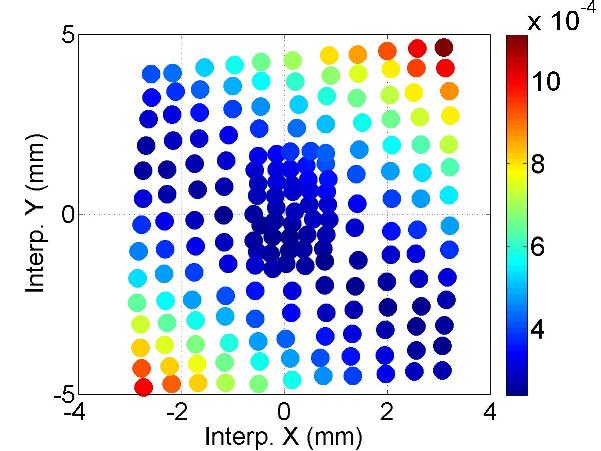}
\label{polar-C2H1-5}
}
\subfigure[\#6 ($f$:9.0553GHz; $Q$:10$^3$)]{
\includegraphics[width=0.23\textwidth]{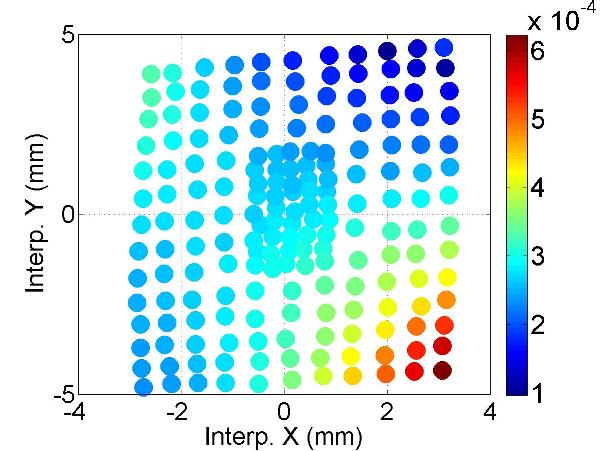}
\label{polar-C2H1-6}
}
\subfigure[\#7 ($f$:9.0604GHz; $Q$:10$^4$)]{
\includegraphics[width=0.23\textwidth]{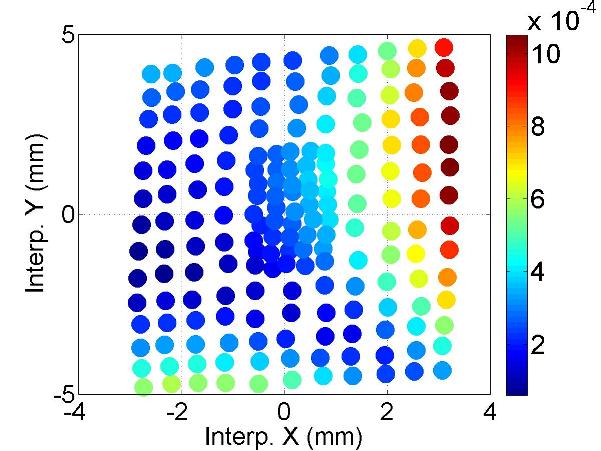}
\label{polar-C2H1-7}
}
\subfigure[\#8 ($f$:9.0615GHz; $Q$:10$^4$)]{
\includegraphics[width=0.23\textwidth]{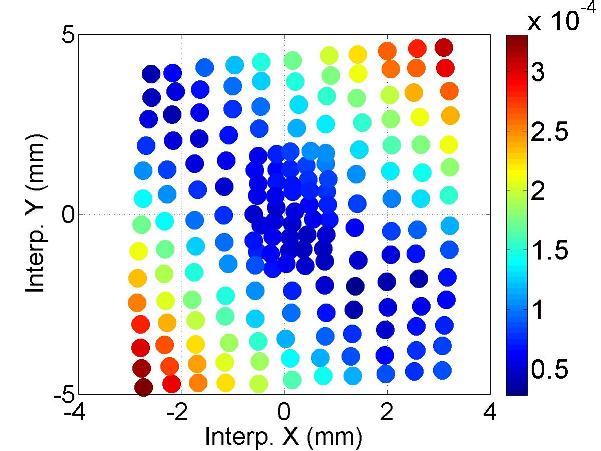}
\label{polar-C2H1-8}
}
\subfigure[\#9 ($f$:9.0648GHz; $Q$:10$^5$)]{
\includegraphics[width=0.23\textwidth]{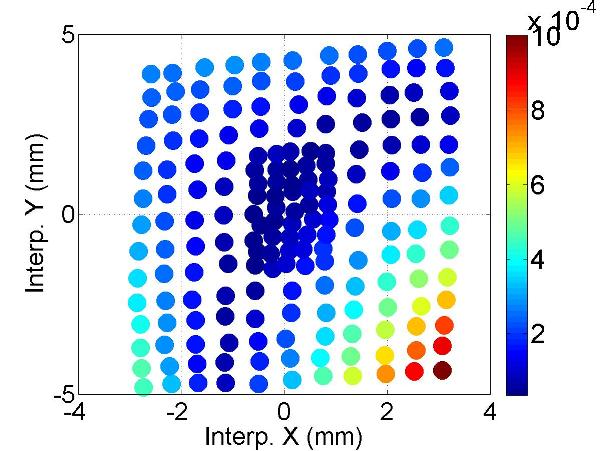}
\label{polar-C2H1-9}
}
\subfigure[\#10 ($f$:9.0660GHz; $Q$:10$^4$)]{
\includegraphics[width=0.23\textwidth]{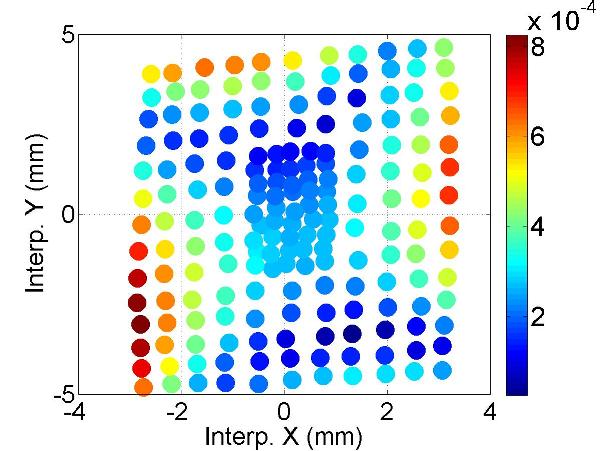}
\label{polar-C2H1-10}
}
\subfigure[\#11 ($f$:9.0694GHz; $Q$:10$^5$)]{
\includegraphics[width=0.23\textwidth]{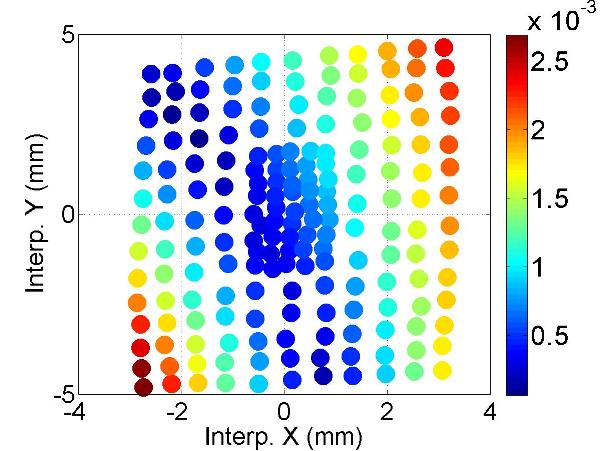}
\label{polar-C2H1-11}
}
\subfigure[\#12 ($f$:9.0709GHz; $Q$:10$^4$)]{
\includegraphics[width=0.23\textwidth]{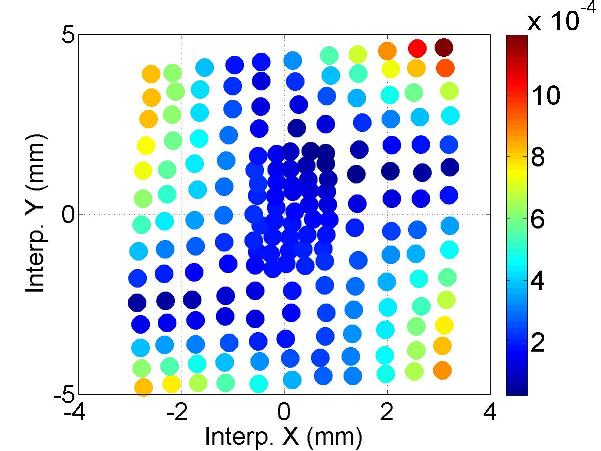}
\label{polar-C2H1-12}
}
\subfigure[\#13 ($f$:9.0720GHz; $Q$:10$^5$)]{
\includegraphics[width=0.23\textwidth]{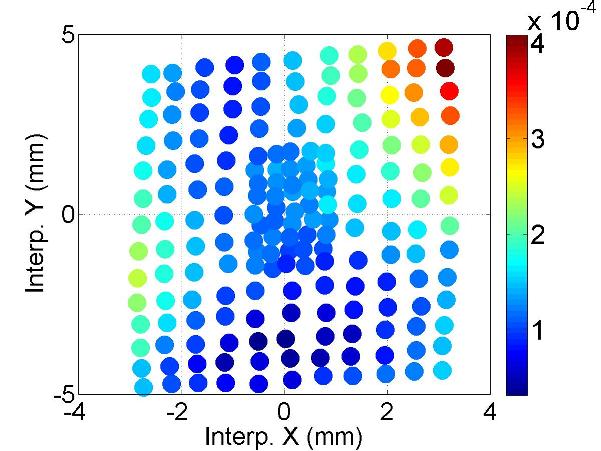}
\label{polar-C2H1-13}
}
\subfigure[\#14 ($f$:9.0756GHz; $Q$:10$^4$)]{
\includegraphics[width=0.23\textwidth]{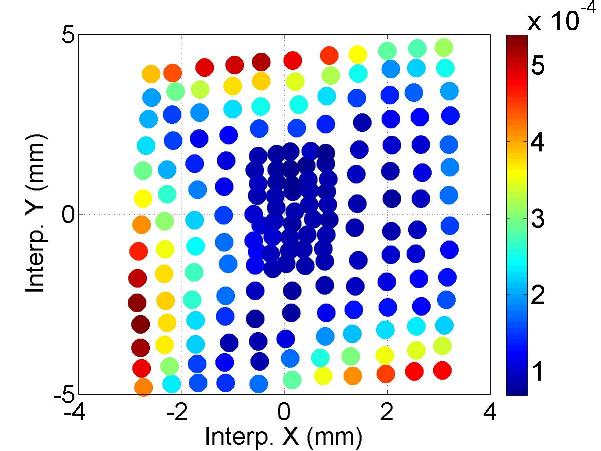}
\label{polar-C2H1-14}
}
\subfigure[\#15 ($f$:9.0812GHz; $Q$:10$^5$)]{
\includegraphics[width=0.23\textwidth]{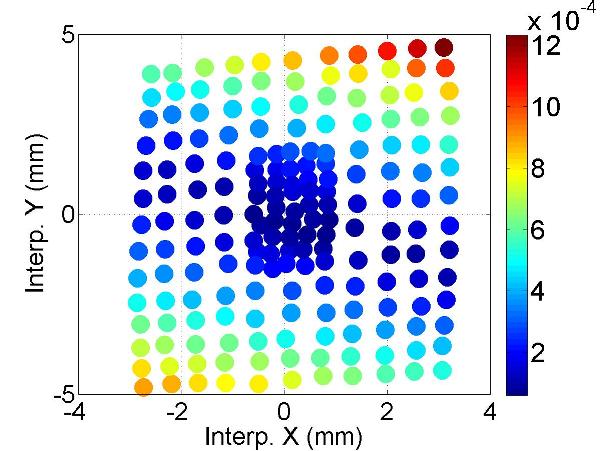}
\label{polar-C2H1-15}
}
\subfigure[\#16 ($f$:9.0821GHz; $Q$:10$^5$)]{
\includegraphics[width=0.23\textwidth]{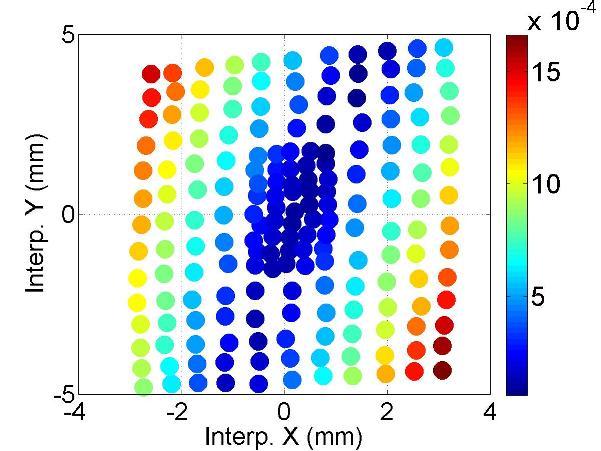}
\label{polar-C2H1-16}
}
\subfigure[\#17 ($f$:9.0860GHz; $Q$:10$^5$)]{
\includegraphics[width=0.23\textwidth]{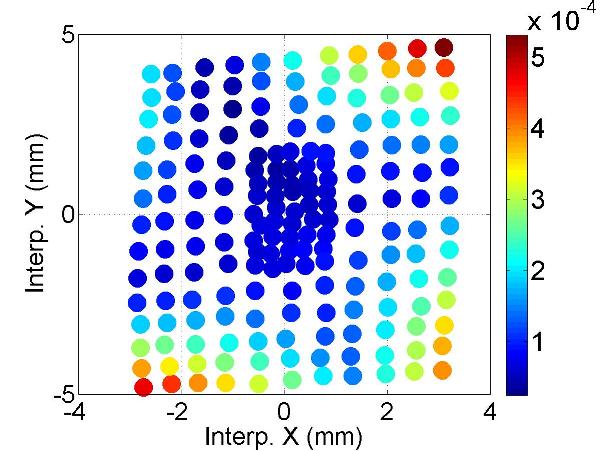}
\label{polar-C2H1-17}
}
\subfigure[\#18 ($f$:9.0962GHz; $Q$:10$^5$)]{
\includegraphics[width=0.23\textwidth]{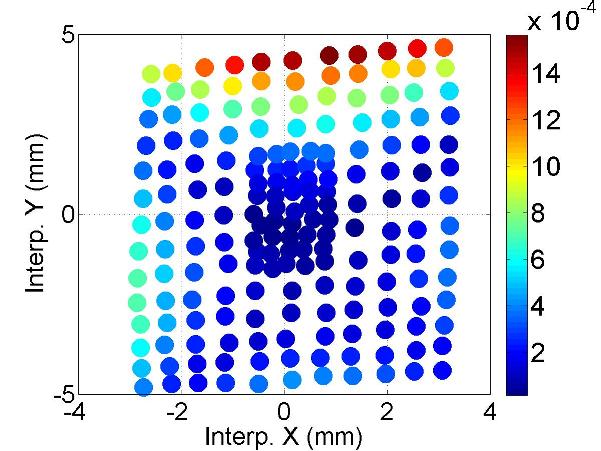}
\label{polar-C2H1-18}
}
\caption{Polarization of the mode.}
\label{spec-polar-C2H1-2}
\end{figure}

\FloatBarrier
\section{D5: HOM Coupler C2H2}
\begin{figure}[h]
\subfigure[Spectrum (C2H2)]{
\includegraphics[width=1\textwidth]{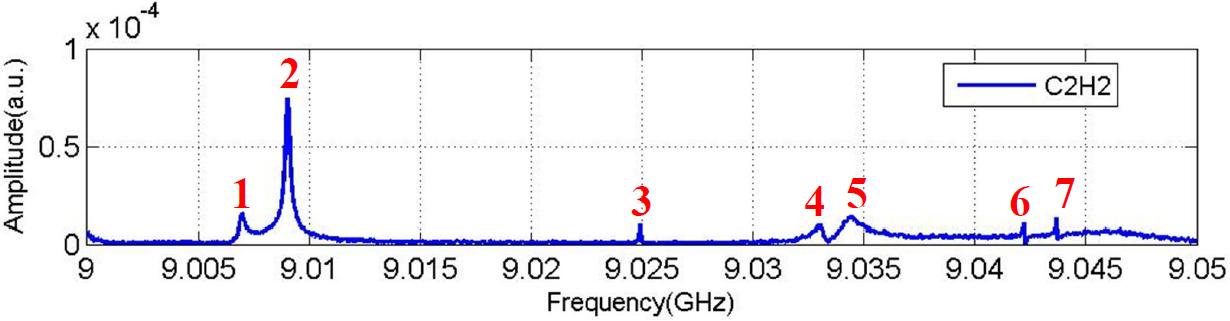}
\label{spec-C2H2-X-1}
}
\subfigure[\#1 ($f$:9.0070GHz; $Q$:10$^4$)]{
\includegraphics[width=0.23\textwidth]{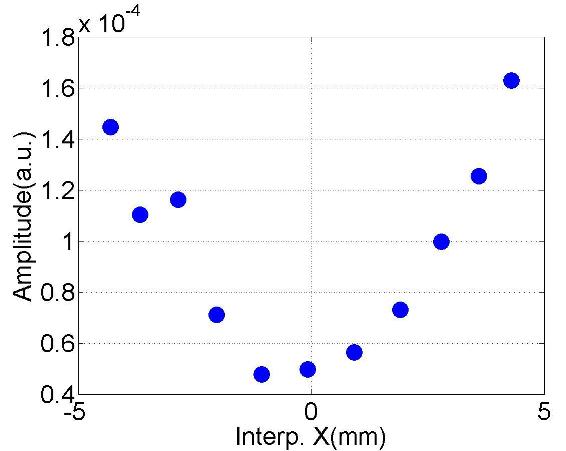}
\label{dep-C2H2-X-1}
}
\subfigure[\#2 ($f$:9.0090GHz; $Q$:10$^4$)]{
\includegraphics[width=0.23\textwidth]{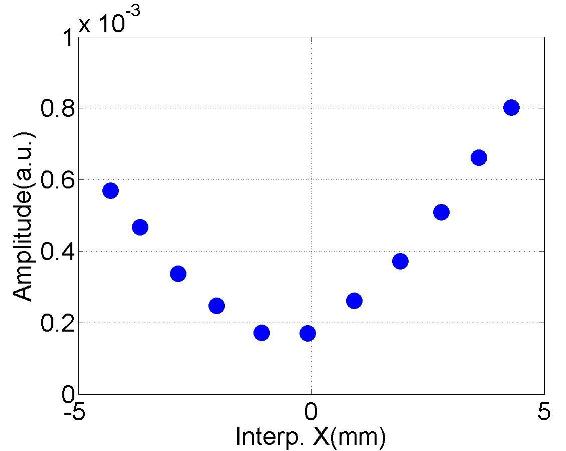}
\label{dep-C2H2-X-2}
}
\subfigure[\#3 ($f$:9.0249GHz; $Q$:10$^5$)]{
\includegraphics[width=0.23\textwidth]{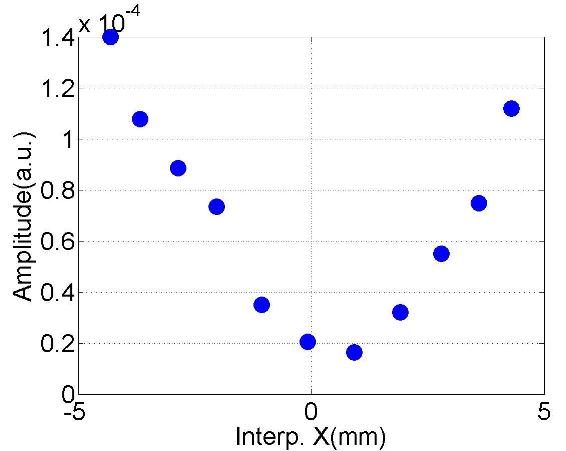}
\label{dep-C2H2-X-3}
}
\subfigure[\#4 ($f$:9.0329GHz; $Q$:10$^4$)]{
\includegraphics[width=0.23\textwidth]{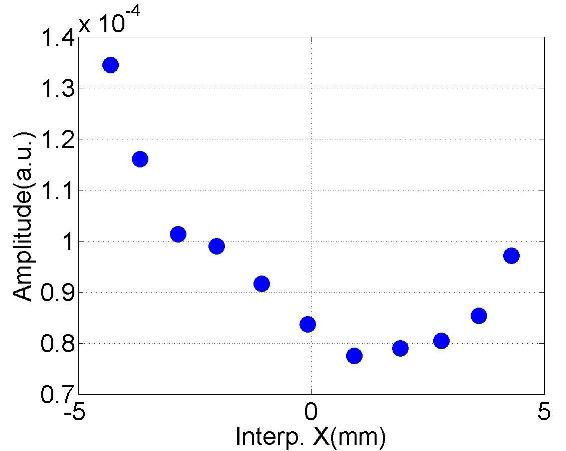}
\label{dep-C2H2-X-4}
}
\subfigure[\#5 ($f$:9.0345GHz; $Q$:10$^4$)]{
\includegraphics[width=0.23\textwidth]{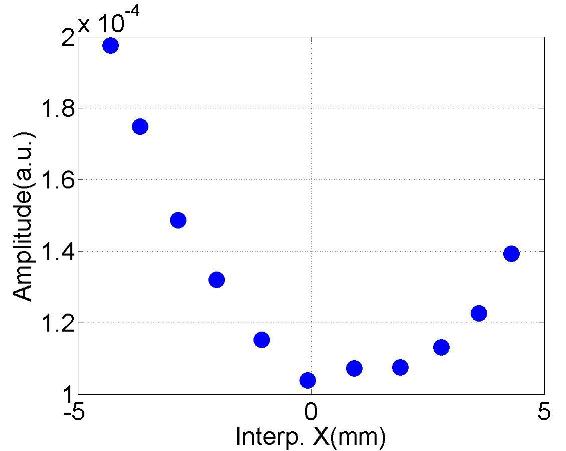}
\label{dep-C2H2-X-5}
}
\subfigure[\#6 ($f$:9.0422GHz; $Q$:10$^5$)]{
\includegraphics[width=0.23\textwidth]{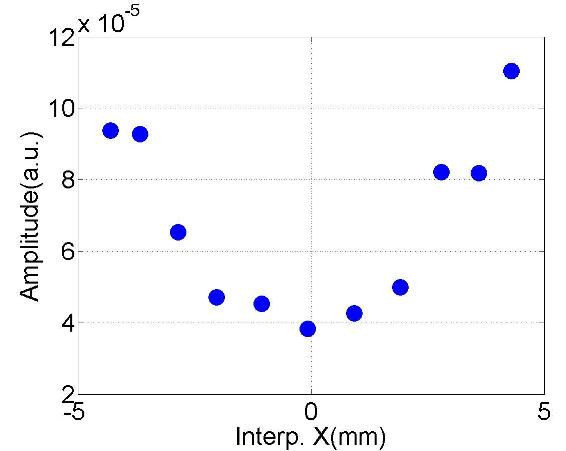}
\label{dep-C2H2-X-6}
}
\subfigure[\#7 ($f$:9.0437GHz; $Q$:10$^5$)]{
\includegraphics[width=0.23\textwidth]{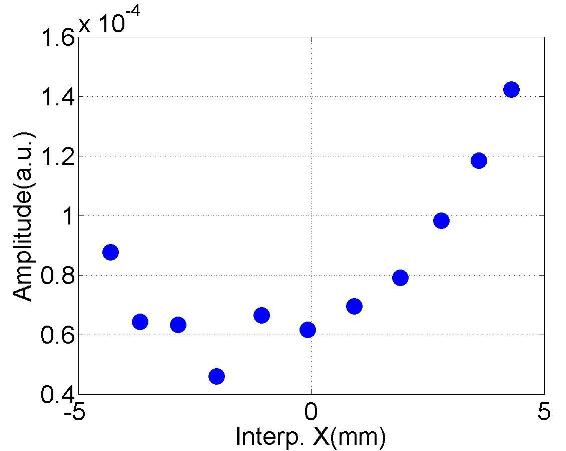}
\label{dep-C2H2-X-7}
}\\
\subfigure[\#1 ($f$:9.0069GHz; $Q$:10$^4$)]{
\includegraphics[width=0.23\textwidth]{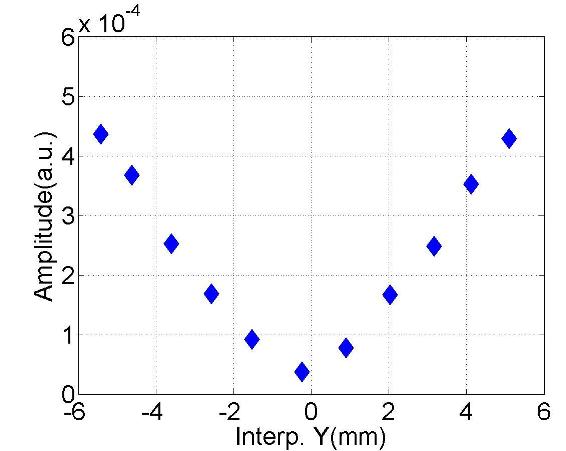}
\label{dep-C2H2-Y-1}
}
\subfigure[\#2 ($f$:9.0090GHz; $Q$:10$^4$)]{
\includegraphics[width=0.23\textwidth]{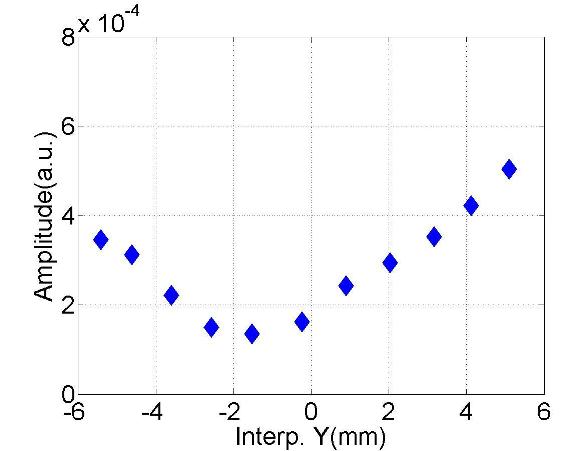}
\label{dep-C2H2-Y-2}
}
\subfigure[\#3 ($f$:9.0249GHz; $Q$:10$^5$)]{
\includegraphics[width=0.23\textwidth]{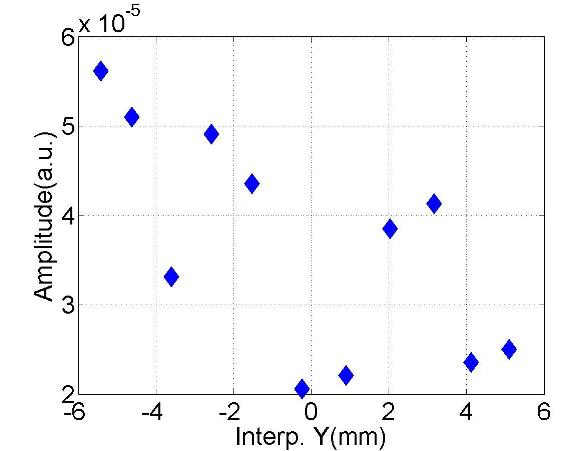}
\label{dep-C2H2-Y-3}
}
\subfigure[\#4 ($f$:9.0330GHz; $Q$:10$^4$)]{
\includegraphics[width=0.23\textwidth]{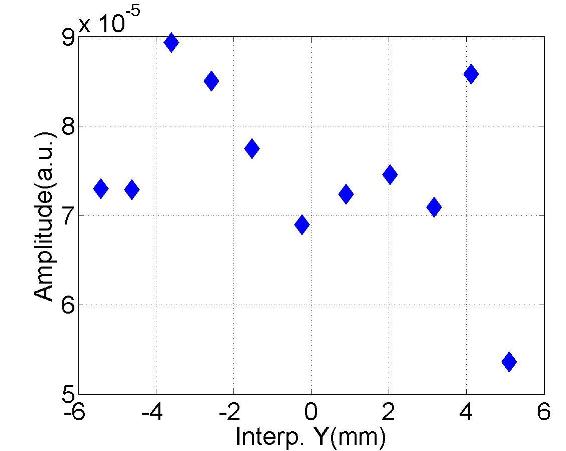}
\label{dep-C2H2-Y-4}
}
\subfigure[\#5 ($f$:9.0346GHz; $Q$:10$^4$)]{
\includegraphics[width=0.23\textwidth]{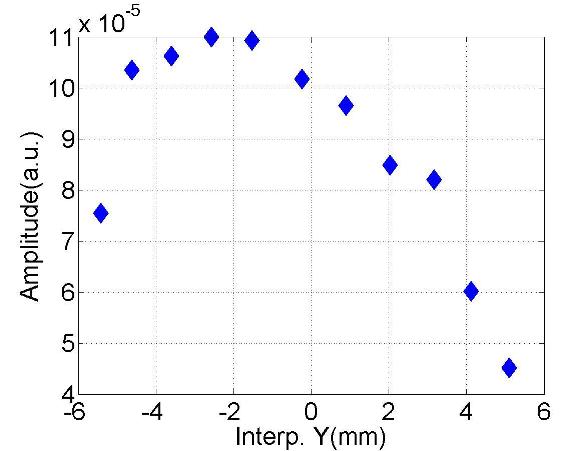}
\label{dep-C2H2-Y-5}
}
\subfigure[\#6 ($f$:9.0422GHz; $Q$:10$^5$)]{
\includegraphics[width=0.23\textwidth]{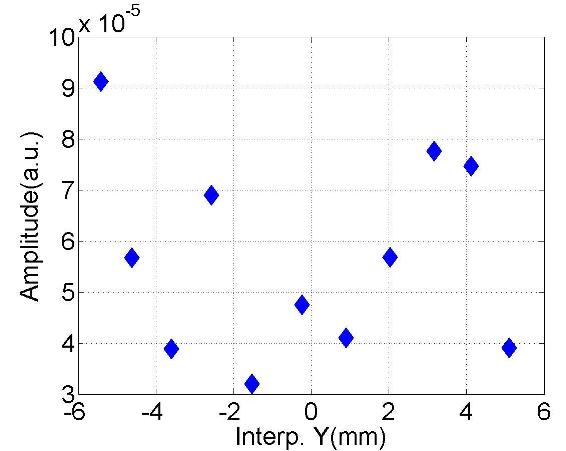}
\label{dep-C2H2-Y-6}
}
\subfigure[\#7 ($f$:9.0437GHz; $Q$:10$^5$)]{
\includegraphics[width=0.23\textwidth]{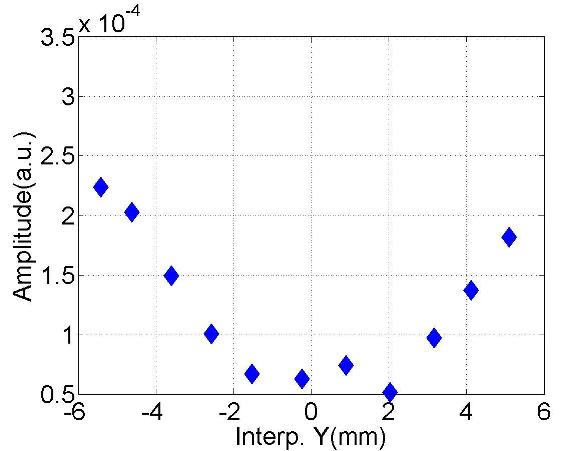}
\label{dep-C2H2-Y-7}
}
\caption{Dependence of the mode amplitude on the transverse beam of{}fset in the cavity.}
\label{spec-dep-C2H2-XY-1}
\end{figure}
\begin{figure}[h]
\subfigure[Spectrum (C2H2)]{
\includegraphics[width=1\textwidth]{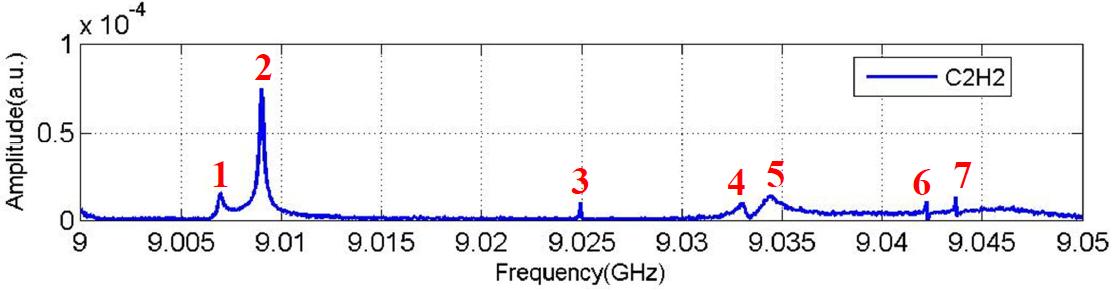}
\label{spec-C2H2-1}
}
\subfigure[\#1 ($f$:9.0069GHz; $Q$:10$^4$)]{
\includegraphics[width=0.3\textwidth]{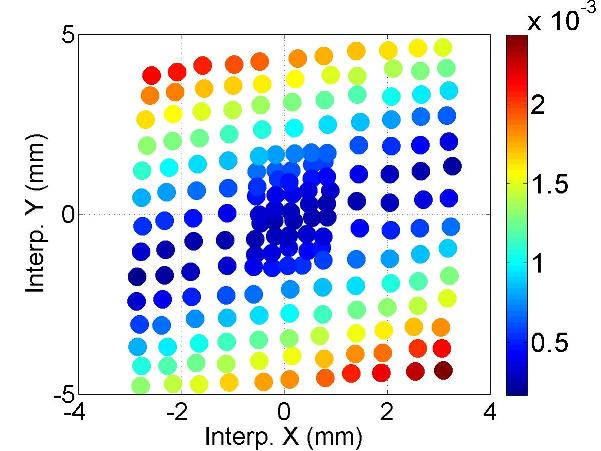}
\label{polar-C2H2-1}
}
\subfigure[\#2 ($f$:9.0090GHz; $Q$:10$^4$)]{
\includegraphics[width=0.3\textwidth]{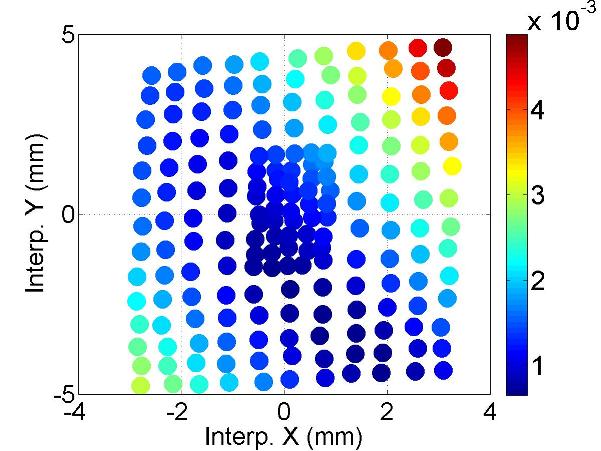}
\label{polar-C2H2-2}
}
\subfigure[\#3 ($f$:9.0249GHz; $Q$:10$^5$)]{
\includegraphics[width=0.3\textwidth]{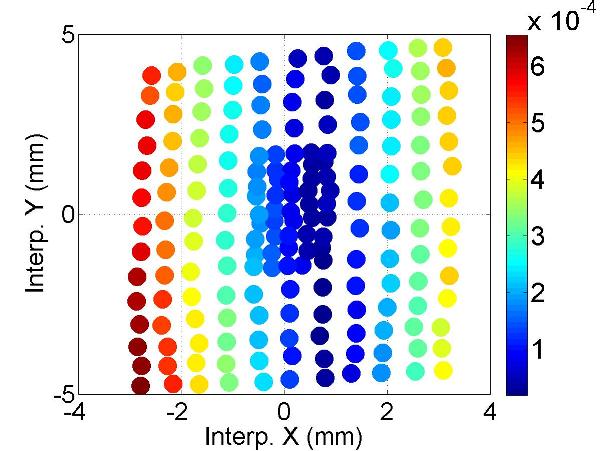}
\label{polar-C2H2-3}
}
\subfigure[\#4 ($f$:9.0330GHz; $Q$:10$^4$)]{
\includegraphics[width=0.3\textwidth]{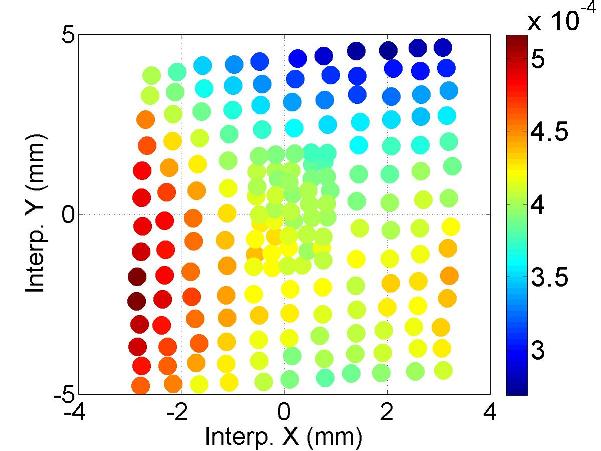}
\label{polar-C2H2-4}
}
\subfigure[\#5 ($f$:9.0345GHz; $Q$:10$^3$)]{
\includegraphics[width=0.3\textwidth]{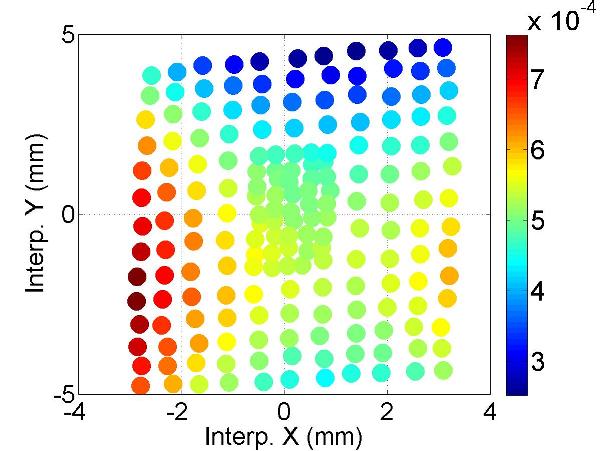}
\label{polar-C2H2-5}
}
\subfigure[\#6 ($f$:9.0422GHz; $Q$:10$^5$)]{
\includegraphics[width=0.3\textwidth]{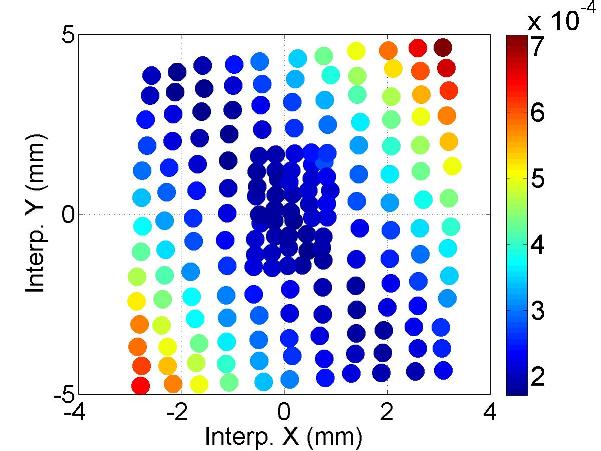}
\label{polar-C2H2-6}
}
\subfigure[\#7 ($f$:9.0437GHz; $Q$:10$^5$)]{
\includegraphics[width=0.3\textwidth]{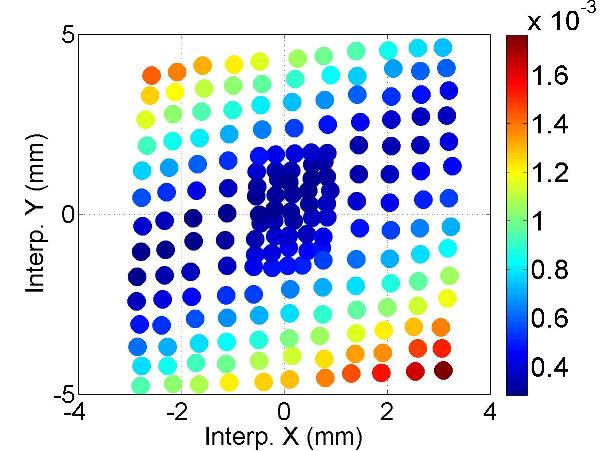}
\label{polar-C2H2-7}
}
\caption{Polarization of the mode.}
\label{spec-polar-C2H2-1}
\end{figure}
\begin{figure}[h]
\subfigure[Spectrum (C2H2)]{
\includegraphics[width=1\textwidth]{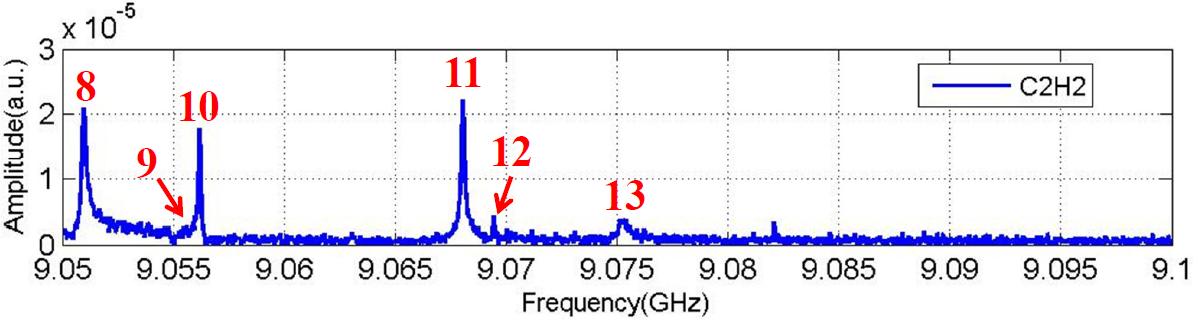}
\label{spec-C2H2-X-2}
}
\subfigure[\#8 ($f$:9.0510GHz; $Q$:10$^4$)]{
\includegraphics[width=0.23\textwidth]{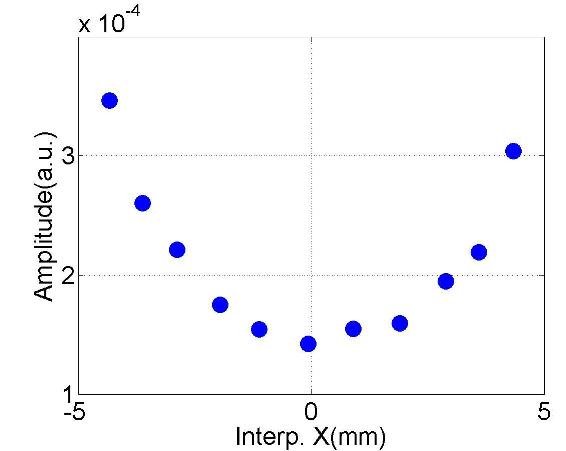}
\label{dep-C2H2-X-8}
}
\subfigure[\#9 ($f$:9.0551GHz; $Q$:10$^4$)]{
\includegraphics[width=0.23\textwidth]{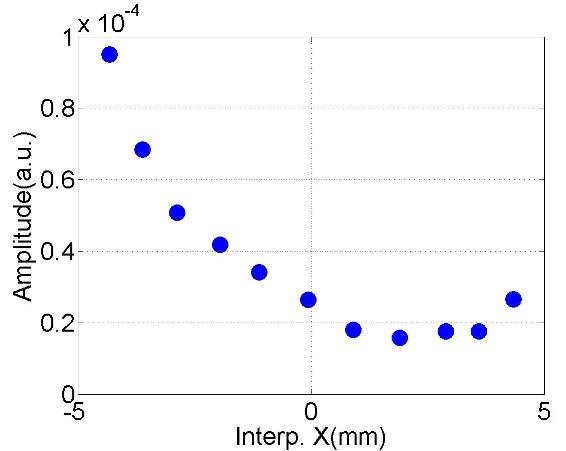}
\label{dep-C2H2-X-9}
}
\subfigure[\#10 ($f$:9.0562GHz; $Q$:10$^5$)]{
\includegraphics[width=0.23\textwidth]{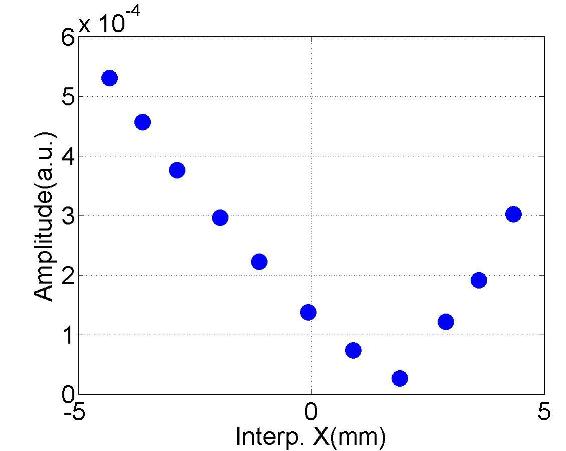}
\label{dep-C2H2-X-10}
}
\subfigure[\#11 ($f$:9.0681GHz; $Q$:10$^5$)]{
\includegraphics[width=0.23\textwidth]{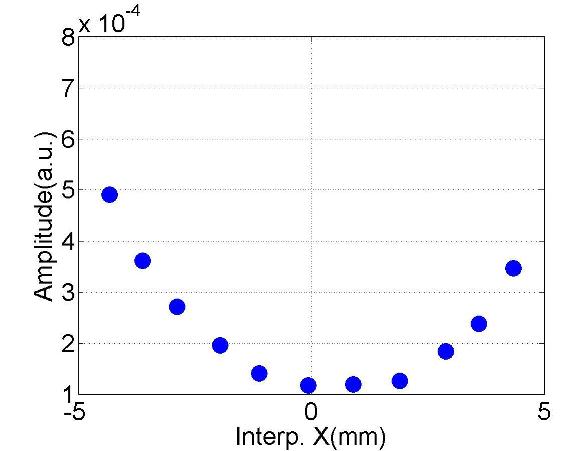}
\label{dep-C2H2-X-11}
}
\subfigure[\#12 ($f$:9.0695GHz; $Q$:10$^5$)]{
\includegraphics[width=0.23\textwidth]{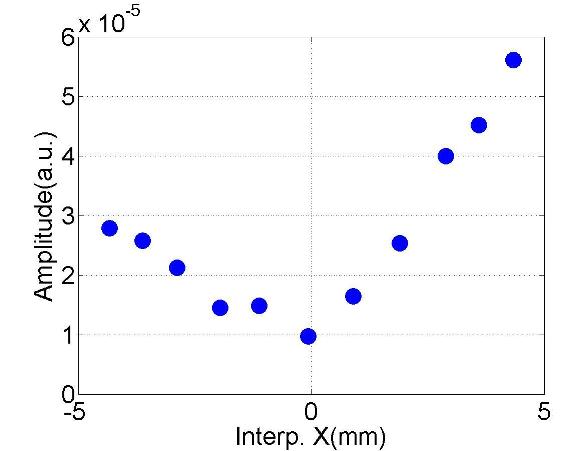}
\label{dep-C2H2-X-12}
}
\subfigure[\#13 ($f$:9.0754GHz; $Q$:10$^4$)]{
\includegraphics[width=0.23\textwidth]{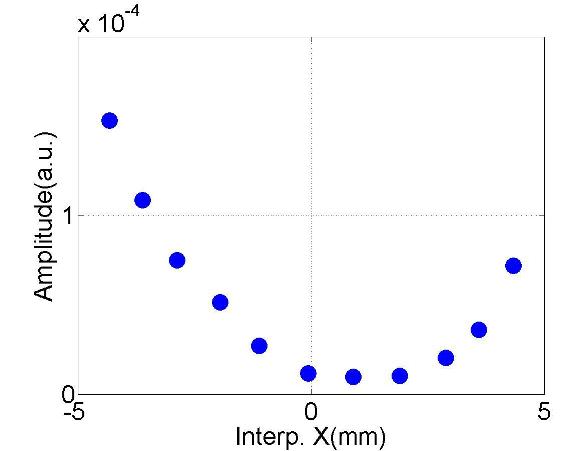}
\label{dep-C2H2-X-13}
}\\
\subfigure[\#8 ($f$:9.0510GHz; $Q$:10$^4$)]{
\includegraphics[width=0.23\textwidth]{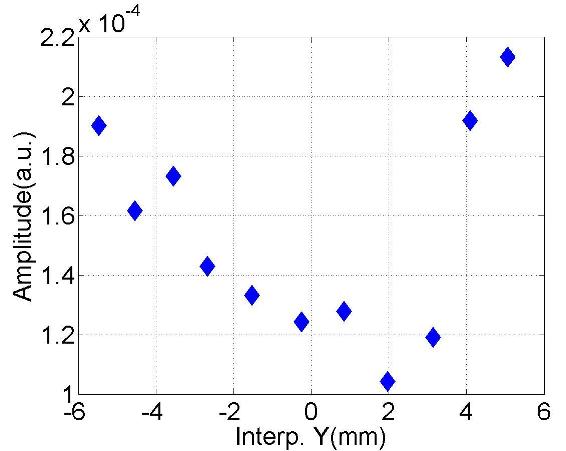}
\label{dep-C2H2-Y-8}
}
\subfigure[\#9 ($f$:9.0551GHz; $Q$:10$^4$)]{
\includegraphics[width=0.23\textwidth]{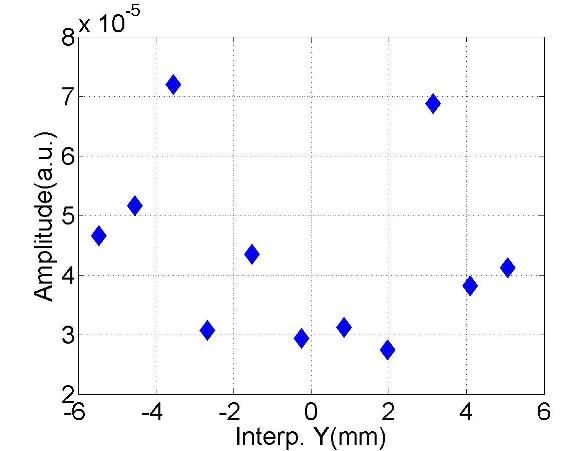}
\label{dep-C2H2-Y-9}
}
\subfigure[\#10 ($f$:9.0562GHz; $Q$:10$^5$)]{
\includegraphics[width=0.23\textwidth]{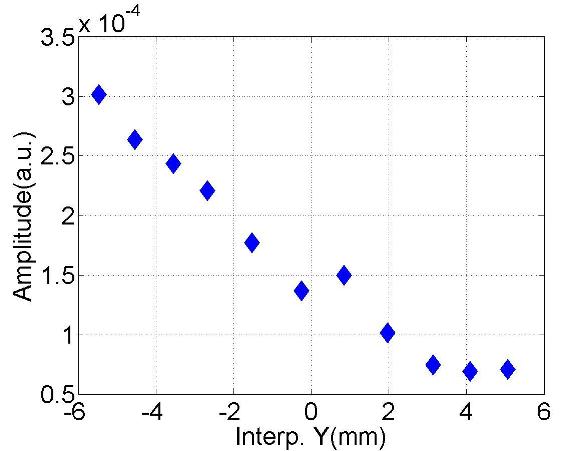}
\label{dep-C2H2-Y-10}
}
\subfigure[\#11 ($f$:9.0680GHz; $Q$:10$^5$)]{
\includegraphics[width=0.23\textwidth]{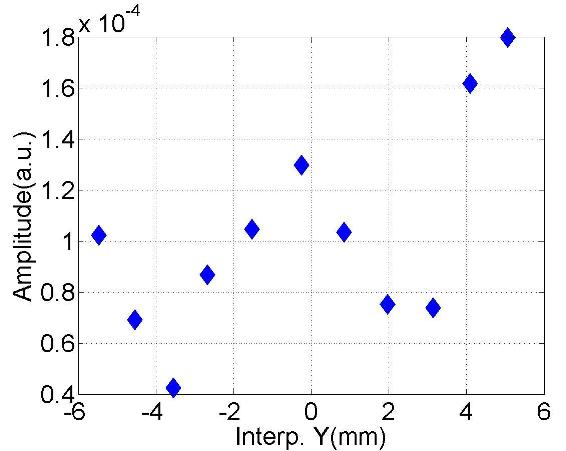}
\label{dep-C2H2-Y-11}
}
\subfigure[\#12 ($f$:9.0695GHz; $Q$:10$^5$)]{
\includegraphics[width=0.23\textwidth]{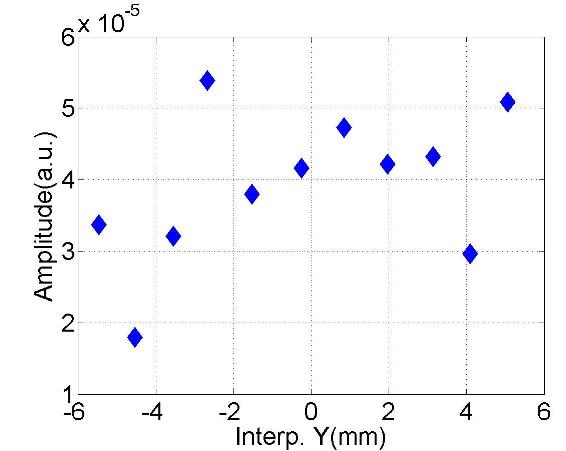}
\label{dep-C2H2-Y-12}
}
\subfigure[\#13 ($f$:9.0755GHz; $Q$:10$^4$)]{
\includegraphics[width=0.23\textwidth]{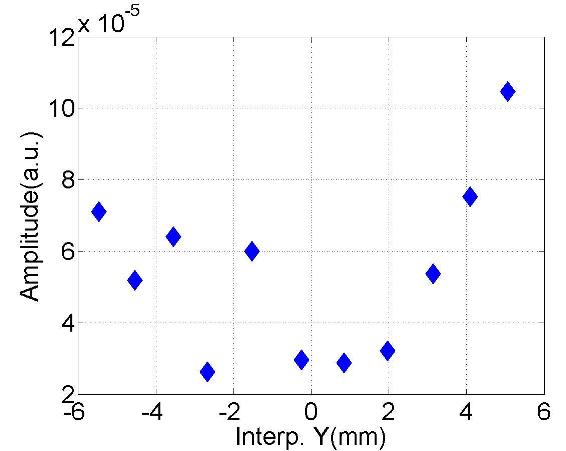}
\label{dep-C2H2-Y-13}
}
\caption{Dependence of the mode amplitude on the transverse beam of{}fset in the cavity.}
\label{spec-dep-C2H2-XY-2}
\end{figure}
\begin{figure}[h]
\subfigure[Spectrum (C2H2)]{
\includegraphics[width=1\textwidth]{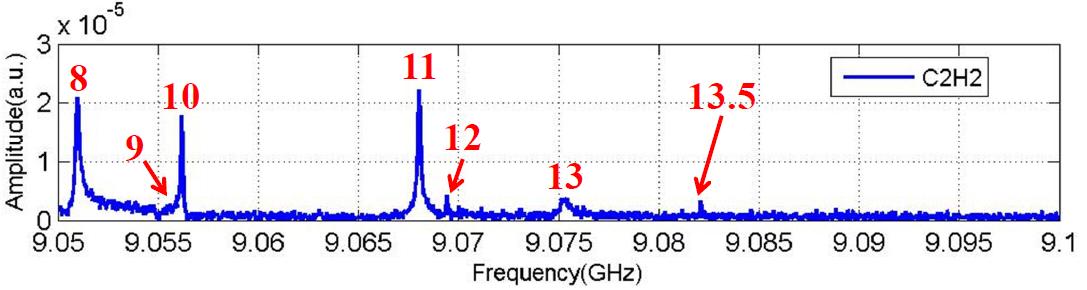}
\label{spec-C2H2-2}
}
\subfigure[\#8 ($f$:9.0510GHz; $Q$:10$^4$)]{
\includegraphics[width=0.31\textwidth]{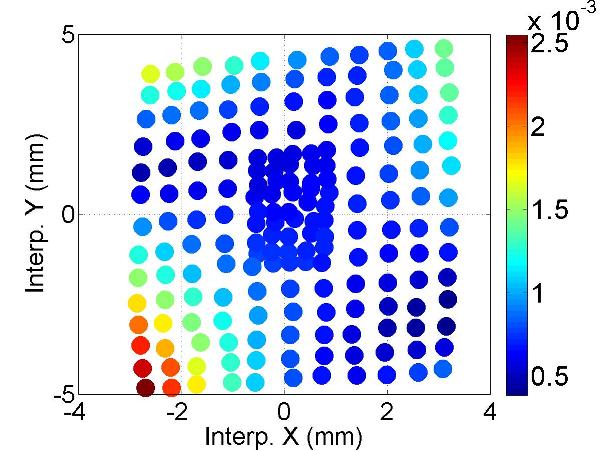}
\label{polar-C2H2-8}
}
\subfigure[\#9 ($f$:9.0550GHz; $Q$:10$^4$)]{
\includegraphics[width=0.31\textwidth]{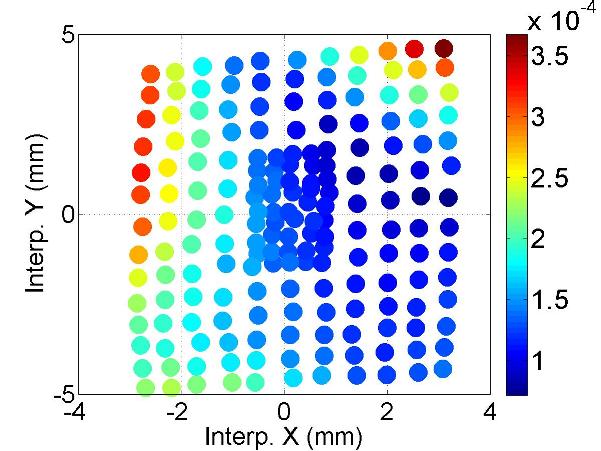}
\label{polar-C2H2-9}
}
\subfigure[\#10 ($f$:9.0562GHz; $Q$:10$^5$)]{
\includegraphics[width=0.31\textwidth]{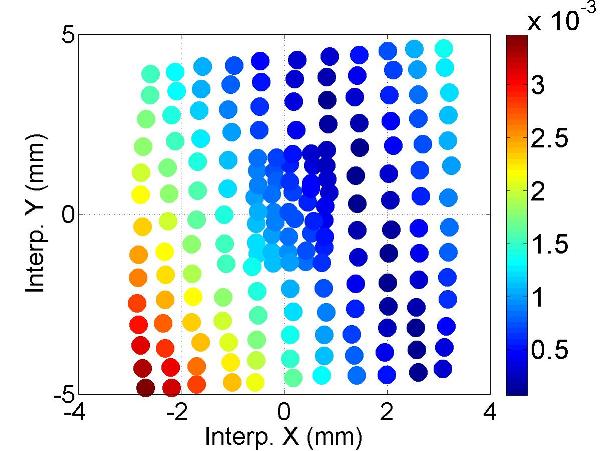}
\label{polar-C2H2-10}
}
\subfigure[\#11 ($f$:9.0681GHz; $Q$:10$^5$)]{
\includegraphics[width=0.31\textwidth]{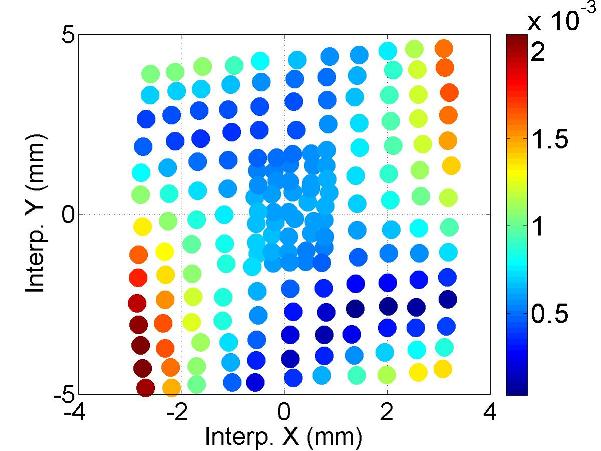}
\label{polar-C2H2-11}
}
\subfigure[\#12 ($f$:9.0695GHz; $Q$:10$^5$)]{
\includegraphics[width=0.31\textwidth]{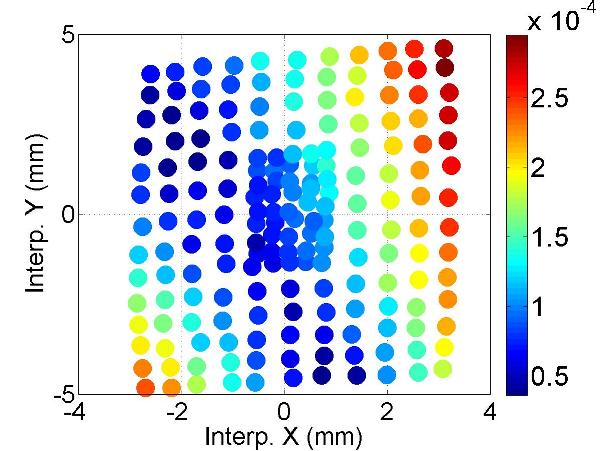}
\label{polar-C2H2-12}
}
\subfigure[\#13 ($f$:9.0754GHz; $Q$:10$^4$)]{
\includegraphics[width=0.31\textwidth]{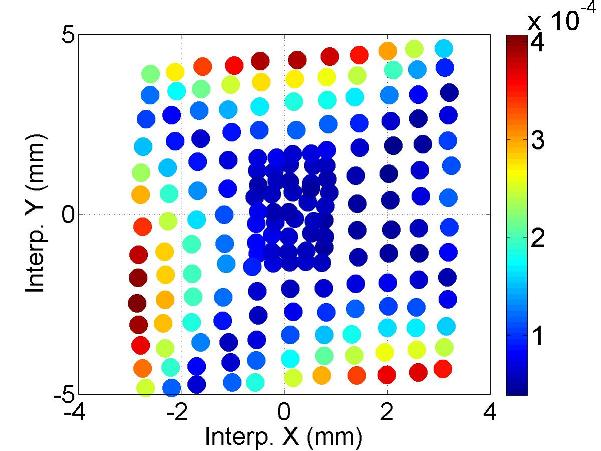}
\label{polar-C2H2-13}
}
\subfigure[\#13.5 ($f$:9.0821GHz; $Q$:10$^5$)]{
\includegraphics[width=0.31\textwidth]{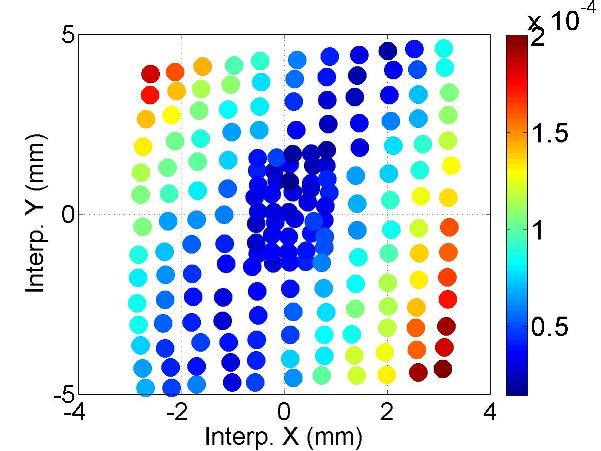}
\label{polar-C2H2-13_5}
}
\caption{Polarization of the mode.}
\label{spec-polar-C2H2-2}
\end{figure}

\FloatBarrier
\section{D5: HOM Coupler C3H1}
\begin{figure}[h]
\subfigure[Spectrum (C3H1)]{
\includegraphics[width=1\textwidth]{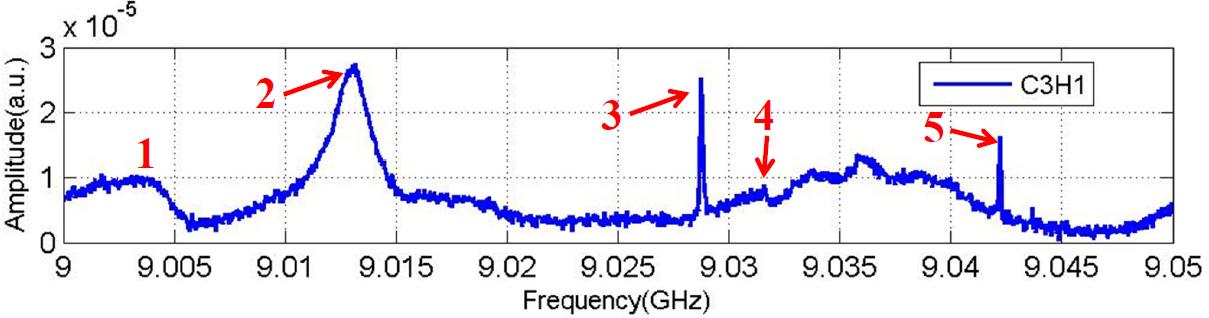}
\label{spec-C3H1-X-1}
}
\subfigure[\#1 ($f$:9.0020GHz; $Q$:10$^3$)]{
\includegraphics[width=0.23\textwidth]{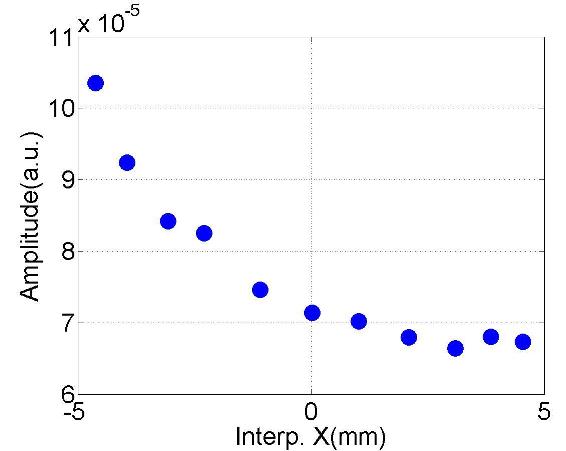}
\label{dep-C3H1-X-1}
}
\subfigure[\#2 ($f$:9.0132GHz; $Q$:10$^3$)]{
\includegraphics[width=0.23\textwidth]{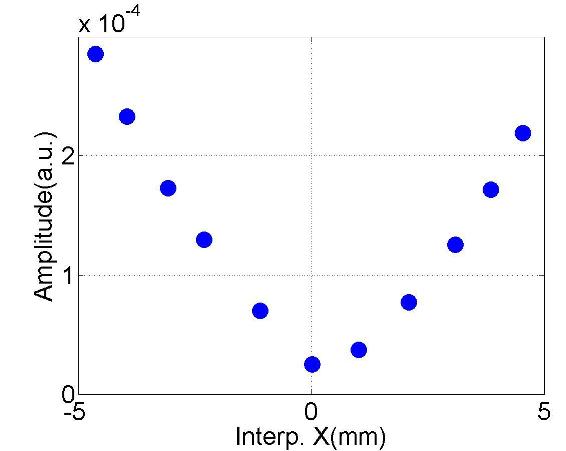}
\label{dep-C3H1-X-2}
}
\subfigure[\#3 ($f$:9.0288GHz; $Q$:10$^5$)]{
\includegraphics[width=0.23\textwidth]{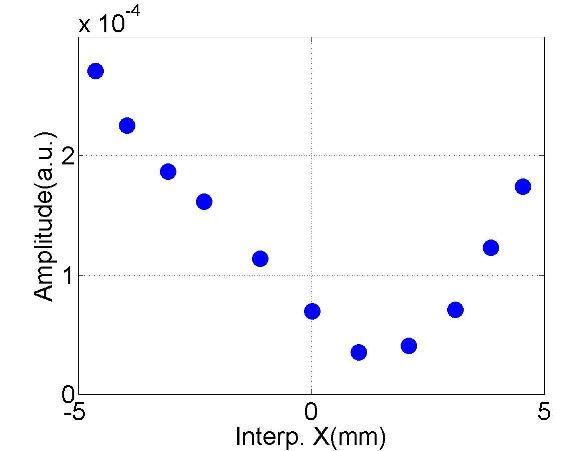}
\label{dep-C3H1-X-3}
}
\subfigure[\#4 ($f$:9.0316GHz; $Q$:10$^4$)]{
\includegraphics[width=0.23\textwidth]{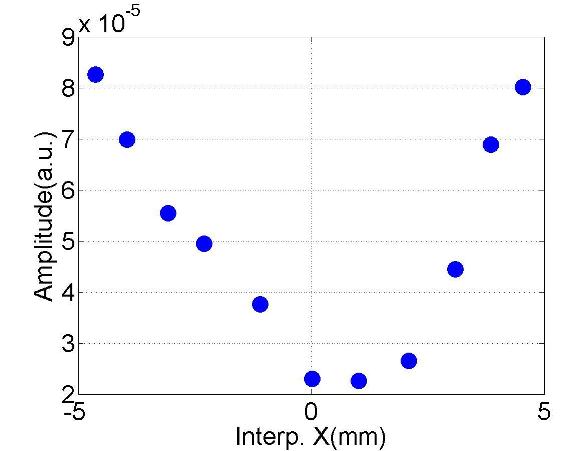}
\label{dep-C3H1-X-4}
}
\subfigure[\#5 ($f$:9.0422GHz; $Q$:10$^5$)]{
\includegraphics[width=0.23\textwidth]{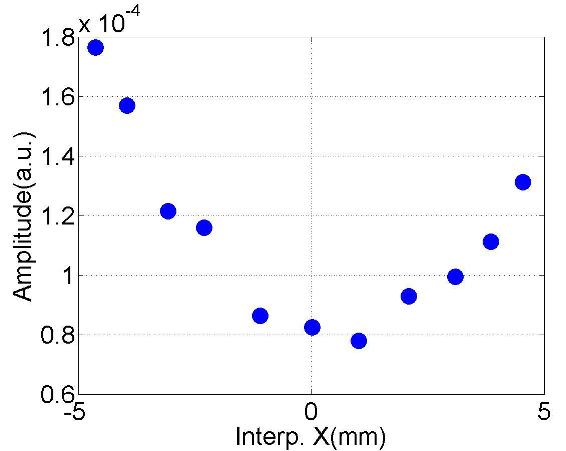}
\label{dep-C3H1-X-5}
}\\
\subfigure[\#1 ($f$:9.0034GHz; $Q$:10$^3$)]{
\includegraphics[width=0.23\textwidth]{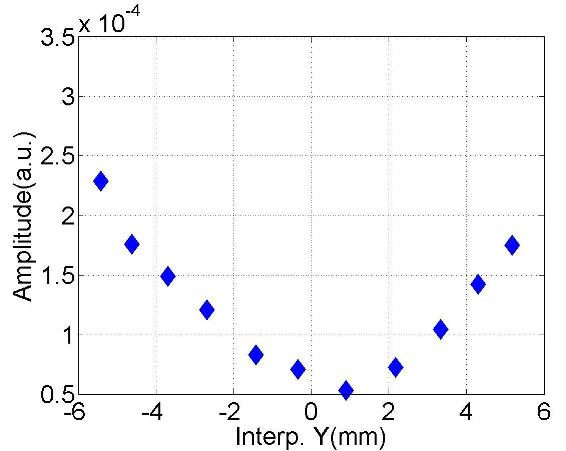}
\label{dep-C3H1-Y-1}
}
\subfigure[\#2 ($f$:9.0123GHz; $Q$:10$^3$)]{
\includegraphics[width=0.23\textwidth]{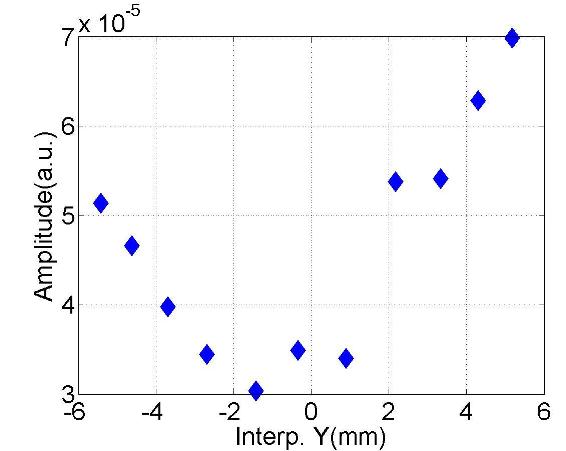}
\label{dep-C3H1-Y-2}
}
\subfigure[\#3 ($f$:9.0287GHz; $Q$:10$^4$)]{
\includegraphics[width=0.23\textwidth]{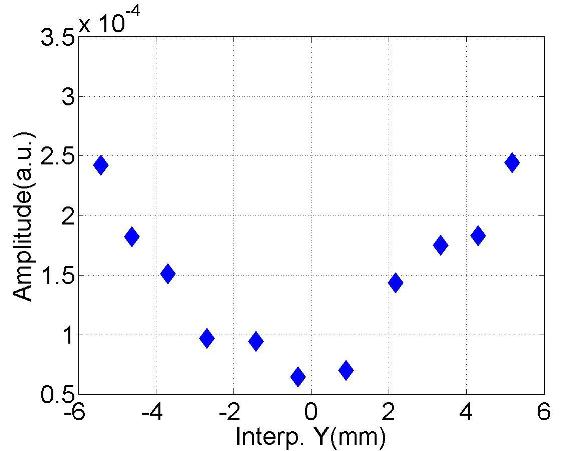}
\label{dep-C3H1-Y-3}
}
\subfigure[\#4 ($f$:9.0317GHz; $Q$:10$^4$)]{
\includegraphics[width=0.23\textwidth]{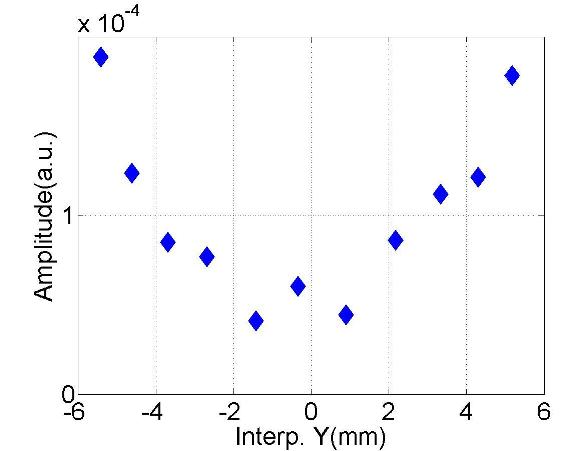}
\label{dep-C3H1-Y-4}
}
\subfigure[\#5 ($f$:9.0422GHz; $Q$:10$^5$)]{
\includegraphics[width=0.23\textwidth]{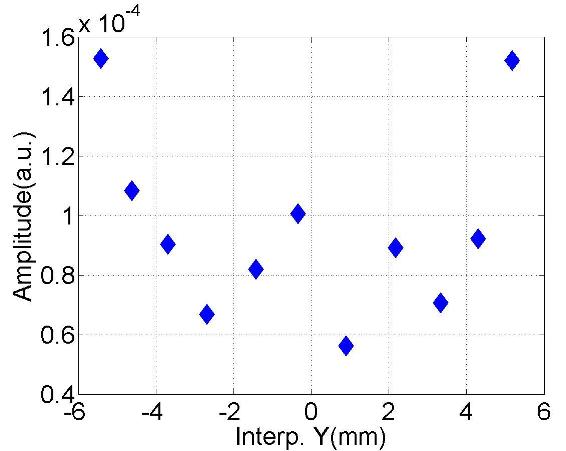}
\label{dep-C3H1-Y-5}
}
\caption{Dependence of the mode amplitude on the transverse beam of{}fset in the cavity.}
\label{spec-dep-C3H1-XY-1}
\end{figure}
\begin{figure}[h]
\subfigure[Spectrum (C3H1)]{
\includegraphics[width=1\textwidth]{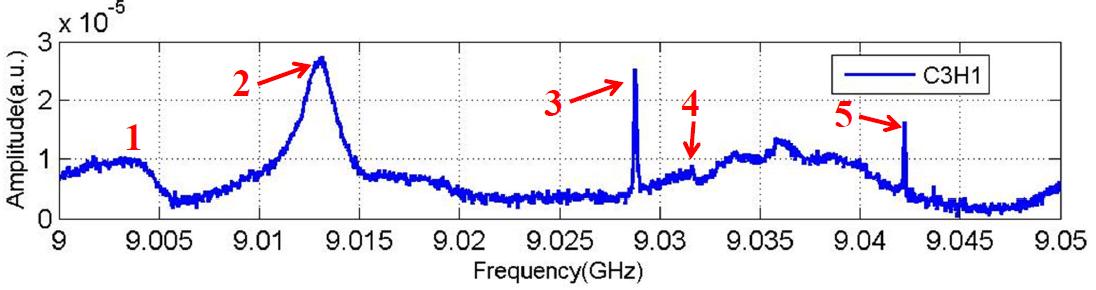}
\label{spec-C3H1-1}
}
\subfigure[\#1 ($f$:9.032GHz; $Q$:10$^3$)]{
\includegraphics[width=0.31\textwidth]{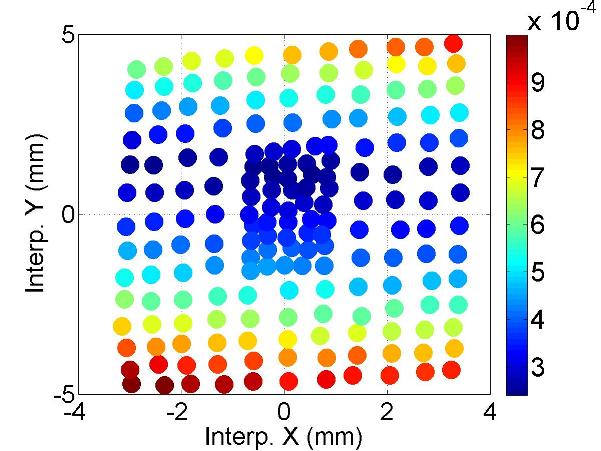}
\label{polar-C3H1-1}
}
\subfigure[\#2 ($f$:9.0129GHz; $Q$:10$^3$)]{
\includegraphics[width=0.31\textwidth]{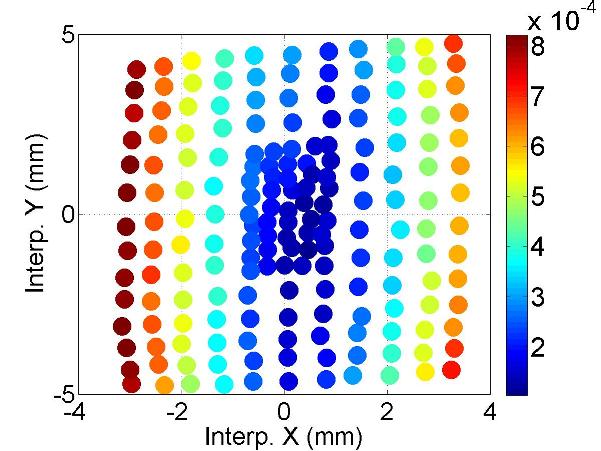}
\label{polar-C3H1-2}
}
\subfigure[\#3 ($f$:9.0287GHz; $Q$:10$^5$)]{
\includegraphics[width=0.31\textwidth]{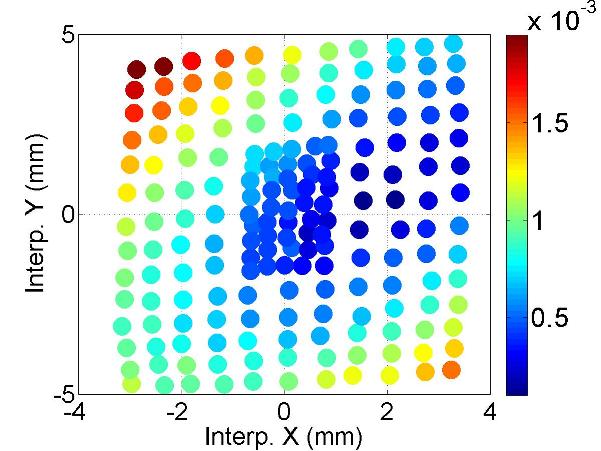}
\label{polar-C3H1-3}
}
\subfigure[\#4 ($f$:9.0317GHz; $Q$:10$^4$)]{
\includegraphics[width=0.31\textwidth]{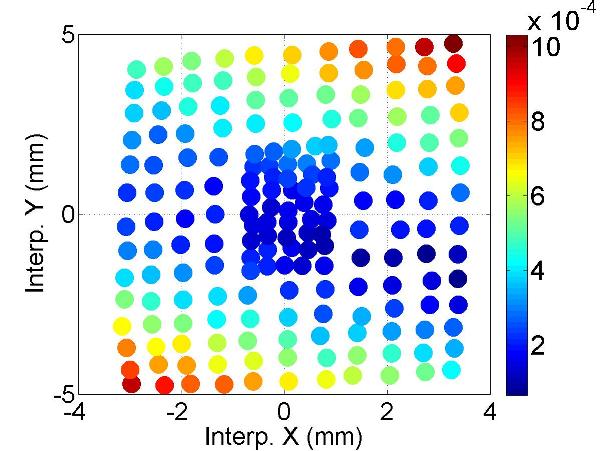}
\label{polar-C3H1-4}
}
\subfigure[\#5 ($f$:9.0422GHz; $Q$:10$^5$)]{
\includegraphics[width=0.31\textwidth]{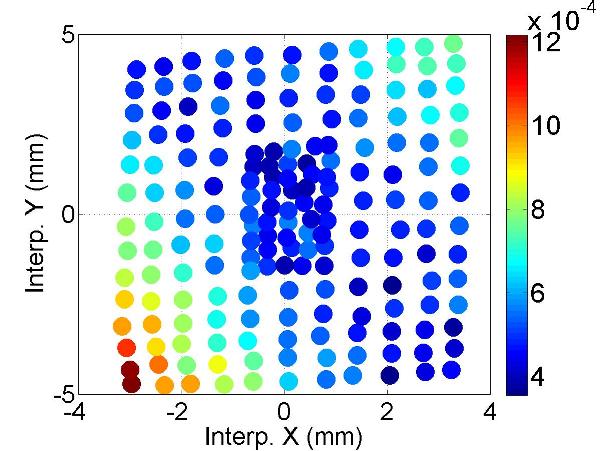}
\label{polar-C3H1-5}
}
\caption{Polarization of the mode.}
\label{spec-polar-C3H1-1}
\end{figure}
\begin{figure}[h]
\subfigure[Spectrum (C3H1)]{
\includegraphics[width=1\textwidth]{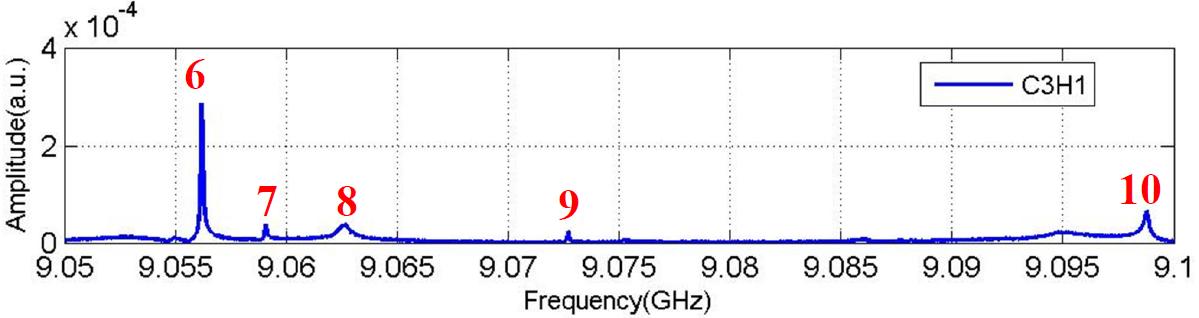}
\label{spec-C3H1-X-2}
}
\subfigure[\#6 ($f$:9.0562GHz; $Q$:10$^5$)]{
\includegraphics[width=0.23\textwidth]{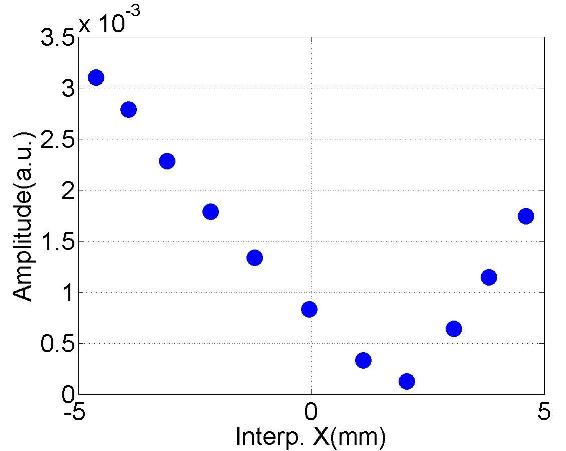}
\label{dep-C3H1-X-6}
}
\subfigure[\#7 ($f$:9.0591GHz; $Q$:10$^5$)]{
\includegraphics[width=0.23\textwidth]{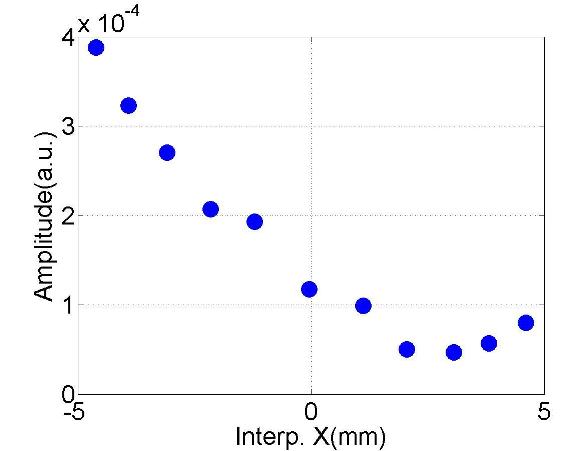}
\label{dep-C3H1-X-7}
}
\subfigure[\#8 ($f$:9.0626GHz; $Q$:10$^4$)]{
\includegraphics[width=0.23\textwidth]{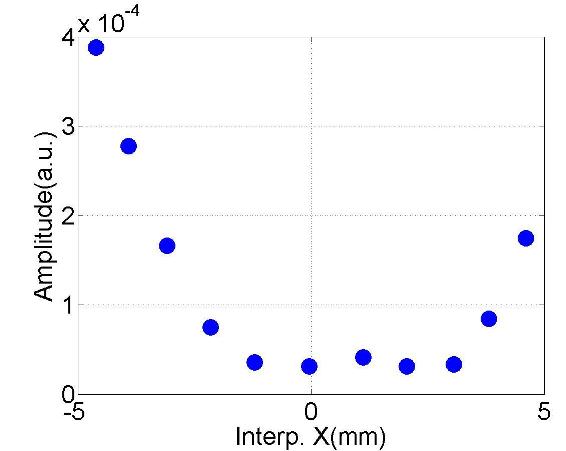}
\label{dep-C3H1-X-8}
}
\subfigure[\#9 ($f$:9.0727GHz; $Q$:10$^5$)]{
\includegraphics[width=0.23\textwidth]{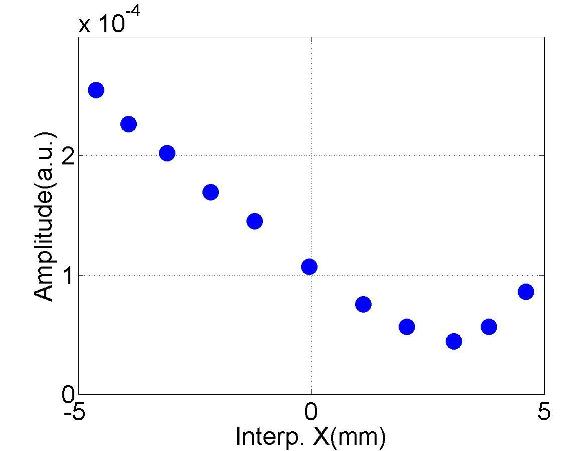}
\label{dep-C3H1-X-9}
}
\subfigure[\#10 ($f$:9.0986GHz; $Q$:10$^4$)]{
\includegraphics[width=0.23\textwidth]{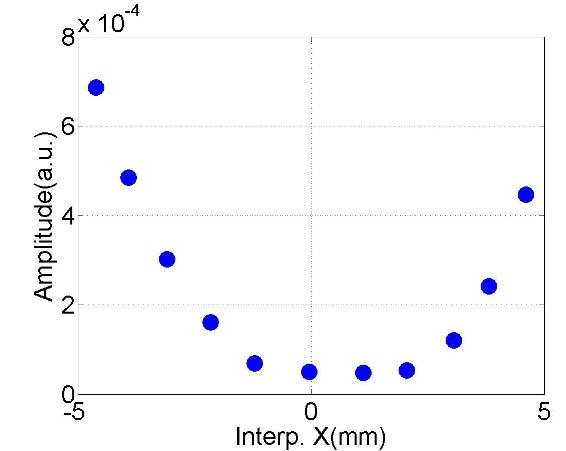}
\label{dep-C3H1-X-10}
}\\
\subfigure[\#6 ($f$:9.0562GHz; $Q$:10$^5$)]{
\includegraphics[width=0.23\textwidth]{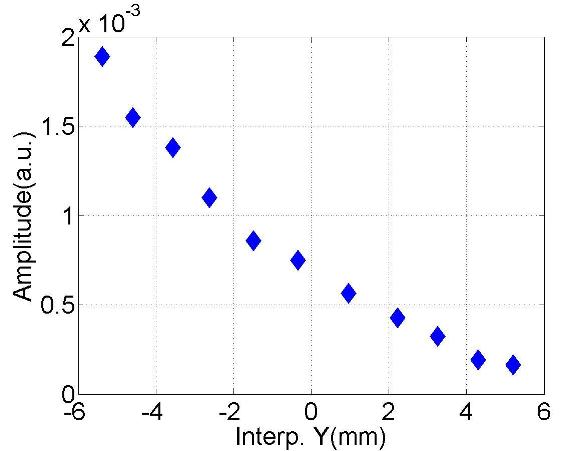}
\label{dep-C3H1-Y-6}
}
\subfigure[\#7 ($f$:9.0591GHz; $Q$:10$^5$)]{
\includegraphics[width=0.23\textwidth]{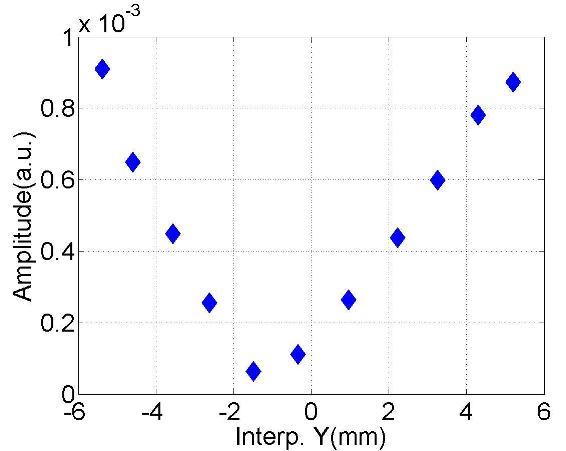}
\label{dep-C3H1-Y-7}
}
\subfigure[\#8 ($f$:9.0627GHz; $Q$:10$^4$)]{
\includegraphics[width=0.23\textwidth]{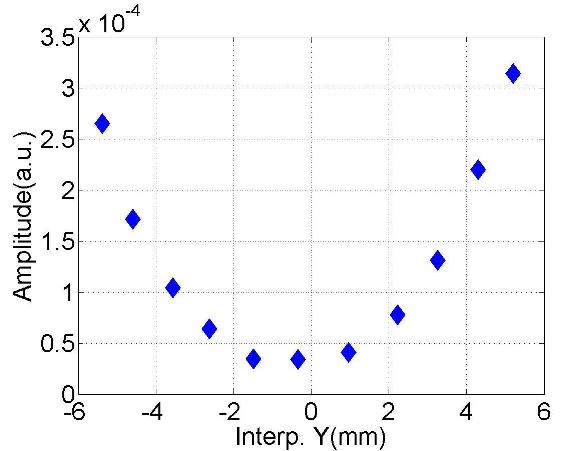}
\label{dep-C3H1-Y-8}
}
\subfigure[\#9 ($f$:9.0727GHz; $Q$:10$^5$)]{
\includegraphics[width=0.23\textwidth]{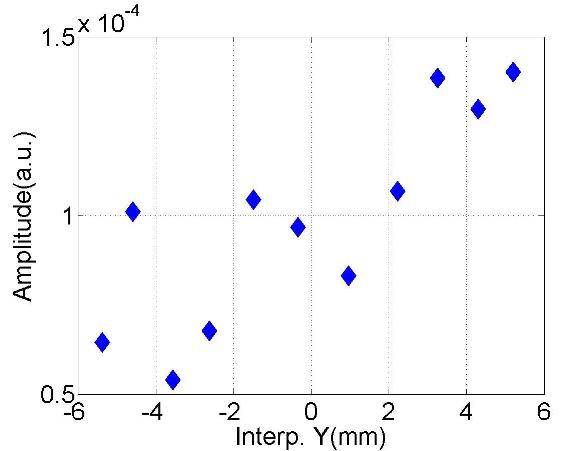}
\label{dep-C3H1-Y-9}
}
\subfigure[\#10 ($f$:9.0986GHz; $Q$:10$^4$)]{
\includegraphics[width=0.23\textwidth]{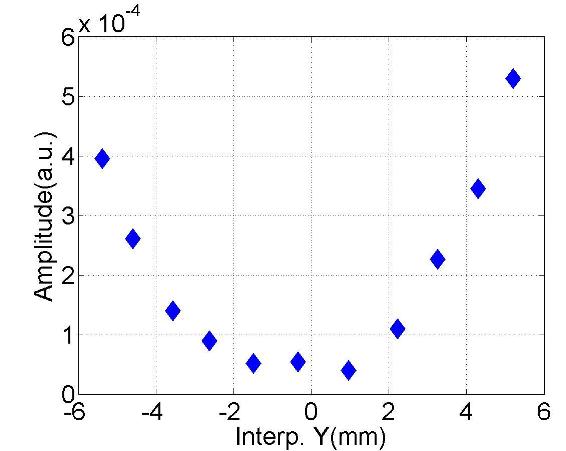}
\label{dep-C3H1-Y-10}
}
\caption{Dependence of the mode amplitude on the transverse beam of{}fset in the cavity.}
\label{spec-dep-C3H1-XY-2}
\end{figure}
\begin{figure}[h]
\subfigure[Spectrum (C3H1)]{
\includegraphics[width=1\textwidth]{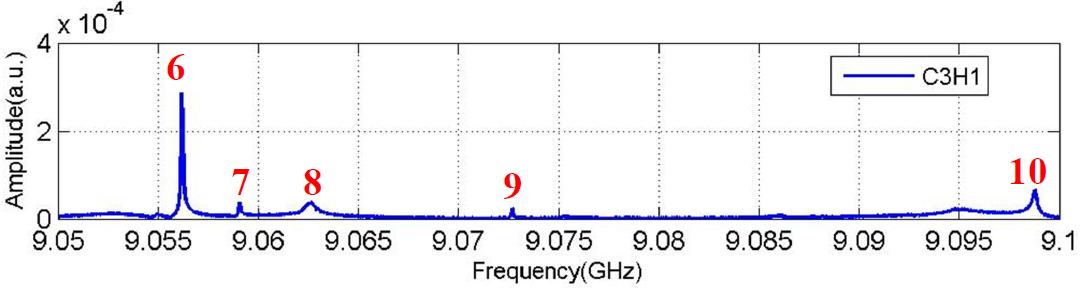}
\label{spec-C3H1-2}
}
\subfigure[\#6 ($f$:9.0562GHz; $Q$:10$^5$)]{
\includegraphics[width=0.31\textwidth]{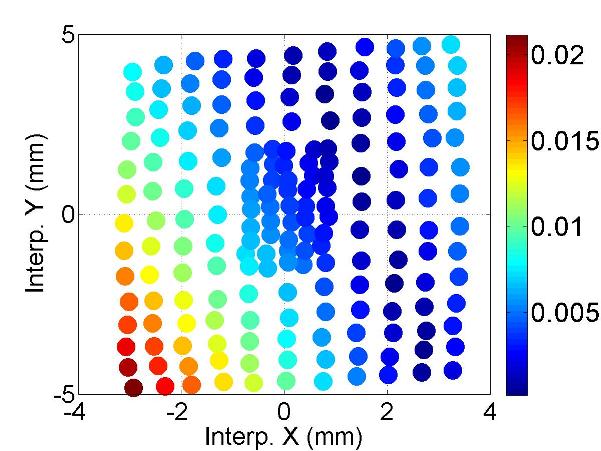}
\label{polar-C3H1-6}
}
\subfigure[\#7 ($f$:9.0591GHz; $Q$:10$^5$)]{
\includegraphics[width=0.31\textwidth]{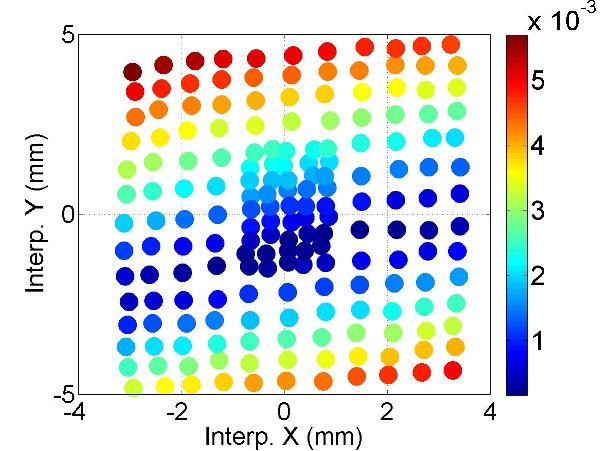}
\label{polar-C3H1-7}
}
\subfigure[\#8 ($f$:9.0628GHz; $Q$:10$^4$)]{
\includegraphics[width=0.31\textwidth]{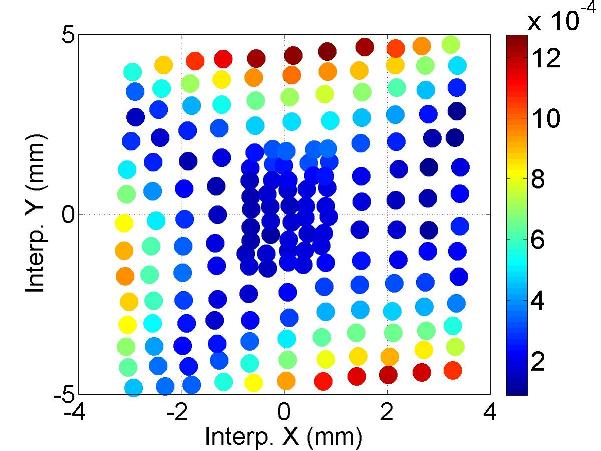}
\label{polar-C3H1-8}
}
\subfigure[\#9 ($f$:9.0727GHz; $Q$:10$^5$)]{
\includegraphics[width=0.31\textwidth]{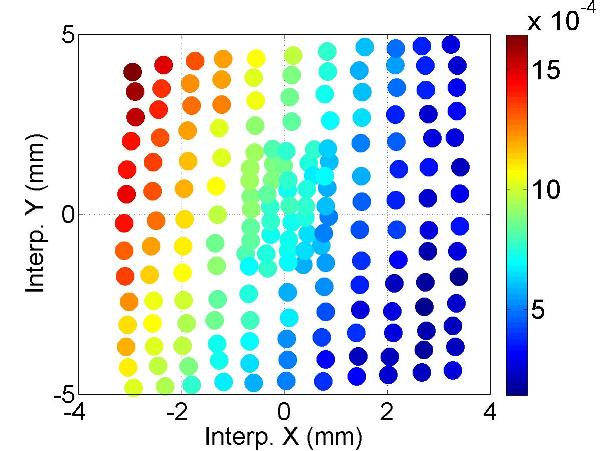}
\label{polar-C3H1-9}
}
\subfigure[\#10 ($f$:9.0986GHz; $Q$:10$^4$)]{
\includegraphics[width=0.31\textwidth]{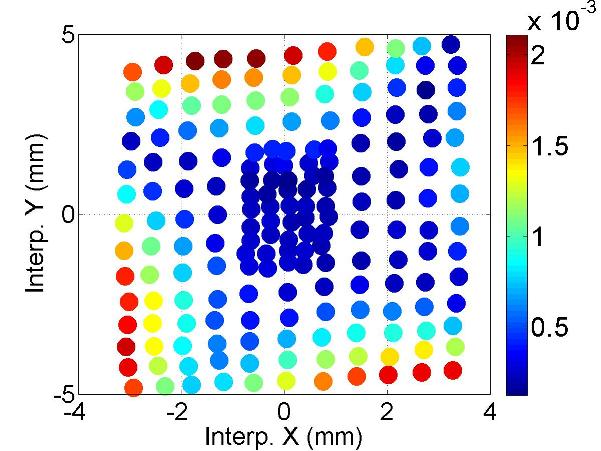}
\label{polar-C3H1-10}
}
\caption{Polarization of the mode.}
\label{spec-polar-C3H1-2}
\end{figure}

\FloatBarrier
\section{D5: HOM Coupler C3H2}
\begin{figure}[h]
\subfigure[Spectrum (C3H2)]{
\includegraphics[width=1\textwidth]{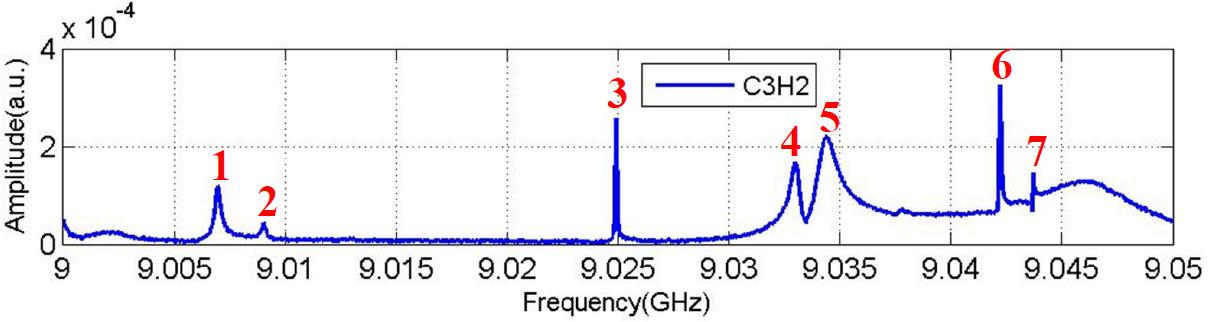}
\label{spec-C3H2-X-1}
}
\subfigure[\#1 ($f$:9.0070GHz; $Q$:10$^4$)]{
\includegraphics[width=0.23\textwidth]{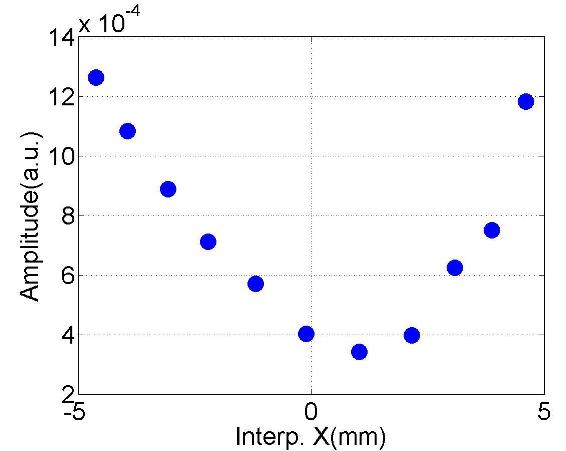}
\label{dep-C3H2-X-1}
}
\subfigure[\#2 ($f$:9.0090GHz; $Q$:10$^4$)]{
\includegraphics[width=0.23\textwidth]{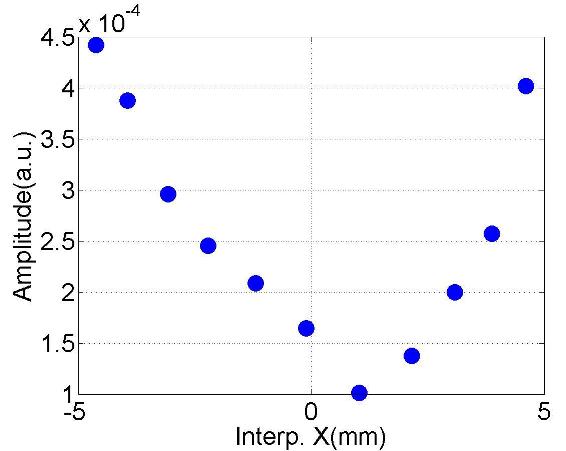}
\label{dep-C3H2-X-2}
}
\subfigure[\#3 ($f$:9.0249GHz; $Q$:10$^5$)]{
\includegraphics[width=0.23\textwidth]{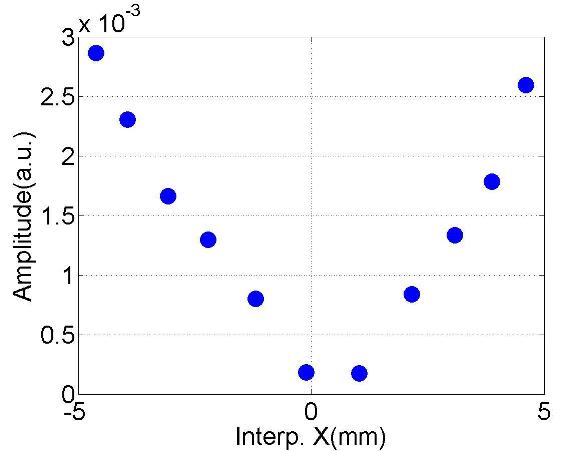}
\label{dep-C3H2-X-3}
}
\subfigure[\#4 ($f$:9.0330GHz; $Q$:10$^4$)]{
\includegraphics[width=0.23\textwidth]{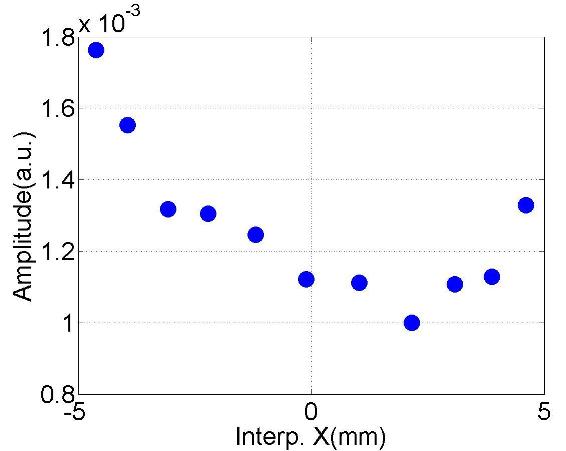}
\label{dep-C3H2-X-4}
}
\subfigure[\#5 ($f$:9.0346GHz; $Q$:10$^4$)]{
\includegraphics[width=0.23\textwidth]{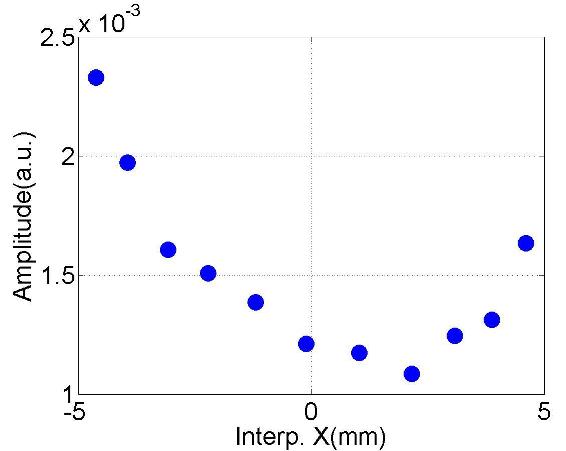}
\label{dep-C3H2-X-5}
}
\subfigure[\#6 ($f$:9.0422GHz; $Q$:10$^5$)]{
\includegraphics[width=0.23\textwidth]{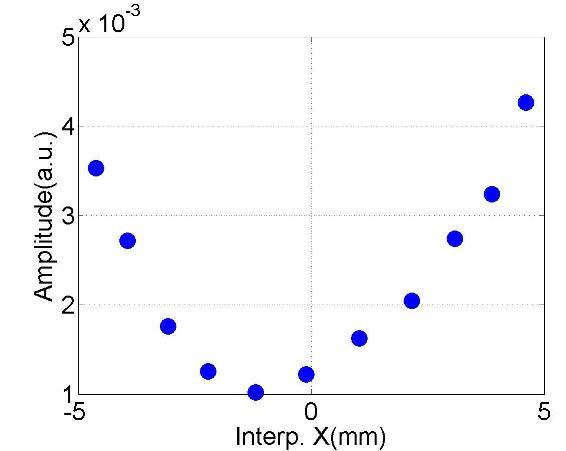}
\label{dep-C3H2-X-6}
}
\subfigure[\#7 ($f$:9.0437GHz; $Q$:10$^5$)]{
\includegraphics[width=0.23\textwidth]{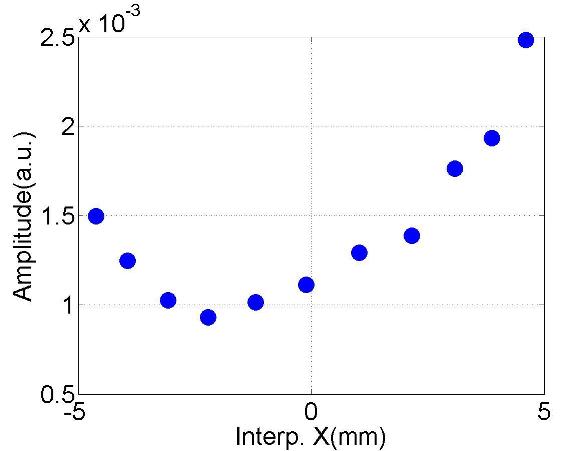}
\label{dep-C3H2-X-7}
}\\
\subfigure[\#1 ($f$:9.0070GHz; $Q$:10$^4$)]{
\includegraphics[width=0.23\textwidth]{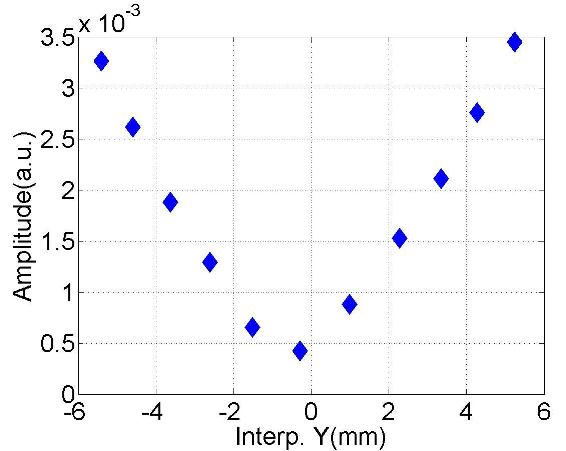}
\label{dep-C3H2-Y-1}
}
\subfigure[\#2 ($f$:9.0089GHz; $Q$:10$^4$)]{
\includegraphics[width=0.23\textwidth]{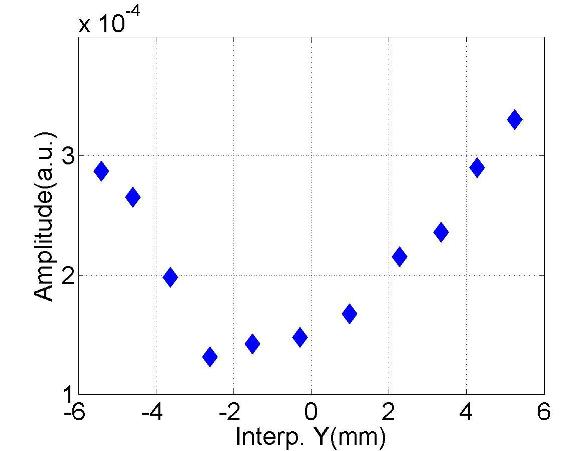}
\label{dep-C3H2-Y-2}
}
\subfigure[\#3 ($f$:9.0249GHz; $Q$:10$^5$)]{
\includegraphics[width=0.23\textwidth]{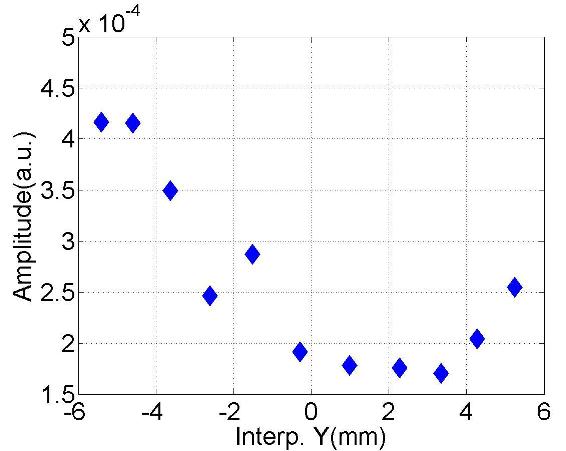}
\label{dep-C3H2-Y-3}
}
\subfigure[\#4 ($f$:9.0330GHz; $Q$:10$^4$)]{
\includegraphics[width=0.23\textwidth]{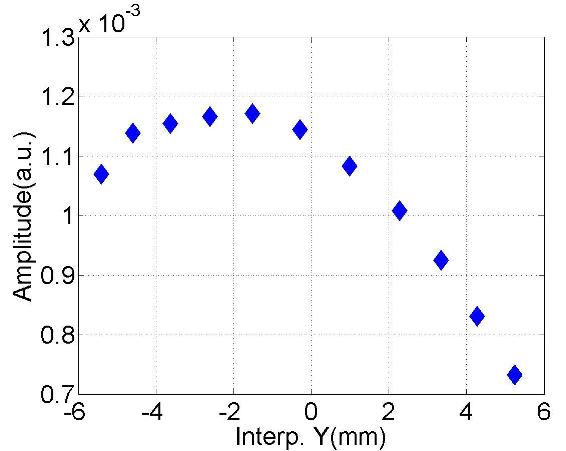}
\label{dep-C3H2-Y-4}
}
\subfigure[\#5 ($f$:9.0347GHz; $Q$:10$^3$)]{
\includegraphics[width=0.23\textwidth]{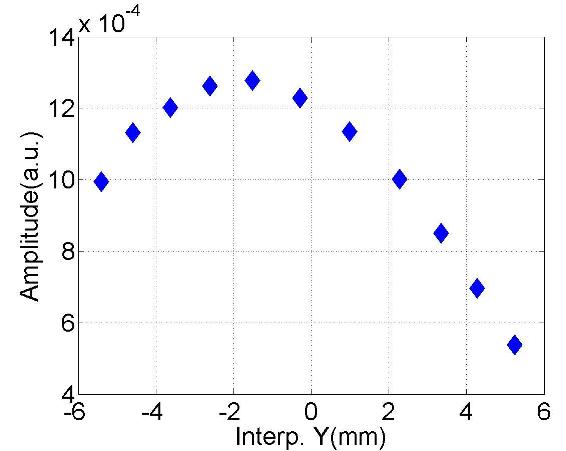}
\label{dep-C3H2-Y-5}
}
\subfigure[\#6 ($f$:9.0422GHz; $Q$:10$^5$)]{
\includegraphics[width=0.23\textwidth]{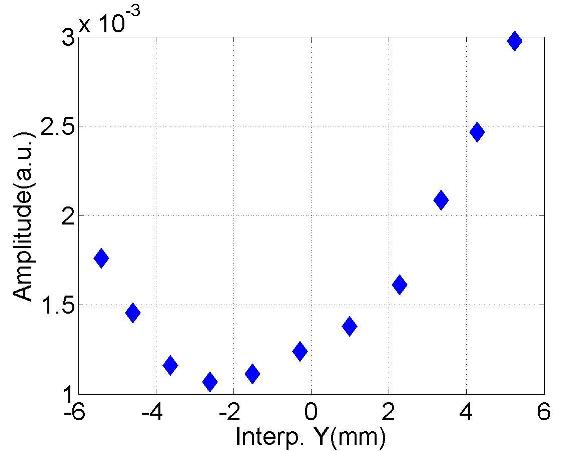}
\label{dep-C3H2-Y-6}
}
\subfigure[\#7 ($f$:9.0437GHz; $Q$:10$^5$)]{
\includegraphics[width=0.23\textwidth]{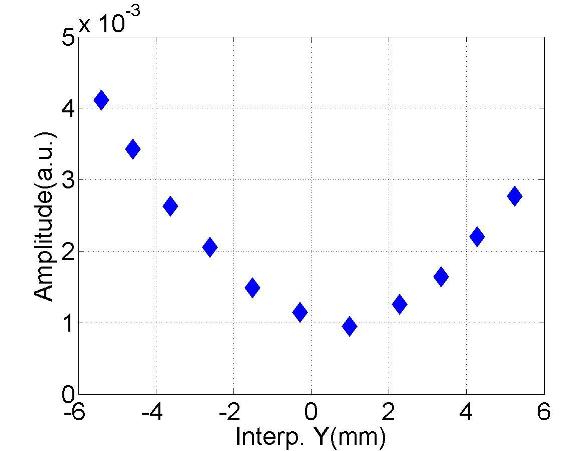}
\label{dep-C3H2-Y-7}
}
\caption{Dependence of the mode amplitude on the transverse beam of{}fset in the cavity.}
\label{spec-dep-C3H2-XY-1}
\end{figure}
\begin{figure}[h]
\subfigure[Spectrum (C3H2)]{
\includegraphics[width=1\textwidth]{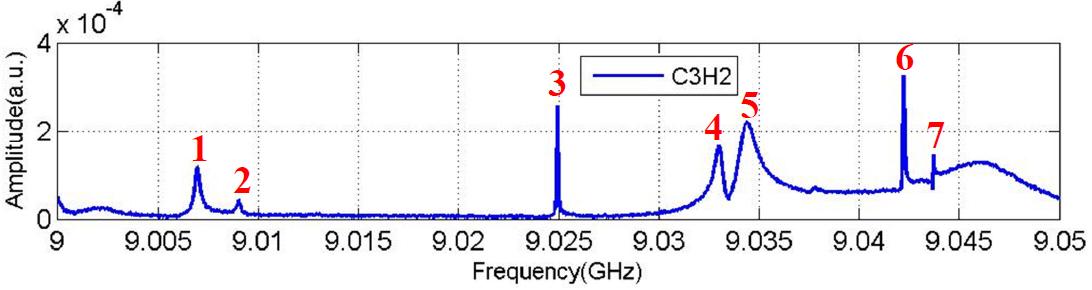}
\label{spec-C3H2-1}
}
\subfigure[\#1 ($f$:9.070GHz; $Q$:10$^4$)]{
\includegraphics[width=0.31\textwidth]{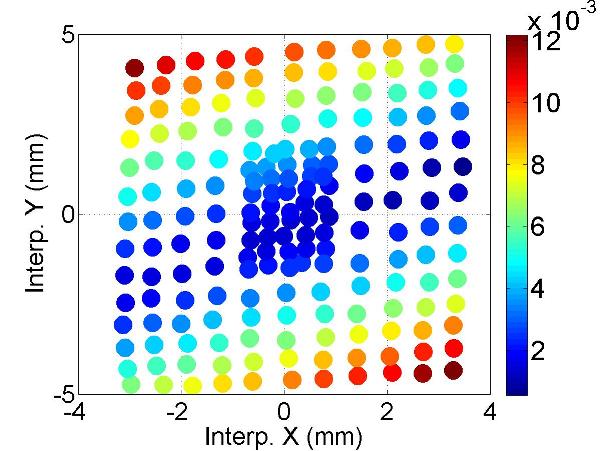}
\label{polar-C3H2-1}
}
\subfigure[\#2 ($f$:9.0090GHz; $Q$:10$^4$)]{
\includegraphics[width=0.31\textwidth]{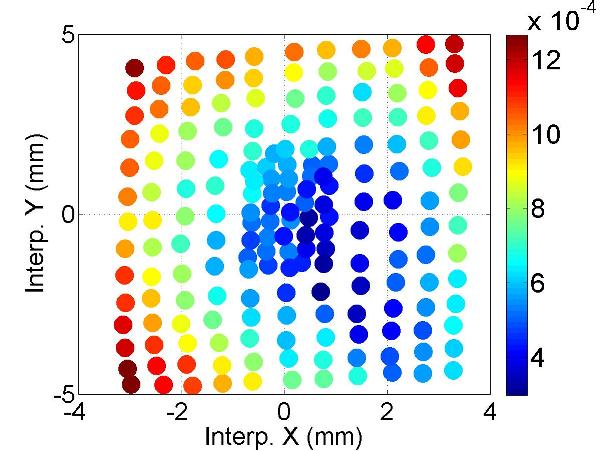}
\label{polar-C3H2-2}
}
\subfigure[\#3 ($f$:9.0249GHz; $Q$:10$^5$)]{
\includegraphics[width=0.31\textwidth]{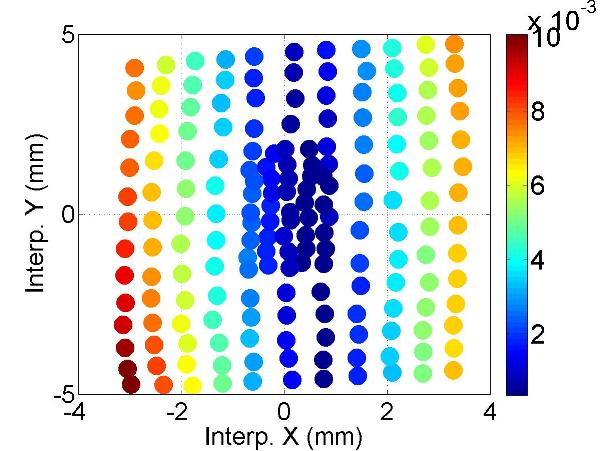}
\label{polar-C3H2-3}
}
\subfigure[\#4 ($f$:9.0330GHz; $Q$:10$^4$)]{
\includegraphics[width=0.31\textwidth]{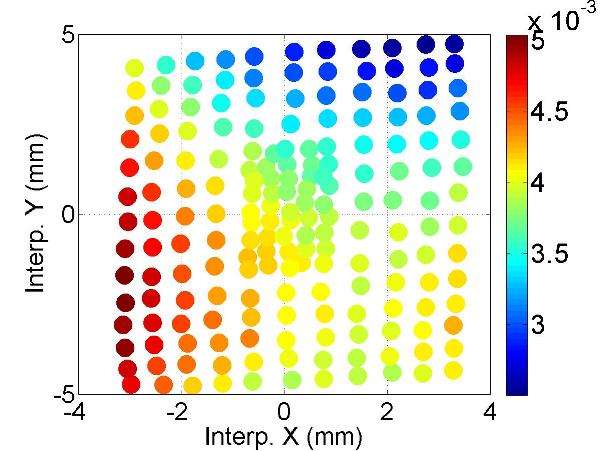}
\label{polar-C3H2-4}
}
\subfigure[\#5 ($f$:9.0347GHz; $Q$:10$^3$)]{
\includegraphics[width=0.31\textwidth]{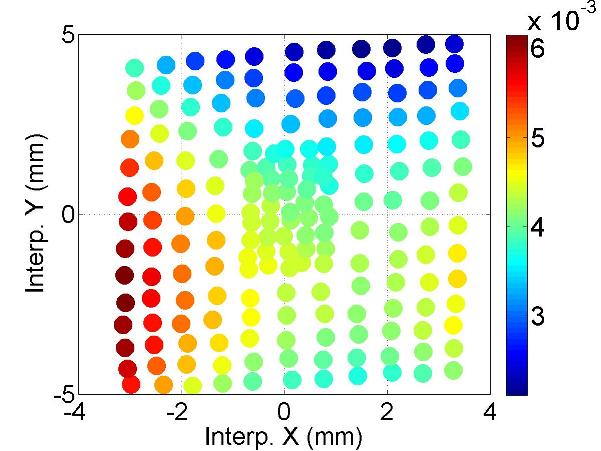}
\label{polar-C3H2-5}
}
\subfigure[\#6 ($f$:9.0422GHz; $Q$:10$^5$)]{
\includegraphics[width=0.31\textwidth]{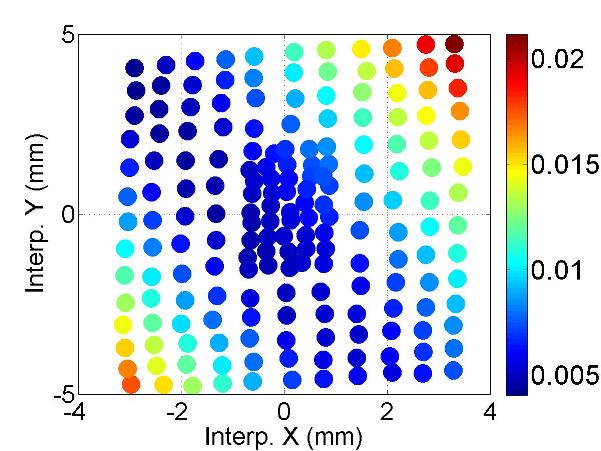}
\label{polar-C3H2-6}
}
\subfigure[\#7 ($f$:9.0437GHz; $Q$:10$^5$)]{
\includegraphics[width=0.31\textwidth]{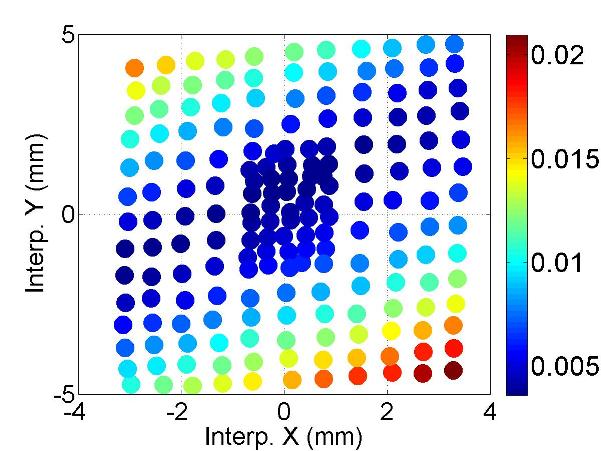}
\label{polar-C3H2-7}
}
\caption{Polarization of the mode.}
\label{spec-polar-C3H2-1}
\end{figure}
\begin{figure}[h]
\subfigure[Spectrum (C3H2)]{
\includegraphics[width=1\textwidth]{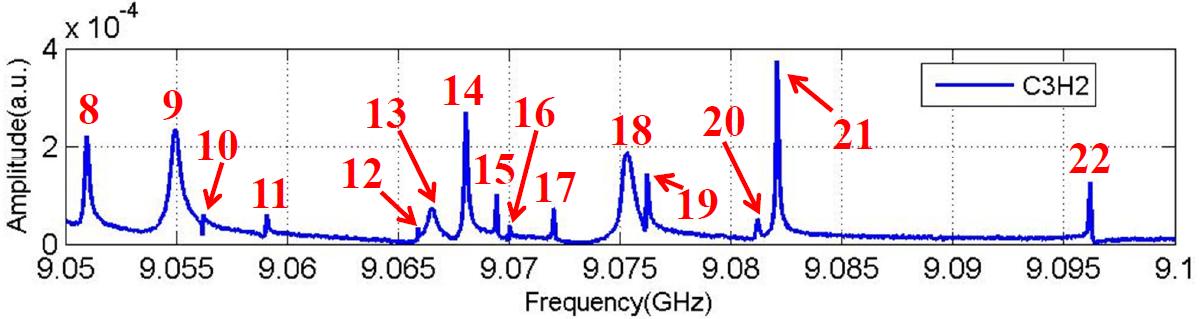}
\label{spec-C3H2-X-2}
}
\subfigure[\#8 ($f$:9.0510GHz; $Q$:10$^4$)]{
\includegraphics[width=0.23\textwidth]{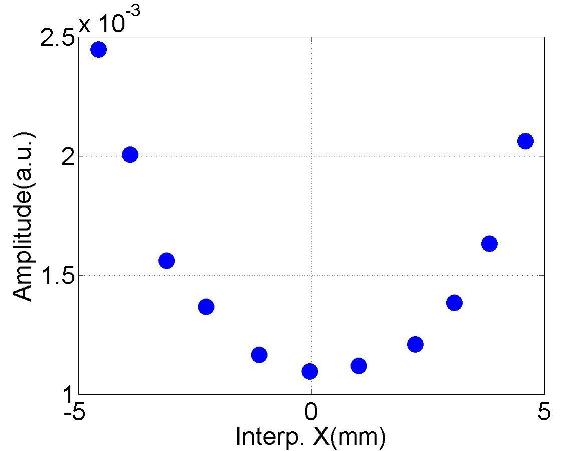}
\label{dep-C3H2-X-8}
}
\subfigure[\#9 ($f$:9.0553GHz; $Q$:10$^4$)]{
\includegraphics[width=0.23\textwidth]{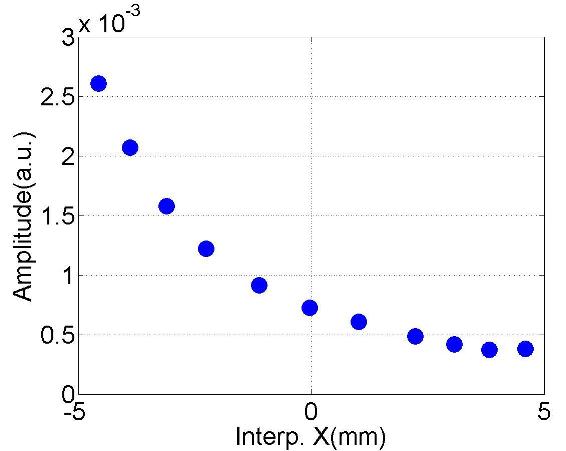}
\label{dep-C3H2-X-9}
}
\subfigure[\#10 ($f$:9.0562GHz; $Q$:10$^4$)]{
\includegraphics[width=0.23\textwidth]{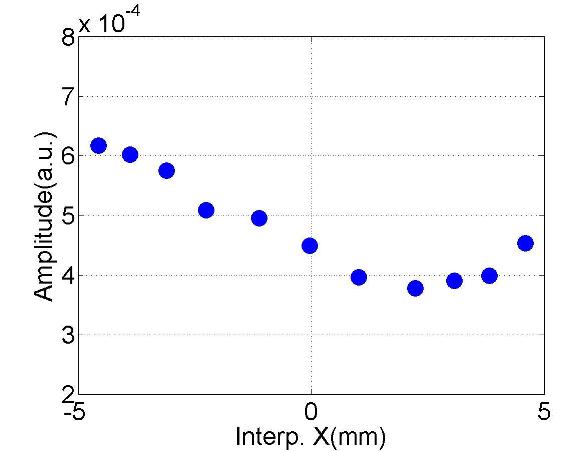}
\label{dep-C3H2-X-10}
}
\subfigure[\#11 ($f$:9.0591GHz; $Q$:10$^4$)]{
\includegraphics[width=0.23\textwidth]{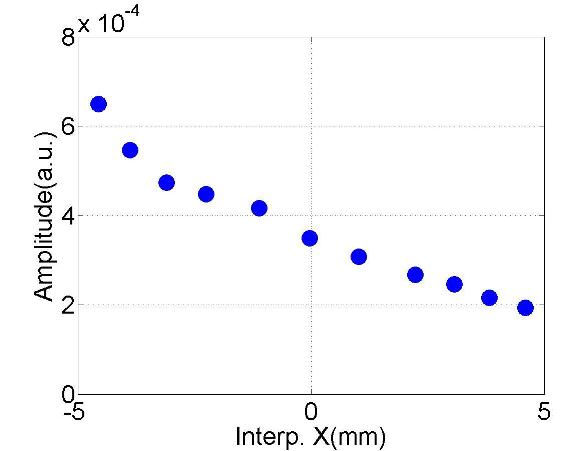}
\label{dep-C3H2-X-11}
}
\subfigure[\#12 ($f$:9.0659GHz; $Q$:10$^5$)]{
\includegraphics[width=0.23\textwidth]{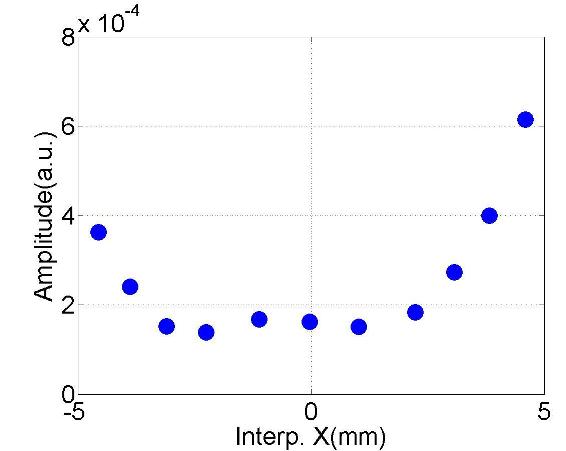}
\label{dep-C3H2-X-12}
}
\subfigure[\#13 ($f$:9.0667GHz; $Q$:10$^4$)]{
\includegraphics[width=0.23\textwidth]{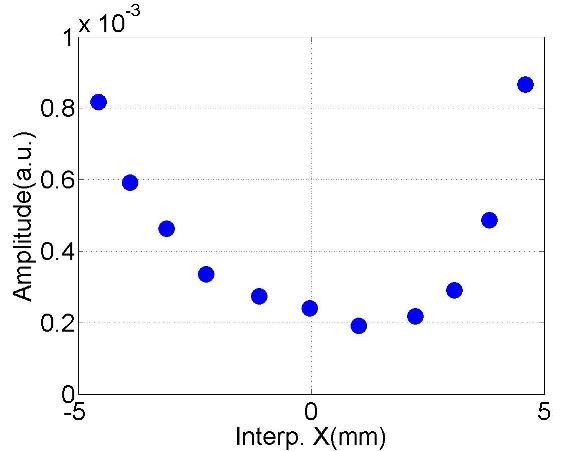}
\label{dep-C3H2-X-13}
}
\subfigure[\#14 ($f$:9.0681GHz; $Q$:10$^4$)]{
\includegraphics[width=0.23\textwidth]{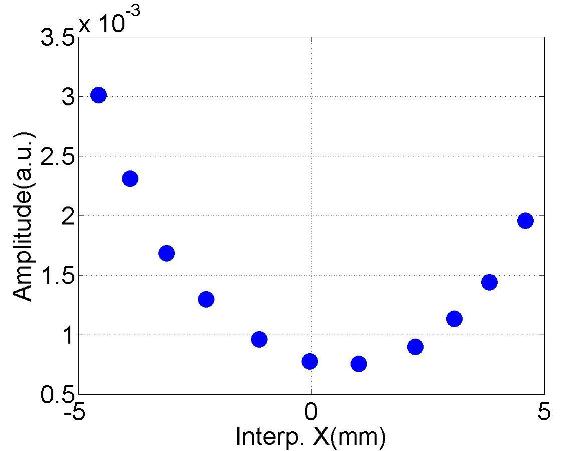}
\label{dep-C3H2-X-14}
}
\subfigure[\#15 ($f$:9.0694GHz; $Q$:10$^5$)]{
\includegraphics[width=0.23\textwidth]{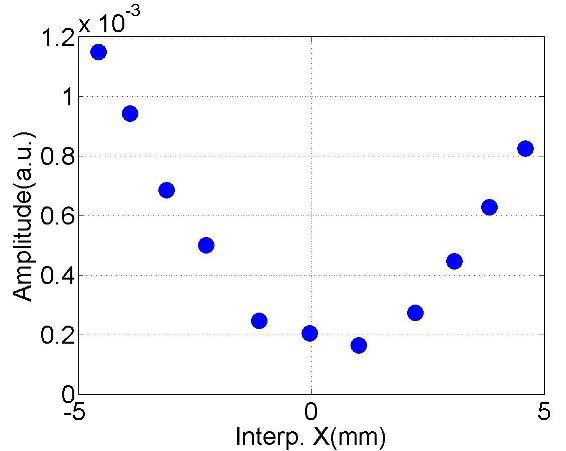}
\label{dep-C3H2-X-15}
}
\subfigure[\#16 ($f$:9.0701GHz; $Q$:10$^5$)]{
\includegraphics[width=0.23\textwidth]{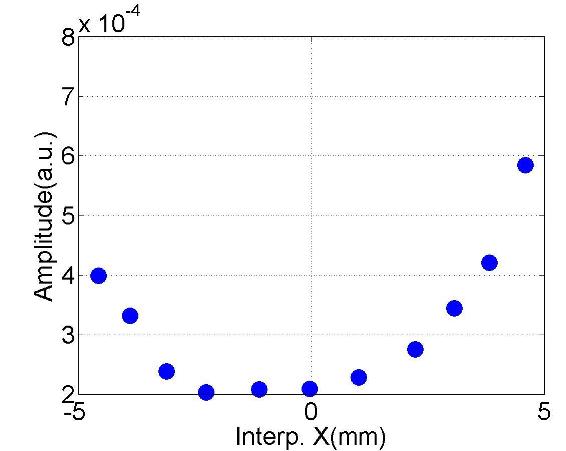}
\label{dep-C3H2-X-16}
}
\subfigure[\#17 ($f$:9.0720GHz; $Q$:10$^5$)]{
\includegraphics[width=0.23\textwidth]{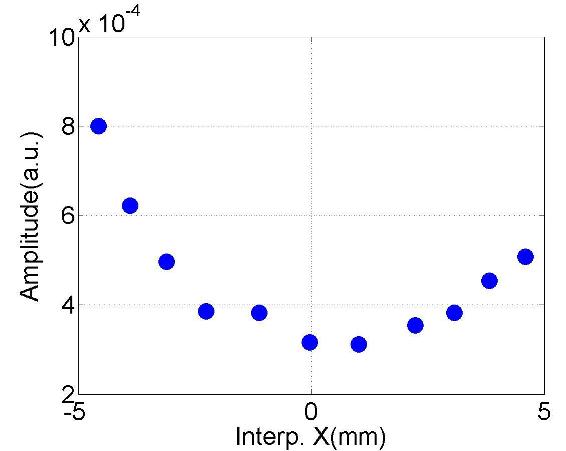}
\label{dep-C3H2-X-17}
}
\subfigure[\#18 ($f$:9.0754GHz; $Q$:10$^4$)]{
\includegraphics[width=0.23\textwidth]{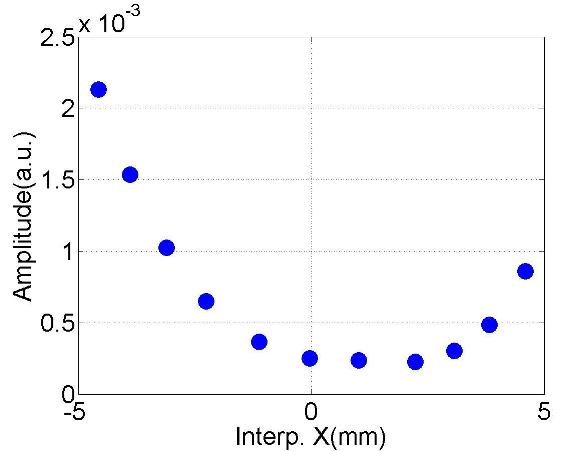}
\label{dep-C3H2-X-18}
}
\subfigure[\#19 ($f$:9.0762GHz; $Q$:10$^5$)]{
\includegraphics[width=0.23\textwidth]{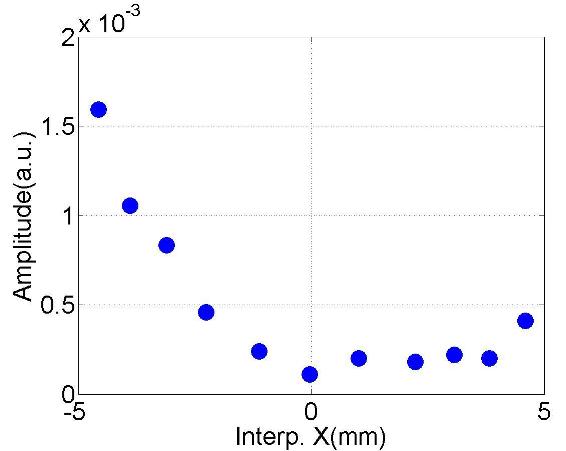}
\label{dep-C3H2-X-19}
}
\subfigure[\#20 ($f$:9.0812GHz; $Q$:10$^5$)]{
\includegraphics[width=0.23\textwidth]{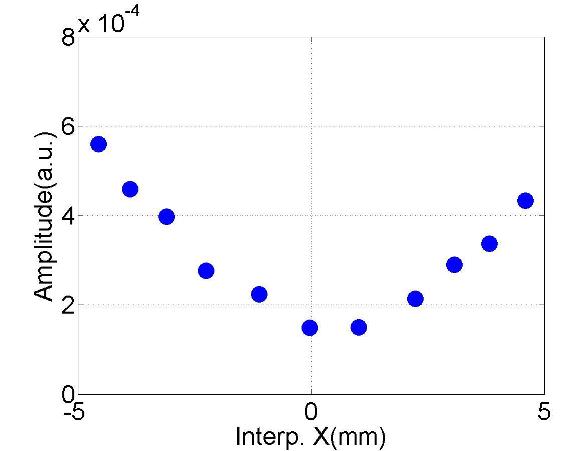}
\label{dep-C3H2-X-20}
}
\subfigure[\#21 ($f$:9.0821GHz; $Q$:10$^5$)]{
\includegraphics[width=0.23\textwidth]{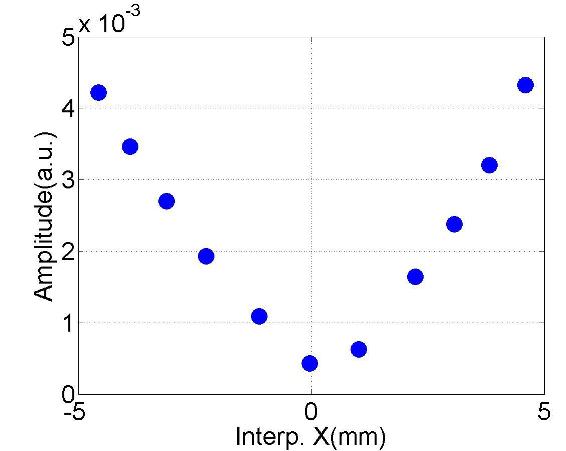}
\label{dep-C3H2-X-21}
}
\subfigure[\#22 ($f$:9.0962GHz; $Q$:10$^5$)]{
\includegraphics[width=0.23\textwidth]{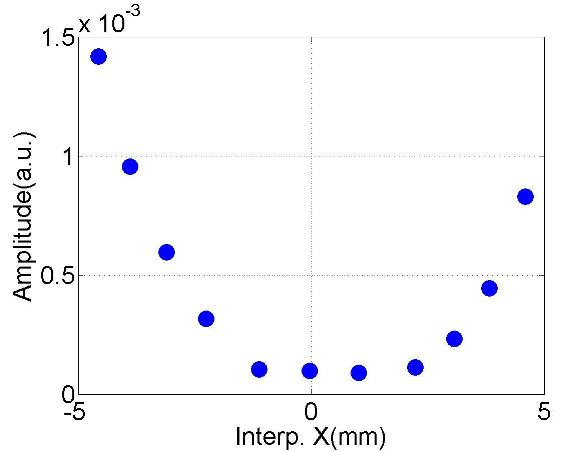}
\label{dep-C3H2-X-22}
}
\caption{Dependence of the mode amplitude on the horizontal beam of{}fset in the cavity.}
\label{spec-dep-C3H2-X-2}
\end{figure}
\begin{figure}[h]
\subfigure[Spectrum (C3H2)]{
\includegraphics[width=1\textwidth]{D5Xmove-Spec-C3H2-2}
\label{spec-C3H2-X-2}
}
\subfigure[\#8 ($f$:9.0511GHz; $Q$:10$^4$)]{
\includegraphics[width=0.23\textwidth]{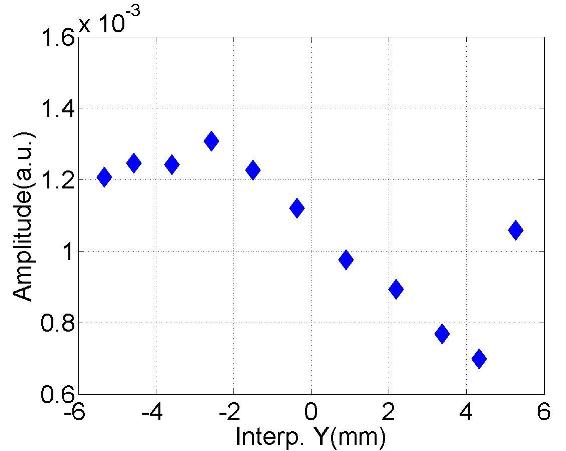}
\label{dep-C3H2-Y-8}
}
\subfigure[\#9 ($f$:9.0551GHz; $Q$:10$^4$)]{
\includegraphics[width=0.23\textwidth]{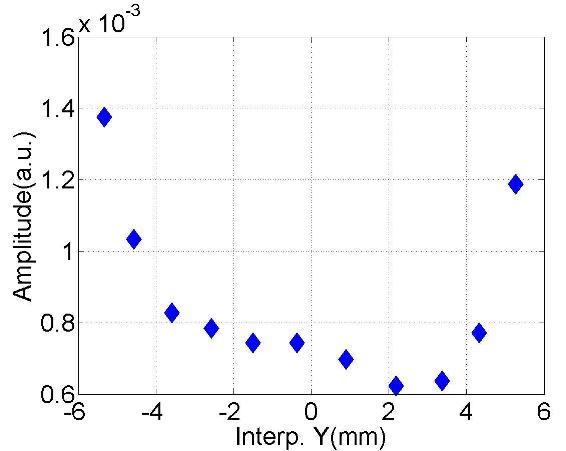}
\label{dep-C3H2-Y-9}
}
\subfigure[\#10 ($f$:9.0563GHz; $Q$:10$^4$)]{
\includegraphics[width=0.23\textwidth]{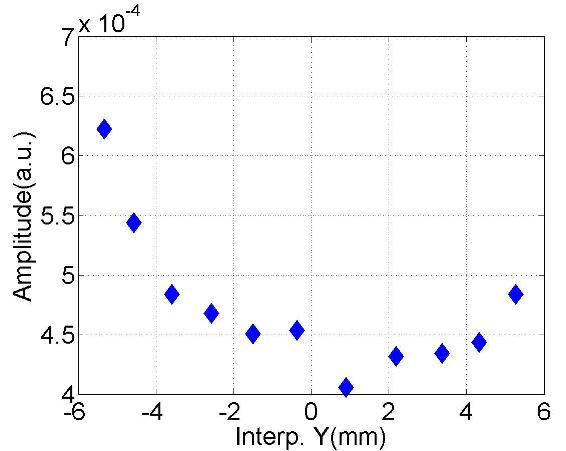}
\label{dep-C3H2-Y-10}
}
\subfigure[\#11 ($f$:9.0591GHz; $Q$:10$^5$)]{
\includegraphics[width=0.23\textwidth]{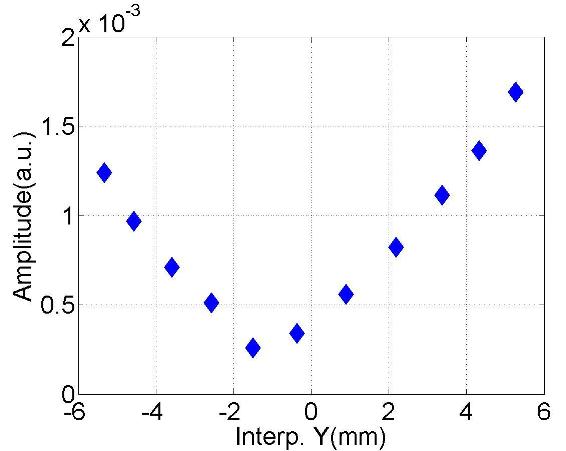}
\label{dep-C3H2-Y-11}
}
\subfigure[\#12 ($f$:9.0659GHz; $Q$:10$^4$)]{
\includegraphics[width=0.23\textwidth]{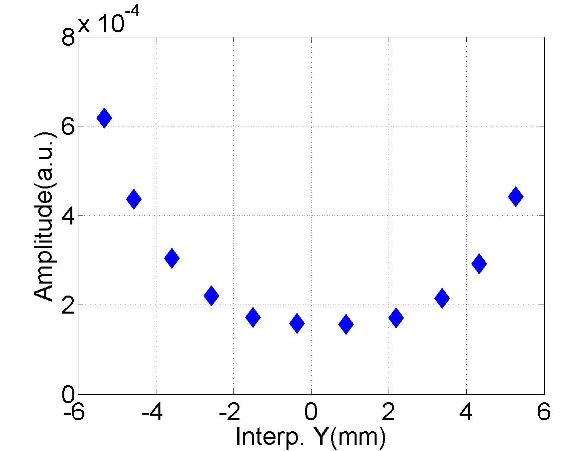}
\label{dep-C3H2-Y-12}
}
\subfigure[\#13 ($f$:9.0665GHz; $Q$:10$^4$)]{
\includegraphics[width=0.23\textwidth]{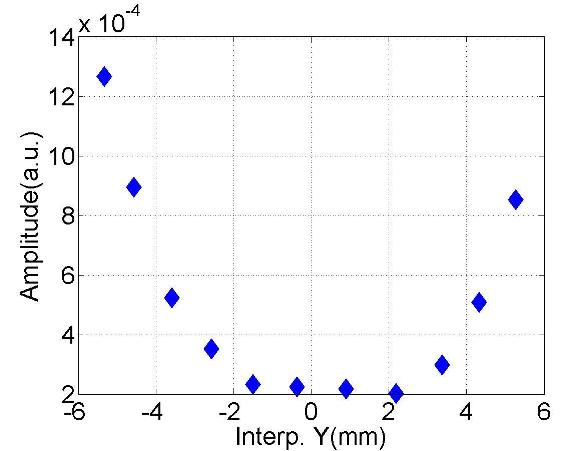}
\label{dep-C3H2-Y-13}
}
\subfigure[\#14 ($f$:9.0681GHz; $Q$:10$^4$)]{
\includegraphics[width=0.23\textwidth]{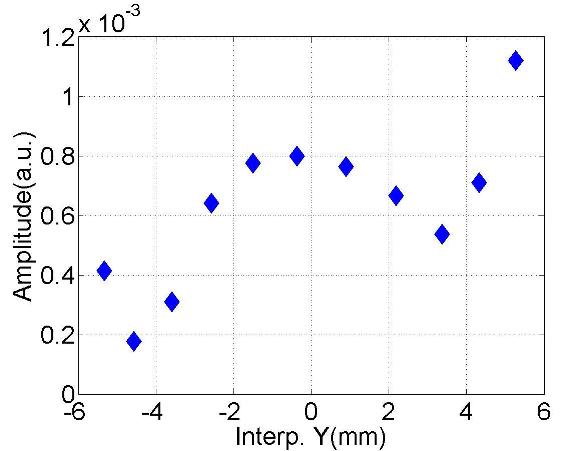}
\label{dep-C3H2-Y-14}
}
\subfigure[\#15 ($f$:9.0694GHz; $Q$:10$^5$)]{
\includegraphics[width=0.23\textwidth]{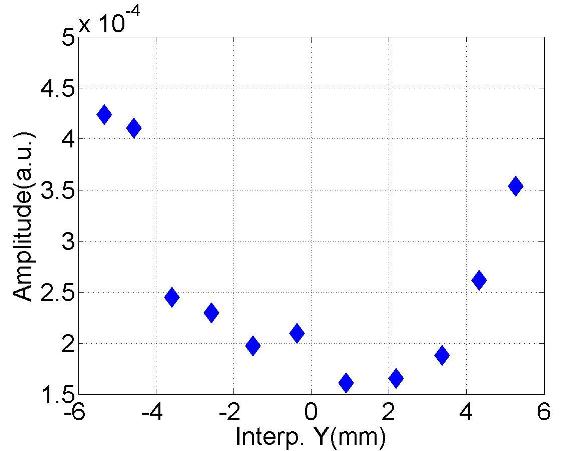}
\label{dep-C3H2-Y-15}
}
\subfigure[\#16 ($f$:9.0700GHz; $Q$:10$^5$)]{
\includegraphics[width=0.23\textwidth]{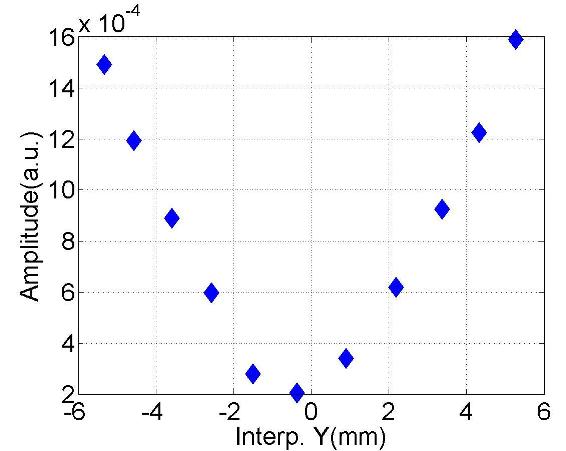}
\label{dep-C3H2-Y-16}
}
\subfigure[\#17 ($f$:9.0720GHz; $Q$:10$^5$)]{
\includegraphics[width=0.23\textwidth]{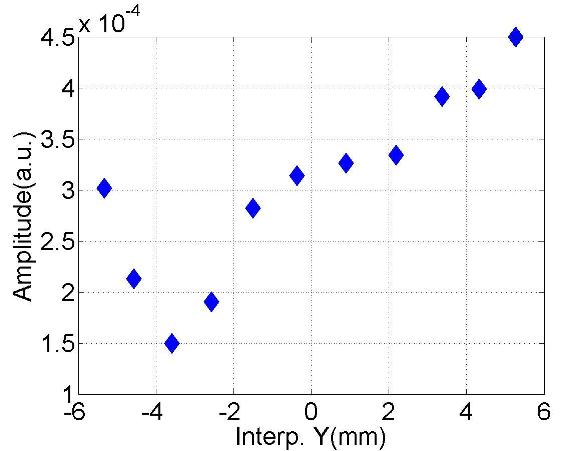}
\label{dep-C3H2-Y-17}
}
\subfigure[\#18 ($f$:9.0753GHz; $Q$:10$^4$)]{
\includegraphics[width=0.23\textwidth]{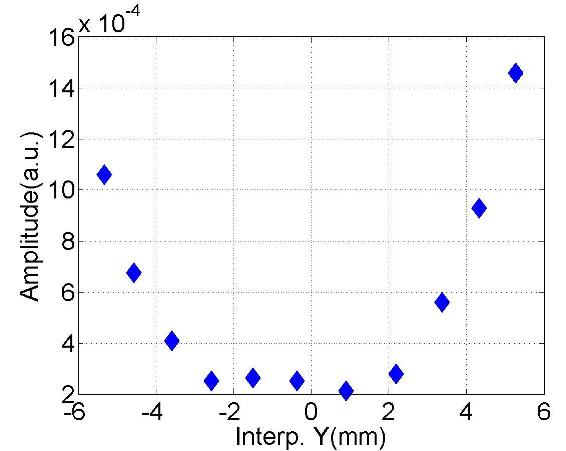}
\label{dep-C3H2-Y-18}
}
\subfigure[\#19 ($f$:9.0762GHz; $Q$:10$^5$)]{
\includegraphics[width=0.23\textwidth]{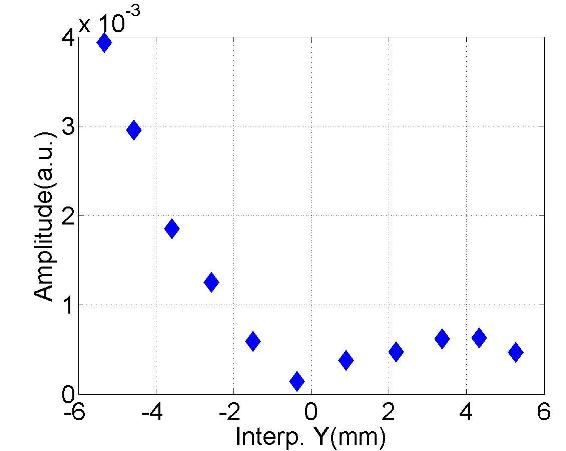}
\label{dep-C3H2-Y-19}
}
\subfigure[\#20 ($f$:9.0812GHz; $Q$:10$^5$)]{
\includegraphics[width=0.23\textwidth]{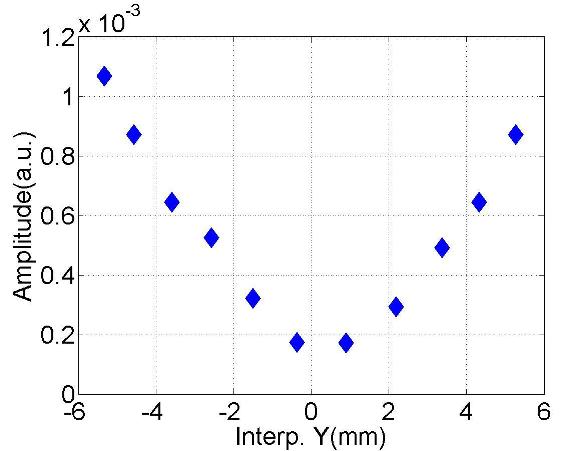}
\label{dep-C3H2-Y-20}
}
\subfigure[\#21 ($f$:9.0821GHz; $Q$:10$^5$)]{
\includegraphics[width=0.23\textwidth]{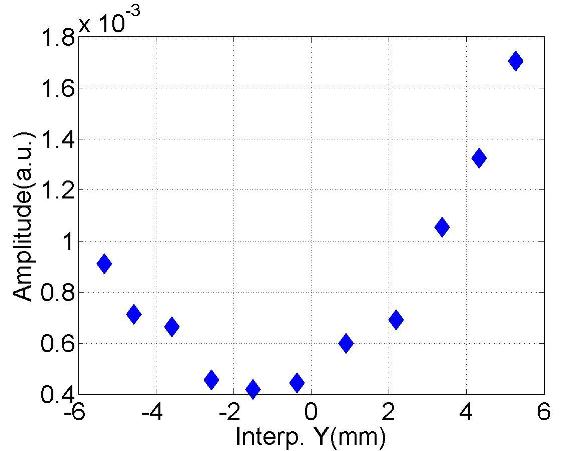}
\label{dep-C3H2-Y-21}
}
\subfigure[\#22 ($f$:9.0962GHz; $Q$:10$^5$)]{
\includegraphics[width=0.23\textwidth]{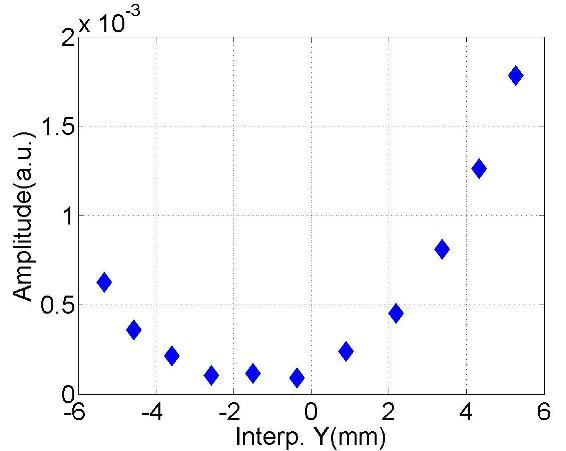}
\label{dep-C3H2-Y-22}
}
\caption{Dependence of the mode amplitude on the vertical beam of{}fset in the cavity.}
\label{spec-dep-C3H2-Y-2}
\end{figure}
\begin{figure}[h]
\subfigure[Spectrum (C3H2)]{
\includegraphics[width=1\textwidth]{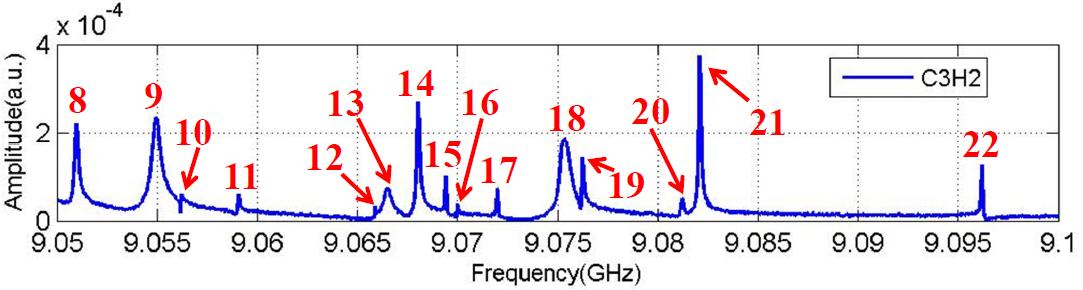}
\label{spec-C3H2-2}
}
\subfigure[\#8 ($f$:9.0510GHz; $Q$:10$^4$)]{
\includegraphics[width=0.23\textwidth]{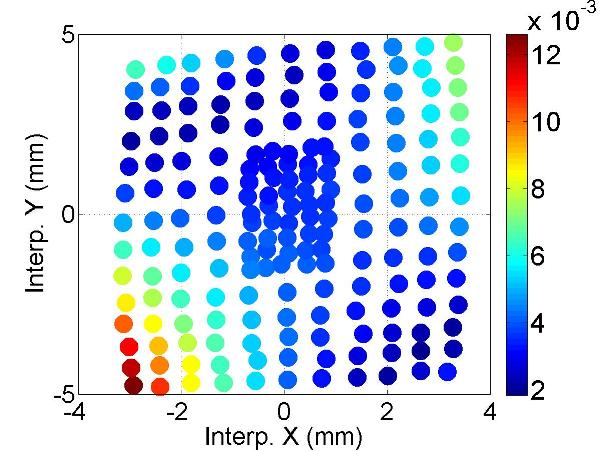}
\label{polar-C3H2-8}
}
\subfigure[\#9 ($f$:9.0551GHz; $Q$:10$^4$)]{
\includegraphics[width=0.23\textwidth]{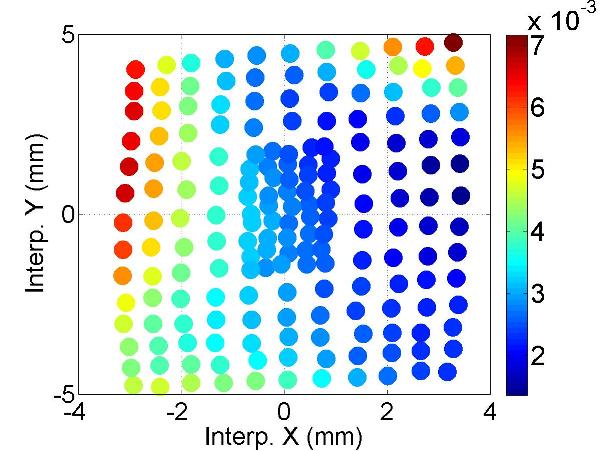}
\label{polar-C3H2-9}
}
\subfigure[\#10 ($f$:9.0562GHz; $Q$:10$^4$)]{
\includegraphics[width=0.23\textwidth]{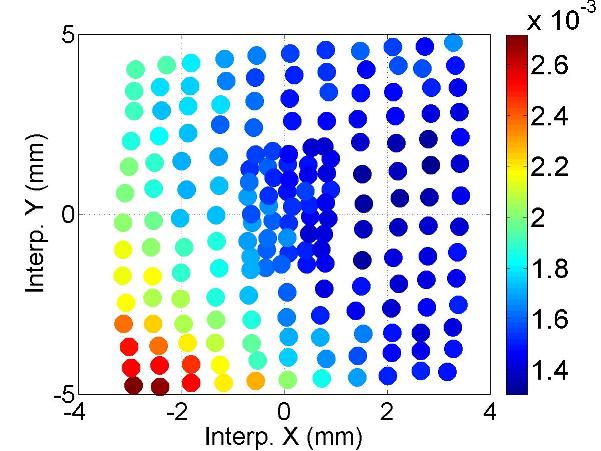}
\label{polar-C3H2-10}
}
\subfigure[\#11 ($f$:9.0591GHz; $Q$:10$^5$)]{
\includegraphics[width=0.23\textwidth]{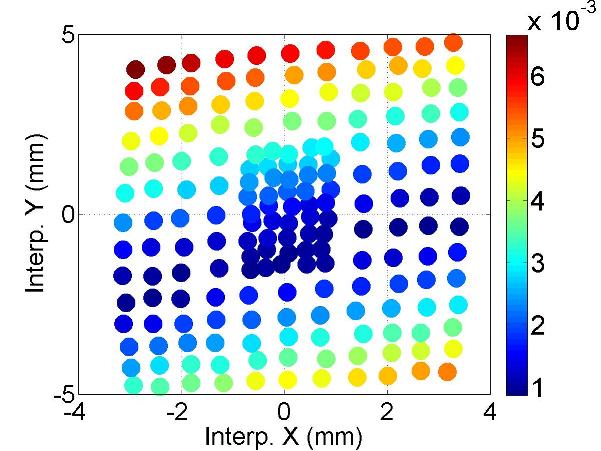}
\label{polar-C3H2-11}
}
\subfigure[\#12 ($f$:9.0659GHz; $Q$:10$^5$)]{
\includegraphics[width=0.23\textwidth]{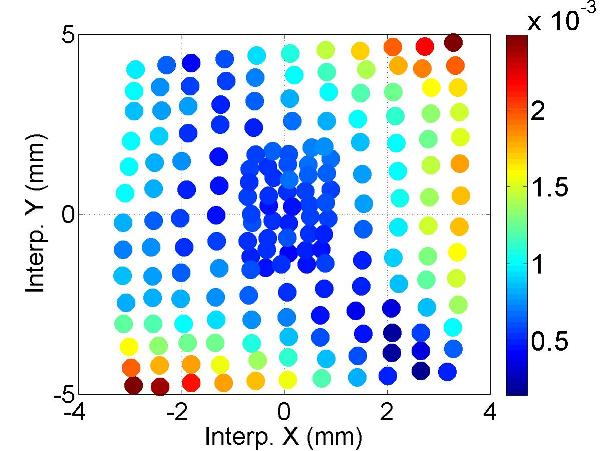}
\label{polar-C3H2-12}
}
\subfigure[\#13 ($f$:9.0666GHz; $Q$:10$^4$)]{
\includegraphics[width=0.23\textwidth]{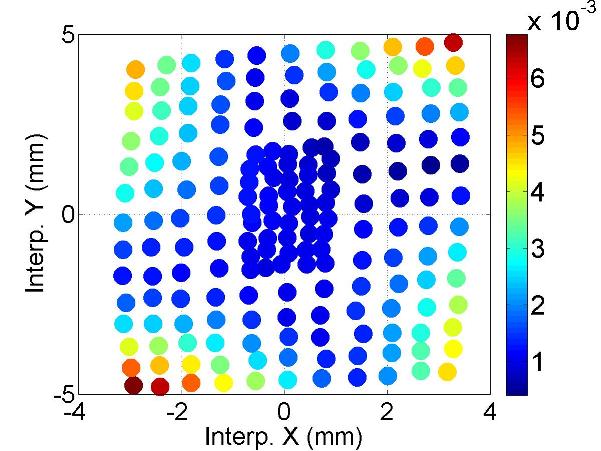}
\label{polar-C3H2-13}
}
\subfigure[\#14 ($f$:9.0681GHz; $Q$:10$^5$)]{
\includegraphics[width=0.23\textwidth]{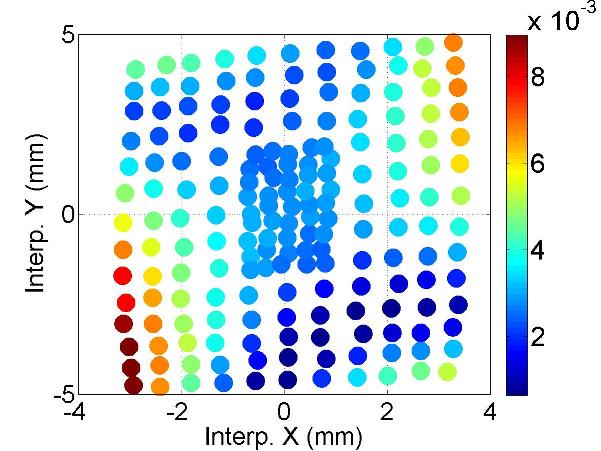}
\label{polar-C3H2-14}
}
\subfigure[\#15 ($f$:9.0694GHz; $Q$:10$^5$)]{
\includegraphics[width=0.23\textwidth]{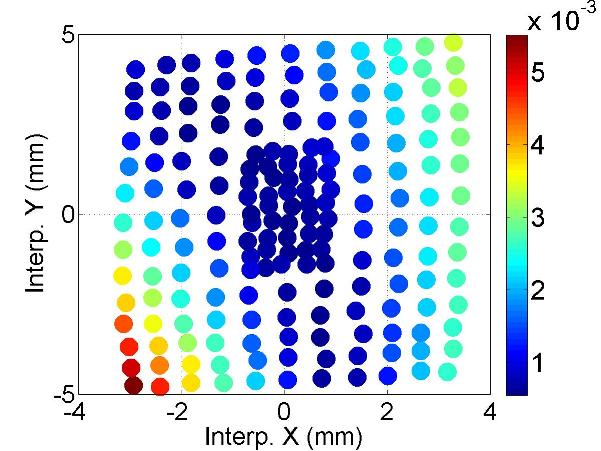}
\label{polar-C3H2-15}
}
\subfigure[\#16 ($f$:9.0700GHz; $Q$:10$^5$)]{
\includegraphics[width=0.23\textwidth]{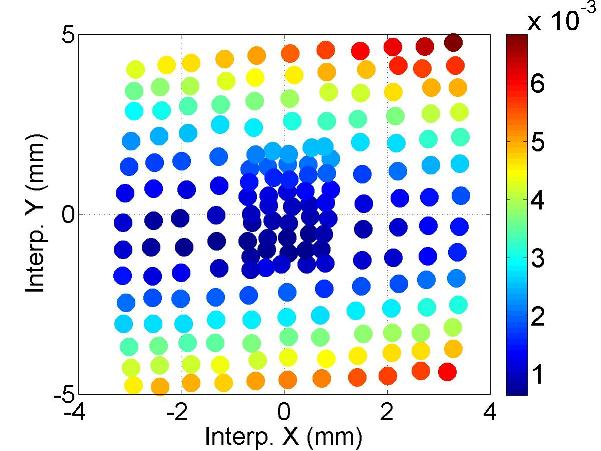}
\label{polar-C3H2-16}
}
\subfigure[\#17 ($f$:9.0720GHz; $Q$:10$^5$)]{
\includegraphics[width=0.23\textwidth]{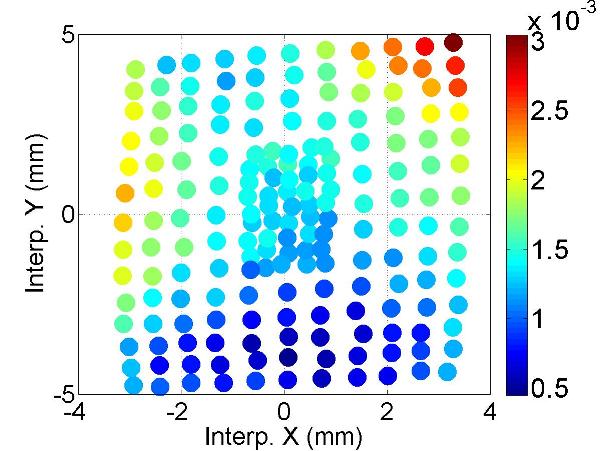}
\label{polar-C3H2-17}
}
\subfigure[\#18 ($f$:9.0754GHz; $Q$:10$^4$)]{
\includegraphics[width=0.23\textwidth]{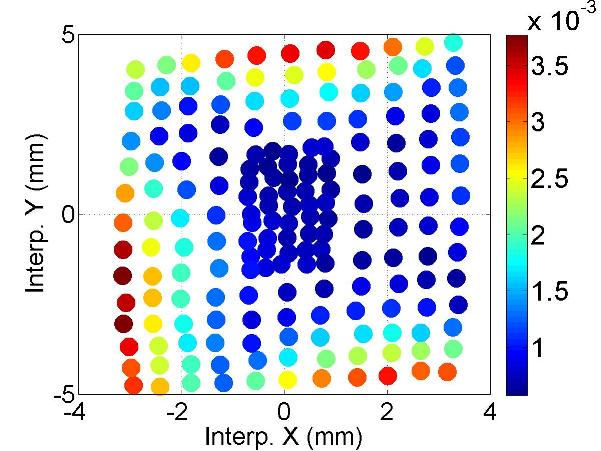}
\label{polar-C3H2-18}
}
\subfigure[\#19 ($f$:9.0762GHz; $Q$:10$^5$)]{
\includegraphics[width=0.23\textwidth]{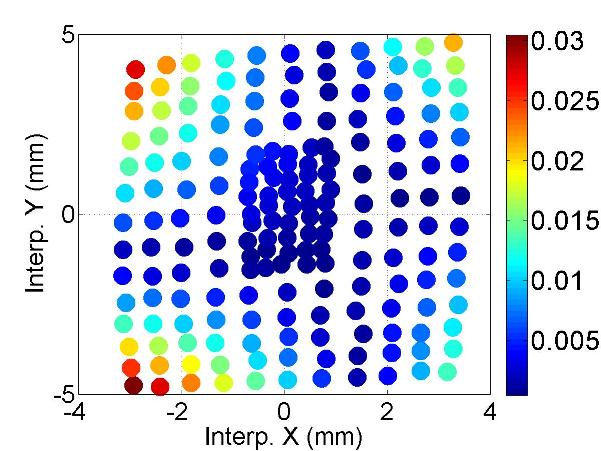}
\label{polar-C3H2-19}
}
\subfigure[\#20 ($f$:9.0812GHz; $Q$:10$^5$)]{
\includegraphics[width=0.23\textwidth]{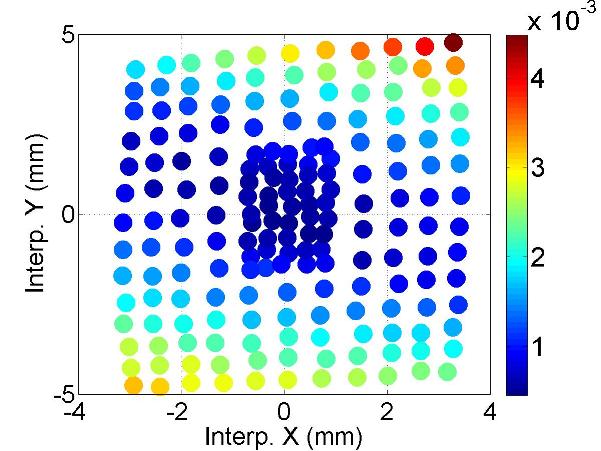}
\label{polar-C3H2-20}
}
\subfigure[\#21 ($f$:9.0821GHz; $Q$:10$^5$)]{
\includegraphics[width=0.23\textwidth]{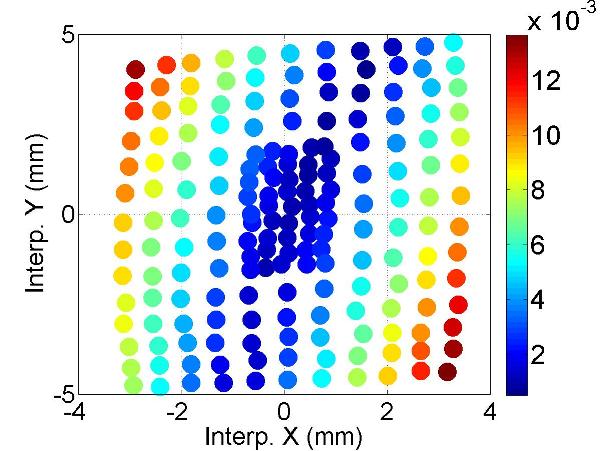}
\label{polar-C3H2-21}
}
\subfigure[\#22 ($f$:9.0962GHz; $Q$:10$^5$)]{
\includegraphics[width=0.23\textwidth]{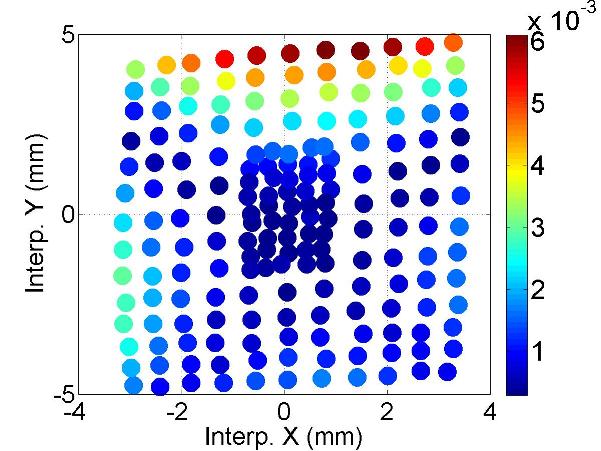}
\label{polar-C3H2-22}
}
\caption{Polarization of the mode.}
\label{spec-polar-C3H2-2}
\end{figure}

\FloatBarrier
\section{D5: HOM Coupler C4H1}
\begin{figure}[h]
\subfigure[Spectrum (C4H1)]{
\includegraphics[width=1\textwidth]{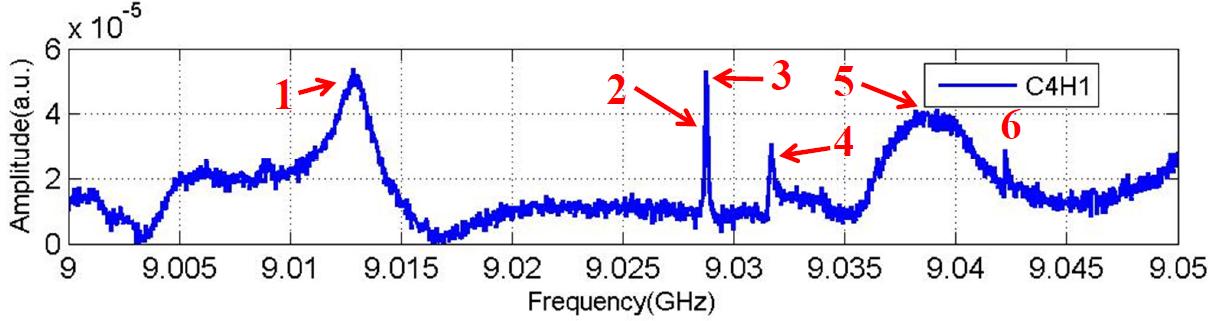}
\label{spec-C4H1-X-1}
}
\subfigure[\#1 ($f$:9.0131GHz; $Q$:10$^3$)]{
\includegraphics[width=0.23\textwidth]{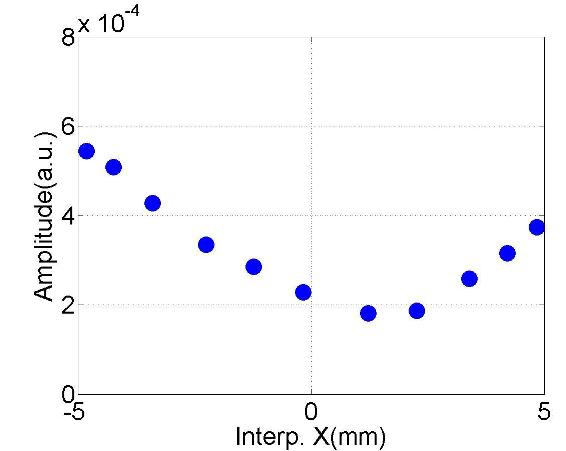}
\label{dep-C4H1-X-1}
}
\subfigure[\#2 ($f$:9.0276GHz; $Q$:10$^4$)]{
\includegraphics[width=0.23\textwidth]{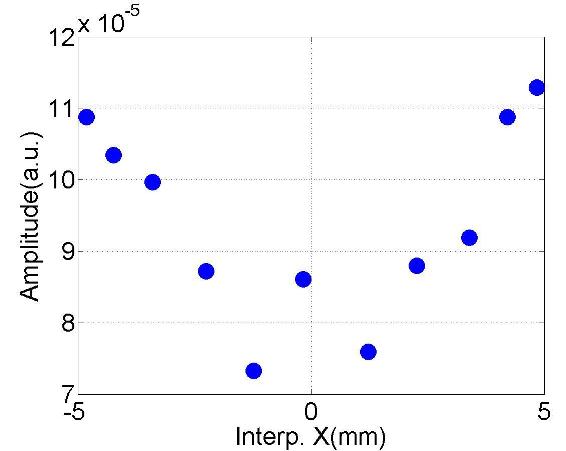}
\label{dep-C4H1-X-2}
}
\subfigure[\#3 ($f$:9.0288GHz; $Q$:10$^5$)]{
\includegraphics[width=0.23\textwidth]{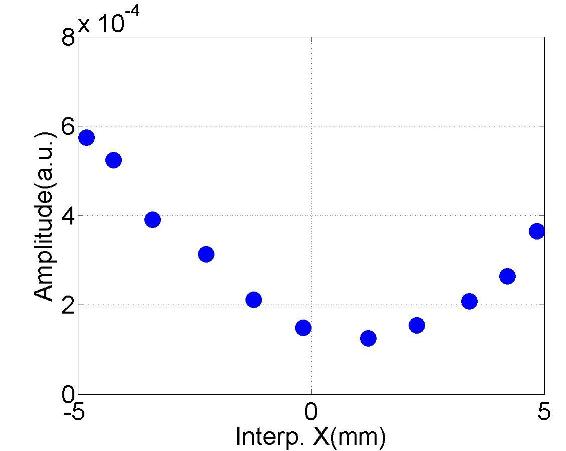}
\label{dep-C4H1-X-3}
}
\subfigure[\#4 ($f$:9.0317GHz; $Q$:10$^4$)]{
\includegraphics[width=0.23\textwidth]{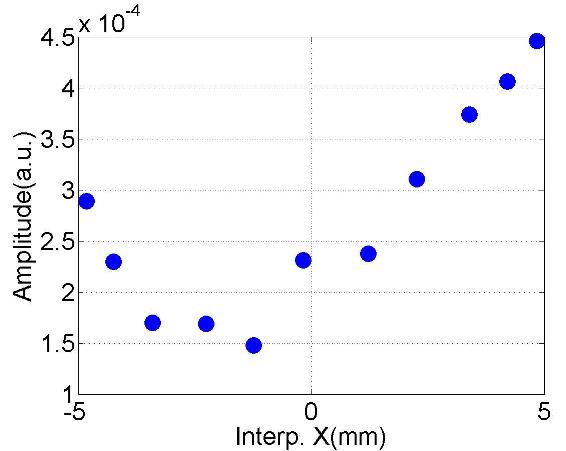}
\label{dep-C4H1-X-4}
}
\subfigure[\#5 ($f$:9.0399GHz; $Q$:10$^3$)]{
\includegraphics[width=0.23\textwidth]{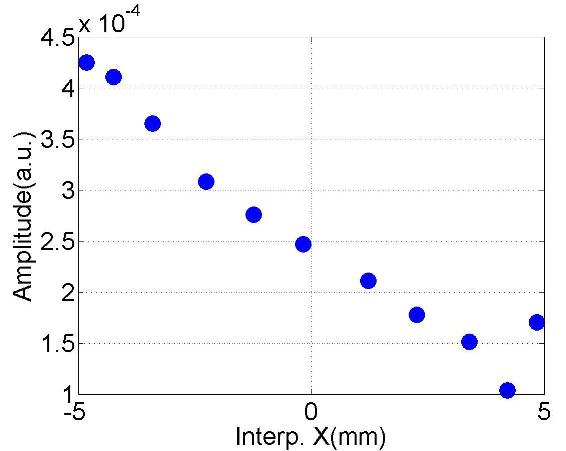}
\label{dep-C4H1-X-5}
}
\subfigure[\#6 ($f$:9.0423GHz; $Q$:10$^4$)]{
\includegraphics[width=0.23\textwidth]{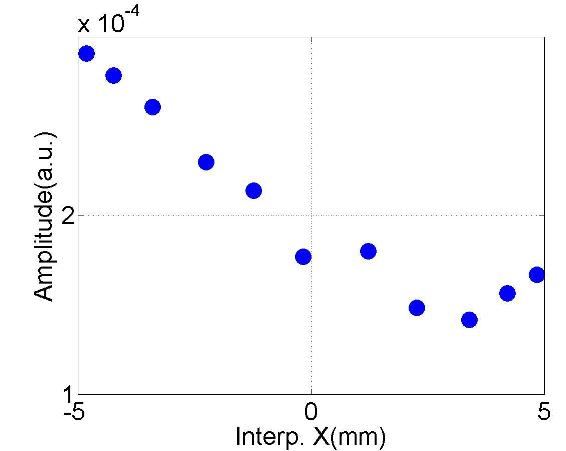}
\label{dep-C4H1-X-6}
}\\
\subfigure[\#1 ($f$:9.0121GHz; $Q$:10$^3$)]{
\includegraphics[width=0.23\textwidth]{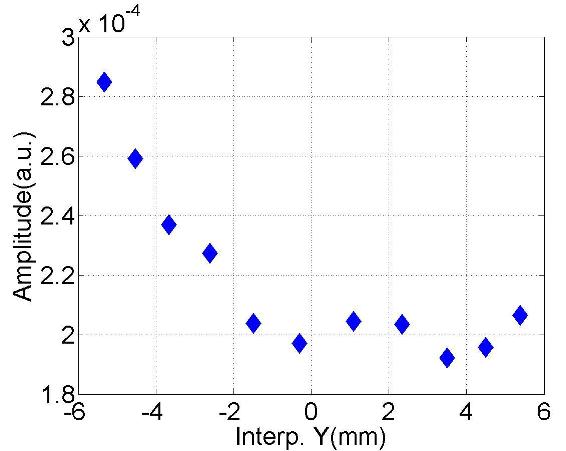}
\label{dep-C4H1-Y-1}
}
\subfigure[\#2 ($f$:9.0283GHz; $Q$:10$^4$)]{
\includegraphics[width=0.23\textwidth]{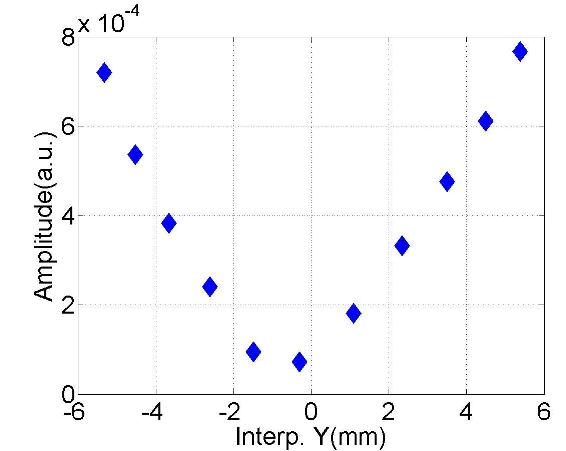}
\label{dep-C4H1-Y-2}
}
\subfigure[\#3 ($f$:9.0289GHz; $Q$:10$^5$)]{
\includegraphics[width=0.23\textwidth]{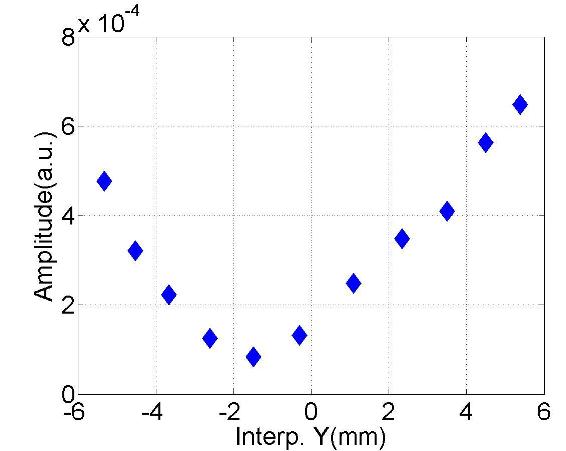}
\label{dep-C4H1-Y-3}
}
\subfigure[\#4 ($f$:9.0317GHz; $Q$:10$^4$)]{
\includegraphics[width=0.23\textwidth]{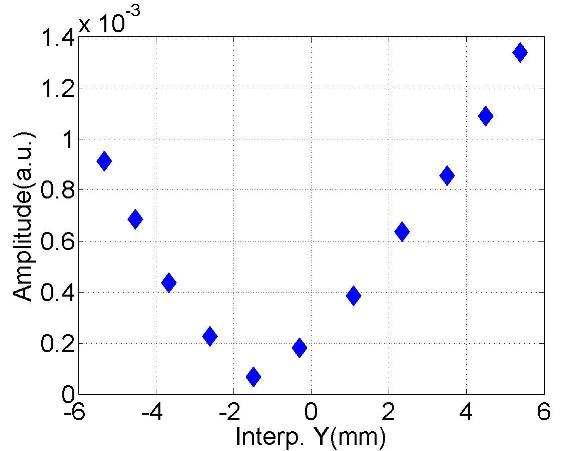}
\label{dep-C4H1-Y-4}
}
\subfigure[\#5 ($f$:9.0400GHz; $Q$:10$^3$)]{
\includegraphics[width=0.23\textwidth]{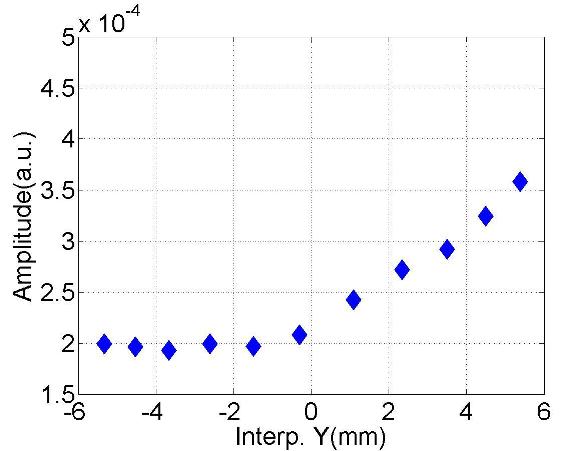}
\label{dep-C4H1-Y-5}
}
\subfigure[\#6 ($f$:9.0423GHz; $Q$:10$^4$)]{
\includegraphics[width=0.23\textwidth]{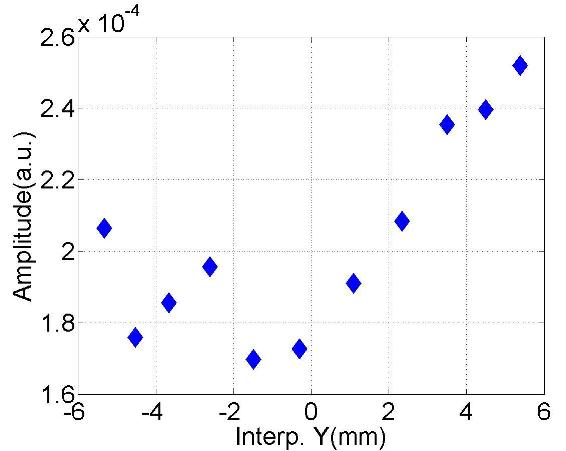}
\label{dep-C4H1-Y-6}
}
\caption{Dependence of the mode amplitude on the transverse beam of{}fset in the cavity.}
\label{spec-dep-C4H1-XY-1}
\end{figure}
\begin{figure}[h]
\subfigure[Spectrum (C4H1)]{
\includegraphics[width=1\textwidth]{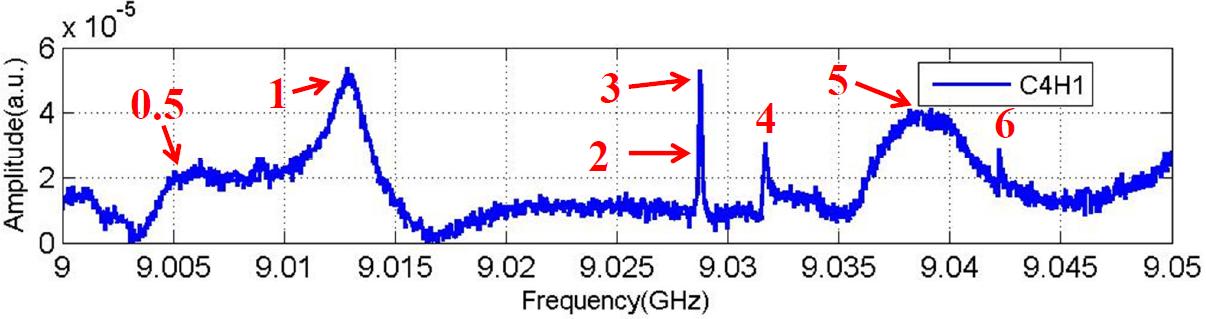}
\label{spec-C4H1-1}
}
\subfigure[\#0.5 ($f$:9.0047GHz; $Q$:10$^3$)]{
\includegraphics[width=0.31\textwidth]{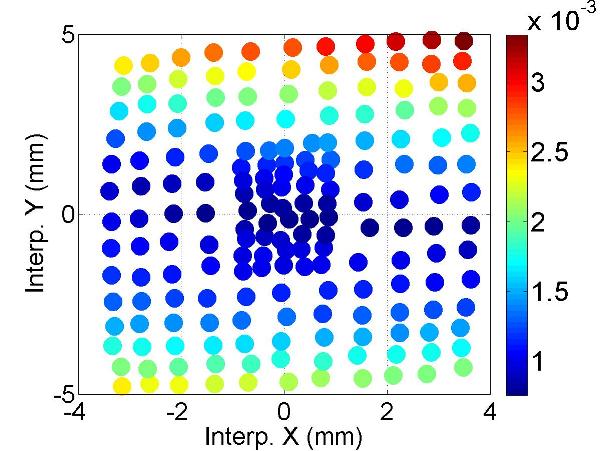}
\label{polar-C4H1-0_5}
}
\subfigure[\#1 ($f$:9.0127GHz; $Q$:10$^3$)]{
\includegraphics[width=0.31\textwidth]{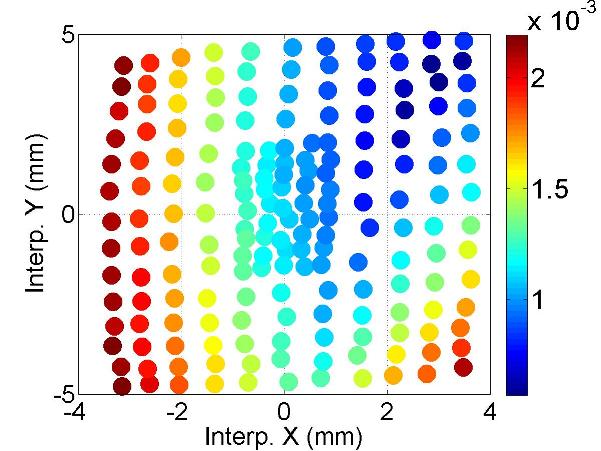}
\label{polar-C4H1-1}
}
\subfigure[\#2 ($f$:9.0282GHz; $Q$:10$^4$)]{
\includegraphics[width=0.31\textwidth]{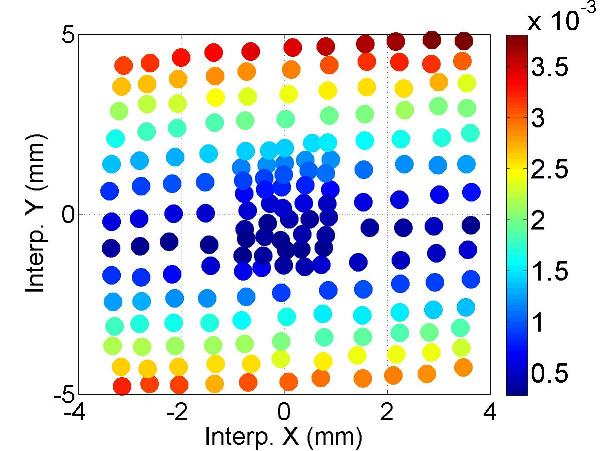}
\label{polar-C4H1-2}
}
\subfigure[\#3 ($f$:9.0288GHz; $Q$:10$^5$)]{
\includegraphics[width=0.31\textwidth]{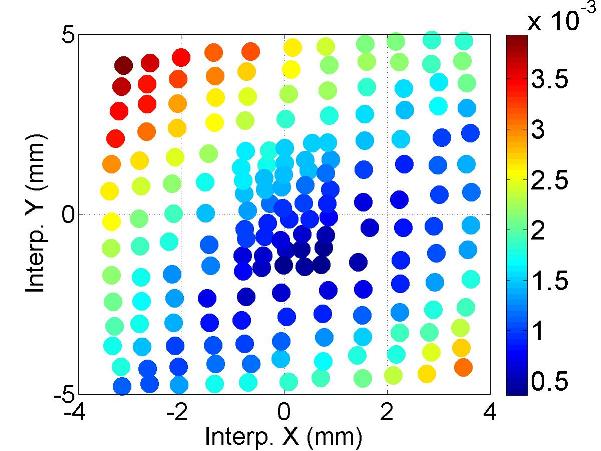}
\label{polar-C4H1-3}
}
\subfigure[\#4 ($f$:9.0317GHz; $Q$:10$^4$)]{
\includegraphics[width=0.31\textwidth]{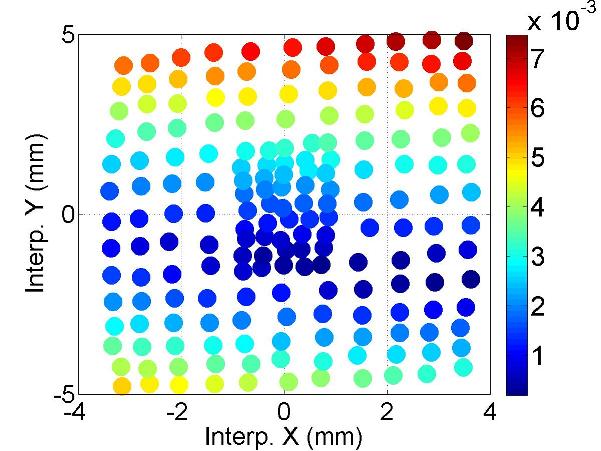}
\label{polar-C4H1-4}
}
\subfigure[\#5 ($f$:9.0400GHz; $Q$:10$^3$)]{
\includegraphics[width=0.31\textwidth]{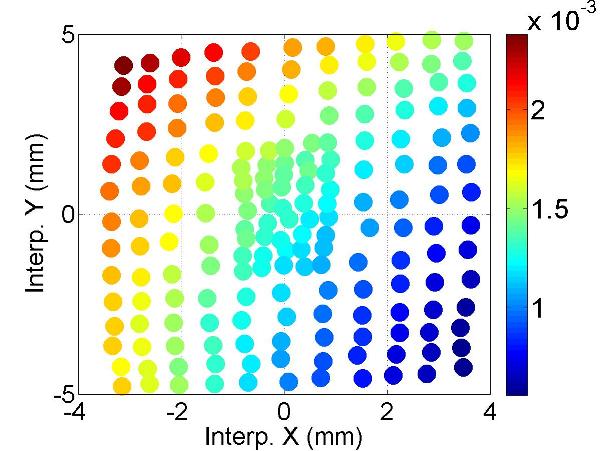}
\label{polar-C4H1-5}
}
\subfigure[\#6 ($f$:9.0422GHz; $Q$:10$^5$)]{
\includegraphics[width=0.31\textwidth]{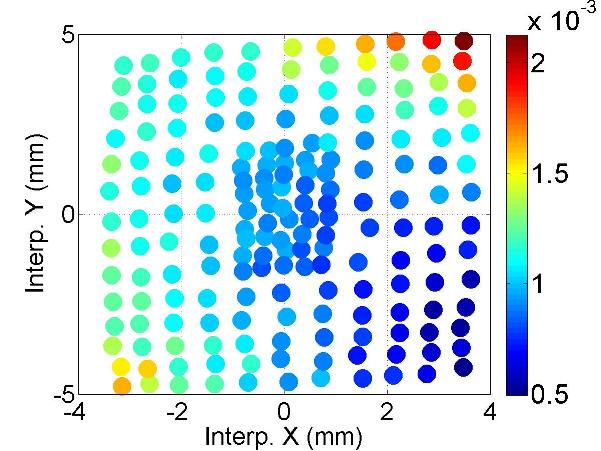}
\label{polar-C4H1-6}
}
\caption{Polarization of the mode.}
\label{spec-polar-C4H1-1}
\end{figure}
\begin{figure}[h]
\subfigure[Spectrum (C4H1)]{
\includegraphics[width=1\textwidth]{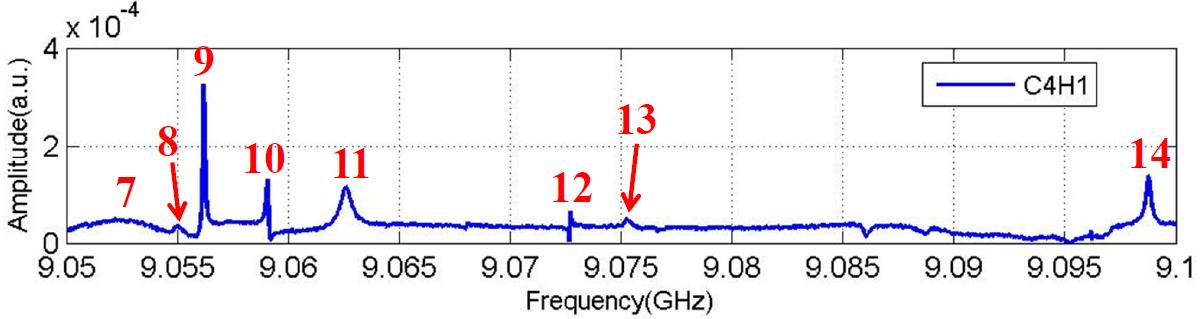}
\label{spec-C4H1-X-2}
}
\subfigure[\#7 ($f$:9.0532GHz; $Q$:10$^3$)]{
\includegraphics[width=0.23\textwidth]{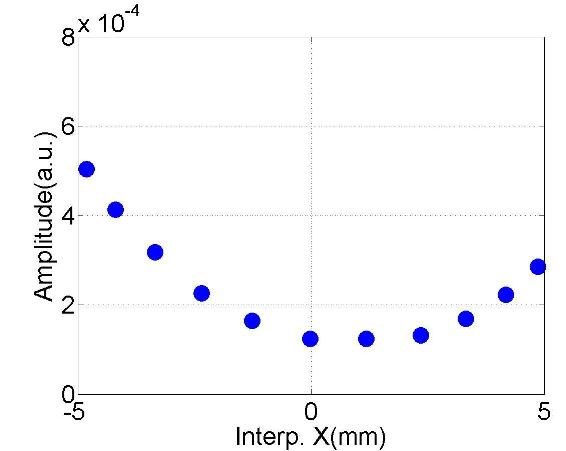}
\label{dep-C4H1-X-7}
}
\subfigure[\#8 ($f$:9.0551GHz; $Q$:10$^4$)]{
\includegraphics[width=0.23\textwidth]{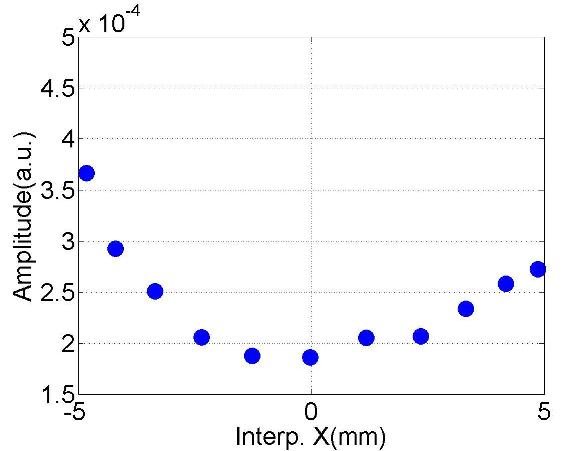}
\label{dep-C4H1-X-8}
}
\subfigure[\#9 ($f$:9.0562GHz; $Q$:10$^5$)]{
\includegraphics[width=0.23\textwidth]{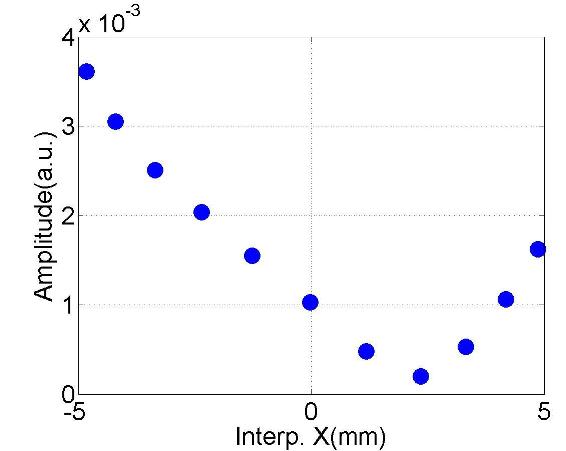}
\label{dep-C4H1-X-9}
}
\subfigure[\#10 ($f$:9.0590GHz; $Q$:10$^5$)]{
\includegraphics[width=0.23\textwidth]{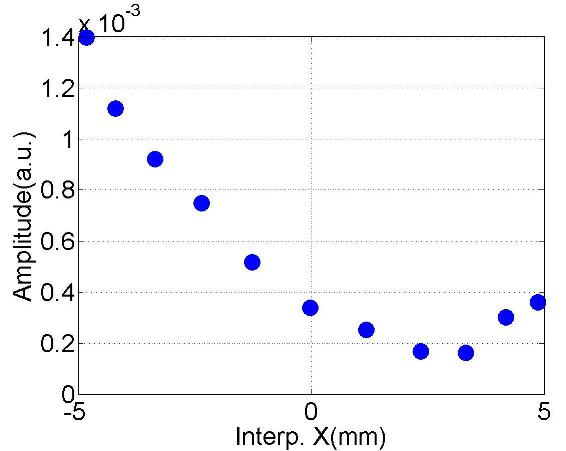}
\label{dep-C4H1-X-10}
}
\subfigure[\#11 ($f$:9.0628GHz; $Q$:10$^4$)]{
\includegraphics[width=0.23\textwidth]{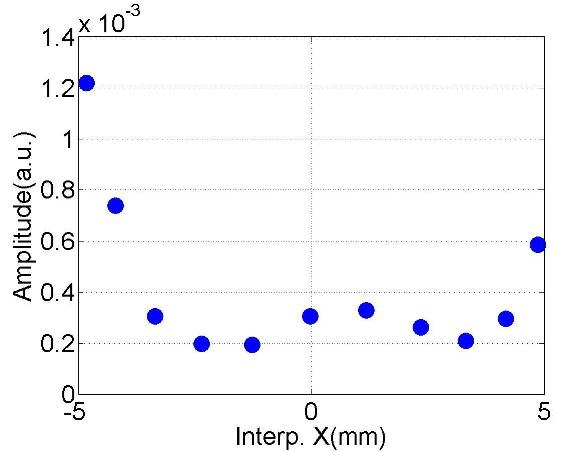}
\label{dep-C4H1-X-11}
}
\subfigure[\#12 ($f$:9.0727GHz; $Q$:10$^5$)]{
\includegraphics[width=0.23\textwidth]{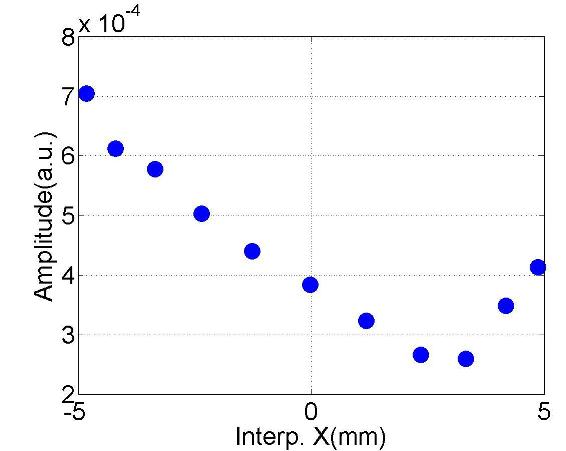}
\label{dep-C4H1-X-12}
}
\subfigure[\#13 ($f$:9.0752GHz; $Q$:10$^4$)]{
\includegraphics[width=0.23\textwidth]{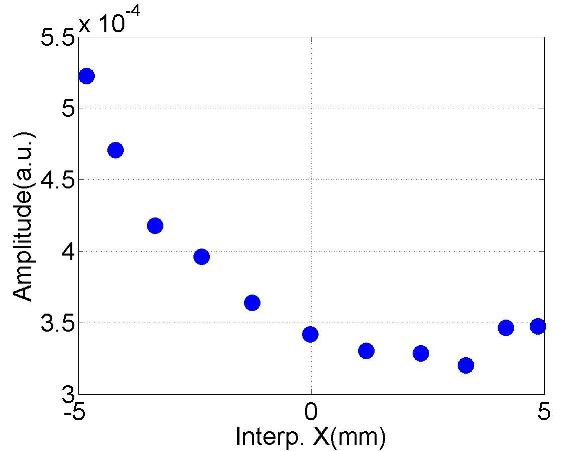}
\label{dep-C4H1-X-13}
}
\subfigure[\#14 ($f$:9.0988GHz; $Q$:10$^4$)]{
\includegraphics[width=0.23\textwidth]{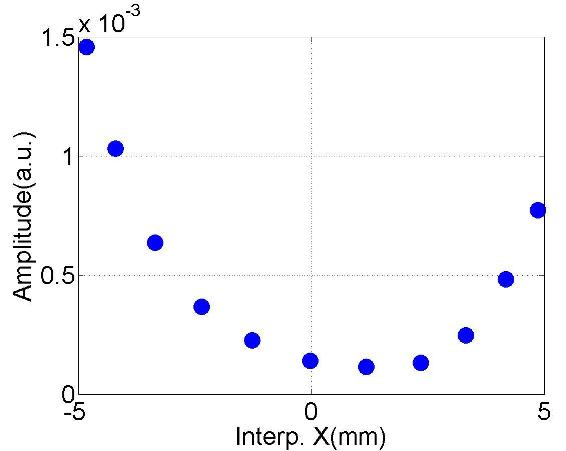}
\label{dep-C4H1-X-14}
}\\
\subfigure[\#7 ($f$:9.0526GHz; $Q$:10$^3$)]{
\includegraphics[width=0.23\textwidth]{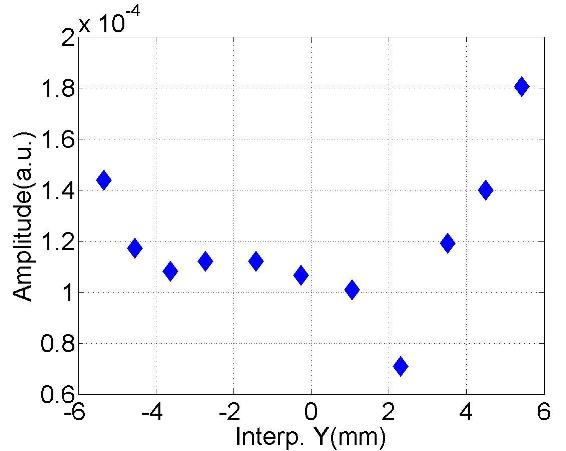}
\label{dep-C4H1-Y-7}
}
\subfigure[\#8 ($f$:9.0552GHz; $Q$:10$^4$)]{
\includegraphics[width=0.23\textwidth]{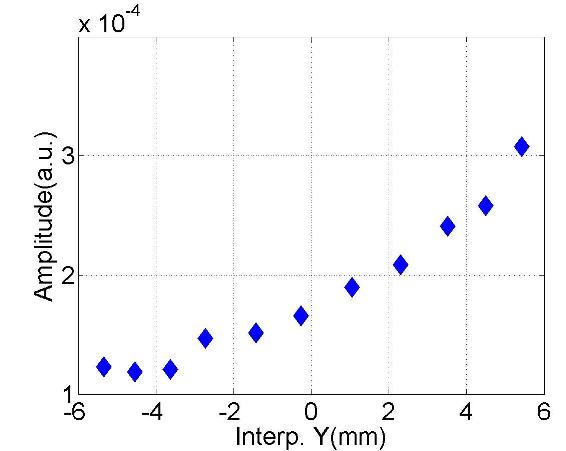}
\label{dep-C4H1-Y-8}
}
\subfigure[\#9 ($f$:9.0562GHz; $Q$:10$^5$)]{
\includegraphics[width=0.23\textwidth]{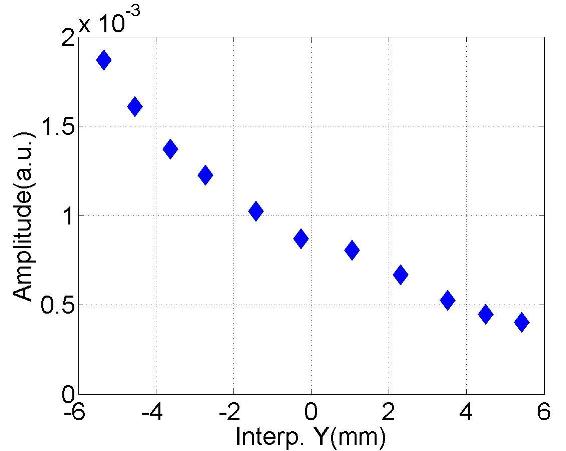}
\label{dep-C4H1-Y-9}
}
\subfigure[\#10 ($f$:9.0591GHz; $Q$:10$^5$)]{
\includegraphics[width=0.23\textwidth]{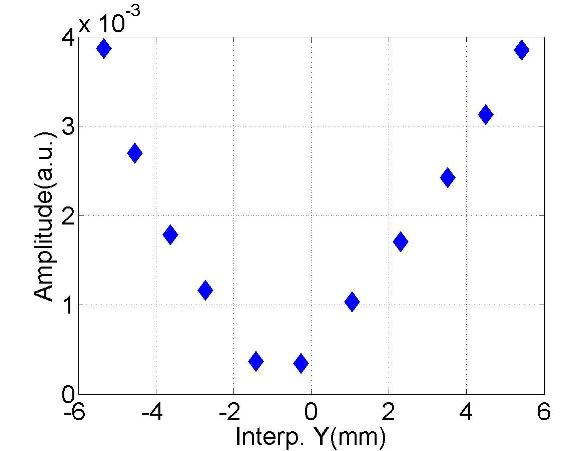}
\label{dep-C4H1-Y-10}
}
\subfigure[\#11 ($f$:9.0626GHz; $Q$:10$^4$)]{
\includegraphics[width=0.23\textwidth]{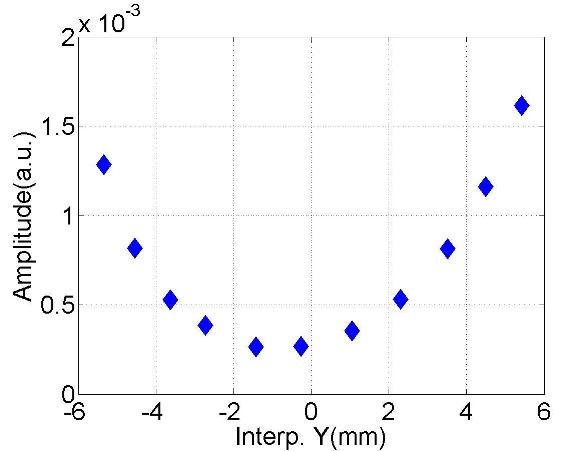}
\label{dep-C4H1-Y-11}
}
\subfigure[\#12 ($f$:9.0727GHz; $Q$:10$^5$)]{
\includegraphics[width=0.23\textwidth]{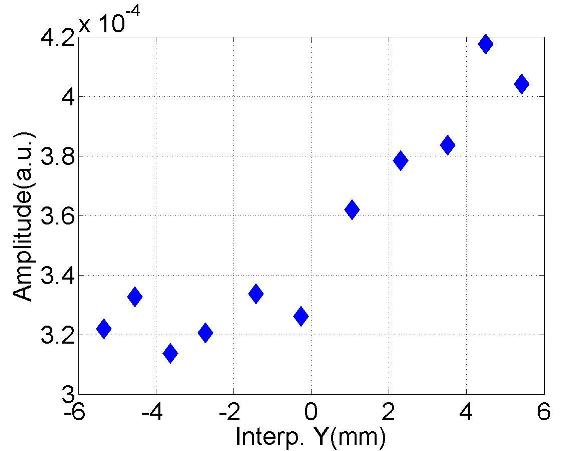}
\label{dep-C4H1-Y-12}
}
\subfigure[\#13 ($f$:9.0752GHz; $Q$:10$^3$)]{
\includegraphics[width=0.23\textwidth]{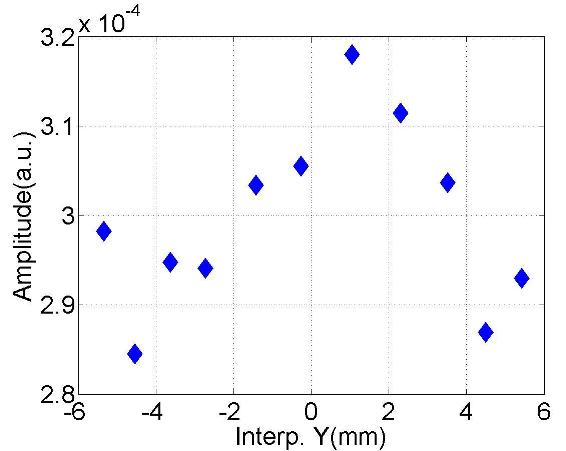}
\label{dep-C4H1-Y-13}
}
\subfigure[\#14 ($f$:9.0988GHz; $Q$:10$^4$)]{
\includegraphics[width=0.23\textwidth]{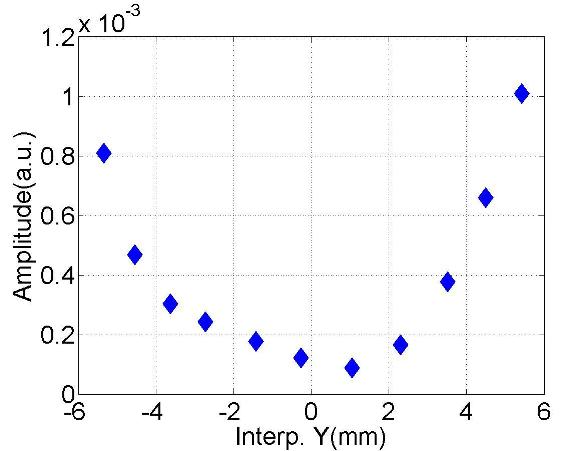}
\label{dep-C4H1-Y-14}
}
\caption{Dependence of the mode amplitude on the transverse beam of{}fset in the cavity.}
\label{spec-dep-C4H1-XY-2}
\end{figure}
\begin{figure}[h]
\subfigure[Spectrum (C4H1)]{
\includegraphics[width=1\textwidth]{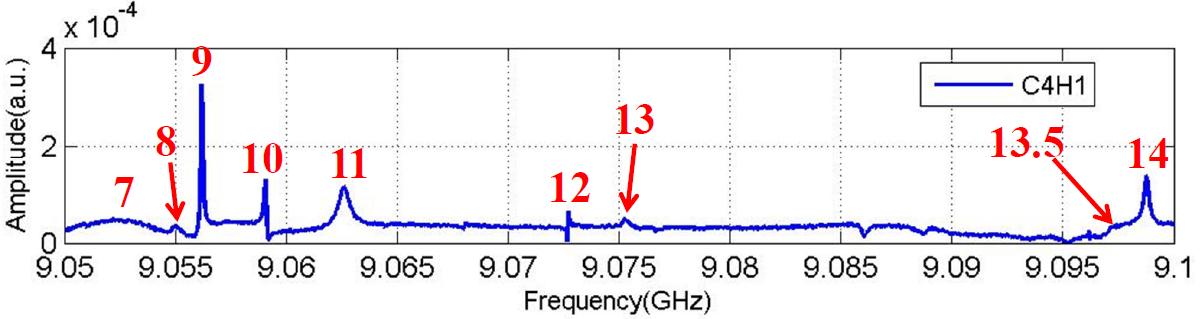}
\label{spec-C4H1-2}
}
\subfigure[\#7 ($f$:9.0530GHz; $Q$:10$^3$)]{
\includegraphics[width=0.31\textwidth]{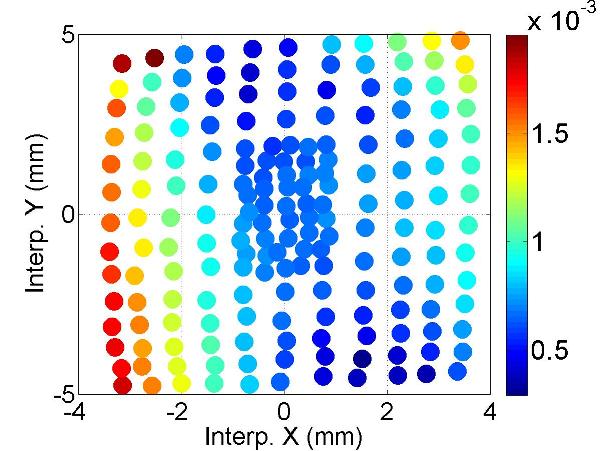}
\label{polar-C4H1-7}
}
\subfigure[\#8 ($f$:9.0552GHz; $Q$:10$^4$)]{
\includegraphics[width=0.31\textwidth]{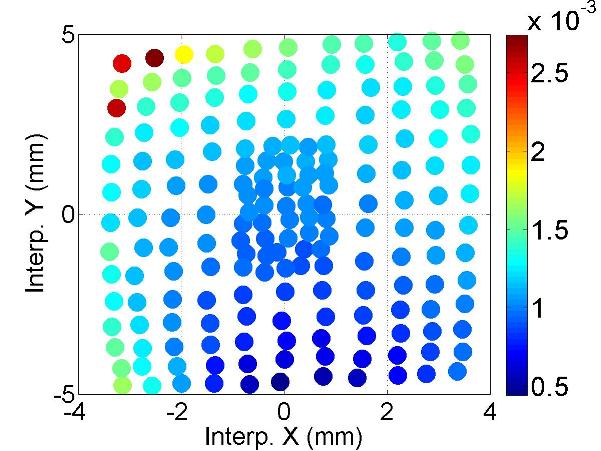}
\label{polar-C4H1-8}
}
\subfigure[\#9 ($f$:9.0562GHz; $Q$:10$^5$)]{
\includegraphics[width=0.31\textwidth]{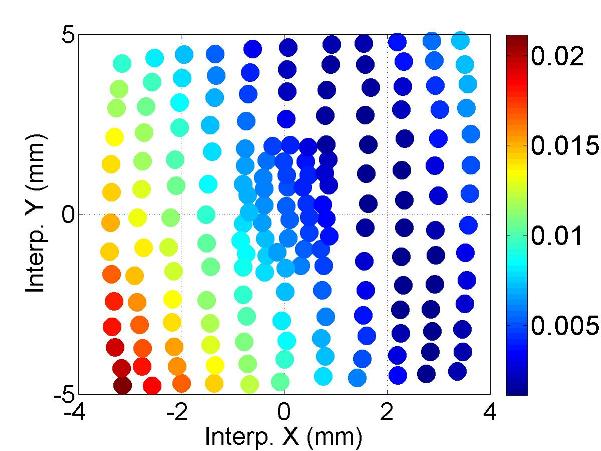}
\label{polar-C4H1-9}
}
\subfigure[\#10 ($f$:9.0591GHz; $Q$:10$^5$)]{
\includegraphics[width=0.31\textwidth]{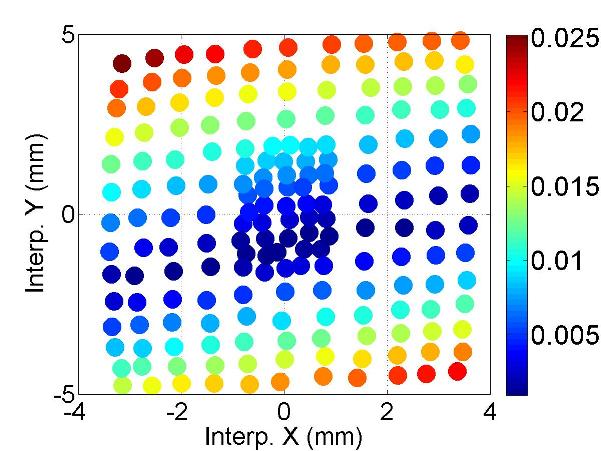}
\label{polar-C4H1-10}
}
\subfigure[\#11 ($f$:9.0627GHz; $Q$:10$^4$)]{
\includegraphics[width=0.31\textwidth]{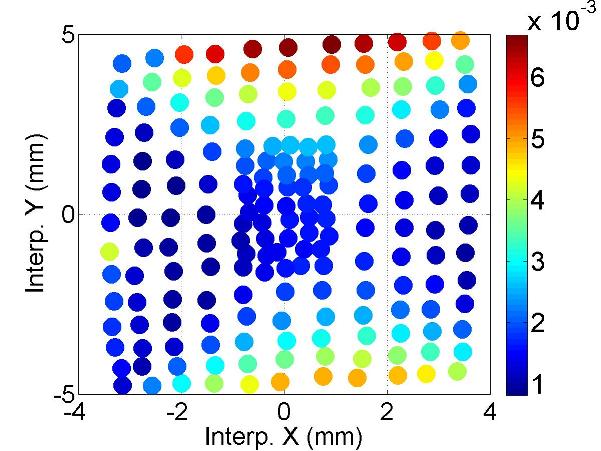}
\label{polar-C4H1-11}
}
\subfigure[\#12 ($f$:9.0727GHz; $Q$:10$^5$)]{
\includegraphics[width=0.31\textwidth]{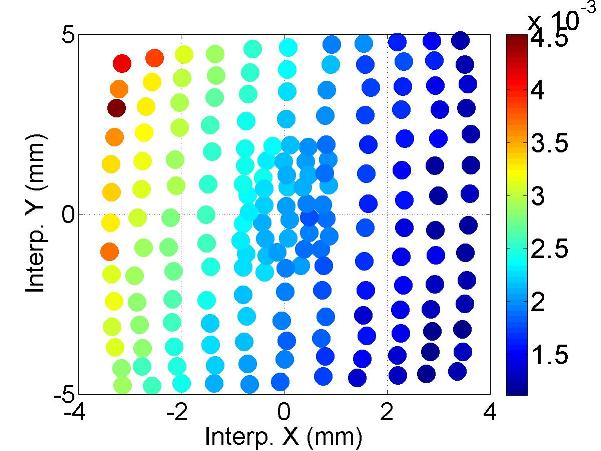}
\label{polar-C4H1-12}
}
\subfigure[\#13 ($f$:9.0752GHz; $Q$:10$^3$)]{
\includegraphics[width=0.31\textwidth]{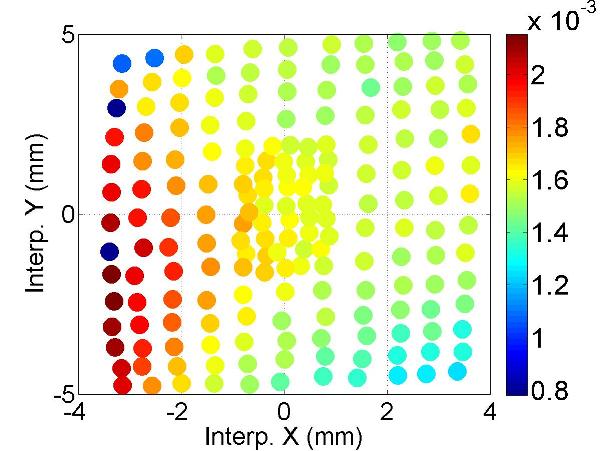}
\label{polar-C4H1-13}
}
\subfigure[\#13.5 ($f$:9.0970GHz; $Q$:10$^4$)]{
\includegraphics[width=0.31\textwidth]{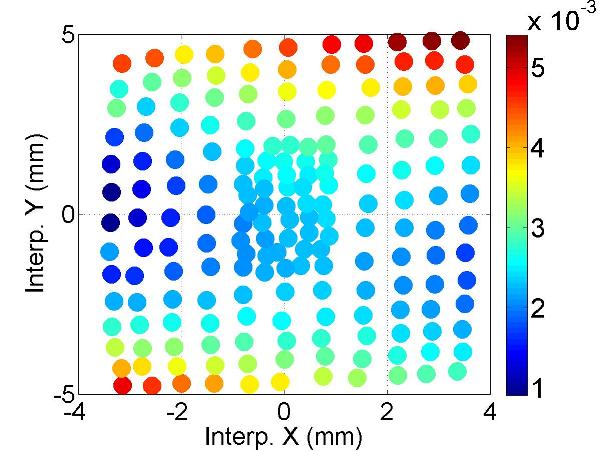}
\label{polar-C4H1-13_5}
}
\subfigure[\#14 ($f$:9.0988GHz; $Q$:10$^4$)]{
\includegraphics[width=0.31\textwidth]{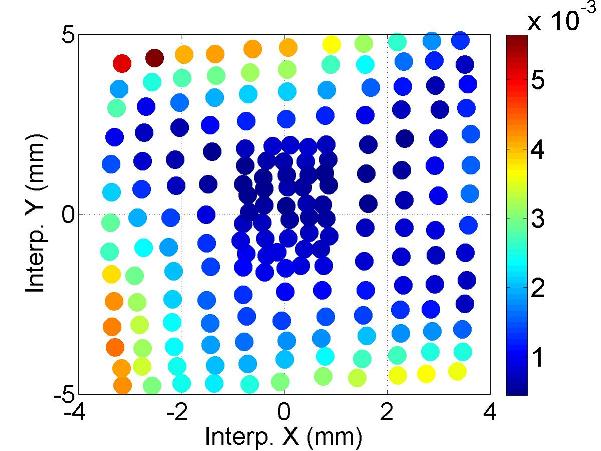}
\label{polar-C4H1-14}
}
\caption{Polarization of the mode.}
\label{spec-polar-C4H1-2}
\end{figure}

\FloatBarrier
\section{D5: HOM Coupler C4H2}
\begin{figure}[h]\center
\subfigure[Spectrum (C4H2)]{
\includegraphics[width=0.85\textwidth]{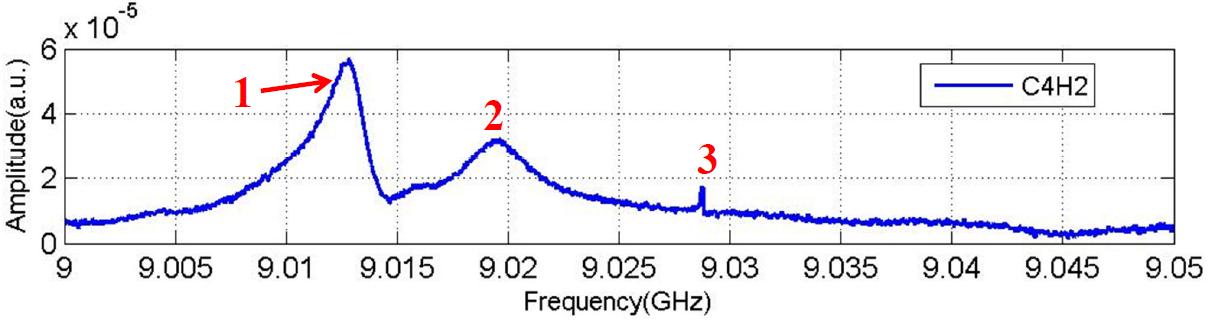}
\label{spec-C4H2-X-1}
}
\subfigure[\#1 ($f$:9.0121GHz; $Q$:10$^3$)]{
\includegraphics[width=0.26\textwidth]{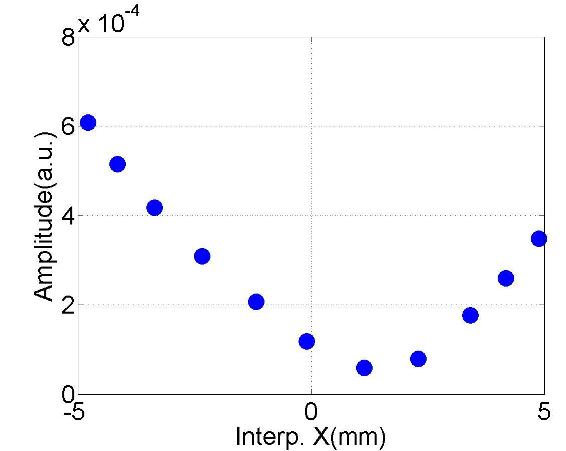}
\label{dep-C4H2-X-1}
}
\subfigure[\#2 ($f$:9.0198GHz; $Q$:10$^3$)]{
\includegraphics[width=0.26\textwidth]{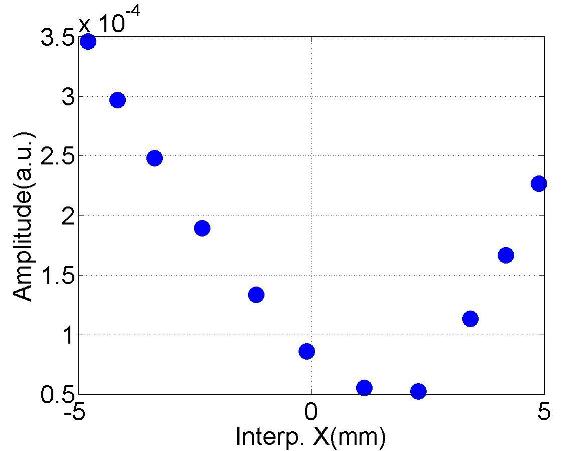}
\label{dep-C4H2-X-2}
}
\subfigure[\#3 ($f$:9.0287GHz; $Q$:10$^4$)]{
\includegraphics[width=0.26\textwidth]{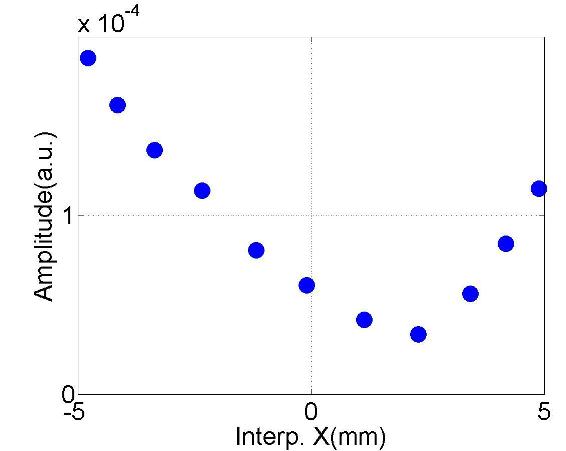}
\label{dep-C4H2-X-3}
}
\subfigure[\#1 ($f$:9.0122GHz; $Q$:10$^3$)]{
\includegraphics[width=0.26\textwidth]{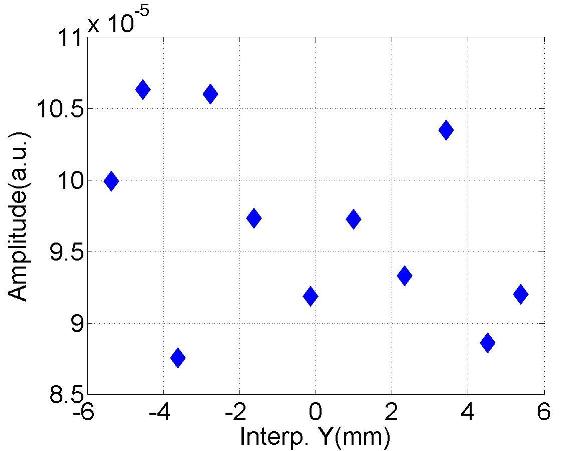}
\label{dep-C4H2-Y-1}
}
\subfigure[\#2 ($f$:9.0198GHz; $Q$:10$^3$)]{
\includegraphics[width=0.26\textwidth]{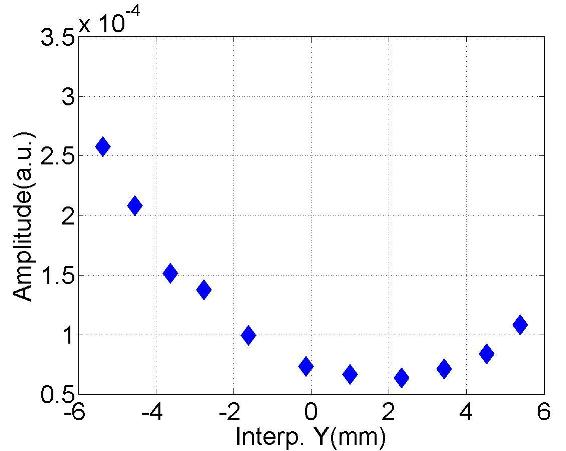}
\label{dep-C4H2-Y-2}
}
\subfigure[\#3 ($f$:9.0287GHz; $Q$:10$^4$)]{
\includegraphics[width=0.26\textwidth]{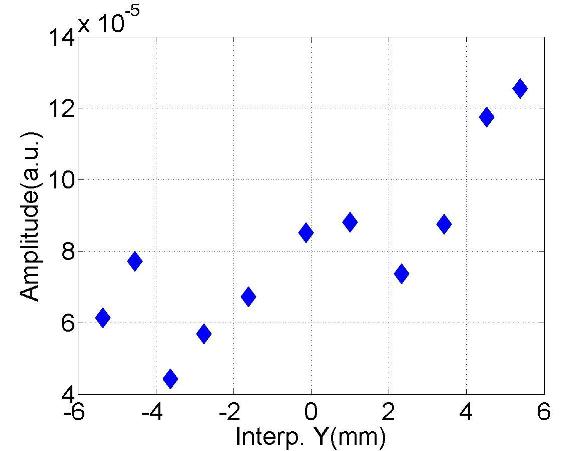}
\label{dep-C4H2-Y-3}
}
\subfigure[\#1 ($f$:9.0121GHz; $Q$:10$^3$)]{
\includegraphics[width=0.27\textwidth]{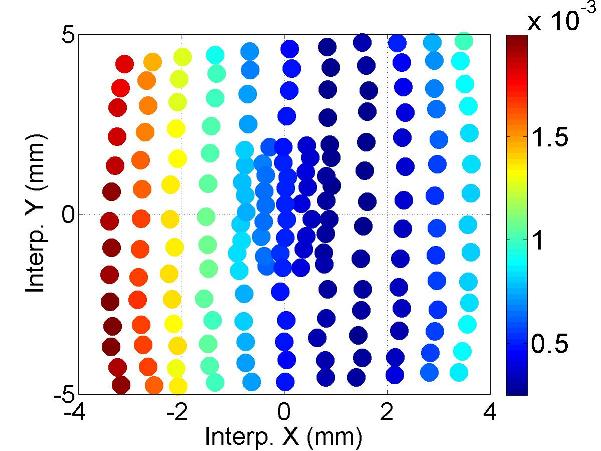}
\label{polar-C4H2-1}
}
\subfigure[\#2 ($f$:9.0199GHz; $Q$:10$^3$)]{
\includegraphics[width=0.27\textwidth]{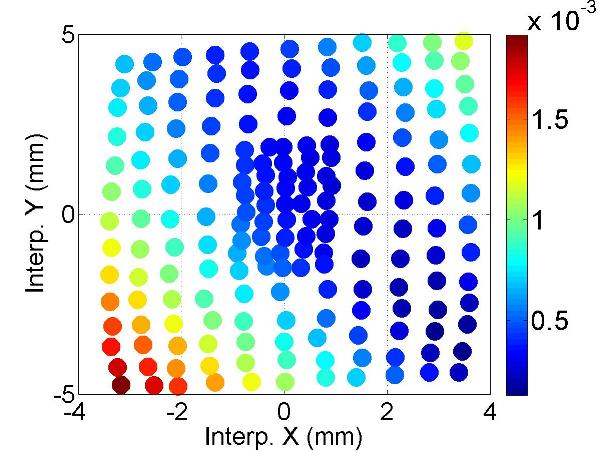}
\label{polar-C4H2-2}
}
\subfigure[\#3 ($f$:9.0287GHz; $Q$:10$^4$)]{
\includegraphics[width=0.27\textwidth]{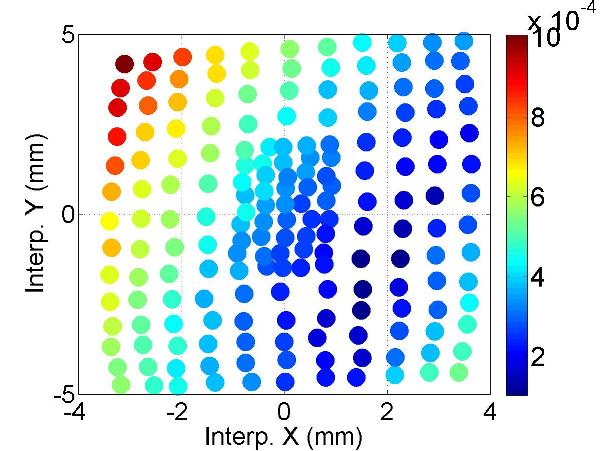}
\label{polar-C4H2-3}
}
\caption{Dependence of the mode amplitude on the transverse beam of{}fset in the cavity.}
\label{spec-dep-C4H2-XY-1}
\end{figure}
\begin{figure}[h]
\subfigure[Spectrum (C4H2)]{
\includegraphics[width=1\textwidth]{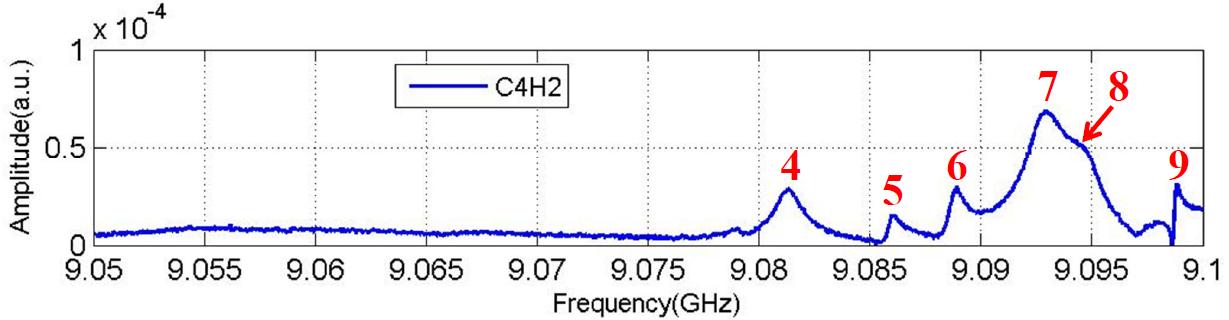}
\label{spec-C4H2-X-2}
}
\subfigure[\#4 ($f$:9.0816GHz; $Q$:10$^3$)]{
\includegraphics[width=0.23\textwidth]{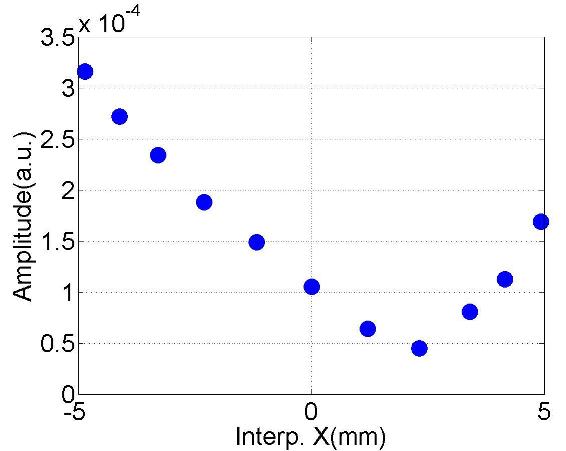}
\label{dep-C4H2-X-4}
}
\subfigure[\#5 ($f$:9.0863GHz; $Q$:10$^4$)]{
\includegraphics[width=0.23\textwidth]{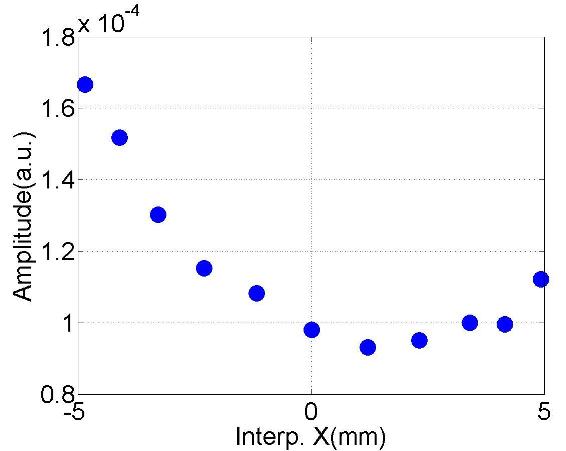}
\label{dep-C4H2-X-5}
}
\subfigure[\#6 ($f$:9.0889GHz; $Q$:10$^4$)]{
\includegraphics[width=0.23\textwidth]{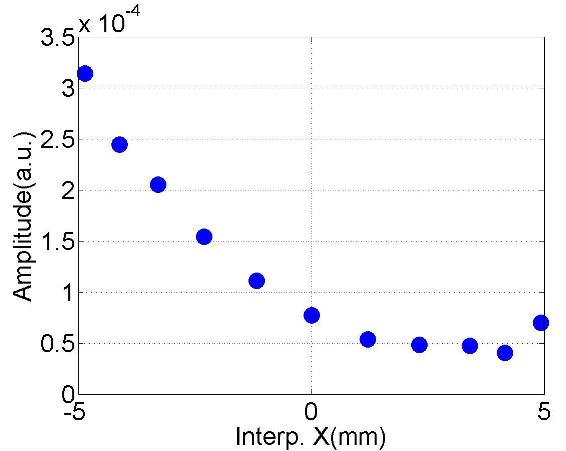}
\label{dep-C4H2-X-6}
}
\subfigure[\#7 ($f$:9.0929GHz; $Q$:10$^3$)]{
\includegraphics[width=0.23\textwidth]{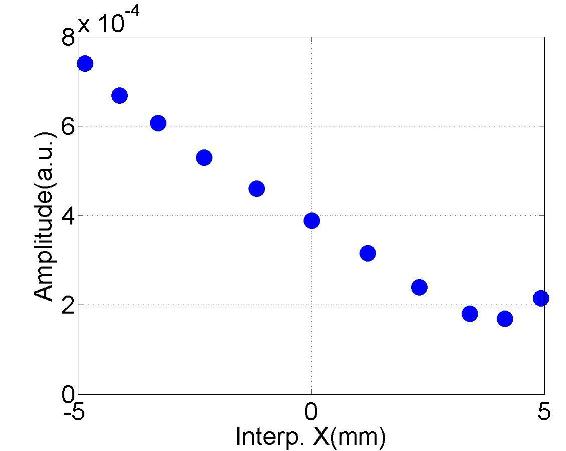}
\label{dep-C4H2-X-7}
}
\subfigure[\#8 ($f$:9.0945GHz; $Q$:10$^3$)]{
\includegraphics[width=0.23\textwidth]{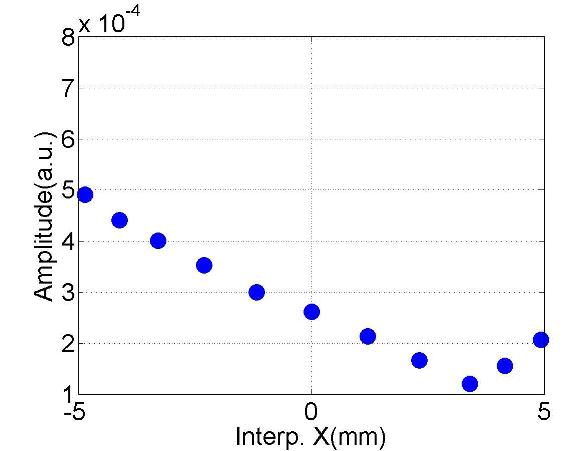}
\label{dep-C4H2-X-8}
}
\subfigure[\#9 ($f$:9.0988GHz; $Q$:10$^4$)]{
\includegraphics[width=0.23\textwidth]{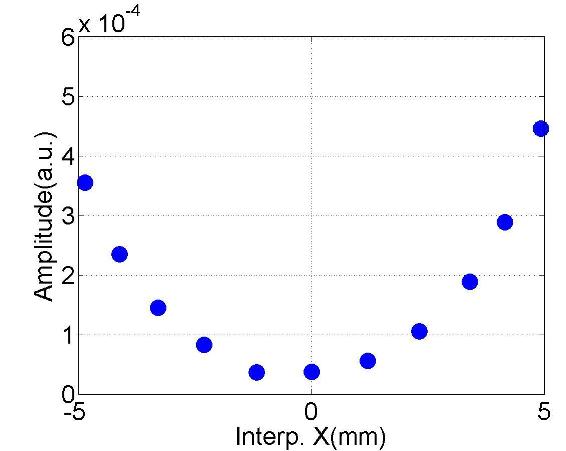}
\label{dep-C4H2-X-9}
}\\
\subfigure[\#4 ($f$:9.0811GHz; $Q$:10$^3$)]{
\includegraphics[width=0.23\textwidth]{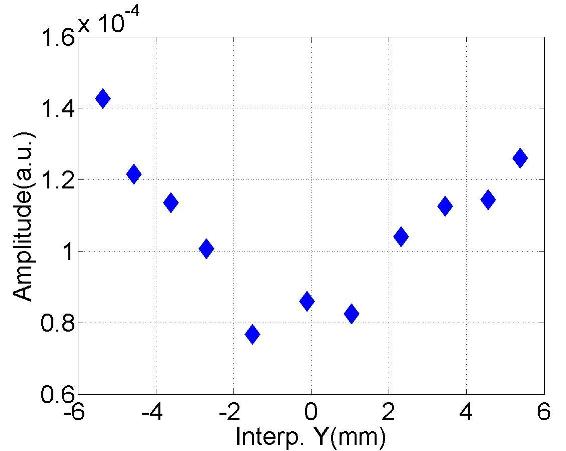}
\label{dep-C4H2-Y-4}
}
\subfigure[\#5 ($f$:9.0862GHz; $Q$:10$^4$)]{
\includegraphics[width=0.23\textwidth]{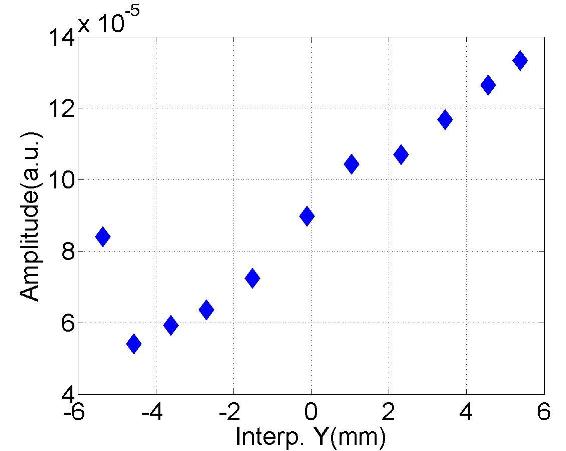}
\label{dep-C4H2-Y-5}
}
\subfigure[\#6 ($f$:9.0887GHz; $Q$:10$^4$)]{
\includegraphics[width=0.23\textwidth]{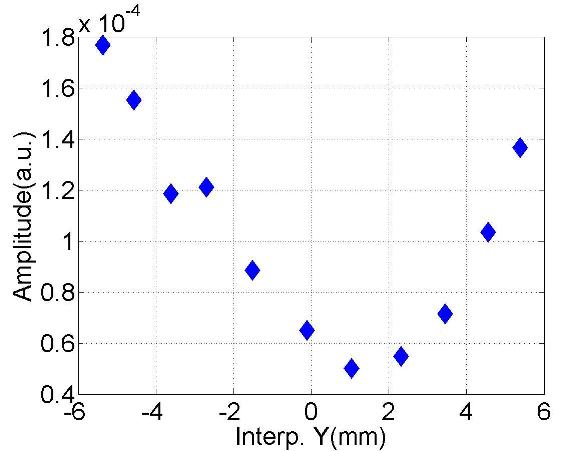}
\label{dep-C4H2-Y-6}
}
\subfigure[\#7 ($f$:9.0930GHz; $Q$:10$^3$)]{
\includegraphics[width=0.23\textwidth]{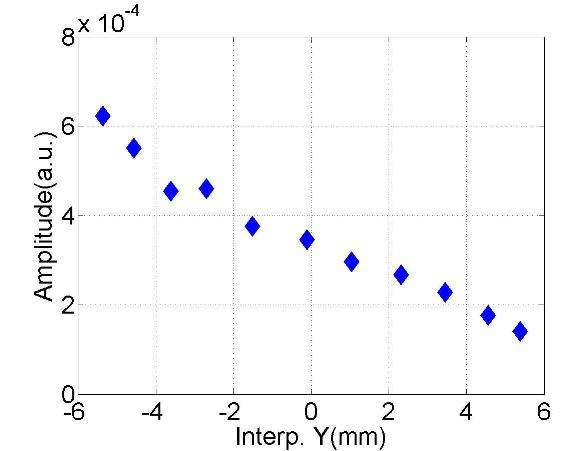}
\label{dep-C4H2-Y-7}
}
\subfigure[\#8 ($f$:9.0944GHz; $Q$:10$^3$)]{
\includegraphics[width=0.23\textwidth]{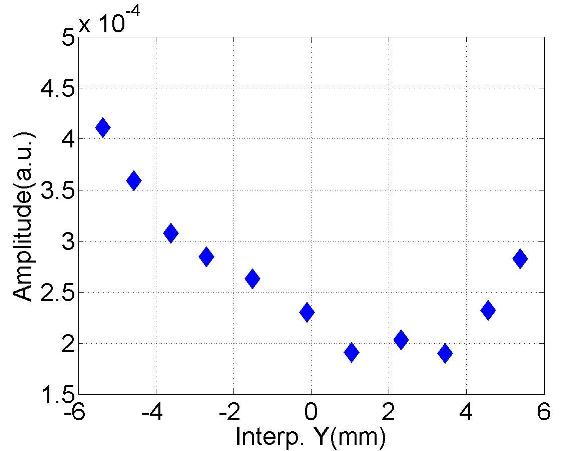}
\label{dep-C4H2-Y-8}
}
\subfigure[\#9 ($f$:9.0988GHz; $Q$:10$^4$)]{
\includegraphics[width=0.23\textwidth]{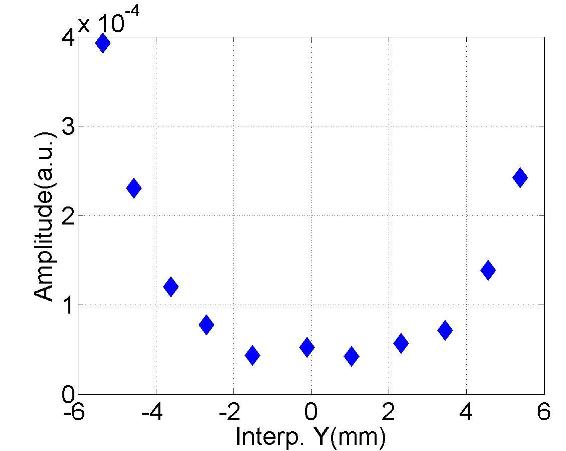}
\label{dep-C4H2-Y-9}
}
\caption{Dependence of the mode amplitude on the transverse beam of{}fset in the cavity.}
\label{spec-dep-C4H2-XY-2}
\end{figure}
\begin{figure}[h]
\subfigure[Spectrum (C4H2)]{
\includegraphics[width=1\textwidth]{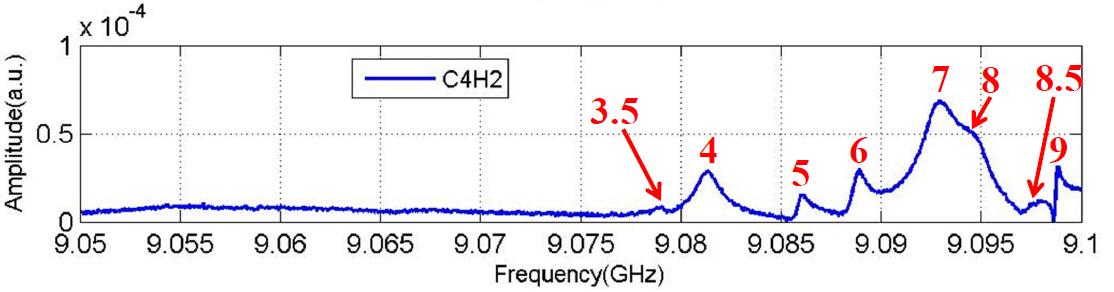}
\label{spec-C4H2-2}
}
\subfigure[\#3.5 ($f$:9.0793GHz; $Q$:10$^4$)]{
\includegraphics[width=0.31\textwidth]{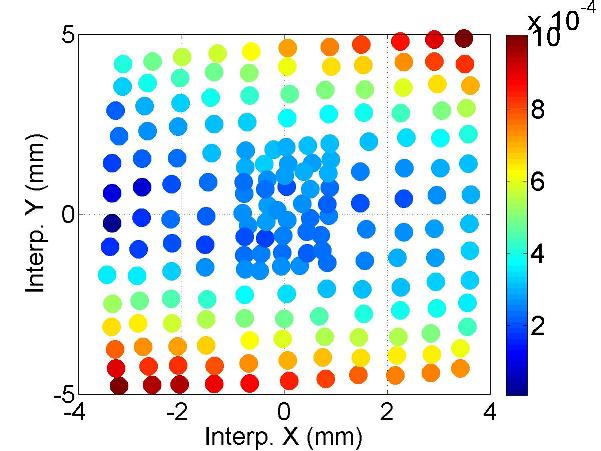}
\label{polar-C4H2-3_5}
}
\subfigure[\#4 ($f$:9.0812GHz; $Q$:10$^3$)]{
\includegraphics[width=0.31\textwidth]{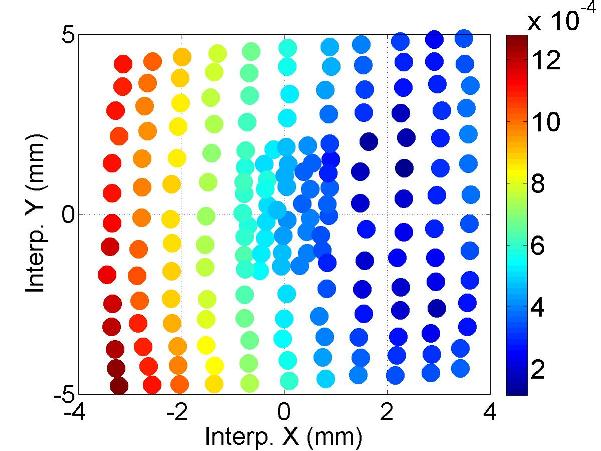}
\label{polar-C4H2-4}
}
\subfigure[\#5 ($f$:9.0862GHz; $Q$:10$^4$)]{
\includegraphics[width=0.31\textwidth]{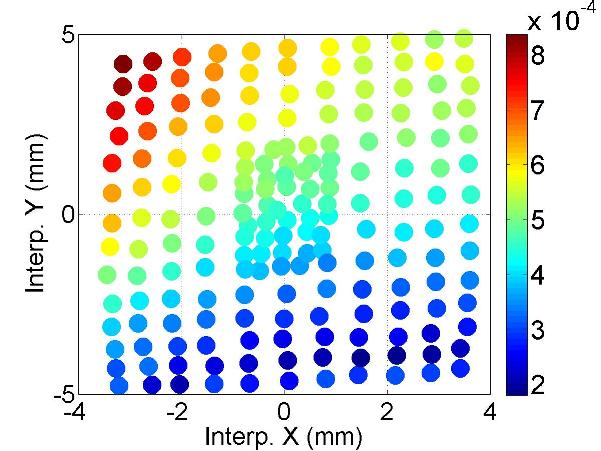}
\label{polar-C4H2-5}
}
\subfigure[\#6 ($f$:9.0887GHz; $Q$:10$^4$)]{
\includegraphics[width=0.31\textwidth]{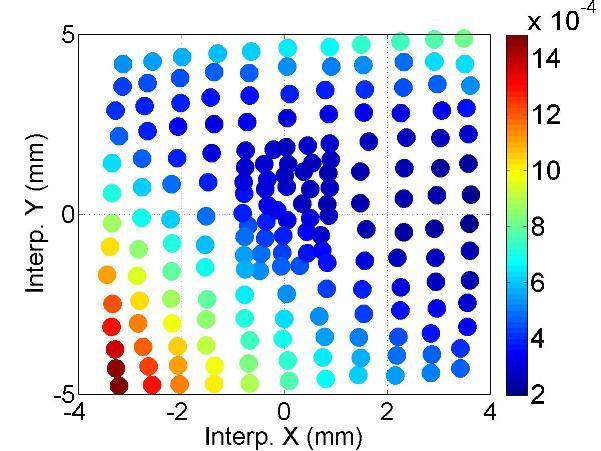}
\label{polar-C4H2-6}
}
\subfigure[\#7 ($f$:9.0929GHz; $Q$:10$^4$)]{
\includegraphics[width=0.31\textwidth]{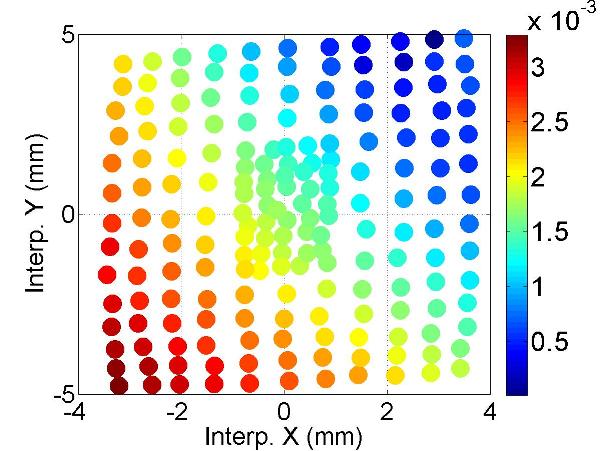}
\label{polar-C4H2-7}
}
\subfigure[\#8 ($f$:9.0943GHz; $Q$:10$^3$)]{
\includegraphics[width=0.31\textwidth]{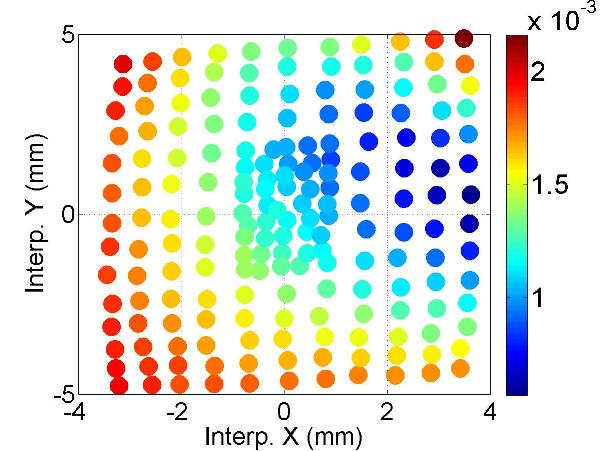}
\label{polar-C4H2-8}
}
\subfigure[\#8.5 ($f$:9.0971GHz; $Q$:10$^3$)]{
\includegraphics[width=0.31\textwidth]{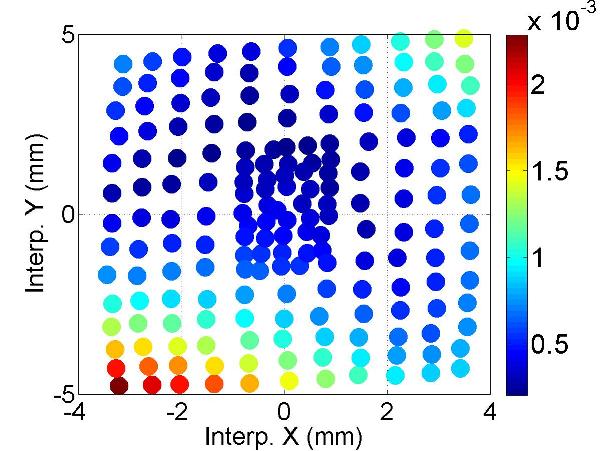}
\label{polar-C4H2-8_5}
}
\subfigure[\#9 ($f$:9.0988GHz; $Q$:10$^4$)]{
\includegraphics[width=0.31\textwidth]{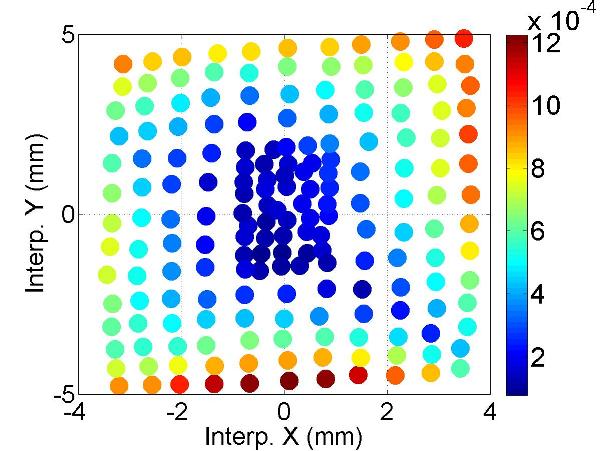}
\label{polar-C4H2-9}
}
\caption{Polarization of the mode.}
\label{spec-polar-C4H2-2}
\end{figure}


\begin{thebibliography}{99}

\bibitem{rwake} K.~Yokoya, ``Cumulative Beam Breakup in Large Scale LINACS'', DESY Report: 86-084, 1986.

\bibitem{rsc} J.~Sekutowicz, ``HOM Damping and Power Extraction from Superconducting Cavities'', \emph{Proceedings of LINAC 2006}, Knoxville, Tennessee, USA, 2006.

\bibitem{rnc} R.M.~Jones, ``Wake f\mbox{}ield Suppression in High Gradient Linacs for Lepton Linear Colliders'', \emph{Phys. Rev. ST Accel. Beams} \textbf{12}, 104801 (2009).

\bibitem{rhombpm-1} G.~Devanz, \emph{et al.}, ``HOM Beam Coupling Measurements at the TESLA Test Facility (TTF)'', \emph{Proceedings of EPAC2002}, Paris, France, 2002.

\bibitem{rhombpm-2} N.~Baboi, \emph{et al.}, ``Preliminary Study on HOM-Based Beam Alignment in the TESLA Test Facility'', \emph{Proceedings of LINAC 2004}, L\"ubeck, Germany, 2004.

\bibitem{rhombpm-3} S.~Molloy, \emph{et al.}, ``High precision superconducting cavity diagnostics with higher order mode measurements'', \emph{Phys. Rev. ST Accel. Beams} \textbf{9}, 112802 (2006).

\bibitem{rflash} S.~Schreiber, \emph{et al.}, ``Status of the FEL User Facility FLASH'', \emph{FEL2011}, Shanghai, China, 2011.

\bibitem{racc39-p1}
K.~Fl$\ddot{o}$ttmann, T.~Limberg, Ph.~Piot, \emph{Generation of Ultrashort Electron Bunches by Cancellation of Nonlinear Distortions in the Longitudinal Phase Space}, TESLA-FEL 2001-06 (2001).

\bibitem{racc39-p2}
J.~Sekutowicz, R.~Wanzenberg, W.F.O.~M$\ddot{u}$ller, T.~Weiland, \emph{A Design of a 3rd Harmonic Cavity for the TTF 2 Photoinjector}, TESLA-FEL 2002-05 (2002).

\bibitem{racc39-p3} E.~Harms, \emph{et al.}, ``Third Harmonic System at Fermilab/FLASH'', \emph{Proceedings of SRF2009}, Berlin, Germany, 2009.

\bibitem{racc39-p4} E.~Vogel, \emph{et al.}, ``Test and Commissioning of the Third Harmonic RF System for FLASH'', \emph{Proceedings of IPAC'10}, Kyoto, Japan, 2010.

\bibitem{racc39-hom} N.~Baboi, \emph{et al.}, ``Using Cavity Higher Order Modes for Beam Diagnostics in Third Harmonic 3.9~GHz Accelerating Modules'', \emph{SRF2011}, Chicago, USA, 2011.

\bibitem{rhommeas-2} I.R.R.~Shinton, \emph{et al.}, ``Higher Order Modes in Third Harmonic Cavities for XFEL/FLASH'', \emph{Proceedings of IPAC'10}, Kyoto, Japan, 2010.

\bibitem{rhommeas-3} P.~Zhang, \emph{et al.}, ``First Beam Spectra of SC Third Harmonic Cavity at FLASH'', \emph{Proceedings of Linear Accelerator Conference LINAC2010}, Tsukuba, Japan, 2010.

\bibitem{rhommeas-4} P.~Zhang, \emph{et al.}, ``Beam-Based HOM Study in Third Harmonic SC Cavities for Beam Alignment at FLASH'', \emph{Proceedings of DIPAC2011}, Hamburg, Germany, 2011.

\bibitem{rhommeas-5} P.~Zhang \emph{et al.}, ``Study of Beam Diagnostics with Trapped Modes in Third Harmonic Superconducting Cavities at FLASH'', \emph{IPAC'11}, San Sebastian, Spain, 2011.

\bibitem{racc39-1} T.~Khabibouline, \emph{et al.}, ``Higher Order Modes of a 3$^{rd}$ Harmonic Cavity with an Increased End-cup Iris'', TESLA-FEL Report: TESLA-FEL 2003-01, 2003.

\bibitem{rgsm-2} I.R.R.~Shinton, \emph{et al.}, ``Higher Order Modes in Third Harmonic Cavities at FLASH'', \emph{Proceedings of Linear Accelerator Conference LINAC2010}, Tsukuba, Japan, 2010.

\bibitem{race3p} I.R.R.~Shinton, \emph{et al.}, ``HOMs in Coupled Cavities of the FLASH Module ACC39'', \emph{IPAC'11}, San Sebastian, Spain, 2011.

\bibitem{rcsc-1} T.~Flisgen \emph{et al.}, ``A Concatenation Scheme for the Computation of Beam Excited HOM Port Signals'', \emph{IPAC'11}, San Sebastian, Spain, 2011.

\bibitem{rtesla-1} J.~Sekutowicz, \emph{Multi-cell Superconducting Structures for High Energy e$^+$e$^-$ Colliders and Free Electron Laser Linacs} (Warsaw University of Technology Publishing House, Poland, 2008), 1st~ed., Chap.~3, p.~27.

\bibitem{rtesla-2} R.~Wanzenberg, ``Monopole, Dipole and Quadrupole Passbands of the TESLA 9-cell Cavity'', TESLA Report: TESLA 2001-33, 2001.

\bibitem{racc39-4} T.~Khabibouline, \emph{et al.}, ``New HOM Coupler Design for 3.9~GHz Superconducting Cavities at FNAL'', \emph{Proceedings of PAC07}, Albuquerque, New Mexico, USA, 2007.

\bibitem{racc39-fnal-1} B.~Hanna \emph{et al.}, ``3.9~GHz Cavities RF and Test Measurements'', unpublished note, 2009.

\bibitem{racc39-fnal-2} T.~Khabibouline (private communication).

\bibitem{rscale-law} K.L.F.~Bane, ``Wake f\mbox{}ield Ef\mbox{}fects in a Linear Collider'', SLAC-PUB-4169, 1986. 

\bibitem{rheav} R.~Bracewell, \emph{The Fourier Transform and Its Applications} (McGraw-Hill, New York, 2000), 3rd~ed., p.~61.

\bibitem{rstat-3} S.S.M.~Wong, \emph{Computational methods in physics and engineering} (World Scientif\mbox{}ic Publishing Co., 1997), 2nd~ed., Chap.~6-1, p.~245.

\bibitem{rpeakfit} PeakFit\textregistered, Ver.~4.12, Systat Software Inc., San Jose, California, USA.

\bibitem{rstat-1} R.L.~Ott and M.T.~Longnecker, \emph{An Introduction to Statistical Methods and Data Analysis} (Duxbury Press, 2008), 6th~ed., Chap.~11.7, p.~611.


\bibitem{racc39-hfss} I.R.R.~Shinton and N.~Juntong, ``Compendium of Eigenmodes in Third Harmonic Cavities for FLASH and the XFEL'', DESY Report: DESY 12-053, 2012.

\bibitem{racc39-cst} P.~Zhang, N.~Baboi and R.M.~Jones, ``Eigenmode simulations of third harmonic superconducting accelerating cavities for FLASH and the European XFEL'', DESY Report: DESY 12-101, 2012.

\bibitem{rrtsa} ``Fundamentals of Real-Time Spectrum Analysis'', Tektronix Application Note, 2010.



\end{thebibliography}
\end{document}